\documentclass[final,3p,times,sort&compress]{elsarticle}

\usepackage{txfonts}
\usepackage{array}
\usepackage{bm}
\usepackage{graphicx}
\usepackage{epsfig}
\usepackage{float}
\usepackage{amsfonts}
\usepackage{color}
\newcommand{\cmark}{\ding{51}}%
\newcommand{\xmark}{\ding{55}}%

\begin{document}


\begin{center}

{\bf \large Hyperuniform  States of Matter}
\bigskip

{Salvatore Torquato}

{\small \it{Department of Chemistry, Department of Physics, Princeton Institute for the Science and Technology of Materials,
and Program in Applied and Computational Mathematics,Princeton University,Princeton, New Jersey 08544, USA}}
\end{center}
\bigskip

\noindent\rule{16.5cm}{0.4pt}
\smallskip

\noindent{\bf Abstract}
\smallskip

Hyperuniform states of matter are correlated systems that are characterized by an
anomalous suppression of long-wavelength (i.e., large-length-scale) density fluctuations compared
to those found in garden-variety disordered systems, such as ordinary fluids and amorphous solids.   
All perfect crystals, perfect quasicrystals and special disordered systems
are hyperuniform. Thus, the hyperuniformity concept enables a unified framework
to classify and structurally characterize crystals, quasicrystals and the exotic disordered varieties.
While disordered hyperuniform systems were largely unknown in the scientific community over a decade ago, now there
is a realization that such systems arise in a host of contexts across the physical, materials, chemical, mathematical, engineering, and biological sciences,
including disordered ground states, glass formation, jamming, Coulomb systems, spin systems, photonic and electronic band structure, localization
of waves and excitations, self-organization,  fluid dynamics, number theory, stochastic point processes,
integral and stochastic geometry, the immune system, and photoreceptor cells.
Such unusual amorphous states  can be obtained via equilibrium or nonequilibrium routes,
and come in both quantum-mechanical and classical varieties.
The connections of hyperuniform states of matter to many different areas of fundamental science  appear to be profound and yet our theoretical
understanding of these unusual systems is only in its infancy. 
The purpose of this review article is to introduce the
reader to the theoretical foundations of hyperuniform ordered and disordered 
systems.  Special focus will be placed on fundamental and practical aspects of the disordered kinds,
including our current state of knowledge of these exotic amorphous systems as well as
their formation and novel physical properties. 

\noindent\rule{16.5cm}{0.4pt}


\tableofcontents
\newpage
\section{Introduction}

The quantitative characterization of density fluctuations in many-particle
systems is a fundamental and practical problem of great interest in the physical, materials,
mathematical and biological sciences.  It is well known that
density fluctuations contain crucial thermodynamic
and structural information about many-particle system both away from \cite{Ve75,Zi77,La80,Chaik95,Tr98b,Han13}
and near critical points \cite{Wi65,Ka66,Fi67,St71,Wi74,Bi92}.  The one-dimensional point patterns associated with the eigenvalues of random matrices and energy levels of integrable quantum systems have been characterized by their  density fluctuations \cite{Berry86,Me91,Bl93}.
The quantification of density fluctuations has been
used to reveal the fractal nature of structures within
living cells \cite{Wa02}. The measurement of galaxy density fluctuations is a powerful way to quantify and study
the large-scale structure of the Universe \cite{Pe93,Ga05}.
Structural fluctuations play a critical role in charge transfer in DNA \cite{Sch08},
thermodynamics of polyelectrolytes \cite{Ye03,Mu06}, and dynamics in glass formation \cite{Be07}.
Depletion phenomena that occur in various molecular systems near interfaces are due to
density fluctuations \cite{Chand99,Sc99}.
Knowledge of density fluctuations in vibrated granular media has been used
to probe the structure and collective motions of the grains \cite{Wa96}.
Local density fluctuations have been recently used to distinguish the spatial
distribution of cancer cells from that of normal, healthy cells \cite{Ji11d}.

\begin{figure}[H]
\centerline{\includegraphics*[width=2in,clip=keepaspectratio]{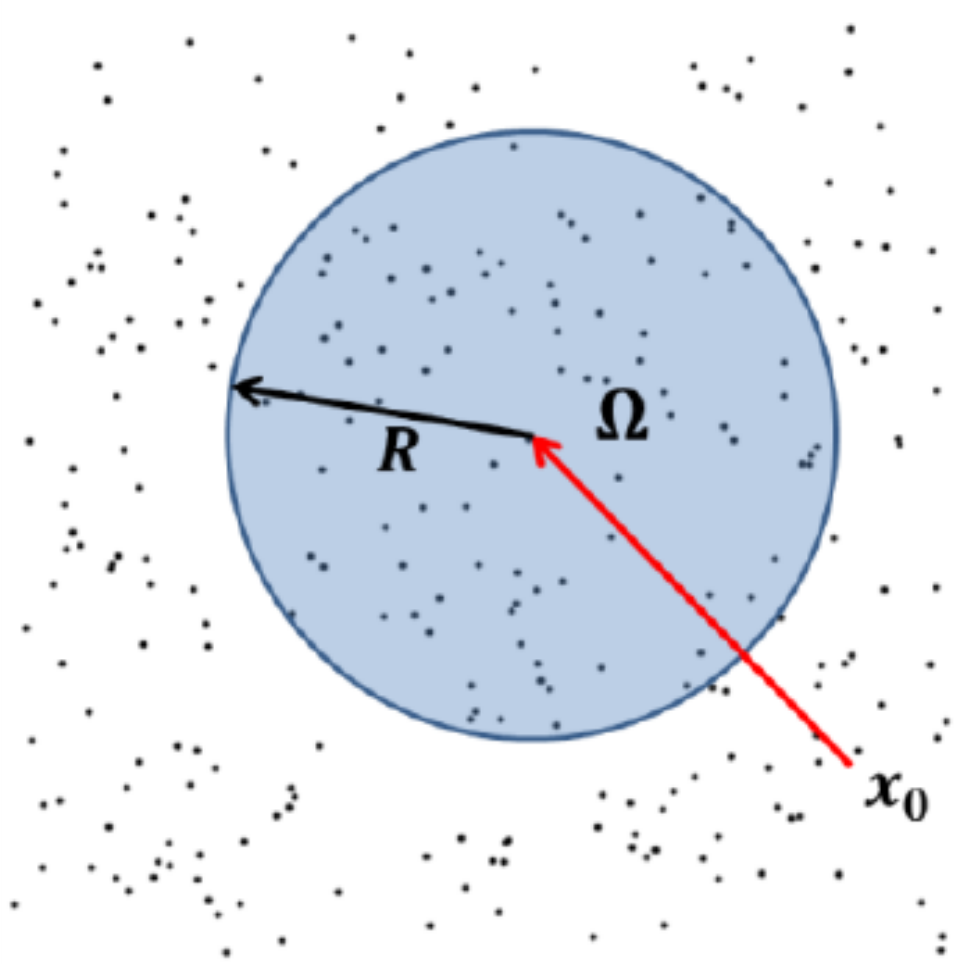}
\includegraphics*[width=2in,clip=keepaspectratio]{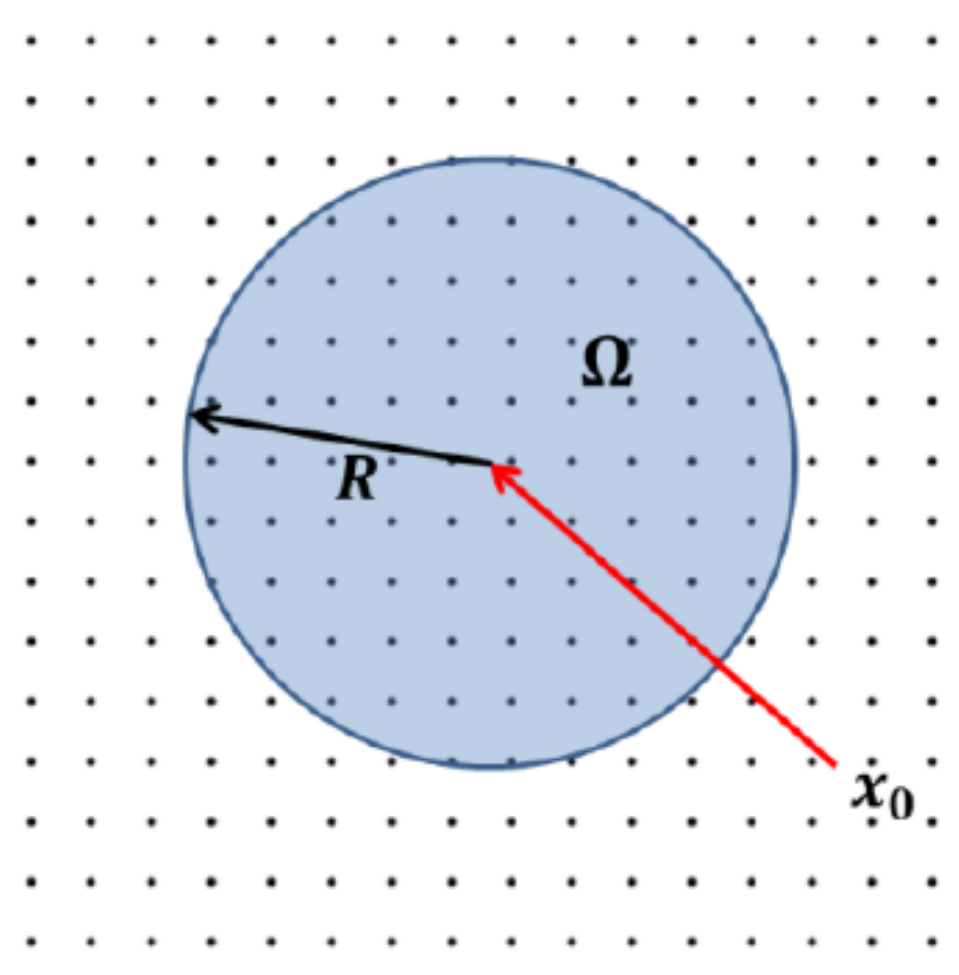}
\includegraphics*[width=2in,clip=keepaspectratio]{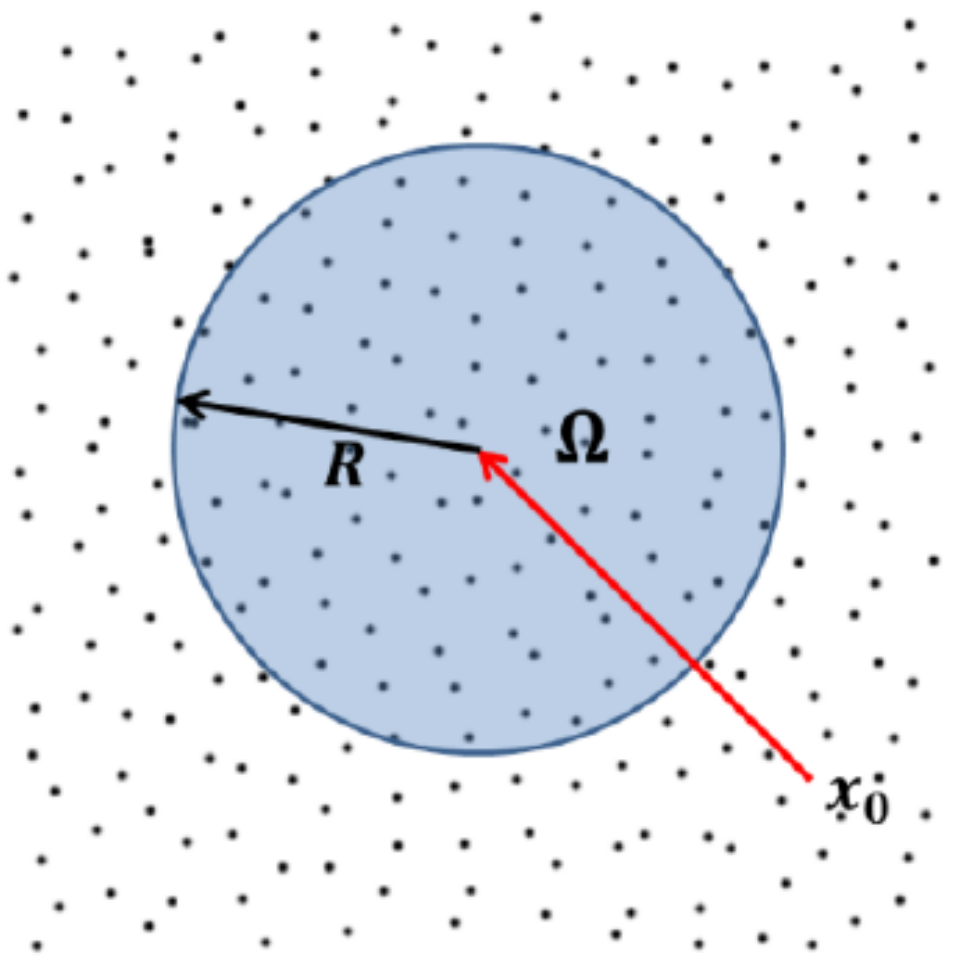}}
\caption{Schematics indicating an observation window $\Omega$, in this case a circular window
in two dimensions,  and
its centroid $\bf x_0$ for a disordered nonhyperuniform (left panel),
periodic (middle panel), and  disordered hyperuniform (right panel)
point configurations. In each of these examples, the density of points within a window will fluctuate as the
window position varies. \ref{gauss} briefly describes the related problem of 
determining how the number of points grows with window size when the window position is fixed and centered at one of the 
points of the point configuration.}
\label{patterns}
\end{figure}

The unusually large suppression of density fluctuations at long wavelengths (large length scales) is central
to the hyperuniformity concept, whose broad importance  for condensed matter physics
and materials science was brought to the fore  only about a decade ago
in a study that focused on fundamental theoretical aspects, including how it provides
a unified means to classify and structurally characterize crystals, quasicrystals and special
disordered point configurations \cite{To03a}.   Hyperuniform systems are poised at an exotic critical point
in which the direct correlation function, defined via the Ornstein-Zernike relation \cite{Han13}, is long-ranged \cite{To03a}, in
diametric contrast to standard thermal critical points in which the total correlation function is long-ranged 
and the corresponding structure factor $S({\bf k})$ diverges to infinity as the wavenumber $|\bf k|$ goes to zero \cite{Wi65,Ka66,Fi67,St71,Wi74,Bi92}.
A hyperuniform (or superhomogeneous \cite{Ga02}) many-particle system in $d$-dimensional Euclidean space
$\mathbb{R}^d$ is one in which (normalized)
density fluctuations are completely suppressed at very large length scales,
implying that the structure factor $S({\bf k})$ tends to zero in the  limit $|{\bf k}| \to 0$.
Equivalently, a hyperuniform system is one in which the number variance $\sigma^2_{_N}(R) \equiv \langle N(R)^2 \rangle -\langle N(R) \rangle^2$ of particles within a spherical observation window of radius $R$ grows more slowly than the window volume in the large-$R$ limit, i.e.,
slower than $R^d$.  Typical disordered systems, such
as liquids and structural glasses, have the standard asymptotic volume
scaling $\sigma^2_{N}(R) \sim R^d$ and hence are not hyperuniform (see the left panel of Fig. \ref{patterns}).
Periodic point configurations (such as the one shown in the middle panel of
Fig. \ref{patterns}) are an obvious class of hyperuniform systems, 
since the number fluctuations are concentrated near the window boundary
and hence have the surface-area scaling $\sigma^2_{_N}(R) \sim R^{d-1}$.  
Less trivially, many perfect quasicrystals \cite{Sh84,Le84,Lev86} possess
the same surface-area  scaling as perfect crystals \cite{Za09,Og17}. Surprisingly,
there is a special  class of disordered particle configurations that have the same asymptotic number-variance scaling behavior as crystals (see the right panel of Fig. \ref{patterns}),
as well as those that have asymptotic scalings that lie between surface-area and volume growth rates (as discussed in Sec. \ref{classes}).

\begin{figure}[H]
\centerline{\includegraphics*[  width=2.8in,clip=keepaspectratio]{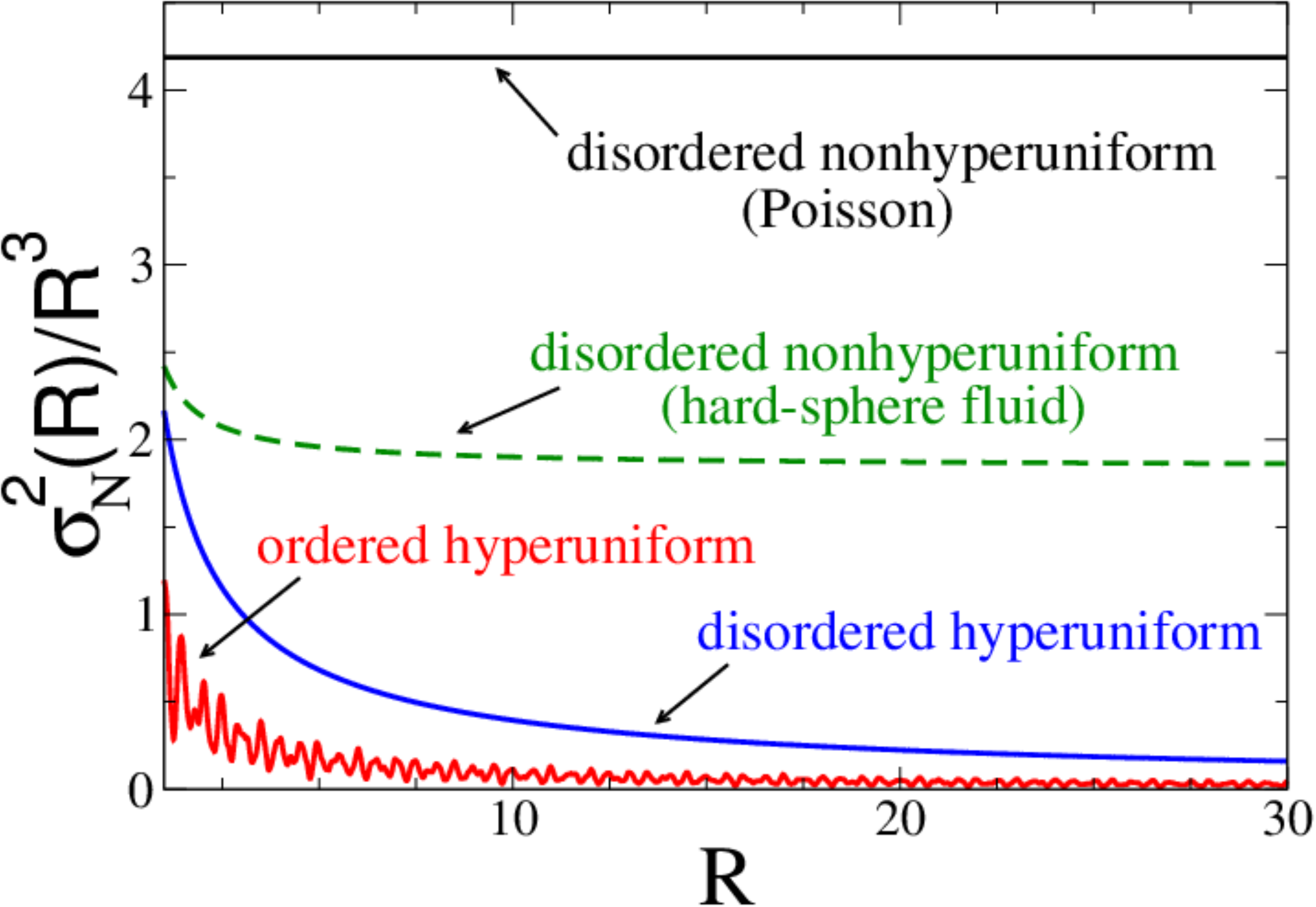}}
\caption{Number variance $\sigma^2_{_N}(R)$, scaled by $R^3$, versus $R$ for four different
many-particle systems in three dimensions at unity density: disordered nonhyperuniform
Poisson (uncorrelated), disordered nonhyperuniform (hard-sphere fluid), disordered hyperuniform (fermionic) \cite{To08b},
and ordered (simple cubic lattice) hyperuniform point configurations. }
\label{scaled-var-3D}
\end{figure}

Figure \ref{scaled-var-3D} shows the number variance (scaled by $R^3$) as a function of $R$  
for four different point configurations in three dimensions at unit density: two disordered nonhyperuniform
systems, the Poisson (uncorrelated)
point process and a low-density equilibrium hard-sphere fluid,  and 
two different hyperuniform systems, one ordered and the other disordered.
For a three-dimensional Poisson point process at unit density, 
the scaled variance $\sigma^2_{_N}(R)/R^3$ is the constant $4\pi/3$ for all $R$. 
The scaled variance for a hard-sphere fluid decreases as $R$ increases but quickly plateaus to
a constant asymptotically. By contrast, for a hyperuniform system,
this scaled number variance tends to decrease as $R$ increases,  apart from small-scale variations, 
 and completely vanishes in the large-$R$ asymptotic limit. 
For a large class of nonspherical convex windows, hyperuniform point configurations
are characterized by a  number variance $\sigma^2_{_N}({\bf R})$ that when scaled
by the window volume $v_1({\bf R})$ tends to zero in the large-window limit, i.e.,
\begin{equation}
\lim_{v_1({\bf R}) \rightarrow \infty} \frac{\sigma^2_{_N}({\bf R})}{v_1({\bf R})} =0,
\label{cond1}
\end{equation}
where $\bf R$ represents the geometrical parameters that define the window shape.
In Sec. \ref{nonspherical}, we will discuss some anomalous situations in which the number variance
for a hyperuniform system can grow even more slowly than the window surface area 
or as fast or faster than the window volume for some window shapes.

Disordered hyperuniform systems and their manifestations were largely unknown in the scientific community
about a decade  and a half ago; only a few examples were known then \cite{Le00,To03a,Ga02}.
Now there is a realization that these systems play a vital role in a number of
problems across the physical, materials, mathematical, and biological sciences.
Specifically, we now know that these exotic states of matter can exist as both {\it equilibrium} and {\it nonequilibrium} phases, including maximally random jammed (MRJ) hard-particle packings \cite{Do05d,Sk06,Za11a,Ji11c,Ho12b,Ch14a,Kl16,At16a,At16b},  jammed athermal soft-sphere
models of granular media~\cite{Si09,Be11}, jammed thermal colloidal packings~\cite{Ku11,Dr15}, 
jammed bidisperse emulsions \cite{Ri17}, dynamical processes in ultracold atoms~\cite{Le14}, 
nonequilibrium phase transitions  \cite{Ja15,He15,We15,Tj15,He17a,He17b,We17,Kw17}, avian photoreceptor patterns \cite{Ji14}, 
receptor organization in the immune system \cite{Ma15}, certain quantum ground states (both fermionic and bosonic) \cite{To08b,Fe56},  classical disordered
(noncrystalline) ground states \cite{Uc04b,Ba08,Ba09a,To15,Zh15a,Zh15b,Zh16a,Zh17a,Zh17b}, the 
distribution of the nontrivial zeros of the Riemann zeta function \cite{To08b,Mon73}, and the eigenvalues of various random matrices \cite{Dy70,Me91}. 

\subsection{Disordered Hyperuniform Systems Are Distinguishable States of Matter}

What qualifies as a distinguishable state of matter? Traditional criteria
include, but are not limited to the following characteristics: (1) it is a homogeneous phase in thermodynamic equilibrium; 
(2) interacting entities are microscopic objects, {\it e.g.}, atoms, molecules or spins;
and (3) often, phases are distinguished by symmetry-breaking
and/or some qualitative change in some bulk property. The liquid-gas
phase transition is an example in which there is concomitant 
latent heat but no symmetry breaking. Modern developments demand broader criteria
to identify a distinguishable state of matter,
which inevitably involves some level of subjectivity. Additional criteria 
include the following attributes: (1) reproducible long-lived metastable or nonequilibrium phases (e.g., 
spin glasses and structural glasses); (2) interacting entities need not be microscopic, but can include building blocks 
that can interact across a wide range of length scales, e.g., colloids, polymers and DNA,
to produce mesoscale materials; and (3) endowed with unique properties.
New states of matter become more compelling if they
give rise to or require new ideas and/or experimental/theoretical tools
and are technologically important.

\begin{figure}[H]
\begin{center}
{\includegraphics[  width=2.in, keepaspectratio,clip=]{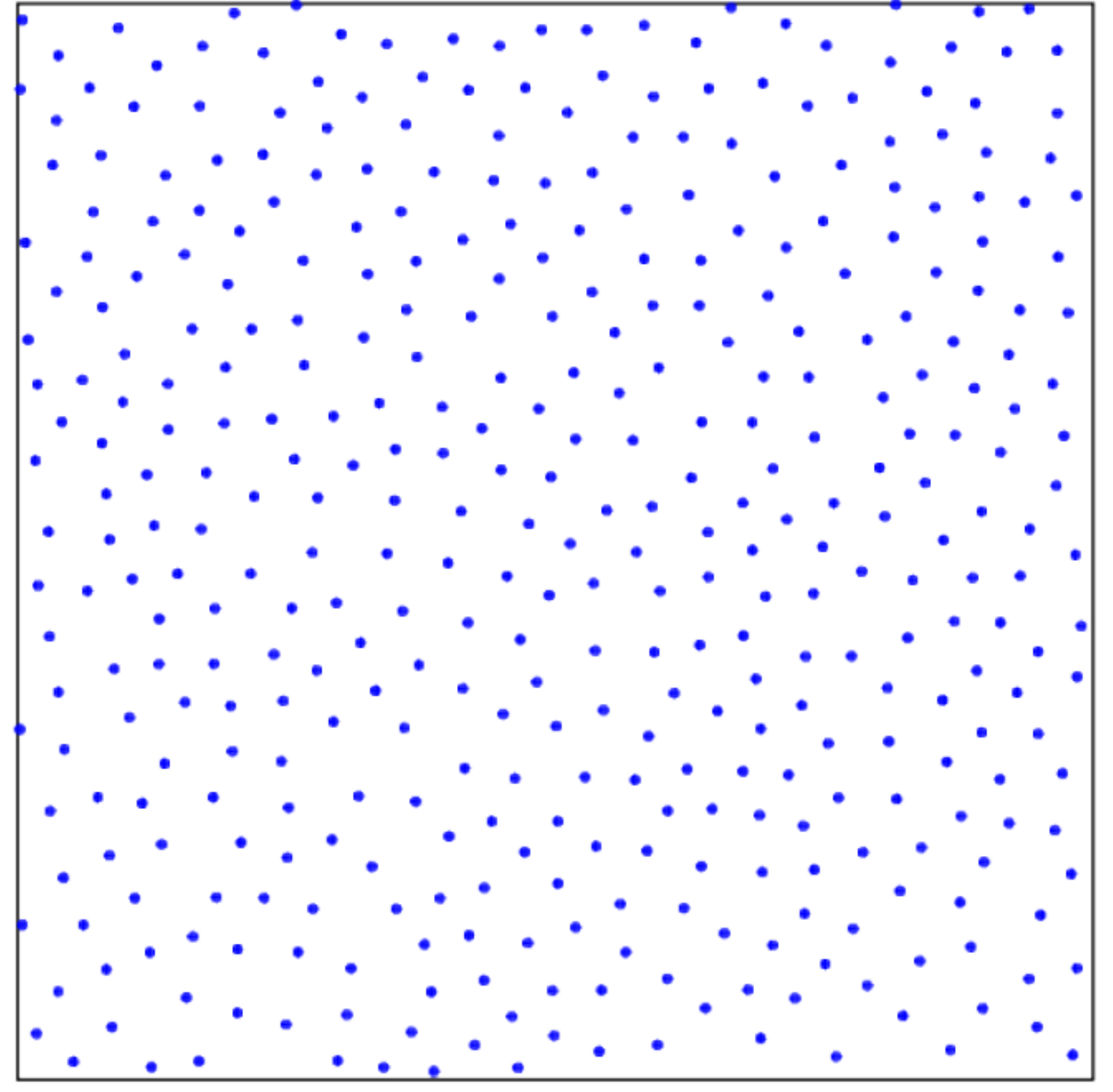}
\hspace{0.55in}\includegraphics[  width=2.in, keepaspectratio,clip=]{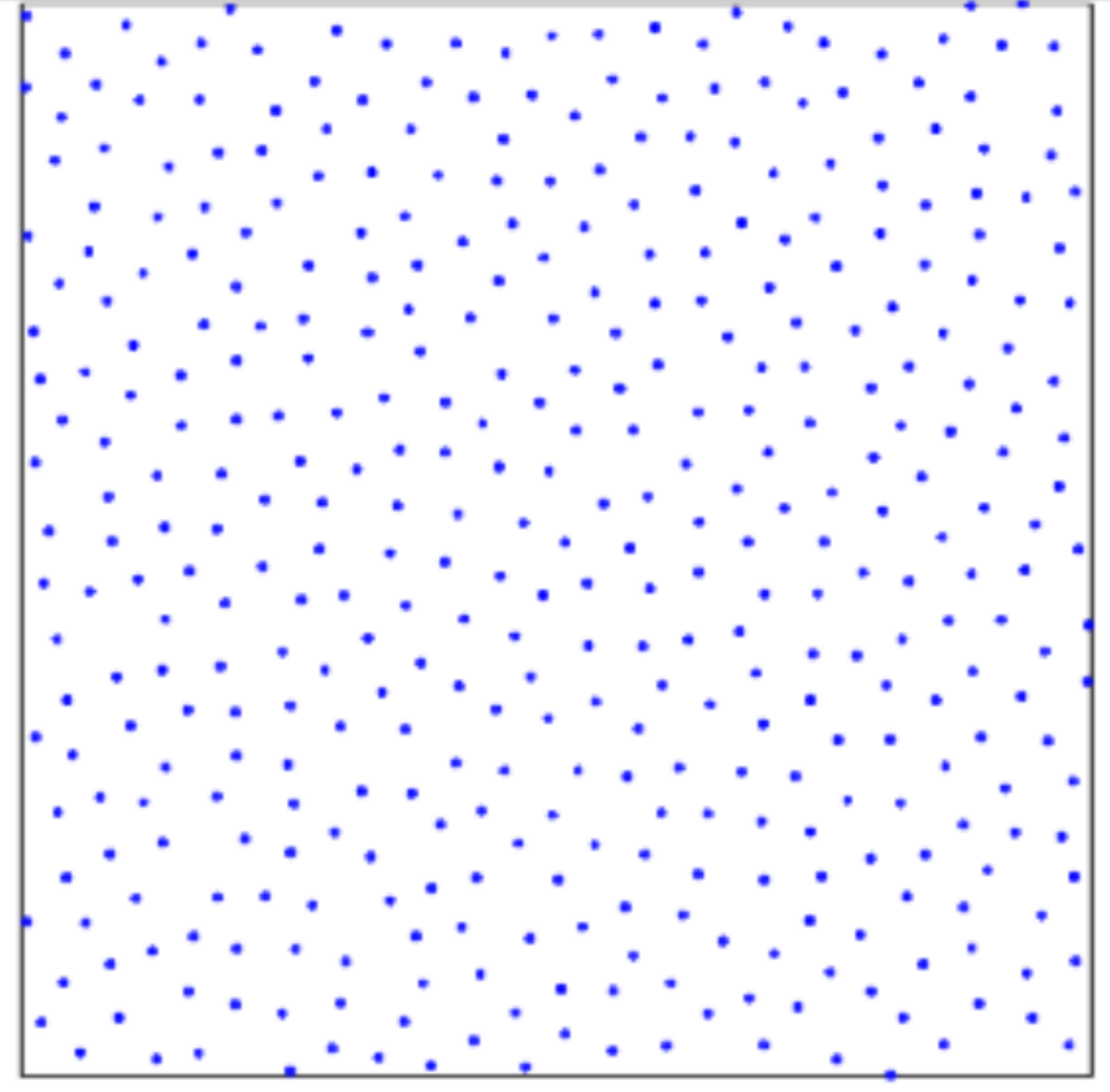}}\vspace{-0.2in}
\caption{A disordered {\it nonhyperuniform} many-particle configuration (left) that
is derived from a dense packing of hard disks generated via the random sequential addition process at saturation
with  $S(0) \approx 0.059$ \cite{To06d,Zh13b}
and a disordered {\it hyperuniform} many-particle configuration (right) taken from Ref. \cite{To16b}.
The latter is obtained by very tiny {\it collective} displacements of the particles in the left panel using
the collective-coordinate procedure reviewed in Sec. \ref{STEALTHY}.
Although the large-scale density fluctuations of these two systems are dramatically different,
the distinctions are very difficult to detect by eye, since
there is a tendency to focus on the structural similarities at short length scales.
Thus, it can often be said that disordered hyperuniform point processes have a ``hidden" order
on large length scales.
}
\label{stealthy}
\end{center}
\end{figure}

\begin{figure}[H]
\centerline{\includegraphics*[  width=1.8in,clip=keepaspectratio]{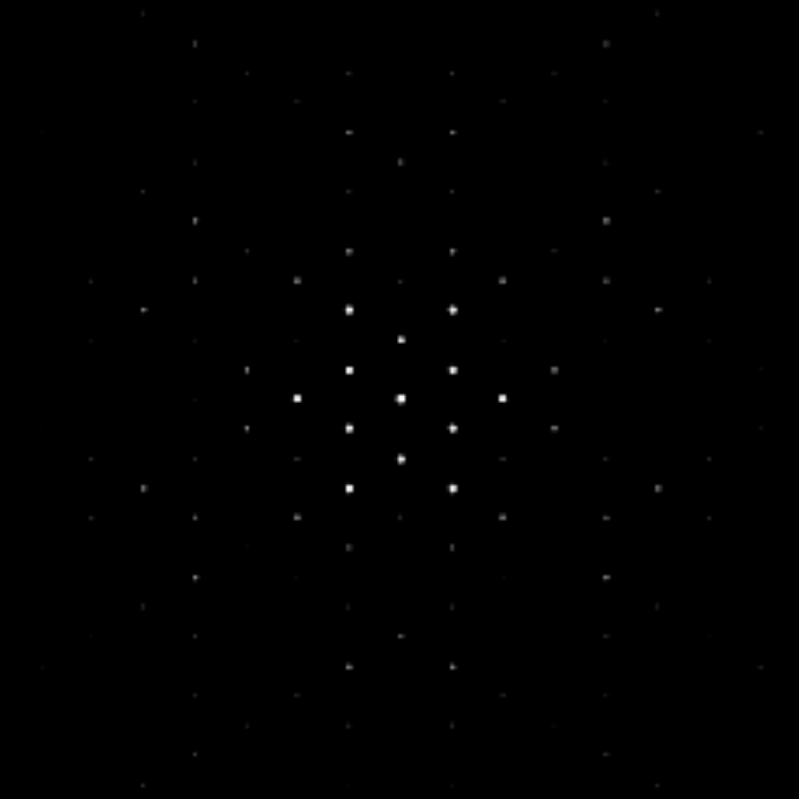}\hspace{0.2in}
\includegraphics*[  width=1.8in,clip=keepaspectratio]{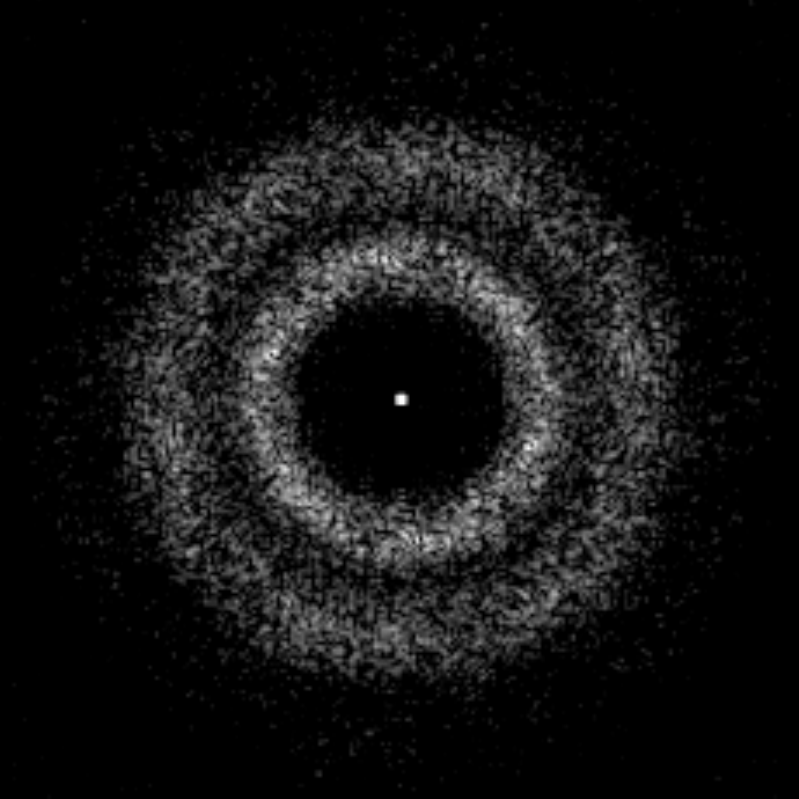}}
\caption{Scattering patterns for two distinctly different hyperuniform point configurations: a six-fold symmetric crystal  (left)
and a disordered ``stealthy" hyperuniform many-particle system (right) \cite{Uc04b,Ba08,To15}.
Observe that in the stealthy disordered case, apart from forward scattering, there is a circular region
around the origin in which there is no scattering, a highly exotic situation for an amorphous state of matter.}
\label{pattern}
\end{figure}

Isotropic disordered hyperuniform materials meet all of these criteria. They are  exotic ideal states
of matter that lie between a crystal and liquid: they are like perfect crystals in the way they suppress large-scale density fluctuations,
a special type of long-range order and yet are like liquids or glasses in that they are statistically isotropic with no Bragg peaks and hence lack any {\it conventional} long-range order. These unusual attributes appear to 
endow such materials with novel  physical properties, as described below.
Importantly, disordered hyperuniform systems can be obtained via equilibrium or nonequilibrium routes,
and come in both quantum-mechanical and classical varieties. Disordered hyperuniform systems  
can have a {\it hidden order not apparent on large length scales}; see Fig. \ref{stealthy} for a vivid example.
 Figure \ref{pattern} shows a typical scattering pattern for a crystal and another for a disordered isotropic hyperuniform system in which there is a circular region around the
origin where there is no scattering, an extraordinary pattern for an amorphous material.
Having become aware of what features to look  for in scattering patterns, disordered hyperuniform systems are now being identified in various contexts across
the fields of  physics, materials science, chemistry, mathematics, engineering and biology \cite{Uc04b,Do05d,Ba08,To08b,Si09,Za11a,Be11,Ku11,Ji11c,Ch14a,
Le14,He15,We15,Tj15,Dr15,Ma15,To15,Zh15a,Zh15b,At16a,Zh16a,Xu16,To16a,To16b,Ch16a,Chr17,He17a,He17b,We17,Kw17,Mar17,Xu17,Zh17a,Zh17b,Wu17,Ch18,Kl18}.

The practical import of the hyperuniformity concept in the context of condensed matter physics 
started to become apparent soon after  it was shown that two- and three-dimensional classical systems of particles interacting with certain soft
long-ranged pair potentials could counterintuitively freeze into highly-degenerate disordered hyperuniform states
at absolute zero temperature with  ``stealthy" scattering patterns, such as the one shown 
in the right panel of Fig. \ref{pattern} \cite{Uc04b,Ba08}. These exotic situations run counter to the common expectation
that liquids  freeze into crystal structures with high symmetry (e.g., face-centered cubic and diamond crystals).
In a subsequent computer simulation study \cite{Fl09b}, it was shown that cellular network solids derived from such stealthy
hyperuniform configurations of particles possess large, {\it isotropic} photonic band gaps comparable in
size to photonic crystals, which was previously thought to be impossible.
This enabled investigators to design disordered hyperuniform cellular solids with unprecedented 
waveguide geometries unhindered by crystallinity and anisotropy, and robust to defects \cite{Fl13,Man13b}.
 A variety of groups have recently fabricated various disordered hyperuniform materials at the 
micro- and nano-scales, including those for photonic applications \cite{Man13a,Ha13,Man13b,Ma16,Zh16},
surface-enhanced Raman spectroscopy \cite{De15}, the
realization of a terahertz quantum cascade laser \cite{De16}, self-assembly
of diblock  copolymers \cite{Zi15b}, periodically-driven emulsions \cite{We15}, and self-assembled
two-dimensional jammed packings of bidisperse droplets \cite{Ri17}.
Recent computational investigations on stealthy hyperuniform systems
have been directed toward optical \cite{Le16}, photonic \cite{Yu15,Fr16,Fr17},  phononic \cite{Gk17},
diffusion and conduction transport \cite{Zh16b,Ch18}, and frequency-dependent dielectric constant behaviors
\cite{Wu17,Ch18,Kl18} and applications. Moreover, another computational study revealed that 
the electronic bandgap of amorphous silicon
widens as it tends toward a hyperuniform state \cite{He13}. Recent
X-ray scattering  measurements indicate that amorphous-silicon samples
can be made to be nearly hyperuniform   \cite{Xie13}.
Thus, an ability to control and design disordered hyperuniform states of matter could lead to the discovery
of novel materials. 

The hyperuniformity concept has brought new attention to the importance of quantifying 
the large-scale structural correlations in amorphous systems, regardless of whether they are hyperuniform.  
This has been  borne out in recent studies concerning molecular jamming processes that underlie glass-formation in polymeric materials \cite{Xu16,Chr17},
and quantifying the large-scale structure of amorphous ices and transitions between their different forms
\cite{Mar17}.
Understanding structural and physical properties of  a system as it approaches a hyperuniform state
or whether  near hyperuniformity  is signaling crucial large-scale structural changes in a system will be shown
to be fundamentally important and is expected to lead to new insights about condensed phases of matter.
 Indeed,  the hyperuniformity concept has
suggested a ``nonequilibrium index" for glasses \cite{Ho12b} as well as new correlation functions
from which one can extract relevant growing length scales as a function of temperature
as a liquid is supercooled below its glass transition temperature \cite{Ho12b,Ma13a},
a problem of intense interest in the glass physics community \cite{Lu07,Be07,Sc07,Chand10,Be12,Hock12,Ch13,Con17}.




\subsection{Hyperuniformity of Heterogeneous Materials}

\begin{figure}[!htp]
\centering
\includegraphics[width=0.3\textwidth]{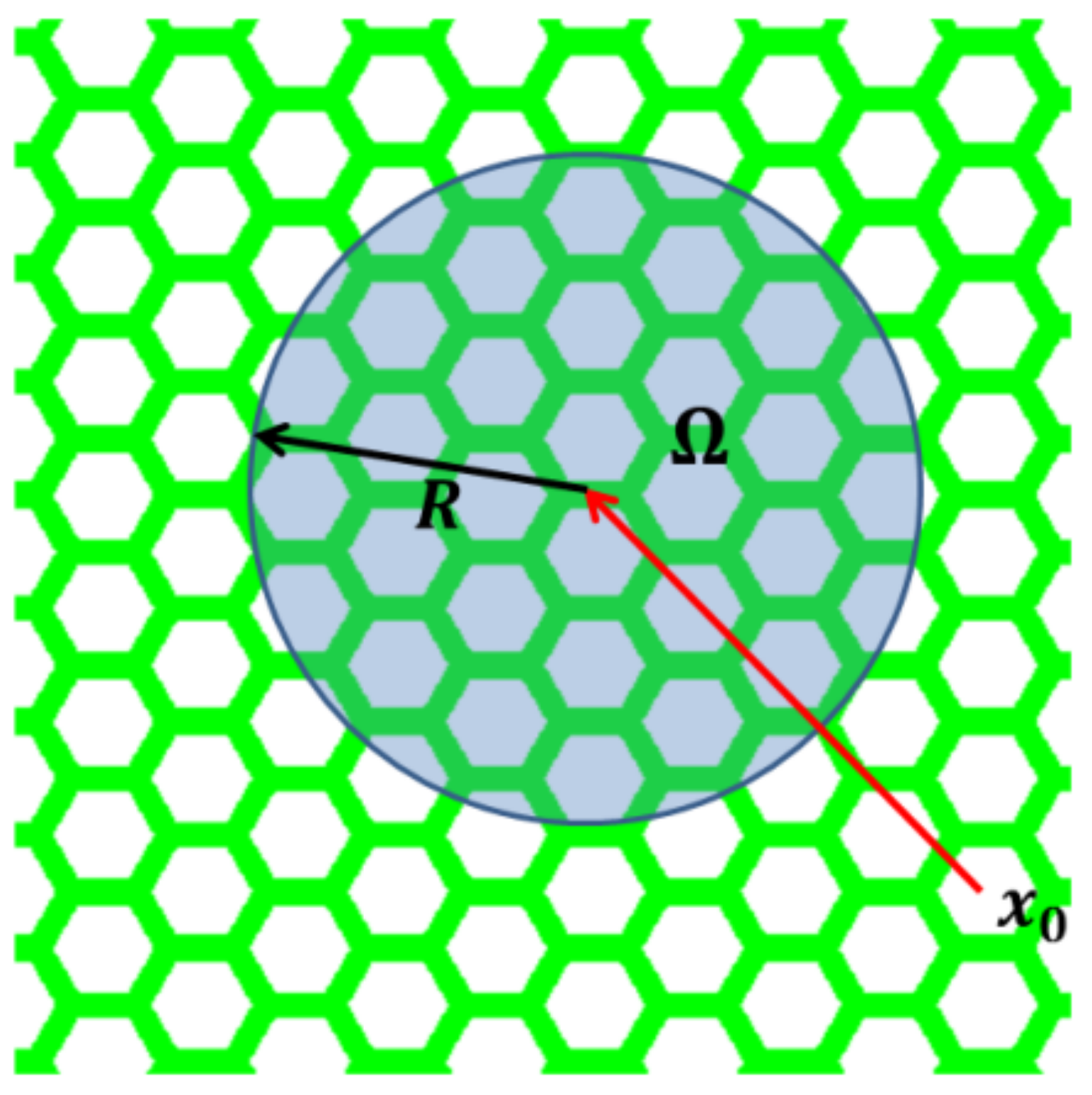}\hspace{0.2in}
\includegraphics[width=0.3\textwidth]{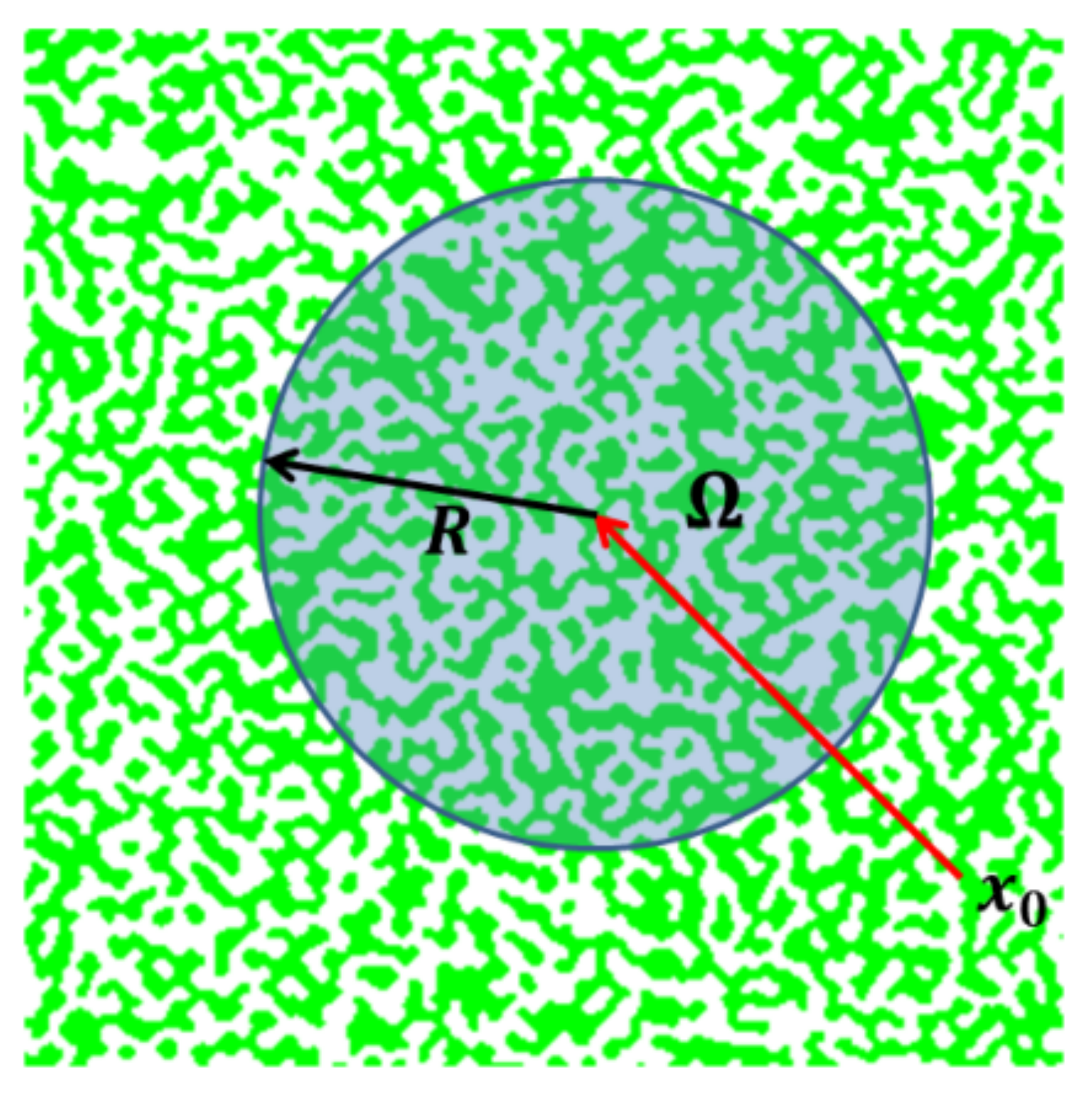}
\caption{Schematics indicating an observation window $\Omega$, in this case a circular window
in two dimensions,  and
its centroid $\bf x_0$ for a periodic heterogeneous medium (left) and 
a general disordered heterogeneous medium (right). In each of these examples, the phase volume fraction or interfacial area within a window will fluctuate as the
window position varies.}\label{cartoon-1}
\end{figure}

The hyperuniformity concept was generalized to the case of heterogeneous materials \cite{Za09}, which are 
composed of domains of different 
materials (``phases") \cite{To02a,Sa03}. Heterogeneous materials abound in Nature and synthetic
situations. Examples include composite and porous media, metamaterials, biological media (e.g., plant and animal tissue),
foams, polymer blends, suspensions, granular media, cellular solids, colloids \cite{To02a}. In the case of two-phase media
(defined  in Sec. \ref{hetero}), one relevant fluctuating quantity is the local phase volume fraction
within an observation window; see Fig. \ref{cartoon-1}. The simplest characterization of such fluctuations is 
the local volume-fraction variance  $\sigma_{_V}^2(R)$ associated with a $d$-dimensional spherical window of radius $R$ that uniformly samples the space \cite{Lu90a,Lu90b,Qu97b,Qu99,To02a}, which is the counterpart to the number variance $\sigma^2_{_N}(R)$ for point
configurations.
The  hyperuniformity condition in the context of
volume-fraction fluctuations in a two-phase heterogeneous system   is
one in which the variance  $\sigma_{_V}^2(R)$ for large $R$ goes to zero more
rapidly than the inverse of the window volume, i.e.,  $R^{-d}$ \cite{Za09}. This  is equivalent
to the  condition that the relevant spectral function, called the spectral density ${\tilde \chi}_{_V}({\bf k})$ (defined in Sec. \ref{hetero})
tends to zero as the wavenumber $|{\bf k}|$ goes to zero.
This generalization of the hyperuniformity concept
has been fruitfully applied to characterize a variety of disordered two-phase systems \cite{Za11a,Za11c, Za11d,Dr15,Ch15,Kl16}
and to the rational design of digitized hyperuniform two-phase media (or, equivalently, two-state spin 
systems) with tunable disorder \cite{Ch18,Di18}.

\subsection{Further Generalizations of the Hyperuniformity Concept}

The hyperuniformity concept has very recently been broadened along four different directions. This includes generalizations to treat fluctuations
in the interfacial area (one of the Minkowski functionals \cite{St95,Sc08}) in heterogeneous media and surface-area driven
evolving microstructures, random scalar fields (e.g., spinodal decomposition), random vector fields (e.g., velocity fields  in turbulence
and transport in random media), and statistically anisotropic
many-particle systems (e.g., structures that respond to external fields) and two-phase media \cite{To16a}. 

\subsection{Overview}

The connections of hyperuniform states of matter to many different areas of fundamental science  appear to be profound and yet our theoretical
understanding of these unusual states of matter is still its nascent stages of development.
The purpose of this review article is directed toward introducing the
reader to the theoretical foundations of hyperuniform ordered and disordered 
systems and their implications.  Special focus will be placed on the disordered variety,
describing our current state of knowledge of disordered hyperuniform systems and their formation,
stressing both fundamental and practical aspects of such exotic states of amorphous matter.

This review article is organized as follows: in Sec. \ref{defs},
we summarize basic definitions and concepts. Formulas  for the local number variance
of point configurations in different settings are derived and/or presented in Sec. \ref{local-N}.
In Sec. \ref{local-V}, we derive formulas for the local volume-fraction
variance of two-phase media. Sections \ref{found-1} and \ref{found-2} focus on the mathematical
foundations of hyperuniformity for point configurations and two-phase media, respectively,
including the three possible hyperuniformity classes.
In Sec. \ref{inverted}, we demonstrate that hyperuniform point configurations
are poised at exotic critical points and describe the critical behaviors
of so-called $g_2$-invariant hyperuniform point processes. A Fourier-based 
optimization procedure to generate a wide class of hyperuniform point configurations
with targeted structure factors that are the classical ground states of certain 
soft long-ranged interactions is described in Sec. \ref{STEALTHY}. In Sec. \ref{Matrices}, 
we discuss the one-dimensional disordered hyperuniform point patterns derived from  random matrices and the
Riemann zeta function. Disordered hyperuniform determinantal point processes are 
described in Sec. \ref{Det}. In Sec. \ref{non-eq}, we report on a variety of disordered nonequilibrium systems
that are putatively hyperuniform, including disordered jammed particle packings
and absorbing-state models. A description of disordered hyperuniform patterns that arise
in biological systems, including photoreceptor cells in avian retina and the immune system
is presented in Sec. \ref{natural}. The generalizations of the hyperuniformity concept 
to treat fluctuations in the interfacial area in  heterogeneous media and surface-area driven
evolving microstructures, random scalar fields, random vector fields, and statistically anisotropic
many-particle systems and two-phase media is taken up in Sec. \ref{GEN}. In Sec. \ref{properties},
we describe novel bulk physical properties of disordered hyperuniform materials that have recently come to light.
We discuss how the degree of hyperuniformity of a perfect hyperuniform system is affected by the introduction of imperfections in Sec. \ref{defects}. 
Section \ref{nearly} addresses the importance of quantifying large-scale density fluctuations of 
many-particle systems, regardless of whether they are hyperuniform, and discusses nearly hyperuniform systems.
Concluding remarks and an outlook for the field are given in Sec. \ref{conclusions}.
\ref{gauss} briefly discusses the Gauss circle problem and its generalizations,
and links to integrable quantum systems.

\section{Basic Definitions and Preliminaries}
\label{defs}

\subsection{Point Processes}
\label{Points}

A stochastic point process  in $d$-dimensional Euclidean space $\mathbb{R}^d$ 
is defined as a mapping from a probability space
to configurations of points ${\bf x}_1, {\bf x}_2, {\bf x}_3, \ldots$. The reader is referred
to Refs. \cite{Le73,St95} for a more detailed mathematical description.
 It is noteworthy that this random setting is quite general.
It incorporates cases in which the location of the points are deterministically
known, such as a lattice or crystal. We will often call a particular realization of a point process
a {\it point configuration}.

A point process is completely statistically characterized by specifying
the countably 
infinite set of $n$-particle probability density functions $\rho_n({\bf r}_1,{\bf r}_2,\ldots,{\bf r}_n)$ 
($n=1,2,3\ldots$) \cite{To06b}.
The distribution-valued function $\rho_n({\bf r}_1,{\bf r}_2,\ldots,{\bf r}_n)$ 
has a probabilistic interpretation: apart from trivial constants,
it is the probability density function
associated with finding $n$ different points at positions $\mathbf{r}_1, \ldots, \mathbf{r}_n$ 
and hence has the nonnegativity property
\begin{eqnarray}
\rho_n({\bf r}_1,{\bf r}_2,\ldots,{\bf r}_n) \ge 0 \qquad \forall {\bf r}_i \in \mathbb{R}^d \quad (i=1,2,\ldots n).
\label{positive}
\end{eqnarray}
The point process is {\it statistically homogeneous} or {\it translationally invariant} if for every constant vector $\bf y$ in $\mathbb{R}^d$, 
$\rho_n({\bf r}_1,{\bf r}_2,\ldots,{\bf r}_n)=\rho_n({\bf r}_1+{\bf y},{\bf r}_2+{\bf y},\ldots,{\bf r}_n+{\bf y})$,
which implies that the point process occupies all of  space and
\begin{eqnarray}
\rho_n({\bf r}_1,{\bf r}_2,\ldots,{\bf r}_n)=\rho^ng_n({\bf r}_{12},\ldots, {\bf r}_{1n}),
\label{nbody}
\end{eqnarray}
where $\rho$ is the {\it number density} (number of points per unit volume in the infinite-volume limit)
and $g_n({\bf r}_{12},\ldots, {\bf r}_{1n})$ is the {\it $n$-particle correlation function},
which depends on the relative positions ${\bf r}_{12}, {\bf r}_{13}, \ldots$,
where ${\bf r}_{ij} \equiv {\bf r}_j -{\bf r}_i$. Unless otherwise stated,
we will assume that statistically homogeneous point processes are also
{\it ergodic}, i.e., any single realization of the ensemble is representative
of the ensemble in the infinite-volume limit and hence a {\it volume} average in the infinite-volume
limit is equal to the ensemble average \cite{To02a}.

For translationally invariant point processes without {\it long-range order}, 
$g_n({\bf r}_{12},\ldots, {\bf r}_{1n}) \rightarrow 1$ when
the points (or ``particles") are mutually far from one another, i.e.,  
as $|{\bf r}_{ij}| \rightarrow\infty$ 
($1\leq i < j < n$), $\rho_n({\bf r}_{1}, {\bf r}_2, \ldots, {\bf r}_{n}) \rightarrow \rho^n$.
Thus, the deviation of $g_n$ from unity  provides a
measure of the degree of spatial correlations between the particles.
Note that for a translationally invariant {\it Poisson} (spatially uncorrelated) point process, 
$g_n=1$ is unity for all values of its argument. 

We call $g_2({\bf r})=g_2(-{\bf r})$ the pair correlation function. 
It is useful to introduce the total correlation function $h({\bf r})$ of a translationally invariant point process, which
is related to the pair correlation function via
\begin{eqnarray}
h({\bf r})\equiv g_2({\bf r})-1
\label{total}
\end{eqnarray}
and decays to zero for large $|{\bf r}|$ in the absence of long-range order. 
Note that $h({\bf r})=0$ for
all $\bf r$ for a translationally invariant Poisson point process.
If the point process is also statistically isotropic (rotational invariant), these pair statistics
are functions only of the radial distance $r=|\bf r|$, i.e., $g_2({\bf r})=g_2(r)$ 
and $h({\bf r})=h(r)$. The {\it cumulative coordination number} $Z(r)$, defined to be the expected
number of points found in a sphere of radius $r$ centered
at an arbitrary point of the point process, is a monotonically increasing function of $r$ that is related to the radial
pair correlation function as follows:
\begin{eqnarray}
Z(r)=\rho s_1(1) \int_0^r x^{d-1} g_2(x) d x,
\label{Z}
\end{eqnarray}
where
\begin{eqnarray}
s_1(r)  =  \frac{2\pi^{d/2}r^{d-1}}{\Gamma(d/2)}
\label{area-sph}
\end{eqnarray}
is the surface area of a  $d$-dimensional sphere of radius $r$ and $\Gamma(x)$ is the gamma
function. 

A collection of spheres in $d$-dimensional Euclidean
space $\mathbb{R}^d$ is called a {\it sphere packing} if no two spheres
overlap. Clearly, the positions of the sphere centers of a packing constitute
an infinite point configuration when the spheres uniformly occupy the space. An important characteristic of a 
packing is its packing fraction $\phi$, which is the fraction of space $\mathbb{R}^d$ covered by the spheres.

Before describing relevant spectral functions, it is useful to  introduce the following definition of the
Fourier transform of a function $f({\bf r})$ that depends on the vector $\bf r$ in $d$-dimensional Euclidean space
$\mathbb{R}^d$:
\begin{eqnarray}
	\tilde{f}(\mathbf{k}) = \int_{\mathbb{R}^d} f(\mathbf{r}) \exp\left[-i(\mathbf{k}\cdot  \mathbf{r})\right] d\mathbf{r},
\end{eqnarray}
where $\bf k$ is a wave vector and  $(\mathbf{k} \cdot \mathbf{r}) = \sum_{i=1}^d k_i r_i$ is the conventional Euclidean inner product of two real-valued vectors.
The function $f({\bf r})$ can generally represent a tensor of arbitrary rank. When it is well-defined, the corresponding inverse Fourier transform is given by
\begin{eqnarray}
f(\mathbf{r}) = \left(\frac{1}{2\pi}\right)^d \int_{\mathbb{R}^d} 	\tilde{f}(\mathbf{k}) \exp\left[i(\mathbf{k}\cdot  \mathbf{r})\right] d\mathbf{k}.
\end{eqnarray}
If  $f$ is a radial function, i.e., $f$ depends only
on the modulus $r=|\mathbf{r}|$ of the vector $\bf r$, 
its Fourier transform is given by
\begin{eqnarray}
{\tilde f}(k) =\left(2\pi\right)^{\frac{d}{2}}\int_{0}^{\infty}r^{d-1}f(r)
\frac{J_{\left(d/2\right)-1}\!\left(kr\right)}{\left(kr\right)^{\left(d/2\right
)-1}} \,d r,
\label{fourier}
\end{eqnarray}
where  $k=|{\bf k}|$ is wavenumber or modulus of the wave vector $\bf k$
and $J_{\nu}(x)$ is the Bessel function of the first kind of order $\nu$.
The inverse transform of $\tilde{f}(k)$ is given by
\begin{eqnarray}
f(r) =\frac{1}{\left(2\pi\right)^{\frac{d}{2}}}\int_{0}^{\infty}k^{d-1}\tilde{f}(k)
\frac{J_{\left(d/2\right)-1}\!\left(kr\right)}{\left(kr\right)^{\left(d/2\right
)-1}} d k.
\label{inverse}
\end{eqnarray}

An important nonnegative spectral function $S({\bf k})$, called the {\it structure factor} (or power
spectrum), is defined as follows: 
\begin{equation}
S({\bf k}) =1+\rho{\tilde h}({\bf k}),
\label{factor}
\end{equation}
where ${\tilde h}({\bf k})$ is the Fourier transform of $h(\bf r)$.  This quantity
is directly related to the scattering-intensity function ${\cal S}({\bf k})$, defined in (\ref{scatter}),
when  forward scattering is excluded. 
While $S({\bf k})$ is a nontrivial function for  spatially correlated point processes,
it is trivially given by
\begin{equation}
S({\bf k}) =1 \qquad \mbox{for all} \quad {\bf k}
\label{factor-P}
\end{equation}
for a translationally invariant Poisson point process. Correlated disordered systems 
are characterized by structure factors that deviate from unity for some wave vectors; 
well-known examples include typical liquids \cite{Han13} as well as disordered equilibrium 
\cite{Han13,To02a} and nonequilibrium \cite{Zh13b} sphere packings. However, the preponderance
of such disordered systems are not hyperuniform.

We note that the value of the total correlation function at the origin, $h({\bf r=0})$,
provides a simple sum rule on its Fourier transform or $S({\bf k})$ must obey, namely, $\rho h({\bf r=0})=(2\pi)^{-d} \int_{\mathbb{R}^d}
[S({\bf k})-1] \,d{\bf k}$. For any point configuration
in which the minimal pair distance is some positive number, $g_2({\bf r=0})=0$ or $h({\bf r=0})=-1$,
and hence the sum rule becomes
\begin{equation}
\frac{1}{(2\pi)^{d}}\int_{\mathbb{R}^d} [S({\bf k})-1]\, d{\bf k}=-\rho.
\label{SUM}
\end{equation}
Examples of point configurations that satisfy (\ref{SUM}) include the centers of
spheres in a sphere packing or the vertices of polyhedral tiles of space.

A hyperuniform \cite{To03a} or superhomogeneous \cite{Ga02} point configuration is one in which the structure factor $S({\bf k})$ tends to zero 
as the wavenumber $k\equiv |\bf k|$ tends to zero, i.e.,
\begin{equation}
\lim_{|{\bf k}| \rightarrow 0} S({\bf k}) = 0,
\label{hyper}
\end{equation}
implying that single scattering of incident radiation at infinite wavelengths is completely suppressed.
As we will see, this class of point configurations
includes perfect crystals,  a large class of perfect quasicrystals \cite{Za09,Og17}
and special disordered many-particle systems. Note that structure-factor definition (\ref{factor}) 
and the hyperuniformity requirement (\ref{hyper}) dictate that the
volume integral of  $\rho h({\bf r})$ over all space is exactly
equal to $-1$, i.e.,
\begin{equation}
\rho \int_{\mathbb{R}^d} h({\bf r}) d{\bf r}=-1,
\label{sum-1}
\end{equation}
which is a direct-space {\it sum rule} that a hyperuniform point process must obey. This means that
$h({\bf r})$ must exhibit negative correlations, i.e., {\it anticorrelations}, for some values of $\bf r$.

{\it Stealthy} point processes are those in which
the structure factor is exactly zero for a subset of wave vectors \cite{Uc04b,Ba08,To15}, meaning that they completely suppress
single scattering  for these wave vectors.
{\it Stealthy hyperuniform} patterns  are a subclass of hyperuniform
systems in which $S({\bf k})$ is zero for a range
of wave vectors around the origin, i.e.,
\begin{equation}
S({\bf k})= 0 \qquad \mbox{for}\; 0 \le |{\bf k}| \le K,
\label{stealth}
\end{equation}
where $K$ is some positive number. As we will see, perfect crystals are stealthy hyperuniform patterns. An example of a disordered  stealthy  and hyperuniform scattering pattern
is shown in the right panel of Fig. \ref{pattern}.

\begin{figure}[H]
\begin{center}
{\includegraphics[  width=1.5in, keepaspectratio,clip=]{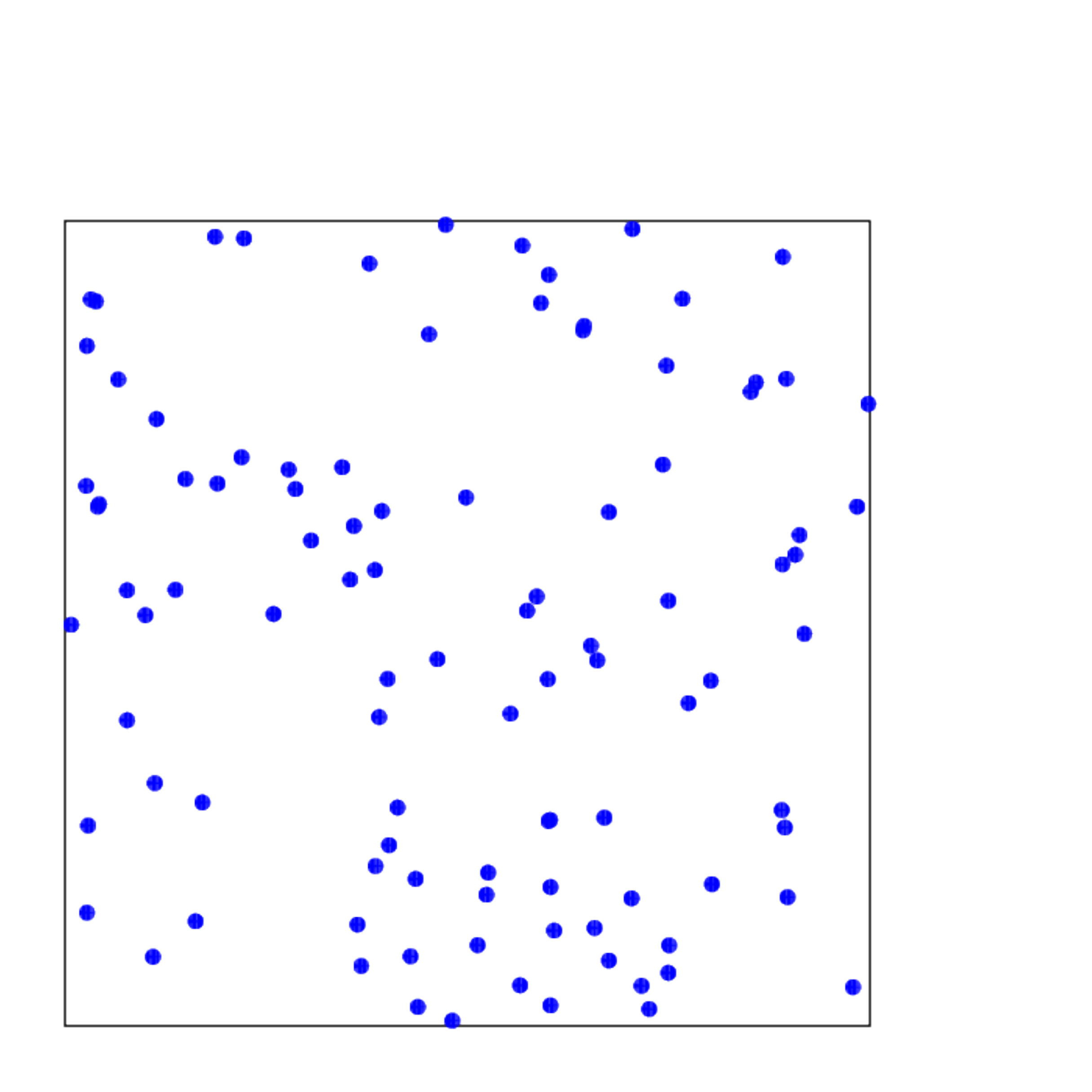}
\hspace{0.15in}
\includegraphics[  width=1.5in, keepaspectratio,clip=]{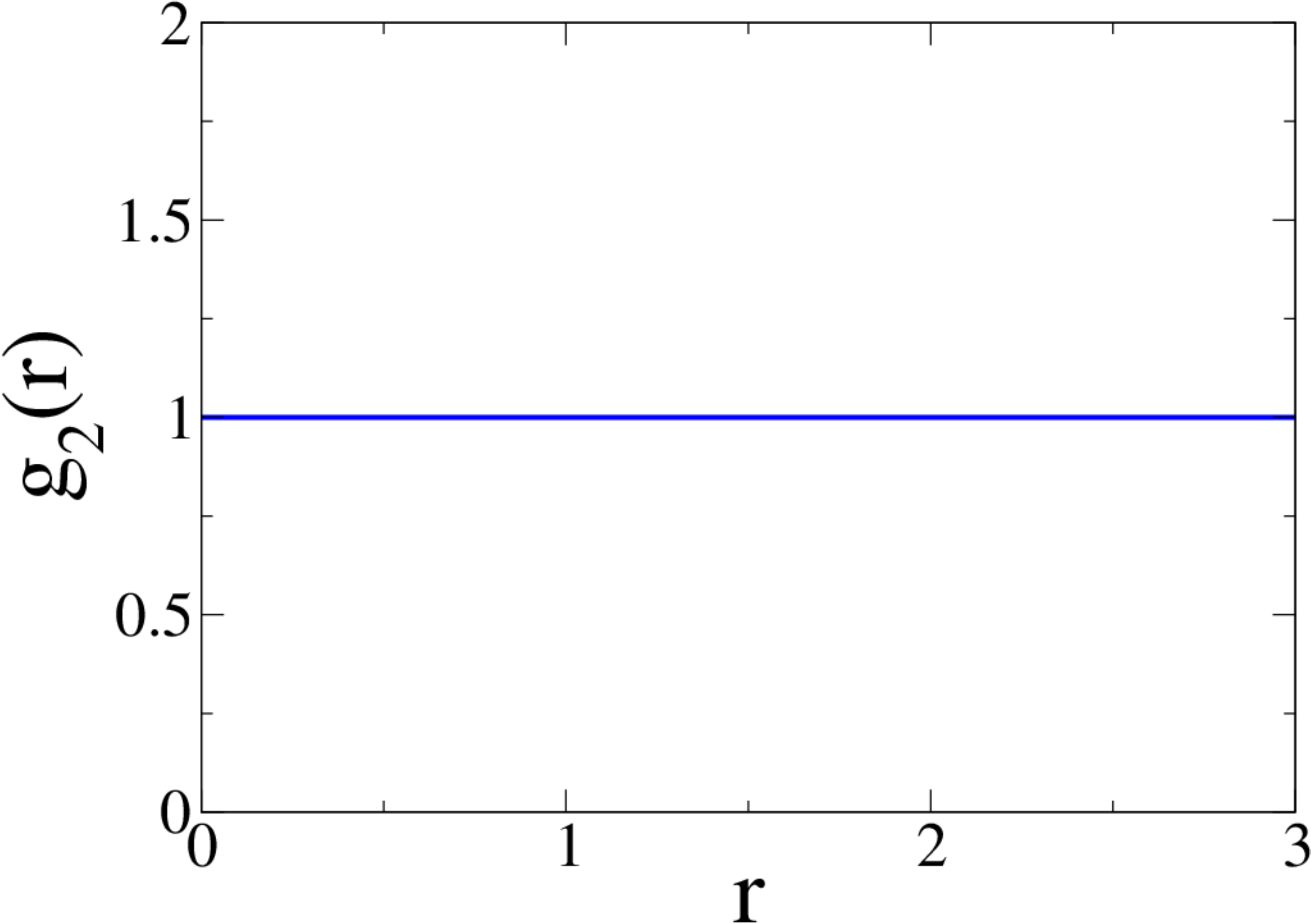}\hspace{0.15in}
\includegraphics[  width=1.5in, keepaspectratio,clip=]{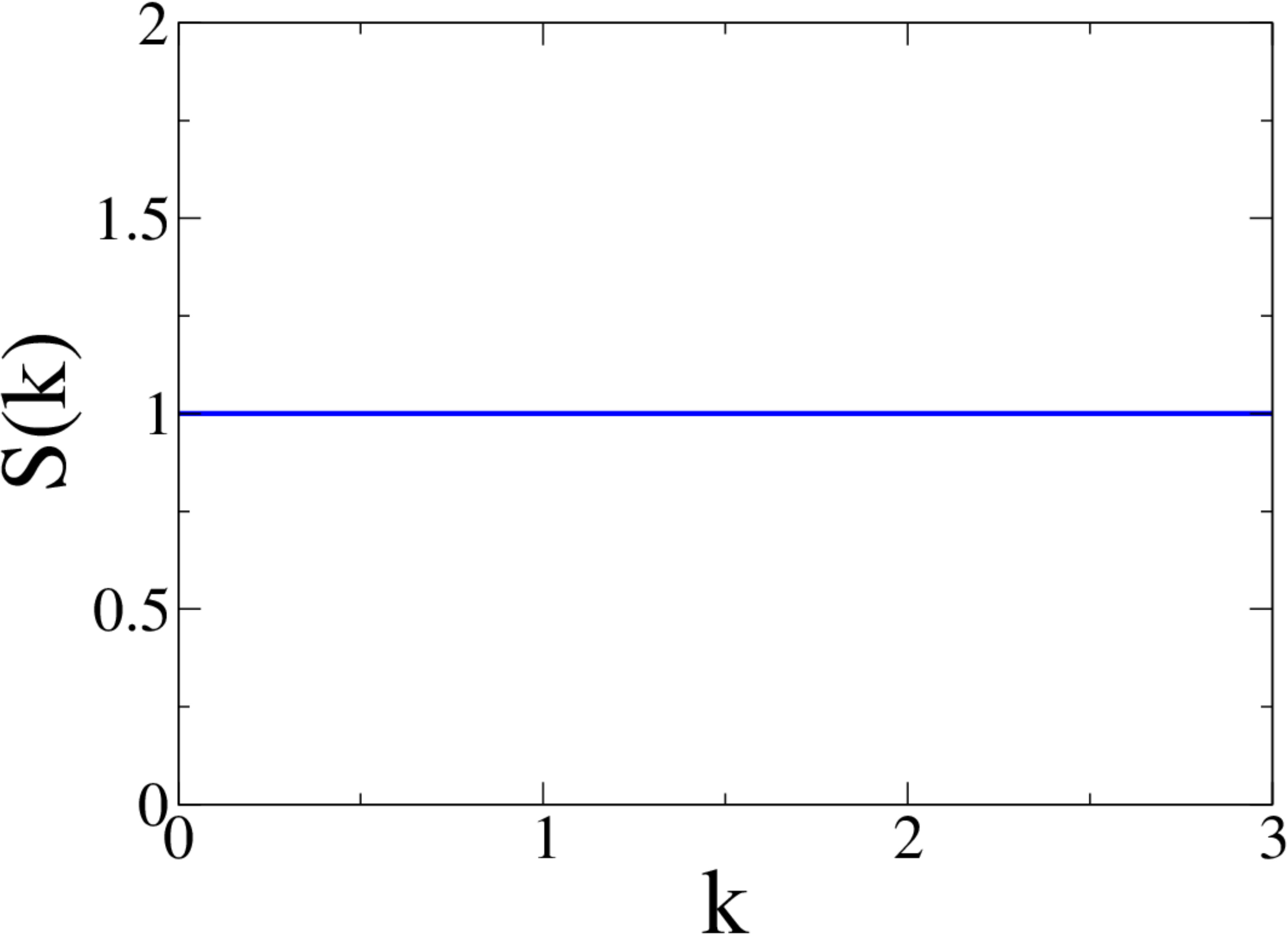}}
\caption{(Left) Portion of a Poisson point configuration. (Middle) Corresponding ensemble-averaged pair correlation function
$g_2(r)$ versus pair distance $r$ in the infinite-volume limit. (Right) Corresponding ensemble-averaged structure factor $S(k)$
versus wavenumber $k$ in the infinite-volume limit. }
\label{Poisson}
\end{center}
\end{figure}

\begin{figure}[H]
\begin{center}
{\includegraphics[  width=1.5in, keepaspectratio,clip=]{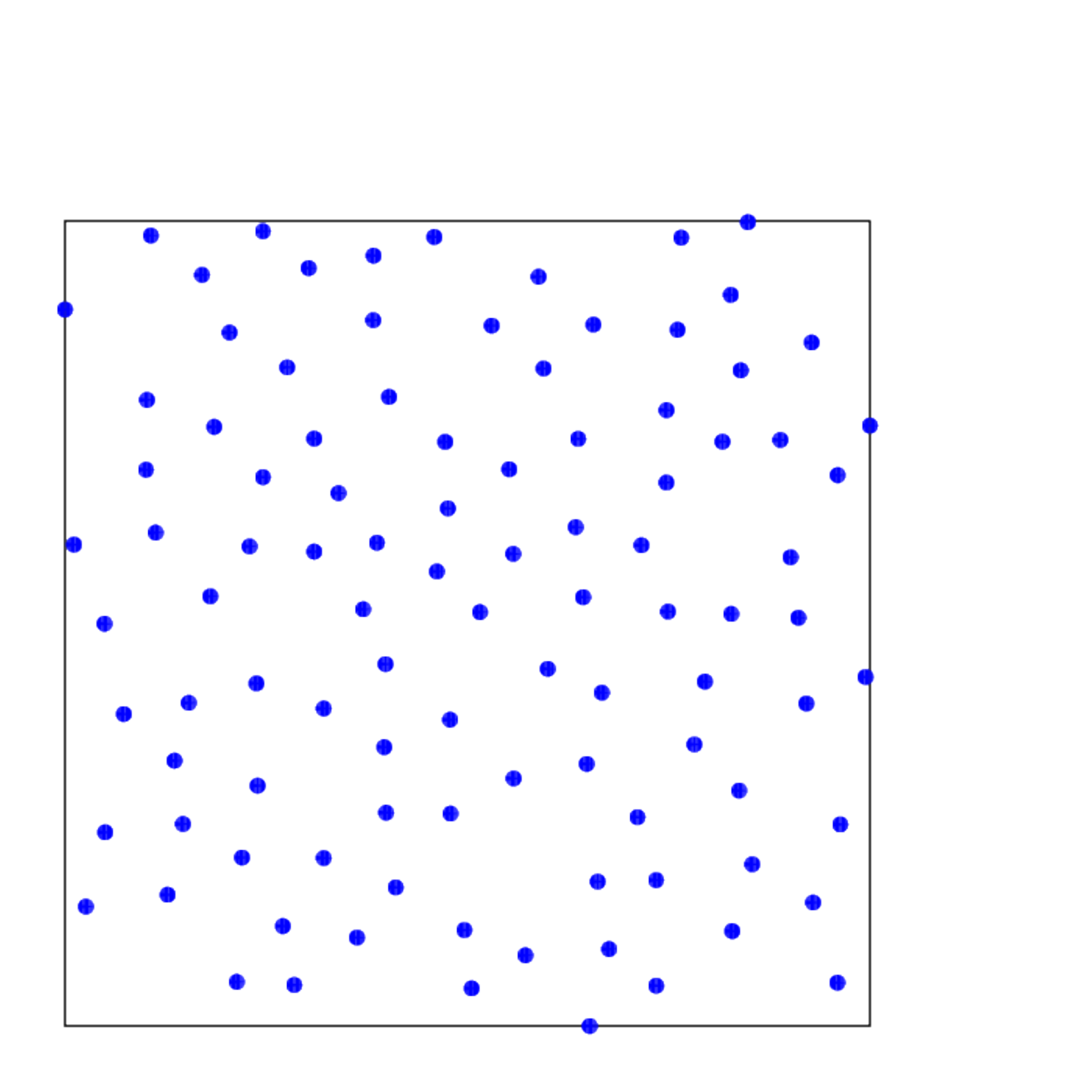}
\hspace{0.15in}
\includegraphics[  width=1.5in, keepaspectratio,clip=]{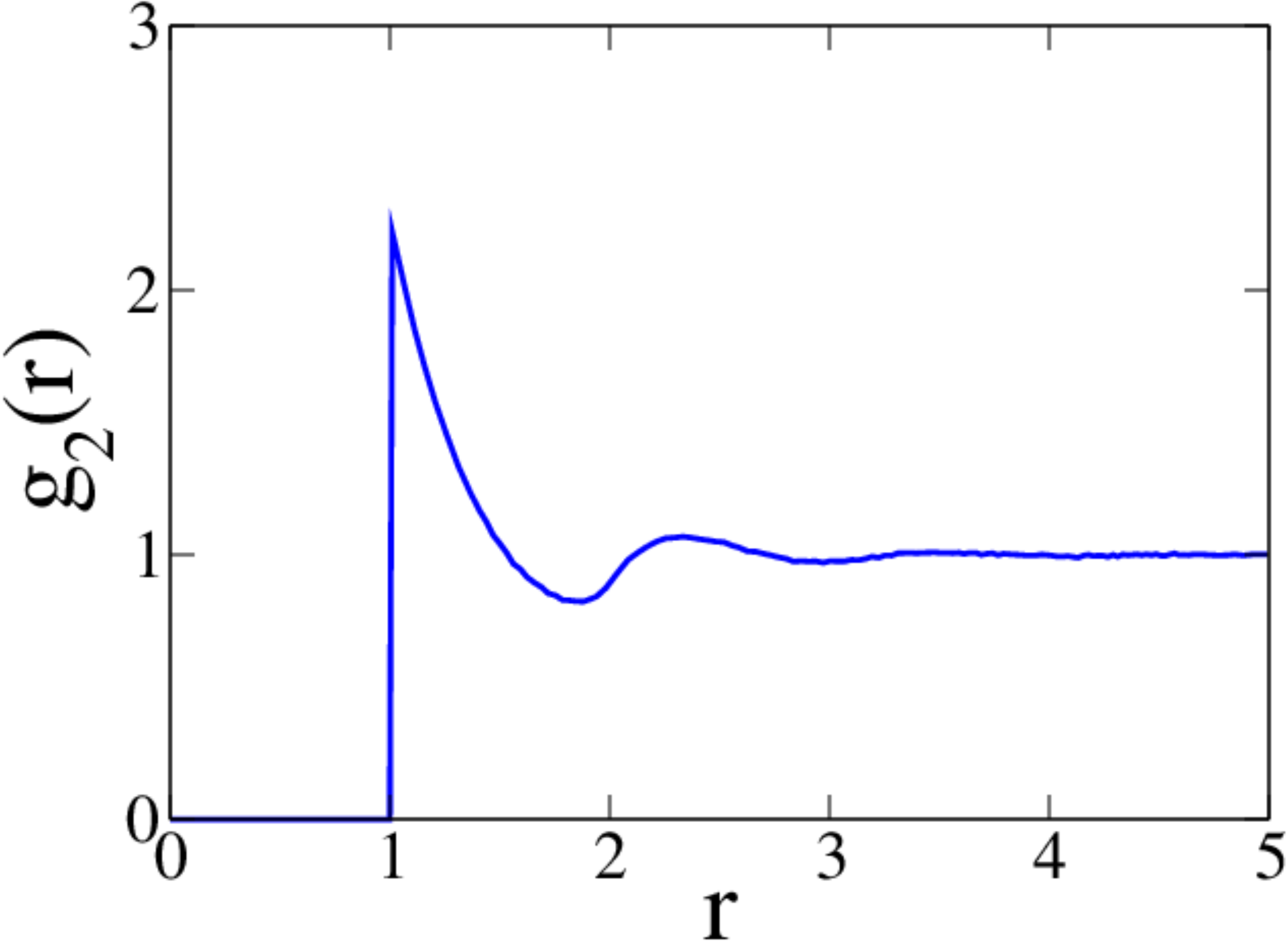}\hspace{0.15in}
\includegraphics[  width=1.5in, keepaspectratio,clip=]{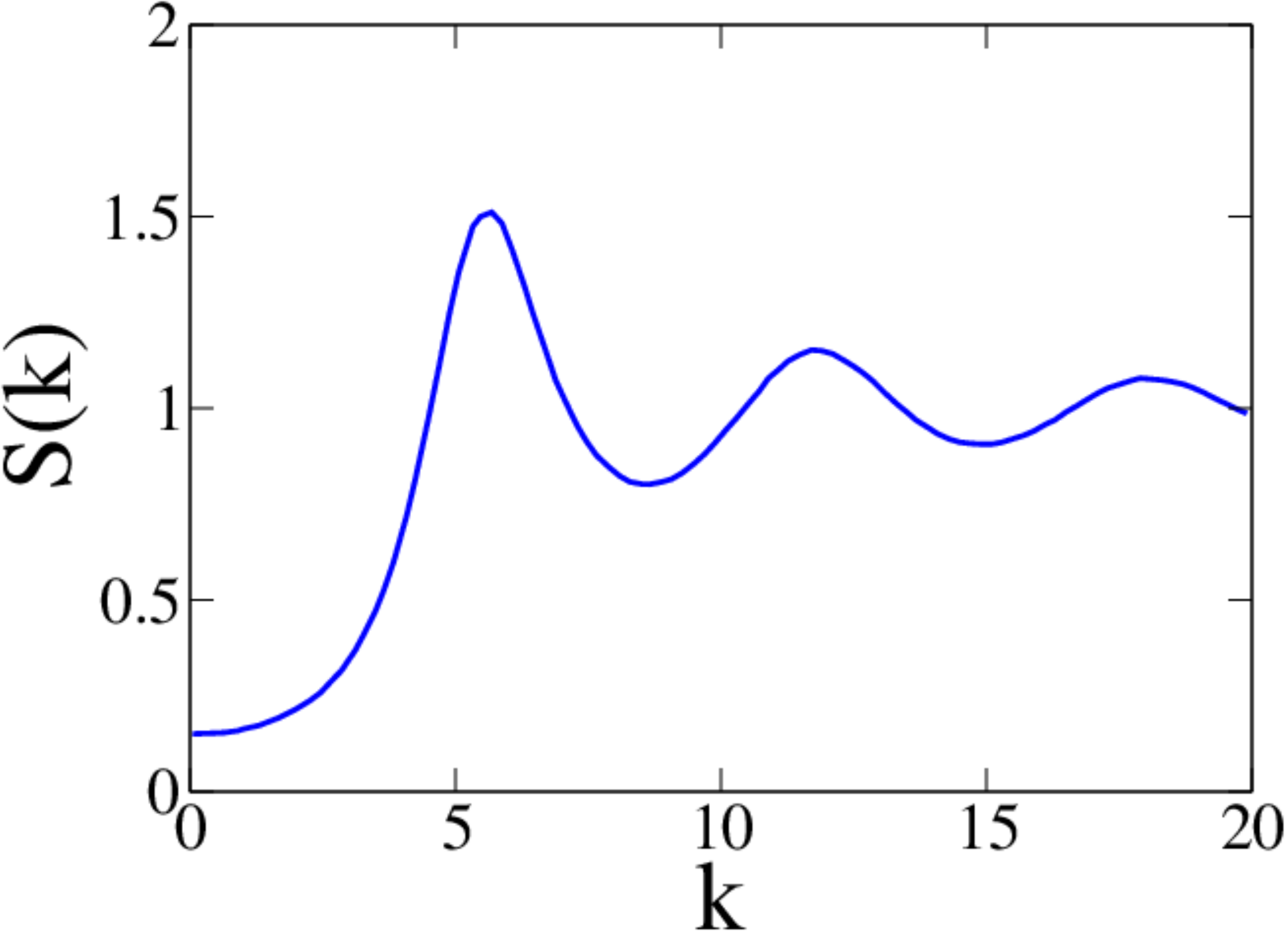}}
\caption{(Left) Portion of a configuration of an equilibrium distribution of hard circular disks
with packing fraction $\phi=0.4$. Only the centers of the disks are shown. (Middle) Corresponding ensemble-averaged pair correlation function
$g_2(r)$ versus pair distance $r$ in the infinite-volume limit in units of the disk diameter.  (Right) Corresponding ensemble-averaged structure factor $S(k)$
versus wavenumber $k$ in the infinite-volume limit. }
\label{Equi}
\end{center}
\end{figure}

\begin{figure}[H]
\begin{center}
{\includegraphics[  width=1.1in, keepaspectratio,clip=]{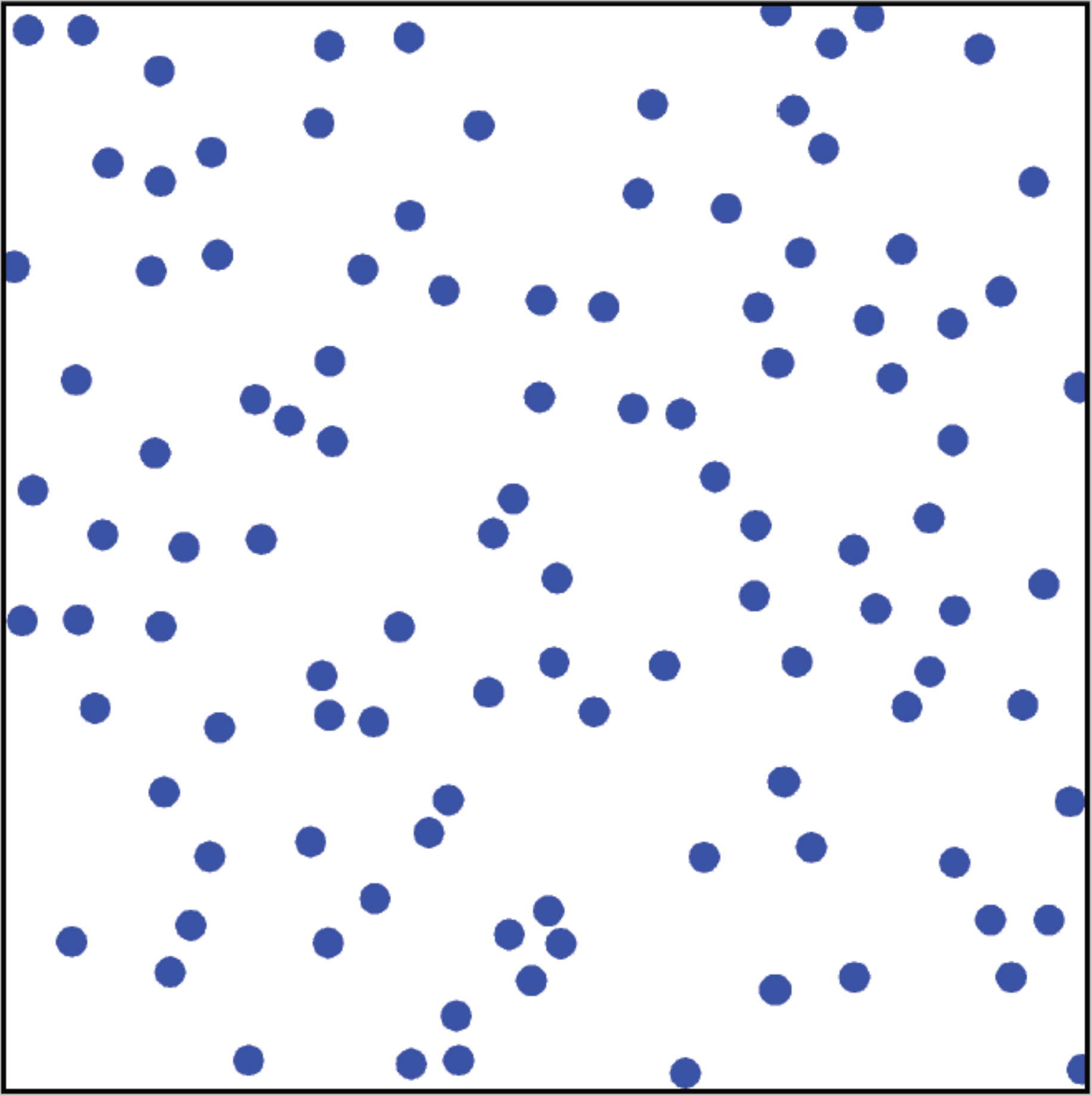}
\hspace{0.15in}
\includegraphics[  width=1.5in, keepaspectratio,clip=]{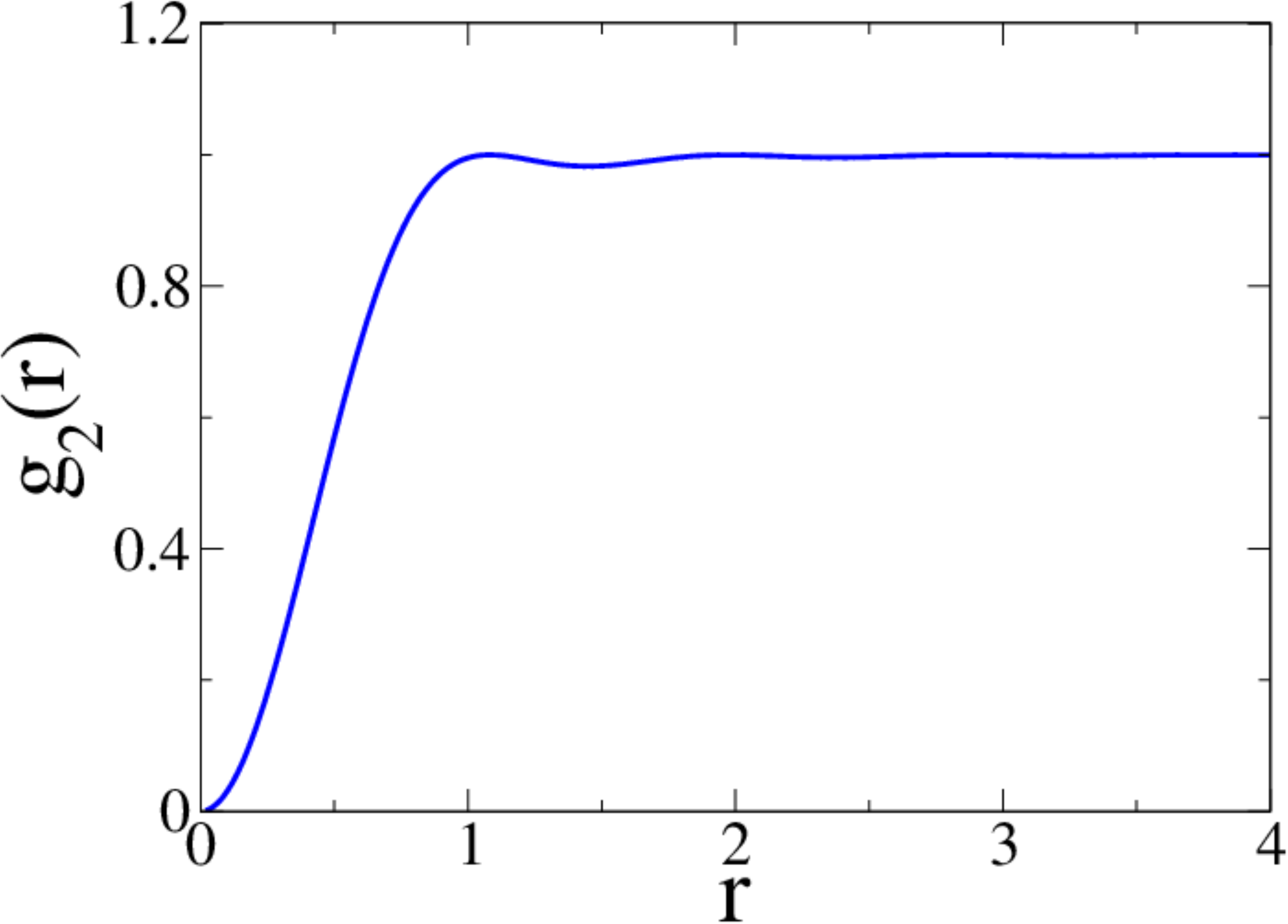}\hspace{0.15in}
\includegraphics[  width=1.5in, keepaspectratio,clip=]{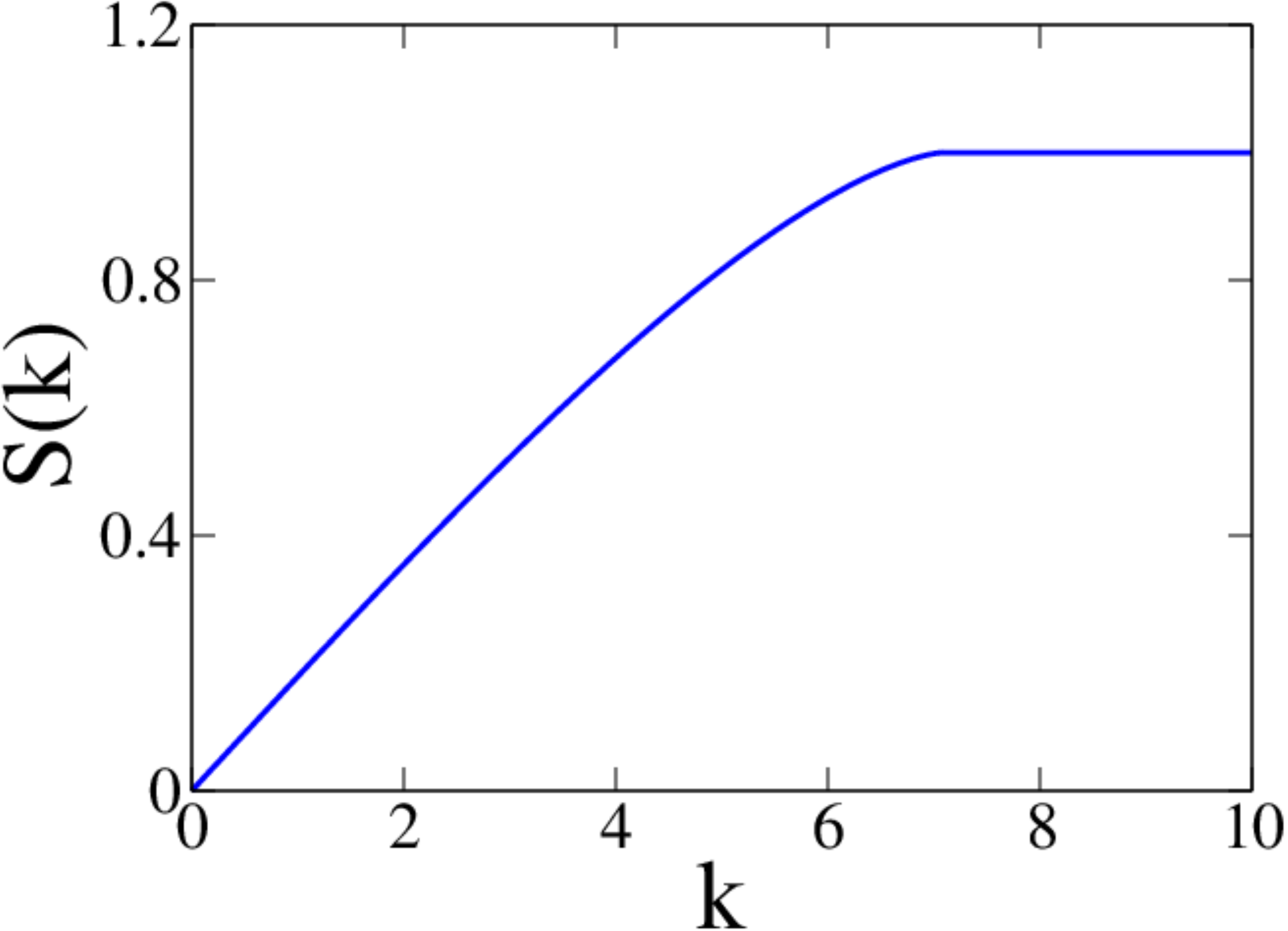}}
\caption{(Left) Portion of a configuration of a fermionic (determinantal) point process at unit density \cite{To08b}. 
(Middle) Corresponding ensemble-averaged pair correlation function
$g_2(r)$ versus pair distance $r$ in the infinite-volume limit.  (Right) Corresponding ensemble-averaged structure factor $S(k)$
versus wavenumber $k$ in the infinite-volume limit. }
\label{Fermi}
\end{center}
\end{figure}

\begin{figure}[H]
\begin{center}
{\includegraphics[  width=1.65in, keepaspectratio,clip=]{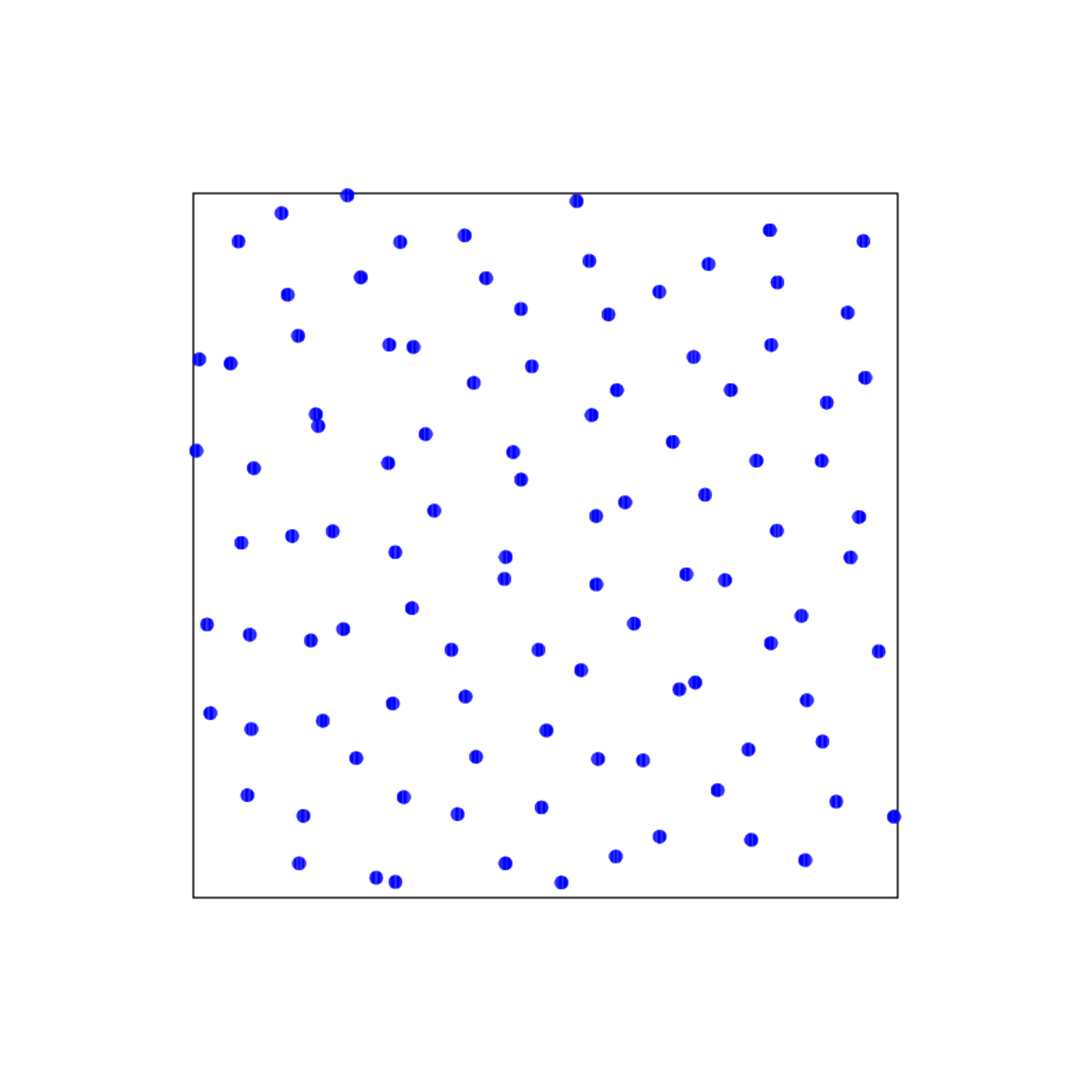}
\hspace{0.15in}
\includegraphics[  width=1.5in, keepaspectratio,clip=]{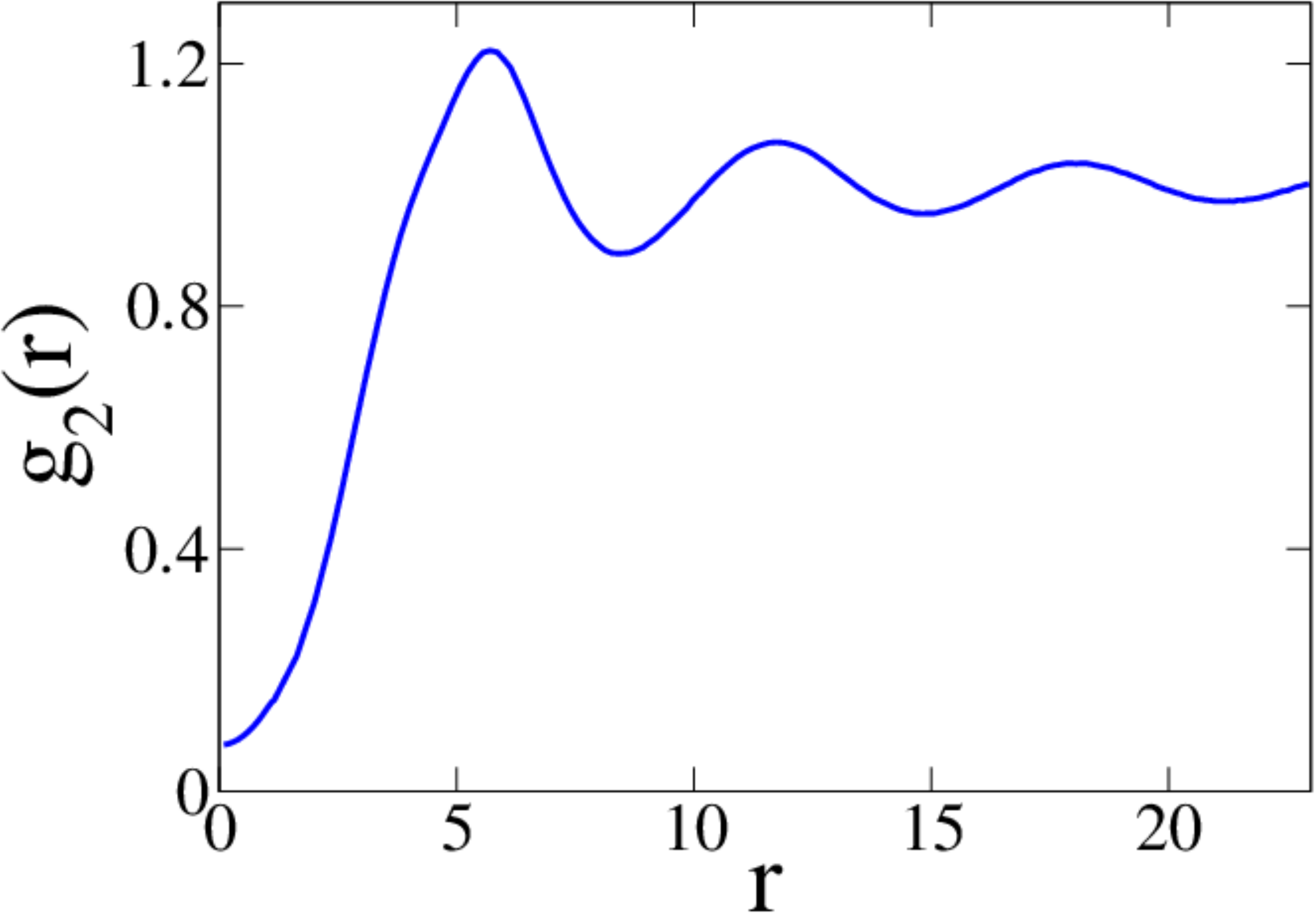}\hspace{0.15in}
\includegraphics[  width=1.5in, keepaspectratio,clip=]{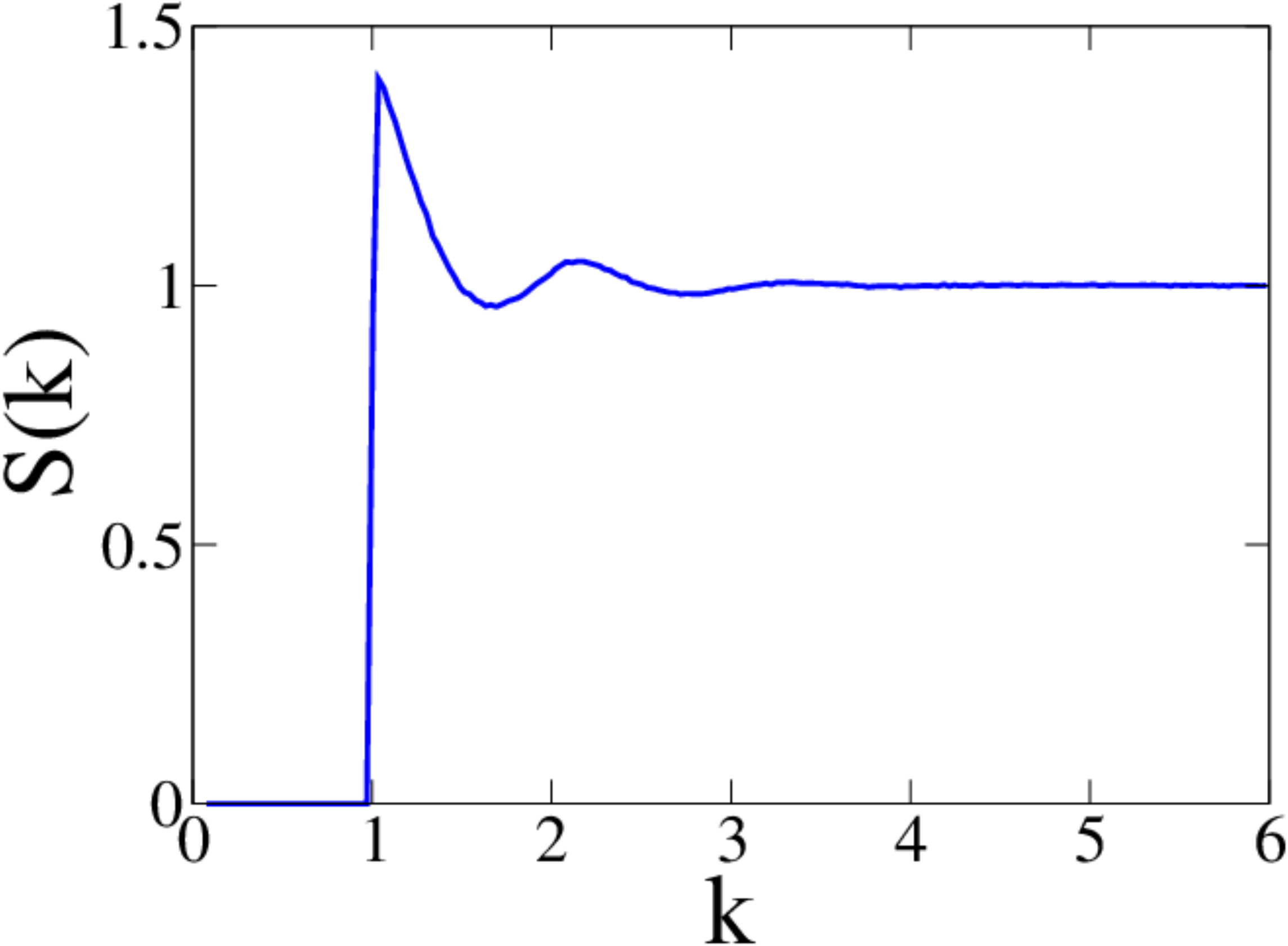}}
\caption{(Left) Portion of a configuration of a disordered stealthy hyperuniform
ground state at unit density \cite{To15}. (Middle) Corresponding ensemble-averaged pair correlation function
$g_2(r)$ versus pair distance $r$ in the infinite-volume limit.  (Right) Corresponding ensemble-averaged structure factor $S(k)$
versus wavenumber $k$ in the infinite-volume limit. }
\label{Stealthy}
\end{center}
\end{figure}

\begin{figure}[H]
\begin{center}
{\includegraphics[  width=1.5in, keepaspectratio,clip=]{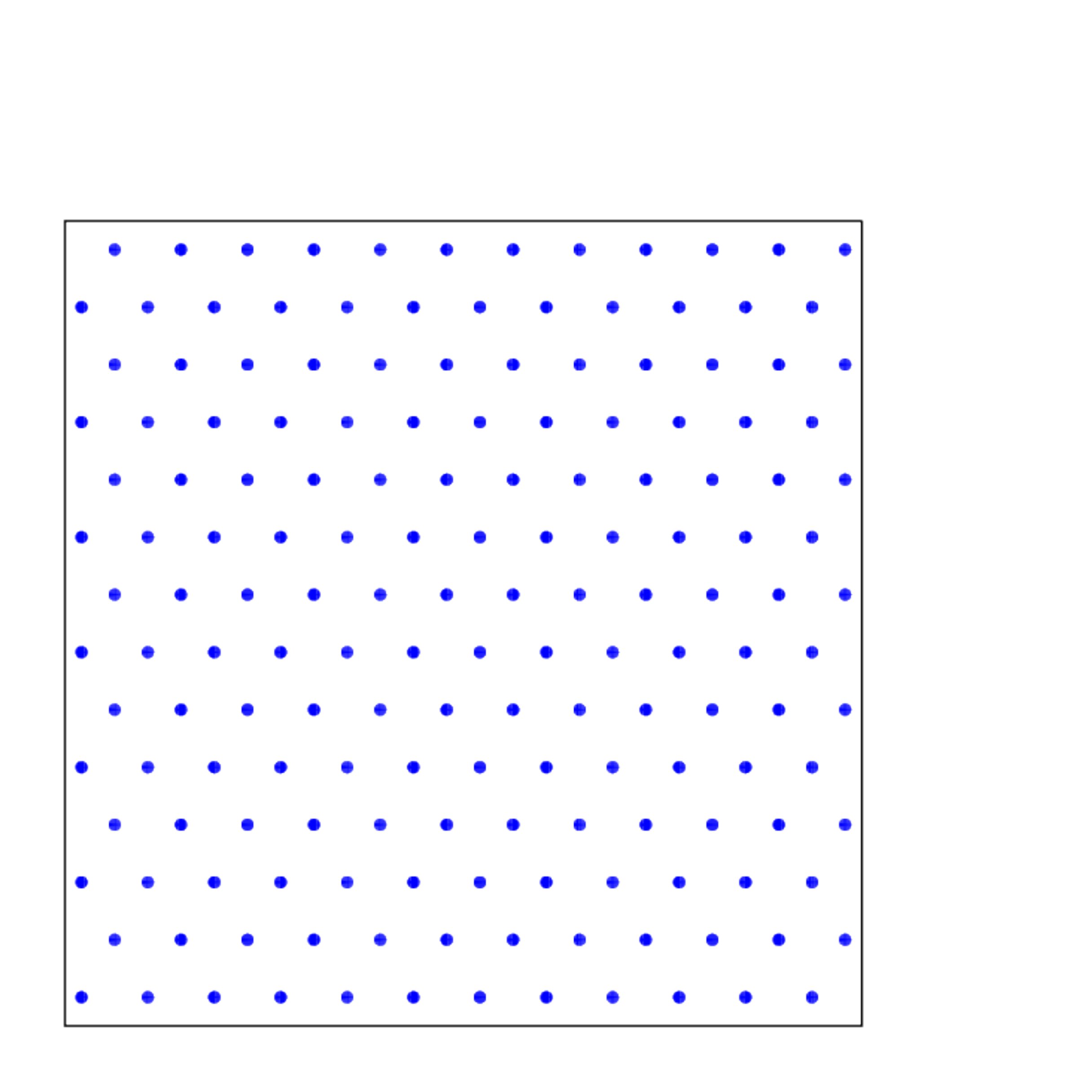}
\hspace{0.15in}
\includegraphics[  width=1.5in, keepaspectratio,clip=]{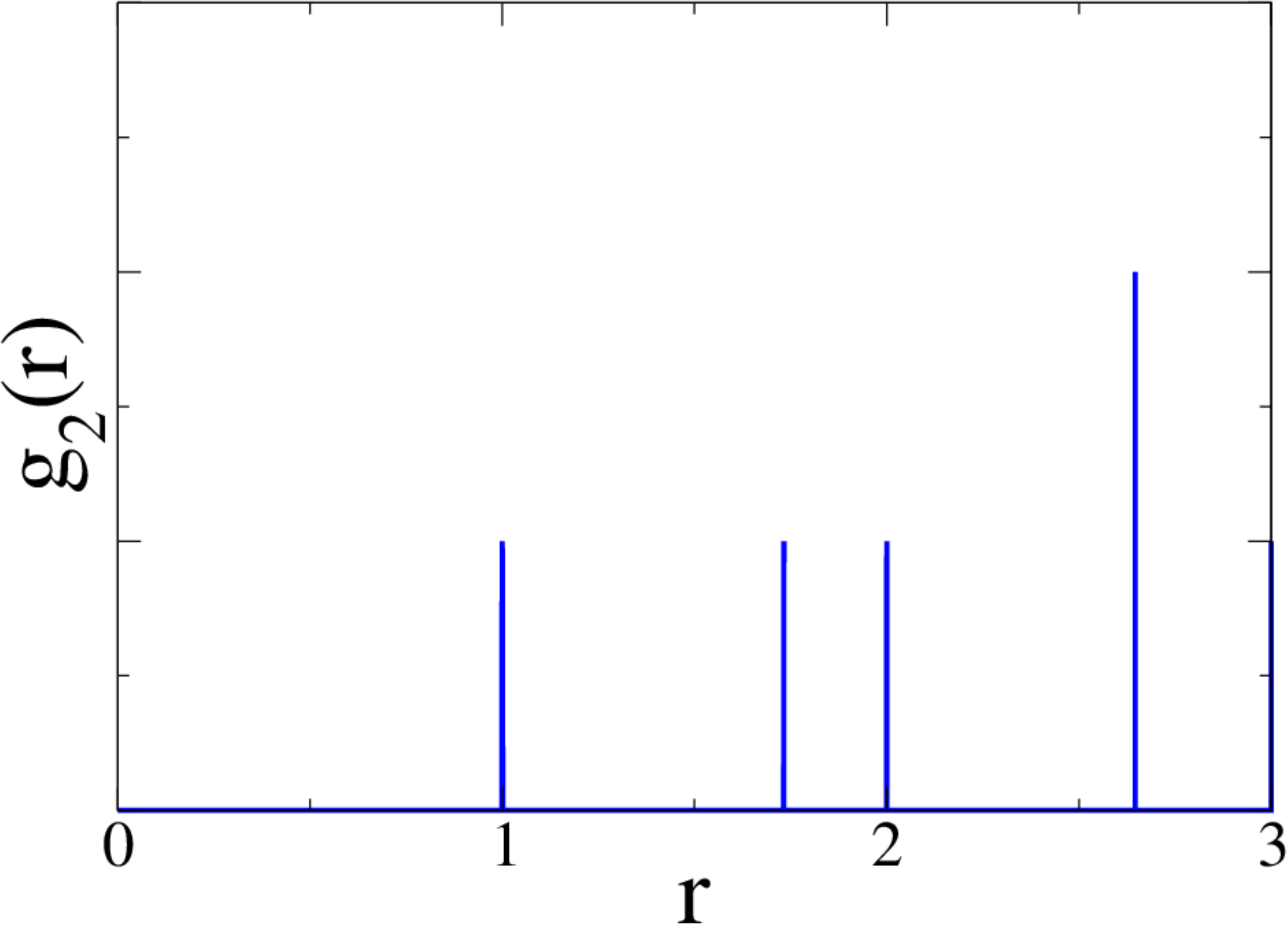}\hspace{0.15in}
\includegraphics[  width=1.5in, keepaspectratio,clip=]{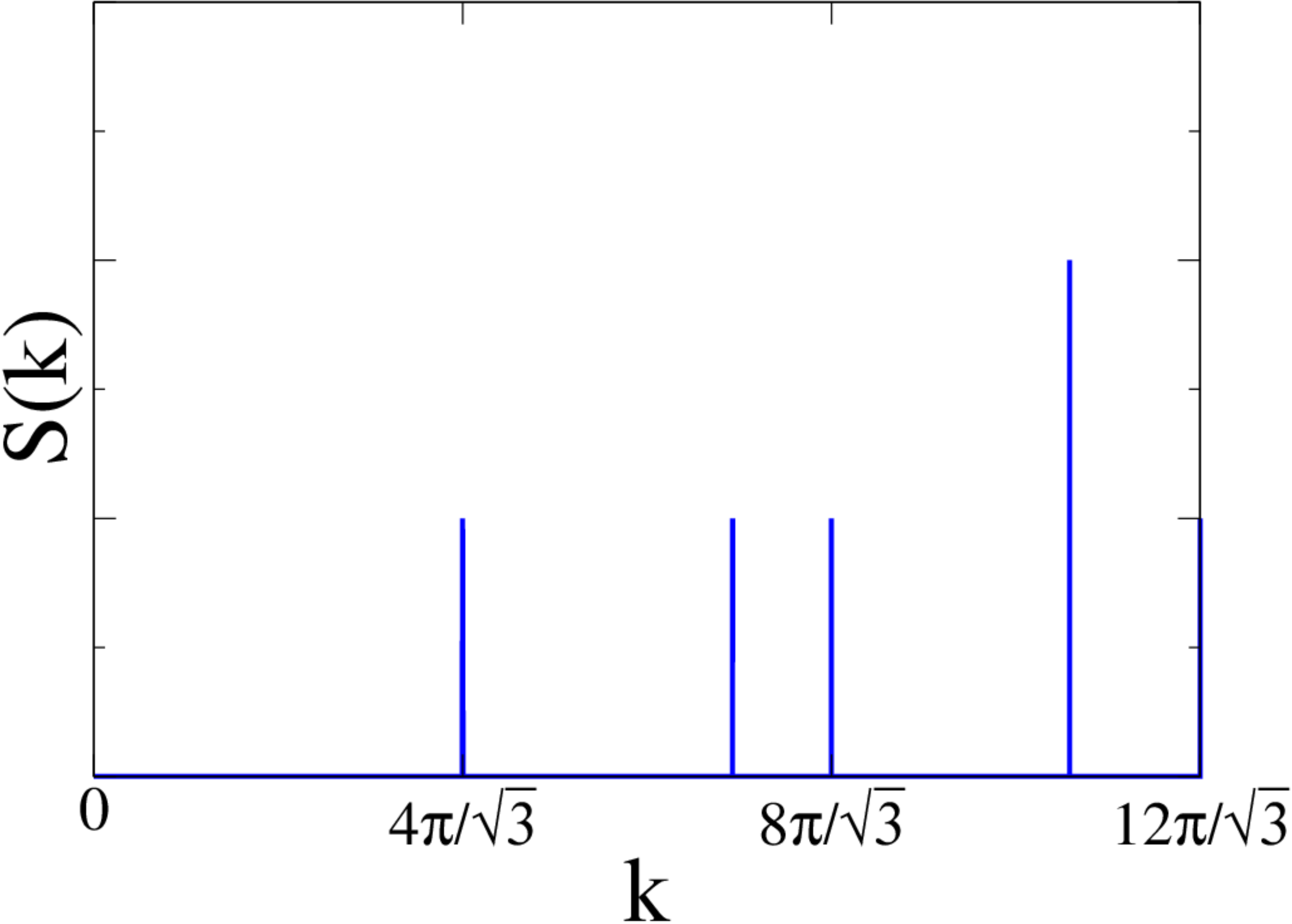}}
\caption{(Left) Portion of an infinite triangular lattice. (Middle) Corresponding angular-averaged pair correlation function
$g_2(r)$ versus pair distance $r$  in units of the nearest-neighbor distance.  (Right) Corresponding angular-averaged structure factor $S(k)$
versus wavenumber $k$.}
\label{Tri}
\end{center}
\end{figure}

It is instructive to compare pair statistics in direct and Fourier spaces to one another for
several different two-dimensional point configurations. Figures \ref{Poisson} and \ref{Equi}
show these pair statistics for the  Poisson point process
and an equilibrium hard-disk fluid, respectively, both of which are not hyperuniform. It is readily seen that 
both structure factors are positive at $k=0$. Figures \ref{Fermi} and \ref{Stealthy}
depict corresponding functions for two different disordered hyperuniform systems: a fermionic
point process and a stealthy hyperuniform point configuration, respectively. These systems will be described more fully
in Secs. \ref{STEALTHY} and \ref{Det}. It suffices to state here that whereas $S({\bf k})$ goes
to zero linearly in $k\equiv |{\bf k}|$ in the limit $k \rightarrow 0$ for the fermionic
example, $S({\bf k})$  is not only zero at the origin, but is ``gapped", i.e.,
it is zero for a contiguous range of wavenumbers around the origin, for the stealthy pattern.
Finally, Fig. \ref{Tri} shows the angular-averaged
pair statistics for the triangular lattice. Like the disordered stealthy example,
$S({\bf k})$ is gapped for the lattice, but unlike the former, possesses
long-range order as reflected by the presence of Bragg peaks out to infinity.

\vspace{-0.5in}

\vspace{-0.3in}

\vspace{-0.3in}

\vspace{-0.3in}

\subsection{Lattices and Periodic Point Configurations}
\label{lattices}

A {\it lattice} $\Lambda$ in $\mathbb{R}^d$ is a subgroup
consisting of integer linear combinations of vectors that constitute a basis for $\mathbb{R}^d$,
and thus, it represents  a special subset of point processes.
Here, the space  can be geometrically divided into identical regions $F$ called {fundamental cells}, each of which contains
just one point  specified by the lattice vector
\begin{equation}
{\bf p}= n_1 {\bf a}_1+ n_2 {\bf a}_2+ \cdots + n_{d-1} {\bf a}_{d-1}+n_d {\bf a}_d,
\label{lattice1}
\end{equation}
where ${\bf a}_i$ are the basis vectors for a fundamental cell
and $n_i$ spans all the integers for $i=1,2,\cdots, d$. We denote by
$v_F$  the volume of $F$.  A lattice is 
called  a {\it Bravais} lattice in the physical sciences.  Every lattice has a dual (or reciprocal) lattice $\Lambda^*$
in which the lattice sites are specified by the dual (reciprocal) lattice vector
 ${\bf q}\cdot {\bf p}=2\pi m$  {for all $\mathbf p$} , where $m=0, \pm 1, \pm 2, \pm 3 \cdots$.
The dual fundamental cell $F^*$ has volume $v_{F^*}=(2\pi)^d/v_F$.
This implies that the number density $\rho_{\Lambda}$ of $\Lambda$
is related to the number  density $\rho_{\Lambda^*}$ of the dual lattice $\Lambda^*$
via the expression 
\begin{equation}
\rho_{\Lambda} \rho_{\Lambda^*}=1/(2\pi)^d.
\label{rho*}
\end{equation}

Common $d$-dimensional lattices include the {\it hypercubic} {$\mathbb{Z}^d$,
}{\it checkerboard} {$D_d$, and }{\it root} {$A_d$ lattices,
defined, respectively, by
}\begin{equation}{
\mathbb{Z}^d=\{(x_1,\ldots,x_d): x_i \in }{{\mathbb{ Z}}}{\} \quad \mbox{for}\; d\ge 1
}\end{equation}
\begin{equation}{
D_d=\{(x_1,\ldots,x_d)\in \mathbb{Z}^d: x_1+ \cdots +x_d ~~\mbox{even}\} \quad \mbox{for}\; d\ge 3
}\end{equation}
\begin{equation}
A_d=\{(x_0,x_1,\ldots,x_d)\in \mathbb{Z}^{d+1}: x_0+ x_1+ \cdots +x_d =0\}  \quad \mbox{for}\; d\ge 1,
\end{equation}
{where $\mathbb{Z}$ is the set of integers;
$x_1,\ldots,x_d$ denote the components of a lattice vector
of either $\mathbb{Z}^d$ or $D_d$; and $x_0,x_1,\ldots,x_d$ denote
the components a lattice vector of $A_d$. The $d$-dimensional lattices $\mathbb{Z}^d_*$, $D_d^*$ and $A_d^*$ are
the corresponding dual lattices.
Following Conway and Sloane \cite{Co93}, we say that
two lattices are }{\it {equivalent}} {or }{\it {similar}} {if one becomes identical
to the other possibly by a rotation, reflection, and change of scale,
for which we use the symbol $\equiv$.  
The $A_d$ and $D_d$ lattices are $d$-dimensional generalizations
of the face-centered-cubic (FCC) lattice defined by $A_3 \equiv D_3$; however, for $d\ge 4$, they are no longer 
equivalent. In two dimensions, $A_2 \equiv A_2^*$ defines the triangular lattice.
In three dimensions, $A_3^* \equiv D_3^*$ defines the body-centered-cubic (BCC)
lattice. 
In four dimensions, the checkerboard lattice and its dual are equivalent,
i.e., $D_4\equiv D_4^*$. The hypercubic lattice $\mathbb{Z}^d\equiv \mathbb{Z}^d_*$ and its dual
lattice are equivalent for all $d$.
}

\begin{figure}
\centerline{\includegraphics*[  width=2.8in,keepaspectratio]{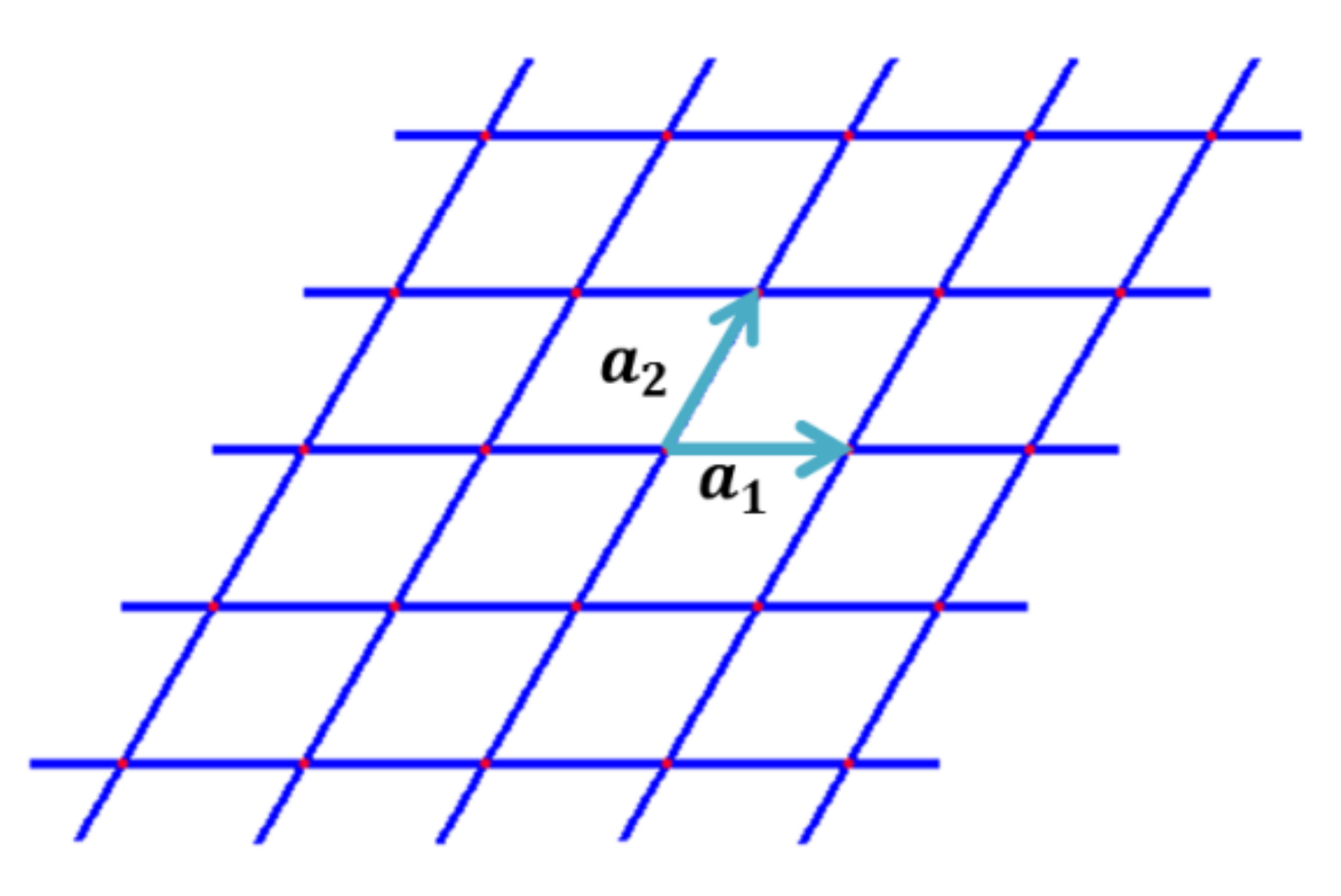} \hspace{-0.2in}
\includegraphics*[  width=2.8in,keepaspectratio]{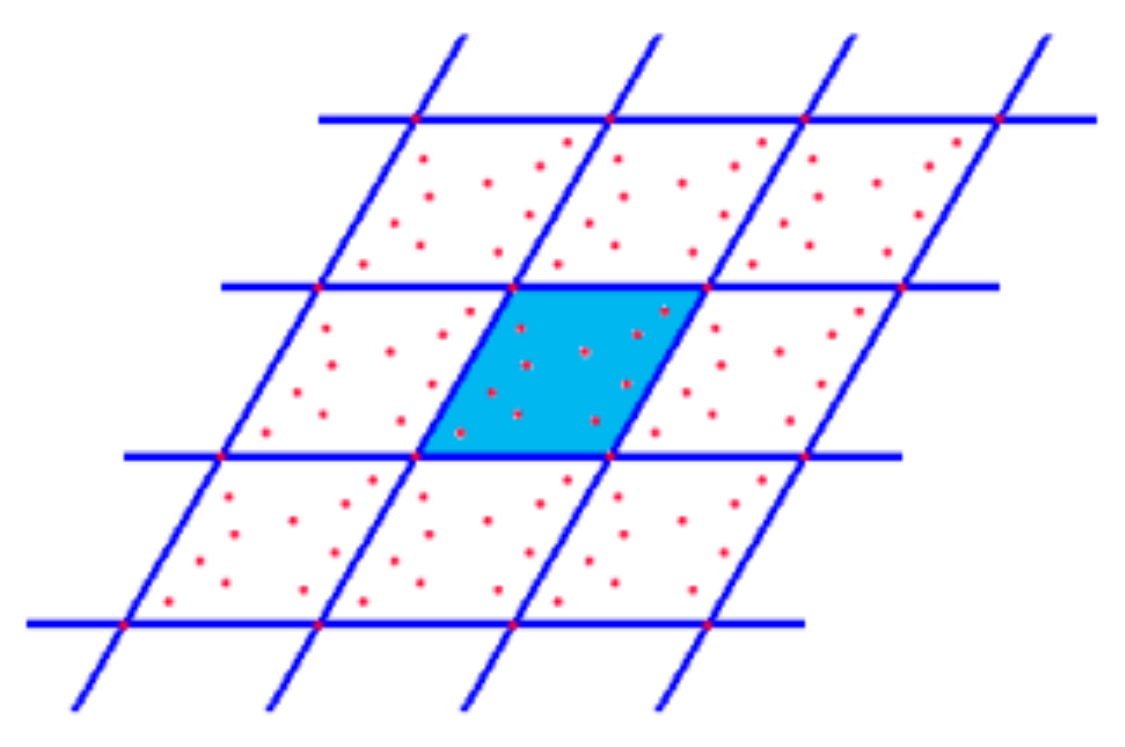}}
\caption{(Bravais) lattice with one particle per fundamental cell (left panel)
and a periodic crystal with multiple particles per fundamental cell (right panel).}
\label{lattice}
\end{figure}

A periodic point configuration (crystal) is a
more general notion than a lattice, since it is
obtained by placing a fixed configuration of $N$ points (where $N\ge 1$)
within a fundamental cell $F$ of a lattice $\Lambda$, which
is then periodically replicated. Thus, the point configuration is still
periodic under translations by $\Lambda$, but the $N$ points can occur
anywhere in $F$; see Fig. \ref{lattice}. Crystals
are characterized by long-range translational and orientational
order with crystallographic symmetries.

A quasicrystal lacks translational periodicity but possesses long-range
orientational order with prohibited crystallographic symmetries, including
five-fold symmetry and three-dimensional icosahedral symmetry \cite{Sh84,Le84,Lev86}. Levine
and Steinhardt correctly predicted that the diffraction pattern of a quasicrystal 
would show a dense set of Bragg  peaks with such prohibited crystallographic symmetries \cite{Le84,Lev86}.

An arbitrary  single periodic point configuration of $N$ points
within $F$ of a lattice $\Lambda$ is specified by its {\it microscopic density} $n({\bf r})$ at 
position $\bf r$, which is defined by
\begin{equation}
n({\bf r})=\sum_{j=1}^N \delta({\bf r}-{\bf r}_j),
\label{local-den}
\end{equation}
where $\delta({\bf r})$ is a $d$-dimensional Dirac delta function.
It is useful for both theoretical and practical reasons to introduce
the {\it complex collective density variable} ${\tilde n}({\bf k })$,
which is simply the Fourier transform of the microscopic density (\ref{local-den}):
\begin{equation}
{\tilde n}({\bf k }) = \sum_{j=1}^{N} \exp(-i{\bf k \cdot r}_j),
\end{equation}
where $\bf k$ is a wave vector in $\Lambda^*$.
This quantity enables one to determine the {\it scattering-intensity} function ${\cal S}({\bf k})$, 
which is proportional to the total intensity measured in elastic scattering radiation experiments \cite{Chaik95,Han13}.
Specifically, the scattering-intensity function is mathematically defined as
\begin{equation}
{\cal S}({\bf k})= \frac{|{\tilde n}({\bf k})|^2}{N},
\label{scatter}
\end{equation}
which is a nonnegative real function with inversion-symmetry, i.e., 
\begin{equation}
{\cal S}({\bf k}) = {\cal S}(-{\bf k})
\label{inv}
\end{equation}
that obeys the bounds
\begin{equation}
0 \le {\cal S}({\bf k}) \le N   \qquad ({\bf k} \neq {\bf 0})
\end{equation}
with ${\cal S}({\bf 0})=N $. 

For a single periodic configuration with a finite number of
$N$ points within a fundamental cell, the scattering intensity  ${\cal S}({\bf k})$
is identical to the structure factor $S({\bf k})$ [cf. (\ref{factor})], 
excluding its value at $\bf k=0$ (forward
scattering). Any such configuration contains a spherical region around the
origin $\bf k=0$ in whose interior there is no scattering for $|{\bf k}| < |{\bf k}_{Bragg}|$,
where ${\bf k}_{Bragg}$ is the minimal positive wave vector, and hence is stealthy,  
as defined by the structure factor condition (\ref{stealth}).

For an ensemble of periodic point configurations generated from a point process within $F$, 
the ensemble average of ${\cal S}({\bf k})$ in a certain infinite-size limit is directly
related to the structure factor $S({\bf k})$ defined in Eq. (\ref{factor})
via  
\begin{equation}
 \lim_{N,v_F\rightarrow \infty}\langle {\cal S}({\bf k}) \rangle = (2\pi)^d \rho \delta({\bf k}) + S({\bf k}),
\end{equation}
where
\begin{equation}
\rho \equiv \lim_{N\rightarrow \infty, v_F \rightarrow \infty}\frac{N}{v_F}
\label{limit-1}
\end{equation}
is the constant {\it number density}. This is the so-called {\it thermodynamic limit}, where it is assumed (unless
otherwise stated) that  the limiting point process is ergodic.

Note that for a single periodic point configuration at number density $\rho$, the radial 
(angular-averaged) pair correlation function can be written as
\begin{equation}
g_2(r)=  \sum_{i=1}^{\infty}\frac{Z_i}{\rho s_1(r_i)}\;\delta(r-r_i),
\label{g2-period}
\end{equation}
where $Z_i$ is the expected coordination number at radial distance $r_i$
(number of points that are exactly at a distance $r=r_i$ from
a point of the point process)
such that $r_{i+1} > r_i$ and $\delta(r)$ is a radial Dirac delta function. 

\subsection{Two-Phase Heterogeneous Media}
\label{hetero}

A random heterogeneous medium is a partition of $d$-dimensional
Euclidean space $\mathbb{R}^d$ into disjoint regions (phases) with interfaces that are known only
probabilistically \cite{To02a}. Such media are ubiquitous in nature and synthetic situations; examples include 
composites, sandstones, granular media, polymer blends, colloids, animal and plant tissue,
gels, foams, and concrete. Throughout this article,  we will consider {\it two-phase} 
heterogeneous media for simplicity.
Since one can always construct a heterogeneous medium by ``decorating" a  point process
with possibly overlapping closed sets (e.g., spheres and polyhedra) or by some other mapping derived from a point
process, characterizing the statistical
properties of a heterogeneous medium represents a more general problem than the study
of point configurations; namely, the latter systems are recovered as a special limit of the former.

Consider a two-phase random heterogeneous  medium, which is a domain of space $\mathcal{V} \subseteq \mathbb{R}^d$ of volume $V$ 
that is partitioned into two disjoint regions:
a phase 1 region $\mathcal{V}_1(\omega)$ of volume fraction $\phi_1$ and a phase 2 region $\mathcal{V}_2(\omega)$ of volume fraction $\phi_2$ such that $\mathcal{V}= \mathcal{V}_1 \cup \mathcal{V}_2$  for some realization $\omega$ of an underlying probability space \cite{To02a}.
The statistical properties of each phase of the system are specified by the countably infinite set of \emph{$n$-point probability functions} $S_n^{(i)}$, 
which are defined by \cite{To82b, To83a, St95,To02a}:
\begin{eqnarray}\label{Sndef}
S_n^{(i)}(\mathbf{x}_1, \ldots, \mathbf{x}_n) = \left\langle\prod_{i=1}^n {\cal I}^{(i)}(\mathbf{x}_i;\omega)\right\rangle,
\end{eqnarray}
where angular brackets denote an ensemble average amd  ${\cal I}^{(i)}(\mathbf{x};\omega)$ is the indicator function for phase $i$ defined by
\begin{eqnarray}
{\cal I}^{(i)}(\mathbf{x};\omega) = \left\{
\begin{array}{lr}
1, \quad \mathbf{x} \in \mbox{phase } i\\
0, \quad \mbox{otherwise},
\end{array}\right.
\end{eqnarray}

The $n$-point probability function $S_n^{(i)}(\mathbf{x}_1, \ldots, \mathbf{x}_n)$
gives the probability of finding the vectors $\mathbf{x}_1, \ldots, \mathbf{x}_n$
all in phase $i$. If the medium is statistically homogeneous, $S_n^{(i)}(\mathbf{x}_1, \ldots, \mathbf{x}_n)$ is translationally invariant
and, in particular, the one-point probability function is independent of position and equal to  the volume fraction of phase $i$:
\begin{eqnarray}
S_1^{(i)}(\mathbf{x}) = \phi_i,
\label{S1}
\end{eqnarray}
while the two-point probability function $S_2^{(i)}({\bf r})$ depends on the displacement vector
${\bf r} \equiv \mathbf{x}_2 -\mathbf{x}_1$. If the medium is also statistically isotropic, then
$S_2^{(i)}(r)$ is a radial function (where $r =|\bf r|$), which gives the probability of finding the
end points of a line segment of length $r$ in phase $i$.

For statistically homogeneous media, the two-point correlation function for phase 2 is simply related
to that for phase 1 via the expression $S^{(2)}_2({\bf r}) = S^{(1)}_2({\bf r}) - 2\phi_1 + 1$,
and hence the \textit{autocovariance} function is given by
\begin{equation}
\label{Auto}
\chi_{_V}({\bf r}) \equiv  S^{(1)}_2({\bf r}) - {\phi_1}^2 =  S^{(2)}_2({\bf r}) - {\phi_2}^2,
\end{equation}
for phase 1 is equal to that for phase 2. Generally, for ${\bf r}=0$,
\begin{equation}
\label{eq109}
 S^{(i)}_2({\bf 0}) = \phi_i , \qquad \chi_{_V}({\bf 0})=\phi_1\phi_2
\label{two}
\end{equation}
and in the absence of any \textit{long-range} order,
\begin{equation}
\label{eq110}
\lim_{|{\bf r}|\rightarrow \infty}  S^{(i)}_2({\bf r}) \rightarrow {\phi_i}^2 , \qquad \lim_{|{\bf r}|\rightarrow \infty}  \chi_{_V}({\bf r}) \rightarrow 0.
\end{equation}

We call the Fourier transform of $\chi_{_V}({\bf r})$,
\begin{equation}
{\tilde \chi}_{_V}({\bf k})=\int_{\mathbb{R}^d} \chi_{_V}({\bf r})\exp[-i ({\bf k}\cdot {\bf r})] d{\bf r},
\label{SPEC}
\end{equation}
the {\it spectral density}, which again can be measured in the laboratory from 
scattering experiments \cite{De49,De57}. This quantity is the heterogeneous-medium analog of the structure factor $S({\bf k})$
for point configurations [cf. (\ref{factor})]. In light of the second equality in (\ref{two}), it immediately
follows that the spectral density generally obeys the following sum rule:
\begin{equation}
\frac{1}{(2\pi)^{d}\,\phi_1\phi_2} \int_{\mathbb{R}^d}
{\tilde \chi}_{_V}({\bf k})\,d{\bf k} =1\,.
\end{equation}

A hyperuniform two-phase medium is one in which the spectral density ${\tilde \chi}_{_V}({\bf k})$ tends to zero as the wavenumber $k\equiv |\bf k|$ tends to zero \cite{Za09}, i.e.,
\begin{equation}
\lim_{|{\bf k}| \rightarrow 0} {\tilde \chi}_{_V}({\bf k}) = 0.
\label{hyper-2}
\end{equation}
This hyperuniformity condition together with definition (\ref{SPEC}) imply that the
volume integral of  $\chi_{_V}({\bf r})$ over all space vanishes identically, i.e.,
\begin{equation}
\int_{\mathbb{R}^d} \chi_{_V}({\bf r}) d{\bf r}=0,
\label{sum-2}
\end{equation}
which is a direct-space {\it sum rule} that a hyperuniform two-phase system  must obey. 
This means that  $\chi_{_V}({\bf r})$ cannot be strictly positive and hence
must exhibit anticorrelations (i.e., negative correlations) for some values of $\bf r$.

\section{Local Number Fluctuations}
\label{local-N}

\subsection{Ensemble-Average Formulation of the Number Variance}

Consider general statistically homogeneous point processes in $d$-dimensional Euclidean space $\mathbb{R}^d$.
Density fluctuations as measured by the number variance within a window of a given shape 
is entirely determined by pair statistics and a certain geometric property of the window.
Here we provide a derivation of the local number variance formula in the ensemble setting
following Torquato and Stillinger \cite{To03a}. Consider a $d$-dimensional observation window $\Omega$ of convex shape and let 
$\bf R$ symbolize the parameters that characterize its geometry and its fixed
orientation with respect to the point process. For example, in
the case of a $d$-dimensional ellipsoidal (hyperellipsoid) or $d$-dimensional cubical 
(hypercubic) window, $\bf R$ would represent the semi-axes of the ellipsoid
or side length of the cube, respectively, as well as the window orientation. For statistical anisotropic
point processes, it can be judicious to choose a nonspherical
window to be consistent with symmetries of the underlying point processes (see Sec. \ref{anisotropy})
but, often, hyperspherical windows are convenient choices. Let us introduce the window indicator function
\begin{equation}
w({\bf x}- {\bf x}_0;{\bf R})=\Bigg\{{1, \quad {\bf x} - {\bf x}_0\in \Omega,
\atop{0, \quad {\bf x}- {\bf x}_0 \notin \Omega,}}
\label{window}
\end{equation}
for a window occupying region $\Omega \subset {\mathbb{R}^d}$ with centroidal position ${\bf x}_0$.
Let $N({\bf x}_0;{\bf R})$ denote the number of points 
within the window  in a realization 
of the ensemble, which is given explicitly by
\begin{eqnarray}
N({\bf x}_0;{\bf R}) &=& \sum_{i=1} w({\bf r}_i-{\bf x}_0;{\bf R}).
\label{N}
\end{eqnarray}
 The ensemble-average number of points contained within the window
in the thermodynamic (infinite-volume) limit is simply
\begin{eqnarray}
\langle N({\bf R}) \rangle &=& \rho v_1({\bf R}),
\label{N(R)}
\end{eqnarray}
where 
\begin{equation}
v_1({\bf R}) = \int_{\mathbb{R}^d} w({\bf r}; {\bf R}) d{\bf r}
\label{V}
\end{equation} 
is the window volume.
The translational invariance (statistical homogeneity) of the point configuration renders the ensemble average
$\langle N({\bf x}_0;{\bf R}) \rangle$ independent of the window coordinate
${\bf x}_0$, and so we suppress  it in (\ref{N(R)}).

Similarly, ensemble averaging the square of $N({\bf x}_0;{\bf R})$  and using
(\ref{N(R)}) gives the {\it local number variance} as
\begin{eqnarray*}
\sigma^2_{_N}({\bf R}) &\equiv& \langle N^2({\bf R}) \rangle-  \langle N({\bf R}) \rangle^2 \nonumber\\
&=&\int_{\mathbb{R}^d} \rho_1({\bf r}_1) w({\bf r}_1-{\bf x}_0;{\bf R}) d{\bf r}_1
+\int_{\mathbb{R}^d} \int_{\mathbb{R}^d}[\rho_2({\bf r}_1,{\bf r}_2) - \rho_1({\bf r}_1)\rho_1({\bf r}_2)]
w({\bf r}_1-{\bf x}_0;{\bf R}) 
w({\bf r}_2-{\bf x}_0;{\bf R}) d{\bf r}_1 d{\bf r}_2.
\end{eqnarray*}
where $\rho_1({\bf r})$ and $\rho_2({\bf r}_1,{\bf r}_2)$ are the one- and two-particle
probability density functions, respectively (see Sec. \ref{Points}). Invoking statistical homogeneity,
 leads to the following simple expression for the number variance in terms
of the  the total correlation function $h({\bf r})$ [cf. (\ref{total})] that
is independent of ${\bf x}_0$:
\begin{equation}
\sigma^2_{_N}({\bf R}) = \langle N({\bf R}) \rangle \Bigg[ 1+\rho\int_{\mathbb{R}^d}  h({\bf r})
\alpha_2({\bf r};{\bf R}) d{\bf r}\Bigg],
\label{N2}
\end{equation}
where
\begin{equation}
\alpha_2({\bf r};{\bf R})= \frac{v_2^{\mbox{\scriptsize int}}({\bf r},
{\bf R})}{v_1({\bf R})}
\label{I}
\end{equation}
is called the {\it scaled intersection volume} function and
\begin{equation}
v_2^{\mbox{\scriptsize int}}({\bf r};{\bf R})= \int_{\mathbb{R}^d} w({\bf x}_0;{\bf R})
w({\bf x}_0+{\bf r};{\bf R}) d{\bf x}_0
\label{int-v}
\end{equation}
is the intersection volume
$v_2^{\mbox{\scriptsize int}}({\bf r};{\bf R})$  of two windows (with the same orientations) 
whose centroids  are separated by the displacement vector  ${\bf r}$. 
Note that  $\alpha_2({\bf r};{\bf R})$ has compact support (non-zero only when two windows overlap) and, by definition, $\alpha_2({\bf r}={\bf 0};{\bf R})=1$.  
Formula (\ref{N2}) was also derived by Landau and Lifschitz \cite{La80},
although they did not explicitly indicate the scaled intersection
volume function $\alpha_2({\bf r};{\bf R})$. Martin and Yalcin \cite{Ma80}
derived the corresponding formula for charge fluctuations in classical
Coulombic systems.

For finite size windows, the integral in (\ref{N2}) will exist
for bounded $h(\bf r)$ because $\alpha_2({\bf r};{\bf R})$ has finite support. For infinitely
large windows, $\alpha_2({\bf r};{\bf R})=1$, and integrability
requires that $h(\bf r)$ decays faster than $|{\bf r}|^{-d}$. For systems in thermal equilibrium, this
will be the case for pure phases away from standard critical points, e.g., liquid-gas and magnetic
varieties. The structure factor S({\bf k})  [defined by (\ref{factor})] at ${\bf k}=\bf 0$ tends
to $+ \infty$  as a thermal critical point
is approached, implying that $h({\bf r})$ becomes long-ranged with a power-law tail that
decays slower than $|{\bf r}|^{-d}$ \cite{Ka66,Fi67,St71,Wi74,Bi92}, which is a manifestation of the underlying
fractal structure  \cite{Wi74,Bi92}. Such fractal systems can be thought of as {\it anti-hyperuniform} with {\it hyperfluctuations},
since they are the antithesis of a hyperuniform system in which $\lim_{|{\bf k}| \rightarrow 0} S({\bf k}) = 0$; see Sec. \ref{vanishing}
for further discussion.

There is a dual Fourier representation of the local number variance formula \cite{To03a}.
Specifically, using Parseval's theorem for Fourier transforms, formula
(\ref{N2}) for an arbitrarily shaped (regular) window can be expressed in terms
of the structure factor [defined by (\ref{factor})] and the nonnegative function
${\tilde \alpha}_2({\bf k};{\bf R})$, the Fourier transform of $\alpha_2({\bf r};{\bf R})$ \cite{To03a}:
\begin{equation}
\sigma^2_{_N}({\bf R})=
\langle N({\bf R}) \rangle \Bigg[\frac{1}{(2\pi)^d} \int_{\mathbb{R}^d} S({\bf k})
{\tilde \alpha}_2({\bf k};{\bf R}) d{\bf k}\Bigg].
\label{var-par}
\end{equation}
Note that
\begin{equation}
{\tilde \alpha}_2({\bf k};{\bf R})=\frac{{\tilde w^2({\bf k};{\bf R})}}{v_1({\bf R})} \ge 0,
\label{I-f}
\end{equation}
where ${\tilde w}({\bf k};{\bf R})$ is the Fourier transform of the window indicator function (\ref{window}), and
\begin{equation}
\frac{1}{(2\pi)^d} \int_{\mathbb{R}^d}  {\tilde \alpha}_2({\bf k};{\bf R}) d{\bf k}= \alpha_2({\bf r}=0;{\bf R})=1.
\label{X}
\end{equation}
It immediately follows from (\ref{var-par}) 
that the number variance is strictly positive whenever $v_1({\bf R}) > 0$  \cite{To03a}.

\subsubsection{Hyperspherical Windows}

Many of the subsequent results will be given for the case of a 
$d$-dimensional spherical (hyperspherical) window of radius $R$ centered
at position ${\bf x}_0$. The general window indicator function (\ref{window}) becomes
$w({\bf x}-{\bf x}_0;R)=\Theta(R-|{\bf x}-{\bf x}_0|)$,
where $\Theta(x)$ is the Heaviside step function
\begin{equation}
\Theta(x) =\Bigg\{{0, \quad x<0,\atop{1, \quad x \ge 0.}}
\label{heaviside}
\end{equation}
Thus, the function $v_1({\bf R})$, defined by (\ref{V}),
becomes the volume of a $d$-dimensional spherical window 
of radius $R$:
\begin{equation}
v_1(R)= \frac{\pi^{d/2}}{\Gamma(1+d/2)}R^d.
\label{v1}
\end{equation}

\begin{figure}
\centerline{ \includegraphics[width=2.7in]{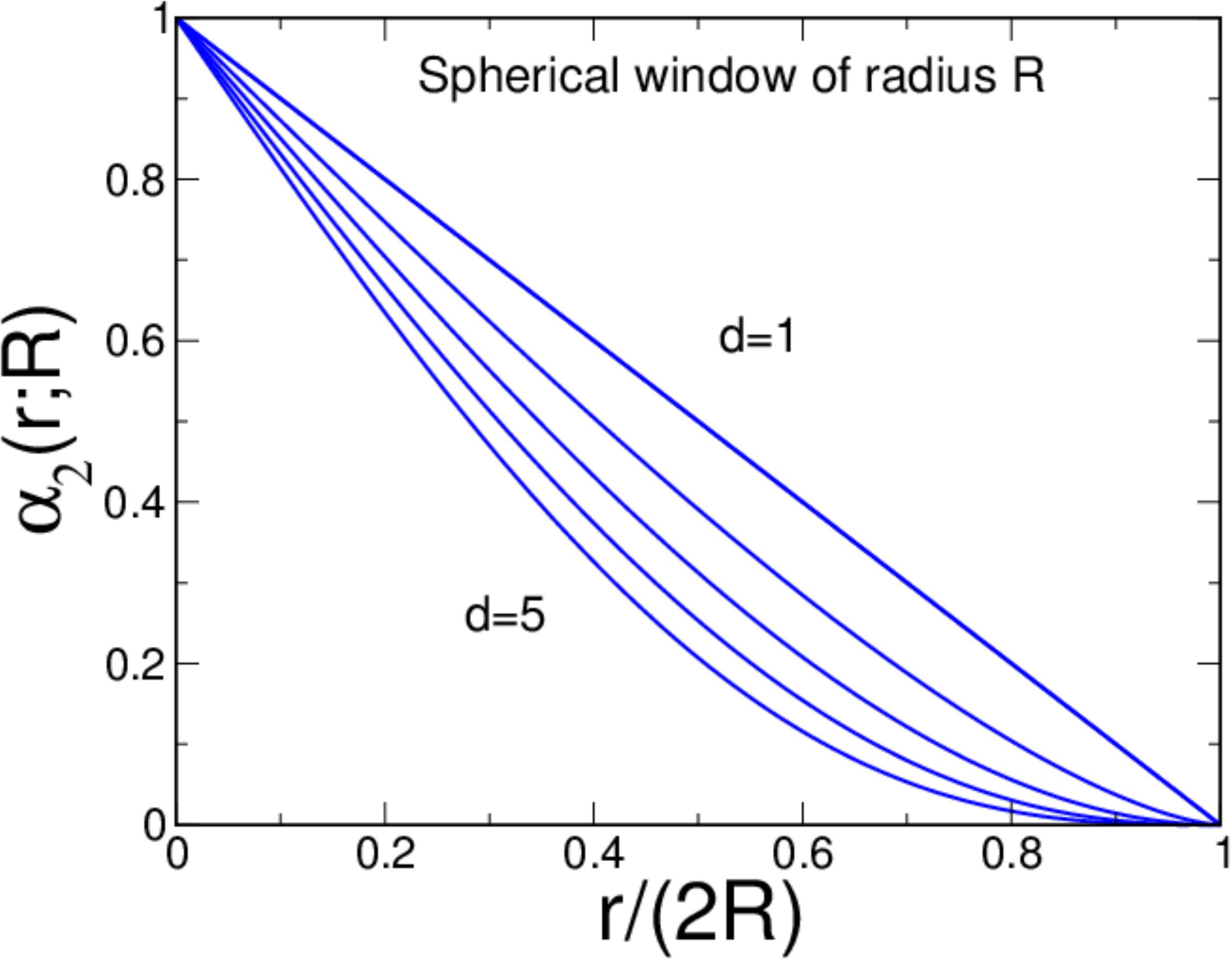} \hspace{0.2in}
\includegraphics[width=2.7in]{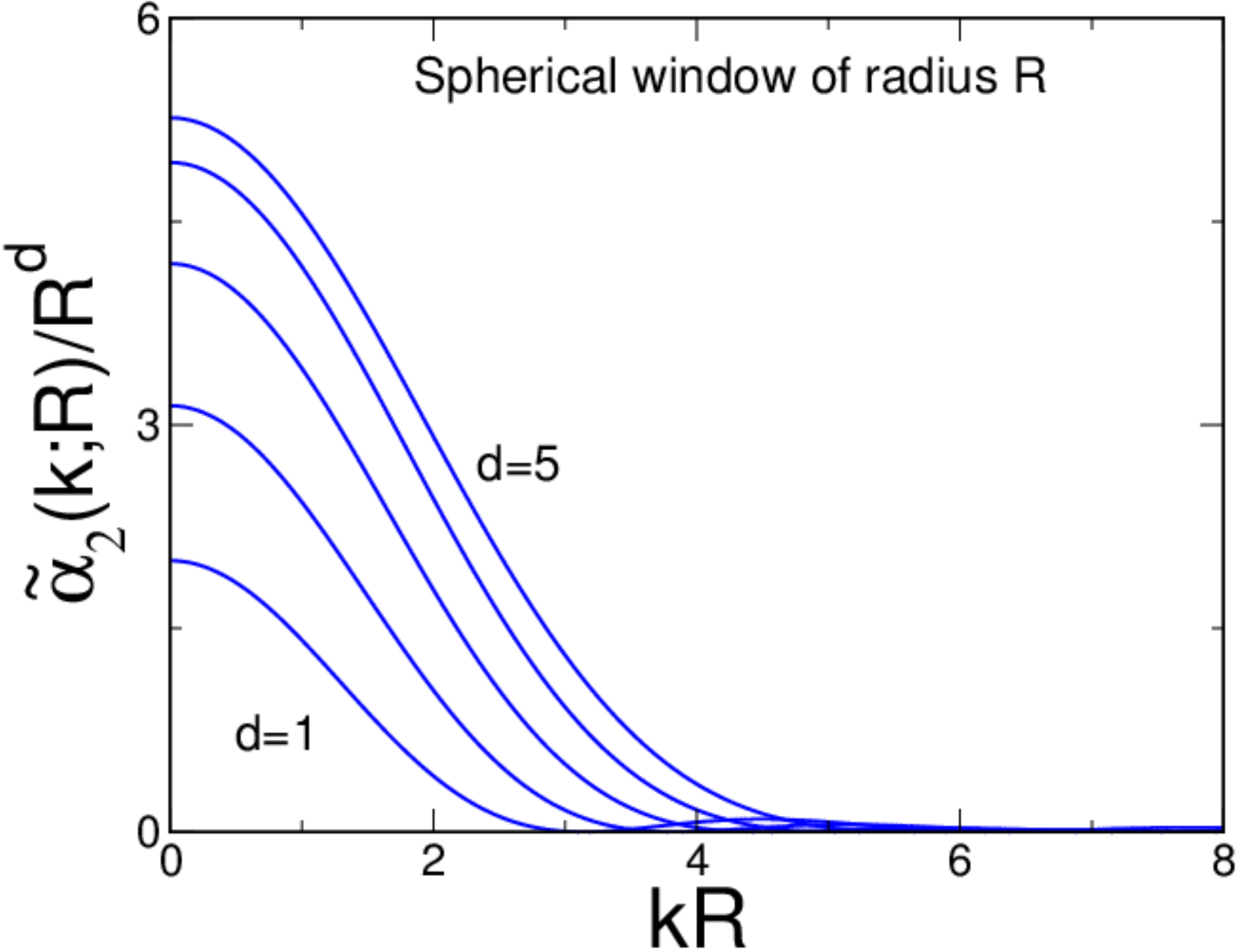} }
\caption{ Left panel: The scaled intersection
volume $\alpha_2(r;R)$ for spherical windows of radius $R$
as a function of $r$ for the first
five space dimensions,  the uppermost curve corresponding to $d=1$. Right panel:  The 
corresponding Fourier transform normalized by $R^d$, ${\tilde \alpha}_2(k;R)/R^d$, as a function of 
$k$,  the uppermost curve corresponding to $d=5$. }
\label{intersection}
\end{figure}

The scaled intersection volume $\alpha_2(r;R)$, which
has support in the interval $[0,2R]$ in the range $[0,1]$, has
the following integral representation in any dimension:
\begin{eqnarray}
\alpha_2(r;R) = c(d) \int_0^{\cos^{-1}[r/(2R)]} \sin^d(\theta) \, d\theta,
\label{alpha}
\end{eqnarray}
where $c(d)$ is the $d$-dimensional constant given by
\begin{eqnarray}
c(d)= \frac{2 \Gamma(1+d/2)}{\pi^{1/2} \Gamma[(d+1)/2]},
\label{C}
\end{eqnarray}
Torquato and Stillinger \cite{To06b} found the following  series representation
of $\alpha_2(r;R)$ for $r \le 2R$
and for any $d$:
\begin{eqnarray}
\alpha_2(r;R)=1- c(d) x+ 
c(d) \sum_{n=2}^{\infty}
(-1)^n \frac{(d-1)(d-3) \cdots (d-2n+3)}
{(2n-1)[2 \cdot 4 \cdot 6 \cdots (2n-2)]} x^{2n-1},
\label{series}
\end{eqnarray}
where $x=r/(2R)$. For even dimensions, relation (\ref{series}) is
an infinite series because it involves transcendental functions, but for odd dimensions, it truncates such
that $\alpha_2(r;R)$ is a univariate polynomial of degree $d$.
For example, for the first three dimensions, the scaled intersection
volumes are respectively given by
\begin{eqnarray} 
\alpha_2(r;R) &=&   \left[ 1- \frac{r}{2R} \right] \Theta(2R-r) \quad (d=1), \\
\alpha_2(r;R) &=&   \frac{2}{\pi} \left[ \cos^{-1}\left(\frac{r}{2R}\right) - \frac{r}{2R}
\left(1 - \frac{r^2}{4R^2}\right)^{1/2} \right]\Theta(2R-r) \quad (d=2),\\
\alpha_2(r;R) &=& \left[ 1 -\frac{3}{4}\frac{r}{R}+\frac{1}{16} \left(\frac{r}{R}\right)^3 \right] \Theta(2R-r) \quad (d=3).
\end{eqnarray}
The Fourier transform of $\alpha_2(r;R)$ is given by 
\begin{equation}
{\tilde \alpha}_2(k;R)= 2^d \pi^{d/2} \Gamma(1+d/2)\frac{[J_{d/2}(kR)]^2}{k^d}.
\label{alpha-k}
\end{equation}
The left panel of Fig. \ref{intersection} shows plots of $\alpha_2(r;R)$ as a function of $r$ for the first five space dimensions. For any space dimension, $\alpha_2(r;R)$
is a monotonically decreasing function of $r$. At a fixed
value of $r$ in the interval $(0,2R)$, $\alpha_2(r;R)$
is a monotonically decreasing function of the dimension $d$. The right panel of Fig. \ref{intersection} depicts
corresponding plots of the normalized Fourier transform of the scaled intersection volume, ${\tilde \alpha}_2(k;R)/R^d$, as a function of the wavenumber.
For any $d$, this is always a nonnegative decaying oscillating function of $k$.

For statistically homogeneous and isotropic point processes, the two-point statistical descriptors
in both direct and Fourier spaces, $h({\bf r})$ and $S({\bf k})$, become radial functions,
and hence for hyperspherical windows,
the number-variance formulas (\ref{N2}) and (\ref{var-par}) simplify as follows:
\begin{eqnarray}
\sigma^2_{_N}(R) &=&  \rho v_1(R)  \Bigg[ 1+\rho s_1(1)\int_0^\infty r^{d-1}  h(r)
\alpha_2(r;R) dr\Bigg] \label{N3} \\
&=&  \rho v_1(R)  \Bigg[\frac{s_1(1)}{(2\pi)^d} \int_0^\infty k^{d-1} S(k) 
{\tilde \alpha}_2(k;R) dk\Bigg],
\label{N4}
\end{eqnarray}
where $s_1(r)$ is the surface area of a $d$-dimensional sphere of radius $r$ [cf. (\ref{area-sph})],
and $\alpha_2(r;R)$ and  ${\tilde \alpha}_2(k;R)$ are given by (\ref{alpha}) and  (\ref{alpha-k}), respectively.
Importantly, these relations are also applicable for periodic or quasiperiodic structure
when the pair statistics are appropriately averaged over angles. After integration by parts, relation
(\ref{N4}) leads to an alternative representation of the number variance \cite{Og17}:
\begin{equation}
\sigma_{_N}^2(R)=
-\frac{\rho v_1(R)}{(2\pi)^d} \int_0^\infty  Z(k) 
\frac{\partial {\tilde \alpha}_2(k;R)}{\partial k} dk,
\label{eqn:local-1}
\end{equation}
where
\begin{equation}
Z(k) = s_1(1) \int_0^k  S(q)  q^{d-1}dq
\label{Zk}
\end{equation}
is the {\it integrated} or {\it cumulative} intensity function within a sphere of radius $k$
of the origin in reciprocal space. The function $Z(k)$ 
has advantages over the structure factor $S(k)$ in the characterization of 
the hyperuniformity of quasicrystals and other point processes with dense Bragg peaks \cite{Og17}.

\subsection{Number Variance for a Single Point Configuration}

Following Torquato and Stillinger \cite{To03a}, here we consider the local number variance 
of a single point configuration consisting of a large
number of points $N$  in a large region of $\mathbb{R}^d$ of volume $V$, 
which is necessarily a volume-average formulation. Fluctuations for a fixed window size arise because
we let the window uniformly sample the space.
Since one is always concerned with infinite-point configurations in $\mathbb{R}^d$,
we ultimately take the thermodynamic limit. However, even in this limit, the volume-average
formulation can lead to different results from those obtained in the ensemble-average formulation when the
system is {\it non-ergodic}. Thus, strictly speaking, the corresponding variances
should be notationally distinguished (e.g., Ref. \cite{To03a} designates the volume-average
variance as  ${\overline {N^2(R)}}-{\overline {N(R)}}^2$). For simplicity, we will henceforth 
invoke erogdicity, enabling us to equate volume-averages with ensemble averages,
and hence continue to designate $\sigma^2_{_N}(R)$ to be the number variance.
We consider a $d$-dimensional spherical window of radius
$R$, keeping in mind that the results below
apply as well (with obvious notational changes) to convex windows
of arbitrary shape. Invoking ergodicity, the local number variance for
a single infinite configuration is given by
\begin{eqnarray}
\sigma^2_{_N}(R) &=& \rho v_1(R) \Bigg[ 1 - \rho v_1(R) +
\frac{1}{N}\sum_{i \neq j} \alpha_2(r_{ij};R) \Bigg]\,.
\label{vol-var}
\end{eqnarray}
Observe that the last term within the brackets is a pairwise sum, the summand of which is exactly zero
for $r_{ij} > 2R$, even for infinitely large systems.

The scaled intersection volume $\alpha_2(r{ij};R)$ appearing in (\ref{vol-var})
is a nonnegative monotonically decreasing function of $r_{ij}$ with compact support (see Fig. \ref{intersection}) and 
hence can be viewed as a {\it repulsive} pair potential between a point $i$ and a point $j$.
It is noteworthy that finding the global minimum of $\sigma_{_N}^2(R)$ is equivalent
to determining the ground state for the pair potential function  $\alpha_2(r;R)$ \cite{To03a}.
More generally, one could devise an optimization
scheme in which a {\it targeted} value of the variance (rather than
an extremal value) is sought.

\subsection{Number Variance for a Single Periodic Point Configurations}

\begin{figure}[bthp]
\centering{\includegraphics[width=0.4\textwidth,clip=]{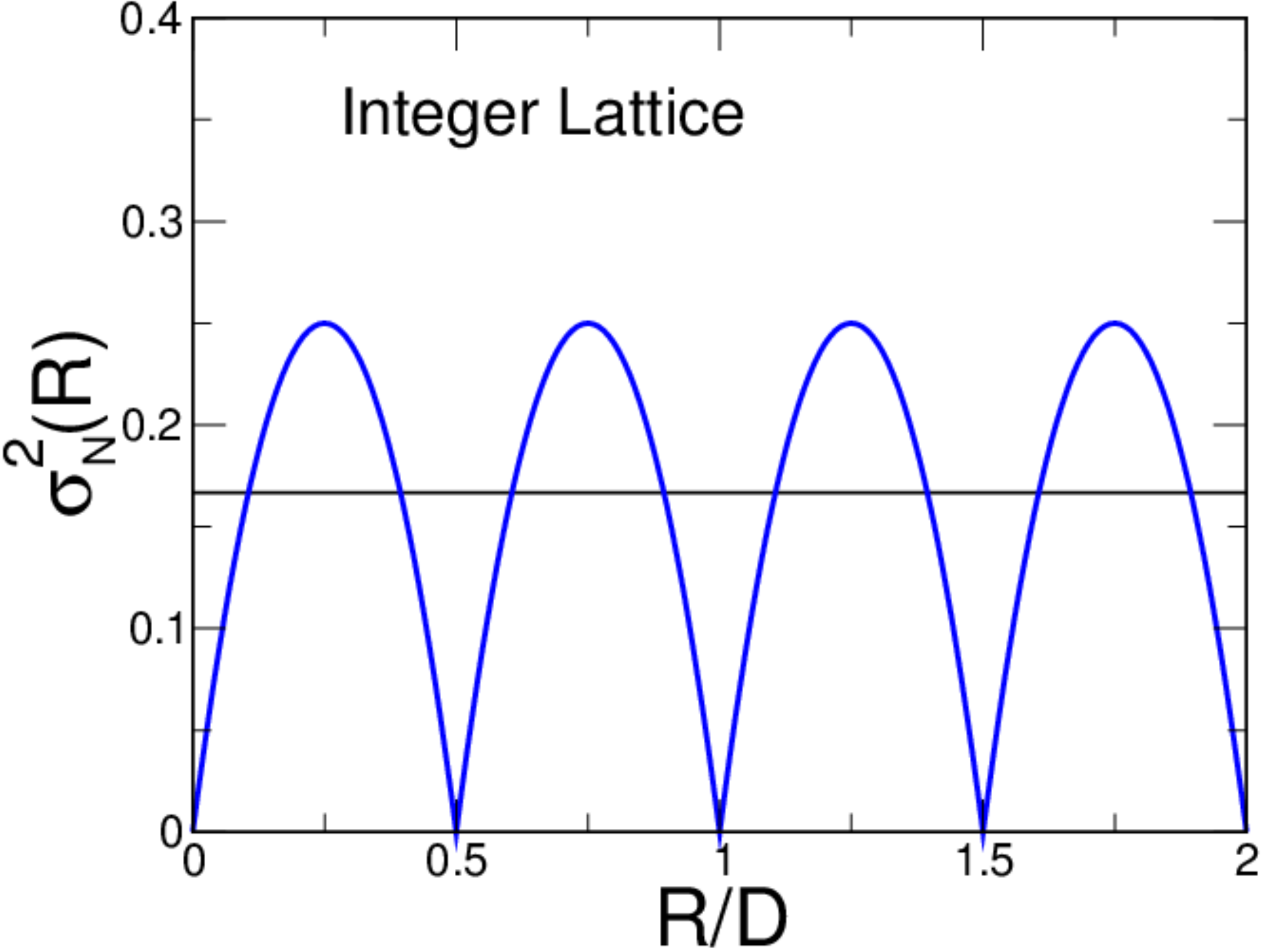}\hspace{0.3in}\includegraphics[width=0.4\textwidth,clip=]{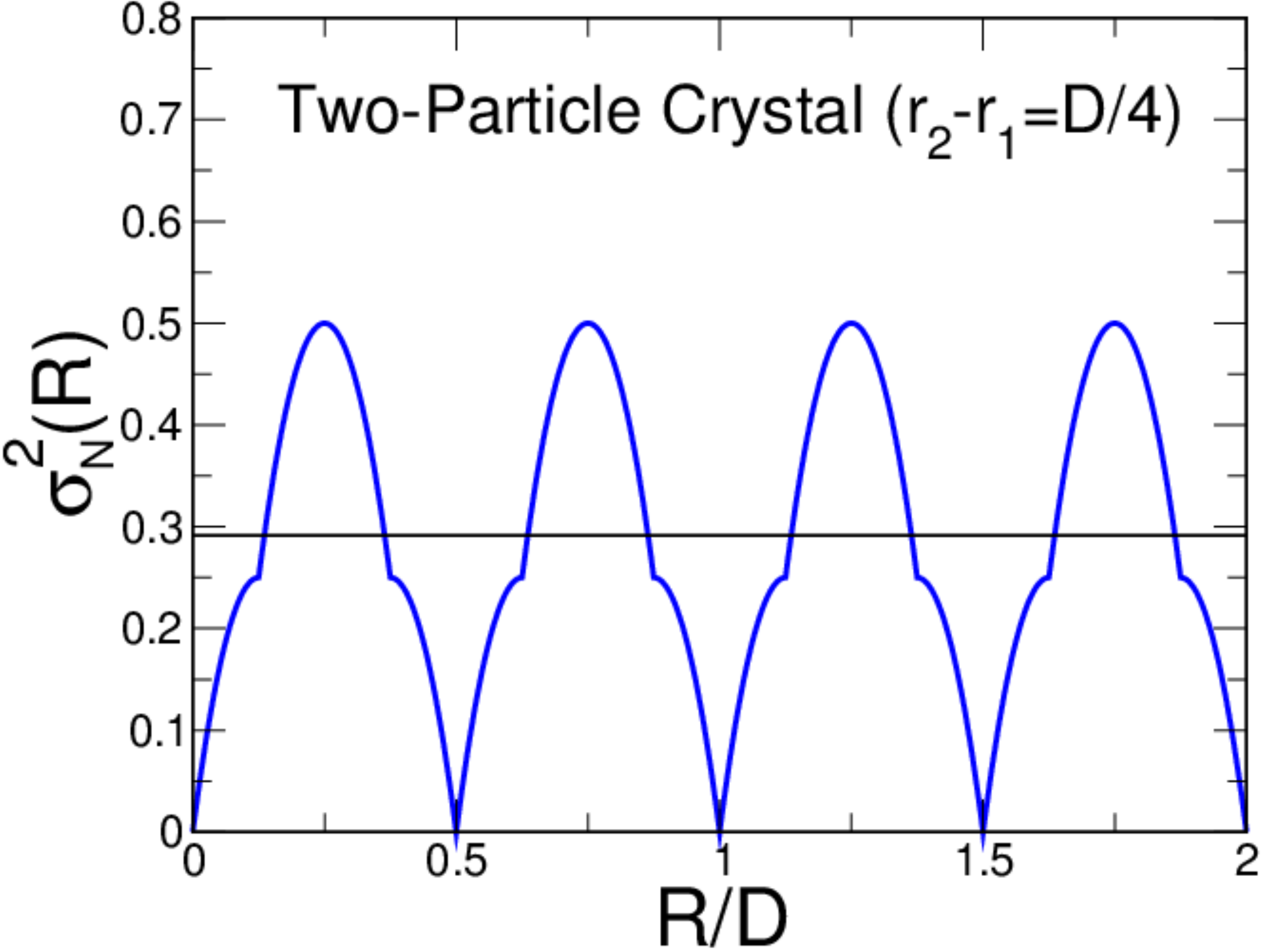}}
\caption{Left panel: The quadratic periodic variance function $\sigma^2_{_N}(R)$  for the
one-dimensional integer lattice given by (\ref{var-single}), where $D$ is the length of the fundamental cell. The horizontal
line is the average ${\overline \Lambda}=1/6$, as predicted by (\ref{var-2b}). Right panel:
The piecewise-quadratic periodic variance function $\sigma^2_{_N}(R)$ for the one-dimensional 
two-particle crystal  for the case $r_2-r_1=D/4$, as predicted  by (\ref{var-2}). The horizontal
line is the average ${\overline \Lambda}=7/24$, as predicted by (\ref{var-2b}).}
\label{1D}
\end{figure}

For periodic point configurations, Fourier analysis leads to an alternative
representation of the local number variance. We begin by  sketching the derivation
for the case of a (Bravais) lattice $\Lambda$ in $\mathbb{R}^d$ and 
consider a $d$-dimensional spherical window of radius $R$ \cite{Ke53b,To03a}.  
For this situation  the number of points $N({\bf x}_0; R)$
 within a window centered at ${\bf x}_0$ [cf. (\ref{N})] within a fundamental cell $F$ of volume $v_F$ is clearly a periodic function of the window position
${\bf x}_0$ and thus it can be expanded in a Fourier series as
\begin{equation}
N({\bf x}_0; R)=\rho v_1(R)+ \sum_{\bf q \neq 0} a({\bf q})
e^{\displaystyle i {\bf q} 
\cdot {\bf x}_0}
\label{N-per}
\end{equation}
where $\bf q$ represents the  reciprocal lattice vectors that specifies the dual 
lattice $\Lambda^*$, defined in Sec. \ref{lattices}, and
\begin{equation}
a({\bf q})= \frac{1}{v_F}\left(\frac{2\pi}{|{\bf q}|R}\right)^{d/2} R^d J_{d/2}(|{\bf q}|R)
\label{a}
\end{equation}
are the corresponding Fourier coefficients.
The number variance $\sigma_{_N}^2(R)$ is obtained by uniformly sampling
the centroid of the window over the fundamental cell and hence
is defined as follows: 
\begin{eqnarray}
\sigma_{_N}^2(R) &\equiv& \frac{1}{v_F} \int_{F} [N({\bf x}_0;R)-\rho v_1(R)]^2
d{\bf x}_0.
\end{eqnarray}
Application of Parseval's theorem for Fourier series to the right-hand
side of this equation \cite{To03a} yields
\begin{equation}
\sigma_{_N}^2(R) = \sum_{\bf q \neq 0} a^2({\bf q}) =\frac{R^d}{v_F^2} \sum_{\bf q \neq 0}
\left(\frac{2\pi}{|{\bf q}|}\right)^{d}  J^2_{d/2}(|{\bf q}|R).
\label{parseval}
\end{equation}
Since the summand in the infinite sum (\ref{parseval}) is nonnegative, any corresponding partial
sum yields a lower bound on $\sigma_{_N}^2(R)$. In one dimension, 
the integer lattice $\mathbb{Z}$  is the unique lattice, 
and from (\ref{parseval}) it follows that its number variance takes the simple form
\begin{equation}
\sigma^2_{_N}(R)=\frac{2}{\pi^2} \sum_{m=1}^\infty \frac{\sin^2(2\pi m R/D)}{m^2},
\label{var-single}
\end{equation}
where $D=v_F$ is the lattice spacing. This integer-lattice variance is a bounded periodic function with period $D/2$ and is 
equal to the quadratic function  $2x(1-2x)$ for $0\le x \le 1/2$, where $x=R/D$
 (see the left panel of Fig. \ref{1D}).

The generalization of the variance formula (\ref{parseval}) for a lattice to 
a periodic point configuration (crystal) in $\mathbb{R}^d$ in which there are $M$ points (where $M\ge 1$)
at position vectors ${\bf r}_1,{\bf r}_2,\ldots,{\bf r}_M$
within a fundamental cell $F$ of a lattice $\Lambda$  is straightforward \cite{To03a}. We simply
state this expression for such a $d$-dimensional crystal:
\begin{equation}
\sigma_{_N}^2(R) = \frac{R^d}{v_F^2} \sum_{\bf q \neq 0}
\left(\frac{2\pi}{|{\bf q}|}\right)^{d}  J^2_{d/2}(|{\bf q}|R) \left[M  +2\sum_{j<k}^{M} \cos[{\bf q}\cdot ({\bf r}_k -{\bf r}_j)]\right],
\label{par-periodic}
\end{equation}
where again  $\bf q$ denotes the reciprocal lattice vectors of the dual 
lattice $\Lambda^*$.
In the special case of one dimension with a fundamental cell of length $D$, this formula reduces to relation (87)
of Torquato and Stillinger \cite{To03a}:
\begin{equation}
\sigma^2_{_N}(R)= \frac{2}{\pi^2} \sum_{m=1}^\infty 
\frac{\sin^2(2\pi m R/D)}{m^2}\Big[M+2\sum_{j<k}^{M}\cos [2\pi m (r_k-r_j)/D]  \Big],
\label{var-2}
\end{equation}
where, without loss of generality, the particle positions are ordered such that $r_{j} \le r_{j+1}$ ($j=1,2,\ldots,M-1$).
In the corresponding expression given in Ref. \cite{To03a}  [Eq. (85)], one of the points is chosen
to be at the origin, resulting in two sums, both of which are missing a factor of 2.
The right panel of Fig. \ref{1D} shows the number variance $\sigma^2_{_N}(R)$ of a one-dimensional crystal with a two-particle basis  in which $M=2$ and $r_2-r_1=D/4$.
Again, we see that the variance is a bounded function, as it must be for any one-dimensional periodic point pattern (crystal).

\section{Local Volume-Fraction Fluctuations in Two-Phase Media}
\label{local-V}

As noted earlier, a heterogeneous medium is a spatial set that can be regarded to be more general than  a point configuration; see Sec. \ref{hetero}.
Of particular interest to us is local volume-fraction fluctuations in two-phase heterogeneous media \cite{Lu90a,Lu90b,Qu97b,Qu99}
and the generalization of hyperuniformity in this more general context \cite{Za09}.
An explicit formula for the former in terms of the relevant two-point descriptor was derived by Lu and Torquato \cite{Lu90b}.
Here we briefly review the derivation of this fluctuation formula.

\begin{figure}[!htp]
\centering
\includegraphics[width=0.40\textwidth]{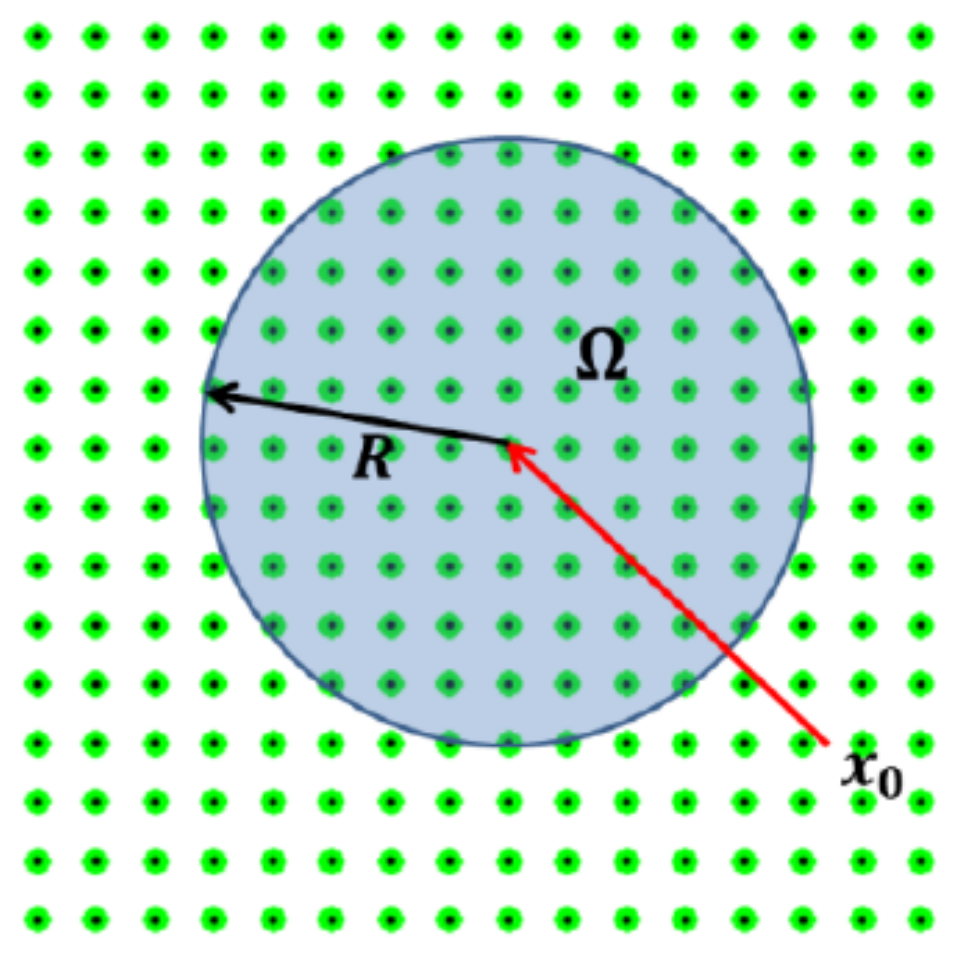}
\includegraphics[width=0.40\textwidth]{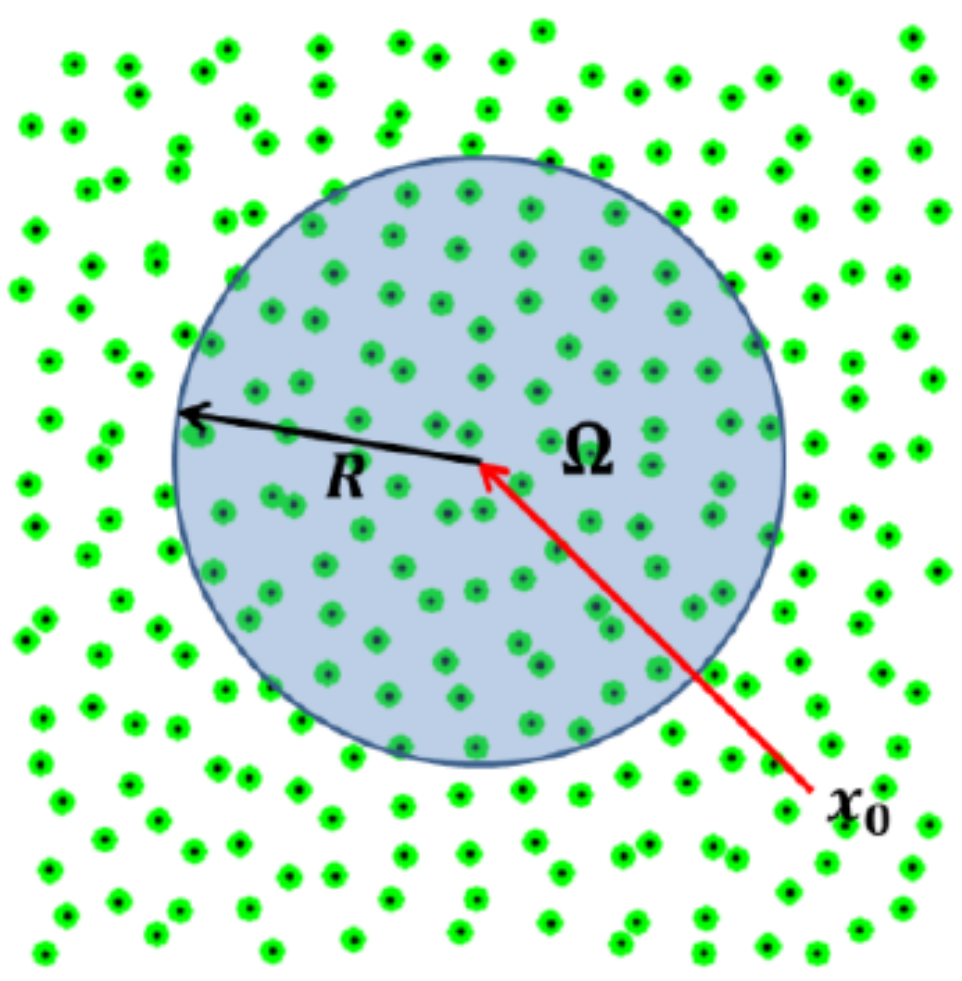}
\caption{Schematics indicating an observation window $\Omega$ for a periodic heterogeneous medium (left) and 
disordered hyperuniform heterogeneous medium (right)
obtained by decorating the point patterns shown in the middle and right panels of Fig. \ref{pattern}, respectively, with nonoverlapping circular disks.}\label{cartoon-2}
\end{figure}

While the global volume fraction, defined by (\ref{S1}), is a fixed constant
for a homogeneous   two-phase medium, the volume fraction within an observation 
window $\Omega \subset \mathcal{V}$  fluctuates as the window samples the space.
For simplicity, we consider a $d$-dimensional spherical window of radius $R$ in two-phase media;
see Fig. \ref{cartoon-2}.  The associated \emph{local volume fraction} $\tau_i(\mathbf{x}_0;R)$ of phase $i$ within 
a window of radius $R$ centered at position ${\bf x}_0$ is specified
explicitly by 
\begin{eqnarray}\label{one}
\tau_i(\mathbf{x}_0; R) = \frac{1}{v_1(R)}\int {\cal I}^{(i)}(\mathbf{x}) w(\mathbf{x}-\mathbf{x}_0; R) d\mathbf{x},
\end{eqnarray}
where $v_1(R)$ is the window volume, ${\cal I}^{(i)}(\mathbf{x})$ 
is the phase indicator function defined by (\ref{I}), and $w$ is the corresponding window indicator function
defined by (\ref{window}).
The variance $\sigma_{_V}(R)^2$ in the local volume fraction is defined by
\begin{eqnarray}\label{three}
\sigma^2_{_V}(R) \equiv \langle\tau_i^2(\mathbf{x}; R) \rangle - \phi_i^2,
\end{eqnarray} 
which trivially related to the so-called {\it coarseness} \cite{Lu90b}.

Following Ref. \cite{Lu90b}, it is straightforward to show that the volume-fraction variance 
can be expressed in terms of the autocovariance function $\chi_{_V}({\bf r})$, defined by (\ref{Auto}), as follows: 
\begin{eqnarray}
\sigma_{_V}^2(R) = \frac{1}{v_1(R)} \int_{\mathbb{R}^d} \chi_{_V}(\mathbf{r})\, \alpha_2(r; R) d\mathbf{r},
\label{phi-var-1}
\end{eqnarray}
where $\alpha_2(r;R)$ is the scaled intersection volume defined by (\ref{alpha}).
The alternative  Fourier representation that is dual to  (\ref{phi-var-1}) is trivially obtained by applying Parseval's theorem to (\ref{phi-var-1}):
\begin{eqnarray}
\sigma_{_V}^2(R) = \frac{1}{v_1(R)(2\pi)^d} \int_{\mathbb{R}^d} {\tilde \chi}_{_V}(\mathbf{k})\, {\tilde \alpha}_2(k; R) d\mathbf{k}.
\label{phi-var-2}
\end{eqnarray}
It should not go unnoticed that the local volume-fraction formulas (\ref{phi-var-1}) and (\ref{phi-var-2})
are  functionally very similar to the number-variance formulas
(\ref{N2}) and (\ref{var-par}) for point configurations, respectively; the integrands 
involve the scaled intersection volume function $\alpha_2$
multiplied by the relevant two-point statistical descriptors for each case.


\section{Mathematical Foundations of Hyperuniformity: Point Configurations}
\label{found-1}

This section is concerned with the mathematical foundations of hyperuniformity
for  point configurations. We discuss hyperuniformity conditions for general
point processes, asymptotic behaviors of the number variance and 
scaling behavior of pair statistics in both direct and Fourier spaces, the three possible
hyperuniformity classes, hyperuniformity
order metrics, and effect of window shape on the number variance.

\subsection{Vanishing of Normalized Infinite-Wavelength Number Fluctuations}
\label{vanishing}

Consider the Fourier-representation of the number variance (\ref{var-par}) for
a general point process in $\mathbb{R}^d$.
Let the window grow infinitely large in a self-similar 
(i.e., shape- and orientation-preserving) fashion. In this limit, which we will denote
simply by $v_1({\bf R}) \rightarrow \infty$, 
the function  ${\tilde \alpha}_2({\bf k};{\bf R})$ appearing in (\ref{var-par})
tends to $(2\pi)^d \delta({\bf k})$, where $\delta({\bf k})$ is a $d$-dimensional Dirac delta function.
For a large class of point configurations and window shapes
(spherical and nonspherical), application of
this limiting result and dividing the variance (\ref{var-par})  by $\langle N({\bf R}) \rangle=\rho v_1({\bf R})$ yields
\begin{equation}
\lim_{v_1({\bf R}) \rightarrow \infty}\frac{\displaystyle \sigma^2_{_N}({\bf R})}{\displaystyle \langle N({\bf R}) \rangle}= 
\lim_{|{\bf k}| \rightarrow 0} S({\bf k})= 1+\rho\int_{\mathbb{R}^d} h({\bf r}) d{\bf r}.
\label{var2}
\end{equation}
Observe that the general variance formula (\ref{var2}) and the hyperuniformity requirement  $\lim_{|{\bf k}| \rightarrow 0} S({\bf k}) = 0$, defined by (\ref{hyper}), then dictate that 
\begin{equation}
\lim_{v_1({\bf R}) \rightarrow \infty}\frac{\displaystyle \sigma^2_{_N}({\bf R})}{\displaystyle 
v_1({\bf R})}= 0,
\label{VAR}
\end{equation}
implying that the variance for a hyperuniform system grows more slowly than the 
window volume. This direct-space hyperuniformity condition provides the justification for the description of relation (\ref{cond1}). 
In other words, formula (\ref{VAR}) states that hyperuniform point processes have vanishing
{\it normalized} density fluctuations at large length scales,
where the normalization factor $v_1({\bf R})$ is proportional to the asymptotic number variance
of a typical disordered system (e.g., Poisson point process or standard liquid state).
In the instance of spherical windows of radius $R$, 
condition (\ref{VAR}) signifies that the number variance $\sigma^2_{_N}(R)$ must grow slower 
than the window volume, i.e., $R^d$, for large $R$. In the case of lattices and special window shapes and orientations, the direct-space
hyperuniformity condition (\ref{var2}) may not apply. Such subtle anomalous situations are
described in Sec. \ref{nonspherical}.

In the case of hyperspherical windows of radius $R$, 
{\it nonhyperuniform} systems possess a normalized variance $ \sigma^2_{_N}(R)/v_1(R)$   that does not vanish
in the limit $R \rightarrow \infty$ such that for large $R$ 
\begin{equation}
\frac{\displaystyle \sigma^2_{_N}(R)}{\displaystyle v_1(R)}= f(R),
\label{NON}
\end{equation}
where $f(R)$ is some positive system-dependent  function of $R$. For typical disordered nonhyperuniform
systems (e.g., Poisson point processes, liquids and glasses), $f(R)$ tends to a
system-dependent constant.  However, for systems in which $S({\bf k})$ becomes
unbounded in the limit $|{\bf k}| \to 0$, $f(R)$ tends to an increasing function of $R$,
and thus can be thought of as {\it anti-hyperuniform} with {\it hyperfluctuations},
since they are the antithesis of a hyperuniform system. 
Examples of such systems include those at thermal critical points \cite{Wi65,Ka66,Fi67,St71,Wi74,Bi92},
all of which have {\it fractal} structures.
Point processes derived from certain Poisson  
hyperplane tessellations for $d\ge 2$ represent a class of hyperfluctuating  systems in which 
it can be rigorously shown that the number variance grows like $R^{2d-1}$ \cite{He06,Hen13};
see also Ref. \cite{Ga08} for related heuristically derived hyperfluctuating scalings.

\subsection{Distinctions Between Equilibrium and Nonequilibrium Infinite-Wavelength Density Fluctuations}

Let us now recall the well-known {\it fluctuation-compressibility theorem} 
that links the isothermal compressibility of
equilibrium  single-component many-particle ensembles at number density $\rho$ and temperature $T$
 to infinite-wavelength density fluctuations \cite{Han13}.
In particular, for  ``open" systems in equilibrium, 
one has
\begin{equation}
\rho k_B T \kappa_T = \frac{ \langle N^2 \rangle_{*} -  \langle N \rangle_{*}^2}{ \langle N \rangle_{*}}
= S({\bf k=  0})= 1+\rho \int_{\mathbb{R}^d} h({\bf r}) d{\bf r},
\label{comp}
\end{equation}
where $k_B$ is Boltzmann's constant, $\langle \rangle_*$ denotes an average 
in the grand canonical ensemble, and $N$ is the fluctuating number of particles in the system
due to equality of the chemical potential with a particle reservoir. 
Observe that the form of the scaled variance (\ref{var2}) 
in the infinite-wavelength limit is identical to that for equilibrium systems
in the infinite-system limit, as given by relation (\ref{comp}). 
The important distinction is that result (\ref{var2}) is derived by considering
fluctuations that arise from moving  an  asymptotically large  window from point to point  
in infinite, ``closed''  generally {\it nonequilibrium} systems.
Sampling density fluctuations associated with such asymptotically large windows  can be viewed
as ensembles of very large ``open'' systems. On the other hand, the ``window"  in the grand canonical ensemble is the infinite system itself
and hence the density fluctuations described in (\ref{comp}) arise from  
density variations in the equilibrium ensemble members. We again stress that (\ref{var2}) is valid
whether the system is in equilibrium or not. When the system is out of equilibrium, then
of course, the global density fluctuations cannot be linked to the isothermal compressibility,
as in the fluctuation-compressibility relation (\ref{comp}).

Importantly,  conditions under which  equilibrium systems are hyperuniform  can be derived 
from this  fluctuation-compressibility theorem (\ref{comp}). For example, any ground state ($T=0)$ in which the isothermal
compressibility $\kappa_T$ is bounded and positive must be hyperuniform
because the structure  factor $S({\bf k =0})$ must be zero according to relation (\ref{comp}).
This includes crystal ground states as well as exotic disordered ground states, such as stealthy ones \cite{Uc04b,Ba08,To15} 
discussed in Sec. \ref{STEALTHY}. More generally, we infer from (\ref{comp}) that if the product
$T \kappa_T$ tends to a nonnegative constant $c$ in the limit $T \rightarrow 0$, 
then the ground-state of this system in this zero-temperature limit must be nonhyperuniform if $c>0$
or  hyperuniform if $c=0$. By the same token, this means that increasing the temperature
by an arbitrarily small positive amount when a system is initially at 
hyperuniform  ground state will destroy perfect hyperuniformity,
since   $S({\bf k =0})$ must deviate from zero by some small amount determined
by the temperature dependence of the product $\kappa_T T$ for small $T$.  This indirectly implies that phonons 
or vibrational modes for sufficiently small $T$ generally destroy the hyperuniformity
of ground states \cite{To15}, as detailed  in Sec. \ref{excited}. Moreover, in order to have a hyperuniform
system that is in equilibrium at any positive $T$, the isothermal compressibility must be zero, i.e.,
the  system must be thermodynamically incompressible \cite{Za11b}; see Refs. \cite{To03a}, \cite{To08b} and \cite{Zh16a}
as well as Secs. \ref{Matrices} and \ref{ocp} for some examples. 

Since supercooled liquids and glasses are systems out of equilibrium, 
the fluctuation-compressibility relation (\ref{comp})  is generally not satisfied.
This motivated Hopkins, Stillinger and Torquato \cite{Ho12b} to introduce and apply
the {\it nonequilibrium index} $X$ defined to be
\begin{equation}
X \equiv \frac{ S({\bf k=0}) }{\rho k_B T \kappa_T} -1.
\label{XX}
\end{equation}
Nonzero values of this index quantifies the degree to which a system under study deviates from thermal equilibrium ($X = 0$).
This index has been applied to study nonequilibrium hard-sphere packings \cite{Ho12b}, well-known atomic models
of glasses \cite{Ma13a}, and ``perfect" glasses \cite{Zh16a}.

\subsection{Asymptotics Via  Ensemble-Average Formulation and Three Hyperuniformity Classes}
\label{classes}

The local number variance $\sigma^2_{_N}(R)$  is generally a function that can be decomposed 
into a global part that grows with the window radius $R$ and a local part that oscillates on 
small length scale  (e.g., mean nearest-neighbor distance) about the global contribution; see Fig. \ref{scaled-var-3D}.
Using the ensemble-average formulation, we will analyze the large-$R$ asymptotic behavior of the variance $\sigma^2_{_N}(R)$
for  a $d$-dimensional spherical window of radius $R$  associated with general hyperuniform systems, which includes 
perfect periodic, perfect quasiperiodic and special disordered point configurations.  Nonspherical windows
will be briefly considered in Sec. \ref{nonspherical}).

Using either of the ensemble-average representations  (\ref{N3}) or (\ref{N4}) of $\sigma^2_{_N}(R)$, 
it is straightforward to show that the global asymptotic growth rate
is determined  by the large-distance behavior of the total
correlation function $h({\bf r})$ [cf. (\ref{total})] or, equivalently, by the small-wavenumber
behavior of the structure factor $S({\bf k})$, respectively. We show that the asymptotic growth behaviors of $\sigma^2_{_N}(R)$  fall into three distinct 
classes: class I in which the growth is proportional to the window surface area $R^{d-1}$, class II in which the growth 
is proportional to $\ln(R)\,R^{d-1}$, and class III
in which the growth is proportional to $R^{d-\alpha}$, where $\alpha \in (0,1)$ is an exponent.

In what follows, we denote by $D$ a characteristic ``microscopic" length scale of
the system. It is useful to introduce the dimensionless density
$\phi$ defined by
\begin{equation}
\phi=\rho v_1(D/2)= \rho \frac{\pi^{d/2}}{2^d\Gamma(1+d/2)}D^d.
\label{phi}
\end{equation}
In some cases, it is convenient to
take $D$ to be the mean nearest-neighbor distance between
the points.  If the point configuration is derived from a
packing of identical nonoverlapping spheres of diameter $D$ or if the minimum pair distance
in a point process is $D$, $\phi$
is simply the fraction of space covered by  nonoverlapping spheres of diameter $D$.
In other instances, it may useful to choose $D=\rho^{-1/d}$, which 
is equivalent to setting the number density to be unity
(i.e., $\rho=1$). This choice will be convenient when comparing the 
number variances for different systems.

\subsubsection{Asymptotics of the Number Variance}

Direct substitution of  the expansion (\ref{series}) of $\alpha_2(r;R)$
into the ensemble-average expression (\ref{N3}), and assuming that the
resulting integrals separately converge, yields the following asymptotic formula for the variance:
\begin{equation}
\sigma^2_{_N}(R)=
2^d\phi\Bigg[ A_{_N}(R)\left(\frac{R}{D}\right)^d+B_{_N}(R)\left(\frac{R}{D}\right)^{d-1}+
o\left(\frac{R}{D}\right)^{d-1}\Bigg]  \qquad (R \rightarrow \infty),
\label{var1}
\end{equation}
where $o(x)$ signifies all terms of order less  than $x$, $D$ is a characteristic microscopic length
mentioned above, and $A_{_N}(R)$ and $B_{_N}(R)$ are, respectively, $d$-dependent asymptotic coefficients
that multiply terms  proportional
to the window  volume ($R^d$) and window surface area ($R^{d-1}$). The ``volume" and ``surface-area" coefficients   are explicitly given by the
following the volume integrals involving the total correlation function $h({\bf r})$, respectively: 
\begin{equation}
A_{_N}(R) =1+\rho\int_{|{\bf r}| \le 2R} h({\bf r}) d{\bf r}=1+\frac{\phi}{v_1(D/2)} 
\int_{|{\bf r}| \le 2R} h({\bf r}) d{\bf r}
\label{A1} 
\end{equation}
and
\begin{equation}
B_{_N}(R)=-\frac{\phi \, c(d)}{2\,D\,v_1(D/2)} \int_{|{\bf r}| \le 2R} h({\bf r})|{\bf r}| d{\bf r},
\label{B1}
\end{equation}
where $c(d)$ is a $d$-dependent constant defined by (\ref{C}).
The restriction on the integration domains to be a spherical region
of radius $2R$ is imposed by the finite support of the scaled intersection volume 
(\ref{alpha})  and hence results in volume and
surface-area coefficients that generally depend on $R$. The asymptotic formula (\ref{var1}) is more
general than the one presented in Ref. \cite{To03a} in that it allows
for coefficients $A_{_N}$ and $B_{_N}$ that depend on $R$ without any assumptions
about their convergence properties.

Observe that the volume coefficient $A_{N}(R)$ in the limit $R \rightarrow \infty$ is equal
to the nonnegative structure factor $S({\bf k})$ [cf. (\ref{factor})] in the 
zero-wavenumber limit i.e.,
\begin{equation}
{\overline A}_{_N} \equiv \lim_{R \rightarrow \infty}A_{N}(R)=\lim_{|{\bf k}| \rightarrow 0}S({\bf k}) \ge 0.
\label{A3}
\end{equation}
We call ${\overline A}_{_N}$ the {\it global} volume coefficient. It is well known that disordered point configurations associated with equilibrium
molecular systems with a wide class of interaction
potentials (e.g., hard-sphere, square-well, and
Lennard-Jones interactions) yield positive values
of the coefficient ${\overline A}_{_N}$ in  gaseous, liquid, and many solid states.
Indeed, ${\overline A}_{_N}$ will be positive for any equilibrium system possessing a positive compressibility
at positive temperatures; see  fluctuation-compressibility theorem (\ref{comp}).
The coefficient ${\overline A}_{_N}$ is positive for a wide
class of nonequilibrium disordered point configurations, including
the random sequential addition (RSA) packing process \cite{To06d,Zh13b}. 
Anti-hyperuniform systems (e.g., standard thermal critical-point states) are at the extreme
end of nonhyperuniformity in which ${\overline A}_{_N}$ is infinitely large.
To summarize, there is an enormous class of  disordered point configurations
in which ${\overline A}_{_N}$  is strictly positive and hence nonhyperuniform; indeed, this is typically true.

According to relations (\ref{hyper}) and (\ref{sum-1}),
a hyperuniform system is one in which the volume coefficient vanishes
in the limit $R \rightarrow \infty$, i.e., 
\begin{equation}
{\overline A}_{_N} = \lim_{|{\bf k}| \rightarrow 0} S({\bf k})=0.
\end{equation}
A {\it class I hyperuniform system} is one in which the 
surface-area coefficient $B_{_N}(R)$ converges to a constant in the limit $R \rightarrow \infty$
and hence, according to relation  (\ref{var1}), $\sigma_{_N} ^2(R)$ grows like the window surface area, as specified by
\begin{equation}
\sigma^2_{_N}(R) \sim 
2^d\phi {\overline B}_{_N}\left(\frac{R}{D}\right)^{d-1} \qquad (R \rightarrow \infty),
\label{var3}
\end{equation}
where
\begin{equation}
{\overline B}_{_N}= \lim_{R \rightarrow \infty} B_{_N}(R)=-\frac{\phi d \Gamma(d/2)}{2Dv_1(D/2) \Gamma(\frac{d+1}{2})
\Gamma(\frac{1}{2})} \int_{\mathbb{R}^d} h({\bf r})|{\bf r}| d{\bf r}
\label{B2}
\end{equation}
is the {\it global} surface-area coefficient, which is positive for a hyperuniform system \cite{To03a}.
Class I systems include disordered point configurations in which 
$h({\bf r})$ decays to zero sufficiently fast for large $|{\bf r}|$; for example,  exponentially
fast or faster (as in $g_2$-invariant systems \cite{To03a} discussed in Sec. \ref{invariant}, one-component
plasmas described in Sec. \ref{ocp} and Weyl-Heisenberg ensembles  mentioned
in Sec. \ref{WH} \cite{Ab17}) or faster than the power-law $1/|{\bf r}|^{d+1}$, as in
some stealthy hyperuniform disordered systems \cite{To15} discussed in Sec. \ref{STEALTHY}).
Moreover, ${\overline B}_{_N}$ converges for statistically homogeneous point configurations that are derived from a
perfect crystal (uniform translations of the entire crystal over the fundamental cell) 
and a large class of perfect quasicrystals, all of which are characterized by Bragg peaks 
in reciprocal space, and hence belong to class I hyperuniform systems. In Sec. \ref{defects}, we
will see that a wide class of ``randomly" perturbed crystal structures \cite{Ga04b,Ga04,Ga08,Ki18a} also belong to
class I.  It is notable that the global surface-area coefficient ${\overline B}_{_N}$ provides a measure of the degree
to which class I hyperuniform systems suppresses large-scale density fluctuations \cite{To03a,Za09}, as we
detail further below.

For one-dimensional class I hyperuniform systems, the number variance is exactly
(not asymptotically) given by
\begin{equation}
\sigma^2_{_N}(R)= 2\phi B_N(R),
\end{equation}
where $B_{_N}(R)$ is given by (\ref{B1}) with $d=1$ \cite{To03a}, implying
that the fluctuations are bounded, i.e., do not grow with $R$.
Aizenman, Goldstein and Lebowitz \cite{Ai01} proved general conditions
under which a one-dimensional system possesses a bounded
variance.

If the point  process in $\mathbb{R}^d$ (hyperuniform or not) can be characterized by radial
total correlation function, i.e., $h({\bf r})=h(r)$, where $r \equiv |{\bf r}|$,
the volume coefficient (\ref{A1}) and surface-area coefficient  (\ref{B1}) 
are expressible in terms of certain moments of the radial
total correlation function, namely,
\begin{eqnarray}
{\overline A}_{_N} &=&1+ d 2^d  \phi \langle x^{d-1} \rangle,
\label{A2} \\
{\overline B}_{_N}&=&-\frac{ d^2 2^{d-1}\Gamma(d/2)}{ \Gamma(\frac{d+1}{2})
\Gamma(\frac{1}{2})} \phi \langle x^{d} \rangle,
\label{B3}
\end{eqnarray}
where
\begin{equation}
\langle x^n \rangle = \int_0^\infty x^n h(Dx) dx
\end{equation}
is the $n$th moment of $h(x)$ and $x=r/D$ is a dimensionless distance. Clearly,
the total correlation function is a radial function
for statistically homogeneous and isotropic point processes as well as
for periodic point configurations in which one averages displacements between
pairs of points over angles; see relation (\ref{g2-period}).
According to the previous analysis, we see that if ${\overline A}_{_N}=0$, the condition for the variance to grow as the surface
area implies that the  the $d$th moment of $h$ must be strictly negative \cite{To03a}, i.e., 
\begin{equation}
\langle x^d \rangle < 0.
\end{equation}

By contrast, point processes
in which the global surface-area coefficient
vanishes (${\overline B}_{_N}=0$) is referred to as  a {\it hyposurficial} system \cite{To03a}. 
Thus, hyposurfical systems obey the following sum rule:
\begin{equation}
\int_{0}^\infty r^d h(r)\; dr =0,
\label{hypo}
\end{equation}
where $h(r)$ represents the radial total correlation function for a statistically
homogeneous and isotropic point process or angular-averaged total correlation function
in the case of a statistically anisotropic system. We see that for the integral (\ref{hypo}) to converge,
$h(r)$ must decay to zero faster than $1/r^{d+1}$ for large $r$ in space dimension $d$.
Since the variance is strictly positive \cite{To03a} and   cannot grow more slowly than the surface area of 
a spherical window \cite{Beck87}, it follows that any such  system
cannot simultaneously be hyperuniform and hyposurficial, i.e.,  the
volume coefficient ${\overline A}_{_N}$ [cf. (\ref{A3})] and 
surface-area coefficient ${\overline B}_{_N}$ [cf. (\ref{B2})] cannot
both be zero.  A homogeneous Poisson point configuration is a simple example of  a hyposurficial system. 
A less trivial example of hyposurficial system is a certain hard-core point process in $\mathbb{R}^d$; see Ref
\cite{To03a} for details. It has recently come to light that hyposurficiality
arises in non-equilibrium phase transitions involving amorphous ices \cite{Ma17}.

Importantly, two other hyperuniform classes are possible if the
coefficient $B_{_N}(R)$ converges to a function of $R$ (not a constant)
in the limit $R \rightarrow \infty$. For example, if the total correlation function
is controlled by the following radial power-law decay:
\begin{equation}
h({\bf r}) \sim \frac{1}{|{\bf r}|^{d+ 1}}  \qquad (|{\bf r}| \rightarrow \infty),
\label{h-decay}
\end{equation}
a similar asymptotic analysis of (\ref{B2}) leads to  a number variance that asymptotically grows
like $\sigma^2_{_N}(R) \sim R^{d-1} \,\ln(R)$, since $B_{_N}(R) \sim \ln(R)$, which we refer to as 
class II hyperuniform systems. Examples include
some quasicrystals \cite{Og17}, classical disordered ground states 
\cite{Uc06b,Zh16a}, zeros of the Riemann zeta function \cite{Mon73,To08b}, 
eigenvalues of random matrices \cite{Me91}, fermionic point processes \cite{To08b}, 
superfluid helium \cite{Fe56,Re67}, maximally random jammed packings \cite{Do05d,Sk06,Za11a,Za11c,Za11d,Ji11c,Ho12b,Ch14a,At16a,At16b}, perturbed lattices
\cite{Ga04b}, 
density fluctuations in early Universe \cite{Pe93,Ga03,Ga05}, and perfect glasses \cite{Zh16a}.
On the other hand, 
\begin{equation}
h({\bf r}) \sim \frac{1}{|{\bf r}|^{d+ \alpha}}  \qquad (|{\bf r}| \rightarrow \infty),
\label{h-decay-2}
\end{equation}
yields a number variance that scales like $R^{d-\alpha}$, where  $0 < \alpha <1$, since $B_{_N}(R) \sim R^{1-\alpha}$. 
We refer to such hyperuniform systems as class III structures. Examples within this class
include classical disordered ground states \cite{Za11b},
random organization models \cite{He15,Tj15}, perfect glasses \cite{Zh16a}, and perturbed lattices
\cite{Ki18a}.

\subsubsection{Asymptotics from Power-Law Structure Factors}
\label{power}

Let us consider hyperuniform systems that are
characterized by a structure factor with a radial power-law form in the vicinity of the origin \cite{To03a,Do05d,Za09}:
\begin{equation}
S({\bf k}) \sim |{\bf k}|^{\alpha} \qquad (|{\bf k}|\rightarrow 0)
\label{S-asy}
\end{equation}
with scaling exponent $\alpha >0$. (Note that it is not possible to construct a hyperuniform system for which $\alpha \leq 0$, 
since the number variance would then grow at least as fast as the volume of the observation window.)
Analysis of  the Fourier-representation of the number variance 
(\ref{var-par}) reveals that  large-$R$ asymptotic behavior of $\sigma_{_N}^2(R)$
is controlled by the  power-law form (\ref{S-asy}), and  depends on the value of the exponent $\alpha$ as follows:
\begin{eqnarray}  
\sigma^2_{_N}(R) \sim \left\{
\begin{array}{lr}
R^{d-1}, \quad \alpha >1 \quad \mbox{(CLASS I)}\\
R^{d-1} \ln R, \quad \alpha = 1 \quad \mbox{(CLASS II)}\\
R^{d-\alpha}, \quad 0 < \alpha < 1 \quad \mbox{(CLASS III)}
\end{array}\right.
\label{sigma-N-asy}
\end{eqnarray}
We see that the scaling regimes in which $\alpha > 1$, $\alpha=1$ and $0 < \alpha < 1$
correspond to class I, II and III hyperuniform systems, respectively. 
Table \ref{CLASS}  provides a summary of model systems
that fall into these three hyperuniformity classes. Almost all of these examples are
expounded upon in Secs. \ref{invariant}, \ref{STEALTHY},  \ref{Matrices},  \ref{Det} and  \ref{non-eq}.

\begin{table}
\caption{Hyperuniform point configurations can exist as both as equilibrium and nonequilibrium phases,
and come in both quantum-mechanical and classical varieties. This table provides a summary of model systems
that fall into the three hyperuniformity classes based on the use of $d$-dimensional spherical
windows of radius $R$ to sample for the number variance $\sigma_{_N}^2(R)$.}
\centering
\begin{tabular}{|c|m{10cm}|} \hline \hline

CLASS  & \hspace{1.5in} MODELS \\ \hline

I: $\sigma_{_N}^2(R) \sim R^{d-1}$ & All crystals \cite{To03a}, many quasicrystals \cite{Za09,Og17}, 
stealthy and other hyperuniform disordered ground states \cite{Uc04b,Uc06b,Ba08,To15,Zh16a}, perturbed lattices
\cite{Ga03,Ga04b,Ga04,Ga08,Ki18a}, $g_2$-invariant disordered point processes \cite{To03a},
one-component plasmas \cite{Ja81,Le00}, hard-sphere plasmas \cite{Lo17,Lo18a}, 
random organization models \cite{He17b}, perfect glasses \cite{Zh16a}, and Weyl-Heisenberg ensembles \cite{Ab17}.\\\hline

II: $\sigma_{_N}^2(R) \sim R^{d-1} \ln(R)$ & Some quasicrystals \cite{Og17}, classical disordered ground states 
\cite{Uc06b,Zh16a}, zeros of the Riemann zeta function \cite{Mon73,To08b}, 
eigenvalues of random matrices \cite{Me91}, fermionic point processes \cite{To08b}, 
superfluid helium \cite{Fe56,Re67}, maximally random jammed packings \cite{Do05d,Za11a,Ji11c,Ch14a,At16a}, perturbed lattices
\cite{Ga04b}.
density fluctuations in early Universe \cite{Pe93,Ga03,Ga05}, and  perfect glasses \cite{Zh16a}.
 \\ \hline

III: $\sigma_{_N}^2(R) \sim R^{d-\alpha}$ ($0 < \alpha <1$)& Classical disordered ground states \cite{Za11b},
random organization models \cite{He15,Tj15}, perfect glasses \cite{Zh16a}, and perturbed lattices
\cite{Ki18a}.
 \\  \hline \hline

\end{tabular}
\label{CLASS}
\end{table}

We stress that the hyperuniform power-law form (\ref{S-asy}) of the structure factor may or may not imply a power-law
decay in the corresponding  total  correlation function $h({\bf r})$ [cf. (\ref{total})]
for large $|{\bf r}|$. The following are three possible outcomes when $\alpha >0$:\smallskip

\noindent{A. Whenever $S({\bf k})$ is analytic at the origin (i.e., admits a Taylor series and so only involves
even integer powers of $|{\bf k}|$ through all orders), the exponent  $\alpha$ in (\ref{S-asy}) must be a positive
even integer and
$h({\bf r})$ must decay to zero exponentially fast (or faster) as $|{\bf r}| \rightarrow \infty$, prohibiting
a power-law decay of $h({\bf r})$. Such hyperuniform systems  are of  class I. }\vspace{0.1in}

\noindent{B. Whenever $S({\bf k})$ is nonanalytic at the origin (including cases in which $\alpha$ is an odd integer)
and sufficiently smooth away from the
origin such that it is $\lfloor \alpha \rfloor+ \lfloor d/2 \rfloor$ times differentiable, where 
$\lfloor x \rfloor$ is the floor function (largest integer less than or equal to $x$),
the asymptotic behavior of the total correlation function  is given by the following inverse power-law form:}
\begin{equation}
\rho h({\bf r}) \sim -\frac{C_1(\alpha,d)}{|{\bf r}|^{d+\alpha}} \qquad (|{\bf r}|\rightarrow \infty),
\label{h-asy}
\end{equation}
which is sufficiently short-ranged such that its volume integral over all space exists. Here
\begin{equation}
C_1(\alpha,d)=\frac{2^{\alpha} \Gamma(1+\alpha/2)\;\Gamma((d+\alpha)/2)\sin(\pi \alpha/2)}{\pi^{1+d/2}}
\end{equation}
is a  constant that depends on the exponent $\alpha$ and $d$, and it is  assumed
that the coefficient multipyling $|{\bf k}|^{\alpha}$ in (\ref{S-asy}) is unity. Note that the presence of the term $\sin(\pi \alpha/2)$
in (\ref{h-asy}) requires the exponent $\alpha$ to lie in one of the intervals
$(0, 2)$, $(4, 6)$, $(8, 10)$, and so forth, in order for the total correlation function
to have the  asymptotic power-law form (\ref{h-asy}). Otherwise, $h({\bf r})$ must decay to zero 
faster than a power law for sufficiently large $|{\bf r}|$.
Therefore, positive even-integer  values of $\alpha$ are types of ``limiting values'' that
overcome the otherwise dominant $|{\bf r}|^{-(d+\alpha)}$ asymptotic scaling of $h({\bf r})$.

This latter case leads to the following natural question: What is the corresponding form
of the small-wavenumber structure factor when $h({\bf r})$ is controlled
by the power law
\begin{equation}
h({\bf r}) \sim -\frac{1}{|{\bf r}|^{d+\alpha}} \qquad (|{\bf r}|\rightarrow \infty),
\end{equation}
where $\alpha$ is a positive even integer? In such situations, it is straightforward to show that
the corresponding structure factor is no longer a pure power law 
but contains a multiplicative logarithmic factor, i.e.,
\begin{equation}
S({\bf k}) \sim -|{\bf k}|^{\alpha}\, \ln(|{\bf k}|)\qquad (|{\bf k}|\rightarrow 0).
\end{equation}
Nonetheless, it is also easy to demonstrate that this still leads to a number variance 
that grows like the window surface area $R^{d-1}$ and hence lies
within class I hyperuniform systems.
\vspace{0.1in}

\noindent{C. Whenever $S({\bf k})$ is nonanalytic at the origin such that it is less than  $\lfloor \alpha \rfloor+ \lfloor d/2 \rfloor$  times differentiable, the scaling form (\ref{S-asy}) corresponds to a total correlation function whose large-$r$ behavior is still controlled by the inverse power-law form (\ref{h-asy}) but modulated by a sinusoidal function of the radial distance
$r$, i.e.,
\begin{equation}
\rho h({\bf r}) \sim -\frac{C_2(\alpha,d)s(r;d)}{|{\bf r}|^{d+\alpha}} \qquad (|{\bf r}|\rightarrow \infty),
\label{h-asy-2}
\end{equation}
where $C_2(\alpha,d)$ and $s(r;d)$ are a positive constant and $d$-dependent sinusoidal function, respectively, whose specific forms depend on the differentiability of the structure factor.

In cases B and C such that $\alpha$ lies in the interval $(0,d)$, we will see in  Sec. \ref{oz} that the direct correlation function $c({\bf r})$, defined via the Ornstein-Zernike relation, becomes long-ranged in the sense that its volume integral is unbounded. Such behavior is in diametric contrast to standard thermal critical points in which $h({\bf r})$ is long-ranged \cite{Wi65,Ka66,Fi67,St71,Wi74,Bi92}, and hence 
a system at a hyperuniform state has been called an ``inverted" critical point \cite{To03a}.

For some infinite point configurations, e.g., lattices, the associated variance oscillates 
around some global average behavior \cite{To03a} (see Fig. \ref{scaled-var-3D}), which  may 
make it difficult to determine  smoothly its asymptotic behavior.
In such cases, it is advantageous to use the {\it cumulative moving average} of the variance ${\overline{\sigma_N ^2}}(R)$ \cite{Ki17}, defined as
\begin{equation}
{\overline{\sigma_{_N}^2}}(R) \equiv \frac{1}{R} \int_0 ^R {\sigma_{_N}^2}(x)\; d{x},
\label{moving}
\end{equation}
to ascertain the large-$R$ asymptotic behavior of $\sigma_{_N}^2(R)$.





\subsection{Asymptotics for a Single Point Configuration}

Consider a single infinite hyperuniform point configuration.
For large $R$, the large-scale variations in $R$ will grow as $R^{d-1}$, and
so we have from (\ref{vol-var}) that \cite{To03a}
\begin{equation}
\sigma^2_{_N}(R) = \Lambda(R) \left(\frac{R}{D}\right)^{d-1}\,, 
\label{sig-asymp}
\end{equation}
where 
\begin{equation}
\Lambda(R)= 2^d \phi \left(\frac{R}{D}\right) \Bigg[ 1-2^d \phi \left(\frac{R}{D}\right)^d + 
\frac{1}{N}\sum_{i \neq j}^N \alpha_2(r_{ij};R) \Bigg]
\label{Lambda(R)}
\end{equation}
is the asymptotic ``surface-area''  function that contains the small-scale variations in $R$.
In the case of class I hyperuniform point configurations, it is useful to average the function $\Lambda(R)$ over $R$, yielding
the constant
\begin{equation}
{\overline \Lambda}=\lim_{L \rightarrow \infty}\frac{1}{L} \int_0^L \Lambda(R) dR.
\label{S-constant}
\end{equation}
This constant is trivially related to the surface-area coefficient $B_N$, 
defined by (\ref{B1}), as follows:
\begin{equation}
{\overline \Lambda} = 2^d \phi B_N=\frac{-~2^{d-1}\phi^2 d \Gamma(d/2)}{Dv_1(D/2)
\Gamma(\frac{d+1}{2})
\Gamma(\frac{1}{2})} \int_{\mathbb{R}^d} h({\bf r})r d{\bf r}.
\label{S-ensemble}
\end{equation}
Because the formula for the
coefficient ${\overline \Lambda}$ is defined 
for a single realization, one can employ it to obtain a particular
point configuration that minimizes it, i.e.,
\begin{equation}
\min_{{\cal C}} {\overline \Lambda},
\end{equation}
where ${\cal C}$ denotes configuration space.

\subsection{Asymptotics for Periodic Point Configurations}

All periodic point configurations that have a finite number
of particles in the fundamental cell belong to class I hyperuniform systems,
as we show below. For a  lattice $\Lambda$, one easily obtains
an asymptotic expression for the variance for large
$R$ by replacing the Bessel function in (\ref{parseval}) by the dominant
term of its asymptotic expansion \cite{To03a}, yielding
\begin{equation}
\sigma^2_{_N}(R) = \Lambda(R) \left(\frac{R}{D}\right)^{d-1} +{o}\left(\frac{R}{D}\right)^{d-1} \,,
\label{lattice-var}
\end{equation}
where $D$ is a characteristic microscopic length scale, say a lattice spacing, and the function
\begin{equation}
\Lambda(R)=\frac{2^{d+1} \pi^{d-1}D^{2d}}{v_F^2} \displaystyle\sum_{\bf q \neq 0} 
\frac{\cos^2\Big[|{\bf q}|R -(d+1)\pi/4\Big]}{(|{\bf q}|D)^{d+1}},
\label{S-period}
\end{equation}
describes small-scale variations in $R$ and ${\bf q}$ is a reciprocal lattice vector of the dual lattice $\Lambda^*$. 
The function $\Lambda(R)$ is bounded and fluctuates around a constant (see Figs. \ref{1D} and \ref{square2}), implying that lattices belong to class I hyperuniform
systems. This constant is equal to the averaged surface-area coefficient ${\overline \Lambda}$ defined
by (\ref{S-constant}) and hence is given by the following convergent sum:
\begin{eqnarray}
{\overline \Lambda} &=&  \frac{2^{d} \pi^{d-1}D^{2d}}{v_F^2}  \sum_{\bf q \neq 0} \frac{1}{(|{\bf q}|D)^{d+1}}.
\label{S-avg}
\end{eqnarray}
We see that finding the lattice in $\mathbb{R}^d$ that minimizes the constant ${\overline \Lambda}$ is equivalent
to finding the dual of the  ground-state lattice associated
with the inverse power-law pair potential $q^{-(d+1)}$ in reciprocal space in dimension $d$.
Note that for the integer lattice $\mathbb{Z}$,  the number variance
$\sigma^2_{_N}(R)$ is exactly equal to the function $\Lambda(R)$, i.e., there are no correction terms
in formula (\ref{lattice-var}) beyond the first term.

\begin{figure}[bthp]
\centering
\includegraphics[width=0.5\textwidth,clip=]{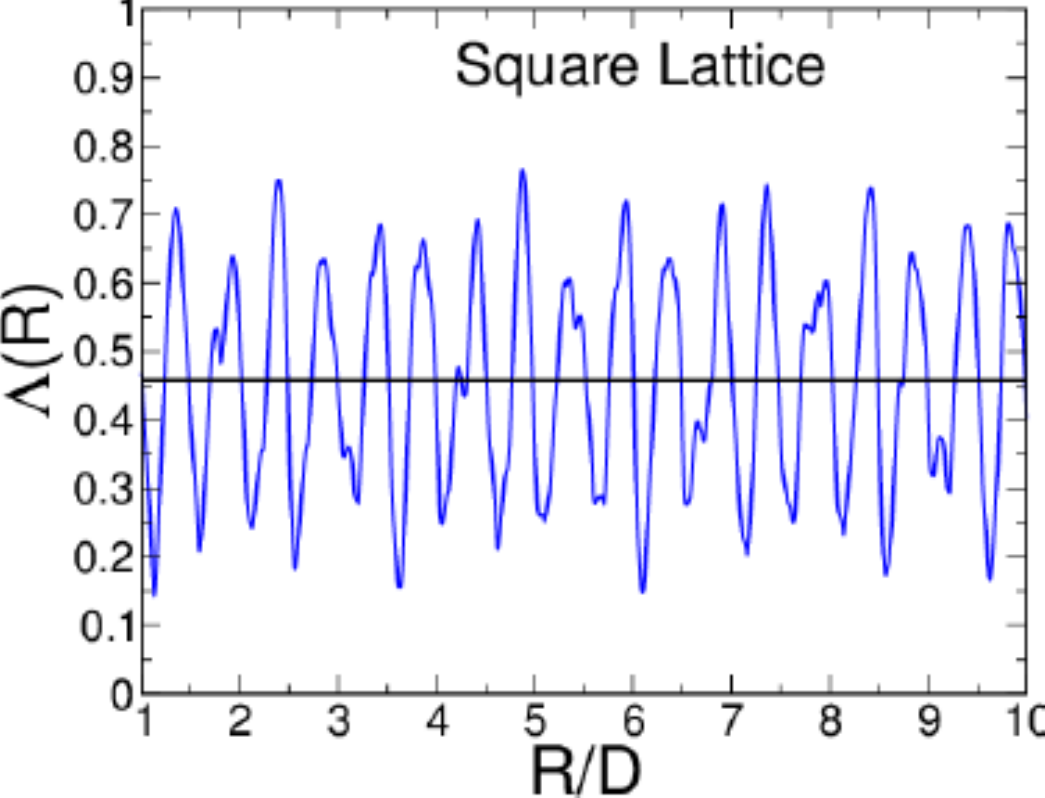}
\caption{The asymptotic surface-area function $\Lambda(R)$ for the square lattice
for $1 \le R \le 10$, where  $D$ is the lattice spacing. The horizontal
line is the asymptotic average value ${\overline \Lambda}=0.457649$ \cite{To03a}.}
\label{square2}
\end{figure}

Remarkably, the averaged surface-area coefficient ${\overline \Lambda}$ is intimately connected to 
the Epstein zeta function $Z_{\Lambda}(s)$ of number theory \cite{Sa06}. For a lattice at unit density,
this  function  is defined as follows \cite{Sa06}:
\begin{eqnarray}\label{epformula}
Z_{\Lambda}(s) = \sum_{\mathbf{p}\neq\mathbf{0}} \frac{1}{|\mathbf{p}|^{2s}} \qquad (\mbox{Re}(s)> d/2),
\end{eqnarray}
where $\mathbf{p}$ is a vector of the lattice $\Lambda$.  It is clear from (\ref{S-avg}) that the dual of the lattice that minimizes the Epstein zeta function $Z_{\Lambda}(s=(d+1)/2)$ among all lattices
will minimize the asymptotic coefficient for the number variance among lattices.  
It is known in two dimensions that the triangular lattice minimizes the Epstein zeta function \cite{Ra53,En64,Sa06} among all lattices,
and in three dimensions, the FCC is at least a local minimum among lattices \cite{En64, Sa06}.
Sarnak and Str{\" o}mbergsson  \cite{Sa06} proved that for dimensions 4, 8, and 24,
the densest known lattice packings (checkerboard  lattice $D_4$, $E_8$ root lattice and Leech lattice $L_{24}$) are at least local minima.  
(Note it was recently proved that the $E_8$ and $L_{24}$ lattices are the densest packings
among all possible packings in dimensions 8 \cite{Vi17} and 24 \cite{Co17}, respectively.) However,
it is almost certainly not true in higher dimensions that the minimizers of the Epstein zeta function are lattice structures, 
since the densest packings are likely to be nonperiodic \cite{To06b}.

For a periodic point pattern in which the fundamental cell contains $M$ points, the asymptotic formula (\ref{lattice-var})
still applies, but where fluctuating surface-area coefficient $\Lambda(R)$ is given by
\begin{equation}
\Lambda(R)=\frac{2^{d+1} \pi^{d-1}D^{2d}}{v_F^2} \displaystyle\sum_{\bf q \neq 0} 
\frac{\cos^2\Big[|{\bf q}|R -(d+1)\pi/4\Big]}{(|{\bf q}|D)^{d+1}}\left[M  +2\sum_{j<k}^{M} \cos[{\bf q}\cdot ({\bf r}_k -{\bf r}_j)]\right],
\label{L-period}
\end{equation}
which immediately follows from relation (\ref{par-periodic}). The fact that the function $\Lambda(R)$ function is bounded
implies that periodic point configurations belong to class I hyperuniform systems. 
The corresponding global average is given by
\begin{eqnarray}
{\overline \Lambda} &=&  \frac{2^{d} \pi^{d-1}D^{2d}}{v_F^2}  \sum_{\bf q \neq 0} \frac{1}{(|{\bf q}|D)^{d+1}} \left[M  +2\sum_{j<k}^{M} \cos[{\bf q}\cdot ({\bf r}_k -{\bf r}_j)]\right].
\label{L}
\end{eqnarray}
This expression simplifies considerably for $d=1$ \cite{To03a}:
\begin{equation}
{\overline \Lambda} = -\frac{M(M-3)}{12}+\sum_{j<k}^{M} f(r_k-r_j),
\label{var-2b}
\end{equation}
where $f(x)$ is the following convex quadratic nonnegative function for $0 \le x \le 1/2$:
\begin{equation}
f(x)=\frac{1}{\pi^2}\sum_{m=1}^{\infty}\frac{1+2 \cos (2\pi m x)}{m^2}=
\frac{1}{2}-2x(1-x).
\label{convex}
\end{equation}
From formula (\ref{var-2b}), it is straightforward to prove that the 
integer lattice $\mathbb{Z}$ yields the global minimum
of $ {\overline \Lambda}=1/6$ among all infinite point patterns \cite{To03a}.

One can also evaluate the asymptotic coefficient ${\overline \Lambda}$
for infinite periodic point configurations using the angular-averaged
pair correlation function relation (\ref{g2-period}) and the ensemble-average formula (\ref{S-ensemble}),
but this is a subtle calculation that must be carried out with care.
The integrand can be appropriately
modified so that a limiting convergent expression for the surface-area coefficient
emerges as follows \cite{To03a}:
\begin{eqnarray}
{\overline \Lambda} &=& \lim_{ \gamma\rightarrow 0^+} 
\frac{2^{d-1}\phi d}{D 
\Gamma(\frac{1}{2})}\left[  \frac{\phi \pi^{d/2}}{v_1(D/2) 
\gamma^{\frac{d+1}{2}}}- \frac{\Gamma(d/2)}{\Gamma(\frac{d+1}{2})}
\sum_{i=1}^{\infty} Z_i r_i e^{\displaystyle-\gamma r_i^2}\right],
\label{S-ensemble2}
\end{eqnarray}
where $Z_i$ is the {\it expected} coordination number at a radial distance $r_i$,
as defined in Eq. (\ref{g2-period}).

\subsection{Hyperuniform Order Metrics}

The global surface-area coefficient ${\overline \Lambda}$, which is trivially related to ${\overline B}_N$, has been shown
to provide a useful measure of the degree to which large-scale density fluctuations is suppressed in class I hyperuniform
systems  \cite{To03a,Za09}. 
In order to compare different hyperuniform systems to one another,
one must choose a way to normalize ${\overline \Lambda}$ so that it is scale-independent.
One choice used in Refs. \cite{To03a} and \cite{Za09} is $\overline{\Lambda}\vert_{D=1}/\phi^{(d-1)/d}$,
where $\phi=\rho v_1(1/2)$ is the dimensionless density defined by (\ref{phi}). Another choice that we will employ
here is the evaluation of the global surface-area coefficient at unit density, which we denote by $\overline{\Lambda}\vert_{\rho=1}$.
In $d$-dimensions, the two normalized surface-are coefficients 
are related by 
\begin{equation}
\overline{\Lambda}\vert_{\rho=1} = (v_1 (1/2))^{(d-1)/d} \frac{\overline{\Lambda}\vert_{D=1}}{\phi^{(d-1)/d}}.
\end{equation}
By normalizing the asymptotic coefficient $\overline{\Lambda}\vert_{\rho=1}$ by the corresponding result
for the variance-minimizing structure, which we denote by $\overline{\Lambda}_{min}\vert_{\rho=1}$, 
Zachary and Torquato \cite{Za09} defined a scalar quantity 
\begin{equation}
\psi_N= \frac{\overline{\Lambda}_{min}\vert_{\rho=1}}{\overline{\Lambda}\vert_{\rho=1}}
\end{equation}
that lies between between 0 and 1.

Table \ref{1d}  lists values of the scale-independent surface-area coefficients for various ordered 
and disordered point processes in one dimension, including point configurations
derived from the quasiperiodic Fibonacci chain by taking the endpoints of each segment. It is seen that the integer
lattice has the lowest hyperuniformity metric in the list. Indeed, it has been proved that
the integer lattice globally minimizes $\overline{\Lambda}\vert_{\rho=1}$   among all point processes \cite{To03a}.
The step-function $g_2$ and step+delta-function $g_2$ are two disordered $g_2$-invariant
point processes described in Sec. \ref{invariant}. The two-particle crystals refer to periodic system with a 2-particle
basis in which $r_2-r_1=1/4$ (see Fig. \ref{1D}) and $r_2-r_1=2/5$. The ``uncorrelated lattice gas" is constructed by tessellating the real line into 
regular intervals and  then a single point is placed in each interval (independently of the others) according 
to a uniform random distribution, a type of ``perturbed lattice" discussed in more detail in Sec. \ref{defects}.
Notice that the quasiperiodic Fibonacci chain (see Ref. \cite{Lev86} for a definition)
has a hyperuniformity order metric that falls between the minimal value
for the integer lattice and that of the uncorrelated lattice gas. Interestingly, it has recently been determined 
that there are some one-dimensional quasicrystals that belong to class II \cite{Og17}.

\begin{table}[bthp]
\caption{Hyperuniformity order metrics $\overline{\Lambda}\vert_{\rho=1}$, $\overline{\Lambda}\vert_{D=1}$ and  and $\psi_N$ for selected
one-dimensional class I hyperuniform point patterns. Except for the Fibonacci-chain
result \cite{Za09}, all of the results are taken from Ref. \cite{To03a}. In the case
of point patterns with a minimal pair separation of $D$, the dimensionless density
represents $\phi$ the packing fraction.}
\centering 
\begin{tabular}{c|c|c|c} \hline \hline
Pattern & $\overline{\Lambda}\vert_{\,\rho=1}$  & $\overline{\Lambda}\vert_{D=1}$ & $\psi_N$\\ \hline
$\mathbb{Z}$ & 0.16667 & 0.16667 & 1.00000\\ \hline
two-particle crystal ($r_2-r_1=2/5$) & 0.18666 & 0.18666 &0.89286 \\ \hline
step+delta-function $g_2$ & 0.18750 & 0.18750 & 0.88889\\  \hline
Fibonacci chain & 0.20110 & 0.20110 & 0.82878\\  \hline
step-function $g_2$ & 0.25 & 0.25 & 0.66667\\  \hline
two-particle crystal  ($r_2-r_1=1/4$) & 0.29167 & 0.29167 & 0.57143\\  \hline
uncorrelated lattice gas & 0.33333  & 0.33333 & 0.50000\\  \hline  \hline
\end{tabular}
\label{1d}
\end{table}

Table  \ref{2d}  lists values of the scale-independent surface-area coefficient for various ordered 
and disordered point processes in two dimensions.  Rankin \cite{Ra53} proved that
the triangular lattice has the smallest normalized surface-area
coefficient for circular windows among all two-dimensional lattices, which is borne out in Table \ref{2d}.
However, there is no proof that the triangular lattice minimizes
$\overline{\Lambda}\vert_{\rho=1}$ among all infinite two-dimensional 
hyperuniform point patterns for circular windows. Nonetheless, it is expected
that the triangular lattice is the global minimizer.
Among the limited set of structures listed
in this table, the one-component plasma has the largest order metric. The point configurations associated with
the two quasicrystal structures, vertices of the Penrose and octagonal tilings,
have order-metric values that lie closer to the minimal value than
that for the one-component plasma. 
There are uncountably many distinct quasicrystals that
have the same symmetry, same fundamental repeating
units (e.g. tiles, clusters of atoms or molecules), and same
support for their diffraction patterns, but which have different
space-filling arrangements of the repeating units and different
peak intensities for their diffraction patterns. These
distinct quasicrystals are said to belong to different {\it local
isomorphism} classes \cite{Le84,Lev86,So86}. It has recently come to light
that the degree of hyperuniformity in the case of 
quasicrystals depends on the local isomorphism class \cite{Li17}.
It was specifically shown that within the local isomorphism class
the Penrose tiling, all of which belong to class I hyperuniform systems, the minimal order metric is achieved by the Penrose tiling
(about  83\% smaller than the maximal value in this set) \cite{Li17}.

\begin{table}[H]
\caption{Hyperuniformity order metrics  $\overline{\Lambda}\vert_{\rho=1}$,  $\overline{\Lambda}\vert_{D=1}/\phi^{1/2}$  
and $\psi_N$ for selected
two-dimensional class I hyperuniform point patterns \cite{To03a,Za09}. In the case
of point patterns with a minimal pair separation of $D$, the dimensionless density $\phi$
represents the packing fraction. The order metrics for the triangular, square, honeycomb, kagom\'e,
step-function $g_2$, step+delta-function $g_2$, and one-component plasma structure were obtained
Ref. \cite{To03a}, while those for the other structures were determined in Ref. \cite{Za09}.}
\centering 
\begin{tabular}{c|c|c|c} \hline \hline
Pattern & $ \overline{\Lambda}\vert_{\,\rho=1}$ & $\overline{\Lambda}\vert_{D=1}/\phi^{1/2}$ & $\psi_N$\\ \hline \hline
$A_2$ (triangular) & 0.450511 & 0.50835  & 1.000000\\ \hline
$\mathbb{Z}^2$ (square) & 0.457648 & 0.51640 & 0.98443\\ \hline
disordered stealthy; $\chi = 0.496$ & 0.46438 & 0.52400 & 0.97015\\ \hline
disordered stealthy; $\chi = 0.402$ & 0.47693 & 0.53816 & 0.94463\\ \hline
honeycomb & 0.502513& 0.56703  & 0.89652\\ \hline
kagom\' e & 0.520206  & 0.58699 & 0.86603\\ \hline
octagonal quasicrystal & 0.52790 & 0.59567 & 0.85340\\ \hline
step+delta-function $g_2$ & 0.531922 & 0.60021 & 0.846949\\ \hline
Penrose tiling & 0.53220 & 0.60052 & 0.84651 \\ \hline
Rectangular kagom\' e  & 0.54051 & 0.60990 & 0.83349 \\ \hline
disordered stealthy; $\chi = 0.302$ & 0.54285 & 0.61254 & 0.82990\\ \hline
$4.8.8$ tessellation & 0.620243 & 0.69987 & 0.72635\\ \hline
step-function $g_2$ & 0.752252 & 0.84883 & 0.598883\\ \hline
one-component plasma & 1.000000 & 1.12838 & 0.45051\\ \hline \hline
\end{tabular}
\label{2d}
\end{table}

Table \ref{3d}  lists values of the scale-independent surface-area coefficients for various ordered 
and disordered point processes in  three dimensions. The previous results for one- and two-dimensional
hyperuniform systems might lead one to conjecture that
the Bravais lattice associated with the densest packing of
congruent spheres in any space dimension $d$ provides the
minimal value of  $\overline{\Lambda}\vert_{\rho=1}$   for spherical
windows. However, in three dimensions, this is definitely not true.
The minimum value of $\overline{\Lambda}\vert_{\rho=1}$ in three
dimensions appears to be achieved for the BCC lattice, which is the lattice dual to the  
to the FCC lattice, corresponding to the densest sphere packing \cite{Ha05}. 
We see that the global surface-area coefficient for the fcc lattice is very slightly larger
than that of the BCC lattice.  Based on our remarks
earlier about the Epstein zeta function, one can only say that the BCC
structure is a local minimum of  $\overline{\Lambda}\vert_{\rho=1}$
among (Bravais) lattices. Note that the ``tunneled" FCC and HCP crystal structures
listed in the table are conjectured to have the lowest density among
all strictly jammed packings of identical spheres \cite{To07}; see Sec. \ref{jamming} for jamming
definitions. However, while there is no proof that BCC is a global minimizer among all infinite three-dimensional hyperuniform point patterns
of $\overline{\Lambda}\vert_{\rho=1}$, it is reasonable to conjecture that this is the case.
Among the limited set of structures listed
in this table, the one associated with the step-function $g_2$ (see Sec. \ref{invariant}) has the largest order metric.

\begin{table}[H]
\caption{ Hyperuniformity order metrics $\overline{\Lambda}\vert_{\rho=1}$, $\overline{\Lambda}\vert_{D=1}/\phi^{2/3}$  
and   $\psi_N$ for selected
three-dimensional class I hyperuniform point patterns \cite{To03a,Za09}. In the case
of point patterns with a minimal pair separation of $D$, the dimensionless density $\phi$
represents the packing fraction. The order metrics for the BCC, FCC, HCP, SC, diamond,
damped-oscillating $g_2$, step-function $g_2$ and step+delta-function $g_2$ were obtained
Ref. \cite{To03a}, while those for the other structures were determined in Ref. \cite{Za09}.}
\centering 
\begin{tabular}{c|c|c|c} \hline \hline
System & $\overline{\Lambda}\vert_{\,\rho=1}$ & $\overline{\Lambda}\vert_{D=1}/\phi^{2/3}$ & $\psi_N$\\  \hline  \hline
$D_3^*$ (BCC) & 0.808633 & 1.24476 & 1.00000 \\  \hline
$D_3$ (FCC) & 0.809127 & 1.24552 & 0.99939 \\  \hline
HCP & 0.809237 & 1.24569 & 0.99926 \\  \hline
$\mathbb{Z}^3$ (SC) & 0.837502 & 1.28920 & 0.96553 \\  \hline
disordered stealthy; $\chi = 0.43$ & 0.842810 & 1.29737  & 0.95949 \\  \hline
$D_3^+$ (diamond) & 0.921772 & 1.41892 & 0.87726 \\  \hline
tunneled FCC &  0.922994 & 1.42080 & 0.87610  \\  \hline
w\" urzite & 0.923669 & 1.42184 & 0.87546 \\  \hline
tunneled HCP & 0.927963 & 1.42845 & 0.87141 \\  \hline
damped-oscillating $g_2$ & 0.940904 & 1.44837 & 0.85942 \\  \hline
step+delta-function $g_2$ & 0.991893& 1.52686 & 0.81524 \\  \hline
step-function $g_2$ & 1.46166 & 2.25000 & 0.55323 \\  \hline \hline
\end{tabular}
\label{3d}
\end{table}

Table \ref{high-d} lists values of the scale-independent surface-area coefficient $\overline{\Lambda}\vert_{\,\rho=1}$  for selected lattices 
across dimensions up to $d=8$ \cite{Za09}. This includes the hypercubic $\mathbb{Z}^d$, checkerboard $D_d$,
root $A_d$, $E_6$ and $E_8$ lattices as well as their corresponding reciprocal (dual) lattices \cite{Co93}. Among the lattices
described in Table \ref{high-d}, the smallest value of  $\overline{\Lambda}\vert_{\,\rho=1}$ is given by $D_d^*$ for 
$d=4$ and $d=5$, and by $E_d^*$ for $d=6,7$ and 8.
For $d=4$ and $d=8$, the best known solutions for the sphere packing and number-variance problems are
identical, namely, $D_4\equiv D_4^*$ and $E_8\equiv E_8*$, respectively. These lattices
are no longer optimal for dimensions in the range  $5 \le d \le 7$ \cite{To10d}.
It is noteworthy that for the first three space dimensions, the best known solutions of the sphere-packing
and number-variance problems (or their ``dual" solutions) are directly related to those of 
two other well-known problems in discrete geometry: the {\it covering}
and {\it quantizer} problems \cite{Co93}, but such relationships may or may not exist for $d>4$, depending on the peculiarities
of the dimensions involved \cite{To10d}.

\begin{table}[H]
\caption{Hyperuniformity order metric $\overline{\Lambda}\vert_{\,\rho=1}$  for selected
lattice families in the first eight space dimensions \cite{Za09}.   Also included
are the maximal packing fractions $\phi$ of the lattices in each dimension.  Note that the $E_d$ and $E_d^*$ lattice families are only uniquely defined
for $d\geq 6$; a hyphen therefore indicates that information for these lattices is not available in lower dimensions.}
\centering
\begin{tabular}{c|c|c|c} \hline \hline
$d$ & $\mathbb{Z}^d$ $(\phi)$ & $A_d$ $(\phi)$ & $A_d^*$ $(\phi)$\\  \hline  \hline
1 & 0.16667  (1) & 0.16667 (1) & 0.16667  (1)\\
2 & 0.457648  ($\pi/4$) & 0.450511  ($\sqrt{3}\pi/6$) &0.450511  ($\sqrt{3}\pi/6$)\\
3 & 0.837502  ($\pi/6$) &  0.809127  ($\sqrt{2}\pi/6$) & 0.808633  ($\sqrt{3}\pi/8$)\\
4 & 1.24273  ($\pi^2/32$) & 1.17426  ($\sqrt{5}\pi^2/40$) & 1.17134  ($\sqrt{5}\pi^2/50$)\\
5 & 1.60656  ($\pi^2/60$) & 1.478629  ($\sqrt{3}\pi^2/45$) &1.46911   ($5\sqrt{5}\pi^2/432$)\\
6 & 1.88060  ($\pi^3/384$) & 1.67639  ($\sqrt{7}\pi^3/336$) & 1.65351  ($9\sqrt{7}\pi^3/5488$)\\
7 & 2.04468  ($\pi^3/840$) & 1.753865  ($\pi^3/210$) & 1.708372  ($49\sqrt{7}\pi^3/61440$)\\
8 & 2.10564 ($\pi^4/6144$) & 1.720951  ($\pi^4/1152$) & 1.64504  ($2\pi^4/6561$)\\ \hline  \hline
$d$ & $D_d$ $(\phi)$ & $D_d^*$ $(\phi)$ & $E_d$ $(\phi)$\\ \hline  \hline
1 & 0.16667  (1) & 0.16667 (1) & - \\
2 & 0.457648  ($\pi/4$) & 0.457648 ($\pi/4$) & -  \\
3 & 0.809127  ($\sqrt{2}\pi/6$)  & 0.808633 ($\sqrt{3}\pi/8$) & - \\
4 & 1.15803  ($\pi^2/16$) & 1.15803 ($\pi^2/16$) & - \\
5 & 1.44268  ($\sqrt{2}\pi^2/30$) & 1.44018  ($\pi^2/30$) & - \\
6 & 1.62341  ($\pi^3/96$) & 1.61020  ($\pi^3/192$) & 1.58945  ($\sqrt{3}\pi^3/144$) \\
7 & 1.69005  ($\sqrt{2}\pi^3/210$) &1.65356   ($\pi^3/420$) &1.60262   ($\pi^3/105$) \\
8 & 1.65765  ($\pi^4/768$) & 1.58318 ($\pi^4/3072$) &1.48166   ($\pi^4/384$) \\ \hline  \hline
$d$ & $E_d^*$ $(\phi)$ &  & \\ \hline  \hline
1 & -  &  & \\
2 & - &  & \\
3 & -  &  & \\
4 & - &  & \\
5 & - &  & \\
6 & 1.587410 ($\sqrt{3}\pi^3/162$) & & \\
7 & 1.59861  ($9\sqrt{3}\pi^3/2240$) & & \\
8 &1.48166   ($\pi^4/384$) & &\\ \hline\hline
\end{tabular}
\label{high-d}  
\end{table}

\subsection{Nonspherical Windows}
\label{nonspherical}

The preponderance of previous theoretical investigations concerning the 
local number variance of point processes have focused on the use of spherical windows. Spherical windows offer many advantages over nonspherical shapes. For example, the mathematical properties of the number variance are easier to derive when spherical windows  are employed, regardless of the symmetries of the underlying point process. This may no longer be true when nonspherical windows are used and the point pattern is  periodic. Moreover, spherical windows have been employed experimentally when direct-space configurational information is available via microscopy 
\cite{Dr15}, for example.

Nonetheless, it is valuable to understand what is the effect of window shape on the number variance and the general conditions under which results for nonspherical windows are not qualitatively similar to those for spherical windows. Nonspherical windows may be good choices to employ to better reflect the symmetries of the underlying point process, e.g., sampling liquid crystals with oriented ellipsoidal windows.
Square- and cubic-shaped windows in two and three dimensions, respectively, are convenient shapes to use when analyzing real material images,
which are necessarily  digitized into square pixels or cubic voxels, respectively.


It is well-known that the large-window asymptotic behavior of the  variance $\sigma_{_N}^2({\bf R})$ for rectangular windows with certain orientations in the special case of  the  square lattice $\mathbb{Z}^2$ can be anomalously different from 
that of circular windows  \cite{Ke48,Be01,Ken63}. 
For example,  for the square lattice and a rectangular window with
a very irrational orientation with respect to the
$x$-axis, the variance grows as $\ln R$ \cite{Be01}.
In these studies, however, the variance was used mainly as a mathematical tool to 
study properties of natural and irrational numbers, and hyperuniformity was not a consideration.

Kim and Torquato  \cite{Ki17} have recently carried out a comprehensive study of the effect of window shape on the 
number variance from the perspective of  hyperuniformity.  Recall that in Sec. \ref{vanishing} it was noted that
 the direct-space hyperuniform condition (\ref{VAR}) that the 
number variance grow asymptotically more slowly than the window volume in usual circumstances is equivalent to the Fourier-space hyperuniformity
condition (\ref{hyper}).
However, for lattices and certain window shapes that share the symmetries of the lattices, the variance growth rate can depend
on the shape as well as the orientation of the window, and in some cases, the
growth rate can be faster than the window volume (see Ref. \cite{Ki17} and references
therein), which may lead one to falsely conclude that accompanying hyperfluctuations is due to 
the system being nonhyperuniform.
Kim and Torquato  \cite{Ki17} showed 
that such anomalous behavior can be completely circumvented by sampling the nonspherical window
uniformly over all window orientations. Specifically, the direct-space hyperuniformity requirement (\ref{VAR}) generalizes in the following way:
\begin{equation}\label{eq:general direct space condition}
\lim_{v_1({\bf R}) \to \infty}\frac{\left<\sigma_N^2 ({\bf R}) \right>_O}{v_1({\bf R})} = 0,
\end{equation}
where $\left<\sigma_N^2 ({\bf R}) \right>_O$ is the orientationally-averaged number variance.
It was also shown that for any hyperuniform point pattern in $\mathbb{R}^d$ (ordered or not), the orientationally-averaged  variance 
has the following common asymptotic behavior: 
\begin{equation}\label{eq:orientation_av_asymp}
\frac{\left<\sigma_N^2 (\bf{R}) \right>_O}{s_1(\bf{R})} \approx -\rho^2 \kappa(d) 
\int_{|{\bf r}| \le 2L_{\bf R}} h({\bf r})|{\bf r}| d{\bf r} ~~~(v_1(\bf{R})\to \infty),
\end{equation}
for any convex window shape.
Here, $L_{\bf R}$ is the largest distance from the centroid of the window to its surface, $s_1(\bf{R})$ is the surface area of the window, and 
$\kappa(d)=\Gamma(d/2)/(2\Gamma(1/2)\Gamma((1+d)/2))$.
This formula generalizes previous results for the square and triangular lattices in two dimensions \cite{Ke48, 
Ma89} and analogous results for charge fluctuations in two- and three-dimensional systems \cite{Ma80}.
The asymptotic expression (\ref{eq:orientation_av_asymp}) also implies that if a point process is statistically isotropic, i.e., $h({\bf r}) = h(|\bf{r}|)$, 
the general expression (\ref{N2}) for the number variance
should exhibit the same asymptotic behavior up to a constant multiplicative factor, regardless of the window shape \cite{Ki17}.

\begin{figure}[ht]
\begin{center}
\includegraphics[width = 0.9\textwidth]{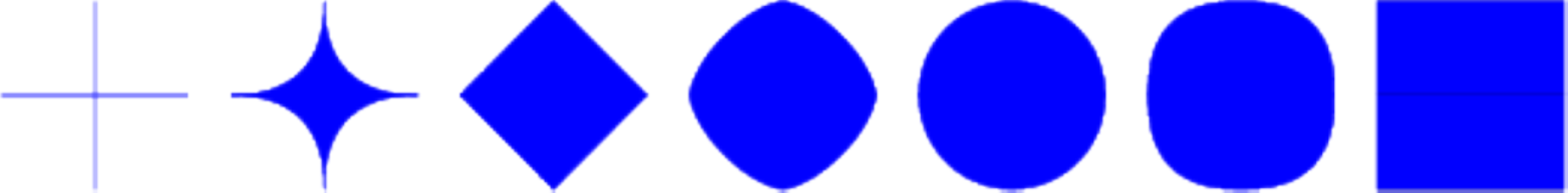}
\end{center}
\caption{Superdisk shapes for various values of the deformation parameter $p$, as adapted from Ref. \cite{Ki17}. Superdisks are defined by the equation $|x|^{2p}+|y|^{2p} = L^{2p}$, where $L$ is a characteristic length and $p$ is the deformation parameter. From the left to the right $p=0,~ 0.25, ~0.5, ~0.75, ~1, ~1.25, ~\mathrm{and}~\infty$. \label{superdisks}}
\end{figure}

To illustrate the richness of the dependence of the window shape on the  number variance, Kim and Torquato
 investigated the behavior of the number variance by sampling the square lattice using ``super disk" windows.
A super disk is the two-dimensional shape  defined by
$|x_1|^{2p}+|x_2|^{2p}  = L^{2p}$
where $p$ is the positive \textit{deformation parameter} and $L$ is a characteristic length scale.
If the deformation parameter $p$ is smaller than 0.5, a super disk is concave, and it interpolates smoothly 
between a cross ($p=0$) and a perfect square ($p=0.5$).
On the other hand, if $p \geq 0.5$, a super disk is convex and  continuously transforms
from a square ($p=0.5$) to the circle ($p=1$) and then back to a square of side length $2L$ in the limit $p\to\infty$, as shown in Fig. \ref{superdisks}.
Superdisk windows with a fixed orientation with respect to the underlying square lattice were studied for cases in which $p\ge 1$.
It was shown that the cumulative moving average of the number variance $\overline{\sigma_{_N}^2} (L)$ [defined by (\ref{moving})]
exhibits the power-law behavior $\overline{\sigma_N^2} (L) \sim L^\gamma$ for large $L$. The exponent $\gamma$ takes on the expected
value of unity at the circle point ($p=1$) and increases continuously and monotonically from this value as the deformation parameter
$p$ increases until it achieves its maximum value of $\gamma=2$ at the perfect square limit  ($p \rightarrow \infty$); see Fig. \ref{super-var}.
Thus, for square windows,  the number variance grows like the window area (proportional to $L^2$) \cite{Ki17}, which is inconsistent with
the usual direct-space hyperuniformity condition (\ref{VAR}).
Referring to the  middle panel in Fig. \ref{super-var},
we see that at rational angles (tangent of which belongs to the rational numbers), the number variance grows as fast as the window area ($L^2$), 
but at  irrational angles, the variance grows slower than the window perimeter, e.g., $\ln{(L)}$ \cite{Ki17}, which is not possible for a circular
window \cite{Be01}.
Moreover, as shown in the right panel of Fig. \ref{super-var}, the orientationally-averaged number 
variance  $\left\langle\sigma_N^2 (L) \right\rangle_O$  exhibit the same asymptotic behavior for any $p\geq 1$, i.e., $\left\langle\sigma_{_N}^2  (L) \right\rangle_O  \propto s_1(L)$, as predicted by (\ref{eq:orientation_av_asymp}),
where $s_1(L)$ is the perimeter of superdisk window.

\begin{figure}[ht]
\begin{center}
\includegraphics[width = 0.31\textwidth,clip=]{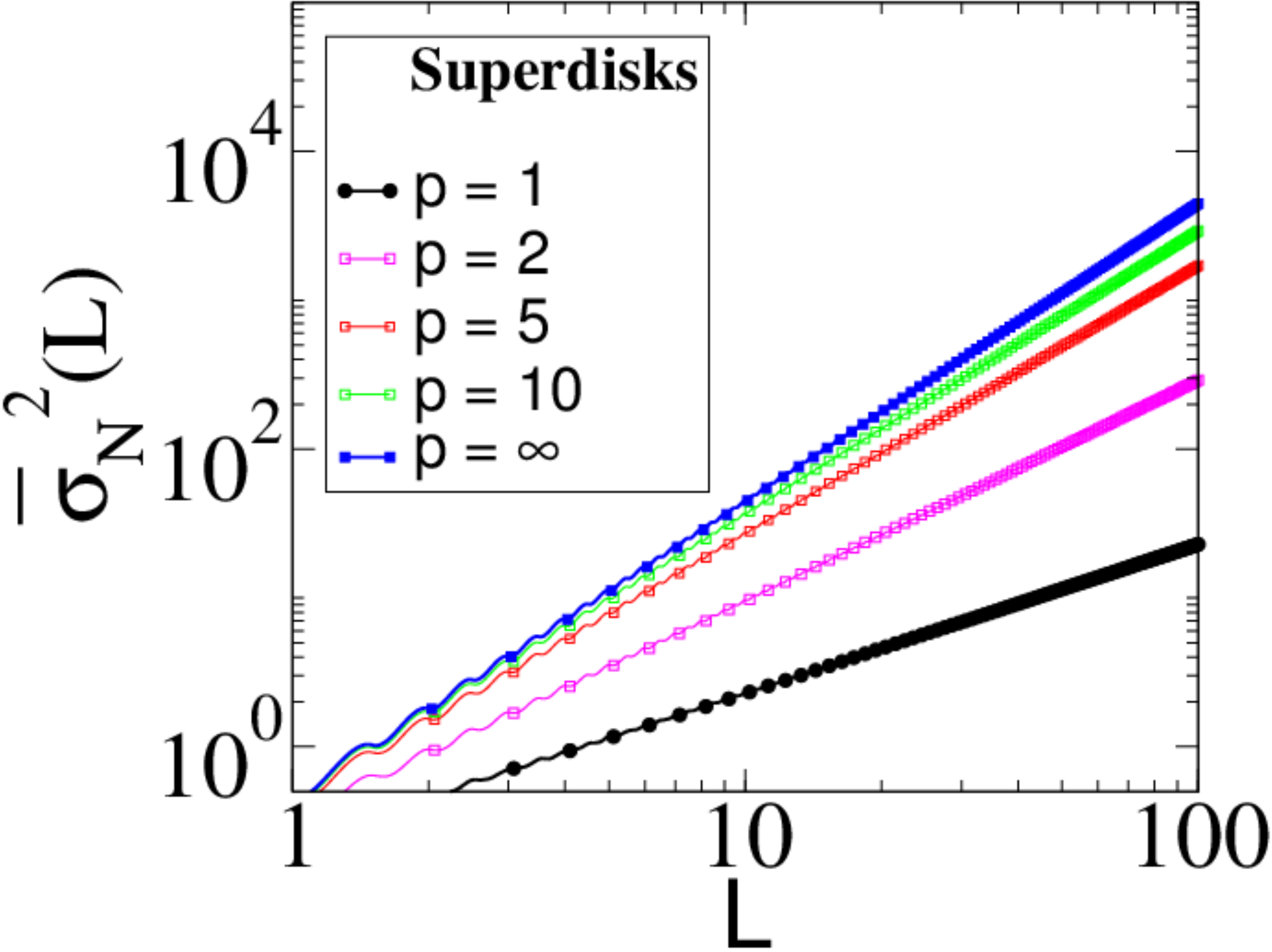}
\includegraphics[width=0.34\textwidth,clip=]{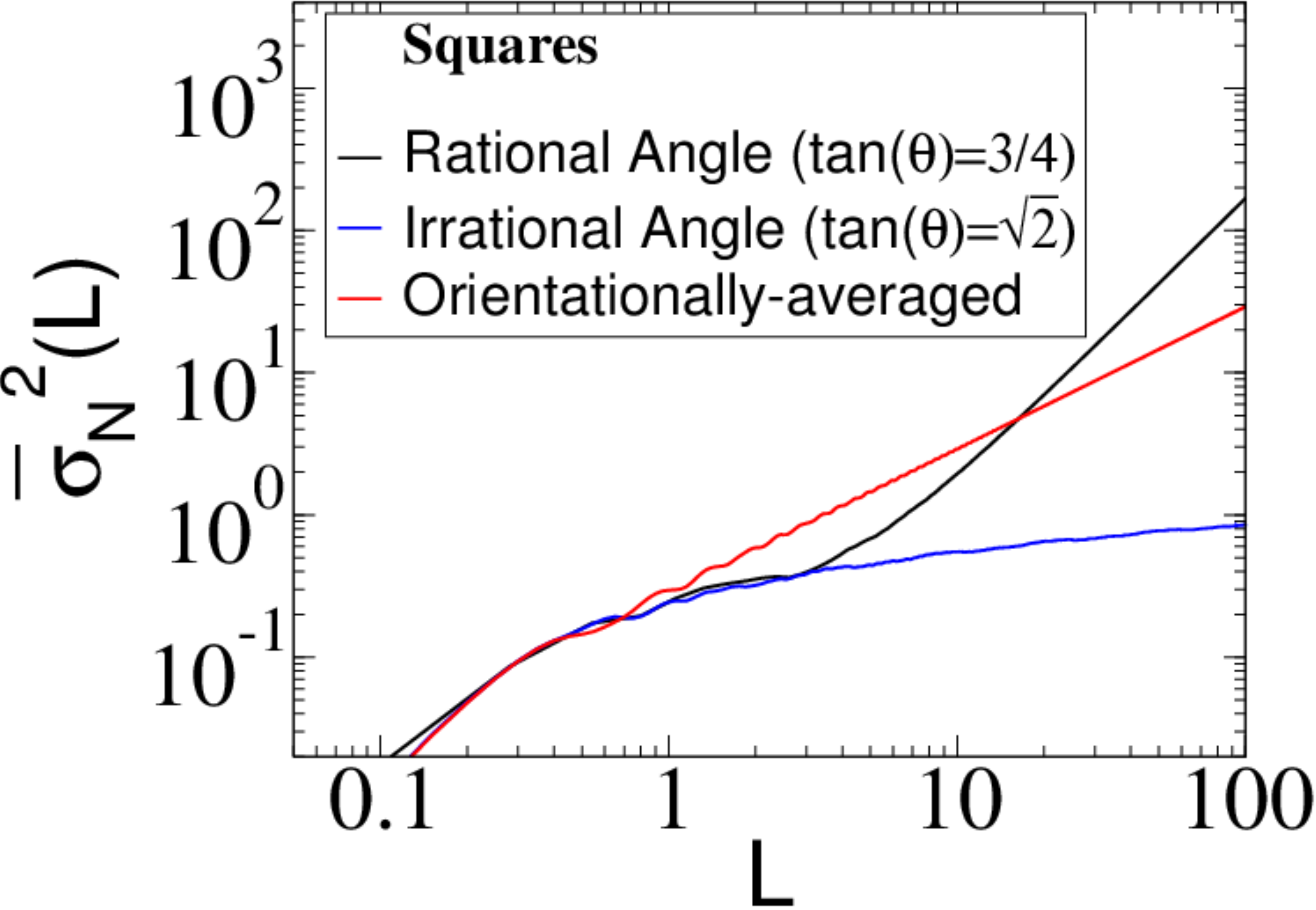}
\includegraphics[width = 0.34\textwidth,clip=]{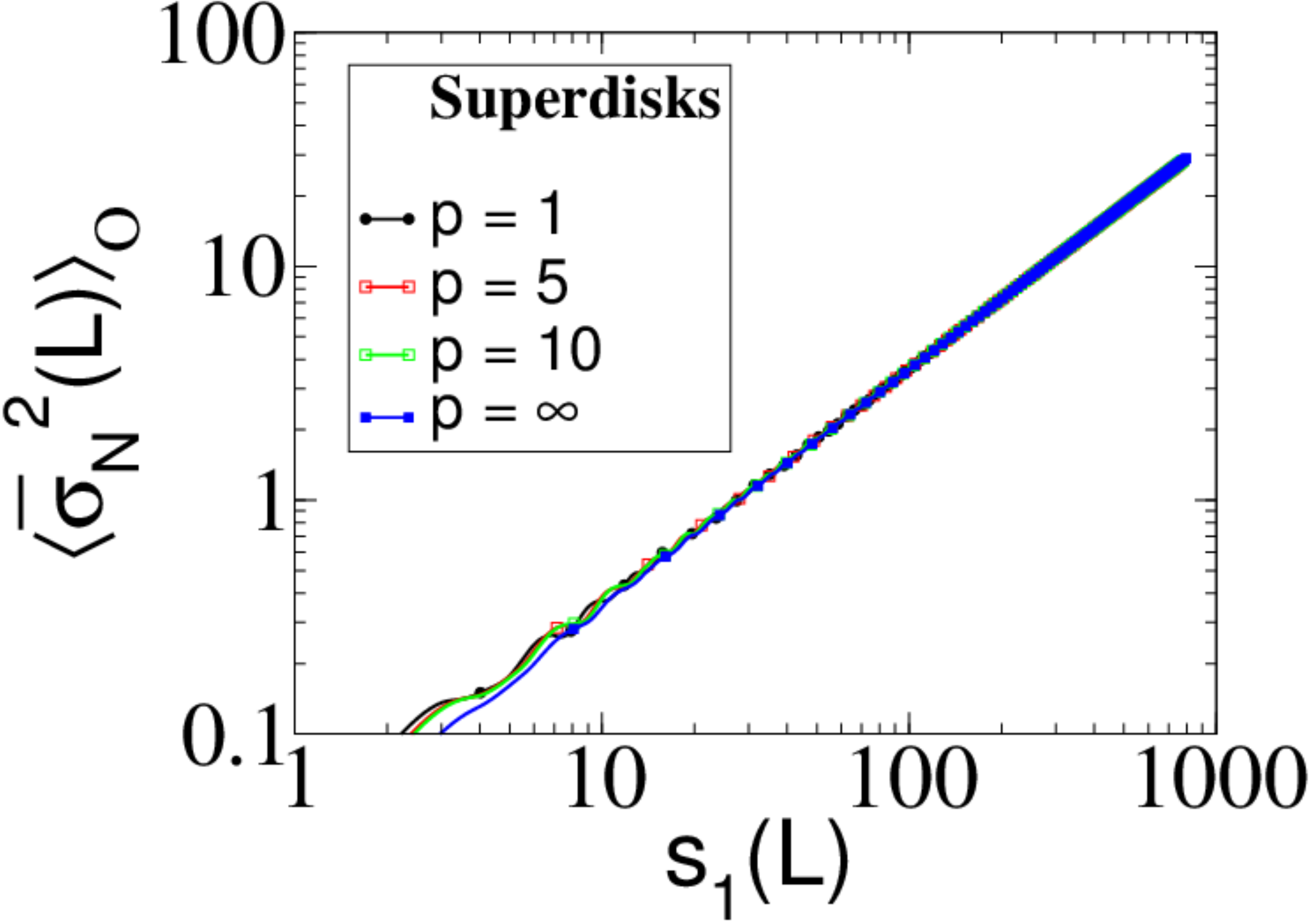}
\end{center}

\caption{Left panel: Log-log plot of the cumulative moving average of the number variance $\overline{\sigma_{_N} ^2} (L)$ of the square lattice as a function of the characteristic length scale $L$ of superdisk windows for selected values of the deformation parameter $p$. The symmetry axes of superdisk window and the square lattice are coincident. The number variance exhibits power-law behavior, i.e., $\overline{\sigma_{_N} ^2} (L) \sim L^\gamma$ where
$\gamma \in [1,2]$.
 For circular windows ($p=1$), the number variance grows like the window perimeter ($\gamma=1$). However, 
as $p$ increases from unity, the exponent $\gamma$ increases continuously and monotonically from unity
until it achieves its maximum value of $\gamma=2$ at the square limit ($p\to\infty$).
Middle panel: Log-log plot of the cumulative moving average of the number variance $\overline{\sigma_N ^2} (L)$ 
versus the side length $2L$ in the case of a square window ($p\to\infty$) at a rational angle
(showing growth rate proportional to the window area), an
irrational angle  (showing growth rate slower than the window perimeter) and the orientationally-averaged case
 (showing growth rate proportional to the window perimeter).  
Right panel: Log-log plot of cumulative moving average of the orientationally-averaged number variance $\left\langle\sigma_N^2 (L)\right\rangle_O $
as a function of the perimeter of superdisk window $s_1(L)$. } \label{super-var}
\end{figure}

It is instructive to understand more mathematically the anomalous variance growth rates.
 We indicated in Sec. \ref{vanishing} that for the preponderance
of point processes and window shapes,  the function  ${\tilde \alpha}_2({\bf k};{\bf R})$ appearing in (\ref{var-par})
tends to $(2\pi)^d \delta({\bf k})$ when the windows grow infinitely large in a self-similar fashion.
However, when the underlying point pattern is a lattice, 
this direct-space condition may no longer apply if $\tilde{\alpha}_2({\bf k};{\bf R})$ can be decomposed into a product of functions
associated with its lower-dimensional forms \cite{Ki17}.
For example, the function $\tilde{\alpha}_2({\bf k};{\bf R})$ for a $d$-dimensional hypercubic window of side length $2L$ can be expressed as 
\begin{equation}\label{eq:scaled intersection volume d-cube}
\tilde{\alpha}_2({\bf k};{L}) = \prod_{i=1}^d \left(\frac{2 \sin(k_iL)}{k_iL} \right)^2,
\end{equation}
where the multiplicand in (\ref{eq:scaled intersection volume d-cube}) is the Fourier transform of the scaled intersection volume of a one-dimensional interval
of length $2L$, ${\tilde \alpha}_2(k;L)$.
The function ${\tilde \alpha}_2({\bf k};L)$ decreases slowly in the direction of any axis of symmetry.
Due to the interplay  between the Bragg peaks of the structure factor $S({\bf k})$ and the slowly decaying function
$\tilde{\alpha}_2({\bf k};L)$ (see Fig. 8 in Ref. \cite{Ki17}), the variance scales like $L^{2(d-1)}$ when the windows are 
perfectly aligned with the lattice and hence grows
faster than the window volume for $d\ge 3$. Of course,  one should not conclude from this result that lattices
are not hyperuniform. 
Indeed, in the preponderance of situations, the structure-factor hyperuniformity condition (\ref{hyper})
implies the direct-space hyperuniformity condition (\ref{VAR}).

\section{Mathematical Foundations of Hyperuniformity: Two-Phase Media}
\label{found-2}

This section is concerned with the mathematical foundations of hyperuniformity
for  two-phase media in  $\mathbb{R}^d$. We discuss hyperuniformity conditions for general
two-phase media, asymptotic behaviors of the volume-fraction variance and 
scaling behavior of two-point statistics in both direct and Fourier spaces, the three possible hyperuniformity
classes, design of general hyperuniform two-phase media, and spectral densities
of packings.

\subsection{Vanishing of Infinite-Wavelength Volume-Fraction Fluctuations}

Consider uniformly sampling a two-phase medium with a $d$-dimensional spherical window of radius $R$.
While the local number variance $\sigma_{_N}^2(R)$ for a general
many-particle system (hyperuniform or not) grows with increasing $R$, the local volume-fraction variance $\sigma_{_V}^2(R)$
for a general two-phase medium decays as $R$ increases \cite{Lu90a,Lu90b,Qu97b,Qu99}, implying that
it vanishes in the limit $R \rightarrow \infty$. For typical disordered
two-phase media, $\sigma_{_V}^2(R)$ decays to zero like the inverse of the window volume, i.e., $R^{-d}$ \cite{Lu90b}.  
It has been shown that for sufficiently large window sizes, the full distribution function of the local volume fraction can be
reasonably approximated by the normal distribution \cite{Qu97b}.

To understand the behavior of the variance $\sigma_{_V}^2(R)$ for a hyperuniform two-phase medium,
consider its Fourier representation given by relation (\ref{phi-var-2}).
In the limit that the spherical window grows infinitely large, the function  ${\tilde \alpha}_2(k; R)$ appearing in (\ref{phi-var-2})
tends to $(2\pi)^d \delta({\bf k})$. Therefore, multiplying the variance (\ref{phi-var-2})  by $v_1(R)$ 
and taking this limit yields
\begin{equation}
\lim_{v_1(R) \rightarrow \infty} v_1(R) \,\sigma^2_{_V}(R)= 
\lim_{|{\bf k}| \rightarrow 0} {\tilde \chi}_{_V}({\bf k})= \int_{\mathbb{R}^d} \chi_{_V}({\bf r}) d{\bf r},
\label{phi-var-3}
\end{equation}
where ${\tilde \chi}_{_V}({\bf k})$ is the spectral density defined by relation (\ref{SPEC}).
Note that the hyperuniformity requirement (\ref{hyper-2}) and relation (\ref{phi-var-3}) dictate that 
\begin{equation}
\lim_{v_1(R) \rightarrow \infty} v_1(R) \,\sigma^2_{_V}(R)= 0,
\label{phi-VAR}
\end{equation}
which signifies that the local volume-fraction variance $\sigma^2_{_V}(R)$ must tend to zero  for large $R$
more rapidly than the inverse of the window volume, i.e., like $R^{-d}$.

Very recently, Chieco {\it et al.} \cite{Chi17} introduced a hyperuniformity disorder length
parameter in pixelized systems that is linked to the volume-fraction variance. 
The continuum limit of point patterns, where pixel size vanishes, was also considered.

\subsubsection{Asymptotic Behavior of the Volume-Fraction Variance for Two-Phase Systems}

For a homogeneous and isotropic two-phase medium and  a spherical observation window of radius $R$, 
substitution of the expansion (\ref{series}) for the scaled intersection volume $\alpha_2(r; R)$ into (\ref{phi-var-1}) implies the following large-$R$ asymptotic expansion for the the volume-fraction variance:
\begin{eqnarray}
\sigma^2_{_V}(R) &=& A_{_V}(R) \left(\frac{D}{R}\right)^d + B_{_V}(R) \left(\frac{D}{R}\right)^{d+1} 
+ o\left(\frac{D}{R}\right)^{d+1},\label{tauasymp}\\
\end{eqnarray}
where
\begin{eqnarray}
A_{_V}(R) &=& \frac{1}{v_1(D)} \int_{|{\bf r}| \le 2R} \chi_{_V}(\mathbf{r}) d\mathbf{r}\label{atau},\\
B_{_V}(R) &=& -\frac{c(d)}{2\;D v_1(D)} \int_{|{\bf r}| \le 2R} \chi_{_V}(\mathbf{r}) |{\bf r}| d\mathbf{r}\label{btau},
\end{eqnarray}
where $A_{_V}(R)$ and  $B_{_V}(R)$ are dimensionless asymptotic coefficients that multiply terms proportional to $R^{-d}$
and $R^{-(d+1)}$, respectively, $D$ represents a characteristic ``microscopic" length scale, and $c(d)$ is a 
$d$-dependent constant defined by (\ref{C}). Note that the coefficient $A_{_V}(R)$ in the limit $R \rightarrow \infty$ is 
proportional to the nonnegative spectral density ${\tilde \chi}_{_V}({\bf k})$ [cf. (\ref{SPEC})] in the 
limit that the wavenumber tends to zero, i.e.,
\begin{equation}
{\overline A}_{_V} \equiv \lim_{R \rightarrow \infty}A_{_V}(R) \propto \lim_{|{\bf k}| \rightarrow 0}{\tilde \chi}_{_V}
({\bf k}).
\label{A_V}
\end{equation}

\subsubsection{Three Classes of Hyperuniform Two-Phase Systems}

Analogous to the classification of hyperuniform point configurations, the 
asymptotic decay behaviors of $\sigma^2_{_V}(R)$ for hyperuniform two-phase media
fall into three distinct  classes: class I in which the decay is like $1/R^{d+1}$, class II in which the decay 
is like $\ln(R)/R^{d+1}$, and class III
in which the decay is like $1/R^{d+\alpha}$, where $\alpha$ is an exponent that
lies in the open interval $(0,1)$. The particular class is determined by
the large-$|\bf r|$ behavior
of the autocovariance function $\chi_{_V}({\bf r})$ or, equivalently, the behavior
of the spectral density ${\tilde \chi}_{_V}({\bf k})$ in the zero-wavenumber limit.

In the case of class I systems, the 
coefficient ${\overline A}_{_V}$, defined in (\ref{A_V}), is exactly zero and the coefficient $B_{_V}(R)$ converges to a constant in the limit $R \rightarrow \infty$
and hence, according to relation  (\ref{tauasymp}), $\sigma_{_V} ^2(R)$ decays like
$1/R^{d+1}$ \cite{Za09}, as specified by 
\begin{equation}
\sigma^2_{_V}(R) \sim {\overline B}_{_V}\left(\frac{D}{R}\right)^{d+1} \qquad (R \rightarrow \infty),
\label{var4}
\end{equation}
where
\begin{equation}
{\overline B}_{_V}= \lim_{R \rightarrow \infty} B_{_V}(R)=-\frac{c(d)}{2\;D v_1(D)} \int_{\mathbb{R}^d} \chi({\bf r})|{\bf r}| d{\bf r},
\label{B4}
\end{equation}
Class I hyperuniform two-phase systems include those in which
$\chi_{_V}({\bf r})$ decays to zero sufficiently fast for large $|{\bf r}|$, including 
two-phase media that are certain decorations of
perfect periodic and disordered hyperuniform point patterns \cite{To16b,Zh16a} (such as the ones depicted in Fig. \ref{cartoon-2}) 
and a large class of perfect quasicrystal point patterns.

Two other hyperuniform classes are possible if the
coefficient $B_{_V}(R)$ does not converge to a  constant
in the limit $R \rightarrow \infty$. For example, if the autocovariance function
is controlled by the following radial power-law decay:
\begin{equation}
\chi_{_V}({\bf r}) \sim \frac{1}{|{\bf r}|^{d+ 1}}  \qquad (|{\bf r}| \rightarrow \infty),
\label{chi-decay}
\end{equation}
an asymptotic analysis of (\ref{btau}) leads to  a volume-fraction variance that asymptotically decays
like $\sigma^2_{_V}(R) \sim R^{-(d+1)} \,\ln(R)$, since $B_{_V}(R) \sim \ln(R)$, which we call
class II hyperuniform two-phase systems. Examples include maximally random jammed packings of disks and spheres \cite{Za11a,Za11c,Kl16,At16a} and of nonspherical particles \cite{Za11c}. On the other hand, if
\begin{equation}
\chi_{_V}({\bf r}) \sim \frac{1}{|{\bf r}|^{d+ \alpha}}  \qquad (|{\bf r}| \rightarrow \infty),
\label{chi-decay-2}
\end{equation}
yields a volume-fraction variance that scales like $R^{-(d+\alpha)}$, 
where  $0 < \alpha <1$, since $B_{_V}(R) \sim R^{1-\alpha}$. 
We refer to such hyperuniform two-phase systems as class III structures.


\subsection{Asymptotics From Power-Law Spectral Densities}

Let us consider the power-law behavior for spectral density in the vicinity
of the origin
\begin{equation}
{\tilde \chi}_{_V}({\bf k}) \sim |{\bf k}|^{\alpha} \qquad (|{\bf k}|\rightarrow 0)
\label{spec-asy}
\end{equation}
for $\alpha >0$. Associated with this scaling form are three different types 
of large-$R$ scaling behaviors of the  volume-fraction variance $\sigma^2_{_V}(R)$ \cite{Za09}:
\begin{eqnarray}  
\sigma^2_{_V}(R) \sim \left\{
\begin{array}{lr}
R^{-(d+1)}, \quad \alpha >1  \qquad \mbox{(CLASS I)}\\
R^{-(d+1)} \ln R, \quad \alpha = 1 \quad \mbox{(CLASS II)}.\\
R^{-(d+\alpha)}, \quad 0 < \alpha < 1 \quad \mbox{(CLASS III)}
\end{array}\right.
\label{sigma-V-asy}
\end{eqnarray}


\subsection{Design of Hyperuniform Two-Phase Materials with Prescribed Spectral Densities}
\label{design}

A Fourier-space based numerical construction procedure has recently been formulated
to design  a wide class of disordered hyperuniform two-phase materials with prescribed spectral densities, which 
enables one to tune the degree and length scales at which the anomalous  suppression 
of volume-fraction fluctuations occur \cite{Ch18}. This is a generalization of the direct-space Yeong-Torquato
construction procedure to generate two-phase media with prescribed autocovariance functions \cite{Ye98a,To02a,Ji09b}.
Clearly, the Fourier-space setting is the most natural one to employ for the purposes
of constructing such disordered hyperuniform materials, since it enables
one to very accurately control the behavior of the spectral density from very long to intermediate
wavelengths. This technique was used to construct a family of phase-inversion-symmetric materials of 
class I with tunable topological  connectedness properties (see Fig. \ref{phase-inv}) as well as  
disordered stealthy hyperuniform dispersions. As we see in Sec. \ref{properties}, these two-phase 
materials have desirable effective physical properties.

\begin{figure}[ht]
\begin{center}
\includegraphics[width = 0.9\textwidth,clip=]{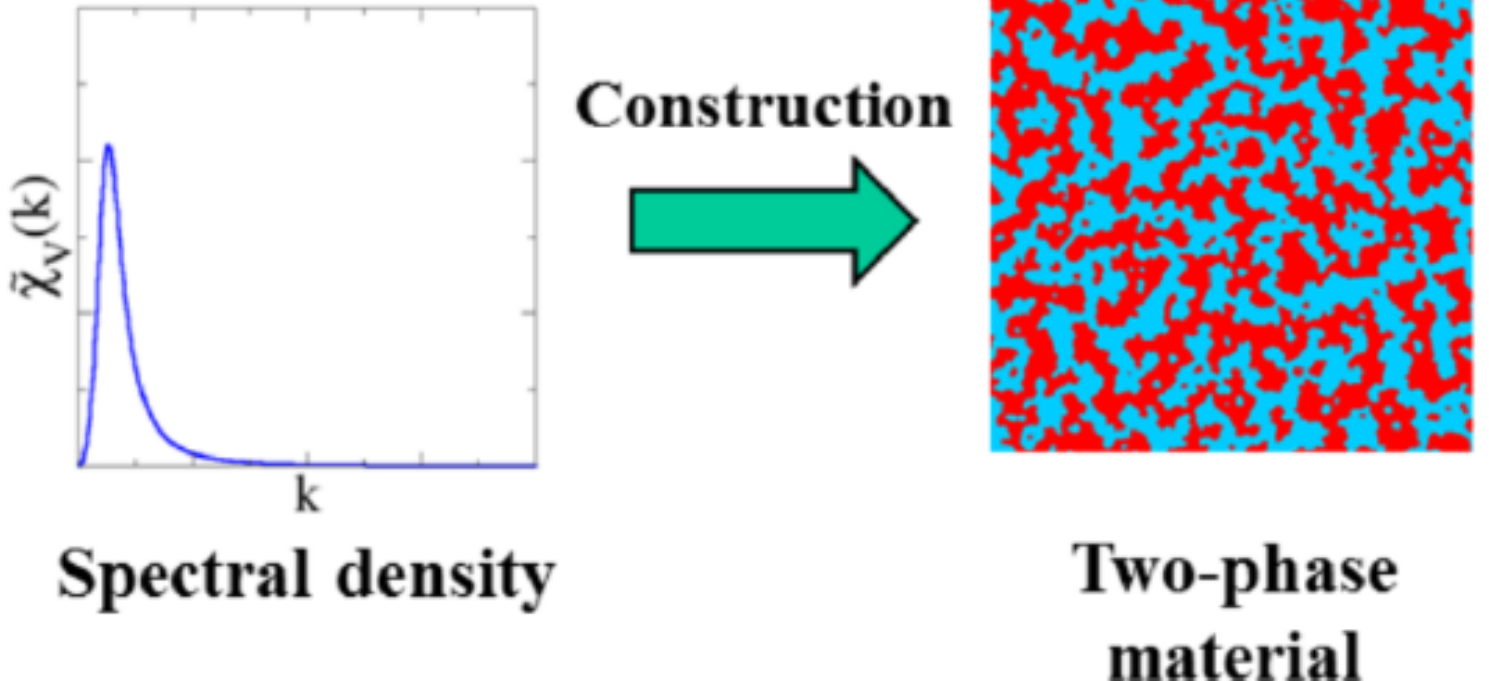}
\end{center}
\caption{A designed disordered hyperuniform spectral density with phase-inversion symmetry of class I and a corresponding
constructed two-phase material in which $\phi_1=\phi_2=1/2$. This figure is adapted
from Ref. \cite{Ch18}.}
\label{phase-inv}
\end{figure}


\subsection{Interrelations Between Number and Volume-Fraction Variances}
\label{packing}

The asymptotic coefficients involved in volume-fraction fluctuations for class I two-phase
systems have been related to those
for number variance fluctuations for class I hyperuniform point processes in the case in which one of the phases
consists of identical spherical inclusions of radius $a = D/2$ \cite{Za09}. This was accomplished
using the exact representation of two-point probability function $S_2$ for such systems
in terms of the $n$-particle correlation functions $g_n$ \cite{To82b, To83a,To02a}.
In particular, for a sphere packing at number density $\rho$, it was found  that the 
coefficient ${\overline A}_{_V}$, defined by (\ref{A_V}), is 
exactly related to ${\overline A}_{_N}$, defined by (\ref{A1}), according to
\begin{equation}
{\overline A}_{_V} = \frac{\phi}{2^d} {\overline A}_{_N} \label{atan},
\end{equation}
where  
\begin{equation}
\phi =\rho v_1(a)
\end{equation}
is the {\it packing fraction}, defined to be the fraction of space covered by the spheres.
 From relation (\ref{atan}), it immediately follows that a system of impenetrable spheres 
derived from a hyperuniform point configuration 
generates a hyperuniform heterogeneous medium with respect to fluctuations in the local volume fraction.
This also follows directly via the spectral density, as will be described below.

A rigorous upper bound on ${\overline B}_{_V}$ in terms of ${\overline B}_{_N}$,
defined by (\ref{B4}) and (\ref{B2}), respectively, has been derived \cite{Za09}, namely,
\begin{equation}
{\overline B}_{_V} \le \frac{ \phi }{2^d} {\overline B}_{_N} \label{BNtau},
\end{equation}
This bound can be obtained  from the more general inequality
\begin{equation}
\sigma^2_{_V}(R) \le \left(\frac{\phi}{\rho v_1(R)}\right)^2 \sigma^2_{_N}(R) \qquad (R \rightarrow \infty).
\end{equation}

When the heterogeneous medium consists of identical {\it overlapping} spheres \cite{To83b,To02a}, 
the volume-fraction variance is generally greater than those 
for impenetrable inclusions at a fixed volume fraction $\phi$ \cite{Lu90b, Qu97b, To02a}.
This behavior arises since the exclusion-volume effects in the latter case
induces a greater degree of uniformity in the
underlying point process. However, at a \emph{fixed reduced density} $\eta = \rho v_1(a)$,
which is greater than the volume fraction of penetrable spheres,  
the general upper bound 
given in (\ref{BNtau}) with the replacement $\phi \rightarrow \eta$ will still hold.

\subsection{Spectral Densities for Packings}

For statistically homogeneous packings of congruent spheres  of radius $a$ in $\mathbb{R}^d$ at number density $\rho$,
the auotcovariance function $\chi_{_V}(r)$ of the particle (sphere) phase is known exactly in terms of the pair correlation function \cite{To85b,To02a},
yielding the autocovariance function as
\begin{eqnarray}
{\chi}_{_V}({\bf r}) &=& \rho \,v_2^{int}(r;a) +\rho^2 v_2^{int}(r;a) \otimes h({\bf r}),
\label{S2-spheres}
\end{eqnarray}
where 
\begin{equation}
m_v(r;a) =\Theta(a-r)=\Bigg\{{1, \quad r \le a,\atop{0, \quad r > a,}}
\label{indicator}
\end{equation}
is a spherical particle indicator function. Previously, this quantity has been 
designated as $m(r;a)$ \cite{To82b,To02a}; we append the subscript $v$ 
here in order to distinguish it from the interface
indicator function of a sphere, denoted by $m_s(r;a)$, as detailed in Sec. \ref{interface}.
Moreover, $v_2^{int}(r;a)=v_1(a)\alpha_2(r;a)$ is the intersection volume of two spheres
of radius $a$ whose centers are separated by a distance $r$, where $v_1(a)$ and $\alpha_2(r;a)$
are defined as in (\ref{phi-var-1}),
Fourier transformation of (\ref{S2-spheres}) gives the corresponding spectral
density in terms of the structure factor \cite{To85b,To02a,Za09}:
\begin{eqnarray}
{\tilde \chi}_{_V}({\bf k})
&=& \phi {\tilde \alpha}_2(k;a) S({\bf k}),
\label{chi_V-S}
\end{eqnarray}
where ${\tilde \alpha}_2(k;a)$ is defined by (\ref{alpha-k}).

It follows immediately from relation (\ref{chi_V-S}) that the hyperuniformity of a sphere packing in terms 
of volume-fraction fluctuations can only arise  if the underlying point configuration (determined by the sphere
centers) is itself hyperuniform, i.e., ${\tilde \chi}_{_V}({\bf k})$ inherits the hyperuniformity property (\ref{hyper-2}) 
as well as its small-wavenumber behavior (i.e., its class) only through the structure factor, not ${\tilde \alpha}_2(k;a)$;
see Ref.  \cite{To16b} for additional details.
Accordingly, note that the spectral densities of maximally random jammed (MRJ) packings of identical  spheres have been computed  
\cite{Kl16} and shown to be class II of hyperuniform two-phase systems, as expected due to the fact
that the MRJ structure factor tends to zero linearly in $|\bf k|$ as  $|{\bf k}|\to 0$ \cite{Do05d,Ho12b,At16a,At16b}.
It is notable that relation (\ref{chi_V-S}) dictates
that ${\tilde \chi}_{_V}({\bf k})$ is zero at those wave vectors where $S({\bf k})$ is zero (i.e., where it is
stealthy) as well as
at the zeros of the function ${\tilde \alpha}_2(k;a)$, which is determined by the zeros of the Bessel function
$J_{d/2}(ka)$. The function  ${\tilde \chi}_{_V}({\bf k})$  will be zero at all
of the zeros of ${\tilde \alpha}_2(k;a)$ for any disordered packing free of any Dirac delta functions (Bragg peaks),
hyperuniform or not.

These results for the pair statistics in both direct and Fourier spaces 
have been generalized to the case of impenetrable spheres
with a continuous or discrete size distribution at overall number density $\rho$ \cite{Lu91,To02a}.
In the case of a continuous distribution in radius ${\cal R}$ characterized by
a probability density function $f({\cal R})$ that normalizes to unity, 
\begin{equation}
\int_{0}^{\infty} f({\cal R}) d{\cal R}=1.
\label{f-normal}
\end{equation}
Let us denote the size average of a function $G({\cal R})$ by
\begin{equation}
\langle G({\cal R}) \rangle_{\footnotesize{\cal R}} \equiv \int_{0}^{\infty} f({\cal R}) G({\cal R}) d{\cal R}.
\label{size}
\end{equation}
It is known that the packing fraction and the autocovariance function  are given respectively by \cite{Lu91,To02a}
\begin{equation}
\phi = \rho \langle v_1({\cal R}) \rangle_{\footnotesize{\cal R}}
\label{pack}
\end{equation}
and 
\begin{equation}
\chi_{_V}({\bf r}) =  \rho   \langle v_2^{int}(r;{\cal R}) \rangle_{\footnotesize{\cal R}} 
+\rho^2 \Big\langle \Big\langle 
 \,m_v(r;{\cal R}_1) \otimes m_v(r;{\cal R}_2) \otimes h({\bf r};{\cal R}_1,{\cal R}_2) \Big\rangle_{\footnotesize{\cal R}_1} \Big\rangle_{\footnotesize{\cal R}_2}, \label{chi-poly-r}
\end{equation}
where $h({\bf r};{\cal R}_1,{\cal R}_2)$ is the appropriate generalization
of the total correlation function for the centers of two spheres of radii ${\cal R}_1$ and ${\cal R}_2$ separated
by a distance $r$. Note that generally $h$ is not symmetric with respect
to interchange of the components, i.e., $h({\bf r};{\cal R}_1,{\cal R}_2) \neq h({\bf r};{\cal R}_2,{\cal R}_1)$.
Fourier transformation of (\ref{chi-poly-r}) gives the corresponding
spectral density
\begin{equation}
{\tilde \chi}_{_V}({\bf k})  =  \rho   \langle {\tilde m}_v^2(k;{\cal R}) \rangle_{\footnotesize{\cal R}} +  \rho^2 \Big\langle  \Big\langle {\tilde m}_v(k;{\cal R}_1)  {\tilde m}_v(k;{\cal R}_2)  {\tilde h}({\bf k};{\cal R}_1,{\cal R}_2)  \Big\rangle_{\footnotesize{\cal R}_1}  \Big\rangle_{\footnotesize{\cal R}_2}.
\label{chi-poly}
\end{equation}
The hyperuniformity
condition for a polydisperse sphere packing  is obtained by setting
the right-hand side of (\ref{chi-poly}) at $\bf k=0$ equal to zero,
implying that the second term involving ${\tilde h}({\bf k};{\cal R}_1,{\cal R}_2)$ must be 
equal to -\,$\rho   \langle {\tilde m}_v^2(k=0;{\cal R}) \rangle_{\footnotesize{\cal R}}=-\,\rho  \langle v^2_1({\cal R}) \rangle_{\footnotesize{\cal R}}$.

One can obtain corresponding results for  spheres
with $M$ different  radii $a_1,a_2,\ldots,a_M$ from the continuous case \cite{To90d,To02a} by letting 
\begin{equation}
f(R) = \sum_{i=1}^{M} \frac{\rho_{i}}{\rho} \delta ({R} -a_{i}), 
\label{f-discrete}
\end{equation}
where $\rho_{i}$ is the number density 
of type-$i$ particles, respectively, and $\rho$ is the {\it total number density}.
Substitution of (\ref{f-discrete}) into (\ref{pack}), (\ref{chi-poly-r}) and (\ref{chi-poly}) 
yields the corresponding packing fraction, autocovariance function and spectral density, respectively, as
\begin{equation}
\phi =\sum_{i=1} \rho_i v_1(a_i),
\end{equation}
\begin{equation}
\chi_{_V}({\bf r})  = \sum_{i=1}^M \rho_i v_2^{int}(r;a_i)  
+  \sum_{i=1}^M  \sum_{j=1}^M  \rho_i \rho_j  
 \,m_v(r;a_i) \otimes m_v(r;a_j) \otimes h({\bf r};a_i,a_j) 
\label{chi-poly-r-dis}
\end{equation}
and
\begin{equation}
{\tilde \chi}_{_V}({\bf k}) = \rho \sum_{i=1}^M {\tilde m}_v^2(k;a_i)  S({\bf k};a_i,a_i)
+ \rho \sum_{i\neq j}^M   
 \,{\tilde m}_v(k;a_i) {\tilde m}_v(k;a_j) S({\bf k};a_i,a_j),
\label{chi-poly-k-dis}
\end{equation}
where
\begin{equation}
 S({\bf k};a_i,a_j)=\frac{\rho_i}{\rho} \delta_{ij} +\frac{\rho_i \rho_j}{\rho}{\tilde h}({\bf k};a_i,a_j) 
\end{equation}
is the so-called {\it partial} structure factor associated with components $i$ and $j$ \cite{Han13}
and  $\delta_{ij}$ is the Kronecker delta.}\\

\noindent{Remarks:}\\

\noindent{1. 
When each subpacking associated with each component is hyperuniform, i.e., $S(0;a_i,a_i)=0$ for all $i$ so that
the first term on the right side of (\ref{chi-poly-k-dis})
is zero at $\bf k=0$], the second term must also be identically zero at $\bf k=0$ (sum of cross terms vanish),
as shown elsewhere \cite{To16b}, leading to the hyperuniformity
of the entire packing, i.e.,  ${\tilde \chi}_{_V}({\bf k=0})$. Such a packing is called {\it multihyperuniform} \cite{Ji14}.
Any decoration of a crystal in which each component
is arranged in a periodic fashion is multihyperuniform. By contrast, constructing disordered multihyperuniform polydisperse
packings is much more challenging. The photoreceptor mosaics in avian retina
\cite{Ji14} and certain multicomponent hard-sphere plasmas \cite{Lo18a} are such examples.

\noindent{2. Generally speaking, examining the structure factor $S({\bf k})$ of the point
configurations derived from the centers of spheres in a polydisperse packing could lead one to incorrectly
conclude that the packing is not hyperuniform. The proper way to ascertain hyperuniformity in this case is through a packing's spectral
density ${\tilde \chi}_{_V}({\bf k})$ \cite{Za11a,Za11c,Za11d,Kl16}. This approach has been profitably
used to diagnose hyperuniformity in experimental studies of disordered jammed polydisperse packings of colloidal spheres and  emulsions \cite{Dr15,Ri17}.}

\noindent{3. The discrete Fourier-space version of the spectral density (\ref{chi-poly-k-dis})
for a multicomponent hard-sphere packing in a fundamental cell
under periodic boundary conditions was given in Ref. \cite{Za11c}.}

\subsection{Nonspherical Windows}

It is noteworthy that there are anomalous circumstances in which the use
of nonspherical observation windows yield a volume-fraction variance $\sigma^2_{_V}({\bf R})$ that do not decrease faster
than the window volume, even though the two-phase system would be deemed to be
hyperuniform according to the spectral condition (\ref{hyper-2}). For example, using square 
windows with fixed orientations, Zachary, Jiao and Torquato \cite{Za11d} studied the
two-dimensional checkerboard model and square lattice decorated by identical squares,
and showed that the corresponding volume-fraction variances decrease with increasing window size 
(while preserving the window shape) as slow as the inverse of the
window volume, despite the fact that these systems are periodic and hence hyperuniform.
Such anomalous fluctuations can arise for periodic two-phase media if nonspherical windows with fixed
orientations share the symmetries of the periodic media. In the case of lattices,
analogous anomalous fluctuations in the number of lattice points within certain
shaped and oriented nonspherical windows
can occur \cite{Be01,Ki17}; see Sec. \ref{nonspherical}. 

\section{Hyperuniformity as a Critical-Point Phenomenon and Scaling Laws}
\label{inverted}

In the ensuing discussion, we return to considering number fluctuations in point configurations.
In their study of density fluctuations  in fluid systems near the critical point,
Ornstein and Zernike  \cite{Or14} defined the
direct correlation function $c({\bf r})$ via an integral
equation that links it to the pair correlation function $g_2({\bf r})$
or, equivalently, the total correlation function $h({\bf r})$.
Specifically, they proposed
a decomposition of $h$ into a ``direct'' part that involves
only $c$ and an ``indirect'' part that involves a convolution of $c$ and $h$:
\begin{equation}
h({\bf r})=c({\bf r})+\rho c({\bf r}) \otimes  h({\bf r}),
\label{OZ}
\end{equation}
where $\otimes$ denotes a convolution integral.
This integral equation has primarily been used to study pair correlations 
of liquids in equilibrium \cite{Han13},
but it is perfectly well-suited to examine correlations in  nonequilibrium
systems, which are of general interest in the study of hyperuniform systems.
Fourier transforming (\ref{OZ}) and solving for ${\tilde h}({\bf k})$ yields
\begin{equation}
{\tilde h}({\bf k})=\frac{{\tilde c}({\bf k})}{1-\rho{\tilde c}({\bf k})},
\label{H}
\end{equation}
where ${\tilde c}({\bf k})$ is the Fourier transform of $c(\bf r)$.

A system at a  thermal critical point, such as a liquid-vapor or magnetic critical point,
has a fractal structure \cite{Wi74,Bi92}, which is characterized by hyperfluctuations, i.e., 
density fluctuations become unbounded, and in this sense is anti-hyperuniform. For general hyperfluctuating
systems,  we see from 
the fluctuation-compressibility theorem (\ref{comp}) that
${\tilde h}({\bf k=0})$ or, equivalently,  the volume integral
of $h({\bf r})$ over all space is unbounded. Thus, $h({\bf r})$ is a long-ranged function characterized by a power-law 
tail that decays to zero slower than $|{\bf r}|^{-d}$ \cite{Or14,Ka66,St71,Wi74,Bi92}. On the other hand,
relation (\ref{H}) dictates that ${\tilde c}({\bf k=0})$ remains bounded
at the critical point density defined by $\rho_c= {\tilde c}({\bf k=0})^{-1}$ and hence $c({\bf r})$ is 
sufficiently short-ranged (roughly, the same range 
as an effective pair potential $v({\bf r})$) in the sense that its volume integral over all space is bounded. Indeed, 
while there is no rigorous proof, there are strong theoretical arguments, using diagrammatic expansions,
that show that the large-distance asymptotic behavior of the direct correlation
of disordered phases is exactly proportional  to the pair potential \cite{Jo64,Stell77}, i.e.,
\begin{equation}
c({\bf r}) \sim -\beta v({\bf r}) \qquad (|{\bf r}| \rightarrow \infty),
\label{c-v}
\end{equation}
where $\beta=(k_B T)^{-1}$ is an inverse temperature. In Fourier space,
this is tantamount to
\begin{equation}
{\tilde c}({\bf k}) \sim -\beta {\tilde v}({\bf k}) \qquad (|{\bf k}| 
\rightarrow 0),
\label{c-v-F}
\end{equation}
where ${\tilde v}({\bf k})$ is the Fourier transform of $v({\bf r})$.

\subsection{Direct Correlation Function for Hyperuniform Systems and New Critical Exponents}
\label{oz}

The direct correlation function $c({\bf r})$ of a hyperuniform
system behaves in an unconventional manner. 
We can express ${\tilde c}({\bf k})$ in terms of ${\tilde h}({\bf k})$ using 
relation (\ref{H}): 
\begin{equation}
{\tilde c}({\bf k})= \frac{{\tilde h}({\bf k})}{S({\bf k})}=\frac{{\tilde h}({\bf k})}{1+\rho{\tilde h}({\bf k})}.
\label{c}
\end{equation}
By definition, a hyperuniform system is one in which ${\tilde h}({\bf k=0})=-1/\rho$,  
i.e., the volume integral of $h(\bf r)$ exists, as, for example, in the case in which
$h({\bf r})$ is sufficiently short-ranged in the sense  that it decays to zero
faster than $|{\bf r}|^{-d}$. Interestingly, this
means that the denominator on the right side of
(\ref{c}) vanishes at ${\bf k=0}$ 
and therefore ${\tilde c}({\bf k=0})$ diverges to $-\infty$. 
This implies that the the volume integral of $c({\bf r})$ does not exist
and hence the real-space direct correlation function
$c(\bf r)$ is long-ranged, i.e., decays slower than $|{\bf r}|^{-d}$.
We see that this stands in diametric contrast to standard thermal critical points 
in which the total correlation function is long-ranged and the direct correlation function
is sufficiently short-ranged such that its volume integral exists \cite{Wi65,Ka66,Fi67,Wi74},
as discussed immediately above. For this reason, it has been said that
hyperuniform systems are at an ``inverted" critical point \cite{To03a}.

Thus, in analogy with thermal critical points,  one expects 
the direct correlation function for a hyperuniform state at critical
reduced density $\phi_c$ to have the following  power-law asymptotic decay for large $|{\bf r}|$ and sufficiently large $d$:
\begin{equation}
c({\bf r}) \sim -\frac{1}{|{\bf r}|^{d-2+\eta}} \qquad (|{\bf r}| \rightarrow \infty),
\label{asymp2}
\end{equation}
 where $(2-d) < \eta <  2$ is a new {\it critical} exponent associated with $c(\bf r)$ for
hyperuniform systems; the upper bound $\eta<2$ ensures
that $c({\bf r})$  is a long-ranged function and the lower bound $\eta> 2-d$
ensures that  $c({\bf r})$  decays at large distances.
(The critical exponent $\eta$ associated with $h({\bf r})$ for thermal systems belonging
to the standard Ising universality class is given exactly
by $1/4$ for $d=2$
and approximately by $0.05$ for $d=3$; see Ref. \cite{Hu87}.)  
This scaling form for $c({\bf r})$ together with the Ornstein-Zernike 
relation (\ref{OZ}) implies the following corresponding power-law
form for the total correlation function at $\phi=\phi_c$:
\begin{equation}
h({\bf r}) \sim -\frac{1}{|{\bf r}|^{d+2-\eta}} \qquad (|{\bf r}| \rightarrow \infty),
\label{asymp3}
\end{equation} 
which  is always a sufficiently short-ranged function such that its volume integral over all space exists whenever $\eta <2$.
We say a system has {\it quasi-long-ranged} (QLR) correlations if $h({\bf r})$ decays 
to zero like a power law $1/|{\bf r}|^a$ for large $|{\bf r}|$  with an exponent $a>d$ so that
the volume integral of $h({\bf r})$ exists.

It is notable that for any hyperuniform particle system derived from an equilibrium ensemble, we
can associate an effective long-ranged interparticle pair potential $v({\bf r})$ whose
asymptotic form is given precisely by (\ref{asymp2}), i.e.,
\begin{equation}
\beta v({\bf r}) \sim \frac{1}{|{\bf r}|^{d-2+\eta}} \qquad (|{\bf r}| \rightarrow \infty),
\label{v-c}
\end{equation}
where we have used relation (\ref{c-v}). This effective repulsive potential can be regarded to be
a {\it generalized} Coulombic interaction between ``like-charged" particles. To maintain
stability, the total potential energy must also include 
a ``background" contribution of equal and opposite ``charge", i.e., the system must have
{\it overall charge neutrality}. Notably, whenever $\eta=0$ so that
$S({\bf k}) \sim |{\bf k}|^2$ as $|{\bf k}| \rightarrow 0$,
the effective pair potential (\ref{v-c}) reduces to the standard Coulombic interaction for $d \ge 3$
and hence the system belongs to class I; see Eq. (\ref{sigma-N-asy}).
In summary, in order to drive an equilibrium many-particle system to a hyperuniform state, 
effective {\it long-ranged repulsive pair interactions} of the form (\ref{v-c}) are required.
On the other hand, long-ranged interactions are not necessary to achieve hyperuniformity
if the system is out of equilibrium; prototypical nonequilibrium examples are MRJ particle
packings that have pure short-ranged hard-particle interactions \cite{Do05d,Sk06,Za11a,Ji11c,Ch14a,At16a,At16b}.
This and other nonequilibrium hyperuniform classes will be described in greater detail in Sec. \ref{non-eq}.

Fourier transformation of the aforementioned large-$|{\bf r}|$ scaling laws for direct-space functions yield
corresponding small-$|\bf k|$ scaling laws for hyperuniform systems. Specifically, 
the Fourier transform of the direct-correlation-function scaling  (\ref{asymp2}) is given by
\begin{equation}
{\tilde c}({\bf k}) \sim -\frac{1}{|{\bf k}|^{\alpha}} \qquad (|{\bf k}| \rightarrow 0),
\label{c-eta}
\end{equation}
where $\alpha$ is the positive exponent introduced in (\ref{S-asy}),
which  determines the hyperuniformity class [cf. (\ref{sigma-N-asy})] and  is related to $\eta$ via
\begin{equation}
\alpha=2 -\eta.
\label{alpha-eta}
\end{equation}
Note that  ${\tilde c}({\bf k})$ contains a singularity at the origin,
the order of which is determined by the exponent $\alpha$. This implies
that a length scale based on the volume integral $-\int_{\mathbb{R}^d} c({\bf r})d{\bf r}$ 
will grow as a hyperuniform state is approached, as specifically described in Sec. \ref{growing}.
The scaling law  (\ref{c-eta})
combined with (\ref{c}) yields the corresponding asymptotic form of the structure factor:
\begin{equation}
S({\bf k}) \sim |{\bf k}|^{\alpha} \qquad (|{\bf k}| \rightarrow 0),
\label{S-eta}
\end{equation}
Importantly, while the exponent $\alpha$ in the structure-factor scaling (\ref{S-eta})
can  generally take on any real positive value up to infinity, it must obey the 
upper bound $\alpha <d$ in the scaling (\ref{c-eta}) in order for the direct correlation
function $c({\bf r})$ to decay at large distances. Note also that while the scaling (\ref{c-eta}) in the special 
case $\alpha=d$ is not integrable in Fourier space, one may sometimes be able to associate a long-range behavior
of direct correlation function $c({\bf r})$ that is not a power-law function; for example, a logarithmic law, i.e.,
\begin{equation}
c({\bf r}) \sim -\beta v({\bf r}) \sim \ln(|{\bf r}|) \qquad (|{\bf r}| \rightarrow \infty),
\label{C-V}
\end{equation}   
which is recognized to be the two-dimensional Coulomb interaction.
We will see in Secs. \ref{invariant}, \ref{Matrices} and \ref{Det} that 
such effective ``log-gas" interactions can arise
in hyperuniform two-dimensional $g_2$-invariant processes of class I, one-dimensional Coulombic systems of 
class II, and two-dimensional Coulombic systems of class I, respectively. 
Observe that when $\alpha=1$  ($\eta=1$), we conclude from (\ref{v-c})
that the effective potential corresponds to a $d$-dimensional charged system
but with Coulombic interactions in a higher $(d+1)$-dimensional space and hence
is a hyperuniform system of class II, which is consistent with results reported in Ref. \cite{Fo96}.

We can define critical exponents associated
with the manner in which certain quantities diverge
as the critical (hyperuniform) point is approached.
Consider a point configuration with a reduced density
$\phi$ that is nearly hyperuniform and which can be
made hyperuniform by increasing and/or decreasing
the density. We denote by $\phi_c$ the reduced density at the hyperuniform
state. The reduced densities $\phi$ and $\phi_c$ play
the same role as temperature $T$ and critical temperature
$T_c$, respectively, in the analogous thermal problem
in the vicinity of a critical point.
In some cases, the direct correlation function of a  many-particle system at a dimensionless
density in the vicinity of
a hyperuniform state, i.e., for $|\phi_c -\phi| \ll 1$, in sufficiently high dimensions 
has  the following large-$r$ asymptotic form:
\begin{equation}
c({\bf r}) \sim \frac{\exp(-|{\bf r}|/\xi)}{|{\bf r}|^{d-2+\eta}},
\end{equation}
where $\xi$ is the {\it correlation length}. If the system approaches
a hyperuniform state from  below the critical density $\phi_c$, the correlation length 
and inverse of the structure factor
at $k=0$, $S^{-1}(0)$, which is proportional to ${\tilde c}(0)$, are described by 
the following scaling laws:
\begin{equation}
\xi  \sim  (1 - \frac{\phi}{\phi_c})^{-\nu} \quad (\phi \rightarrow \phi_c^{-}),
\label{c-power}
\end{equation}
\begin{equation}
S^{-1}(0) \sim  \left(1-\frac{\phi}{\phi_c}\right)^{-\gamma}, \quad 
(\phi \rightarrow \phi_c^{-}), 
\label{S-power}
\end{equation}
where $\nu$ and $\gamma$ are nonnegative critical exponents. Observe
that $S^{-1}(0)$ is a measure of the degree of hyperuniformity of a system away from a
critical point. Combination of three previous scaling laws
leads to the following interrelation between the exponents:
\begin{equation}
\gamma=(2-\eta)\nu.
\label{inter}
\end{equation}
Of course, the specific values of the critical exponents determine the ``universality'' class of
the hyperuniform system. Analogous critical exponents can be defined for densities
near but above $\phi_c$. Table \ref{critical} provides a summary of the scaling laws
and critical exponents.


\begin{table}
\centering
\parbox{3.9in}{\caption{Definitions of the critical exponents
in the vicinity of or at a hyperuniform state. Here $S^{-1}(0)$
is the inverse of the structure factor at ${\bf k=0}$, $\xi$ is the
correlation length, and $c({\bf r})$ is the direct correlation function.
The scaling laws for the latter apply for sufficiently large dimensions. }
\label{critical}
}
\begin{tabular}{cc}\hline\hline
Exponent& Asymptotic behavior\\ \hline 
$\gamma$ & $S^{-1}(0)  \sim  (1 - \frac{\phi}{\phi_c})^{-\gamma}$\quad ($\phi \rightarrow \phi_c^{-}$)   \\
$\gamma^{~\prime}$ & $S^{-1}(0)  \sim
(\frac{\phi}{\phi_c}-1)^{-\gamma^\prime}$\quad ($\phi \rightarrow \phi_c^{+}$)   \\
$\nu$ & $\xi  \sim  (1 - \frac{\phi}{\phi_c})^{-\nu}$\quad ($\phi \rightarrow \phi_c^{-}$)   \\
$\nu^{~\prime}$ & $\xi  \sim  (\frac{\phi}{\phi_c}-1)^{-\nu^\prime}$\quad ($\phi \rightarrow \phi_c^{+}$)   \\
$\eta$ & $c({\bf r})  \sim  |{\bf r}|^{2-d-\eta} $\quad ($\phi =\phi_c$)    \\ 
$\xi$ &\quad $c({\bf r}) \sim \exp(-|{\bf r}|/\xi)/|{\bf r}|^{d-2+\eta}$\quad ($|\phi_c -\phi| \ll 1$) \\ \hline\hline
\end{tabular}
\end{table}

In the following subsection, we describe so-called $g_2$-invariant hyperuniform 
point configurations in which the critical behavior
can be exactly determined in any dimension $d$.

\subsection{Critical behavior of $g_2$-Invariant Hyperuniform Point Configurations}
\label{invariant}

A $g_2$-invariant process
is one in which a chosen nonnegative form for
the pair correlation function $g_2$ remains
invariant over a nonvanishing density range while keeping
all other relevant macroscopic variables fixed \cite{To02b}. The upper
limiting ``terminal'' density is the point above which
the nonnegativity condition on $S({\bf k})$
[cf. (\ref{factor})] would be violated. Thus, at the terminal
or critical density, if $S({\bf k=0})=0$, the system is hyperuniform, if realizable. 
In Ref. \cite{To03a}, a variety of $g_2$-invariant processes in $\mathbb{R}^d$
in which the number variance scales like the window surface area (i.e., belong to class I)  were exactly 
studied. Here we summarize some of those exact results by reporting  the 
corresponding surface-area
coefficients, structure factors,
direct correlation functions,  and  associated critical exponents.

\subsubsection{Step-Function $g_2$}

The simplest  $g_2$-invariant process that was considered by Torquato and Stillinger \cite{To03a} is one
in which a radial pair correlation function is defined by the unit step function, i.e.,
\begin{equation}
g_2(r)=\Theta(r-D)
\label{step-f}
\end{equation}
or, equivalently, a total correlation function $h(r)=-\Theta(D-r)$.
Any system with such pair correlations
corresponds to a packing of identical sphere with hard-core diameter $D$. In the special case of 
identical hard spheres in equilibrium in the limit $\rho \rightarrow 0$,
$g_2$ is exactly given by this step-function form as well as those with finite
densities, as described below.  
Substitution of $h(r)=-\Theta(D-r)$ into (\ref{factor}) gives the structure factor for $\phi$ in the
range $0 \le \phi \le \phi_c$ to be
\begin{equation}
S(k)=1-\Gamma(1+d/2) 
\left(\frac{2}{kD}\right)^{d/2}
\left(\frac{\phi}{\phi_c}\right) J_{d/2}(kD),
\end{equation}
where 
\begin{equation}
\phi_c=\frac{1}{2^d}
\end{equation}
is the terminal or critical density, which is the density at which this  system is hyperuniform. 
The fact that $h(r)$ is exactly zero for all $r>D$ results in a structure factor $S(k)$ that is analytic
not only at the origin, but for all $k$. The Ornstein-Zernike relation (\ref{c}) yields an
exact expression for the Fourier transform of the direct
correlation function:
\begin{equation}
{\tilde c}(k)= \frac{\displaystyle -\left(\frac{2\pi}{kD}\right)^{d/2} D^d
J_{d/2}(kD)}{\displaystyle 1-\Gamma(1+d/2) 
\left(\frac{2}{kD}\right)^{d/2}
\left(\frac{\phi}{\phi_c}\right) J_{d/2}(kD)}.
\end{equation}
Notably, Ref. \cite{Cr03} provides numerical evidence
that the step-function $g_2$ is to a good approximation realizable by systems of impenetrable
$d$-dimensional spheres (with $d=1$ and $d=2$)
for densities up to the terminal density. 
Thus,  satisfying the nonnegativity conditions 
on $g_2(r)$ and $S(k)$ in this instance is sufficient to ensure realizability 
of such a point process to within numerical accuracy. 
Figure \ref{step-g2} shows results for the sphere packing corresponding to the step-function
$g_2$-invariant process in three dimensions at the critical hyperuniform packing
density $\phi_c=1/8$.

\begin{figure}[H]
\begin{center}
{\includegraphics[  width=2.6in, keepaspectratio,clip=]{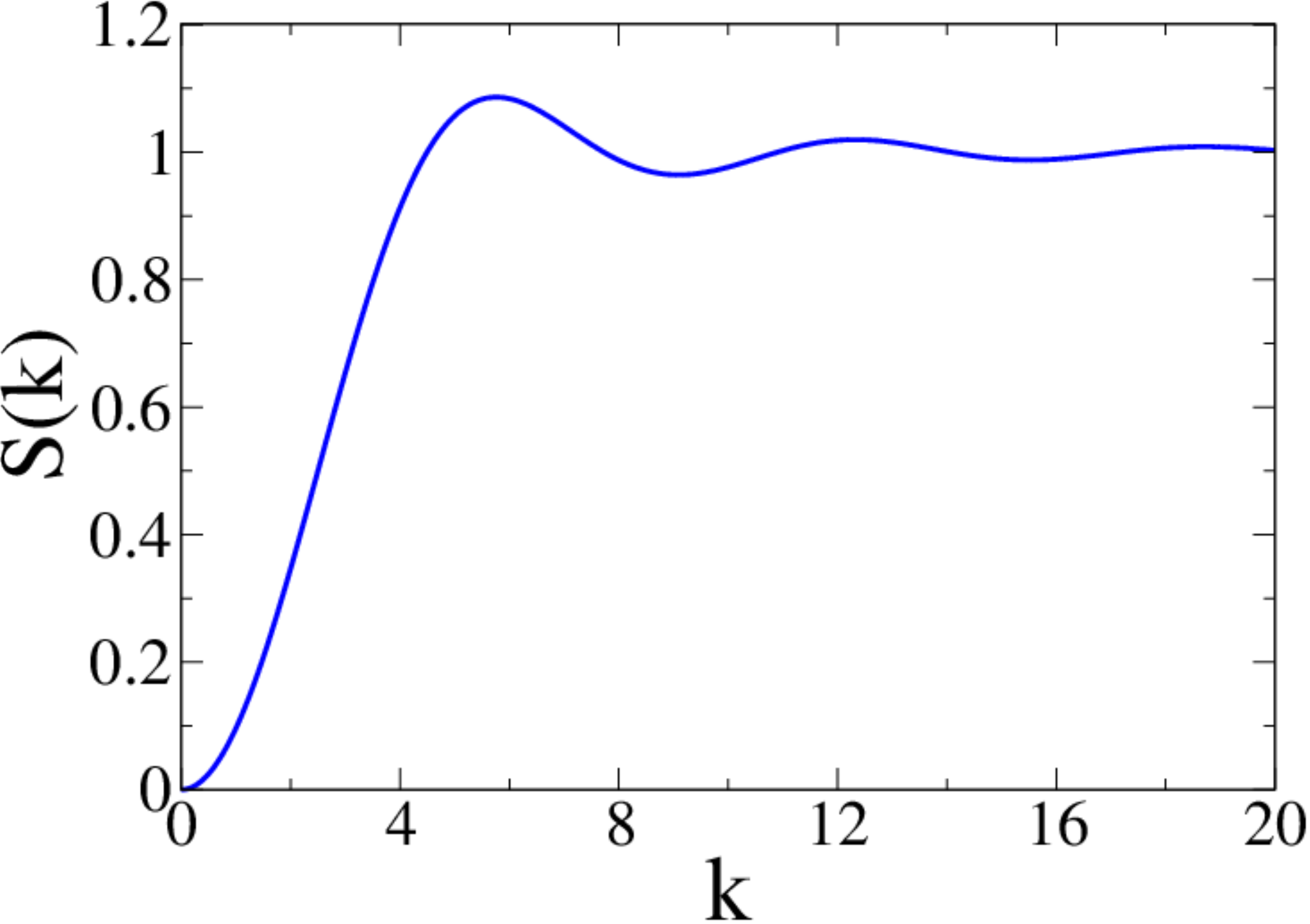}\\ \vspace{0.2in}
\includegraphics[  width=2.6in, keepaspectratio,clip=]{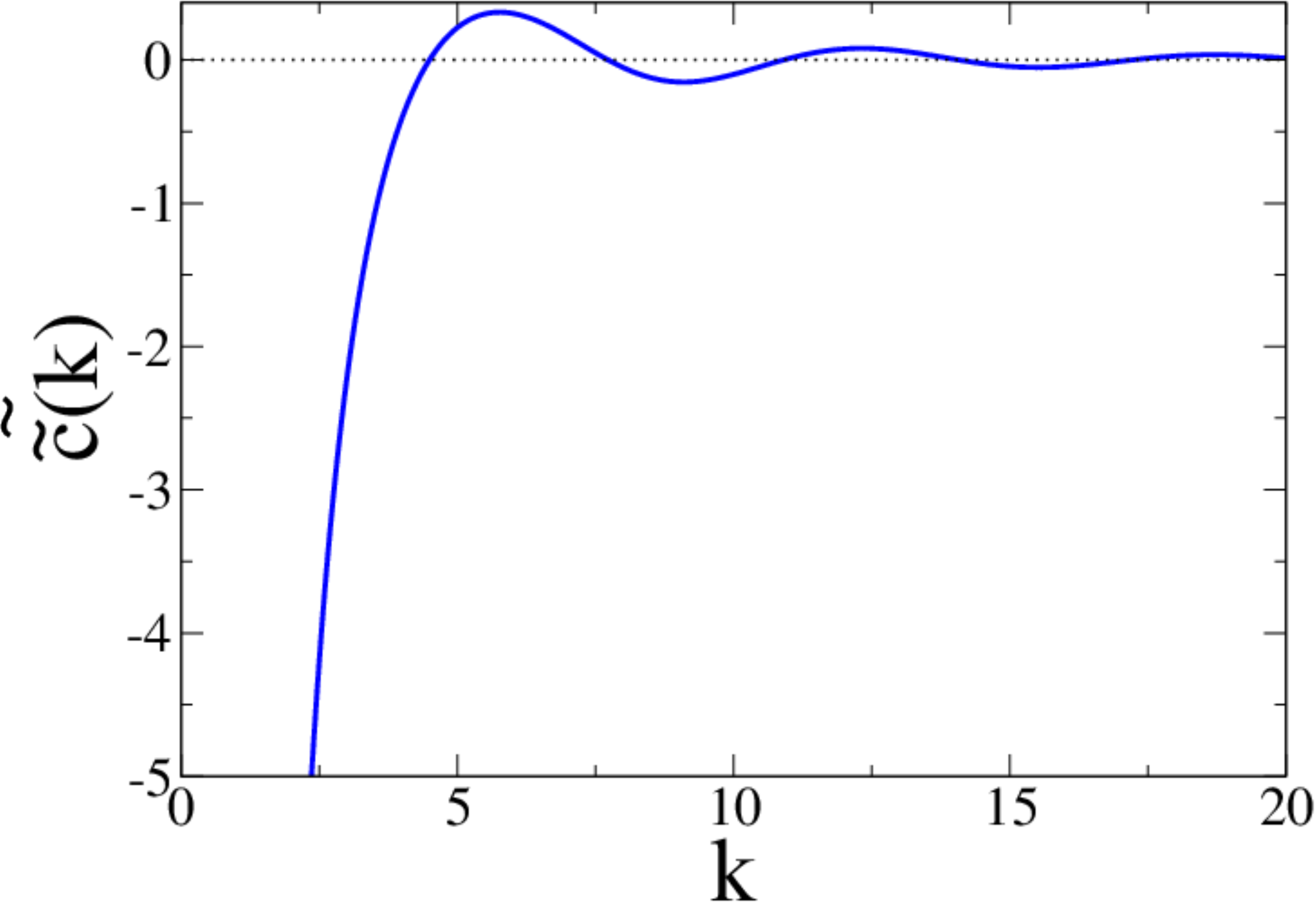}\\ \vspace{0.2in}
\includegraphics[  width=2.6in, keepaspectratio,clip=]{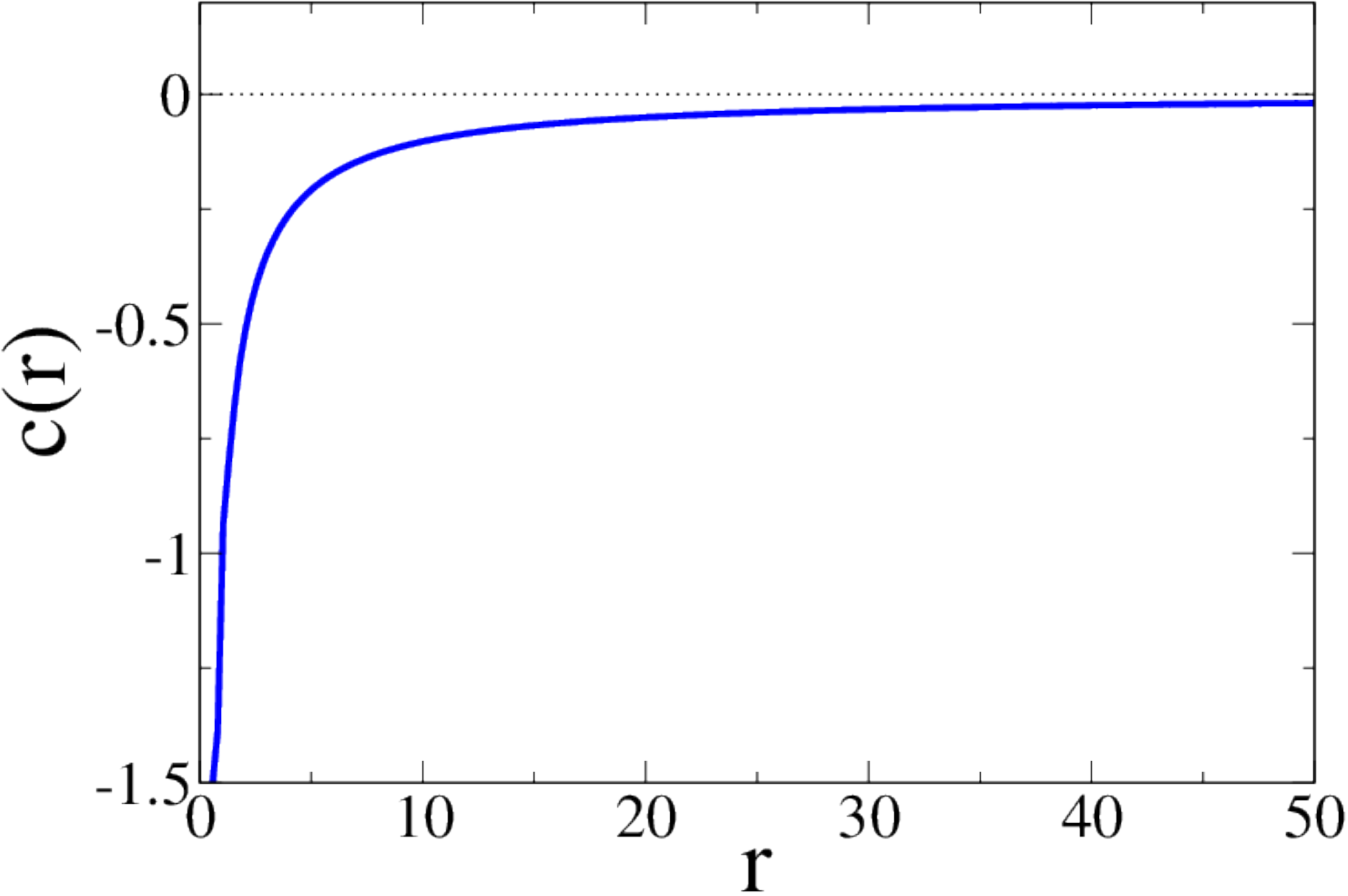}}
\caption{Results for the sphere packing corresponding to the step-function
$g_2$-invariant process in three dimensions at the critical hyperuniform packing
density $\phi_c=1/8$, where the hard-core diameter $D$ is taken to be unity. Top panel: 
Structure factor. Middle panel: Fourier transform
of the direct correlation function. Bottom panel: Direct correlation function. }  \label{step-g2}
\end{center}
\end{figure}

Combination of  (\ref{step-f}), (\ref{A2}) and (\ref{B2})
yields the volume and surface-area coefficients as
\begin{equation}
A_N=S(k=0)=1-2^d\phi, \qquad B_N=\frac{\overline \Lambda}{2^d \phi}=
\frac{ 2^{d-2}d^2\Gamma(d/2)}
{\Gamma((d+3)/2)
\Gamma(1/2)} \phi.
\end{equation}
The reduced density $\phi$ defined by (\ref{phi})
(equivalent to the covering fraction of the hard cores
of diameter $D$) lies
in the range $0 \le \phi \le \phi_c$, where
\begin{equation}
\phi_c=\frac{1}{2^d}
\end{equation}
is the terminal or critical density, i.e., the density
at which the system is hyperuniform, where $A_N=0$ and 
\begin{equation}
B_N={\overline \Lambda}=
\frac{d^2 \Gamma(d/2)}
{4\Gamma((d+3)/2)
\Gamma(1/2)}.
\label{lambda-1}
\end{equation}
The values of the  surface-area
coefficient ${\overline \Lambda}$ for $d=1,2$ and 3 are given
in Tables \ref{1d}, \ref{2d} and \ref{3d}, respectively.

Thus, the small-$k$ expansions of 
$S(k)$ and ${\tilde c}(k)$, which determine their behavior
in the vicinity of the critical point,   are respectively given by
\begin{equation}
S(k)=\displaystyle \left(1-\frac{\phi}{\phi_c} \right)
+ \frac{1}{2(d+2)} \frac{\phi}{\phi_c}(kD)^2 +{\cal O}[(kD)^4]
\label{S-step}
\end{equation}
and 
\begin{equation}
{\tilde c}(k)=\frac{-v_1(D) }{\displaystyle 
 \left(1-\frac{\phi}{\phi_c} \right)
+ \frac{1}{2(d+2)} \frac{\phi}{\phi_c}(kD)^2 +{\cal O}[(kD)^4]},
\label{c-step}
\end{equation}
where $v_1(D)$ is the volume of a $d$-dimensional sphere
of radius $D$ [cf. (\ref{v1})]. Comparison of equations (\ref{S-step}) and (\ref{c-step}) to the scaling relations
(\ref{S-eta}) and  (\ref{S-power})  yield the exponents $\eta=0$ and $\gamma=1$. 
At the critical point $\phi=\phi_c$, we see that 
\begin{equation}
S(k) \sim k^2 \quad \mbox{and} \quad {\tilde c}(k) \sim - k^{-2} \quad (k \rightarrow 0).
\label{step-scalings}
\end{equation}
The correlation length $\xi$ can be extracted from relation (\ref{c-step}), which can be rewritten as
\begin{equation}
k^2 {\tilde c}(k)+ \xi^{-2} {\tilde c}(k)= -G, \qquad k D \ll 1
\label{k-diff1}
\end{equation}
where
\begin{eqnarray}
\xi &=& \frac{D}{[ 2(d+2)\phi_c]^{1/2}} \left(1-\frac{\phi}{\phi_c} \right)^{-1/2}, 
\qquad \phi \rightarrow \phi_c^{-}, \label{xi-step}\\ \nonumber \\
G &=& \frac{2(d+2) v_1(D)}{D^2} \frac{\phi_c}{\phi}, \label{G-step}
\end{eqnarray}
 Comparison of (\ref{xi-step}) to scaling relation (\ref{c-power}) yields the exponent $\nu =1/2$.
The exponent values $\gamma=1$, $\nu=1/2$,
and $\eta=0$ are consistent with the interrelation (\ref{inter}).
Inversion of (\ref{k-diff1}) yields the partial differential equation
\begin{equation}
\nabla^2 c(r)-\xi^{-2} c(r)= G \delta({\bf r}), \qquad r \gg D,
\label{diff1}
\end{equation}
where the spherically symmetric Laplacian operator $\nabla^2$ in any dimension $d$ is given by 
\begin{equation}
\nabla^2=\frac{1}{r^{d-1}} \frac{\partial}{\partial r} \left[ r^{d-1} \frac{\partial }
{\partial r}\right].
\label{Laplace}
\end{equation}
We see that the direct correlation function in real space for large $r$ 
is determined by the Green's function of the {\it linearized Poisson-Boltzmann} equation.

Let us first determine the solutions of (\ref{diff1})  
at the critical point $\phi=\phi_c$, where $\xi$ diverges to
infinity. Thus, the asymptotic behavior of $c(r)$ for $r \gg D$ is given by the infinite-space Green's function for the $d$-dimensional Laplace equation \cite{To02a}, and so we obtain
\begin{equation}
c(r) = \cases{
\displaystyle{-6 \left(\frac{r}{D}\right)}, 
     &$d = 1$, \cr
 \displaystyle{4 \ln \left(\frac{r}{D}\right)}, 
     &$d = 2$, \cr
 \displaystyle{- \frac{2(d+2)}{d(d-2) } \left(\frac{D}{r}\right)^{d-2}},     
     &$d \ge 3$.
}
\label{c-green1}
\end{equation}
Observe that it is only for $d \ge 3$ that $c(r)$ follows the power-law
form (\ref{c-eta}) with an exponent
$\eta=0$. The fact that $\eta$ takes on an integer value is due to the fact
that ${\tilde h}(k)$ is an analytic function.
We will see in Sec. \ref{ocp} that the real-space direct correlation function of 
the one-component plasma \cite{Ba78,Na80,Ch80,Ba80,Ja81,Le00,Se14,Sa15,Se15,Leb16,Pe18} for $d \ge 2$  has precisely the same
asymptotic form as the ones indicated in (\ref{c-green1}), albeit with different
amplitudes (prefactors).

For fixed $r$ and in the limit $\xi \rightarrow \infty$, the solutions of (\ref{diff1}) are
\begin{equation}
c(r) = \cases{
\displaystyle{-6 \frac{\phi_c}{\phi} \left(\frac{\xi}{D}\right) \exp(-r/\xi)}, 
     &$d = 1$, \cr
 \displaystyle{4 \frac{\phi_c}{\phi}\ln \left(\frac{r}{D}\right) \exp(-r/\xi)}, 
     &$d = 2$, \cr
 \displaystyle{- \frac{2(d+2)\phi_c}{d(d-2) \phi}
 \left(\frac{D}{r}\right)^{d-2}\exp(-r/\xi)},     
     &$d \ge 3$.
}
\label{c-green2}
\end{equation}
On the other hand, observe that as $r \rightarrow \infty$
for fixed $\xi$, the asymptotic behavior changes according to the relation
\begin{equation}
c(r)=- \frac{(d+2)\sqrt{2\pi}\phi_c}{\Gamma(1+d/2) \phi}
\left(\frac{D}{\xi}\right)^{(d-3)/2} \left(\frac{D}{r}\right)^{(d-1)/2}
\exp(-r/\xi), \qquad d \ge 1.
\end{equation}

\subsubsection{Step+Delta Function $g_2$}

Here we consider the $g_2$-invariant process defined by
a radial distribution function  that consists 
of the  aforementioned unit step function plus a delta function contribution
that acts at $r=D$:
\begin{equation}
g_2(r)=\Theta(r-D)+ \frac{Z}{\rho s_1(D)}\delta(r-D), 
\label{step-delta}
\end{equation}
where $Z$ is a nonnegative constant and $s_1(D)$ is
the surface area of a sphere of radius $D$ defined by (\ref{area-sph}).
The function (\ref{step-delta}) was one of several examples studied by Torquato
and Stillinger \cite{To02b} to understand the relationship
between short-range order and maximal density in sphere
packings. In that investigation, $Z$ was interpreted
as the average contact coordination number. Here we consider
their case IV in which the condition
\begin{equation}
Z=\frac{2^d d}{d+2}\phi
\end{equation}
is obeyed in order to constrain the location of the minimum
of $S(k)$ to be at $k=0$, thus enforcing hyperuniformity. Here the reduced density $\phi$ lies
in the range $0 \le \phi \le \phi_c$, and
\begin{equation}
\phi_c=\frac{d+2}{2^{d+1}}
\end{equation}
is the critical density, which also results as a special limit
of the pair correlation function corresponding to the 
dilute and narrow limit of the {\it square-well potential} studied by Sakai, Stillinger,
and Torquato \cite{Sa02}.

Substitution of (\ref{step-delta}) into (\ref{A2}) and (\ref{B3})
yields the volume and surface-area coefficients as
\begin{equation}
A_N=S(k=0)=1-\frac{2^{d+1}}{d+2}\phi, \qquad B_N=\frac{\overline \Lambda}{2^d \phi}=
\frac{ 2^{d-2}d^2 \Gamma(d/2)}
{(d+2)\Gamma((d+3)/2)
\Gamma(1/2)}\phi.
\end{equation}
At the critical density, $A_N=0$ and 
\begin{equation}
{\overline \Lambda}=2^d \phi_c B_N=\frac{ d^2 
(d+2)\Gamma(d/2)}
{16\Gamma((d+3)/2)
\Gamma(1/2)}.
\label{lambda-2}
\end{equation}
The values of the scale-independent surface-area
coefficient ${\overline \Lambda}$ for $d=1,2$ and 3 are given
in  Tables  \ref{1d}, \ref{2d} and \ref{3d}, respectively.

The structure factor and Fourier transform of the direct
correlation function can be exactly obtained for $\phi$ in the
range $0 \le \phi \le \phi_c$ \cite{To03a} and yields the following small-$k$ behavior:
\begin{equation}
S(k)=\displaystyle \left(1-\frac{\phi}{\phi_c} \right)
+ \frac{1}{8(d+2)(d+4)} \frac{\phi}{\phi_c}(kD)^4 +{\cal O}[(kD)^6]
\label{S-step-delta}
\end{equation}
and 
\begin{equation}
{\tilde c}(k)=\frac{-2v_1(D) }{\displaystyle 
 \left(1-\frac{\phi}{\phi_c} \right)
+\frac{1}{8(d+2)(d+4)} \frac{\phi}{\phi_c}(kD)^4 +{\cal O}[(kD)^6]}.
\label{c-step-delta}
\end{equation}
Comparison of expression (\ref{S-step-delta}) to the scaling relations
(\ref{S-eta}) and  (\ref{S-power})  yield the exponents $\eta=-2$ and $\gamma=1$. 
The correlation length
$\xi$ is defined via (\ref{c-step-delta}), which we rewrite as
\begin{equation}
k^4 {\tilde c}(k)+ \xi^{-4} {\tilde c}(k)= -G, \qquad k D \ll 1
\label{k-diff2}
\end{equation}
where
\begin{eqnarray}
\xi &=& \frac{D}{[ 8(d+2)(d+4)\phi_c]^{1/4}} \left(1-\frac{\phi}{\phi_c} \right)^{-1/4}, 
\qquad \phi \rightarrow \phi_c^{-}, \label{xi-step-delta}\\ \nonumber\\
G &=& \frac{16(d+2)(d+4)v_1(D)}{D^4} \frac{\phi_c}{\phi} . \label{G-step-delta}
\end{eqnarray}
Comparison of (\ref{xi-step-delta}) to
the power law (\ref{c-power}) yields the exponent $\nu =1/4$.
The exponent values $\gamma=1$, $\nu=1/4$,
and $\eta=-2$ are consistent with the interrelation (\ref{inter}).
Inversion of (\ref{k-diff2}) yields the partial differential
equation
\begin{equation}
\nabla^4 c(r)+\xi^{-4} c(r)= -G \delta({\bf r}), \qquad r \gg D,
\label{diff2}
\end{equation}
where $\nabla^4 \equiv \nabla^2\nabla^2$ is the spherically symmetric 
{\it biharmonic operator}, and $\nabla^2$ is given by (\ref{Laplace}). 

The solutions of (\ref{diff2})  
at the critical point $\phi=\phi_c$ ($\xi \rightarrow \infty$)
are given by the infinite-space
Green's function for the $d$-dimensional biharmonic equation. It is only for
$d \ge 5$ that the solutions admit a power law of the form (\ref{c-power})
with an exponent $\eta=-2$, namely,
\begin{equation}
c(r) = - \frac{8(d+2)(d+4)}{d(d-2)(d-4) }\left(\frac{D}{r}\right)^{d-4}, \qquad d \ge 5.
\label{c-green3}
\end{equation}

\section{Hyperuniform Disordered Classical Ground States}
\label{STEALTHY}

The ``collective-coordinate" optimization procedure represents  a powerful 
{\it reciprocal-space-based} approach to generate disordered hyperuniform classical many-particle systems 
in $d$-dimensional Euclidean space  $\mathbb{R}^d$ with prescribed (or targeted)
structure factors. These particle configurations correspond 
to exotic disordered classical  ground states of many particles 
interacting with certain bounded long-ranged interactions, including isotropic pair potentials 
\cite{Fa91,Uc04b,Uc06b,Ba08,Ba09a,Ba09b,Za11b,To15,Zh15a,Zh15b}, anisotropic pair potentials \cite{Ma13},
and two-, three- and four-body potential functions \cite{Uc06b}. Not only are these ground states
endowed with novel physical properties, such as photonic characteristics \cite{Fl09b,Fl13,Man13a,Man13b,Ha13,Le16,Fr16,Ma16,Fr17}, 
phononic band gaps \cite{Gk17}, dielectric characteristics \cite{Wu17,Ch18,Kl18},
and  transport properties \cite{Zh16b,Ch18} (see Sec. \ref{properties}),
but their corresponding excited (positive temperature) states are characterized by
singular thermodynamic properties, as detailed in Sec. \ref{excited}. In what follows, we first
describe numerical procedures to generate a large class of tunable disordered hyperuniform
ground-state configurations with high precision and 
then we discuss predictive ensemble theories for so-called ``stealthy" ground states.

\subsection{Stealthy Configurations via Collective-Coordinate Optimizations}

The simplest setting involves the consideration of pairwise additive potentials 
$v({\bf r})$ that are bounded and integrable such that their
Fourier transforms ${\tilde v}({\bf k})$ exist. If $N$
identical point particles reside in a fundamental cell $F$ of volume $v_F$ in $d$-dimensional
Euclidean space $\mathbb{R}^d$ 
at positions ${\bf r}^N \equiv {\bf r}_1, \ldots, {\bf r}_N$ under periodic
boundary conditions, the total potential energy $\Phi_N({\bf r}^N)$ [equal to the pairwise 
sum of the potential $v({\bf r})$] also has the following Fourier representation:
\begin{equation}
\Phi_N({\bf r}^N) =\frac{N}{2v_F} \left[\sum_{\bf k} {\tilde v}({\bf k}){\cal S}({\bf k})
-\sum_{\bf k} {\tilde v}({\bf k})\right],
\label{pot}
\end{equation}
where ${\cal S}({\bf k })$ is the single-configuration structure factor, as defined by (\ref{scatter}).
The key idea behind the collective-coordinate approach is to require that ${\tilde v}({\bf k})$
be a bounded, positive function with support for wave vectors $\bf k \in Q$, where $\bf Q$ is a prescribed subregion of Fourier space,
and if the particles are arranged so  that ${\cal S}({\bf k})$ is driven to its minimum value of zero 
for all wave vectors where ${\tilde v}({\bf k})$ has support (except $\bf k=0$), then it is clear from relation (\ref{pot})
that the system must be at its ground state or global energy minimum.
These ground-state configurations
have been called  ``stealthy" \cite{Ba08} because the structure factor  
${\cal S}({\bf k})$  is {\it constrained} to be zero for  $\bf k \in Q$ (except $\bf k=0$), 
meaning that they completely suppress single scattering of incident radiation 
for these wave vectors. 

For purposes of illustration, we will focus on the case in which $\bf Q$ is taken to be a sphere
of radius $K$ centered at the origin, and hence  by construction, such stealthy ground states are also hyperuniform, since the structure factor
${\cal S}({\bf k})$  is constrained to be zero within this ``exclusion" sphere, implying no single scattering events
from infinite wavelength down to a wavelength of the order of $2\pi/K$; see
Fig. \ref{pattern} for an example of such a scattering pattern.
The corresponding class 
of stealthy radial  potential functions  ${\tilde v}(k)$ in $\mathbb{R}^d$ have support
within this sphere and the general functional form 
\begin{equation}
{\tilde v}({\bf k})= V(k) \Theta(K-k),
\label{v-k}
\end{equation}
where, for simplicity, $V(k)$ is infinitely differentiable in the open interval $(0,K)$ 
and $\Theta(x)$ is the Heaviside step function.  
The corresponding direct-space radial pair potential $v(r)$ is 
necessarily a delocalized, long-ranged function that is integrable in $\mathbb{R}^d$.
Two specific families of potentials that fall within the  class of  stealthy interactions
are the ``power-law" and ``overlap" potentials \cite{To15}, examples of which are depicted
in Figs. \ref{plot-power} and \ref{plot-over}.
While the form of the direct-space power-law potential  is
 very similar to the weakly decaying  Friedel oscillations of the electron density 
in a variety of systems, including molten metals  as well as graphene \cite{As76,Bac10}, 
the overlap potential $v(r)$ mimics effective interactions
that arise in certain  polymer systems \cite{Ha14}.
In either family of potentials,  the direct-space potential $v(r)$
is controlled by an envelop that cannot decay slower than the  inverse power law $1/r^{d+1}$;
see Ref. \cite{To15} for additional details.

\begin{figure}[H]
\begin{center}
{\includegraphics[  width=2.5in, keepaspectratio,clip=]{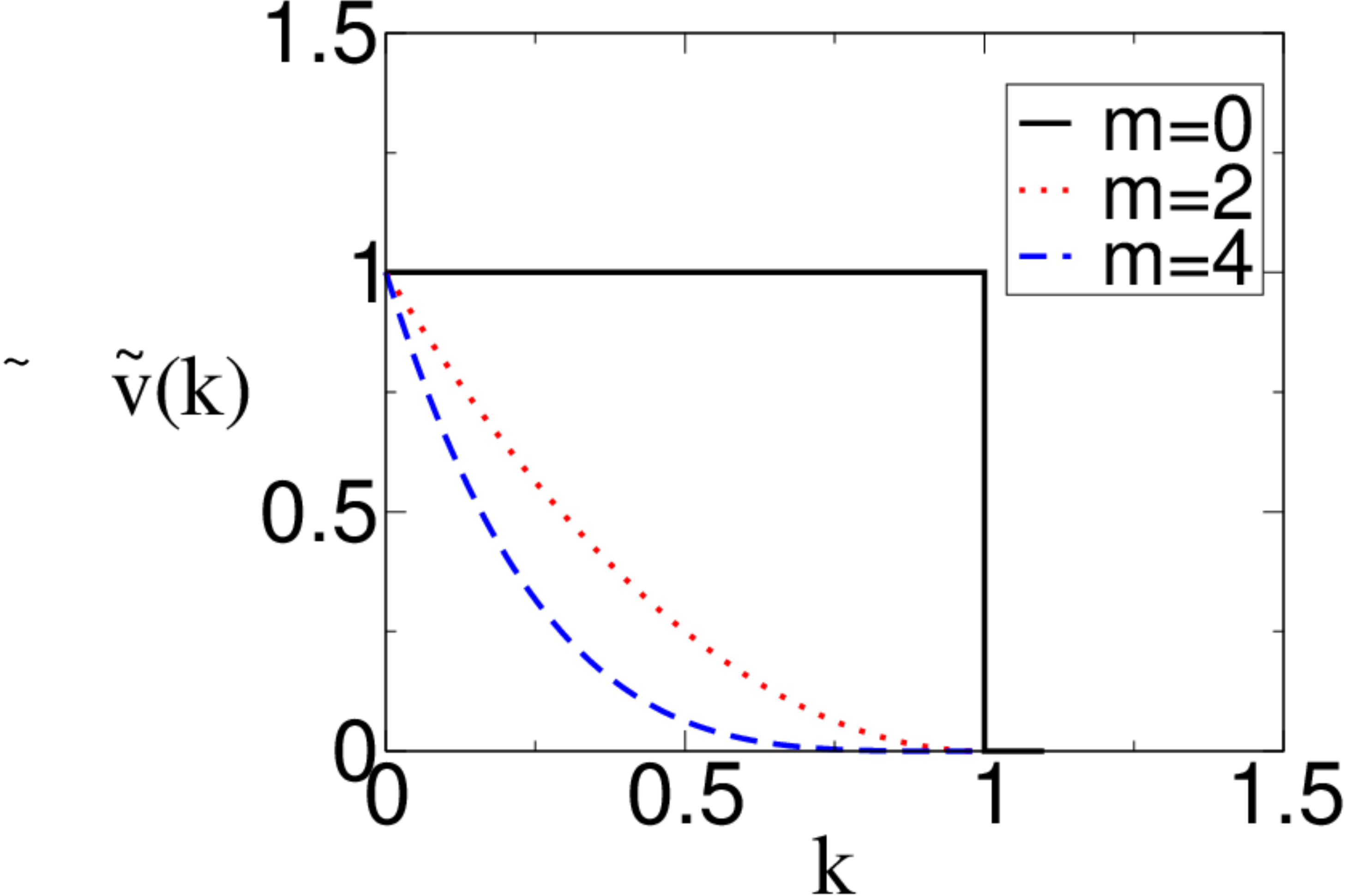} 
\includegraphics[  width=2.5in, keepaspectratio,clip=]{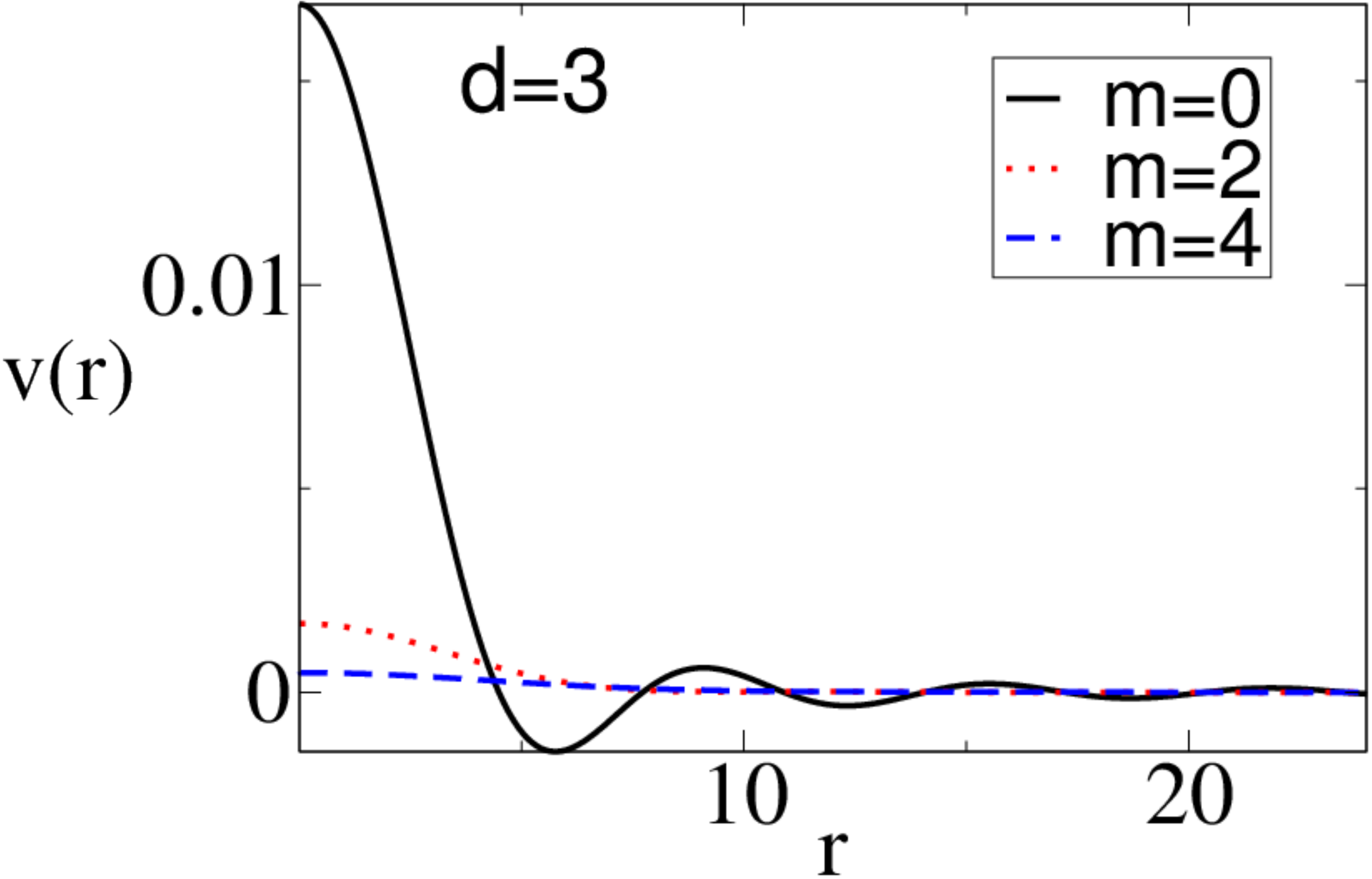}}
\caption{Left panel: Fourier power-law potential ${\tilde v}(k)$ for the special cases
$m=0$, $2$, and $4$ that apply for any $d$ with $V(0)=K=1$. Right panel: Corresponding direct-space power-law potentials
$v(r)$ in the instance $d=3$. }  \label{plot-power}
\end{center}
\end{figure}

\begin{figure}[ht]
\begin{center}
{\includegraphics[  width=2.5in,clip=, keepaspectratio]{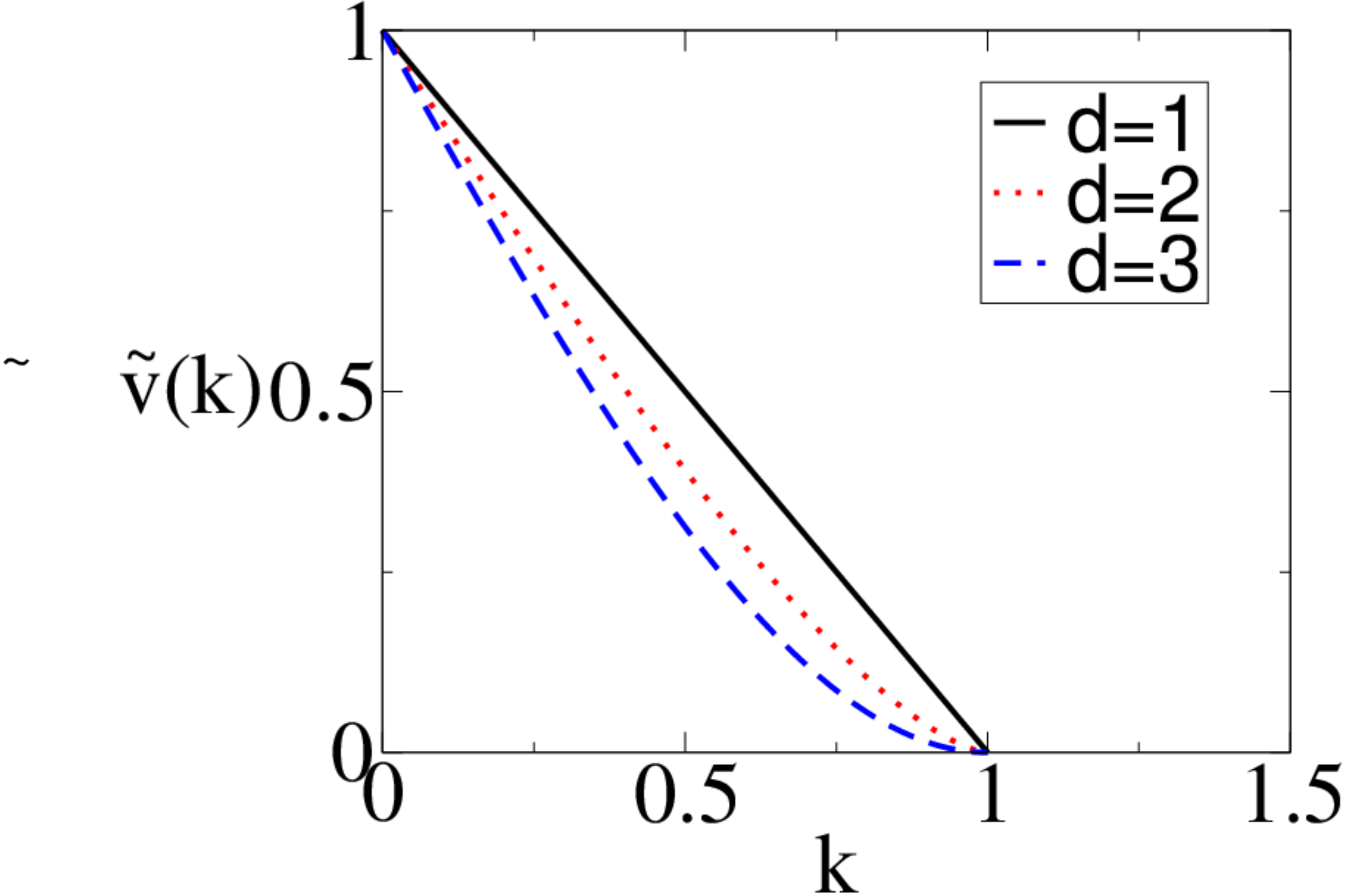} 
\includegraphics[  width=2.5in, clip=,keepaspectratio]{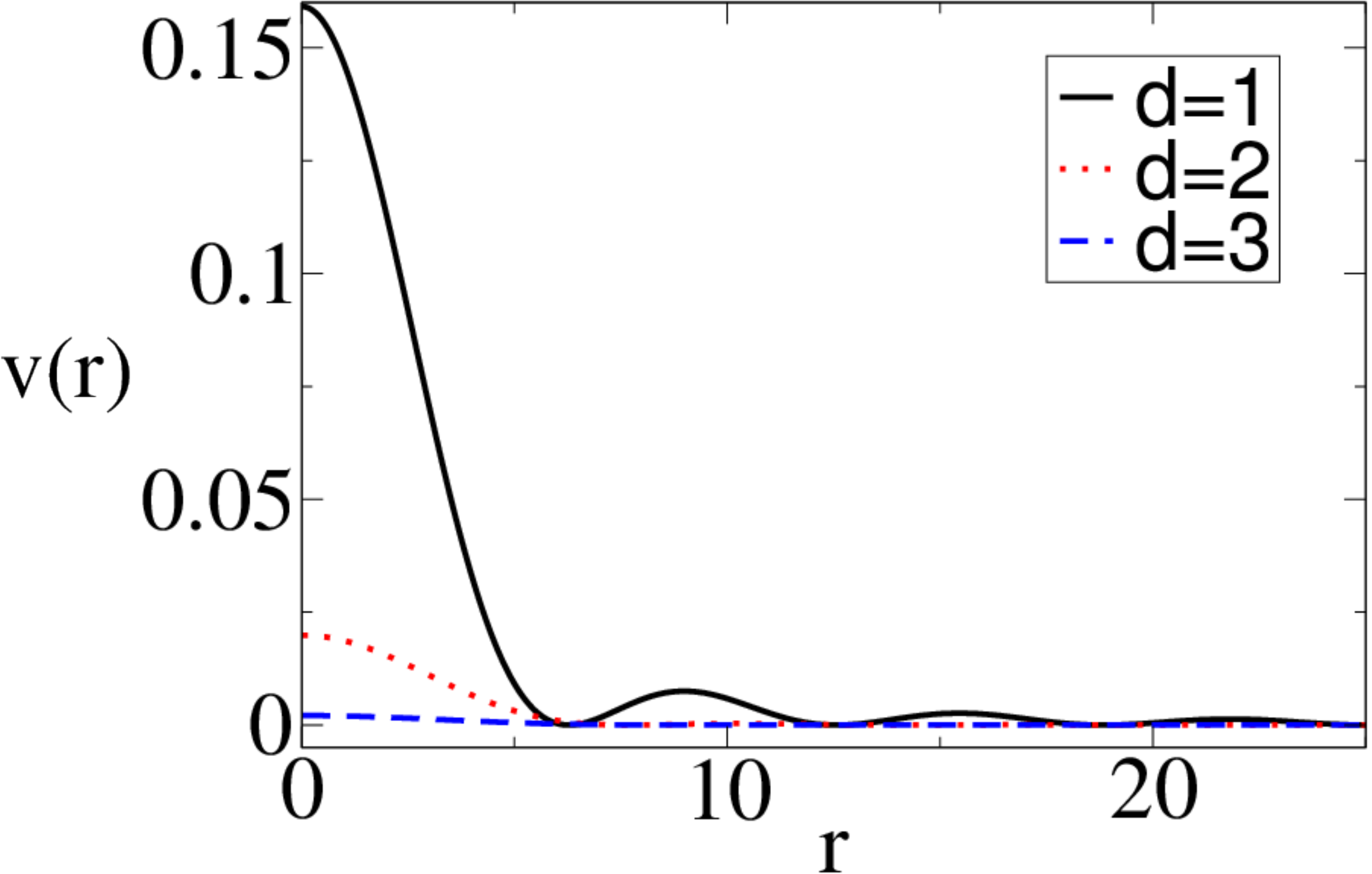}}
\caption{Left panel: Fourier overlap potential ${\tilde v}(k)$ for the first three
space dimensions with $V(0)=K=1$. Right panel: Corresponding direct-space overlap potentials
$v(r)$, which oscillate but are always non-negative.}  
\label{plot-over}
\end{center}
\end{figure}

\begin{figure}[H]
\begin{center}
\includegraphics[  width=2.75in, keepaspectratio,clip=]{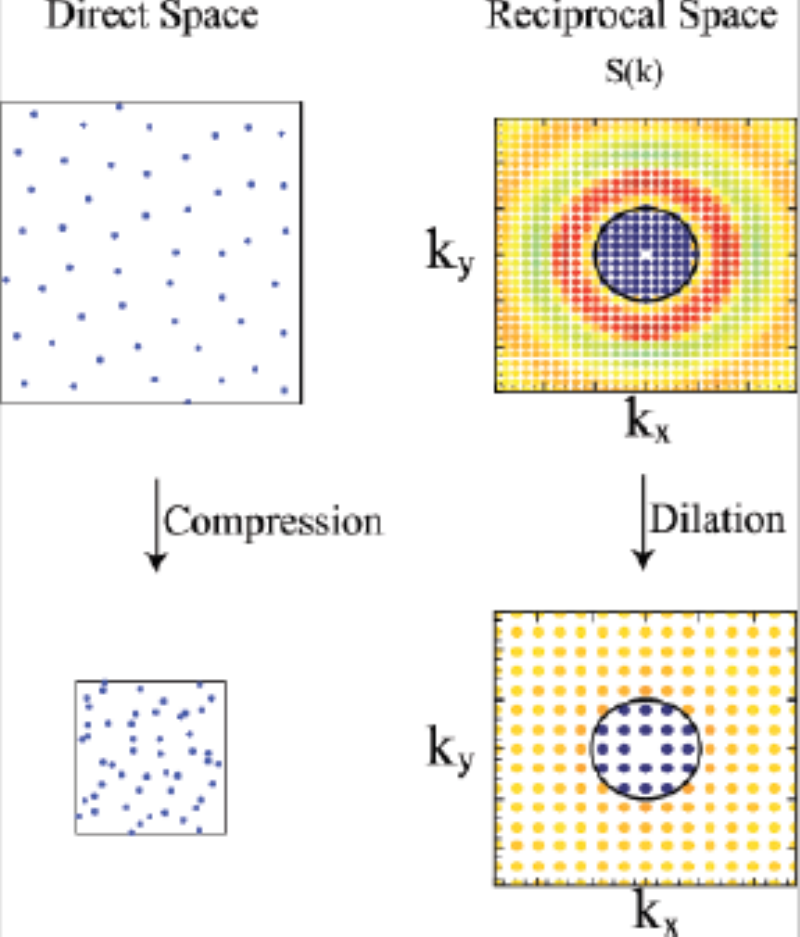}\vspace{-0.1in}
\caption{Schematic taken from Ref. \cite{To15} that illustrates the inverse relationship between the
number density $\rho$ and relative fraction of constrained degrees of freedom $\chi$
for  a fixed reciprocal-space exclusion-sphere of radius $K$ (where dark blue  $\bf k$ points signify zero intensity
with green, yellow, and red points indicating increasingly larger intensities)
for a stealthy ground state. A compression of a disordered ground-state configuration {with a fixed number of particles $N$ }in direct space leads to a dilation of the lattice spacing in reciprocal space,
thus reducing the number of constrained  wave vectors within the exclusion sphere and hence increasing 
the dimensionality of the ground-state manifold per particle $d_c$.}
\label{rho_chi}
\end{center}
\end{figure}

The nature of the ground-state configuration manifold (e.g., the degree of order) depends
on the number of constrained wave vectors.
As the number of $\bf k$ vectors for which ${\cal S}({\bf k})$ is constrained to be zero increases, i.e., as $K$ increases,
the dimensionality of the ground-state configuration manifold per particle, $d_C$, decreases. 
 Let $M(K)$ be the number of  independently
constrained wave vectors.   The parameter $\chi = M(K)/[d(N-1)]$,
gives a measure of the relative fraction of constrained degrees of freedom compared
to the total number of degrees of freedom $d(N-1)$ (subtracting out the system  translational degrees of freedom) \cite{Ba08},
which, in the thermodynamic limit, is given by \cite{To15}
\begin{equation}
\rho \,\chi =\frac{v_1(K)}{2d\,(2\pi)^d},
\label{rho-chi}
\end{equation}
where $v_1(K)$ is the volume of a $d$-dimensional sphere (hypersphere) of radius $K$ [cf. (\ref{v1})].
We see that for fixed $K$ and $d$, which sets the potential, 
the ``tuning" parameter $\chi$ is inversely proportional to the number density $\rho$.
The dimensionality of the configuration space per particle is given by $d_C=d(1-2\chi)$ for $0 \le \chi \le 1/2$ \cite{To15}
and hence depends on $\chi$ or, equivalently, the density $\rho$.
For sufficiently small $\chi$, the ground states are highly degenerate 
and overwhelmingly  disordered.
Clearly, when the system is free of any constraints, i.e., if $\chi=0$, it is structurally
like a noninteracting classical ideal gas
(even if it is not an ideal gas thermodynamically) \cite{To15}. This situation runs counter
to traditional understanding that  ideal-gas configurations correspond to the opposite zero-density limit
of classical systems of particles. The reason for this inversion of limits
is due to the fact that a compression of the system in direct space
leads to a dilation of the lattice spacing in reciprocal space, thus expelling $\bf k$-vectors that
can be constrained within the exclusion zone and thus increasing $d_c$. This compression process is
schematically shown in Fig. \ref{rho_chi} \cite{To15}. 
 If  $\chi$ is a positive but very small number, the ground states remain disordered
and highly degenerate, even if $d_C$ has now been suddenly reduced 
due to the imposed constrained degrees of freedom. While the ground-state manifold contains periodic configurations (e.g., Bravais lattices and lattices
with a basis) in this ``disordered" regime, these are sets of zero measure in the thermodynamic limit.
When the configurational dimensionality collapses to zero in a low-density  regime ($\chi=1/2$), there can be
concomitant  phase transition to a crystal phase that depends on the space dimension and
the ensemble under consideration \cite{Fa91,Uc04b,Su05,Ba08}.

Generally, a numerically obtained ground-state configuration depends on the number of particles $N$ within the fundamental cell, initial particle configuration, shape of the fundamental cell, and particular optimization technique that is used \cite{To15}.
Various optimization techniques have been employed
to find the globally energy-minimizing configurations for stealthy and other prescribed structure factors
within an exceedingly small numerical 
tolerance \cite{Fa91,Uc04b,Uc06b,Ba08,Ba09a,Ba09b,Za11b,Ma13,Zh15a,Zh15b}.
Figures \ref{stealth-1} and \ref{stealth-2} show numerically obtained stealthy  hyperuniform
disordered configurations in both two and three dimensions at selective values
of $\chi$, as obtained from Ref. \cite{Zh15a}. Regardless of the dimension, it is clear that the degree of short-range order 
(tendency for particles to repel one another) increases
as $\chi$ increases \cite{Uc04b,Ba08,To15,Zh15a}.

\begin{figure}[H]
\begin{center}
{\includegraphics[  width=3.in, keepaspectratio,clip=]{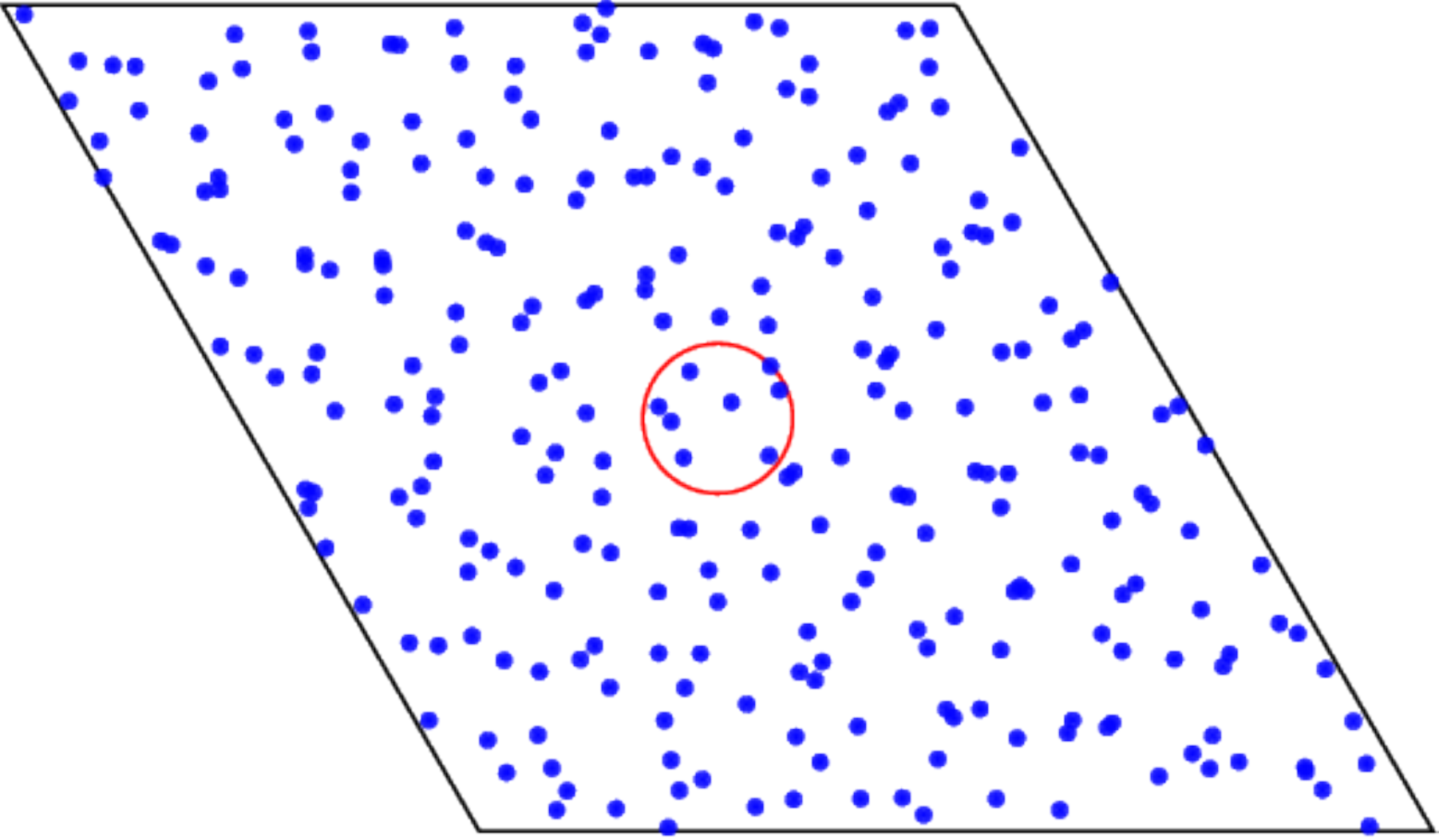} 
\includegraphics[  width=3.in, keepaspectratio,clip=]{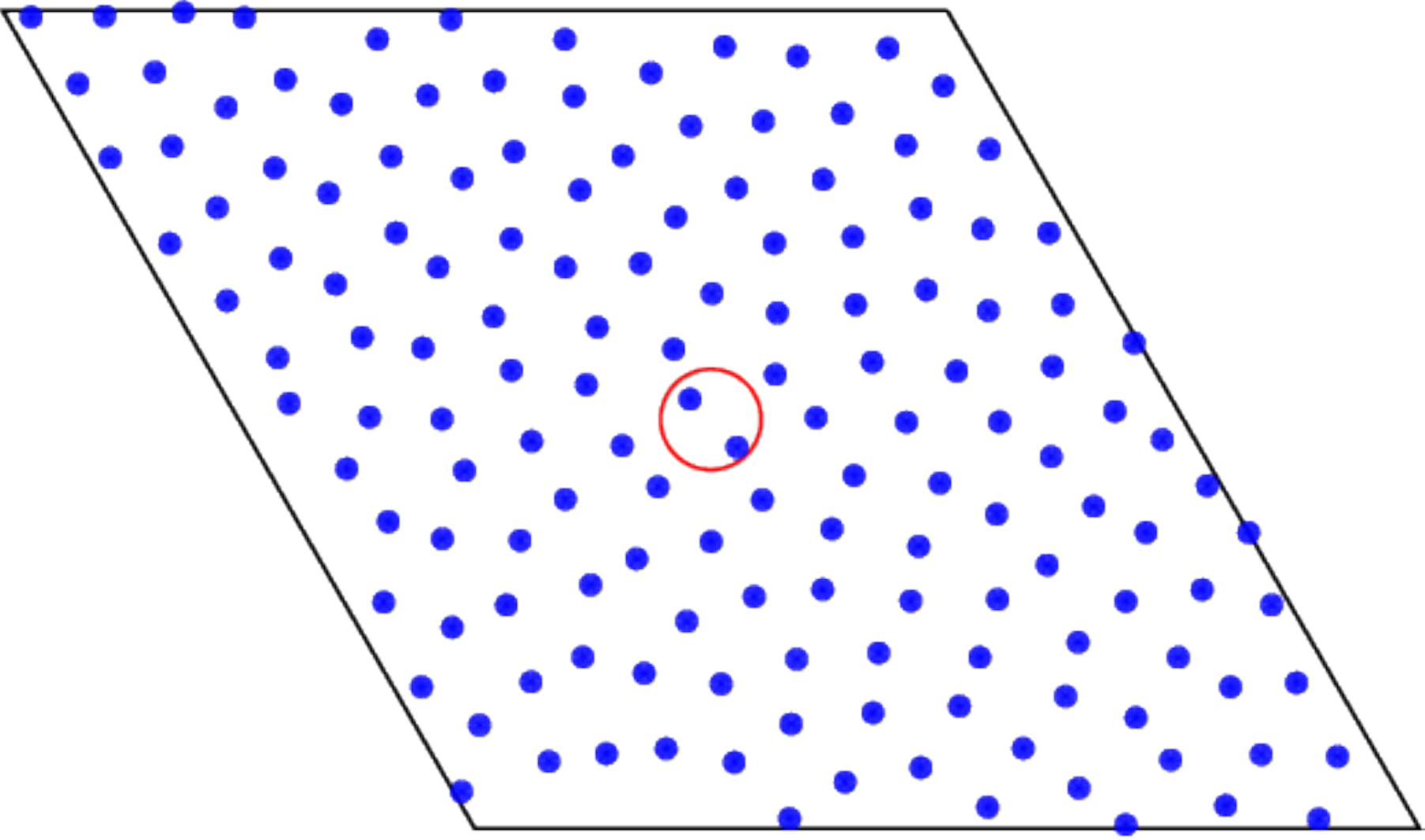}}
\caption{Two stealthy two-dimensional configurations in a rhombical fundamental cell containing $N=271$ 
particles with  $\chi=0.1$ (left panel) and $N=151$ particles with $\chi=0.4$ 
(right panel),  as adapted from Ref. \cite{Zh15a}. It is clear that the system with  $\chi=0.4$ has
substantially more short-range order than  the one with $\chi=0.1$. The diameters of the circular
windows shown in each figure represent the characteristic wavelength $2\pi/K$ above which single-scattering events
are completely suppressed. In the case of $\chi=0.4$, this means that no scattering occurs on length scales 
above that corresponding to approximately the mean-nearest-neighbor spacing.} \label{stealth-1}
\end{center}
\end{figure}

\begin{figure}[H]
\begin{center}
{\includegraphics[  width=2.8in, keepaspectratio,clip=]{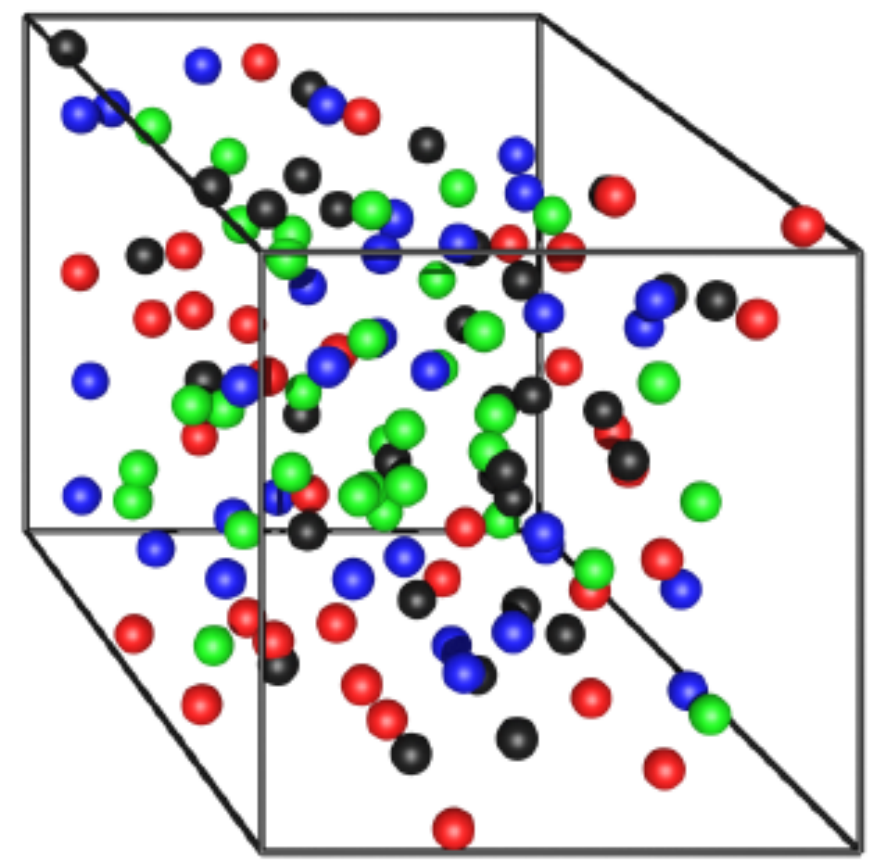} 
\hspace{0.2in}\includegraphics[  width=2.8in, keepaspectratio,clip=]{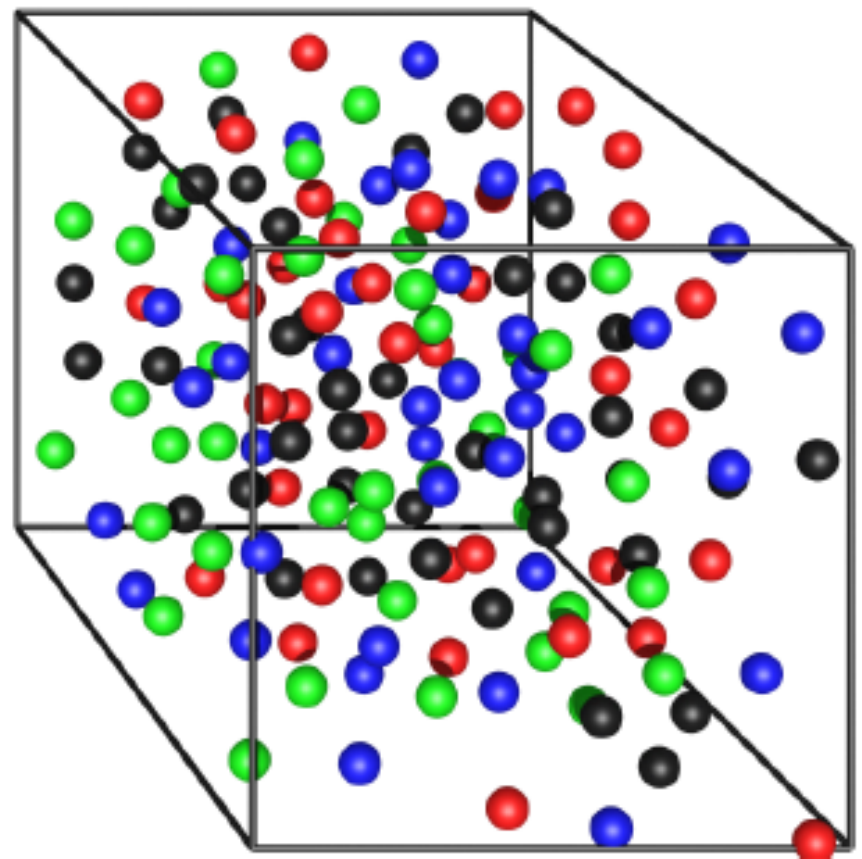}}
\caption{Two stealthy three-dimensional configurations in a cubic fundamental cell containing $N=131$ 
particles with  $\chi=0.1$ (left panel) and $N=161$ particles with $\chi=0.4$ 
(right panel), as adapted from Ref. \cite{Zh15a}. As in the analogous two-dimensional cases exhibited in Fig. \ref{stealth-1}, 
the system with  $\chi=0.4$ has substantially more short-range order than  the one with $\chi=0.1$. The identical point particles are depicted as finite-sized spheres with different colors for visualization purposes.} \label{stealth-2}
\end{center}
\end{figure}

The topography of the energy landscape in the disordered regime
is sufficiently simple such that a variety of different optimization techniques  yield the
globally minimizing energy configurations with a 100\%
percent success rate from random or other initial conditions \cite{Zh15a}. It turns out that
disordered ground-state manifold has many directions 
in which the landscape is absolutely flat  and fully connected \cite{Ba09a,Ba09b}. Continuous perturbations from
these states in configuration space can take the system from one ground state to
another energetically degenerate ground state without any energy
cost. Notably, while disordered stealthy hyperuniform ground states have positive bulk
moduli \cite{To15}, they cannot resist shear \cite{Ba09a,Ba09b}. Thus, their Poisson ratios are
equal to unity. Indeed, this is true for any $\chi < \chi_{max}^*$, including
those ground states in the ordered regime, except at the unique ground state associated
with $\chi =\chi_{max}^*$ \cite{Ba09a,Ba09b}. Here $\chi_{max}^*$ is the largest possible value
of $\chi$ consistent with the stealthy constraint; see Sec. \ref{general} for a more precise
definition. The energy landscape becomes considerably more complex for $\chi > \chi_{max}^*$, 
as discussed in Sec. \ref{ensemble}.

More generally, stealthy configurations can be those ground states that
correspond to minimizing ${\cal S}({\bf k})$ to be zero at other sets of wave vectors, not
necessarily in a connected set around the origin; see Ref. \cite{Ba08} for specific examples. 
Of course, when ${\cal S}({\bf k})$ is not zero around the origin, the ground states
are no longer hyperuniform.

\subsection{Can Disordered Stealthy Particle Configurations Tolerate Arbitrarily Large Holes?}
\label{holes}

A  ``hole" in a many-particle system in  $d$-dimensional Euclidean space 
$\mathbb{R}^d$ is defined to be a spherical region of 
a certain radius that is empty of particle centers. The probability of finding a hole 
of arbitrarily large size in typical disordered many-particle systems in the infinite-system-size limit 
(e.g., equilibrium liquid states) is non-zero.  Disordered hyperuniform systems 
could provide examples of disordered systems with bounded hole sizes
because the formation of large holes might be inconsistent with
hyperuniformity, which suppresses large-scale density fluctuations.
However, we know that not all disordered hyperuniform
systems prohibit arbitrarily large holes. For example, in a
hyperuniform fermionic-point process in $d$ spatial dimensions (see Sec. \ref{fermi}), the hole
probability size  scales as $\exp[-c r (d+1)]$ (where $c$ is a constant) for large $r$
and for the hyperuniform 2D one-component plasma (see Sec. \ref{ocp}) the hole probability scales as 
$\exp(-cr^4)$ for large $r$ \cite{Hough09}. Hence, both of these systems allow arbitrarily large holes.
Therefore, hyperuniformity alone is not a sufficient condition
to guarantee boundedness of the hole size. 

However, it is reasonable to conjecture that disordered stealthy systems have bounded hole sizes, since they
strongly suppress density fluctuations for a finite range of wavelengths.
A recent study presents strong evidence
that disordered stealthy configurations across in any dimension cannot tolerate arbitrarily 
large holes in the infinite-system-size limit, i.e., the hole probability has compact support \cite{Zh17a}.
It was conjectured  that maximum hole size depends inversely on $K$, and estimates
of the corresponding dimension-dependent constant for $d=1,2$ and $3$ were provided.
Ghosh and Lebowitz have provided a rigorous proof of this hole conjecture \cite{Gh17}.
This bounded-hole-size property of disordered stealthy systems
apparently accounts for their novel thermodynamic and physical properties, including desirable band-gap, optical and
transport characteristics \cite{Fl09b,Le16,Zh16b,Wu17,Ch18}; see Sec. \ref{properties}.

\subsection{Configurations with Designed Hyperuniform Structure Factors via Collective Coordinates}

The collective-coordinate technique has been applied to target even more general forms
of the structure factor for a prescribed set of wave vectors such that the structure factor
is not necessarily minimized to be zero in this set, e.g.,  power-law forms and positive constants, called {\it equiluminous} systems \cite{Uc06b,Ba08,Za11b}. 
In these more general situations,  the resulting configurations
are the ground states of interacting many-particle systems with 2-, 3- and  4-body interactions.
Note that unlike stealthy ground states, $\chi=1$ (not $\chi=1/2$) is the critical value when
one runs out of degrees of freedom that can be independently
constrained in the large-system limit \cite{To15}. Here we briefly describe
both underconstrained systems ($\chi <1$) and so-called ``perfect glasses", which
are overconstrained systems ($\chi >1$).

\subsubsection{Underconstrained Systems}

One can create disordered hyperuniform configurations that correspond to targeted structure factors 
with the power-law form $S({\bf k}) = |{\bf k}| ^{\alpha}$ in the radial interval $0 \le |{\bf k}| \le K$ for 
any positive value of the exponent $\alpha$ (see Fig. \ref{powers}) 
thereby spanning the three hyperuniformity classes: I, II and III. Configurations corresponding
to the targeted power-law structure factors depicted in Fig. \ref{powers}  are illustrated in Fig. \ref{collective}.
We see that clustering of particles, characteristic of class III systems \cite{Uc06b,Za11b}), is 
not  inconsistent with hyperuniformity. This provides a vivid counterexample to the
notion that the development of hyperuniformity goes hand in hand with the formation
of short-range order and vice versa, as was suggested in Ref. \cite{Fr16}. It is seen 
that as the exponent $\alpha$ increases, short-range order increases and the tendency for particles
to ``cluster" decreases.  All of the cases shown in Fig. \ref{collective}
are underconstrained systems with $\chi=0.471$, and hence cannot resist
shear stresses \cite{Zh16a}.

\begin{figure}[H]
\begin{center}
\includegraphics[  width=3in, keepaspectratio,clip=]{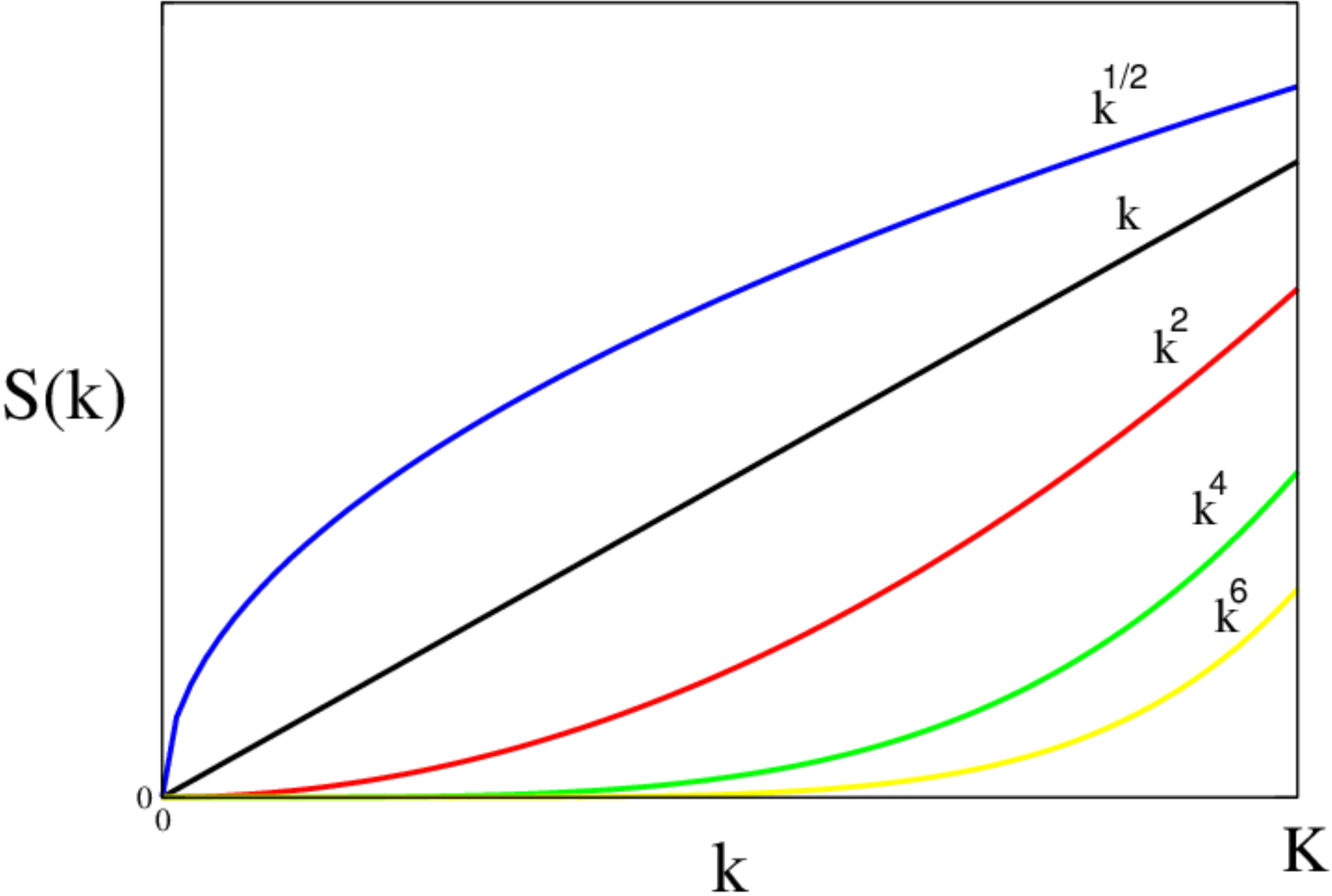} 
\caption{Target power-law structure factor $S({\bf k}) = |{\bf k}| ^{\alpha}$ in the radial interval $0 \le |{\bf k}| \le K$ for selected values
of the exponent $\alpha$ as carried out in Refs. \cite{Uc06b} and \cite{Za11b}. The ``stealthy" case corresponds to $\alpha \rightarrow \infty$ \cite{Uc04b,Ba08}.}  \label{powers}
\end{center}
\end{figure}

\begin{figure}[H]
\begin{center}
{\includegraphics[  width=1.2in, keepaspectratio,clip=]{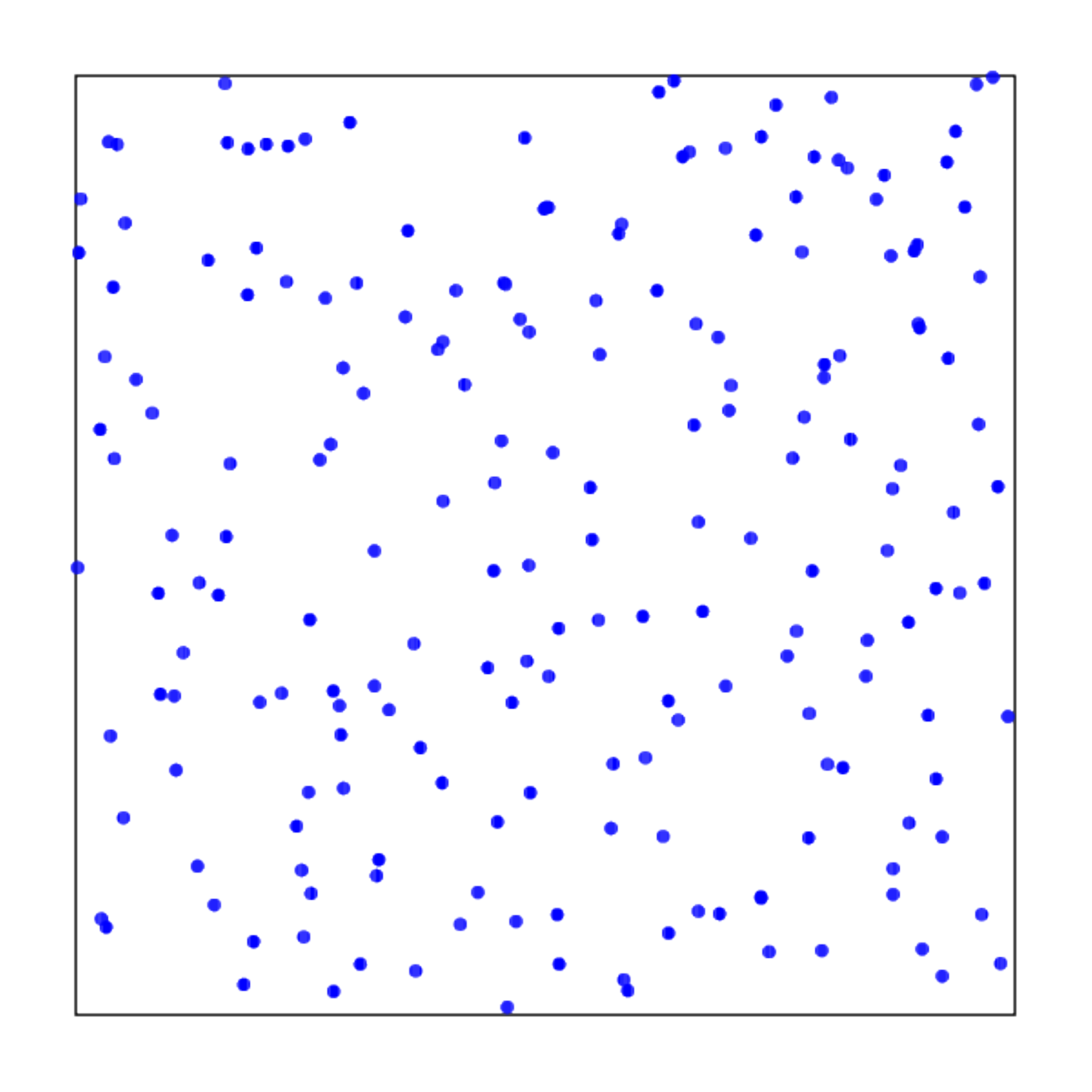} 
\includegraphics[  width=1.2in, keepaspectratio,clip=]{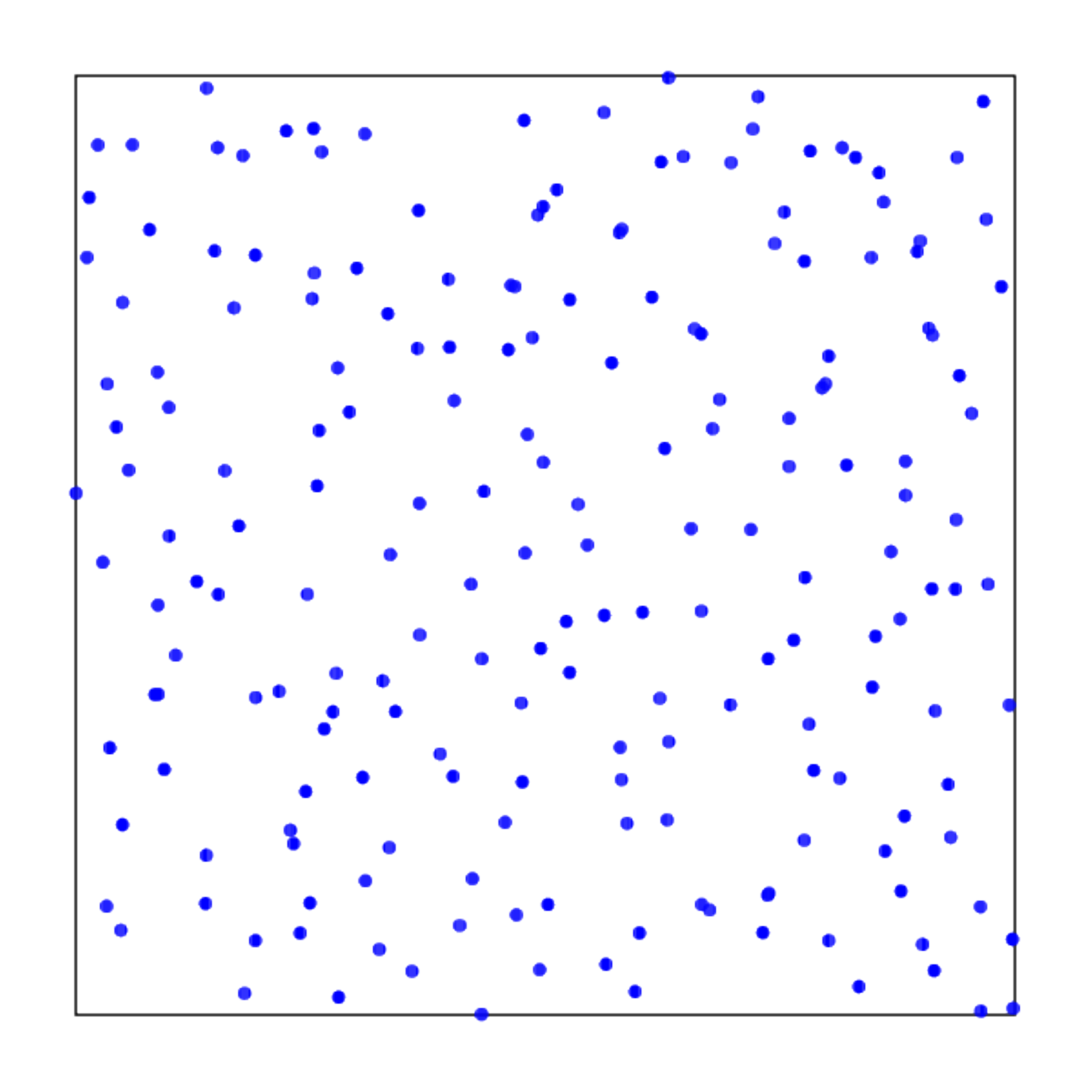}
\includegraphics[  width=1.2in, keepaspectratio,clip=]{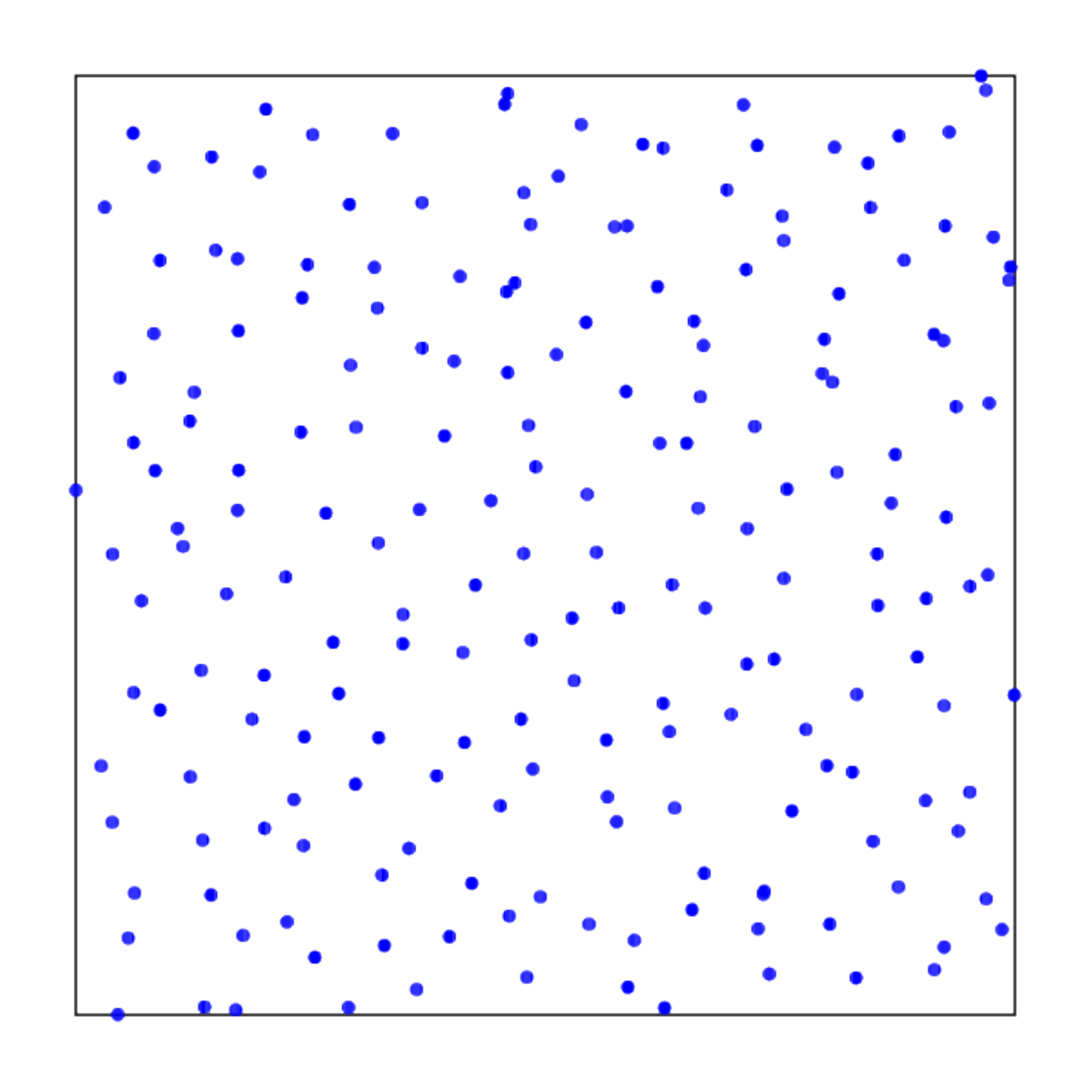}} \\
{\includegraphics[  width=1.2in, keepaspectratio,clip=]{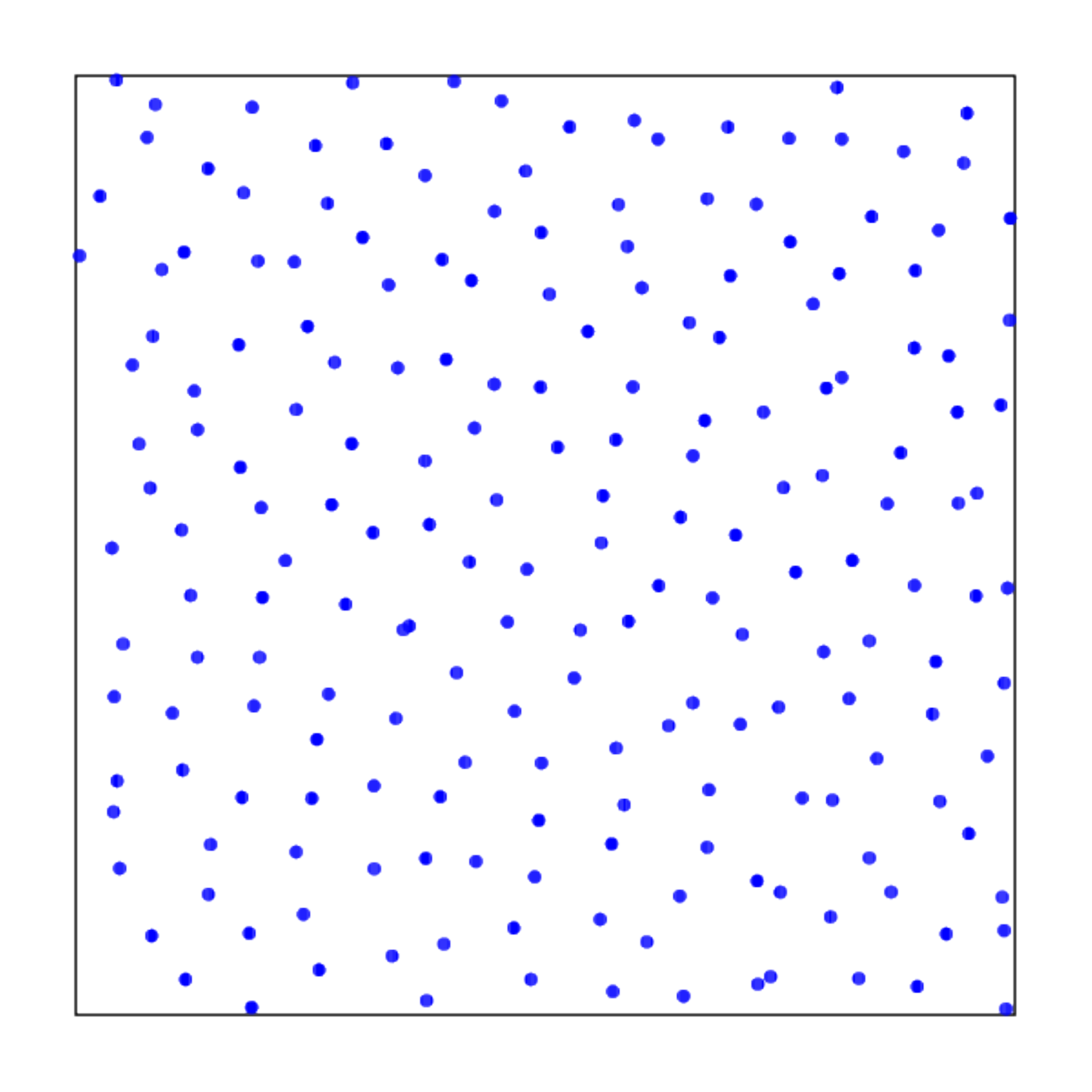}
\includegraphics[  width=1.2in, keepaspectratio,clip=]{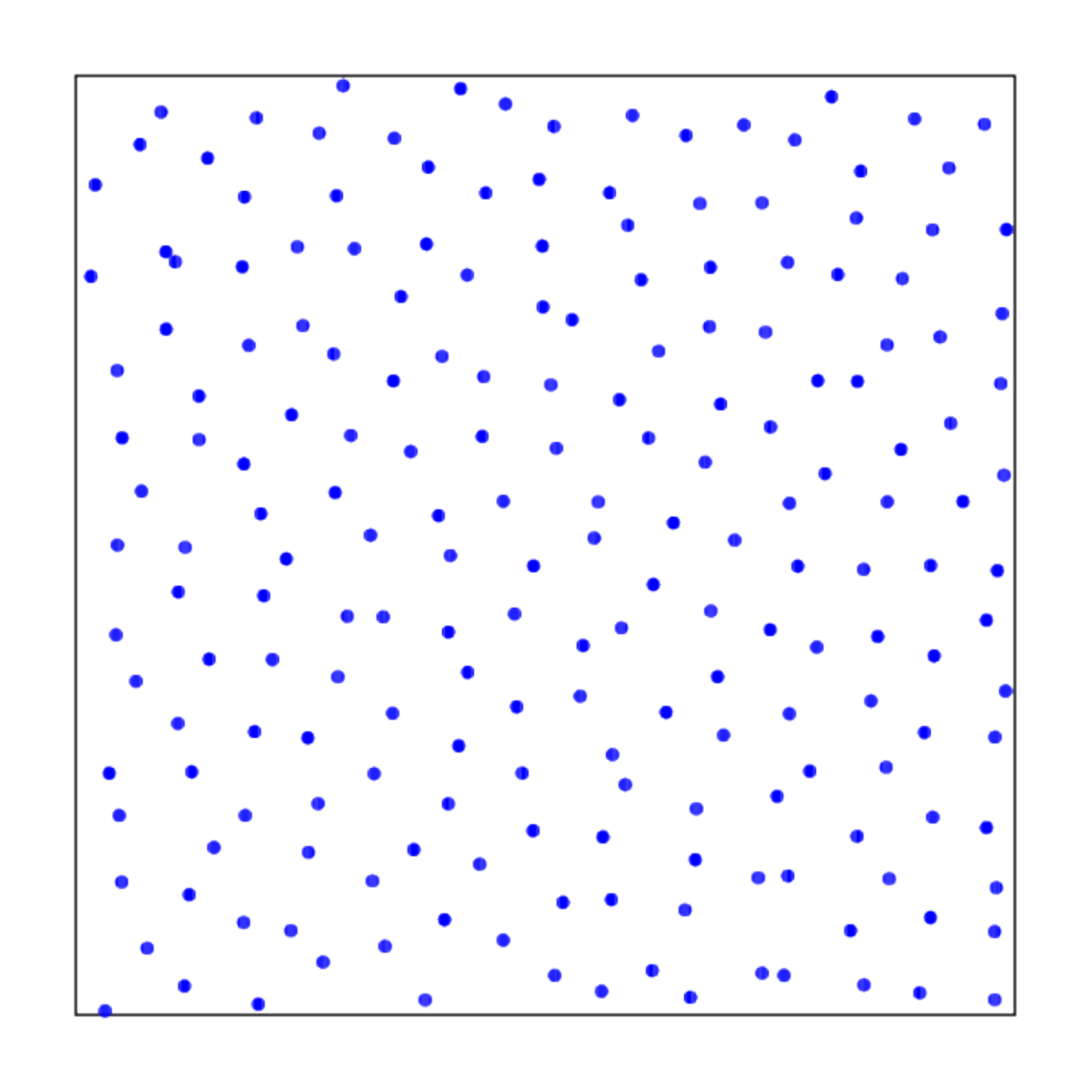} 
\includegraphics[  width=1.2in, keepaspectratio,clip=]{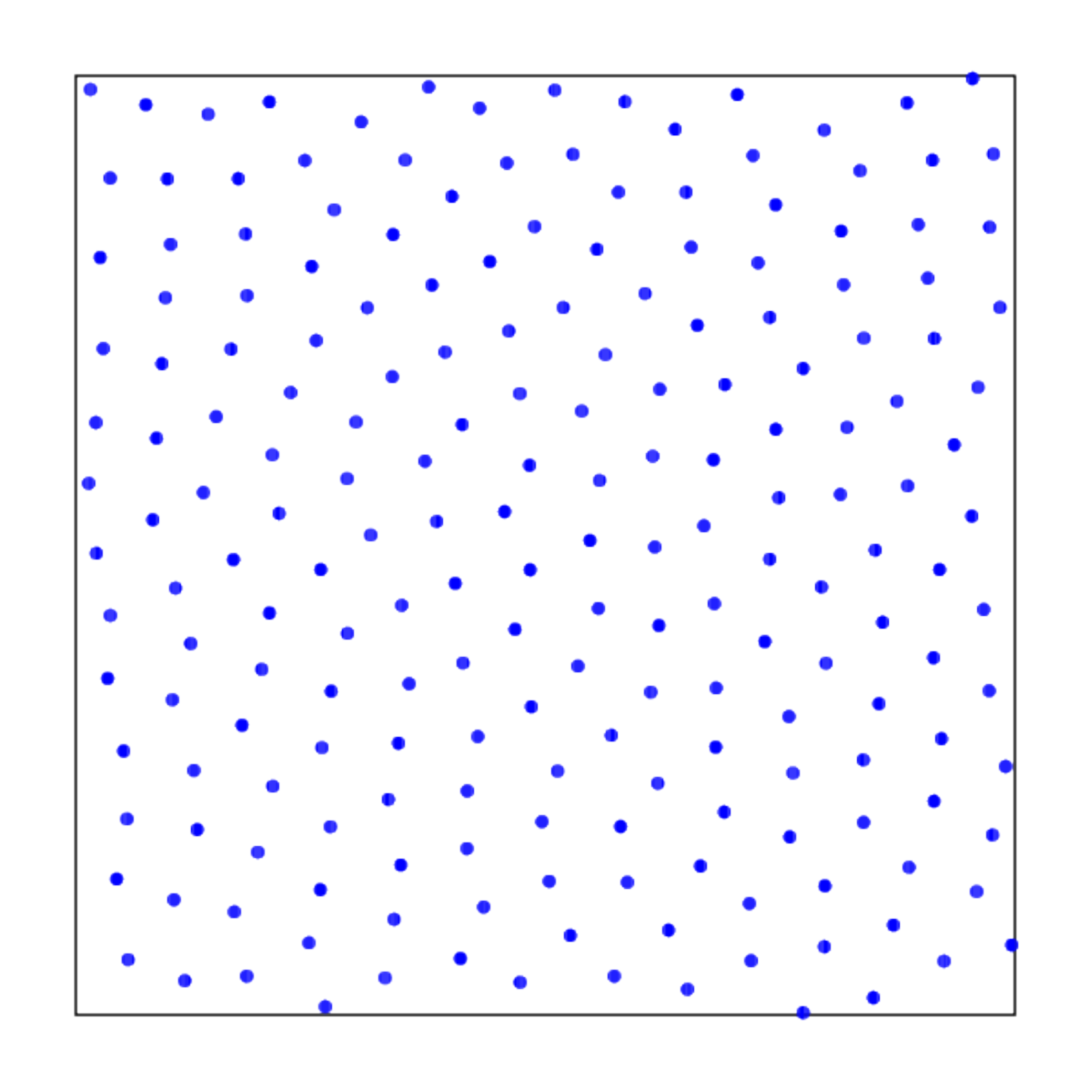}}
\caption{Configurations of particles in two dimensions with prescribed power-law structure factors $S({\bf k}) = |{\bf k}|^{\alpha}$ for $0 \le |{\bf k}| \le K$ corresponding to $\alpha=1/2,1,2,4,6,\infty$
with $N=200$ and $\chi=0.47$. The class III case ($\alpha=1/2$) is characterized by a
large degree of particle clustering. Increasing the exponent $\alpha$ (left to right), increases the short-range order and decreases the tendency for particles
to ``cluster". The limit $\alpha \rightarrow \infty$ corresponds to a stealthy hyperuniform disordered state (bottom right).}  \label{collective}
\end{center}
\end{figure}

\subsubsection{Perfect-Glass Paradigm}

It is well-known that rapid cooling of liquids below a certain temperature range can result in a transition to glassy states
of matter that are structurally disordered with long-range order with the mechanical rigidity of a solid \cite{Chaik95}. The traditional understanding of glasses includes their thermodynamic metastability with respect to crystals.
A singular counterexample to this scenario is the {\it perfect glass} whereby  
interactions are designed to completely eliminate the possibilities of
crystalline and quasicrystalline phases, while creating mechanically stable {\it hyperuniform} amorphous glasses down
to absolute zero temperature \cite{Zh16a}. Since the perfect-glass model completely banishes
crystal and quasicrystal formation, it circumvents the Kauzmann paradox \cite{Ka48}
in which the extrapolated entropy of a metastable supercooled liquid
drops below that of the stable crystal and hence is distinguished from the associated {\it ideal glass}.
Perfect-glass interactions are derived from the collective-coordinate optimization scheme by requiring the structure factor 
to have disordered hyperuniform power-law forms over a very wide range of wavenumbers
around origin (indeed, overconstrained with $\chi> 1$), including values that automatically include all 
possible Bragg peaks. The resulting two-, three- and four-body interactions enable
a perfect glass to resist both compressive and shear deformations.
A perfect glass represents a soft-interaction analog of the maximally
random jammed (MRJ) packings of hard particles \cite{To00a,To10c}; see Sec. \ref{jamming}. 
These latter states can be regarded to be the epitome of
a glass, since they are out of equilibrium, maximally disordered, hyperuniform, mechanically rigid with
infinite bulk and shear moduli, and can never crystallize (infinitely ``frustrated") due to configuration-space trapping.  A novel feature of equilibrium
systems of identical particles interacting with a perfect-glass potential at positive temperature is that,
due to their hyperuniformity, they have a non-relativistic speed of sound that is infinite.

\begin{figure}[H]
\begin{center}
\includegraphics[  width=2.5in, keepaspectratio,clip=]{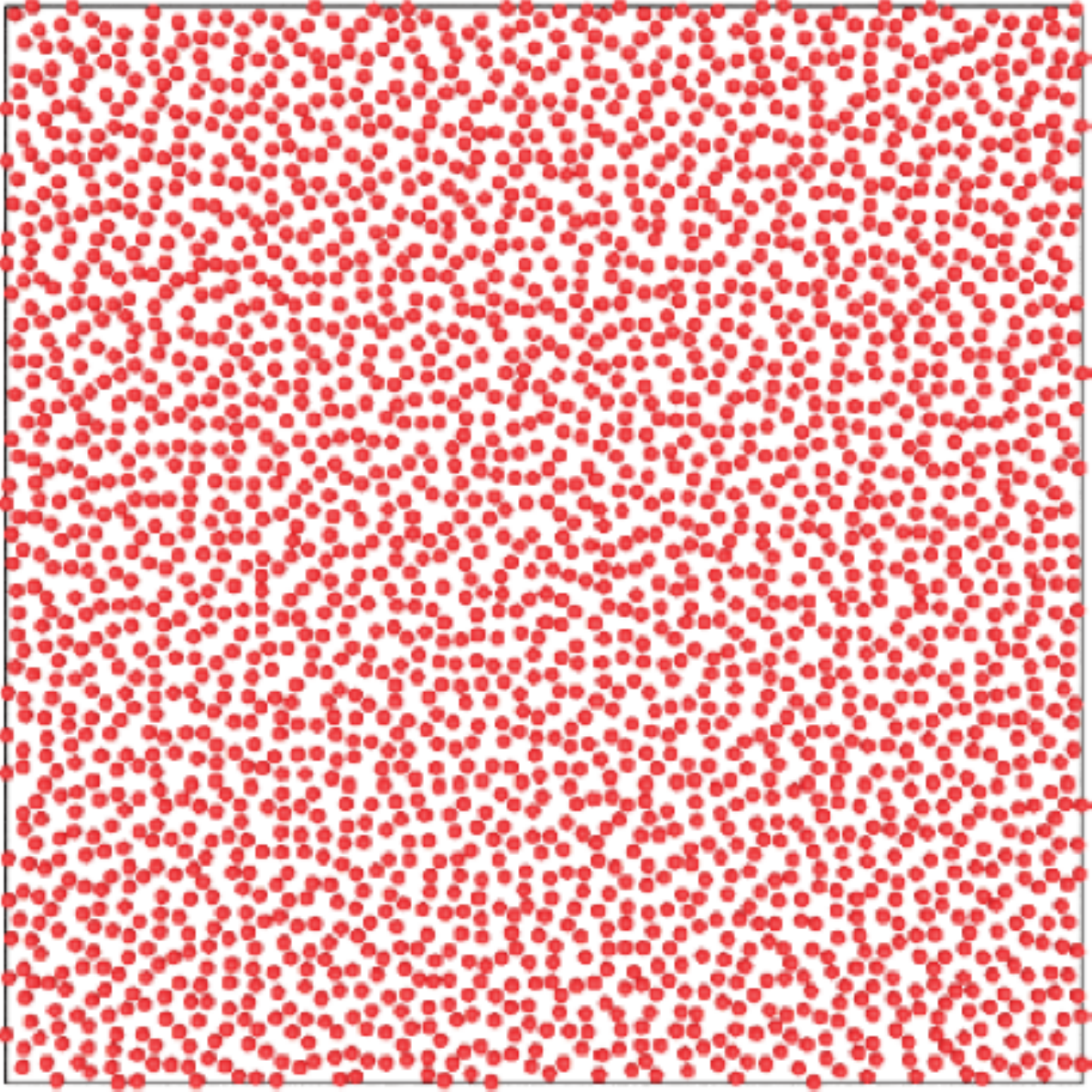} \hspace{0.3in}
\includegraphics[  width=3.5in, keepaspectratio,clip=]{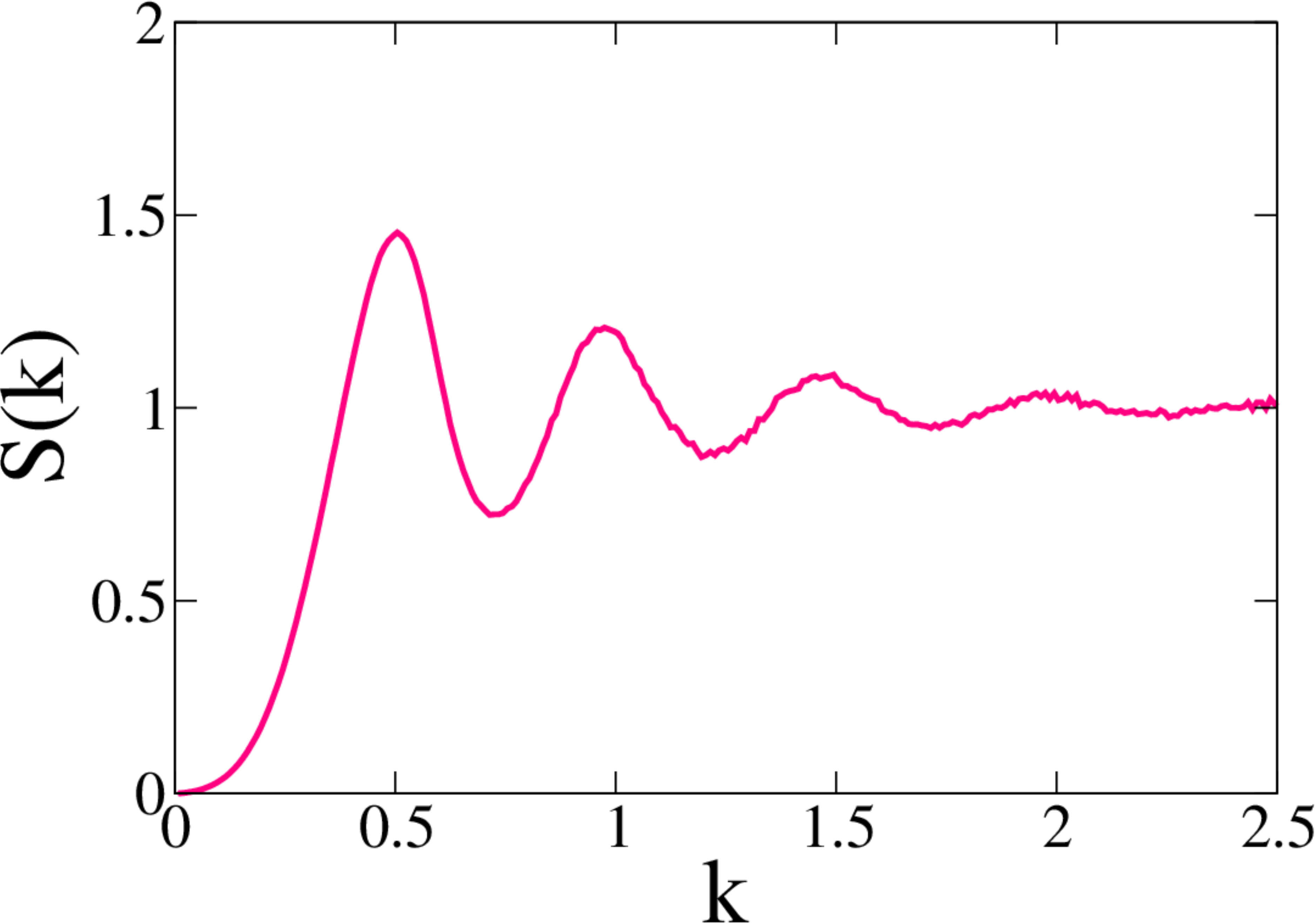} 
\caption{Left panel: Snapshot of a 2D perfect-glass configuration of 2,500 particles 
in which the structure factor is targeted to be  the power-law form 
 $S({\bf k}) = |{\bf k}|^2$  with $\chi=5.1$ and $K=1$ \cite{Zh16a}. Right panel: Resulting optimized structure factor.}  \label{p-glass}
\end{center}
\end{figure}

It has recently been shown that the perfect-glass model possesses disordered classical
ground states with an {\it enumeration entropy} ${\cal S}_E=k_B\ln(\Omega_E)$ that is zero, where
$\Omega_E$ is the number of {\it distinct} accessible structural patterns (aside from trivial symmetry operations) and $k_B$ is the Boltzmann constant.
This means that the disordered ground states of perfect glasses, an example of which is depicted in the left panel of Fig. \ref{p-glass}, are 
configurationally unique such that they can always be superposed onto each other or their mirror image.
Such ``unique" disorder is a highly counterintuitive situation that heretofore had not been identified.
Zero-enumeration entropy is normally associated with crystalline  classical ground states,
and the few previously known disordered classical ground states of many-particle systems are all 
highly degenerate states, including stealthy disordered ground states \cite{To15}.

\subsection{Ensemble Theory for Stealthy Hyperuniform Disordered Ground States}
\label{ensemble}

The task of formulating an ensemble theory that yields
analytical predictions for the structural characteristics and other properties of stealthy degenerate
ground states in $d$-dimensional Euclidean space $\mathbb{R}^d$ is highly nontrivial because the dimensionality of the  configuration
space depends on the number density $\rho$ and there is a multitude of ways
of sampling the ground-state manifold, each with its own probability measure
for finding a particular ground-state configuration. Recently, some initial steps 
have been taken to develop an ensemble theory for stealthy ground states \cite{To15}.
General exact relations for thermodynamic properties
(energy, pressure, and isothermal compressibility) as well as exact conditions 
that both the ensemble-averaged pair correlation function
$g_2({\bf r})$ and  ensemble-averaged structure factor $S({\bf k})$ must obey for any $d$ have been derived, which then
can be applied to the ground-state 
of any well-defined stealthy ensemble as a function of $\rho$ in any $d$.

\subsubsection{General Results}
\label{general}

For a general stable radial pair potential function $v({\bf r})$, it is well-known
that the ensemble average of the energy (\ref{total}) per particle $u$ in the thermodynamic limit can be written in terms of  
the pair correlation function $g_2({\bf r})$:
\begin{eqnarray}
u \equiv \lim_{N \rightarrow \infty} \left\langle \frac{\Phi_N({\bf r}^N)}{N }\right\rangle
   &=&\frac{\rho}{2} \int_{\mathbb{R}^d}  v({\bf r})g_2({\bf r}) d{\bf r} 
\label{u1}
\end{eqnarray}
where angular brackets denote an ensemble average and $\rho$ is the number density in the thermodynamic limit.
Let us now specialize to stealthy potential functions ${\tilde v}({\bf k})$ with support in $0 \le |{\bf k}| \le K$
of the class specified by  Eq. (\ref{v-k}). It has been shown \cite{To15} that whenever particle configurations in 
$\mathbb{R}^d$ exist such that $S({\bf k})$  is constrained
to achieve its minimum value of zero for $0 \le  |{\bf k}| \le K$, the system must be at its ground state or global
energy minimum, and  the average ground-state energy per particle
$u$ in any well-defined ensemble is given exactly by
\begin{equation}
u=v_0\left[\frac{\rho}{2}  - \gamma d\rho \chi\right], \qquad (\rho^*_{min} \le \rho < \infty),
\label{ground2}
\end{equation}
where  $v_0\equiv {\tilde v}({\bf k=0})$ denotes the Fourier-space potential at the origin and and $\gamma = (2\pi)^d v({\bf r=0})/(v_0 v_1(K))$. The relation (\ref{ground2}) 
follows from (\ref{u1}), energy duality relations \cite{To08a,To11b}, and the aforementioned stealthy conditions on
${\tilde v}({\bf k})$  and $S({\bf k})$. 
Importantly, the generally degenerate ground-state manifold  is invariant to the specific choice of the stealthy function ${\tilde v}(k)$ at fixed $\rho$ and $d$.
Here $\rho_{min}^*$ is the minimal density associated with
the dual of the densest Bravais lattice in direct space  (as elaborated below), and
$\gamma \in (0,1]$ is a constant whose value depends on the specific form of the 
stealthy-potential class, but (as noted above) has no effect on the ground-state manifold. 
For stealthy potentials,
the ground-state pressure $p$, as obtained from the average energy [$p=\rho^2 \left(\partial u/\partial \rho\right)_T$] for all possible values of $\rho$ or $\chi$, is given by the
following simple expression:
\begin{equation}
p= \frac{\rho^2}{2}v_0, \qquad (\rho^*_{min} \le \rho < \infty).
\label{p-u}
\end{equation}
Hence, the isothermal compressibility $\kappa_T\equiv \rho^{-1}\left(\partial \rho/\partial p\right)_T$
of such a ground state is $\kappa_T=v_0/\rho^2$. It is seen
that as $\rho$ tends to infinity, the compressibility tends to zero.
Although the pressure obtained via the ``virial relation" \cite{Han13} is generally expected to be equivalent to that obtained from the energy route, for a certain class of stealthy potentials, the pressure obtained via the viral route may either be ill-defined or divergent \cite{To15}.
This situation serves to illustrate the mathematical subtleties that can arise because of the long-ranged nature
of stealthy potentials in direct space.
These general exact relations can be profitably employed to test corresponding
computer simulation results.

It is noteworthy that a particular periodic crystal with a finite basis is a  stealthy hyperuniform ground state
for all positive $\chi$ up to its corresponding  maximum value $\chi_{max}$ (or minimum
value of the number density $\rho_{min}$) determined by its first positive Bragg peak ${\bf k}_{Bragg}$
[minimal positive wave vector for which $S({\bf k})$ is positive]. Reference \cite{To15} 
lists the pair $\chi_{max}$, $\rho_{min}$ for some common
periodic patterns in one, two, three, and four dimensions, all
of which are part of the ground-state manifold. At fixed $d$, we call $\rho_{min}^*$ the smallest possible value of $\rho_{min}$, 
which corresponds to the density associated with the dual of the densest Bravais lattice in direct space, and 
represents the critical density value below which a stealthy ground state
does not exist for all $k \le |{\bf k}_{Bragg}^*|$.  Correspondingly, $\chi_{max}^*$ is 
the largest possible value of $\chi_{max}$.
For $d=1,2,3$ and 4, this critical minimal density
occurs for the integer lattice ($\rho_{min}^*=1/(2\pi)=0.15915\ldots$) \cite{Fa91}, 
triangular lattice ($\rho_{min}^*=\sqrt{3}/(8\pi^2)=0.06581\ldots$) 
\cite{Uc04b,Su05}, body-centered cubic lattice ($\rho_{min}^*=1/(8\sqrt{2}\pi^3)=¼ 0.00285\ldots$) \cite{Su05,Ba08}, 
and $D_4$ checkerboard lattice ($\rho_{min}^*=1/(32\pi^3)=0.0003208\ldots$) \cite{To15}.

While the fact that periodic structures are part of the ground-state manifold does not provide any clues about their occurrence probability in some ensemble, one can show how their existence leads to disordered degenerate ground states arise as part of the ground-state manifold for sufficiently small $\chi$, as we now briefly outline. 
It is notable that in the case $d=1$, there is no non-Bravais lattice (periodic structure
with a basis $n \ge 2$) for which $\chi_{max}$ is greater than $1/2$, implying
that the ground-state manifold is nondegenerate (uniquely the integer lattice) for $1/2 < \chi \le 1$ \cite{To15}.
This case is to be contrasted with the cases $d \ge 2$ where the ground-state manifold must be degenerate 
for $1/2 <\chi < \chi_{max}^*$ and nondegenerate only at the point $\chi=\chi_{max}^*$, 
as implied by the list of periodic structures summarized in Tables II-IV given in Ref. \cite{To15}.  The following Lemma, together with the fact that any periodic crystal with a finite basis
is a stealthy ground state  can be used to demonstrate rigorously how complex 
aperiodic patterns can be ground states, entropically favored or not \cite{To15}.

\noindent{\sl Lemma.---}At fixed $K$, a configuration comprised of the union (superposition) of $m$ different stealthy ground-state
configurations in $\mathbb{R}^d$ with $\chi_1,\chi_2,\ldots,\chi_m$, respectively,
is itself stealthy with a $\chi$ value given by
\begin{equation}
\chi= \left[\sum_{i=1}  \chi_i^{-1}\right]^{-1},
\label{harm}
\end{equation}
which is the harmonic mean of the $\chi_i$ divided by $m$.

\subsubsection{Pseudo Hard Spheres in Fourier Space and Pair Statistics}

Unlike thermodynamic properties and certain general features
of pair statistics, the specific functional forms of $g_2({\bf r})$ and $S({\bf k})$ 
and other structural functions  depend on the stealthy ensemble under consideration \cite{To15}.
Recently, a statistical-mechanical theory has been formulated in the canonical ensemble 
in the zero-temperature limit, i.e.,  
the probability of observing a configuration is proportional
to $\exp[-\Phi_N({\bf r}^N)/(k_B T)]$ in the limit $T \rightarrow 0$. By exploiting an ansatz that 
the entropically favored (most probable) stealthy 
ground states in the canonical ensemble behave like ``pseudo" equilibrium  hard-sphere systems in Fourier space with an ``effective packing fraction"
that is proportional to $\chi$, one can employ well-established integral-equation formulations
for the pair statistics of hard spheres in direct space \cite{Han13,To02a} to obtain
accurate theoretical predictions for $g_2({\bf r})$ and $S({\bf k})$ for a moderate range of $\chi$
about $\chi=0$ \cite{To15}. To get an idea of the large-$r$ asymptotic behavior of the  pair correlations, consider 
the total correlation function $h(r)$ in the limit $\chi \rightarrow 0$ for any $d$, which
is exactly given by
\begin{equation}
\rho h(r) = -\left(\frac{K}{2 \pi r}\right)^{d/2}J_{d/2}(Kr) \qquad (\chi \rightarrow 0),
 \label{eq_hr}
 \end{equation}
and for large $r$ is given asymptotically by
\begin{equation}
\rho h(r) \sim -\frac{1}{r^{(d+1)/2}}\cos(r-(d+1)\pi/4) \qquad (r \rightarrow +\infty).
\end{equation}
Thus, the longed-ranged oscillations of  $h(r)$
are controlled by the power law  $-1/r^{(d+1)/2}$.

Figure \ref{Sk} shows that the structure factor $S({\bf k})$ predicted by this pseudo-hard-sphere theory
is in excellent agreement with the corresponding simulated quantities for $\chi=0.05$, 0.1, and $0.143$ for $d=3$
in the case of radial stealthy potentials \cite{To15}.
In Fig. \ref{g-1}, the theoretical results for the pair correlation function $g_2$
are compared to corresponding  simulation  results
across the first three space dimensions. Again, we see excellent agreement between theory and simulations,
which validates the pseudo-hard-sphere Fourier-space ansatz.
Figure \ref{g-2} depicts the theoretical predictions for $g_2$ for 
 $\chi=0.15$ across the first four space dimensions. It is seen that increasing dimensionality
increases short-range correlations.

\begin{figure}[H]
\begin{center}\includegraphics[  width=2in,
  keepaspectratio,clip=]{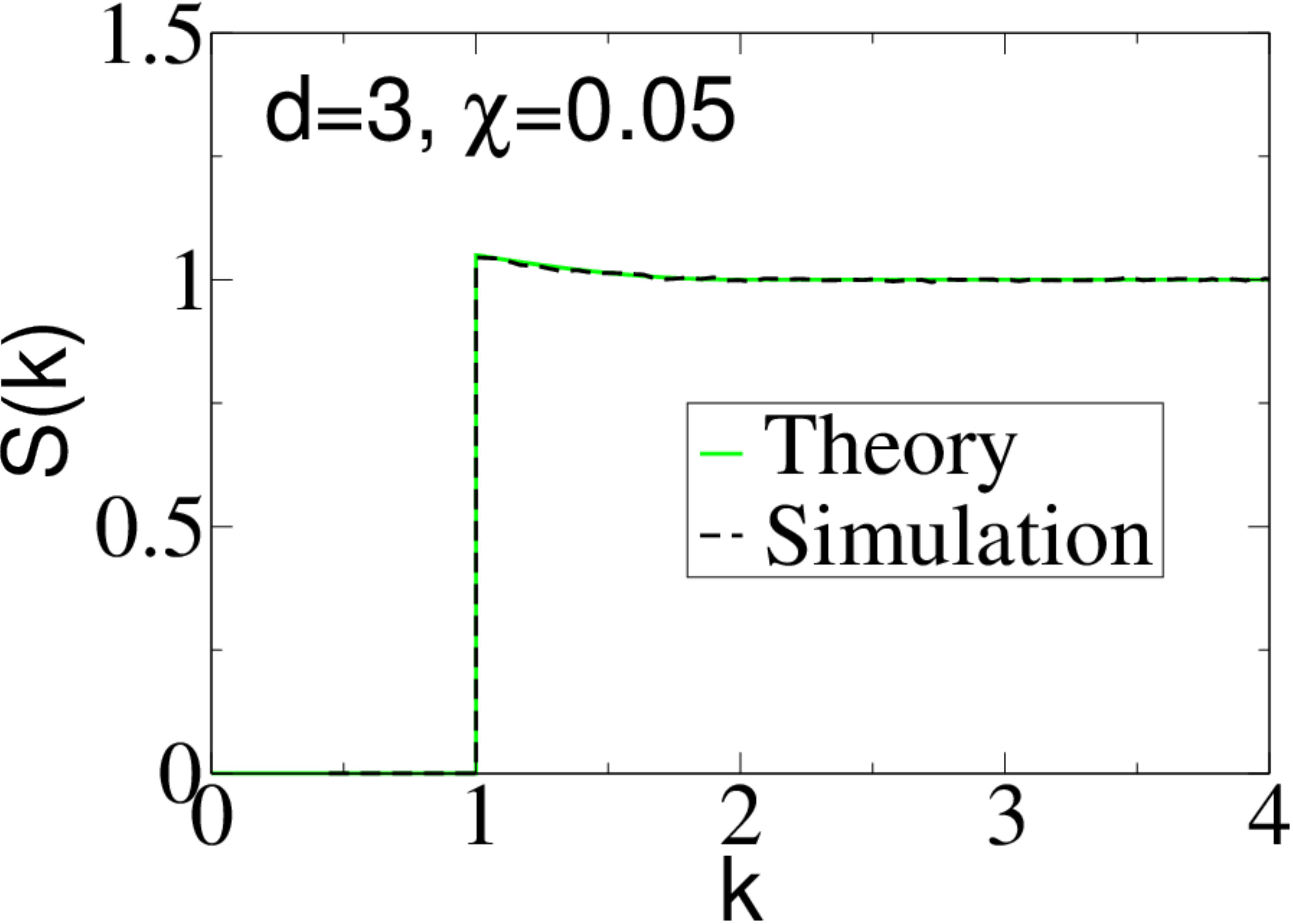}\hspace{0.3in}
\includegraphics[  width=2in,
  keepaspectratio,clip=]{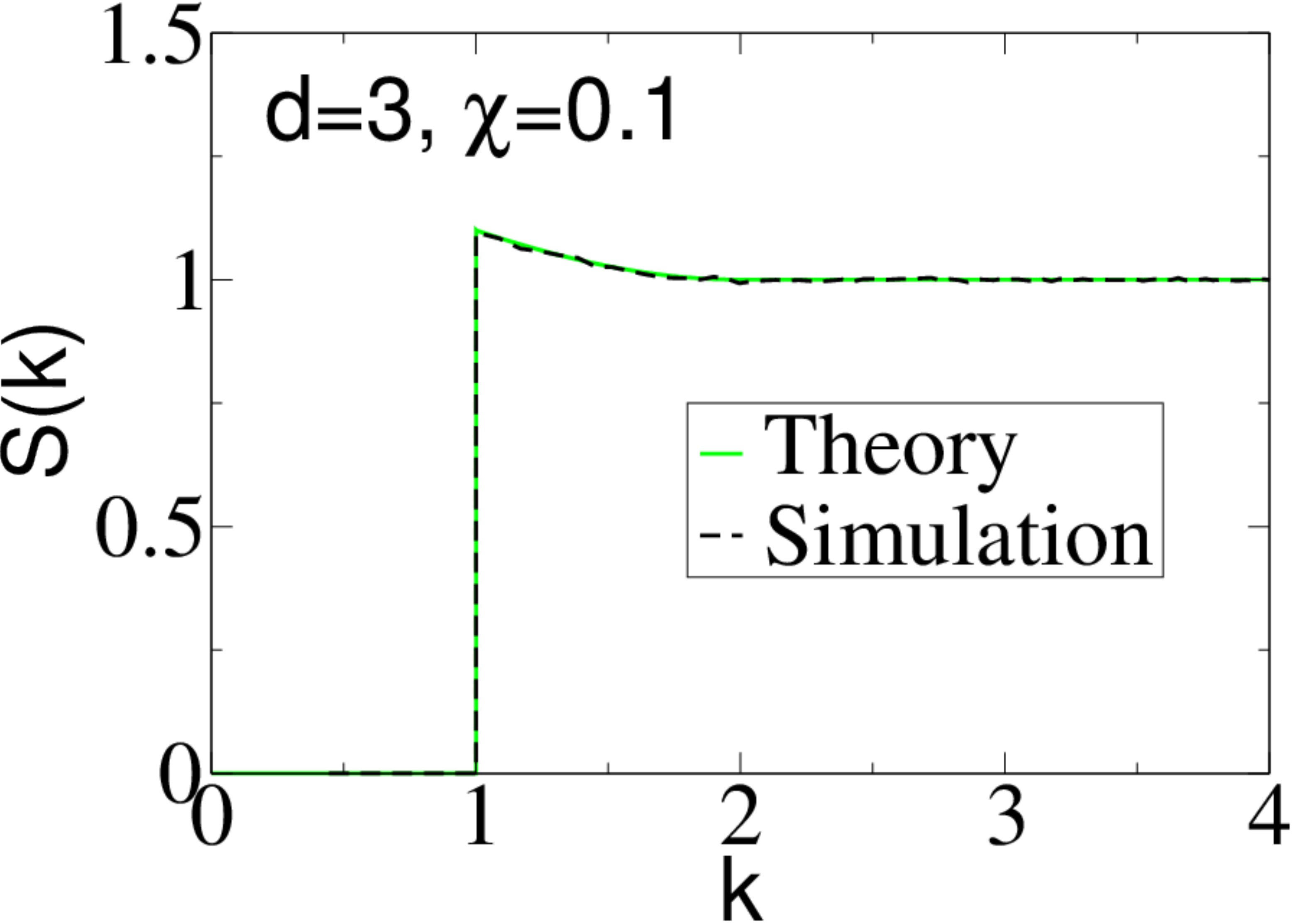}
\includegraphics[  width=2in,
  keepaspectratio,clip=]{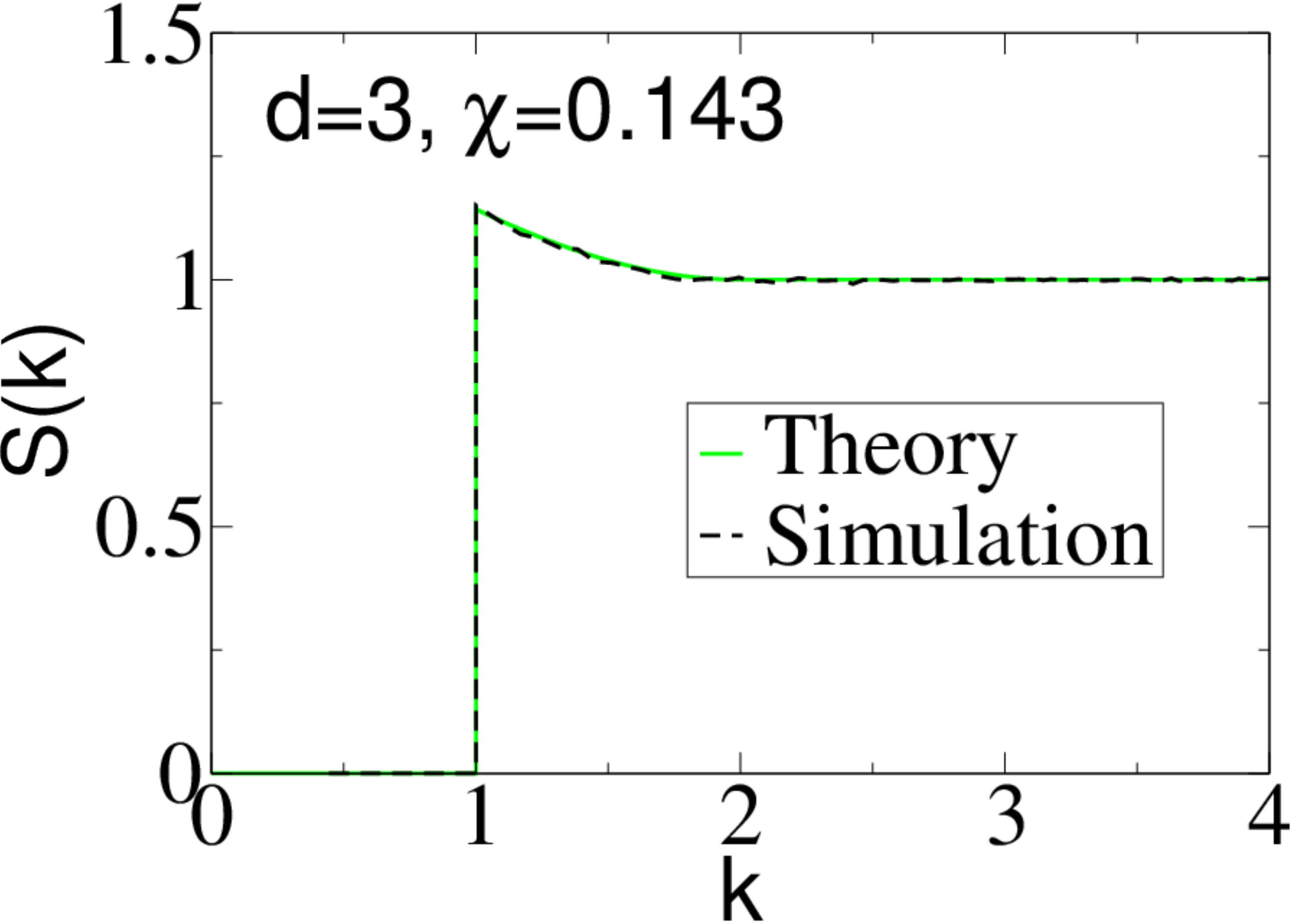}
\caption{Comparison of theoretical and simulation results for the radial structure factor 
$S(k)$ of three-dimensional stealthy systems for $\chi=0.05$, 0.1, and 0.143 in the canonical ensemble, as taken from  Ref. \cite{To15}. Here, $K=1$.}\label{Sk}
\end{center}
\end{figure}
\vspace{-0.1in}
\begin{figure}[H]
\begin{center}\includegraphics[  width=2in,
  keepaspectratio,clip=]{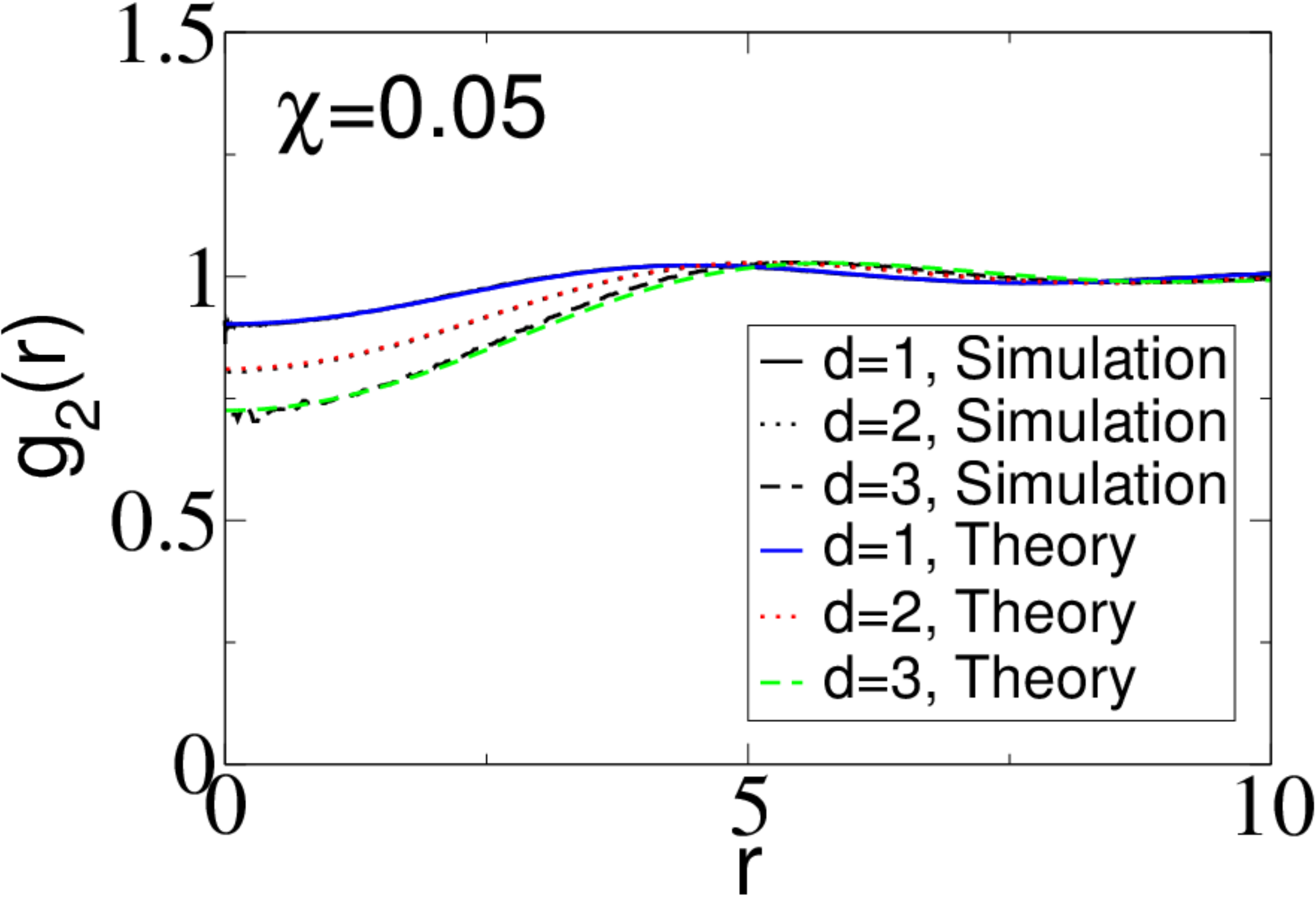}\hspace{0.3in}
\includegraphics[  width=2in,
  keepaspectratio,clip=]{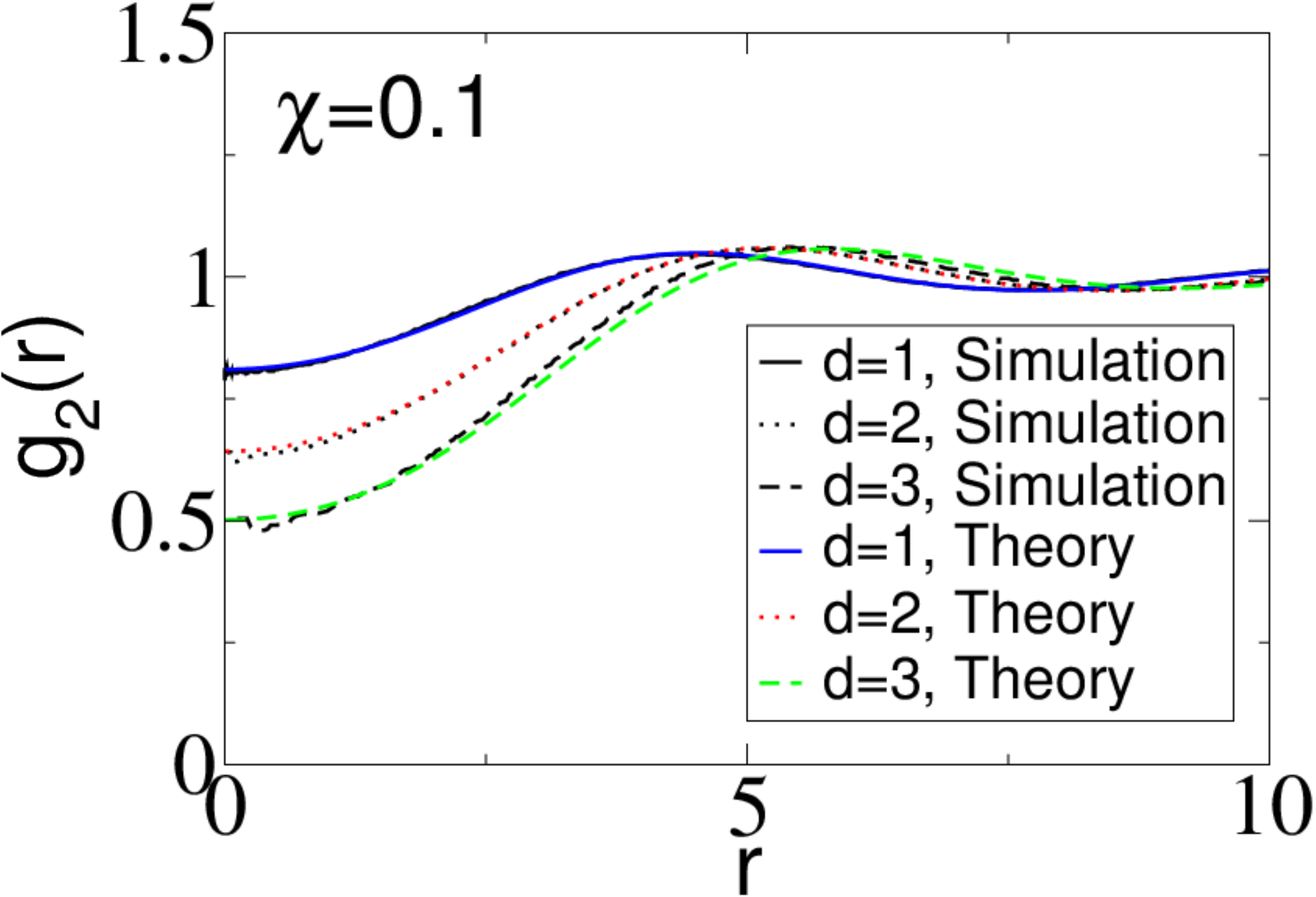}
\includegraphics[  width=2in,
  keepaspectratio,clip=]{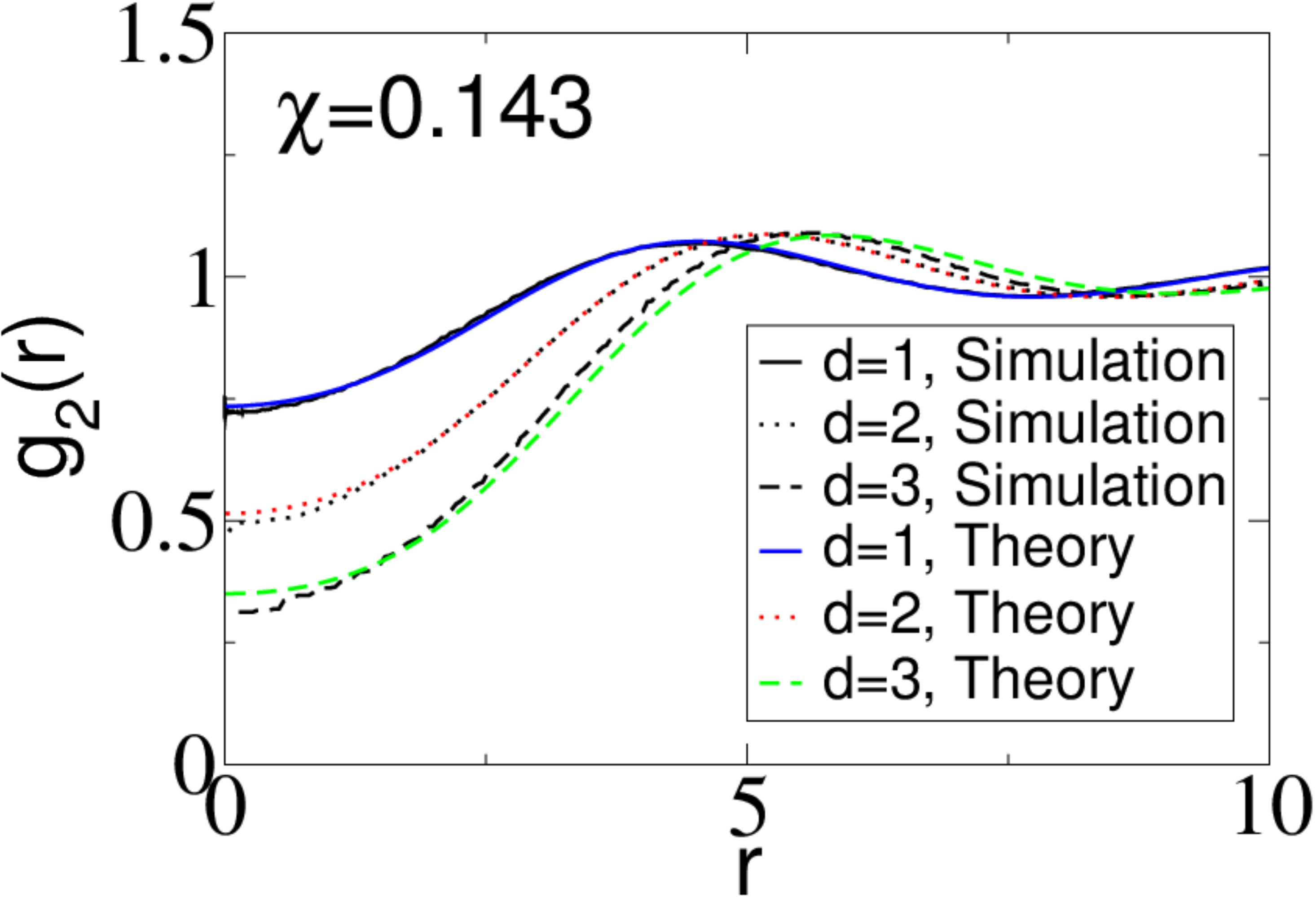}
\caption{Comparison of theoretical and simulation  results for the radial pair correlation function $g_2(r)$ of stealthy systems for 
$\chi=0.05$, 0.1, and 0.143 in the canonical ensemble across the first three space dimensions, as taken from  Ref. \cite{To15}. Here, $K=1$.}\label{g-1}
\end{center}
\end{figure}

\begin{figure}[H]
\begin{center}\includegraphics[  width=2.2in,
  keepaspectratio,clip=]{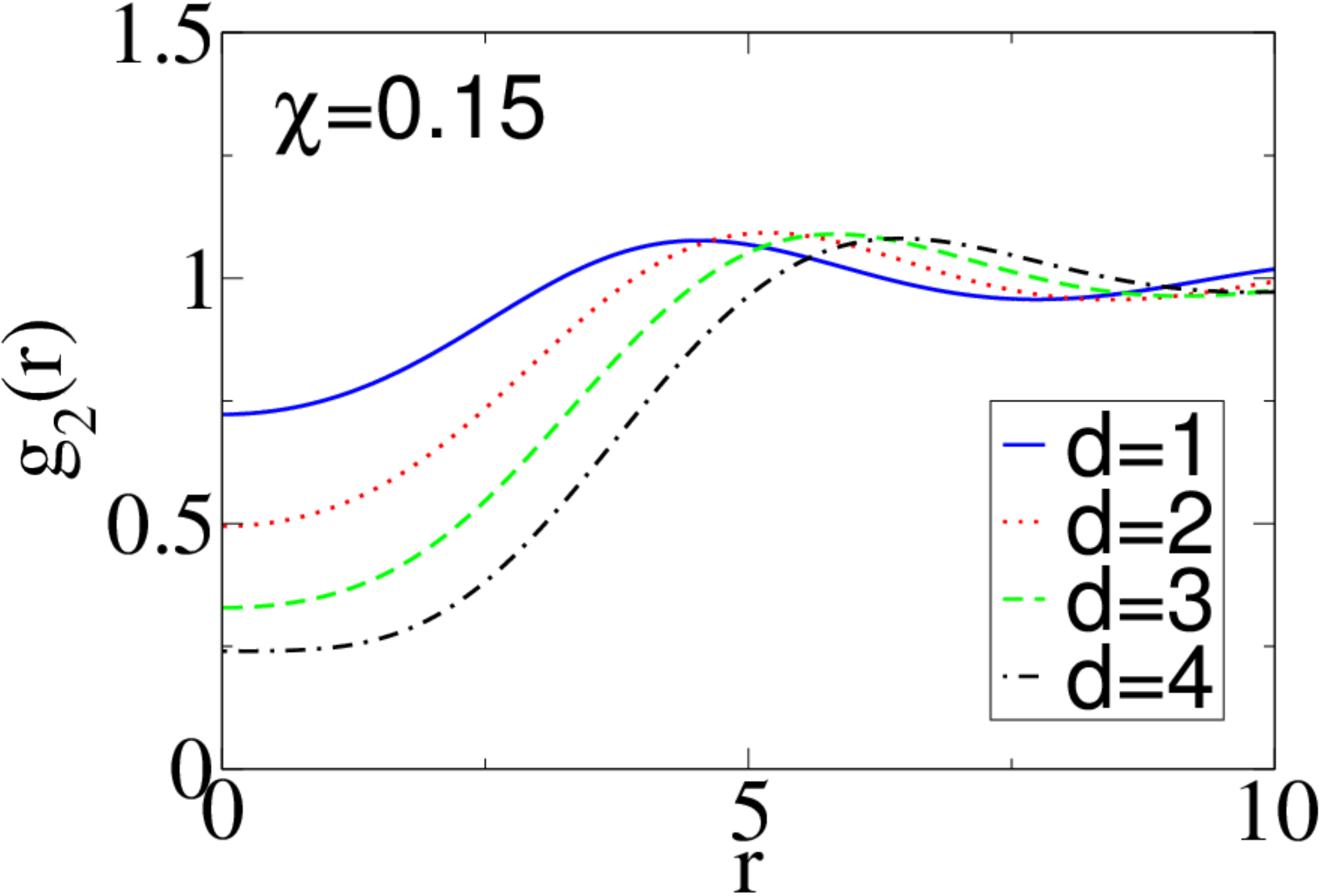}
\caption{Theoretical predictions for the radial pair correlation function $g_2(r)$ of stealthy systems for 
 $\chi=0.15$ in the canonical ensemble across the first four space dimensions, as taken from  Ref. \cite{To15}. Here $K=1$.}\label{g-2}
\end{center}
\end{figure}

\subsubsection{Number Variance and Translational Order Metric $\tau$}

The aforementioned analytical  results for $g_2({\bf r})$ and $S({\bf k})$ were employed
to ascertain other other structural characteristics of the entropically favored stealthy disordered
ground states, including the local number variance, a new  translational order metric $\tau$, and nearest-neighbor
functions across dimensions \cite{To15}.  Figure \ref{number} compares the local number variance
$\sigma^2(R)$ versus $R$ for $\chi=0.2$ across the first
three space dimensions to corresponding simulation results. It is seen that the theoretical predictions
are in good agreement with the simulations.

\begin{figure}
\begin{center}
\includegraphics[  width=3.5in,
  keepaspectratio,clip=]{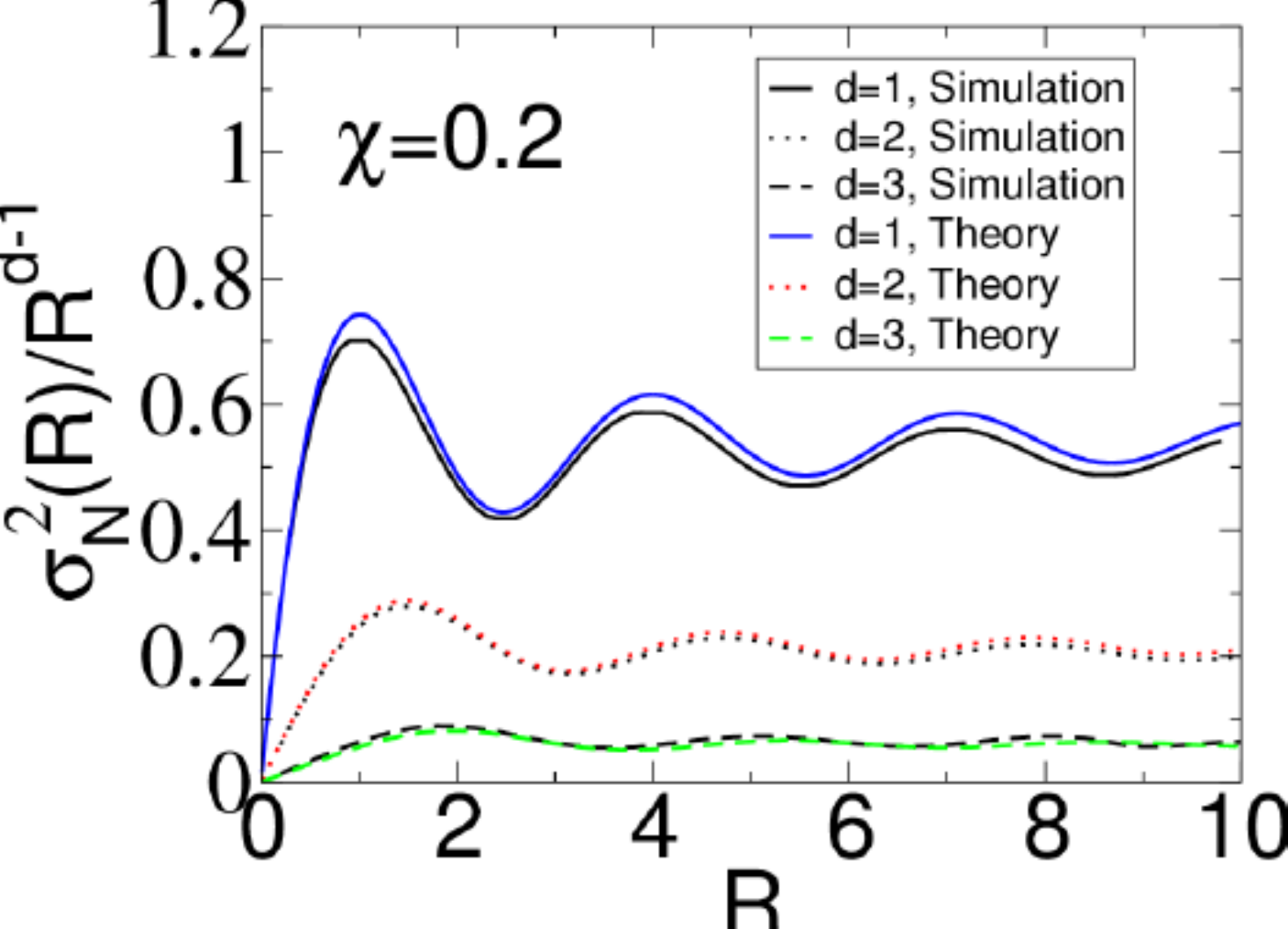}
\caption{Comparison of analytical and numerical results for the local number variance 
$\sigma^2_{_N}(R)$ versus $R$  of stealthy systems for $\chi=0.2$
across the first three space dimensions, as obtained in Ref. \cite{To15}. Here, $K=1$.}
\label{number}
\end{center}
\end{figure}

We have seen that both short- and long-scale correlations increase as $\chi$ increases.
A useful scalar positive order metric $\tau$ that captures the degree to which translational order
increases with $\chi$ across length scales was proposed in Ref. \cite{To15}:
\begin{eqnarray}
\tau &\equiv & \frac{1}{D^d}\int_{\mathbb{R}^d} h^2({\bf r}) d{\bf r} \nonumber \\
&=& \frac{1}{(2\pi)^d D^d} \int_{\mathbb{R}^d} {\tilde h}^2({\bf k}) d{\bf k},
\label{TAU}
\end{eqnarray}
where $D$ denotes some characteristic length scale.
Note that for an ideal gas (spatially uncorrelated Poisson point process), $\tau=0$
because $h(r)=0$ for all $r$. Thus, a  deviation of $\tau$ from zero
measures translational order with respect  to the fully uncorrelated
case. We see that both positive and negative correlations across length scales  will make  a positive contribution
to $\tau$, and hence is expected to be a sensitive detector of ordering in disordered systems.
Because $\tau$ diverges for any {\it infinite} perfect  crystal, it is a quantity
that is better suited to distinguish the degree of pair correlations
in amorphous systems. The metric $\tau$ has been profitably employed
to characterize order in non-stealthy point and spin (or digitized) configurations, including
equilibrium hard-disk and hard-sphere systems \cite{Zh16b}, amorphous
ices \cite{Mar17} and binary digitized systems \cite{Di18}. Note that 
for systems in the vicinity of liquid-gas  and magnetic critical points, the order metric $\tau$ 
is infinitely large due to the fact that the structure factor diverges as ${\bf k} \rightarrow {\bf 0}$  in the infinite-system-size limit. Thus,
while caution should  exercised in interpreting 
$\tau$ as an order metric in the vicinity of a thermal critical point, it can be fruitfully employed to detect
whether a disordered system is approaching a critical point.

Using the leading-order term in the $\chi$-expansion for either $h(r)$ or ${\tilde h}(k)$ [cf. (\ref{eq_hr})] and 
relation (\ref{TAU}) yields the following corresponding asymptotic expansion for $\tau$ for 
entropically-favored stealthy disordered ground states \cite{To15}:
\begin{equation}
\tau = \frac{4 d^2(2\pi)^d}{v_1(1)} \chi^2 + {\cal O}(\chi^3),
\label{formula}
\end{equation}
where we have taken $D=K^{-1}$. Thus, for stealthy ground states in the canonical ensemble, the order metric
$\tau$ grows quadratically with $\chi$ for small $\chi$. Since the error is of
order $\chi^3$, it was argued that the formula (\ref{formula}) will be a very good approximation
of $\tau$ up to relatively large values of $\chi$. Indeed, this is confirmed by simulations
up to $\chi \approx 0.45$, as shown in Fig. \ref{tau}. The dramatic rise of $\tau$ (by many
orders of magnitude) as $\chi$ increases from small values to about $\chi=0.45$
is a testament to its capacity to detect the increase of short-, intermediate
and long-range order as $\chi$ increases.


\begin{figure}[H]
\begin{center}\includegraphics[  width=3.5in,
  keepaspectratio,clip=]{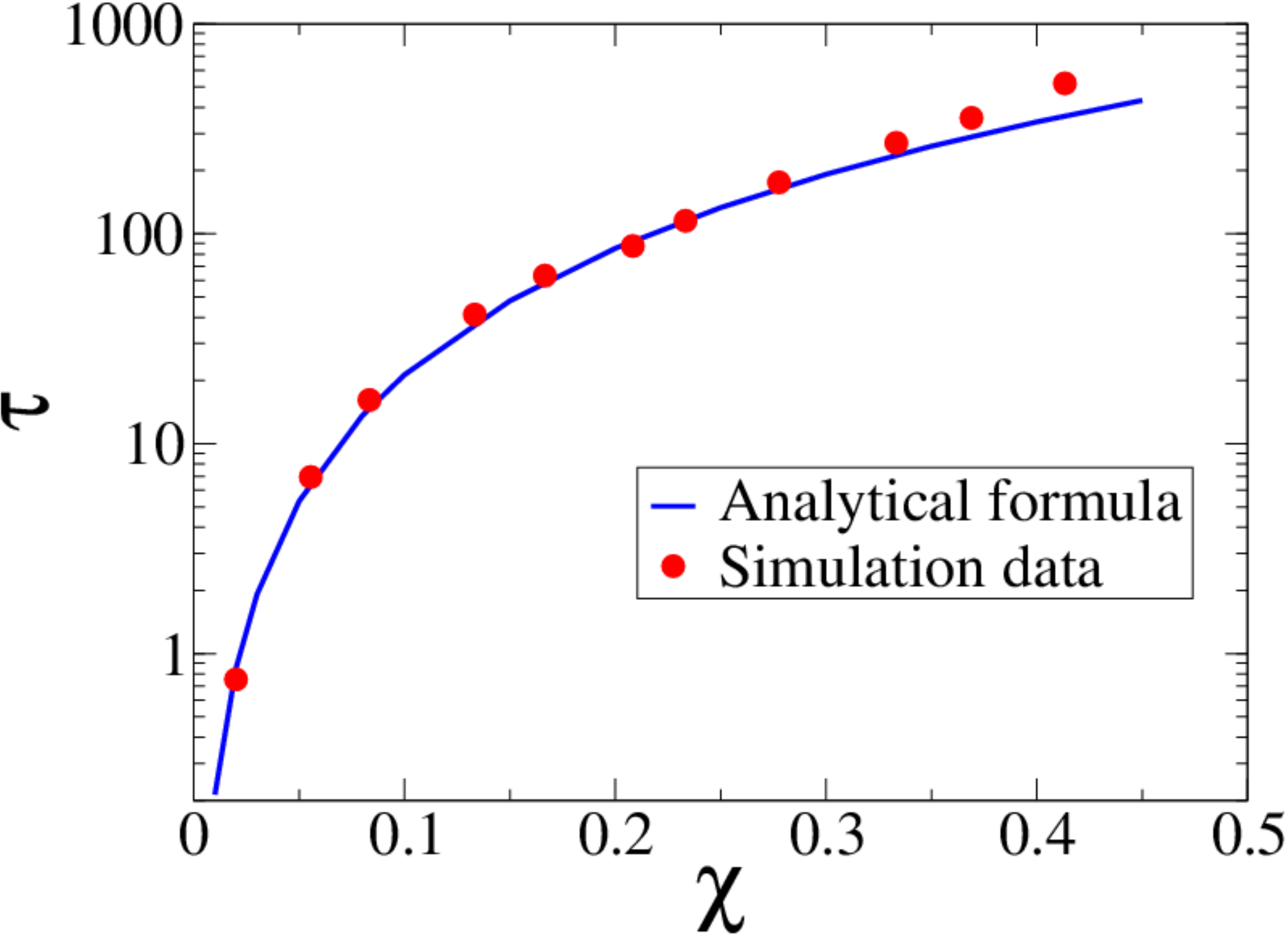}
\caption{Comparison of the analytical formula (\ref{formula}) for the $\tau$ order metric as
a function of $\chi$ in three dimensions for entropically 
favored stealthy disordered ground states to corresponding simulation data 
obtained in Ref. \cite{Zh16b}.}
\label{tau}
\end{center}
\end{figure}

\subsubsection{Entropically Favored (Most Probable) States and Ground-State Manifold}

For $\chi$ between 0 and 1/2, stealthy ground states are disordered and possess
a configurational dimension per particle $d_C=d(1-2\chi)$. At $\chi=1/2$,
the configurational dimensionality per particle collapses to zero, and there is a
concomitant  phase transition to a stable crystal phase, the nature of which depends
on the space dimension \cite{To15}. The entropically favored (most probable) ground-state phase diagram is schematically depicted
in Fig. \ref{phase}, which applies to the first four space dimensions in the canonical ensemble
in the limit $T \to 0$.
For $d=1$, the only crystal phase allowed is the integer lattice,
and hence there can be no phase coexistence. 
For $d=2$, simulations \cite{Ba08,Ba09a,Ba09b,Zh15a} indicate that the stable crystal
phase is the triangular lattice for $1/2 \le \chi \le \chi_{max}^*$. However, for $d=3$,
it is possible that there may be more than one crystal phase \cite{Zh15a,Zh15b}.  
Since four dimensions is more similar to two dimensions in that
the lattice corresponding to  $\chi_{max}^*$ is equivalent to its dual, we would expect
that the $D_4$ lattice is the stable crystal for $1/2 \le \chi \le \chi_{max}^*$, but this
remains to be confirmed. While  the configuration space is fully connected for sufficiently small $\chi$,
quantifying its topology as a function of $\chi$ up to  $\chi_{max}^*$ is an outstanding 
open problem. At some intermediate range of $\chi$, the topology of the ground-state 
manifold undergoes a sequence of one or more disconnection events, but this process
is poorly understood and demands future study. In the limit $\chi \rightarrow \chi_{max}^*$, the 
disconnection becomes complete at the unique crystal ground state.

\begin{figure}[H]
\begin{center}
\includegraphics[  width=3in,  keepaspectratio,clip=]{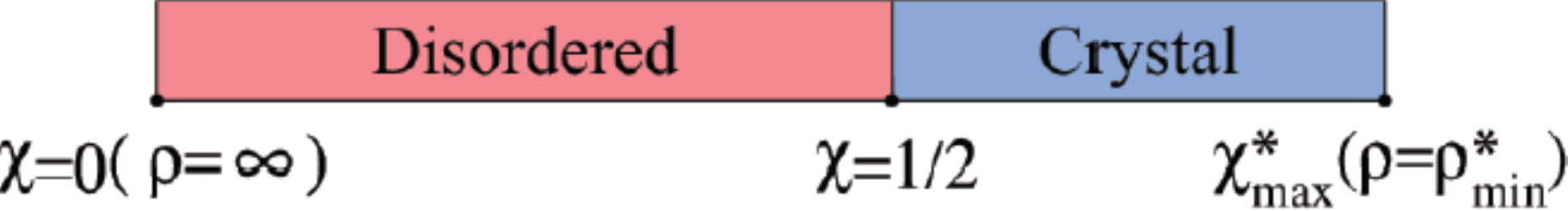}
\caption{Phase diagram for the entropically favored (most probable) stealthy ground states in the canonical ensemble
as a function of $\chi$, which applies to the first four space dimensions \cite{To15}.  }\label{phase}
\end{center}
\end{figure}

The possible configurations that can arise
as part of the ground-state manifold for $\chi > 1/2$, regardless
of their probability of occurrence, not only include periodic crystals for $d\ge 2$,  
but also aperiodic structures, reflecting the complex
nature of the energy landscape. For example, for some $\chi$ above 1/2,
so-called ``stacked-slider" phases  are part of
the stealthy ground-state manifold, even if they are 
not entropically favored. Stacked-slider phases were first
discovered numerically from random (e.g., high-temperature) initial configurations  in two dimensions in Ref. \cite{Uc04b} and were originally
called ``wavy" crystals because they were observed to consist
of particle columns that display a meandering displacement
away from linearity. However,  ``stacked-slider
phases" is a more suitable name for
this phase for arbitrary space dimensions  and this designation will be used henceforth.
The authors of Ref. \cite{Uc04b}  easily distinguished stacked-slider
phases from crystal phases by a lack of periodicity in direct
space and a lack of Bragg peaks with crystallographic symmetry in its 
scattering pattern. On the other hand, stacked-slider phases were distinguished from disordered phases
by the nature of their scattering patterns outside the exclusion region; specifically,
while the structure factors of the former are positive for $|{\bf k}| > K$, those of the  stacked-slider phases
are induced to be zero at some $\bf k$'s outside the exclusion
sphere by virtue of the constraints imposed inside this region \cite{Uc04b}.

Guided by numerical results \cite{Uc04b,Zh15b}, it  has recently been shown that
stacked-slider phases can be constructed analytically  by stacking lower-dimensional stealthy configurations
in a higher-dimensional space such that different lower-dimensional ``layers" (hyperplanes) can
slide (translate) with respect to one another \cite{Zh15b}; see two-dimensional examples
shown in Fig. \ref{stacked-1} and their corresponding scattering patterns in Fig. \ref{stacked-2}.
The fact that the stacking directions
are orthogonal to the sliding directions endow stacked-slider phases with 
unique orientational-order characteristics. More specifically, 
it was demonstrated that stacked-slider
phases are generally nonperiodic, statistically anisotropic structures
that possess long-range orientational order but have
zero shear modulus  \cite{Zh15b}. Therefore, they are  distinguishable states of matter
distinct from stealthy disordered ground states without any Bragg peaks. Since stacked-slider phases are part
of the ground-state manifold of stealthy potentials, they
are also hyperuniform.

\begin{figure}[H]
\centerline{{\includegraphics*[  width=1.4in,clip=
keepaspectratio]{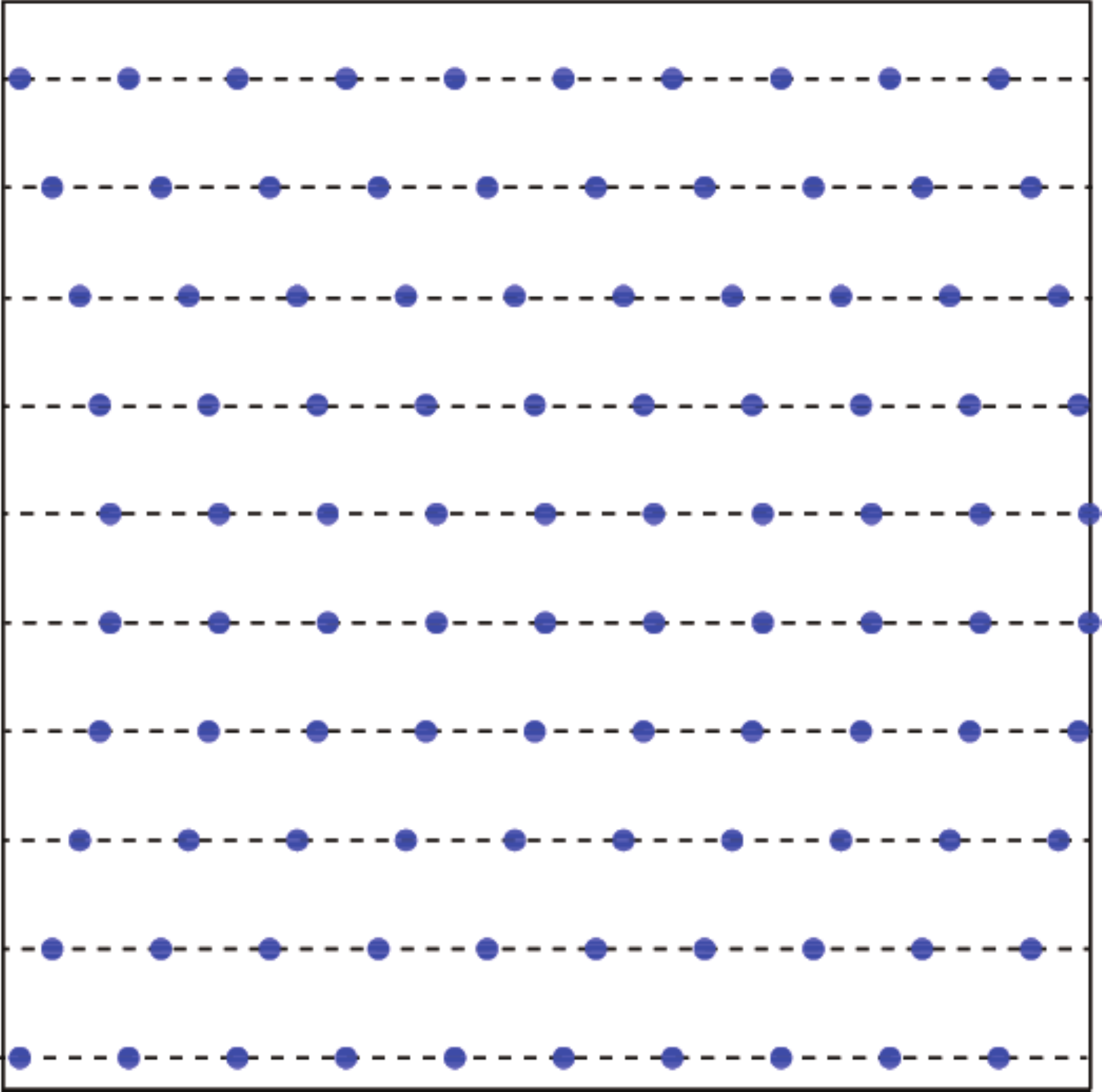}
\hspace{0.2in}\includegraphics*[  width=1.4in,clip=
keepaspectratio]{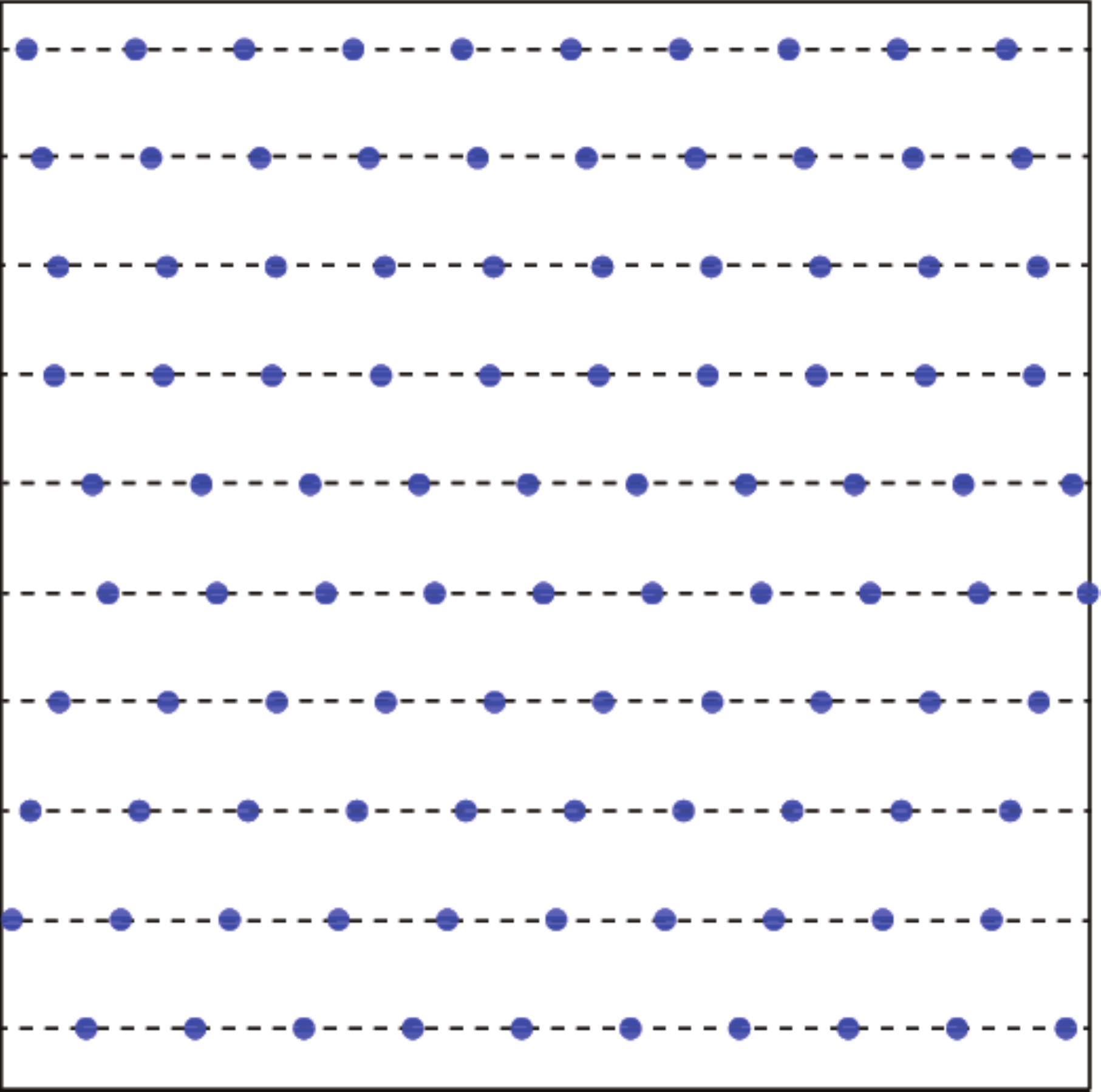}\hspace{0.2in}\includegraphics*[  width=1.4in,clip=
keepaspectratio]{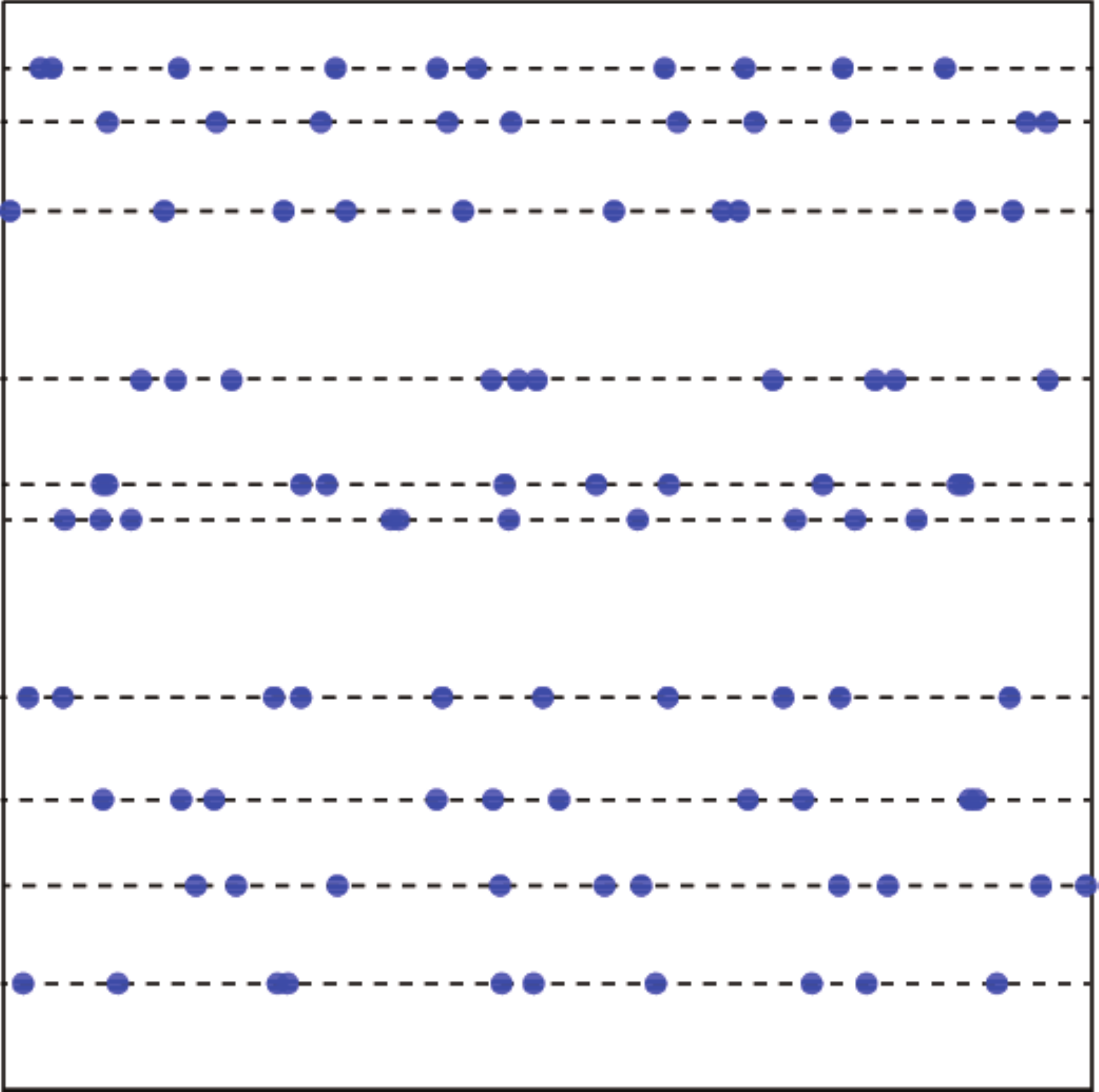}}}
\caption{Three examples of stealthy stacked-slider phases in $\mathbb{R}^2$ with $N=100$ that
are part of the ground-state manifold, even if not entropically favored. Left panel:
Horizontal rows in the square lattice are coherently translated with respect
to one another. Here $\chi=\pi/4$. Middle panel: Horizontal rows in the square lattice are randomly translated with respect
to one another. Here  $\chi=\pi/4$. Right panel: Horizontal stealthy disordered stackings of disordered stealthy 1D configurations.
Here $\chi=\pi/18$ and each horizontal row has $\chi=2/9$.}
\label{stacked-1} 
\end{figure}
\vspace{-0.2in}

\begin{figure}[H]
\centerline{{\includegraphics*[  width=2in,clip=
keepaspectratio]{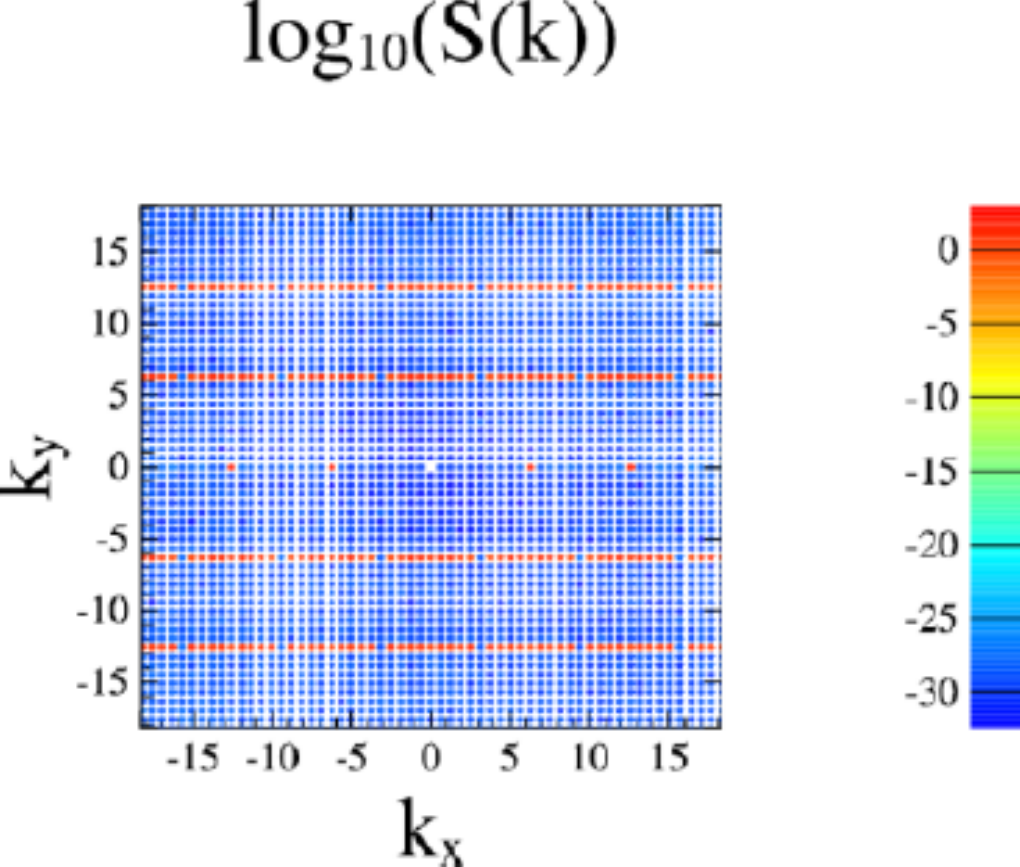}
\hspace{0.2in}\includegraphics*[  width=2in,clip=
keepaspectratio]{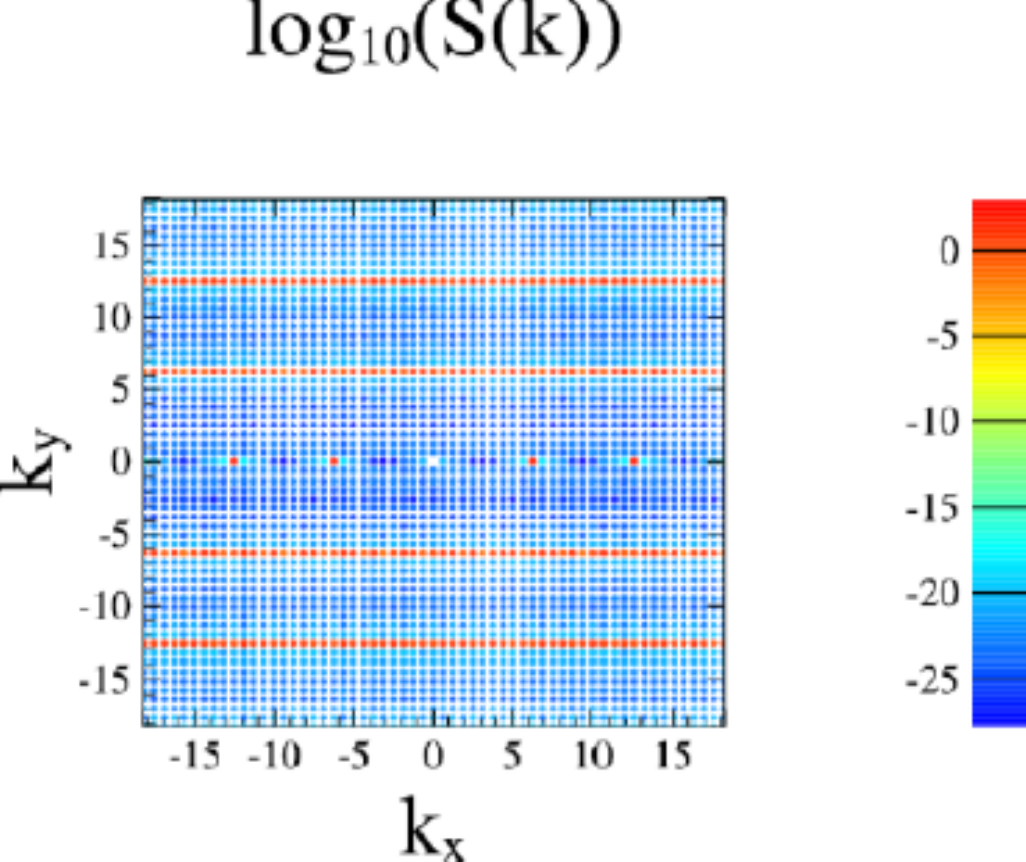}\hspace{0.2in}\includegraphics*[  width=2in,clip=
keepaspectratio]{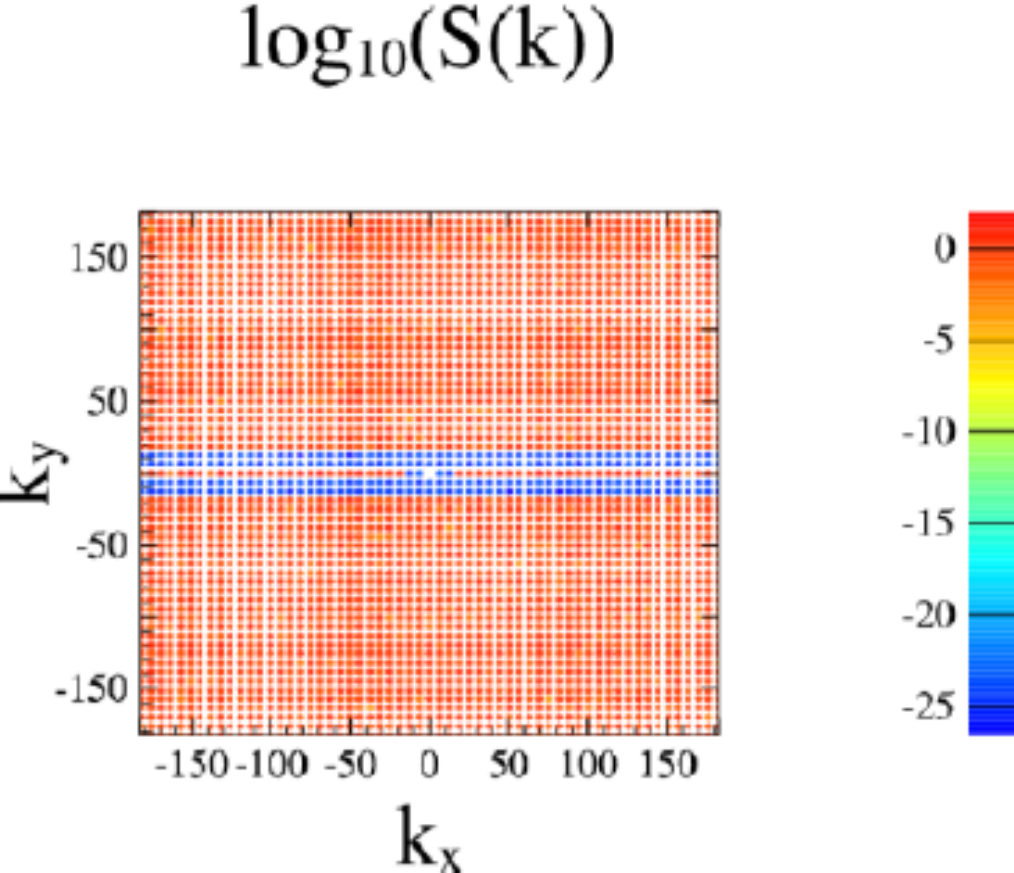}}}
\caption{The scattering patterns corresponding to the configurations to the left, middle and right
panels, respectively, shown in Fig. \ref{stacked-1}. Note that in each example, there are
wave vectors in which scattering is suppressed outside the circular exclusion region
centered at the origin.
}
\label{stacked-2}
\end{figure}

\subsubsection{Comparison of Stealthy Configurations to Some Common States of Matter}

Stealthy disordered  ground states as well as stealthy stacked-slider phases
are distinguishable states of matter that are generally uniquely different
from crystals, quasicrystals and some other common states
matter.  Table \ref{phases} compares these states of matter.
Unlike perfect crystals, perfect quasicrystals to do not possess long-range translational order.
Crystals, quasicrystals, and stacked-slider phases all have long-range orientational
order, but with different symmetries. While crystals can
only have two-fold, three-fold, four-fold, or six-fold rotational
symmetries, quasicrystals have prohibited crystallographic
rotational symmetries (e.g., five-fold symmetry in the plane
or icosahedral symmetry in three dimensions \cite{Le84,Sh84,Lev86}). Disordered stealthy ground states are isotropic and hyperuniform,
cannot support shear stresses, and lack long-range translational and orientational  order. Stacked-slider phases generally do not
have any rotational symmetry, but the fact that their stacking directions
are different from the sliding directions endows them with unique
orientational order.

\begin{table*}[bthp]
\caption{Comparison of the properties of some common states of matter, as summarized in Ref. \cite{Zh15a}. Here ``crystals" and ``quasicrystals" signify perfect crystals and perfect quasicrystals, respectively, without any defects (e.g., phonons and phasons). The checks and crosses indicate whether or not different phases have the attributes listed in the first column. This table is adapted from the one given in Ref. \cite{Zh15b}.}
\renewcommand{\arraystretch}{2}
\begin{tabular}{|c || c| c |c| c| c| c|}
\hline
& \parbox{1.3cm}{Crystals \\ \cite{As76,Chaik95}} & \parbox{2cm}{Quasicrystals \\ \cite{Sh84,Le84,Lev86}}& \parbox{2cm}{Stacked-slider phases \\ \cite{Uc04b,Zh15b}} & \parbox{3cm}{Disordered stealthy  \\ ground states \\ \cite{Uc04b,Uc06b,Ba08,To15,Zh15a}}& \parbox{2.2cm}{Typical liquid crystals \\ \cite{De95}} & \parbox{1.3cm}{Typical liquids \\ \cite{Han13}}\\
\hline
Translational order & \cmark & \xmark & \xmark & \xmark & \xmark & \xmark \\\hline
Orientational order & \cmark & \cmark & \cmark & \xmark & \cmark & \xmark \\\hline
Hyperuniformity & \cmark & \cmark & \cmark & \cmark & \xmark & \xmark \\\hline
Anisotropy & \cmark & \cmark & \cmark & \xmark & \cmark & \xmark \\\hline
Positive shear modulus & \cmark & \cmark & \xmark & \xmark & \xmark & \xmark \\\hline
\end{tabular}
\label{phases}
\end{table*}

\subsubsection{Excited (Vibrational) States}
\label{excited}

 One can also derive accurate analytical formulas for the structure factor 
at the origin $S({\bf k}=0)$, and thermal expansion coefficient 
for the excited (vibrational) states associated with stealthy ground states at sufficiently small temperatures.
Specifically, for small $\chi$ (large $\rho$)
and small $T$, it has been shown that $S(0)$ varies linearly with 
$T$ and proportional to $\chi$ \cite{To15}:
\begin{equation}
S(0) \sim C(d) \,\chi\, T,
\label{S0}
\end{equation}
in units $k_B=v_0=K=1$, where $C(d)= 2d \,(2\pi)^d /v_1(1)$ is a $d$-dependent constant. 
Figure \ref{ex} shows that the prediction of relation (\ref{S0})
is in excellent agreement with  MD simulation results in the case $d=2$.
An interesting conclusion to be drawn from this analysis 
is that $S(0)$ can be arbitrarily close to zero at positive temperatures when $T$ is arbitrarily small.
This means that, for all practical purposes, such systems at positive $T$ are effectively hyperuniform.
We remind the reader that perfect hyperuniformity is not necessarily required in order to achieve
novel physical properties in technological applications. 

\begin{figure}[H]
\begin{center}\includegraphics[  width=3.5in,keepaspectratio,clip=]{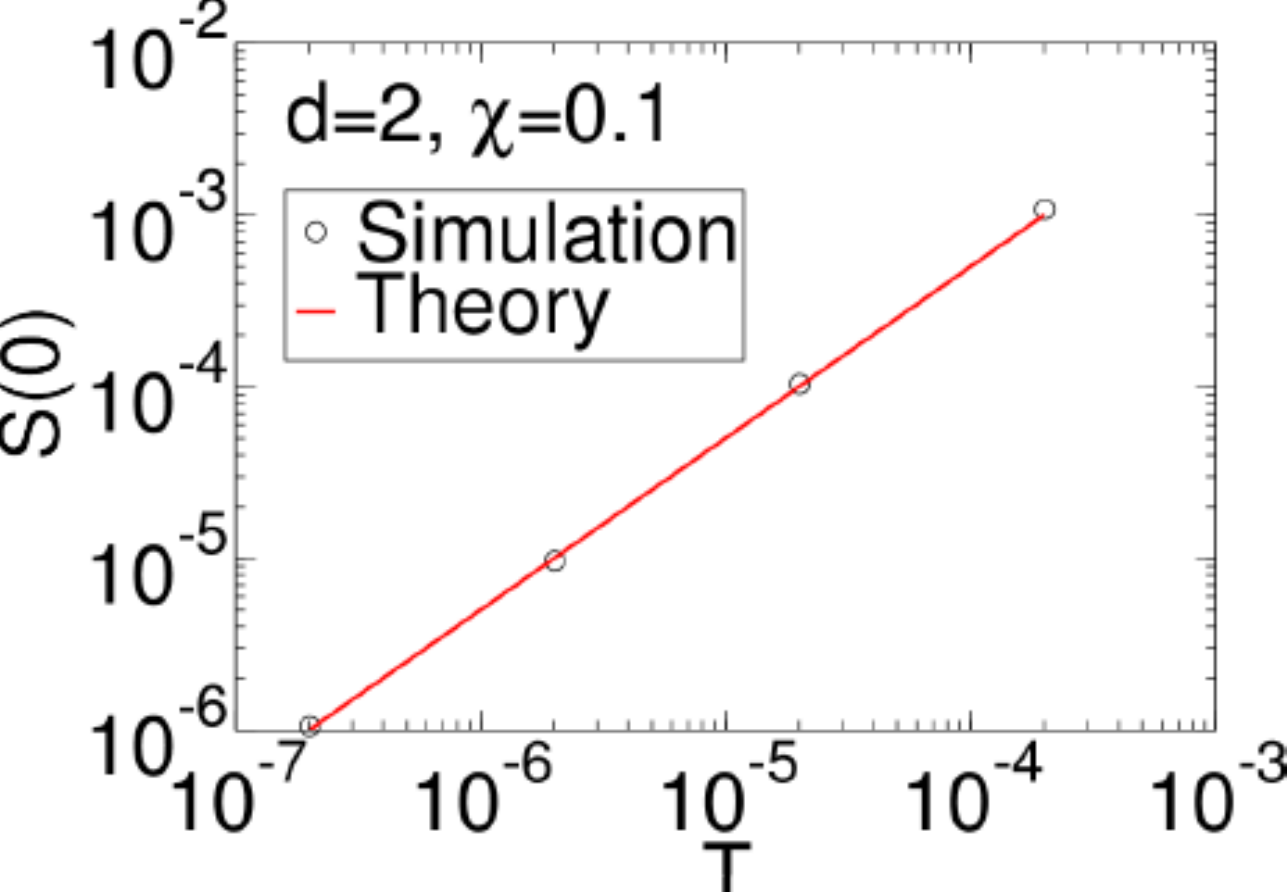}
\caption{Comparison of the theoretically predicted structure factor at the origin $S(0)$ versus absolute temperature $T$,
as obtained from Eq. (\ref{S0}), to our corresponding simulation results for
a two-dimensional stealthy system at $\chi=0.1$. Here we take $k_B=v_0=K=1$. This figure is taken from Ref. \cite{To15}. }\label{ex}
\end{center}
\end{figure}

Interestingly,  at positive temperatures within the harmonic regime, particle systems
interacting with isotropic stealthy pair potentials exhibit negative thermal expansion \cite{Ba09a,Ba09b},
which is an unusual behavior for single-component
systems with isotropic interactions. When
 heated at constant pressure, the system attains a density maximum \cite{Ba09a,Ba09b}.

\section{1D Hyperuniform Patterns Derived from  Random Matrices and the Zeta Function}
\label{Matrices}

It is well known that there are remarkable connections
between the statistical properties of   certain random matrices, the zeros of the Riemann zeta function,
classical Coulomb gases, and fermionic ground states \cite{Wi51,La55,Me60,Mc60,Dy62a,Dy62b,Dy62c,Dy63,Me91,Dy70,Mo73,Od87,Rud96,Fe98,Ka99,To08b,Fo10,Ta11}, all of which
can be regarded to be one-dimensional point configurations.
These correspondences came to light through a chance meeting between Freeman Dyson and Robert Montgomery
during tea time at the Institute for Advanced Study in Princeton, New Jersey in 1972. 
Underlying each of the aforementioned four seemingly disparate systems  are certain one-dimensional disordered point processes 
whose statistical properties (under certain limits) are believed to be identical. Here we will re-examine these
results through the ``hyperuniformity lens," including two other random matrices 
that are hyperuniform but are not connected to the Riemann zeta function.

Random-matrix theory was pioneered by Wigner \cite{Wi51,La55} in the 1950's 
to model the statistics of the eigenvalues and eigenfunctions of complex many-body quantum systems
by studying the spectrum of certain large random matrices.
It was more fully developed by Dyson and Mehta in the 1960's \cite{Me60,Dy62a,Dy62b,Dy62c,Dy63}.
Wigner's original study was concerned with neutron excitation spectra of heavy
nuclei. Since this seminal work, random matrices have found applications 
in a host of fields including quantum chaos \cite{Be85,Berry86},
  nuclear physics \cite{Gu98}, black holes \cite{Ba14,Cot17}, quantitative finance
\cite{La99,Pl02}, complex energy landscapes \cite{Maj09}  and number theory \cite{Mo73,Ka99}. 
There are three prominent theories of random Hermitian matrices that
model the Hamiltonians of a wide class of random dynamical systems; see
the excellent book by Mehta \cite{Me91}. If the Hamiltonian is symmetric under time reversal, 
the relevant theory is that of the Gaussian orthogonal ensemble (GOE), 
which consists of the sets of all $N\times N$ real symmetric matrices with the unique
probability measure that is invariant under orthogonal transformations
such that the individual matrix entries are independent random variables.
The Gaussian symplectic ensemble (GSE) models Hamiltonians that are symmetric under time reversal
but possess no rotational symmetry. The GSE consists of the sets of all $N\times N$ Hermitian quaternionic matrices  with the unique
probability measure that invariant under symplectic transformations
such that the individual matrix entries are independent random variables. On the other hand, the Gaussian 
unitary ensemble (GUE), which models random Hamiltonians without time reversal symmetry, is directly relevant to certain properties of the Riemann zeta function
as described below.
The GUE  consists of the set of all $N\times N$ Hermitian matrices together with a Haar measure.
This is the unique probability measure on the set of $N\times N$ Hermitian matrices that is invariant under
conjugation by unitary matrices such that the individual matrix entries are independent random variables.
The {\it universality conjecture} asserts that the $n$-particle correlations functions of a large 
class of random matrices exhibit a bulk universal behavior
in the large $N$-limit depending only on the symmetry class of the matrix ensemble \cite{Me91,Ka99,Fo10,Ta11}.

Dyson \cite{Dy62a,Me91} showed that the eigenvalue distributions of the GOE, GUE and GSE
are given by the Gibbs canonical distribution function $P_N({\bf r}^N)$ that correspond to 
one-dimensional systems of certain charged point particles at unit number density with  positions ${\bf r}^N$ contained within
an interval of length $L \subset \mathbb{R}$ at certain temperatures, i.e.,
\begin{equation}
P_N({\bf r}^N) =\frac{1}{Z(N,L,T)}  \exp[-\beta \Phi_N({\bf r}^N)],
\end{equation}
where
\begin{equation}
\Phi_N({\bf r}^N)=\frac{1}{2} \sum_{i=1}^N |{\bf r}_i|^2 - \sum_{i < j}^N \ln(|{\bf r}_i -{\bf r}_j|)\,.
\label{phi-G}
\end{equation}
is total potential energy of the system, $Z(N,L,T)$ is the canonical partition function,
and $\beta=(k_BT)^{-1}$ is a reciprocal temperature. The second term in (\ref{phi-G})
corresponds to point particles interacting via the two-dimensional repulsive logarithmic
Coulomb potential, $v(r)=-\ln(r)$, whereas the first term in (\ref{phi-G})
is a harmonic potential that confines the charged particles by attracting them towards the origin. 
In the thermodynamic limit, the eigenvalue distributions of the GOE, GUE and GSE correspond configurationally to these Coulomb gases 
at reciprocal temperatures $\beta=
1, 2, 4$, respectively \cite{Me91}.  The confinement of the gas by a harmonic 
potential is absent in the closely related  {\it circular ensembles} originally considered by Dyson.

In the particular case of the GUE in the limit $N\rightarrow \infty$ such
that the mean gap distance between eigenvalues in the vicinity of the origin is suitably normalized, the
pair correlation function at number density $\rho=1$ is known exactly:
\begin{eqnarray}
g_2(r)=1-\frac{\sin^2(\pi r)}{(\pi r)^2}.
\label{g2-zeros}
\end{eqnarray}
The aforementioned normalization has the effect of magnifying the bulk of the eigenvalue density on $\mathbb{R}$ such that
(\ref{g2-zeros}) is well-defined.
We see that this point process is always {\it negatively correlated}, i.e., $g_2(r)$ never exceeds unity
(see Fig. \ref{zeta-zero}) and is ``repulsive" in the sense
that $g_2(r) \rightarrow 0$ as $r$ tends to zero. In fact, the pair correlation function tends to zero quadratically
in $r$ in this asymptotic limit; specifically,
\begin{equation}
g_2(r)=\frac{\pi^2}{3}r^2-\frac{2\pi^4}{45}r^4+{\cal O}\left(r^6 \right) \qquad (r \rightarrow 0).
\label{gue-g2-2}
\end{equation}
The corresponding structure factor $S(k)$ for the GUE at unit number density is given by
\begin{eqnarray}
S(k) = \left\{
\begin{array}{lr}
\frac{\displaystyle k}{\displaystyle 2\pi}, \quad 0 \le k \le 2\pi\\\\
\displaystyle{1}, \quad k > 2\pi.
\end{array}\right.
\label{spec}
\end{eqnarray}
We see that structure factor goes to zero linearly in $k$ as wavenumber goes to zero and is equal to unity
for $k > 2\pi$. Hence, this one-dimensional point process is perfectly hyperuniform and disordered. 
This small-$k$ behavior of the structure factor is reflected in a pair correlation whose decay envelope
for large $r$ is controlled by the inverse power law $1/r^2$; see the general relations (\ref{S-asy}) 
and (\ref{sigma-N-asy}). Clearly, the physical origin 
of the hyperuniformity of this point process is the long-range Coulombic interactions between the point particles.

It is noteworthy that Dyson \cite{Dy70} proved that the $n$-particle correlation function 
for the GUE in the thermodynamic limit is exactly given by the following
determinant:
\begin{eqnarray}
g_n(r_{12},r_{13},\ldots,r_{1n})=\mbox{det}\left( \frac{\sin(\pi r_{ij})}{\pi r_{ij}}\right)_{i,j=1,\ldots,n}.
\label{gn-zeros}
\end{eqnarray}
Thus, the $g_n$ for $n \ge 3$ are entirely determined by the pair correlation function $g_2$.

\begin{figure}[bthp]
\centerline{\psfig{file=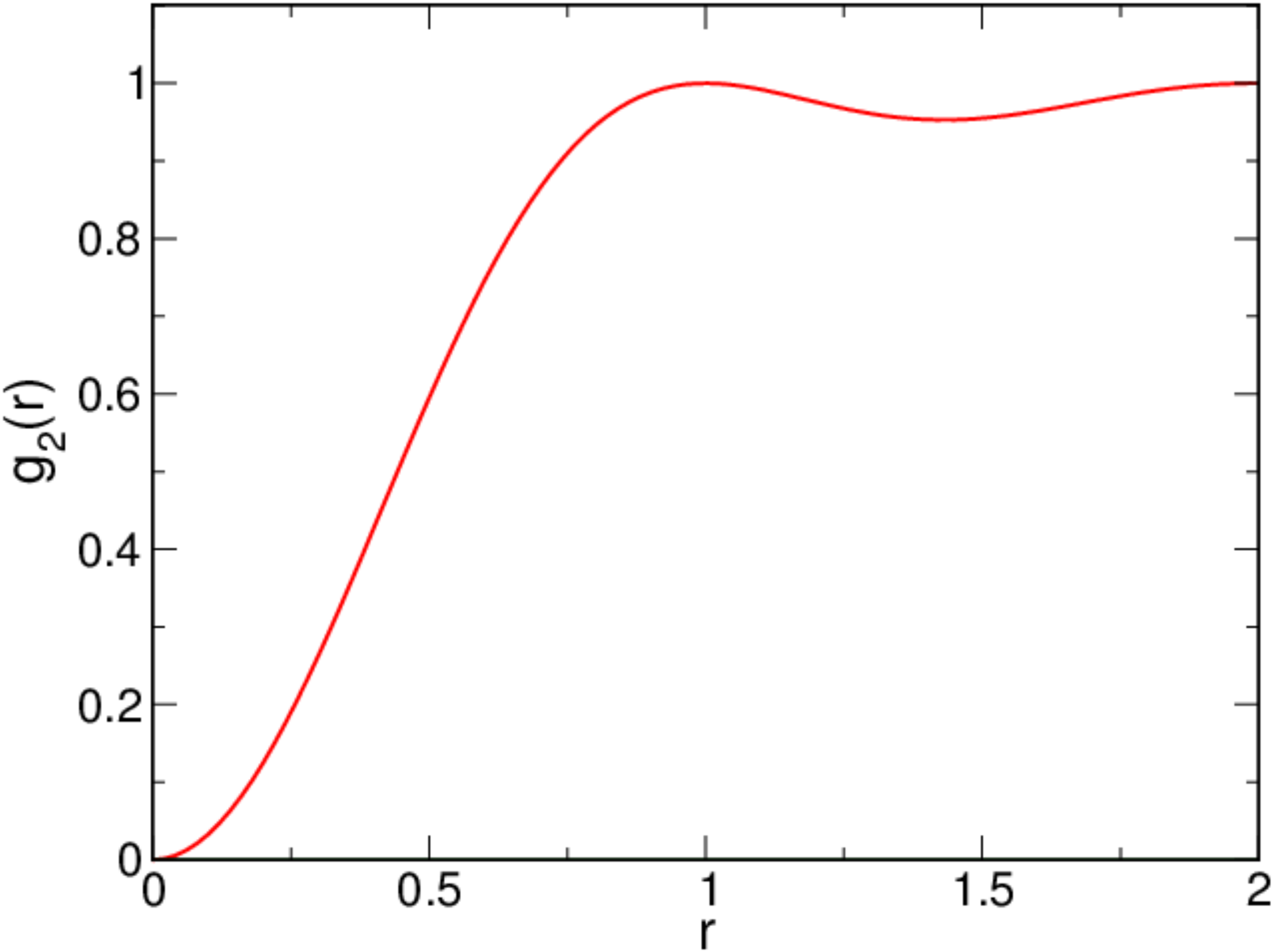,width=2.6in,clip=}\hspace{0.3in}\psfig{file=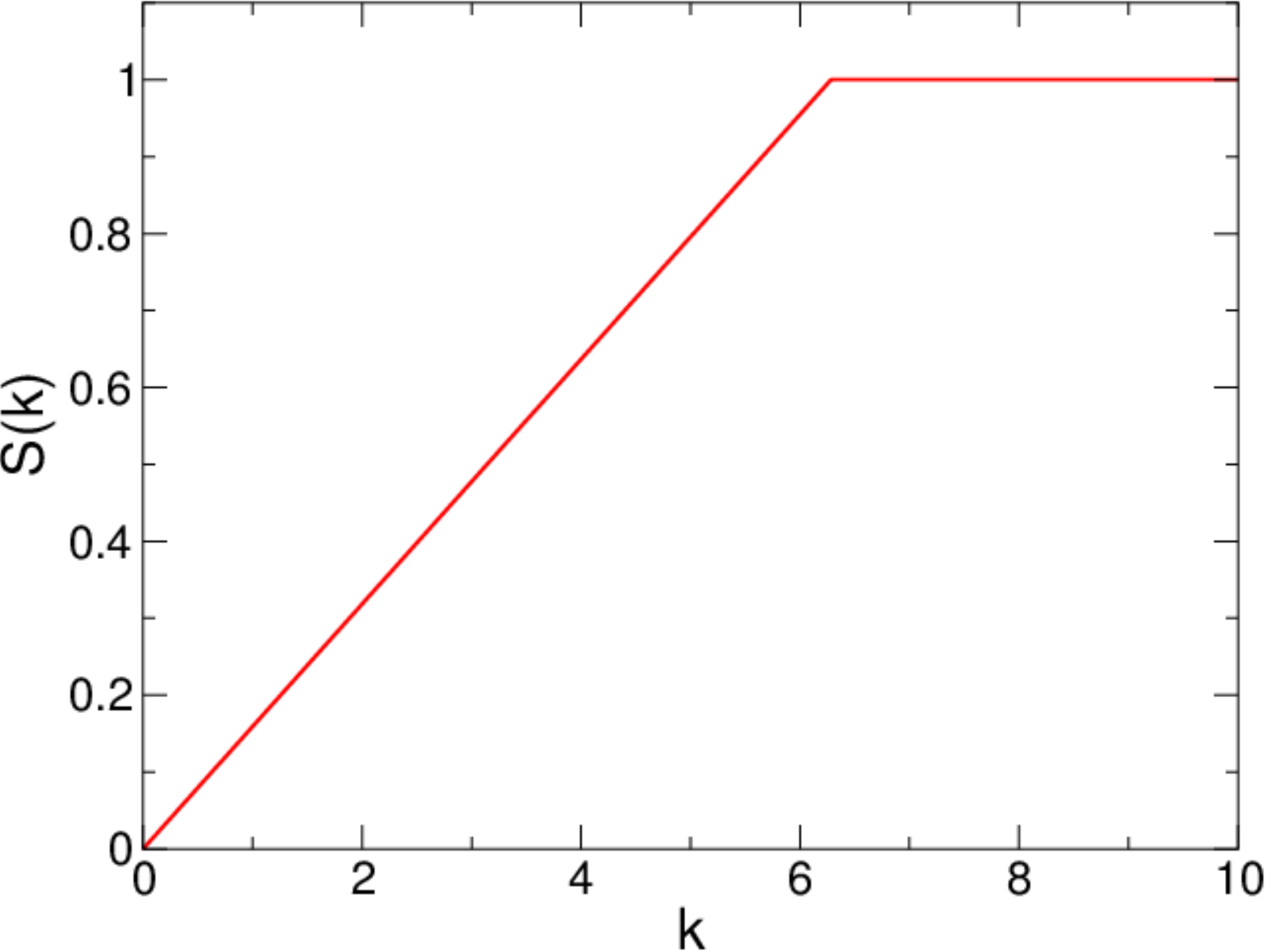,width=2.6in,clip=}}
\caption{Left panel: The pair correlation function $g_2(r)$ versus
distance $r$ for the eigenvalues of the GUE in the large-$N$ limit, as given by (\ref{g2-zeros}),
which is conjectured to be the same as the one characterizing the nontrivial zeros of the Riemann zeta function.
Noninteracting spin-polarized fermions in their ground state in $\mathbb{R}$ have the same pair correlation function, as will
be discussed in Sec. \ref{Det}.
Right panel: The corresponding  structure factor $S(k)$ [cf. (\ref{spec})],
as a function of wavenumber $k$.}
\label{zeta-zero}
\end{figure}

There is a remarkable correspondence between the statistical properties of the eigenvalues
of the GUE and those of the nontrivial zeros of the Riemann zeta function $\zeta(s)$.
The latter is a function of a complex variable $s$
and can be written as the following infinite series:
\begin{equation}
\zeta(s) =\sum _{n=1}^{\infty }{\frac {1}{n^{s}}},
\end{equation}
which converges for $\mbox{Re}(s)>1$.  However, $\zeta(s)$ has a unique analytic continuation to the entire complex plane,
excluding the simple pole at $s=1$.
According to the Riemann hypothesis, the nontrivial
zeros of the zeta function lie along the critical line $s=1/2 + it$ with $t\in\mathbb{R}$ in the complex plane; see Fig. \ref{nontrivial}.
Montgomery \cite{Mo73} conjectured that the pair correlation function associated with the nontrivial zeros
in the asymptotic limit (high on the critical line), when appropriately normalized, is given exactly by (\ref{g2-zeros}),
which he did not realize until his chance meeting with Dyson, was identical to the GUE
pair correlation. This correspondence was further established by Odlyzko \cite{Od87}, who numerically verified the Riemann hypothesis for the first $10^{13}$ nontrivial zeros of the zeta function and at much larger heights, and confirmed that its $g_2(r)$
agrees with (\ref{g2-zeros}). Rudnick and Sarnak \cite{Ru96}
proved that, under the Riemann hypothesis, the nontrivial zeros
have $n$-particle densities for any $n$ given by (\ref{gn-zeros}).
The reader is referred to the excellent review article
by Katz and Sarnak \cite{Ka99}, which discusses the connection
between the zeros of zeta functions and classical symmetric
groups, of which the three canonical random-matrix ensembles
are but special cases.

\begin{figure}[bthp]
\centerline{\includegraphics[  width=4in,keepaspectratio,clip=]{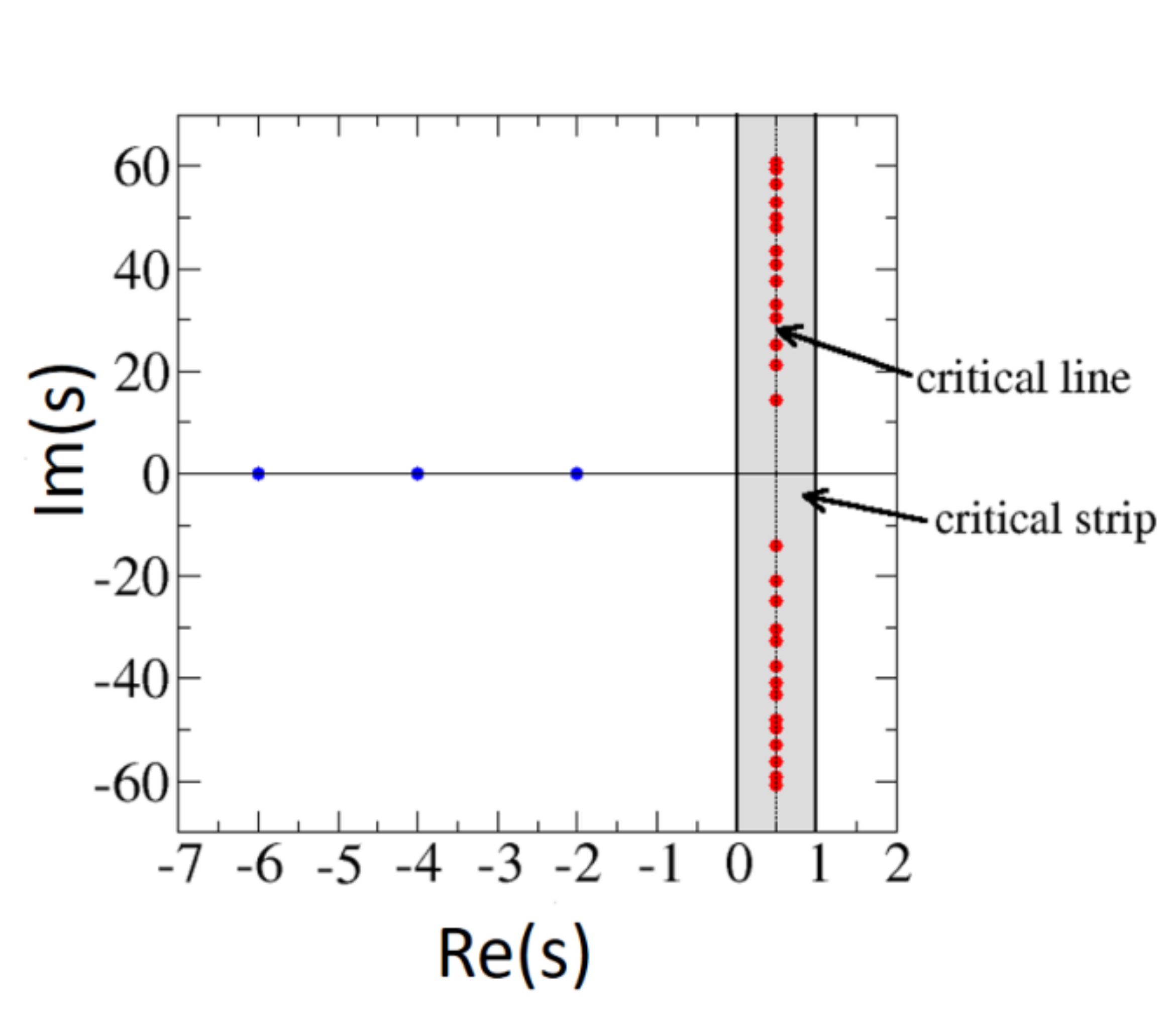}}
\caption{A schematic of a few of the trivial zeros (blue circles)
and nontrivial zeros (red circles) of the Riemann zeta function in the complex plane. 
It is known that all of the nontrivial zeros must lie in the ``critical strip" defined
by the open strip" $\;\{ \zeta \in \mathbb{C}: 0 < \mbox{Re}(\zeta) <1\}$. Riemann conjectured
that all of the nontrivial zeros of the zeta function lie along the critical line $1/2 + it$ with $t\in\mathbb{R}$.}
\label{nontrivial}
\end{figure}

For spin-polarized free fermions in $\mathbb{R}$ (fermion gas) at number density $\rho=1$, it is known that the pair correlation function in the {\it ground state} (i.e., completely filling the Fermi ``sphere") is given by
\begin{eqnarray}
g_2(r)=1-\frac{\sin^2(k_F r)}{(k_F r)^2},
\label{g2-fermion}
\end{eqnarray}
where $k_F$ is the Fermi radius, which is the one-dimensional
analog of the well-known three-dimensional result \cite{Fe98}.
Therefore, we see that when $k_F=\pi$, we obtain
the GUE pair correlation function (\ref{g2-zeros}). The repulsive nature
of the point process in this context arises physically from
the Pauli exclusion principle. Exact generalizations of
these one-dimensional point processes in
$d$-dimensional Euclidean space $\mathbb{R}^d$ for any $d$
are discussed in Sec. \ref{Det}.

The suitably normalized pair correlation functions for the GOE and GSE
are known analytically \cite{Me91}  and given  respectively by
\begin{equation}
g_2(r)=1-\frac{\sin^2(\pi r)}{(\pi r)^2}+\frac{1}{2(\pi r)^2}\Big(\pi r \cos(\pi r)-\sin(\pi r)\Big)\Big(2\, \mbox{Si}(\pi r)-\pi\Big) 
\label{goe-g2-1}
\end{equation}
and 
\begin{equation}
g_2(r)=1-\frac{\sin^2(2\pi r)}{(2\pi r)^2}+\frac{1}{4(\pi r)^2}\Big(2\pi r \cos(2\pi r)-\sin(2\pi r)\Big)
\,\mbox{Si}(2\pi r), 
\label{gse-g2-1}
\end{equation}
where $\mbox{Si}(x)\equiv \int_0^x (\sin(t)/t) \;dt$ is the sine integral. The left panel of Fig. \ref{goe-gse} shows that the pair correlation functions for the GOE and GSE exhibit oscillations that are substantially suppressed and amplified, respectively, compared
to that of the GUE (see Fig. \ref{zeta-zero}). The small-$r$ and large-$r$ asymptotic behaviors
of these pair correlation functions are different from one another and the GUE asymptotic behaviors.
 While the point particles associated with the GOE and GSE
tend to repel one another for small $r$ (as they do in the GUE), $g_2(r)$ tends to zero as $r \rightarrow 0$
linearly in $r$ in the former  and  quartically in $r$ in the latter; more precisely,
\begin{equation}
g_2(r)=\frac{\pi^2}{6}r-\frac{\pi^4}{60}r^3+{\cal O}\left(r^5 \right) \qquad (r \rightarrow 0)
\label{goe-g2-2}
\end{equation}
and
\begin{equation}
g_2(r)=\frac{16\pi^4}{135}r^4-\frac{128\pi^6}{4725}r^6+{\cal O}\left(r^8 \right) \qquad (r \rightarrow 0).
\label{gse-g2-2}
\end{equation}
By contrast, the GUE $g_2(r)$ in the small-$r$ limit tends to zero quadratically, as seen in Eq. (\ref{gue-g2-2}).
Moreover, the large-$r$ asymptotic behaviors for the GOE and GSE are given respectively by
\begin{equation}
g_2(r)=1-\frac{1}{(\pi r)^2}+ \frac{1+\cos^2(\pi r)}{(\pi r)^4} - {\cal O}\left(\frac{1}{r^6}\right) \qquad (r \rightarrow \infty)
\label{goe-g2-3}
\end{equation}
and
\begin{equation}
g_2(r)=1+\frac{\cos(2\pi r)}{4 r}-\frac{2+\pi \sin(\pi r)}{4(\pi r)^2} + {\cal O}\left(\frac{1}{r^4}\right) \qquad (r \rightarrow \infty)
\label{gse-g2-3}
\end{equation}

The closed-form exact expressions for the structure factors for the GOE and GSE corresponding to (\ref{goe-g2-1}) and (\ref{gse-g2-1}) are 
respectively given by \cite{Me91} 
\begin{eqnarray}
S(k) = \left\{
\begin{array}{lr}
{\displaystyle \frac{k}{\pi} -\frac{k}{2\pi}\ln(k/\pi+1)}, \quad 0 \le k \le 2\pi\\\\
\displaystyle{2 -\frac{k}{2\pi} \ln\left(\frac{k/\pi+1}{k/\pi -1}\right)}, \quad k > 2\pi.
\end{array}\right.
\label{goe-S-1}
\end{eqnarray}
and
\begin{eqnarray}
S(k) = \left\{
\begin{array}{lr}
{\displaystyle \frac{k}{4\pi} -\frac{k}{8\pi}\ln\Big(|1-k/(2\pi)|\Big)}, \quad 0 \le k \le 4\pi\\\\
\displaystyle{1}, \quad k > 4\pi.
\end{array}\right.
\label{gse-S-1}
\end{eqnarray}
Observe that the GSE structure factor possesses a cusp at $k=2\pi$, which
is to be contrasted with the smooth GOE structure factor; see the right panel of Fig. \ref{goe-gse}.
Moreover, like the GUE structure factor (\ref{spec}), we see  that the structure factors of both the GOE and GSE tend to zero linearly in $k$ 
in the limit $k \rightarrow 0$; specifically
\begin{equation}
S(k)=\frac{1}{\pi} k- \frac{1}{2\pi^2} k^2 + {\cal O}(k^3) \qquad (k \rightarrow 0)
\label{goe-S-2}
\end{equation}
and
\begin{equation}
S(k)=\frac{1}{4\pi} k+ \frac{1}{8\pi^2} k^2 + {\cal O}(k^3) \qquad (k \rightarrow 0).
\label{gse-S-2}
\end{equation}
In the opposite large-$k$ asymptotic limit, the GOE structure factor decays to unity like $1/k^2$, i.e.,
\begin{equation}
S(k)=1-\frac{\pi^2}{3k^2}+ \frac{\pi^4}{5k^4} + {\cal O}(\frac{1}{k^6}) \qquad (k \rightarrow \infty)
\label{goe-S-3}
\end{equation}
whereas $S(k)=1$ for any $k > 4\pi$ in the case of the GSE.

\begin{figure}[bthp]
\centerline{\psfig{file=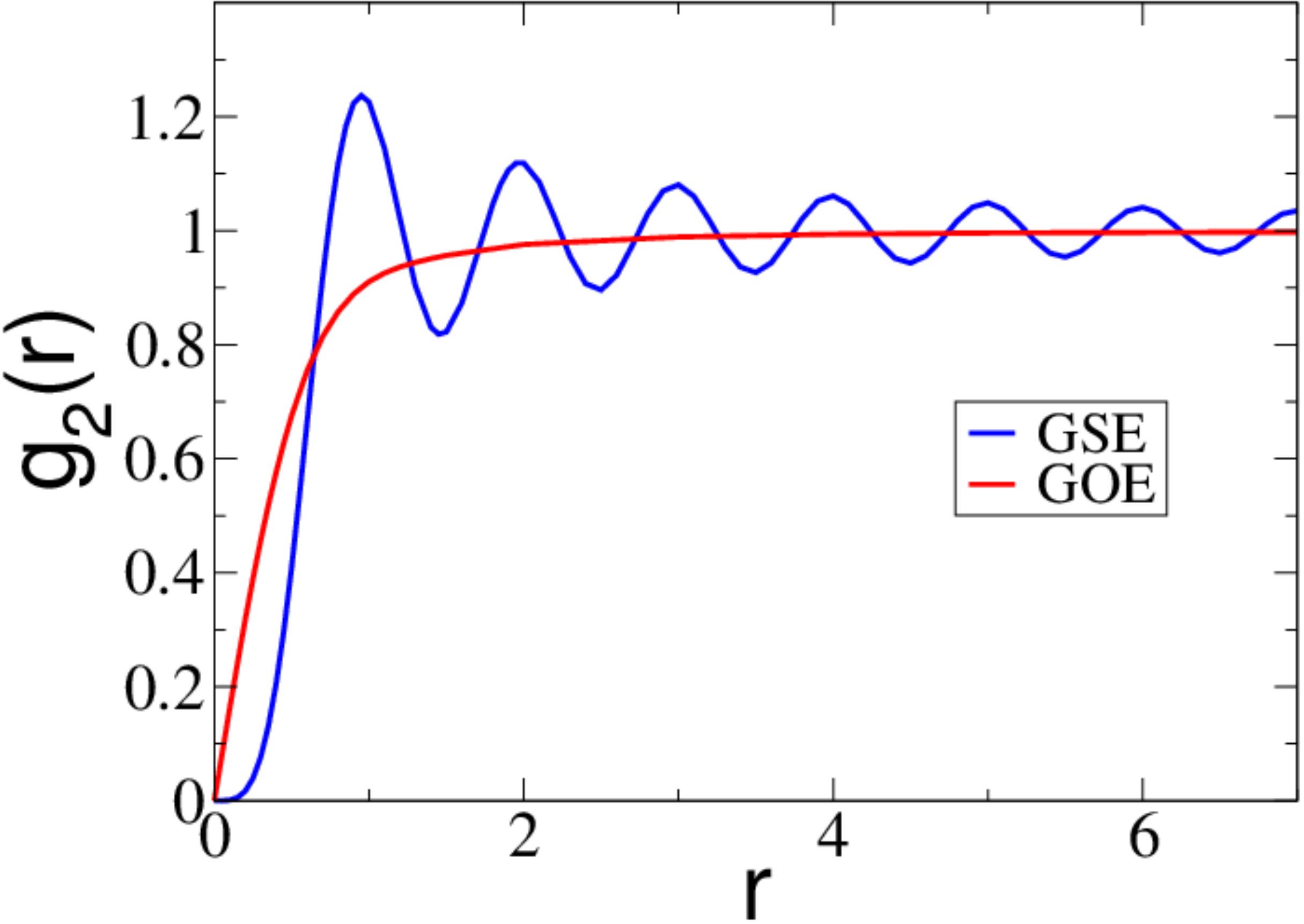,width=2.6in,clip=}\hspace{0.3in}\psfig{file=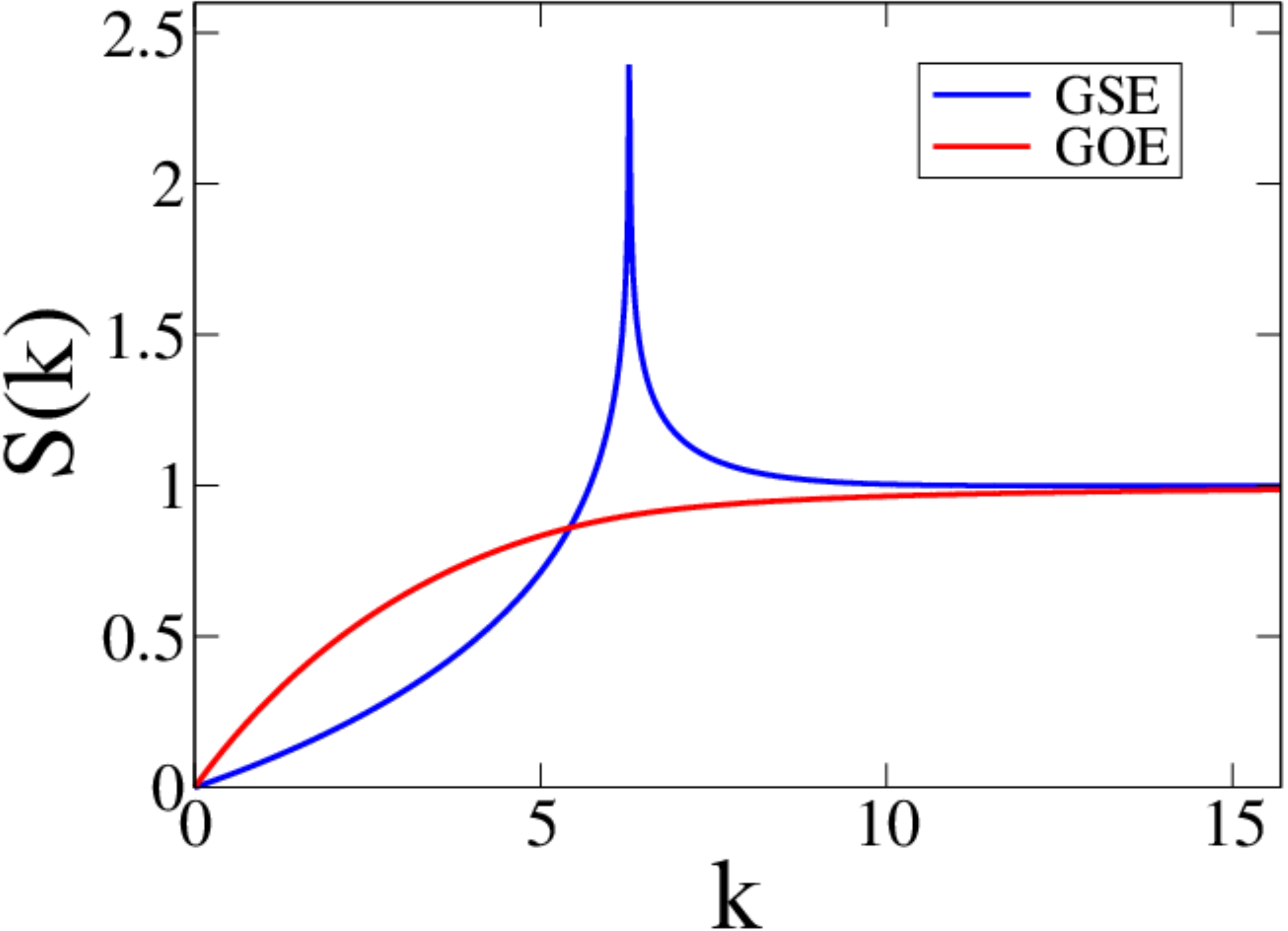,width=2.6in,clip=}}
\caption{Left panel: The pair correlation functions for the eigenvalues of the GOE and GSE in the large-$N$ limit, as given by 
(\ref{goe-g2-1}) and (\ref{gse-g2-1}), respectively.
Right panel: The corresponding  structure factors, as given by 
(\ref{goe-S-1}) and (\ref{gse-S-1}).}
\label{goe-gse}
\end{figure}

The number variance in the large-$R$ limit for any of the three random Hermitian matrices 
is given by
\begin{equation}
\sigma^2_{_N}(R)= a_\beta \ln(R) + {\cal O}(1) \qquad (R \rightarrow \infty),
\label{matrices}
\end{equation}
where $a_\beta = 2/(\beta \pi^2)$, and $\beta =1, 2,4$ for GOE, GUE, and GSE, respectively,
and thus they all belong to class II hyperuniform systems such that GSE 
suppresses density fluctuations to the greatest degree at large $R$.
This logarithmic growth with $R$ is consistent with a structure factor $S(k)$ that tends to zero linearly in $k$
in the small-wavenumber limit; see the general relations (\ref{S-asy}) and (\ref{sigma-N-asy}).

\section{Hyperuniform Determinantal Point Processes}
\label{Det}

One class of disordered point processes that exhibits exact hyperuniformity is  \emph{determinantal point processes},
which are characterized by a joint probability distribution given by the determinant of a finite-rank,
positive, bounded, and self-adjoint operator.  
Determinantal point processes were introduced
by Macchi \cite{Ma75}, who originally called them fermion point processes
because they are rooted in quantum fermionic statistics.
Soshnikov \cite{So00} presented a review of this subject and discussed
applications to random matrix theory, statistical mechanics,
quantum mechanics, and representation theory. Examples of determinantal point processes arise
in  uniform spanning trees \cite{Bu93}, self-avoiding random walks \cite{Jo04} and 
random polynomials \cite{Pe05}, and have been reported 
both on the real line \cite{Cos04, De08, To08b, Sc09} and in
higher Euclidean space dimensions \cite{To08b, Sc09,Ab17}.

\subsection{General Considerations}

In what follows, we briefly review the mathematics of determinantal point processes.
Without loss of generality, the number density is set to unity ($\rho=1$) in
the ensuing discussion.
Let $H({\bf r})=H(-{\bf r})$ be  a translationally invariant Hermitian-symmetric kernel of an integral
operator $\cal H$. A translationally invariant {\it determinantal point process} in $\mathbb{R}^d$ exists if 
the $n$-particle density functions for $n \ge 1$ are given by the following
determinants:
\begin{eqnarray}
\rho_n({\bf r}_{12},{\bf r}_{13},\ldots,{\bf r}_{1n})=\mbox{det}[H({\bf r}_{ij})]_{i,j=1,\ldots,n}.
\label{rhon-det}
\end{eqnarray}
 By virtue of the nonnegativity of the $\rho_n$ in the pointwise sense [cf. (\ref{positive})]
and (\ref{rhon-det}), it follows that $\cal H$ must have nonnegative minors and $\cal H$ must
be a nonnegative operator, which implies that $H({\bf r})$ is positive semidefinite.
The latter implies that the  Fourier transform ${\tilde H}({\bf k})$ of the kernel $H({\bf r})$ is nonnegative.
In particular, let
\begin{eqnarray}
0 \le {\tilde H}({\bf k}) \le 1 \quad \mbox{for all} \; {\bf k},
\label{f-K}
\end{eqnarray}
and $H({\bf 0})=1$, implying the sum rule
\begin{equation}
\frac{1}{(2\pi)^{d}}\int_{\mathbb{R}^d} {\tilde H}({\bf k}) d{\bf k}=1.
\label{sum-rule}
\end{equation}
It follows that a  Fourier transform ${\tilde H}({\bf k})$ that satisfies the inequalities 
(\ref{f-K}) and sum rule (\ref{sum-rule}) describes a determinantal
point process with a pair correlation
function $g_2({\bf r})$ given by
\begin{eqnarray}
g_2({\bf r})=1-|H({\bf r})|^2,
\label{g2-determ}
\end{eqnarray}
such that
\begin{eqnarray}
0 \le g_2({\bf r}) \le 1 \qquad \mbox{and} \qquad g_2({\bf 0})=0,
\label{conditions}
\end{eqnarray}
and a $n$-particle density function given by (\ref{rhon-det}).
We see that the total correlation function 
is given by $h({\bf r})=-|H({\bf r})|^2$ and therefore the corresponding structure factor is given by
\begin{equation}\label{Sdetpphyp}
S({\bf k}) = 1-{\cal F}[|H|^2](k),
\end{equation}
where $\cal{F}$ denotes the Fourier transform.
To obtain a hyperuniform  determinantal point process  specified by some $H({\bf r})$ that satisfies
the aforementioned conditions, it is necessary and sufficient that 
${\tilde H}({\bf k})$ be an indicator  function of a set $\omega$ in $\bf k$-space, i.e., ${\tilde H}({\bf k}) = 1$ for ${\bf k} \in \omega$
and  ${\tilde H}({\bf k})=0$ otherwise \cite{Cos04}. 

\subsection{Fermi-Sphere Point Processes}
\label{fermi}

The so-called Fermi-sphere point process \cite{To08b} provides a $d$-dimensional generalization of 
the one-dimensional   point processes
corresponding to the eigenvalues of the GUE, the zeros of the Riemann zeta function
or the fermionic gas, all of which were described in Sec. \ref{Matrices}. This disordered
hyperuniform point process belongs to class II and corresponds exactly to the 
one associated with the ground state 
of a noninteracting spin-polarized fermions in which the Fermi sphere is completely filled.
The $d$-dimensional Fermi-sphere point process is a special determinantal point process obtained by taking  ${\tilde H}({\bf k})$ 
to be the indicator function
of a (Fermi) sphere of radius $K= 2 \sqrt{\pi}\, [\Gamma(1+d/2)]^{1/d}$  \cite{To08b}.
This means that the $n$-particle densities given by (\ref{rhon-det}) have a squared kernel
$|H({\bf r})|2={\tilde \alpha}_2(r;K)/(2\pi)^{d}$ and hence, using (\ref{g2-determ}),
the pair correlation function of such a point process at unit density is given by
\begin{eqnarray}
g_2({\bf r})= 1-2^d\Gamma(1+d/2)^2\frac{J^2_{d/2}(Kr)}{(Kr)^{d}}.
\label{g2-d-2}
\end{eqnarray}
Moreover, the corresponding structure factor  takes the form
\begin{eqnarray}
S({\bf k})=1-\alpha_2(k;K),
\label{S-d}
\end{eqnarray}
where $\alpha_2(k;K)$ is the scaled intersection volume of two $d$-dimensional spheres
of radius $K$ separated by a distance $k$, i.e., the function $\alpha_2(r;R)$,
specified by (\ref{alpha}), with the replacements $r \rightarrow k$ and $R \rightarrow K$.
It follows from the properties of $\alpha_2(k;K)$  in (\ref{S-d})
that the structure factor $S({\bf k})$ obeys the bounds
$0 \le S({\bf k}) \le 1$  for all $\bf k$,
and  achieves its maximum value of unity for $|{\bf k}| \ge 2K$.

Figure \ref{fermions} shows the pair statistics
in both real and Fourier space for $d=1$ and $d=3$
at unit density. Perfect hyperuniformity is manifest in the small-wavenumber behavior
of the structure factors. In both one and three dimensions, the point particles tend to repel one another,
as reflected by the fact that $g_2(r) \rightarrow 0$ as $r$ tends to zero. 
We also see that the amplitudes of the oscillations
in $g_2(r)$ that are apparent for $d=1$ are significantly
reduced in the corresponding three-dimensional pair correlation function. 
By virtue of the asymptotic properties of the Bessel function of arbitrary order,
the small-$r$ and large-$r$ forms of the pair correlation function (\ref{g2-d-2}) are respectively given by
\begin{eqnarray}
g_2(r)= \frac{K^2}{d+2} r^2 - \frac{(d+3)K^4}{2(d+2)^2(d+4)} r^4 +\; {\cal O}(r^6)  \qquad (r \rightarrow 0)
\label{quad}
\end{eqnarray}
and
\begin{eqnarray}
g_2(r)=1 - \frac{2\Gamma(1+d/2)\cos^{2}\,(rK -\pi(d+1)/4)}{K\, \pi^{d/2+1}\, r^{d+1}} \qquad (r \rightarrow \infty).
\end{eqnarray}
We see that $g_2(r)$ tends to zero quadratically in $r$ in the limit $r \rightarrow 0$,
independent of the dimension.
Moreover, $g_2(r)$ tends to unity  for large pair distances
with a decay rate that is controlled by the power law $-1/r^{d+1}$
for any $d \ge 1$, and hence the corresponding structure factor $S(k)$ at $\rho=1$ tends to zero linearly in $k$ in
the limit $k\rightarrow 0$ such that  
\begin{eqnarray}
S(k)=\frac{c(d)}{2K}k +\; {\cal O}(k^3)  \qquad (k \rightarrow 0),
\label{linear}
\end{eqnarray}
where $c(d)$ is a $d$-dependent positive constant given by (\ref{C}). Hence,  Fermi-sphere point processes
belongs to class II hyperuniform systems (see Sec. \ref{classes}). Generalizations
of this determinantal process, called {\it Fermi-shells} point processes, also belong to class II \cite{To08b}.

\begin{figure}[bthp]
\centerline{\includegraphics[  width=3in,  keepaspectratio,clip=]{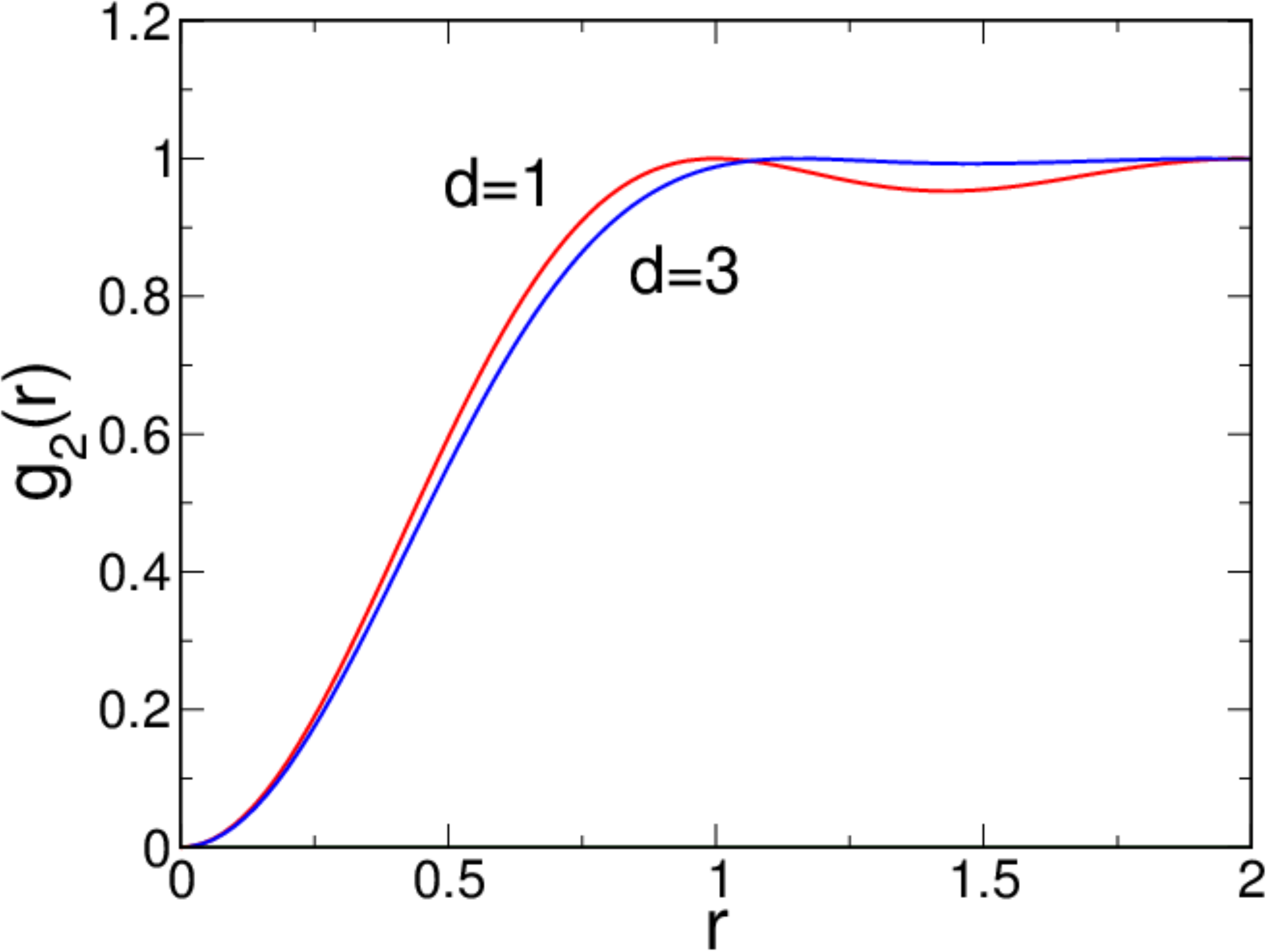}\hspace{0.3in}\includegraphics[  width=3in,  keepaspectratio,clip=]{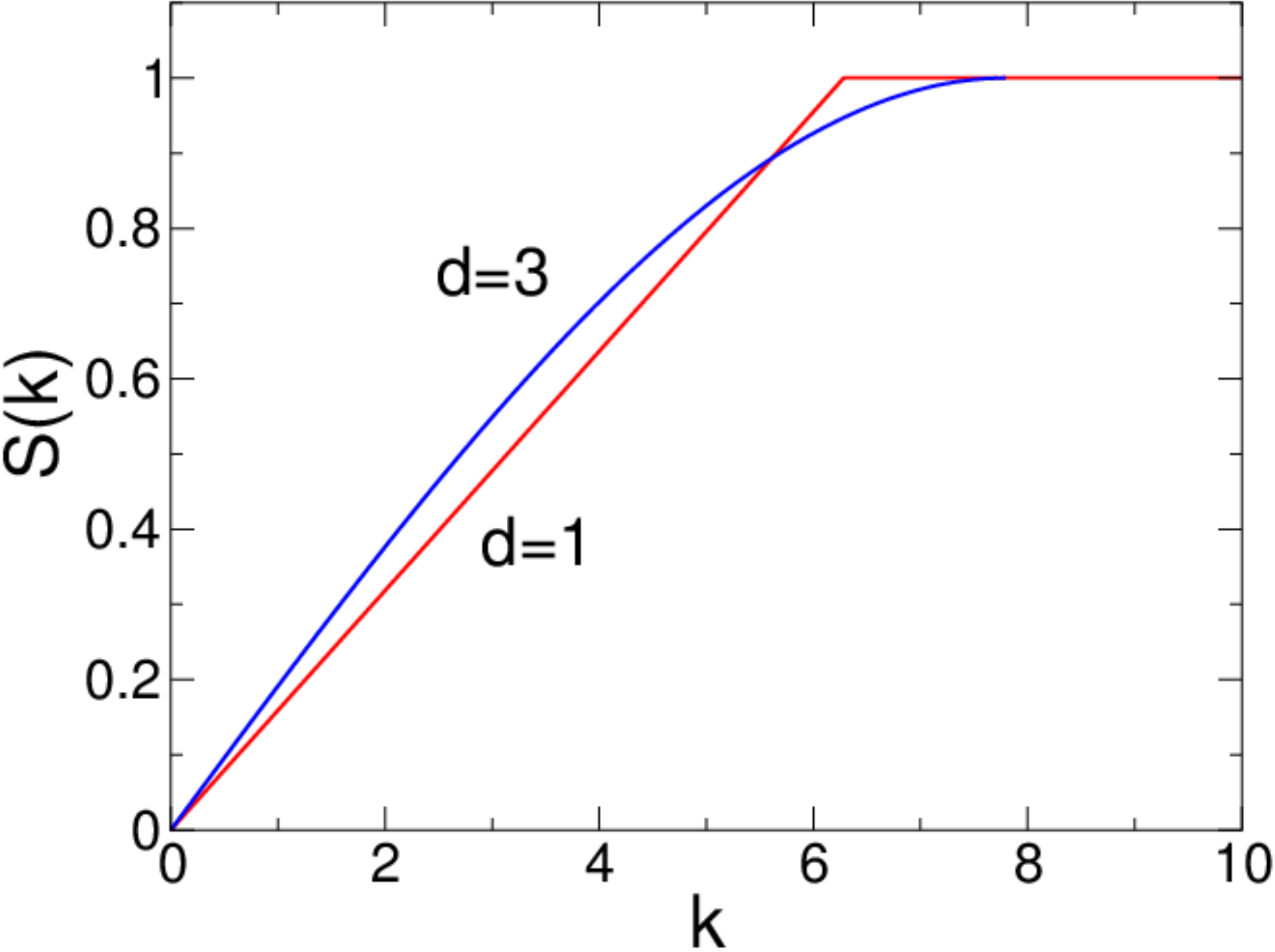}}
\caption{Results for fermionic-sphere point processes. Left panel: Comparison of the pair correlation functions for $d=1$ and $d=3$, as obtained from (\ref{g2-d-2}).
Right panel: The corresponding structure factors, as obtained from (\ref{S-d}), tend to zero as $k \rightarrow 0$ linearly
in $k$, and hence exhibit exact hyperuniformity within class II. These figures are adapted from those given in Ref. \cite{To08b}.}
\label{fermions}
\end{figure}

It is notable  that Fermi-sphere point processes in the high-$d$
asymptotic limit can be characterized by an effective ``hard-core" diameter
that grows like the square root of $d$. \cite{To08b}. Moreover, in the
high-$d$ limit, the point process behaves effectively like  a sphere packing
with a coverage fraction of space that is no denser than $1/2^d$.
This coverage fraction has a special significance in the study of sphere
packings in high-dimensional Euclidean spaces, including
Minkowski's lower bound on the density of the densest lattice sphere packings \cite{Mi05}.
as well as exactly solvable disordered sphere-packing models  \cite{To06a,To06b}.

Using (\ref{g2-d-2}) into (\ref{N2}), it has been shown \cite{To08b} that the number variance
$\sigma^2_{_N}(R)$ for a spherical window of radius $R$ of Fermi-sphere point processes
in any dimension $d$ for large $R$ grows like 
\begin{eqnarray}
 \sigma^2_{_N}(R)= \frac{d\pi^{(d-4)/2}}{2\Gamma[(d+1)/2]\Gamma(1+d/2)^{1/d}} \ln(R)R^{d-1}
+ {\cal O}(R^{d-1}) \qquad (R \rightarrow \infty).
\label{fermion-var}
\end{eqnarray}
This class-II hyperuniform growth law in three dimensions also arises in  maximally random jammed sphere packings \cite{Do05d}, which can be viewed
as  prototypical glasses because they are simultaneously perfectly mechanically rigid,
maximally disordered and perfectly nonergodic. Since the coefficient multiplying
$\ln(R)$ in  (\ref{fermion-var}) decays to zero exponentially fast as $d \rightarrow \infty$,
the surface-area term $R^{d-1}$ increasingly
becomes the dominant one in the large-$d$ limit.

Figure \ref{points} depicts realizations of the Fermi-sphere point processes
in one and two dimensions, which were generated using an algorithm
devised by Scardicchio, Zachary and Torquato \cite{Sc08} that was based on a formal description 
given by Hough et al. \cite{Ho06}. 
The reader is referred to Ref.  \cite{Sc08} for  
details and applications of this algorithm.

\begin{figure}[bthp]
\centerline{\psfig{file=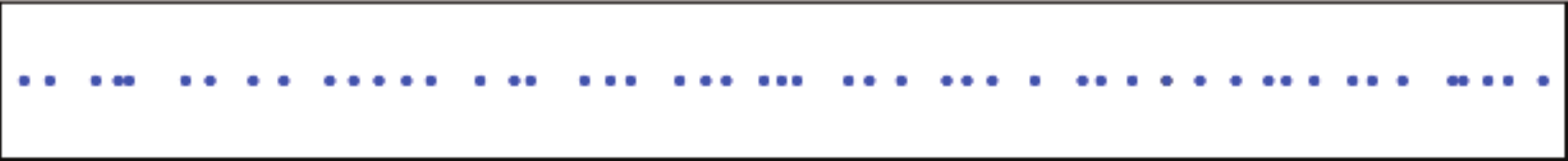,width=4.2in,clip=}}\vspace{0.25in}
\centerline{\psfig{file=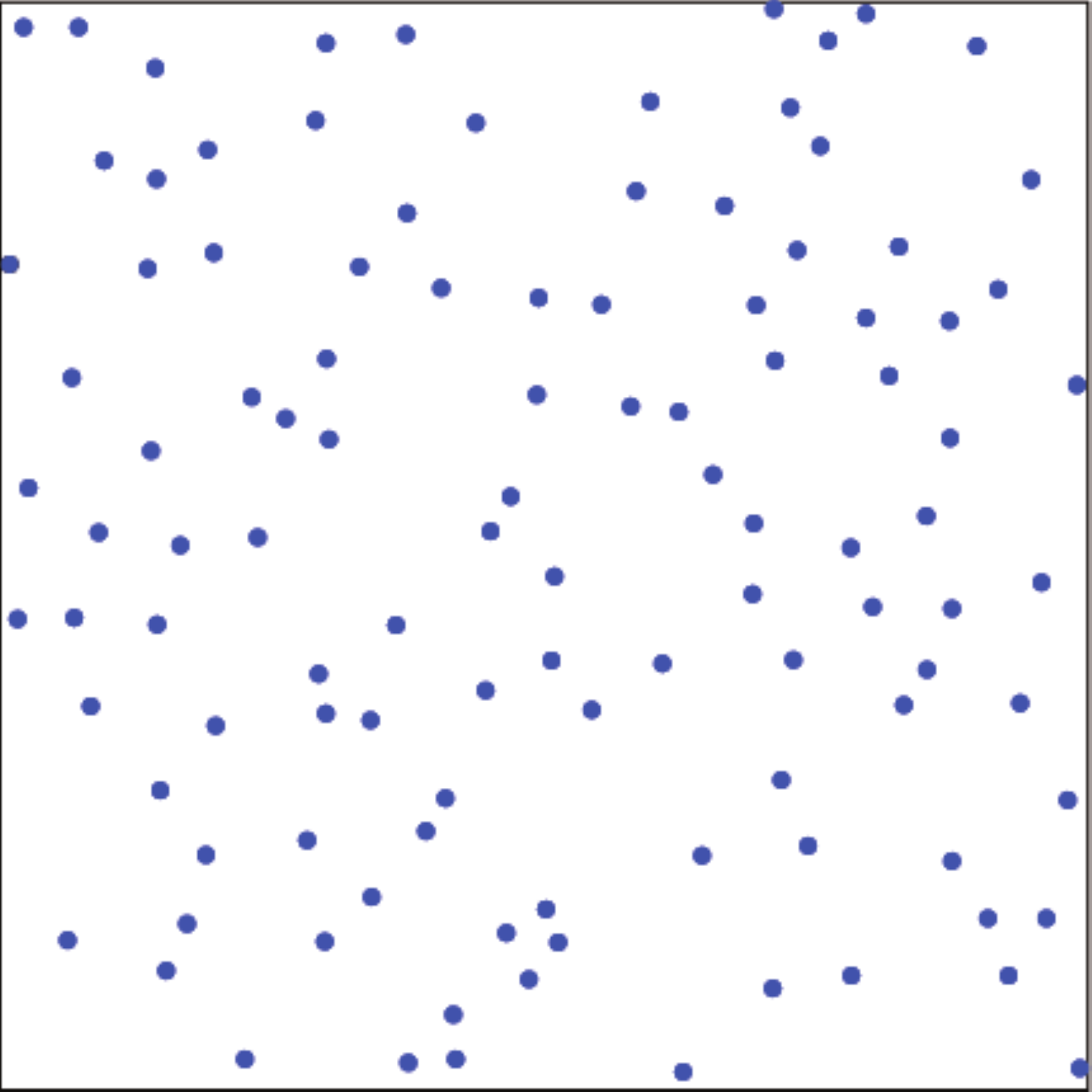,width=2.8in,clip=}}
\caption{Top panel: A realization of 50 points of a Fermi-sphere point
process in a linear ``box" subject to periodic boundary conditions generated from the
algorithm described in Ref. \cite{Sc08}. Bottom panel:  A realization of 100 points of a Fermi-sphere point
process in a square box subject to periodic boundary conditions generated from the
algorithm described in Ref. \cite{Sc08}.}
\label{points}
\end{figure}


It is noteworthy that the nearest-neighbor distribution functions
for Fermi-sphere point processes were also evaluated
and rigorously bounded \cite{To08b}. Nearest-neighbor probability density functions give
the probability of finding a nearest-neighbor point within some differential annulus
a radial distance $r$ around some reference position. The associated complementary cumulative functions give the probability
of finding spheres of radius $r$ centered at some reference position empty of any 
points, which are sometimes referred to as ``hole" probabilities. Among other results, it was shown
that the probability of finding a large spherical cavity of radius $r$ in a $d$-dimensional
Fermi-sphere point process behaves like a Poisson point process but in dimension
$d+1$, i.e., this probability is given by $\exp[-\kappa(d) r^{d+1}]$ for large $r$ and finite $d$,
where $\kappa(d)$ is a positive $d$-dependent constant. 

Finally, we note that superfluid helium at zero temperature
possesses a structure factor $S({\bf k})$ equal to $\hslash\,|{\bf k}|/(m c)$ in the zero-wavenumber limit \cite{Fe56,Re67},
where $m$ is the mass of a helium molecule and $c$ the speed of propagation of the phonons.  Hence, this
strongly interacting system of bosons in its ground state is hyperuniform 
of class II, as are spin-polarized free fermions as well as
 maximally random jammed packings of disks and spheres \cite{Do05d,Za11a,Za11c,Kl16,At16a} and of nonspherical
particles \cite{Za11c,Ji11c,Ch14a} via either $S({\bf k})$ or the spectral density ${\tilde \chi}_{_V}({\bf k})$.

\subsection{Two-Dimensional Ginibre Ensemble: One-Component Plasma}
\label{ocp}

An example of a two-dimensional determinantal point process that exhibits hyperuniform behavior is generated by 
the Ginibre ensemble (corresponding to the complex eigenvalues of random complex matrices with independent Gaussian
entries) \cite{Gi65}, which is a special case of the so-called two-dimensional
one-component plasma  \cite{Ba78,Na80,Ch80,Ba80,Ja81,Le00,Se14,Sa15,Se15,Leb16,Pe18}. A one-component plasma (OCP) is an equilibrium system of 
identical point particles of charge $e$ interacting via the log Coulomb potential and immersed in a rigid, uniform background of opposite charge to ensure overall charge neutrality,
i.e., the total system potential energy is given by
\begin{equation}
\Phi_N({\bf r}^N)=N \sum_{i=1}^N V({\bf r}_i) - \sum_{i < j}^N \ln(|{\bf r}_i -{\bf r}_j|)\,,
\label{phi-OCP}
\end{equation}
where $V({\bf r})$ is the background potential and we set $e=1$.
For $\beta=2$, the total correlation function for the OCP (Ginibre ensemble) in the thermodynamic limit
has been ascertained exactly  by Jancovici \cite{Ja81}:
\begin{eqnarray}
h(r) = -\exp\left(-\rho\pi r^2\right).
\label{h-OCP}
\end{eqnarray}
This faster-than-exponential decay of $h(r)$ demonstrates that 
{\it quasi-long-ranged correlations} (power-law decays faster than $1/r^{d}$) are not required for a disordered system to be hyperuniform.
Figure \ref{ginibrefig} shows a finite configuration (under periodic boundary) for this case
generated in Ref. \cite{Za09} using the algorithm  presented in Ref. \cite{Ho06} and elaborated in Ref. \cite{Sc09}.  
The corresponding structure factor is given by 
\begin{equation}
S(k)=1-\exp\left(-\frac{k^2}{4\pi \rho}\right)
\label{S-OCP}
\end{equation}
(see Fig. \ref{ginibrefig}) and hence is analytic everywhere, in contrast to the fermionic-sphere structure factor,
which is nonanalytic at $k=0$ and $k=K$ in any $d$ (see Sec. \ref{fermi}).
Therefore, this system  is hyperuniform with  the following smooth small-$k$ behavior:
\begin{equation}
S(k) \sim k^2 \qquad (k\rightarrow 0).
\label{S-ocp}
\end{equation}

According to Sec. \ref{oz}, relation (\ref{S-ocp}) implies that the Fourier transform of the direct correlation
function is singular at the origin such that
\begin{equation}
{\tilde c}(k) \sim -\frac{1}{k^2} \qquad (k\rightarrow 0),
\label{C-ocp}
\end{equation}
the nonintegrability of which is a reflection of the long-ranged nature of the Coulomb 
interaction in (\ref{phi-OCP}), which is consistent with asymptotic logarithmic 
large-distance behavior of the direct correlation function $c(r)$ indicated in relation (\ref{C-V}), i.e.,
\begin{equation}
c(r) =-\beta v(r) \sim \ln(r) \qquad (r \rightarrow \infty).
\end{equation} 
 The scaling laws (\ref{S-ocp}) and (\ref{C-ocp}) apply as well to
the OCP in three and higher dimensions, and thus  lead
to the expected large-$r$ Coulombic scaling laws for the direct correlation function
and effective pair interactions for $d \ge 3$:
\begin{equation}
c(r) = -\beta v(r) \sim  - \frac{1}{r^{d-2}} \qquad (r \rightarrow \infty),
\end{equation}
Thus, according to Sec. \ref{classes}, OCP systems for $d \ge 2$,  fall in class I hyperuniform systems.
Notice that since the step-function $g_2$-invariant
processes are characterized by the same scaling laws for $S(k)$ and ${\tilde c}(k)$,
the corresponding asymptotic scaling for the direct correlation function $c(r)$ 
is also purely Coulombic, as indicated in relation (\ref{c-green1}).

\begin{figure}[bthp]
\centering
\includegraphics[width=0.31\textwidth,clip=]{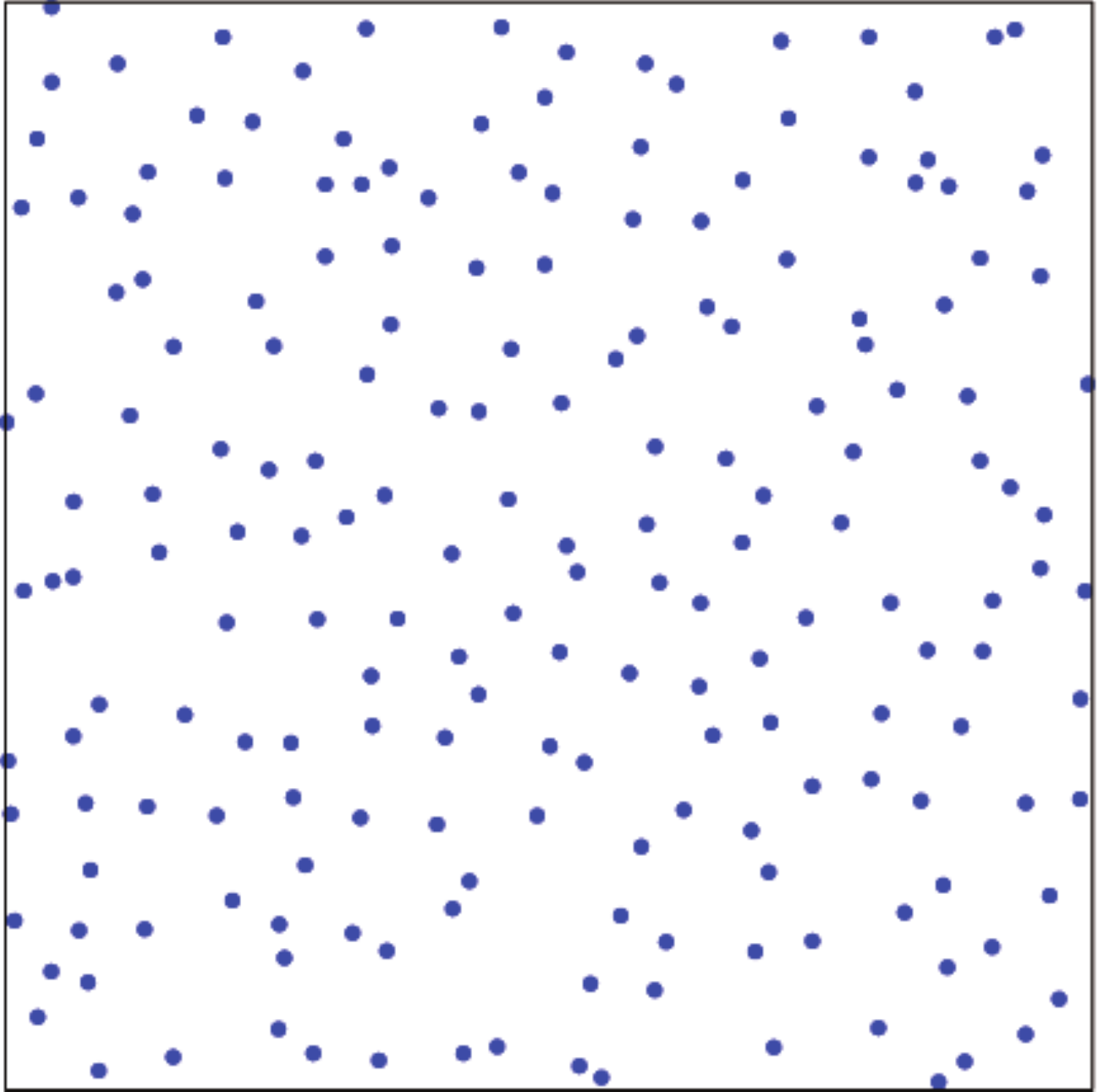} \hspace{0.2in}\includegraphics[width=0.5\textwidth,clip=]{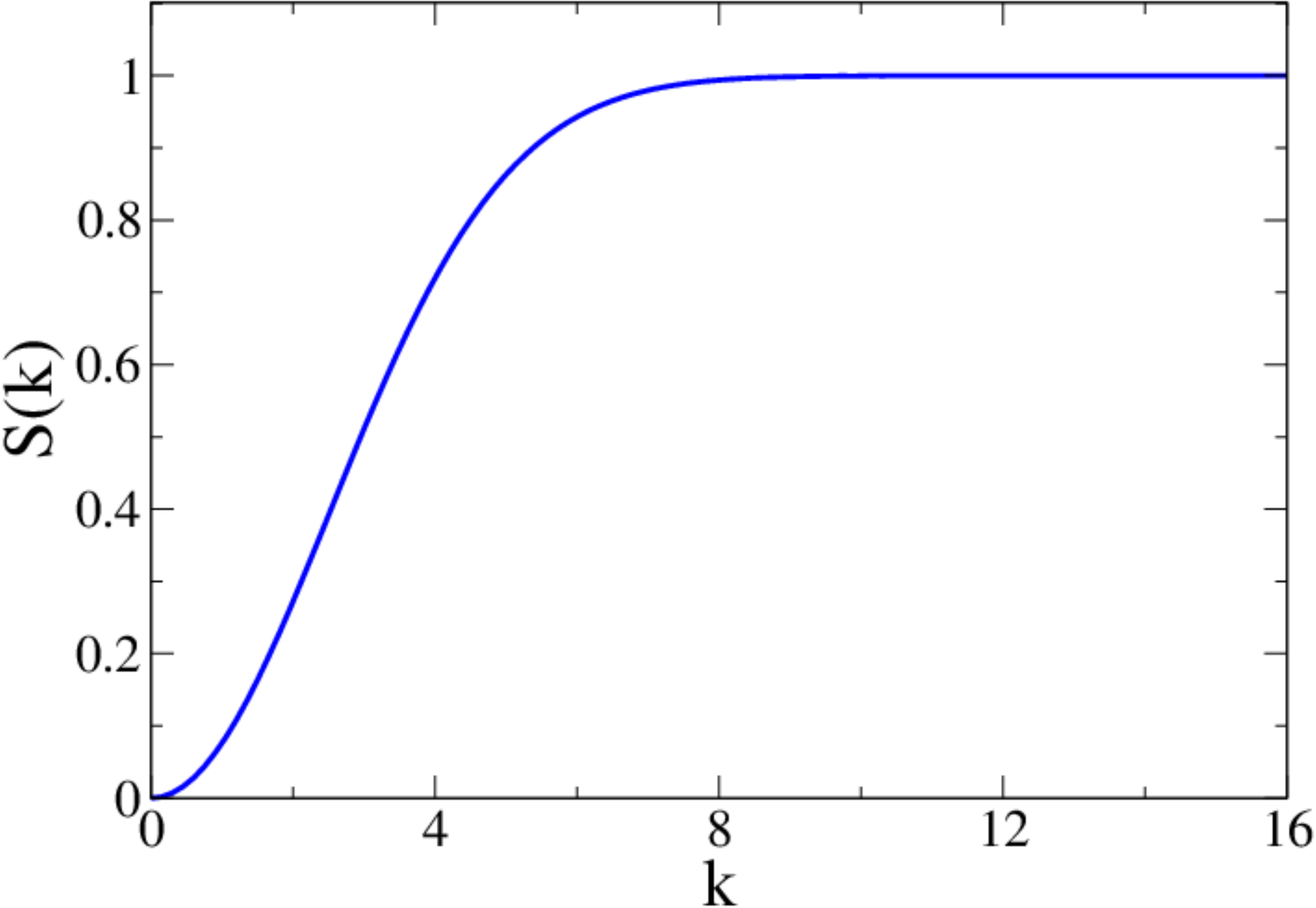}
\caption{Left panel: A realization of a point process generated from the Ginibre ensemble \cite{Gi65}, which corresponds
to the two-dimensional one-component plasma with $\beta=2$
\cite{Za09}. Right panel: The corresponding structure factor in the infinite-size limit at unit number density. Notice that, unlike
the fermionic-sphere point processes (see Fig. \ref{fermions}), the structure factor
is analytic everywhere and, in particular, goes to zero quadratically in $k$ as $k \rightarrow 0$.}\label{ginibrefig}
\end{figure}

It bears re-emphasizing that the OCP  provides a critical lesson, namely, point patterns do not need to
possess  quasi-long-ranged pair  correlations 
in order to be hyperuniform, as evidenced by the Gaussian form of the total correlation
function (\ref{h-OCP}).
Note that OCP fluid phases at other temperatures must  always be hyperuniform because, due
to overall charge neutrality with a rigid background, the reduced charge fluctuations are in correspondence
with number fluctuations, both of which grow like the window surface area.
Interestingly, the probability of finding a large spherical cavity of radius $r$ in the 2D OCP
behaves like a Poisson point process but in dimension four, i.e., this ``hole" probability is given by $\exp(-c r^{4})$ for large $r$
\cite{Hough09}.

Combination of either (\ref{N3}) and (\ref{h-OCP}) or (\ref{N4}) and (\ref{S-OCP}) 
exactly yields the local number variance for the 2D OCP with $\beta=2$ to be 
\begin{equation}
\sigma_{_N}^2(R) = \pi R^2 \exp\left(-2\pi R^2\right)[ I_0(2\pi R^2)+I_1(2\pi R^2)],
\label{OCP-2}
\end{equation}
where $I_{\nu}(x)$ is a modified Bessel function of order $\nu$. The corresponding
large-$R$ asymptotic behavior of the number variance is given by
\begin{equation}
\sigma_{_N}^2(R) = R -\frac{1}{16\pi R} +{\cal O}\left(\frac{1}{R^3}\right) \qquad (R \rightarrow \infty).
\label{OCP-asy}
\end{equation}
The local number variance rapidly saturates to its asymptotic value, as shown  in Fig. \ref{var-OCP}.

\begin{figure}[bthp]
\centering
\includegraphics[width=0.5\textwidth,clip=]{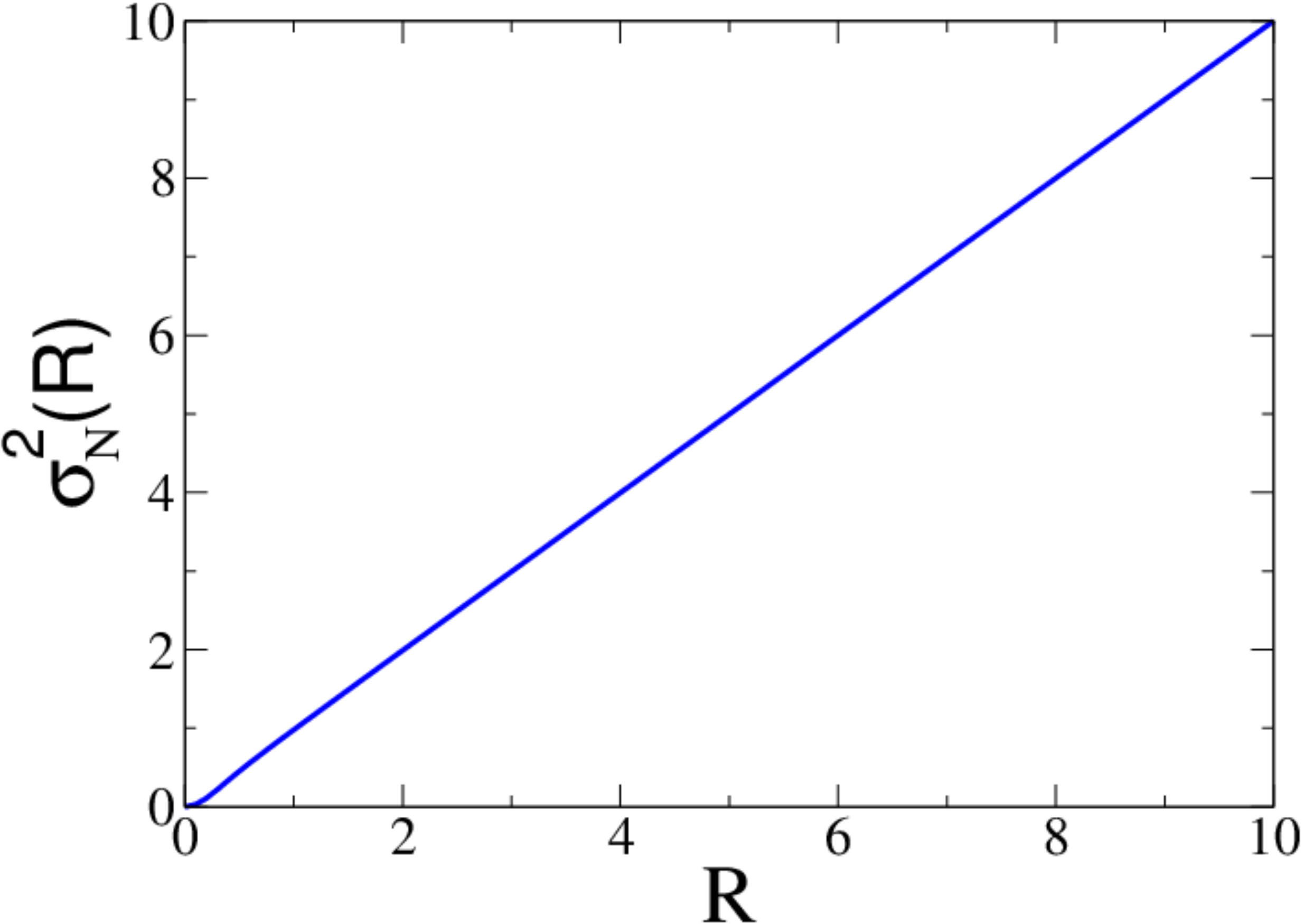} 
\caption{The local number variance $\sigma^2(R)$ versus $R$ for the 2D OCP, as given by
(\ref{OCP-2}). It is noteworthy that the large-$R$ asymptotic limit of $R$ [see relation (\ref{OCP-asy}]
is achieved at very small values of $R$ (say $R> 1$). }\label{var-OCP}
\end{figure}

We remark that Laughlin's celebrated
ansatz for the ground state wave function associated with the  fractional quantum Hall effect 
is tantamount to a mapping to a classical 2D OCP in which the charges depend
on the filling fraction \cite{Laugh87}.
Notably, Sandier and Serfaty \cite{Sa15}
have rigorously established a link between the Ginzburg-Landau model  for vortices
in superconductors  in the critical regime  and the 2D OCP. Here the interaction between the 
vortices is well-described by a 2D OCP in which the background
potential $V({\bf r})=|{\bf r}|^2$.

Finally, we comment on superionic phases, including superionic ice \cite{Ca99}, which is a special 
group of ice phases at high temperatures and pressures that may
exist in ice-rich planets and exoplanets. In superionic ice, ``liquid" hydrogen ions can coexist with a
relatively rigid crystalline oxygen sublattice of opposite charge. Thus, it is
 expected that OCPs can be well-approximated by superionic ice under appropriate conditions.
Recent evidence has emerged that this is indeed the case by virtue of the fact
that the liquid hydrogen phase is nearly hyperuniform as measured by the structure factor \cite{Su15}.
Other solid-state superionic conductors at ambient temperatures and pressure, such as silver iodide (AgI) \cite{Bo79,Mak09} used in batteries
and sensors, should exhibit near hyperuniformity for the same reasons.

\subsection{Weyl-Heisenberg Ensemble}
\label{WH}

The infinite Weyl-Heisenberg ensemble is a class of determinantal point processes
associated with the Schr\"{o}dinger representation of the Heisenberg
group. It has recently been proved that such determinantal point processes
are perfectly  hyperuniform  \cite{Ab17}.  This provides another class of examples of d-dimensional determinantal
point processes that are hyperuniform beyond the aforementioned Fermi-type varieties \cite{To08b,Sc09}. It was also
proved there that the number variance $\sigma^2_{N}(R)$ associated with a spherical observation window of
radius $R$ grows like the surface area of the window, i.e., as $R^{d-1}$, and hence are hyperuniform
of class I. Explicit formulas
for the corresponding total correlation functions $h({\bf r})$ were obtained; they exhibit a large-$|{\bf r}|$ decay
that is faster than exponential; specifically, like a Gaussian \cite{Ab17}. In the radial case,
the structure factor $S({\bf k})$ near the origin tends to zero quadratically in $|\bf k|$
in all space dimensions, and thus exhibits the same behavior as the Ginibre ensemble in two dimensions.
We note that the Weyl-Heisenberg ensemble includes as a special case a multi-layer extension
of the Ginibre ensemble modeling the distribution of electrons in higher
Landau levels, which has recently been the object of study in the realm of the
Ginibre-type ensembles associated with polyanalytic functions \cite{Hai13}. Moreover, this family
of Weyl-Heisenberg ensembles includes new structurally anisotropic processes, and thus 
provides a rigorous means to explore ``directional" hyperuniformity, which is a recent generalization
of the hyperuniformity concept that is reviewed in Sec. \ref{anisotropy}.

\subsection{Multicomponent Hard-Sphere Plasmas}

It has recently been recognized that multicomponent equilibrium plasmas made of {\it nonadditive} hard spheres with Coulombic interactions
enable one to generate a very wide class of disordered hyperuniform as well as multihyperuniform
systems  at positive temperatures \cite{Lo17,Lo18a}. It is  the infinite parameter space (particle
size distribution, composition and non-additivity parameter) and long-ranged interactions
afforded by them that provides greater tunablilty to achieve hyperuniform states. Specific
theoretical and computational results have been obtained for two-component non-additive hard-disk plasmas in two dimensions \cite{Lo17,Lo18a}.
In the case of multihyperuniform two-component hard-disk plasmas, it was shown
that multihyperuniformity competes with phase separation and stabilizes a clustered phase \cite{Lo18a}.

\section{Classical Nonequilibrium Systems}
\label{non-eq}

In this section, we report on a variety of disordered nonequilibrium systems 
that are putatively hyperuniform according to numerical/experimental protocols and analyses.
Specifically, we describe the hyperuniformity characteristics
of ordered and disordered jammed particle systems \cite{To03a,Do04d,Za11a,Za11c,Be11,Ku11,Ho12b,Dr15,Kl16,At16a,At16b,Ri17}
and absorbing-state models \cite{He15,Tj15,We15,He17a,He17b,We17,Kw17}.

\subsection{Ordered and Disordered Jammed Particle Packings}

Understanding the characteristics of jammed particle packings provides basic insights into the
structure and bulk properties of crystals, glasses, and granular media and selected aspects of
biological systems \cite{Pa10,To10c}. A {\it packing} is a large collection of
hard (nonoverlapping) particles  in either a finite-sized container or
in $d$-dimensional Euclidean space $\mathbb{R}^d$ \cite{Co98,To10c}. Recall that the packing fraction $\phi$ is the fraction
of space covered by the hard particles.  ``Jammed" packings are those particle
configurations such that each particle is in contact with
its nearest neighbors so that mechanical stability
of a specific type is conferred to the packing, as defined more precisely below.
Jammed packings  have received considerable attention in the theoretical and experimental literature 
\cite{To00b,To01b,Oh02,Oh03,Do04b,Ma05,Wy05,Ma08,Mail09,Ma09,Si09,Pa10,To10c,Ku11,Za11a,Be11,Ja13,Dr15}.

Preliminary results in 2003 indicated that disordered jammed packings
of spheres exhibited hyperuniform behavior \cite{To03a}, which at that time was regarded
to be a highly exotic large-scale property of a hard-sphere system. This result suggested
a link between a certain class of jammed states and hyperuniformity. The most natural
theoretical framework to understand this connection is the  ``geometric-structure" approach,
which emphasizes the quantitative characterization of
single-packing configurations without regard to their occurrence
frequency in the algorithmic method used to
produce them; see Ref. \cite{To10c} and references therein.
While a comprehensive review of this literature is beyond the scope
of the present article, we briefly review those aspects
of the geometric-structure approach that will aid
in understanding the conditions under which jamming and hyperuniformity
may or may not be linked.

\subsubsection{Jamming Categories}

Much of the ensuing discussion focuses on  packings of frictionless
identical hard spheres in the absence of gravity. Three broad and mathematically precise hierarchical
``jamming" categories of sphere packings can be distinguished depending on the nature of their
mechanical stability \cite{To01b,To03c}.  In order of increasing stringency (stability), for
a finite sphere packing, these are the following:
\begin{enumerate}
\item {\it Local~jamming}: Each particle in the packing is locally trapped by
its neighbors (at least $d+1$ contacting particles,
not all in the same hemisphere), i.e., it cannot be translated while fixing the positions
of all other particles;
\item {\it Collective~jamming:} Any locally jammed configuration in which no
subset of particles can simultaneously be displaced so that its members
move out of contact with one another and with the remainder set fixed;; and
\item {\it Strict~jamming:} Any collectively jammed configuration that disallows
all  uniform volume-decreasing strains of the system
boundary is strictly jammed, implying that their
bulk and shear moduli are infinitely large \cite{To03c}.
\end{enumerate}
Importantly, the jamming category of a given sphere
packing depends on the boundary conditions employed \cite{To01b,To03c,Do04a,Ch15}. For example, hard-wall boundary
conditions \cite{To01b} generally yield different
jamming classifications from periodic boundary conditions \cite{Do04a}.
These jamming categories, which are closely related to the concepts of ``rigid" and ``stable"
packings found in the mathematics literature \cite{Co98},
requires that the packing possesses no ``rattlers", 
which are particles that are free to move about their respective confining cages \cite{Do05c,To10c}.

\begin{figure}[bthp]
\centerline{\includegraphics[height=1.5in,keepaspectratio,clip=]{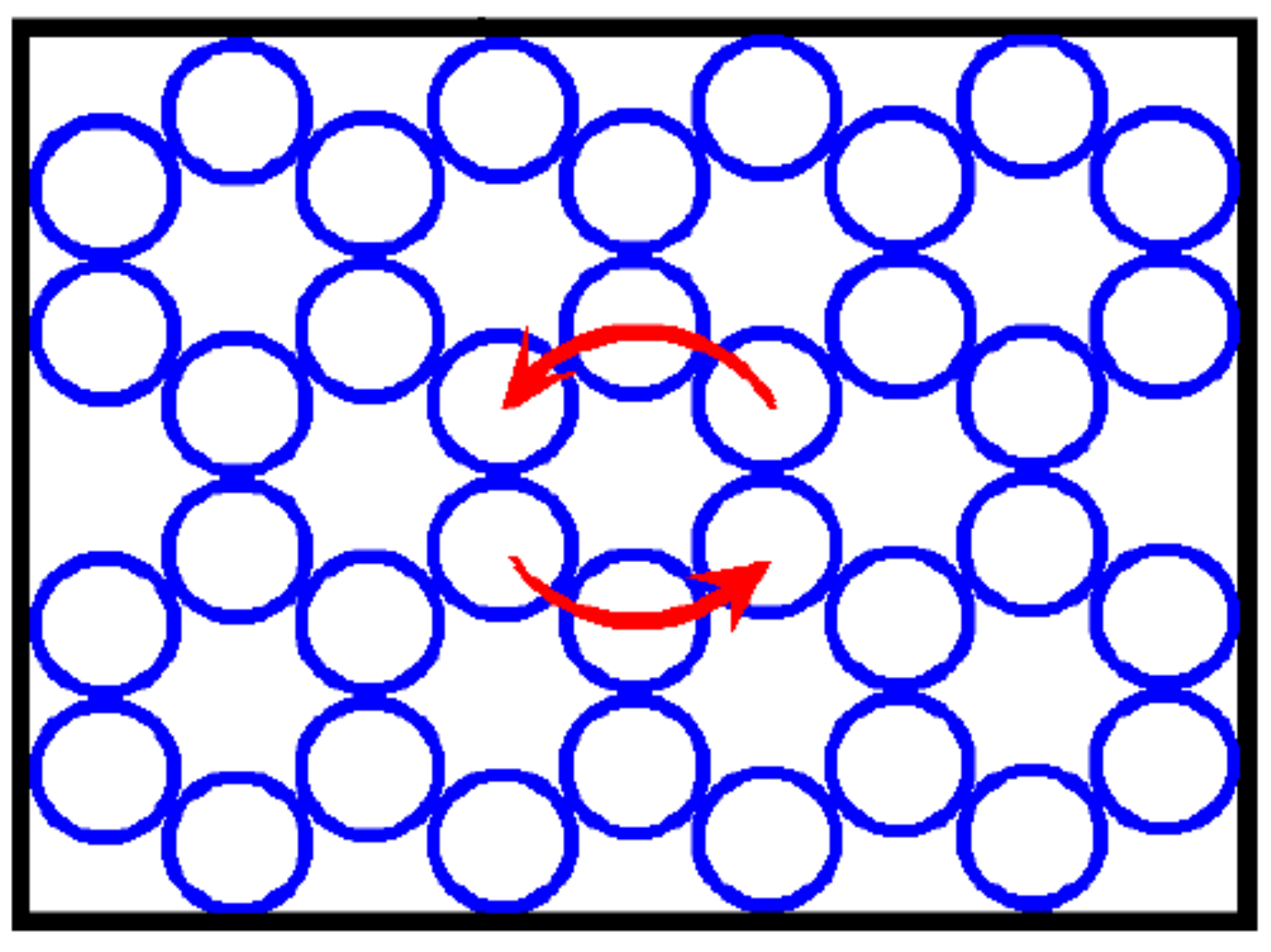}
\hspace{0.3in} \includegraphics[width=1.7in,keepaspectratio,clip=]{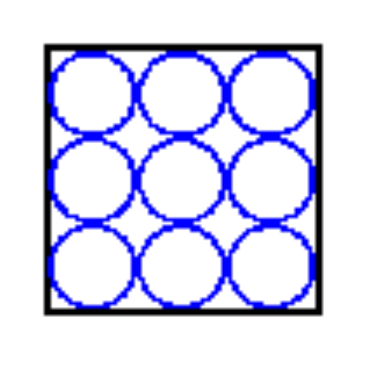}
\hspace{0.3in} \includegraphics[width=1.7in,keepaspectratio,clip=]{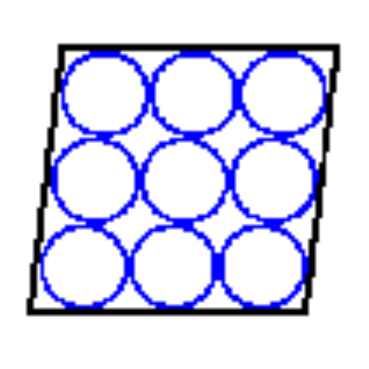}}

\caption{\footnotesize Illustrations of jamming categories taken from Ref. \cite{To10c}.
Leftmost panel: The honeycomb-crystal packing within a rectangular hard-wall container
is locally jammed, but is not collectively jammed (e.g., a collective rotation
of a hexagonal particle cluster, as shown, will unjam the packing). Middle panel:
Square-lattice packing within a square hard-wall container is collectively jammed.
Rightmost panel:
The square-lattice packing shown in the middle panel can be sheared
and hence is not strictly jammed. Thus, we see that
the square-lattice packing with square hard-wall boundary conditions
can only be collectively jammed even in the infinite-volume limit.
}
\label{examples}
\end{figure}

Figure \ref{examples} shows examples of ordered locally and
collectively jammed packings of circular hard disks in two dimensions
within hard-wall containers. The honeycomb-crystal packing ($\phi=\pi/3^{3/2}=0.605\ldots$
in the infinite-size limit)
is only locally jammed because there are collective particle
motions that will lead to its collapse, suggesting that there exist
particle rearrangements  and motions of the container boundary 
that can lead to collectively or strictly packings with a higher density. Such a densification
process starting with the original {\it unsaturated} honeycomb packing
could be accomplished, for example, using an ``adaptive-shrinking-cell" optimization
scheme \cite{To09b,To09c}. A {\it saturated}  packing of hard spheres is one in which there is no space available to add another sphere. 
Figure \ref{examples} also shows that while the square-lattice packing 
within a hard-wall container ($\phi=\pi/4=0.785\ldots$
in the infinite-size limit) is collectively jammed, implying that
it can be sheared to yield a denser packing that is strictly jammed,
e.g., the triangular lattice packing (with $\phi=\pi/\sqrt{12}=0.907\ldots$ in the infinite-size
limit). A portion of this infinite strictly jammed packing is shown in 
Fig. \ref{tri}, which is not only hyperuniform but stealthy for the
reasons noted in Sec. \ref{ensemble}.

\subsubsection{Geometric-Structure Approach and Order Maps}
\label{jamming}

It is also valuable to quantify the degree of ordering in a packing, 
especially any that is jammed. To 
this end, a variety of scalar order metrics $\psi$ have been 
employed and applied that depend on the configurational coordinates of a
packing ~\cite{To00b,To10c} with the normalization 
$0 \le \psi \le 1$, where $\psi=0$ corresponds to  
the most disordered state (i.e., Poisson point process) and 
$\psi=1$ is the most ordered state (e.g., fcc packing). Using the geometric-structure 
approach, one can construct an ``order map" in the 
$\phi$--$\psi$ plane~\cite{To10c}, where 
jammed packings form a subset in this map. The boundaries of the 
jammed region delineate extremal structures, including, for example, the 
maximally dense packings (the face-centered-cubic packing and its stacking 
variants with $\phi_{\mathrm{max}}=\pi/\sqrt{18}\approx 
0.74$~\cite{Ha05}) and the least dense strictly jammed packings 
(conjectured to be the `tunneled crystals' with 
$\phi_{\mathrm{min}}=2\phi_{\mathrm{max}}/3 \approx 0.49$~\cite{To07}).

Among all strictly jammed sphere packings  in  $\mathbb{R}^d$, 
the one that exhibits maximal disorder (minimizes some given order metric $\psi$)
is of special interest. This ideal state is called the maximally random jammed (MRJ) state \cite{To00b}; see Fig.~\ref{map}.
The  MRJ concept is geometric-structure based, since it refers to a single packing 
that is maximally disordered subject to the strict jamming constraint, regardless of its probability of occurrence 
in some packing protocol.  Thus, the MRJ state is conceptually and quantitatively
different from random close packed (RCP) packings \cite{Be60,Ber65}, which have recently been suggested to be the most probable jammed configurations within an ensemble
\cite{Oh03}. The differences between these states are even starker
in two dimensions, e.g.,   MRJ packings of identical circular disks in $\mathbb{R}^2$ have been shown to be dramatically
different from their RCP counterparts, including their respective densities,
average contact numbers, and degree of order \cite{At14}. The MRJ state under the strict-jamming constraint is a prototypical glass
\cite{To07} in that it is maximally disordered (according to a variety of order metrics) without any  long-range order (Bragg peaks)
and perfectly rigid (i.e., the elastic moduli are indeed unbounded \cite{To03c,To10c}). The jammed backbone of the MRJ state is 
{\it isostatic}, i.e., the number of exact contacts in the jamming limit is exactly equal to the
number of degrees of freedom, once rattlers are removed; in the infinite-size  limit, this
implies that the average number of contacts per sphere is $2d$ in $d$ dimensions \cite{Ed94,Do05c}.  Moreover,  MRJ packings in  $\mathbb{R}^d$ are characterized by negative ``quasi-long-range" (QLR) pair correlations 
in which the total correlation function $h({\bf r})$ decay to zero asymptotically 
like $-1/|{\bf r}|^{d+1}$ \cite{Do05d,Sk06,Ho12b,At16a,At16b}.
Equivalently, this means that the structure factor $S({\bf k})$ tends to zero linearly
in $|\bf k|$ in the limit $|{\bf k}| \rightarrow 0$, and hence place putative MRJ sphere
packings in the same hyperuniformity class as Fermi-sphere point processes (see Sec. \ref{fermi}) and
superfluid helium in its ground state \cite{Fe56,Re67}, i.e., they all belong to class II.

It is noteworthy that Minkowski correlation functions associated with the Voronoi cells
of MRJ packings exhibit even stronger anti-correlations than
those shown in the standard pair-correlation function \cite{Kla14}.

\begin{figure}[bthp]
\centerline{
\includegraphics[height=2.8in,keepaspectratio,clip=]{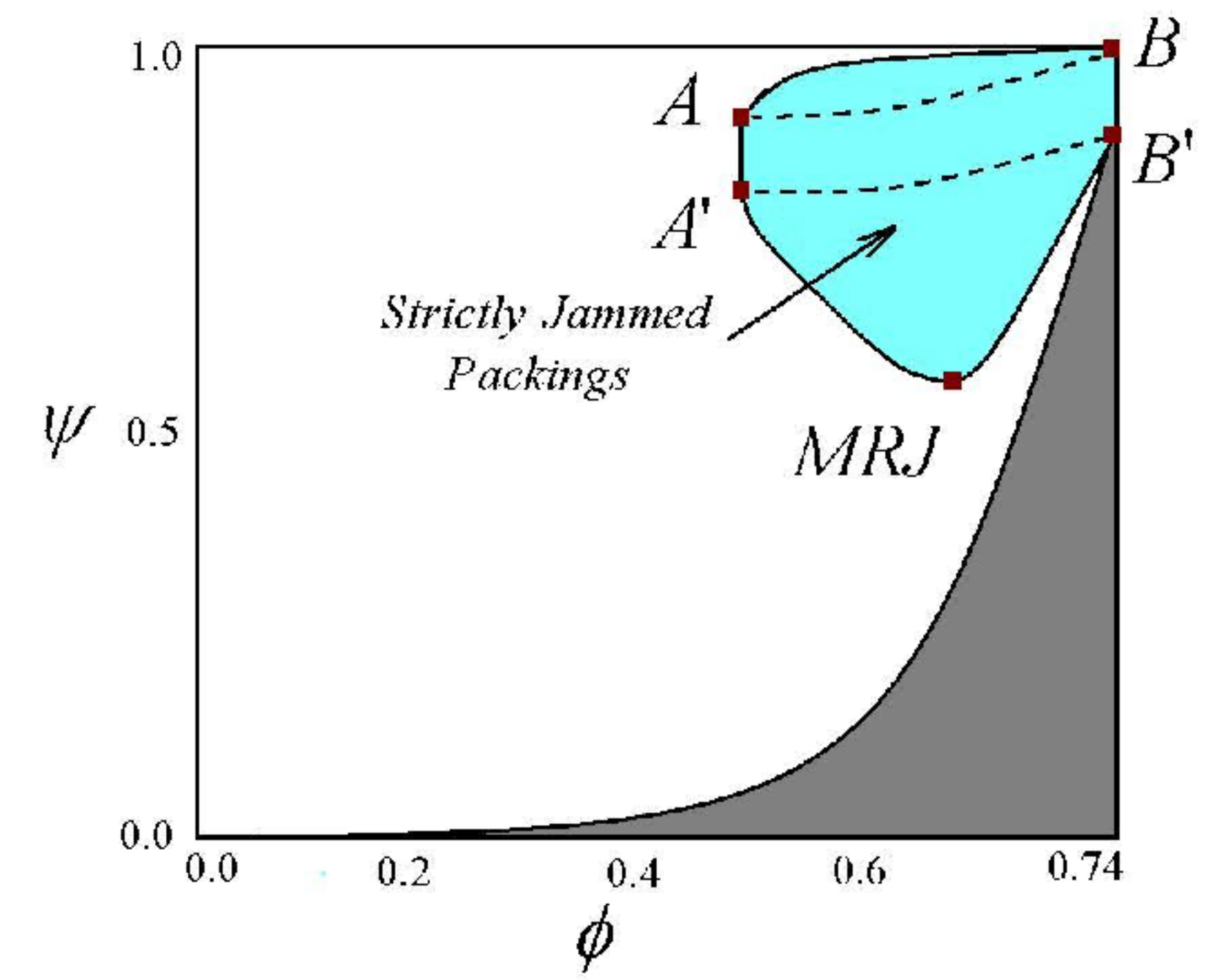}\hspace{0.25in}\includegraphics[height=2.8in,keepaspectratio,clip=]{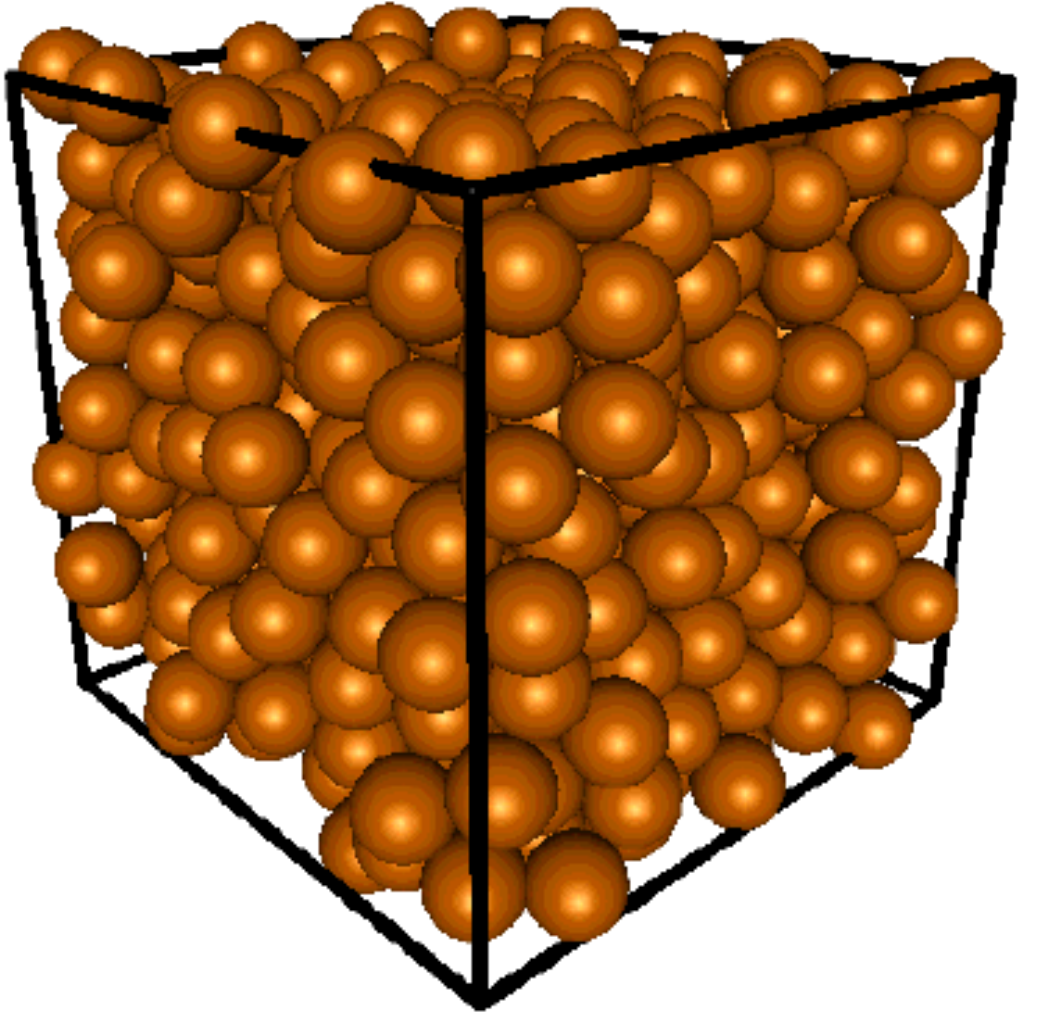}}
\caption{ Left panel: Schematic order map in the density-order
($\phi$-$\psi$) plane for identical hard spheres, including strictly jamming states in $\mathbb{R}^3$
under periodic boundary conditions. White and blue regions contain the attainable
packings, blue regions represent the jammed subspaces, and  dark shaded regions
contain no packings, as taken from Ref. \cite{To10c}. The locus of points along the boundary
of the jammed set are optimal points \cite{To07}.
The locus of points $A$-$A^{\prime}$ correspond to the lowest-density jammed
packings \cite{To07}. The locus of points $B$-$B^\prime$ correspond
to the densest jammed packings (face-centered-cubic packing and its stacking 
variants). The point MRJ represents
the maximally random jammed states, i.e., the most disordered
states subject to the jamming constraint. The
represented packings are not subject to rattler exclusion. Right panel: A three-dimensional MRJ-like configuration
of 500 spheres with $\phi \approx 0.64$ produced
using the Lubachevsky-Stillinger (LS) packing algorithm \cite{Lu90} with a fast expansion rate \cite{To00b}.}
\label{map}
\end{figure}

\subsubsection{Jamming-Hyperuniformity Conjecture}
\label{conjecture}

It is instructive now to state a conjecture due to Torquato and Stillinger
concerning the conditions under which a certain class of strictly jammed packings are hyperuniform. \\

\noindent{\bf Conjecture}: {\sl Any strictly jammed saturated infinite packing of identical spheres in $\mathbb{R}^d$ is hyperuniform \cite{To03a}}.\\

To date, there is no known counterexample to this conjecture, notwithstanding a recent study that calls into question the link between jamming and hyperuniformity \cite{Ik15}. We emphasize that the conjecture eliminates packings that may have a rigid backbone but possess ``rattlers''. Typical numerical packing protocols that have generated disordered jammed packings tend to contain a small concentration of rattlers; because of 
these movable particles, the whole (saturated) packing cannot be deemed to be ``jammed''.  Therefore, the conjecture cannot apply to these packings -- a subtle point that has not been fully appreciated \cite{Ik15,Wu15,Ik17}.

\begin{figure}[bthp]

\centerline{\includegraphics[height=2.6in,keepaspectratio,clip=]{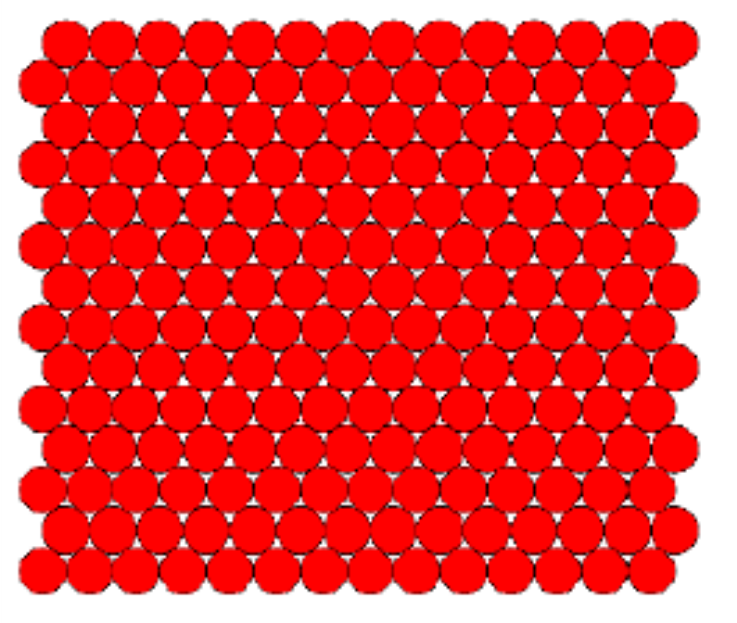} \hspace{-0.2in}
 \includegraphics[height=2.6in,keepaspectratio,clip=]{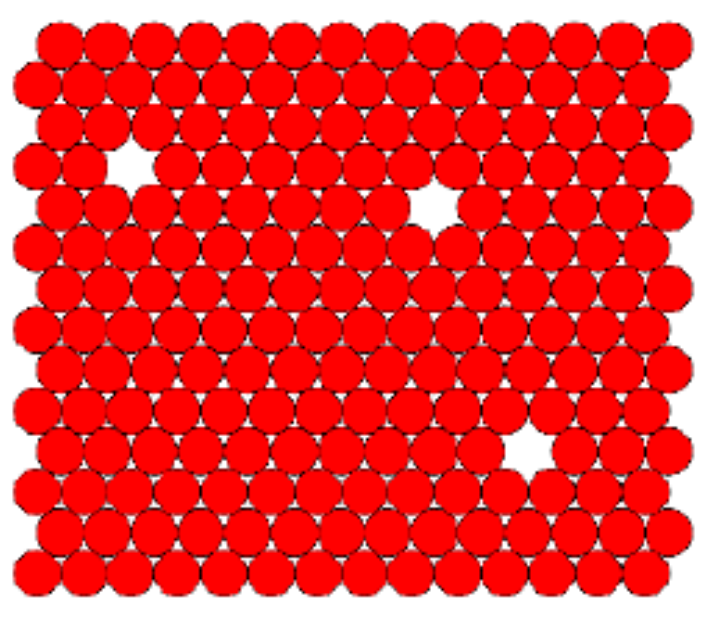}}
\caption{\footnotesize 
Left panel: A portion of the   
infinite triangular-lattice circle packing, which
is both strictly jammed and hyperuniform.
Right panel: A portion of an infinite triangular-lattice
circle packing in which a small fraction of particles
are randomly removed such that there are no di-vacancies.
This defective triangular-lattice packing is strictly
jammed \cite{St03} but not hyperuniform.  
}
\label{tri}
\end{figure}

What is the rationale behind the conjecture? First, the saturation condition is necessary in the conjecture because
hyperuniformity can be degraded by ``imperfections." For example, the random removal
of a finite fraction of circular disks from the perfect strictly jammed triangular-lattice packing 
such that there are no di-vacancies (absence of nearest-neighbor particles), as shown in Fig. \ref{tri}, will result
in a packing that is still strictly jammed \cite{St03} but will no longer
be hyperuniform -- the random defects induce diffuse scattering
that has a nonzero intensity in the zero-wavenumber limit and hence
$S(0) \neq 0$.  Similarly, it is known that randomly removing a finite fraction of particles from the
perfect strictly jammed fcc-lattice sphere packing such that  there are no ``tri-vacancies''  
to maintain strict jamming results in nonhyperuniform packings  \cite{St03}.
Moreover, it is known that collisions in equilibrium hard-sphere configurations along the stable ``crystal" branch
(see Fig. \ref{MRJ}) and on approach to jammed ordered states (such as the triangular lattice and fcc lattice in two and three dimensions, respectively) are not hyperuniform due to large-scale collective vibrational motions and only become exactly hyperuniform when the ideal jammed state without any 
imperfections or defects is attained \cite{At16b}.  One  expects to achieve exact hyperuniformity 
on the approach to disordered jammed states that are defect-free, including the complete absence of rattlers \cite{At16a}.  Thus, based on these considerations, it seems reasonable to conjecture that statistically homogeneous disordered strictly jammed saturated packings of identical spheres are hyperuniform,
since such packings cannot possess rattlers.

\begin{figure}[H]
\centerline{\includegraphics[height=3.5in,keepaspectratio,clip=]{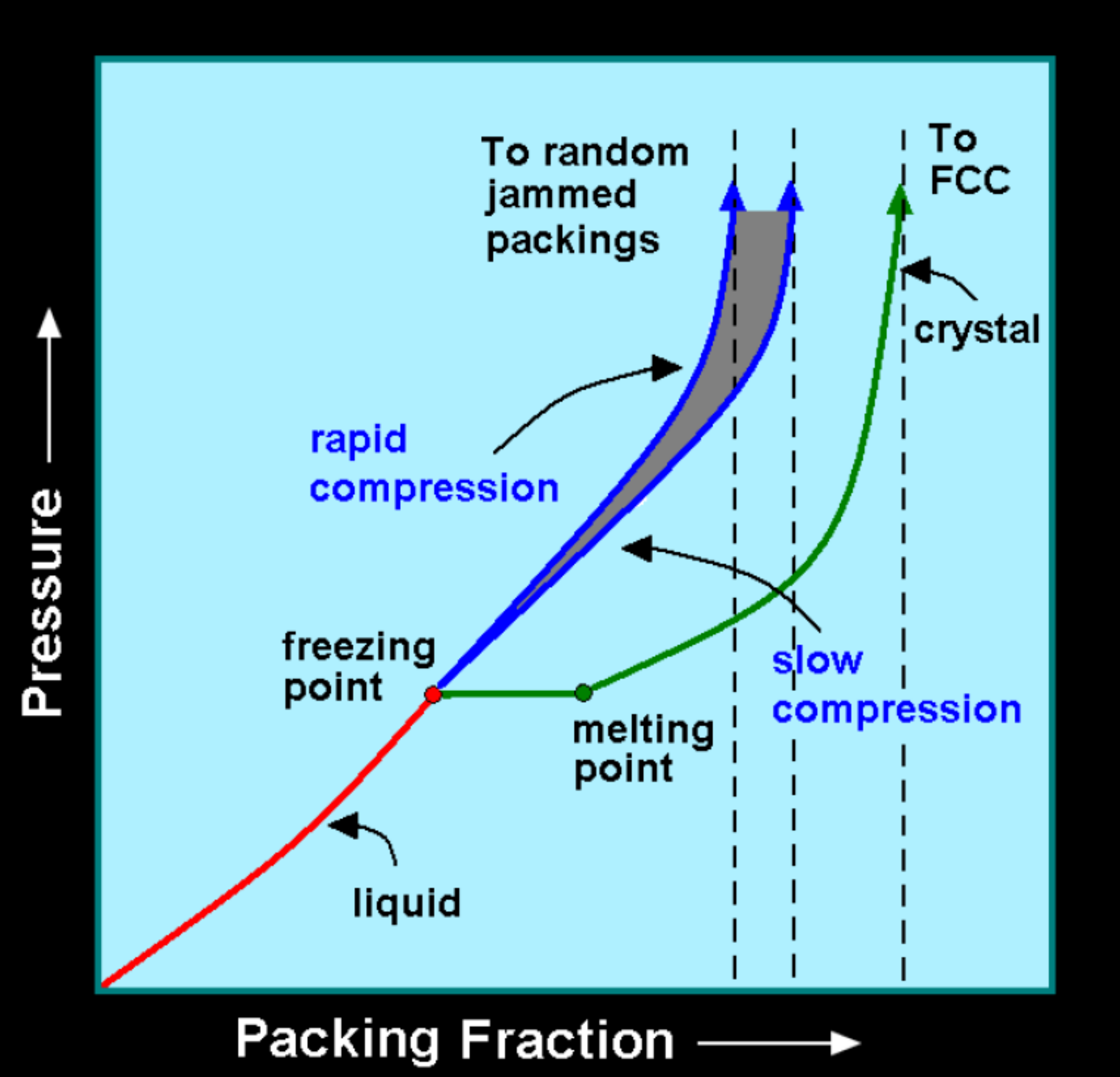} }
\caption{The isothermal phase behavior of three-dimensional
hard-sphere model in the pressure-packing fraction
plane, as taken from Ref. \cite{To10c}.  An infinitesimal compression rate
of the disordered liquid traces out the thermodynamic equilibrium path,
shown in green, including a first-order freezing transition to a crystal branch that
ends at the maximally dense fcc state ($\phi=\pi/\sqrt{18}=0.740\ldots$) , 
which is strictly jammed and hyperuniform.
Importantly, hard-sphere collisions along the stable crystal branch induce large-scale
density fluctuations that destroy hyperuniformity unless the system is exactly
at the close-packed fcc jamming point \cite{At16b}.
Rapid compressions of the liquid produce a range of amorphous metastable extensions
of the liquid branch that jam only at the their respective terminal
densities. Ideally, the MRJ state could be regarded to be the end point of 
a metastable branch with the fastest compression rate consistent with jamming. 
}
\label{MRJ}
\end{figure}

\subsubsection{Numerical Simulations of MRJ-Like States}

Donev, Stillinger and Torquato  \cite{Do05d} were the first to determine the extent to which large MRJ-like sphere packings in $\mathbb{R}^3$ were hyperuniform using a modified Lubachevsky-Stillinger molecular-dynamics packing protocol \cite{Do05a,Do05b}. 
Under such an event-driven molecular dynamics,   particles  in a periodic simulation box (that is allowed to perform volume-preserving deformations) undergo thermal motion
but also (quickly) grow in size at a certain expansion rate starting from a Poisson distribution of points until
it ideally produces a jammed state with a diverging collision rate. Even though the packings contained was a significant rattler fraction (about 2.5\%), precluding their applicability  to the aforementioned conjecture, Donev et al. \cite{Do05d} nonetheless found that a packing of $10^6$ particles 
was to an excellent approximation hyperuniform with a structure factor that 
exhibits an unusual nonanalytic linear dependence near the origin, i.e., $S({\bf k}) \sim |{\bf k}|$ as $|{\bf k}| \rightarrow 0$;
see Fig. \ref{S-MRJ}. This implies that QLR  pair correlations  in which $h(r)$ decays to zero
with the power-law scaling $-1/r^4$ in accordance with the asymptotic analysis presented in Sec. \ref{classes}. When the rattlers were removed from the packing, the structure factor at the origin had a substantially larger value, showing that the backbone alone is far from hyperuniform.
This numerical finding supporting the link between ``effective" hyperuniformity of a disordered packing 
and mechanical rigidity
spurred a number of subsequent numerical and experimental investigations that reached similar conclusions 
either via $S(k)$ or the spectral density ${\tilde \chi}_{_V}(k)$
\cite{Za11a,Za11c,Be11,Ku11,Ho12b,Dr15,Kl16,At16a}.
In all cases, effective or near hyperuniformity is conferred because the majority of the particles are contained in the strictly-jammed backbone.
Indeed, it has been systematically shown that as a hard-sphere system, substantially away from a jammed state, is driven toward strict jamming through densification, $S(0)$ monotonically decreases until effective hyperuniformity is achieved at the putative MRJ state.  Specifically, $S(0)$ was found to approach zero approximately linearly as a function of density from 93\% to 99\% of the jamming density, where extrapolating the linear trend in $S(0)$ to jamming density yielded $S(0) = -1 \times 10^{-4}$ \cite{Ho12b}.  This study clearly established a correlation between distance to jamming and hyperuniformity, and additionally introduced a ``nonequilibrium index'' $X$, defined by relation (\ref{XX}), 
describing the interplay between hyperuniformity and a dynamic measure of distance to jamming. 

\begin{figure}[bthp]
\begin{center}
\includegraphics[width=3.in,keepaspectratio,clip=]{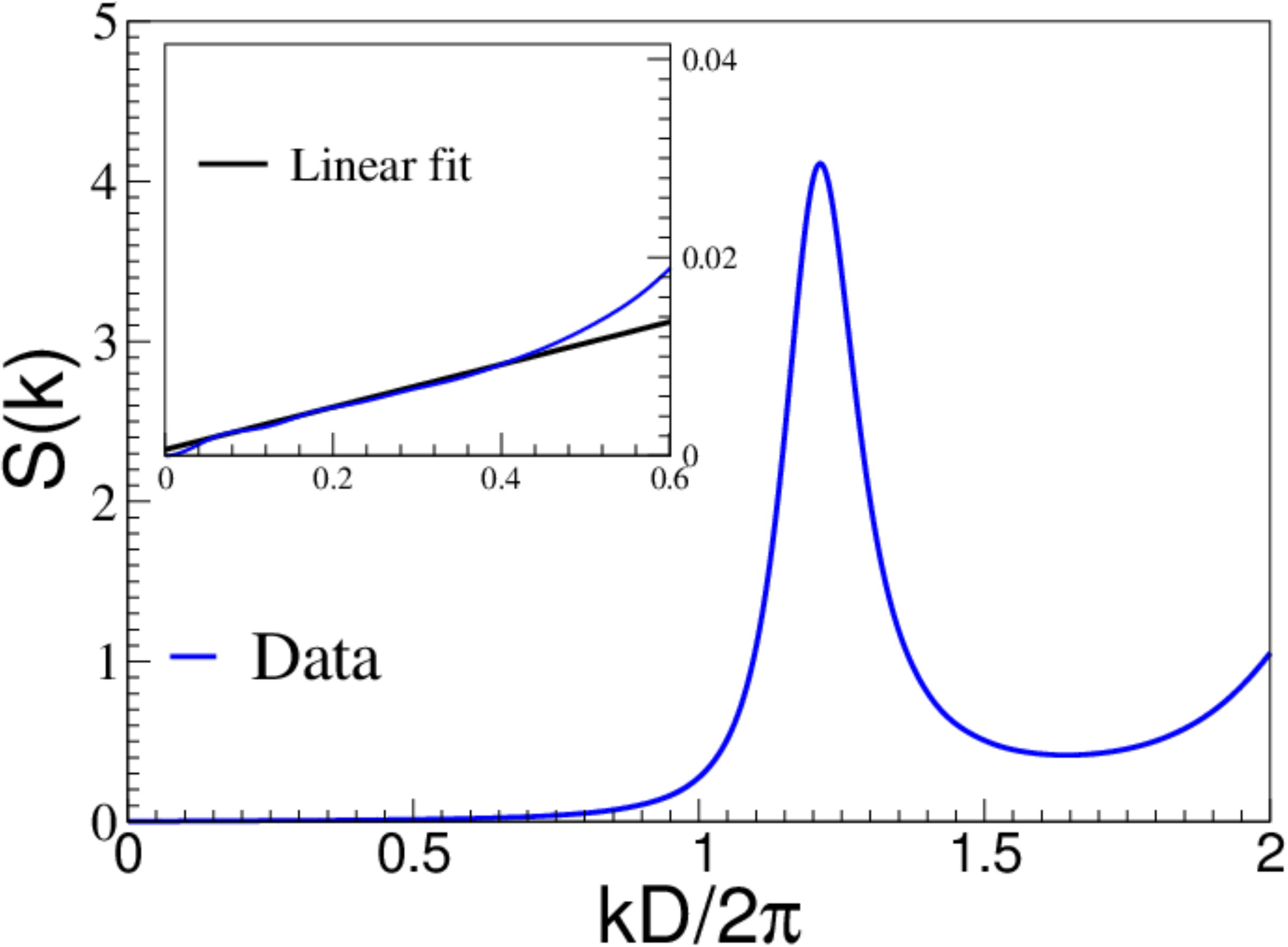}\hspace{0.15in}\includegraphics[width=3.in,keepaspectratio,clip=]{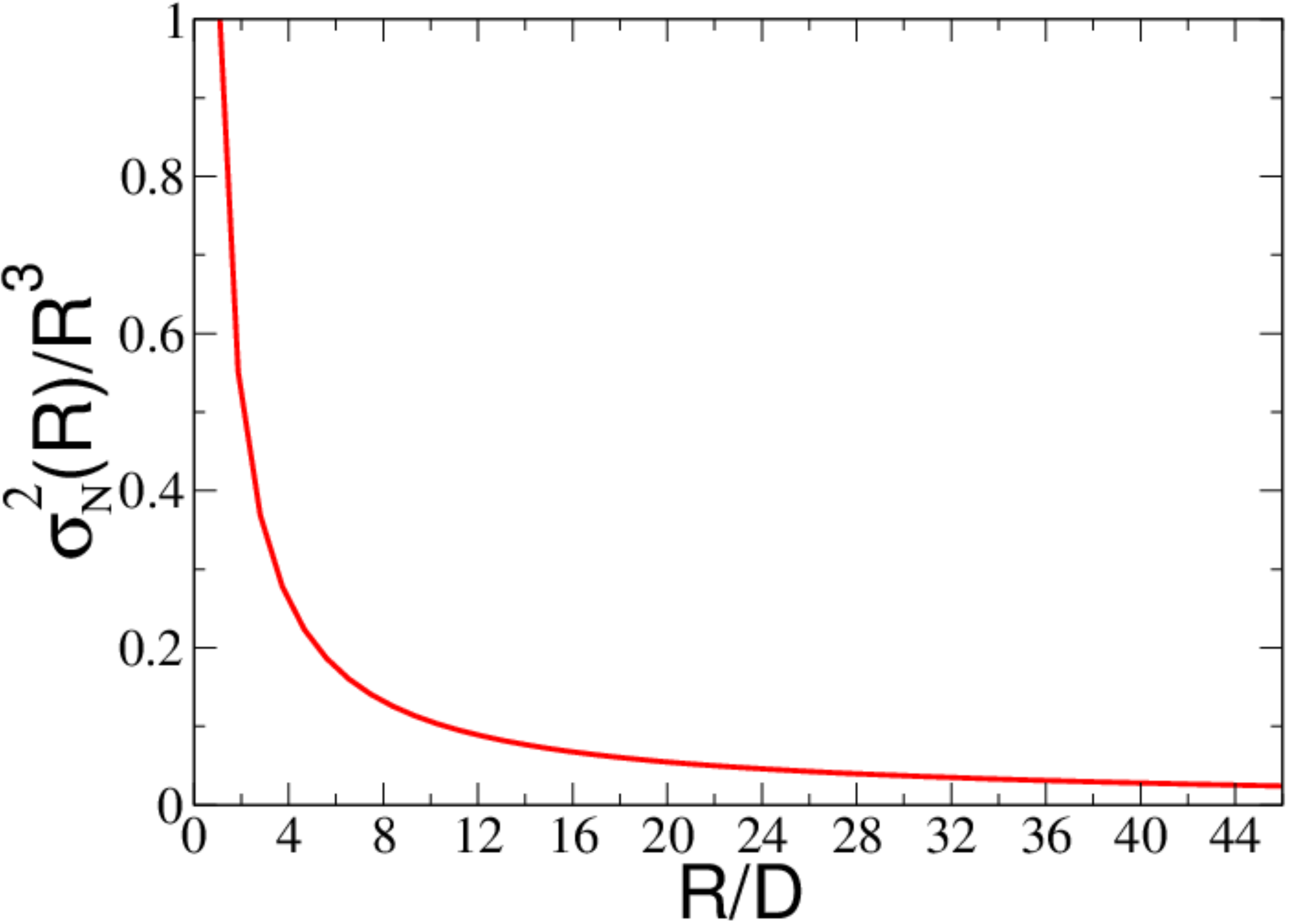}
\end{center}
\caption{ Left panel: The  structure factor $S(k)$ as a function
of the dimensionless wavenumber $kD/(2\pi)$ for a million-particle packing of identical
three-dimensional spheres at a putative MRJ state, where $D$ is the hard-sphere diameter \cite{Do05d}. The inset shows the linear in $|\bf k|$ 
nonanalytic behavior at ${\bf k=0}$ with $S(0) = 6 \times 10^{-4}$.
Right panel: The corresponding scaled number variance $\sigma_N^2(R)/R^3$ versus $R$. The fact that $S({\bf k})$ goes to zero
linearly in $|\bf k|$ implies that $\sigma_N^2(R)$ grows like $R^2\ln(R)$ for large $R$ and hence $\sigma_N^2(R)/R^3$ is a decaying function
of $R$, as it should be for a hyperuniform system.}
\label{S-MRJ}
\end{figure}

It is instructive to contrast the effective hyperuniform behavior of these nonequilibrium packings 
to that of nonhyperuniform identical hard spheres in equilibrium along the stable disordered (fluid) branch
for packing fractions in the range $0 \le \phi \le \phi_F$, where the freezing point is $\phi_F \approx 0.494$
in three dimensions; see Fig. \ref{MRJ}. At these packing-fraction extremes, $S(0)=1$ and $S(0)=0.02$ \cite{Han13,Do05c}, respectively.
The latter minimal value is still about two orders of magnitude larger than $S(0)$
reported for MRJ packings  \cite{Do05c,Ho12b}. Moreover, the
QLR  behavior of $g_2(r)$ of MRJ states  distinguishes it from that 
of the equilibrium hard-sphere fluid \cite{Han13}, which possesses a structure factor that is analytic at $\bf k
=0$ and thus has a pair-correlation function that decays
exponentially fast to unity for large $r$.

Even though polydisperse packings are part of the original Torquato-Stillinger conjecture \cite{To03a},
effective hyperuniformity can be observed in such
packings \cite{Za11a,Za11c,Be11,Ku11,Dr15} provided that the size distribution is suitably
constrained and the rattler concentration is sufficiently small. It is even possible that the conjecture can be
extended to  strictly jammed saturated packings of polydisperse spheres;
however, the determination of the additional conditions for such an extension
is highly nontrivial. Nonetheless, jamming is again a crucial
necessary property to attain near-hyperuniformity in two and three dimensions. Note that for packings
of spheres with a size distribution and of nonspherical particles, 
it is the spectral density ${\tilde \chi}_{_V}({\bf k})$
(not the structure factor $S({\bf k})$ associated with the points
that define the particles centroidal positions) that must be computed \cite{Za11a}. Of course, the 
hyperuniformity condition is that ${\tilde \chi}_{_V}({\bf k})$ must vanish
in the small-wavenumber limit, as specified by (\ref{hyper-2}); see also Sec. \ref{packing}.

As a practical matter, is is important to note that ascertaining the hyperuniformity of limited samples sizes  often encountered in simulations 
and laboratory experiments can often be done more accurately in
direct space via the local variance than in reciprocal space via the appropriate
spectral function \cite{Dr15}.

\subsubsection{Critical Slowing Down as a Disordered Jammed State is Approached}
\label{slowing}

A fascinating open question remains as to whether putative MRJ packings can be made to be even more hyperuniform than established to date or exactly hyperuniform with improved numerical protocols as the system size is made large enough.
This is an extremely delicate question to answer because one must be able to ensure that true jamming is achieved to within a controlled tolerance as the system size increases  without bound. However,
any packing algorithm necessarily must treat a finite system and hence the smallest accessible positive wavenumber at which $S(k)$ or ${\tilde \chi}(k)$ can be measured is of the order of $2\pi/N^{1/d}$, where $N$ is the number of particles.
The situation is further complicated by noise at the smallest wavenumbers, numerical and protocol-dependent errors, and the reliance on extrapolations of such uncertain data to the zero-wavenumber limit.

It has been recently shown that various standard packing protocols struggle to reliably create packings that are jammed for even modest system sizes of $N \approx 10^3$  bidisperse disks in two dimensions \cite{At16a}. Importantly, while these packings {\it appear} to be jammed by conventional tests, 
 rigorous linear-programming jamming tests \cite{Do04a,Do04c} reveal that they are not.
Evidence  suggests that deviations from hyperuniformity in putative MRJ packings can in part be explained by a shortcoming of the numerical protocols to generate exactly-jammed configurations as a result of a type of  ``critical slowing down'' \cite{Bi92} as the packing's collective rearrangements in configuration space become locally confined by high-dimensional ``bottlenecks'' through which escape is a rare event.
Thus, a critical slowing down implies that it becomes increasingly difficult numerically to drive the value of $S(0)$ down to its minimum
value of zero if a true jammed critical state could  be attained.  Moreover, the inevitable presence of even a small fraction of
rattlers generated by current packing algorithms destroys perfect hyperuniformity. 
In this regard, one should note that  nearly hyperuniform point configurations  can be made to be exactly hyperuniform by very tiny collective displacements via the collective-coordinate approach \cite{Uc06b}, which by construction enables the structure factor to be constrained to take exact targeted values at a range of wave vectors, as shown recently in Refs. \cite{To16b} (see also Fig. \ref{stealthy}) and \cite{At16a}.

In summary, the difficulty of ensuring jamming
as $N$ becomes sufficiently large to access the small-wavenumber regime in the structure factor
as well as the presence of rattlers that degrade hyperuniformity makes it virtually
impossible to test the Torquato-Stillinger jamming-hyperuniformity conjecture via current
numerical packing protocols.

\subsubsection{``Effective" Hyperuniformity Criterion}

Since hyperuniformity is an infinite-wavelength property of a point configuration in $\mathbb{R}^d$ and numerical
simulations as well as lab experiments are limited by system size and subject to error/noise, it is desirable 
from a practical viewpoint to devise a rough criterion 
for what one considers to be ``effective" or ``near" hyperuniformity. While this is ultimately subjective, 
an empirical operational definition that has been proposed \cite{At16a} for such behavior is
that the first dominant peak value of the structure factor    relative to its
estimated value at the origin is roughly of the order of $10^4$ or larger; equivalently, the  ratio
\begin{equation}
H \equiv \frac{S({\bf k=0})}{S({\bf k}_{peak})}
\label{Hyp}
\end{equation}
is of the order of $10^{-4}$ or smaller, where ${\bf k}_{peak}$ is the location of the first dominant peak
of the structure factor. In the case of heterogeneous media, a similar ``hyperuniformity metric"
applies, except where $S(\bf k)$ is replaced with the spectral density ${\tilde \chi}_{_V}({\bf k})$, as
defined in Sec. \ref{hetero}.

\subsubsection{Toward the Ideal MRJ State}
\label{ideal-MRJ}

While one should not generally expect exact hyperuniformity for disordered packings with rattlers, it has been demonstrated  that when jamming is ensured, disordered packings 
with a small fraction of rattlers can be very nearly hyperuniform, and deviations from hyperuniformity correlate with an inability to ensure jamming, suggesting that strict jamming and hyperuniformity are indeed linked \cite{At16a}. This raises the possibility that the ideal MRJ packing 
possesses no rattlers such that the jammed backbone contains every sphere in the packing.
This would imply that disordered isostatic sphere packings with rattlers possess a higher degree of order (than ideal MRJ packings
without rattlers), even when the rattlers are included,
because rattler cages require the caged particles and their neighbors be more correlated
in order to house the rattlers. This possibility provides the impetus for the development of packing algorithms that
produce large disordered strictly jammed packings that are rattler free.  
Typical packing algorithms, to a good approximation, will produce disordered isostatic sphere
packings with approximately 3\% rattlers \cite{Lu90,Do05c,Cha10,To10c}. The Torquato-Jiao (TJ) sphere packing 
algorithm~\cite{To10e}, which employs linear-programming techniques that become exact
as jamming is approached, efficiently  produces disordered, isostatic strictly jammed packings with unsurpassed 
numerical fidelity with a substantially lower rattler fraction of 1.5\,\% when the
number of spheres $N$ is sufficiently small \cite{At13}. Thus,  an outstanding, challenging task is the formulation
of numerical packing protocols that generate not only rattler-free disordered
isostatic packings but ones in which $N$ is large enough to access very small wavenumbers
in the structure factor to ascertain the degree of hyperuniformity as accurately as possible. 

In the subsequent subsection on absorbing-state models, the ideal MRJ state will be interpreted
as the end point of a nonequilibrium dynamical process of a  many-particle system 
in which jamming is tantamount to being at a critical-absorbing state.

\subsubsection{Is Hyperuniformity Also a Signature of  MRJ Packings of Aspherical Particles?}

Over the past decade there has been increasing interest in the
effects of particle shapes on the characteristics of disordered
jammed packings, since deviations from sphericity introduce 
rotational degrees of freedom. This implies that a nonspherical particle requires more contacts to 
stabilize mechanically than a sphere \cite{Do04b,Do05a,Do05b} and hence packings of the former 
are generally denser than  sphere packings. Packings of aspherical particles 
are useful models of heterogeneous materials,  granular
media and structural glasses \cite{To02a,Do04b}. Nonspherical particles  that have been studied include
ellipsoids \cite{Do04b,Ma05,Mail09,Za11a}, ``superballs" \cite{Ji10b,Za11a,Ti15}, ``superellipsoids" \cite{De10}, and
polyhedra \cite{Li08,Gl09,To09c,Ja10,Sm10,Bak10}. Numerical simulations have been performed
that to a very good approximation produce MRJ-like isostatic packings
of identical non-spherical particles in two and three dimensions, including
ellipsoids and superballs \cite{Za11a,Za11c,Za11d} as well as
polyhedra, such as the Platonic solids \cite{Ji11c} and truncated tetrahedra \cite{Ch14a}.
In all of these cases, the spectral function (either the 
structure factor or spectral density) appears to go zero linearly
in the wavenumber with a slope that depends on the particle shape 
\cite{Za11a,Za11c,Za11d}. Hence, these jammed disordered packings possess QLR pair correlations in which
$h(r)$ or $\chi_{_V}(r)$ decays asymptotically to zero with the scaling $-1/r^{d+1}$, showing that they
 belong to the same universality class as MRJ sphere
packings.  Thus, at least for this limited class of aspherical shapes, 
hyperuniformity (to a good approximation) appears to be a signature
of the MRJ state.

\subsection{Driven Nonequilibrium Systems and Critical Absorbing States}

It is  well-established that many-particle systems far from equilibrium 
typically possess long-range correlations; see, for example,  \cite{Sp83,Ga90,Ch91,De07,Bo08}.   
Absorbing-state models are far-from-equilibrium many-body systems that
provide excellent examples of nonequilibrium phase transitions between two distinct phases 
-- an active phase, which is a steady state with never-ending dynamics, and an absorbing state where the dynamics 
cease - with a well-defined critical point \cite{Hi00}. 
Following an initial transient, any system below its critical state
ultimately evolves to an absorbing state, in which each of the particles satisfies some local criterion for the cessation of its evolution under the dynamics. The details of what is this criterion and the nature of the dynamics defines the specific absorbing-state model. Whether the system is in the active or the absorbing phase is determined by the value of the control parameter (e.g., fraction of space covered by the particles), as will be detailed below. 
\begin{figure}[bthp]
\centerline{
\includegraphics[height=2.4in,keepaspectratio,clip=]{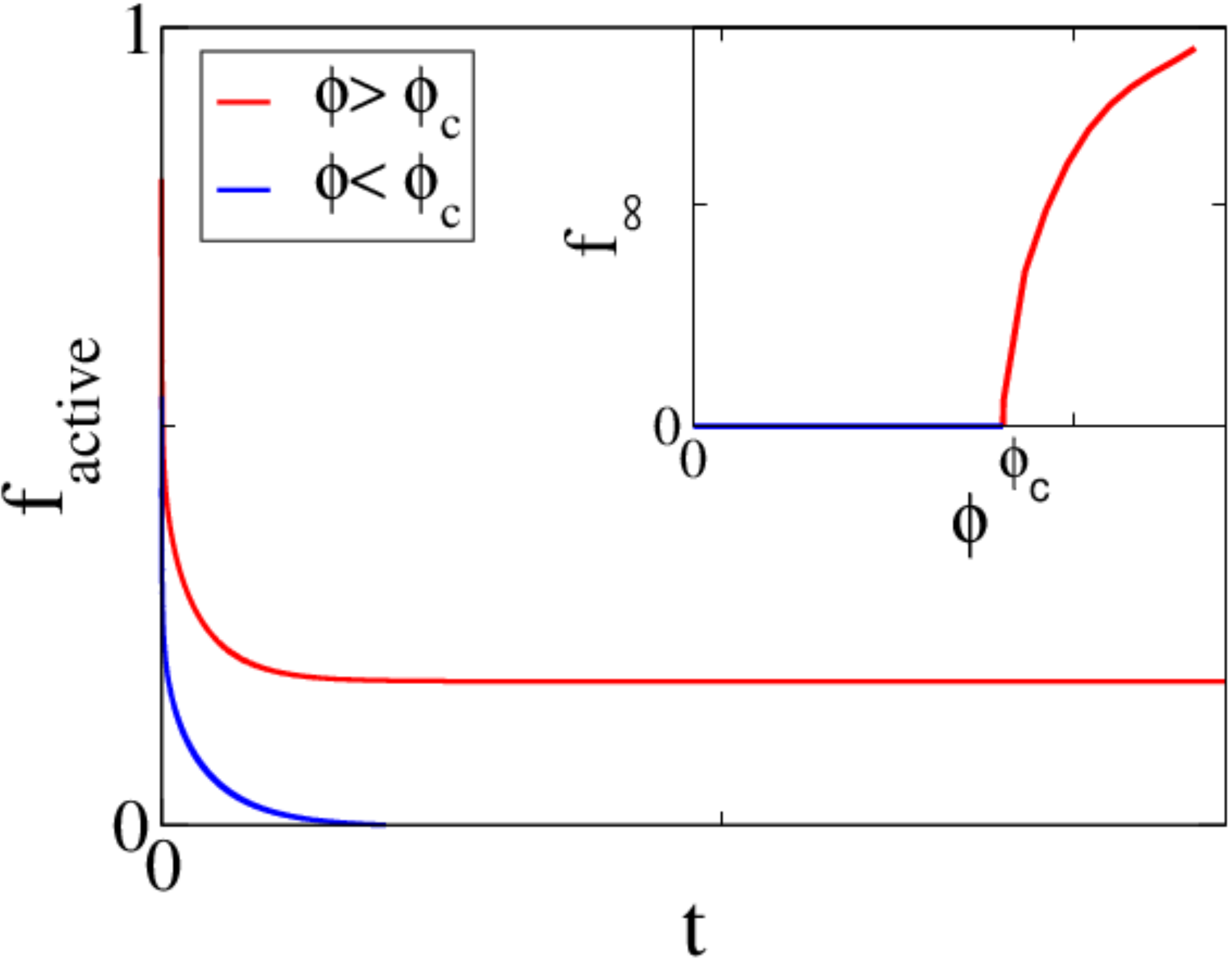}}
\caption{Schematic illustrating the time ($t$) evolution of the fraction of active particles $f_{active}(t)$
below the critical packing fraction $\phi_c$ (blue curve)
and above $\phi_c$ (red curve). The inset shows the behavior of the active fraction
at infinite time $f_{\infty}$ as a function of the packing fraction $\phi$. A
second-order phase transition occurs at $\phi=\phi_c$, where the system is hyperuniform.}
\label{decay}
\end{figure}

At a given instant of time $t$, a system has fraction of active particles $f_{active}(t)$, and the active
fraction decreases up to a steady-state value $f_\infty \equiv \lim_{t\to \infty} f_{active}(t)$ over time; see
Fig. \ref{decay}. If a control parameter, such as the packing fraction $\phi$, is smaller than a critical value $\phi_c$, the
steady-state fraction $f_\infty$ becomes identically zero. 
As $\phi$ approaches the critical point $\phi_c$ from below, both the lifetime of the transient
and appropriately defined correlation length diverge.
When the parameter $\phi$ is larger than the critical point $\phi_c$, the steady-state fraction $f_\infty$ is
positive, and the system lies in the active phase.
The transition  bears a great resemblance to a second-order equilibrium phase transition \cite{Hi00,Lu04,He08}; see 
the inset of Fig.  \ref{decay}. 
It is natural to ask about the nature of the critical absorbing states, which is obtained by studying the absorbing phase as the control parameter approaches its critical value, $\phi_c$. The recent numerical observations that  several absorbing state models at criticality are effectively hyperuniform long-range pair correlations \cite{He15} together with the fact that such disordered systems may be realized experimentally \cite{Pi05,Chaik08} are significant developments.

\subsubsection{Hyperuniformity in Absorbing-State Models}
\label{slowing-2}

The properties of several absorbing-state  models belonging to the conserved directed
percolation universality class were numerically studied by Hexner and Levine \cite{He15}.  The models examined
include the conserved lattice gas in two and three dimensions \cite{Ro00}, the Manna model in one dimension \cite{Hi00,Le04} and random organization
models in one and two dimensions \cite{Chaik08}.  They found  that at the critical point the absorbing
states are effectively hyperuniform. The various models differ in the criteria that define which
of the particles are active and the particle displacement rules.  Regardless of their details,
they found that the nonequilibrium phase transitions are characterized by a set of universal
critical exponents that depend only on the dimensionality. Specifically, they found
that the structure-factor exponent $\alpha$ defined in Eq. (\ref{S-asy}) is given
by $\alpha \approx 0.425$, $\alpha \approx 0.45$ and $\alpha \approx 0.24$
in one, two and three dimensions, respectively. Therefore, according to the asymptotic number-variance
scaling relation (\ref{sigma-N-asy}), these hyperuniform systems belong to class III; specifically,
$\sigma^2_{_N}(R) \sim R^{0.575}$ for $d=1$, $\sigma^2_{_N}(R) \sim R^{1.55}$ for $d=2$,
and $\sigma^2_{_N}(R) \sim R^{2.76}$ for $d=3$.

\begin{figure}[bthp]
\centerline{
\includegraphics[height=2.2in,keepaspectratio,clip=]{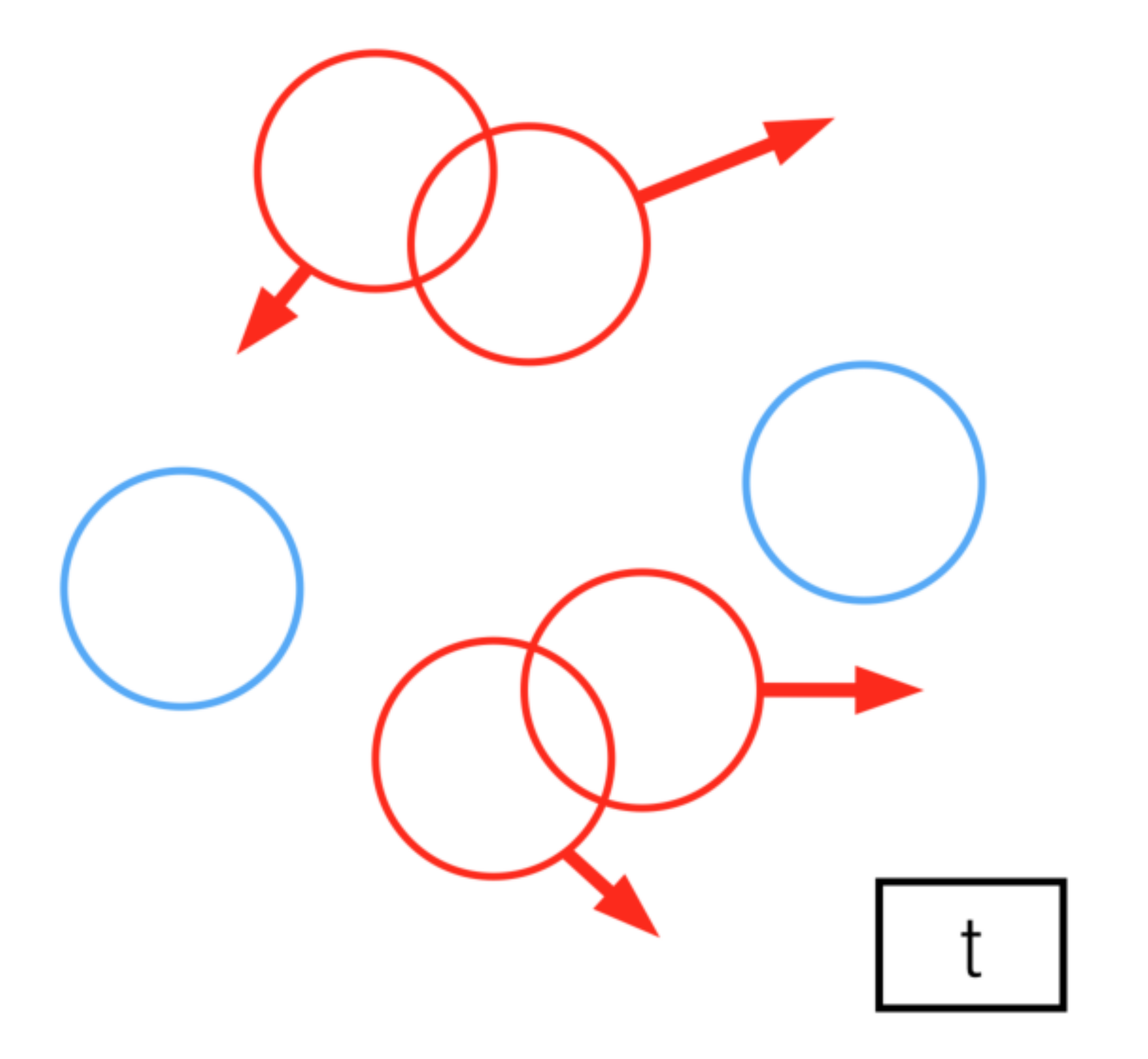} \hspace{0.2in}
\includegraphics[height=2.2in,keepaspectratio,clip=]{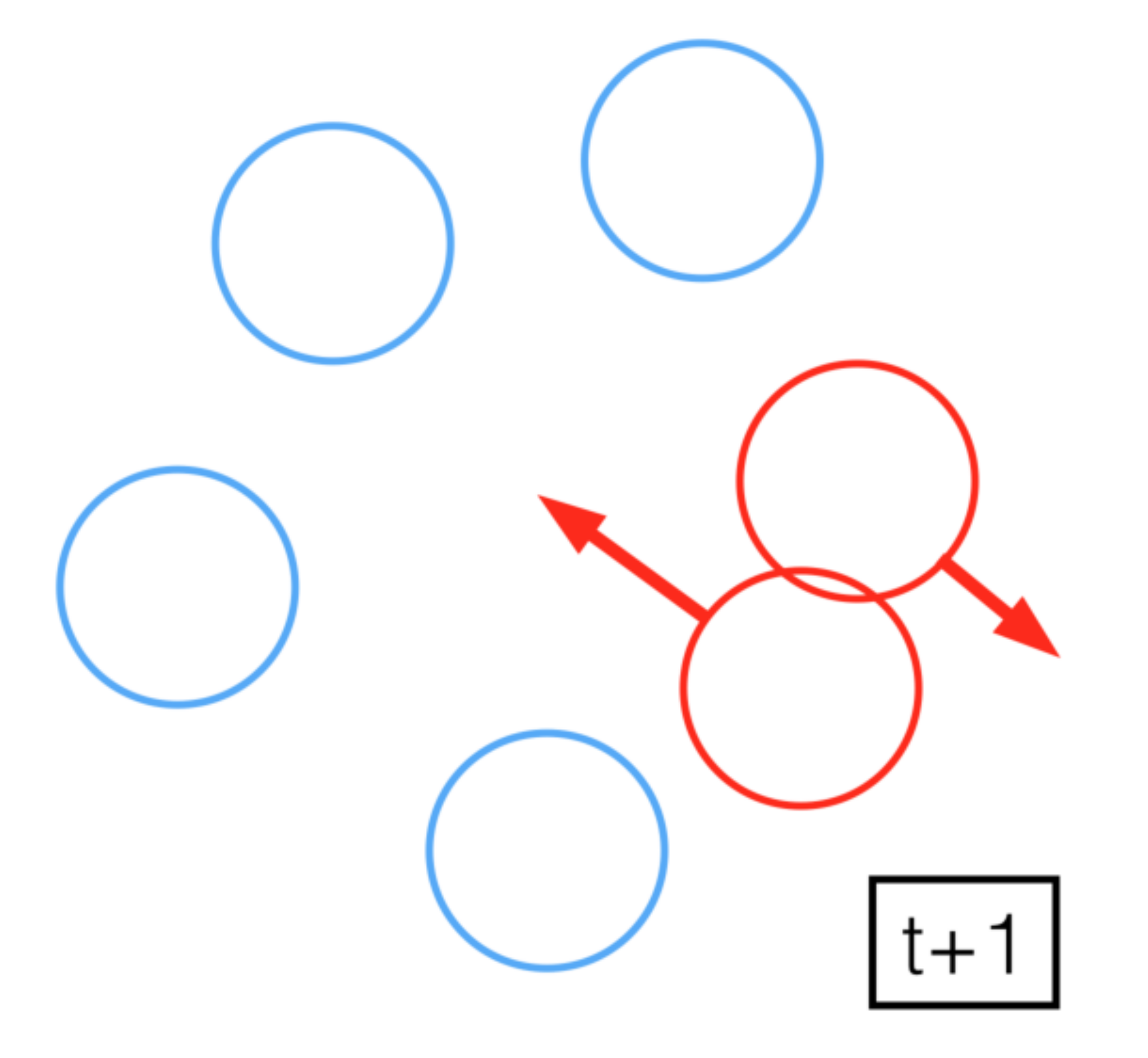} \hspace{0.2in}}
\caption{Schematic indicating overlapping active particles (red) and nonoverlapping
inactive particles  (blue) at times $t$ (left panel) and $t+1$ (right panel). At time $t$,
the maximum possible displacement vectors  of active particles are indicated with arrows (whose
directions and lengths are chosen according to some stochastic rule)
and determine which of the particles are active or inactive at the next time step, i.e., at time $t+1$.
 }
\label{cartoons}
\end{figure}
  
It is instructive to review briefly aspects of the random organization model that was put forward by Cort{\` e} et al. \cite{Chaik08} to understand how the irreversible collisions  that generally produce diffusive chaotic dynamics in periodically sheared suspensions 
at low Reynolds number can also cause systems to self-organize to avoid future collisions. Such dynamics can lead to a non-fluctuating quiescent state
with a dynamical phase transition separating it from fluctuating diffusive states.  The model begins from an initial distribution of points 
(e.g., Poisson point process) in some region of $d$-dimensional Euclidean space $\mathbb{R}^d$, often taken to be a fundamental cell
under periodic boundary conditions.   Next, each point is surrounded by an ``influence region" of some well-defined shape and size.  
Particles whose influence regions overlap with those of other particles are deemed active, and at the next time step, all active particles 
are translated by some distance (``kick") in random directions; see Fig. \ref{cartoons}.  The kick sizes may be drawn randomly from a given distribution, which is taken to be short-ranged.  This process is iterated until the system achieves a steady state (active phase) or the dynamics ends (absorbing phase).
An anisotropically-shaped influence region is an appropriate choice for sheared systems, which was the original motivation for the
model \cite{Pi05,Chaik08}.

\begin{figure}[bthp]
\centerline{\includegraphics[height=2.7in,keepaspectratio,clip=]{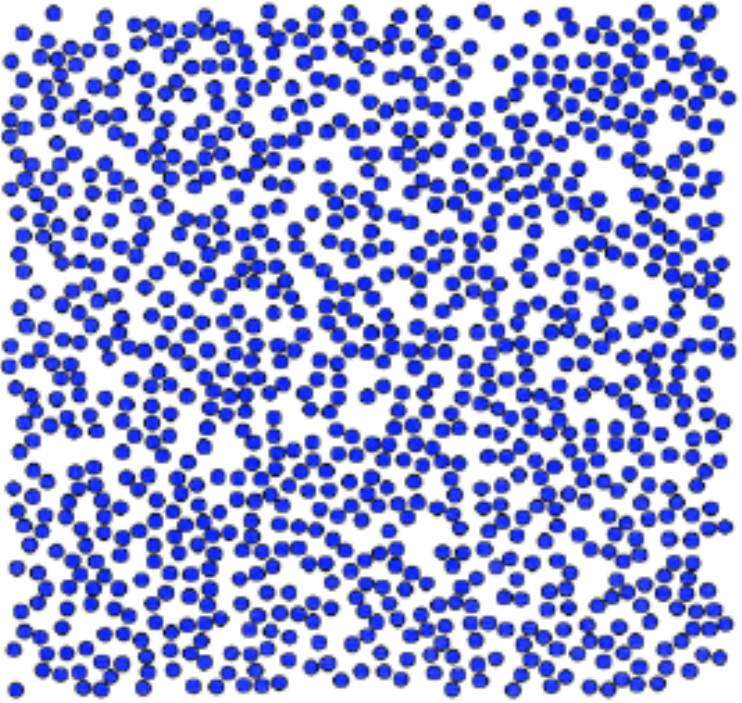}}
\centerline{\includegraphics[height=2.7in,keepaspectratio,clip=]{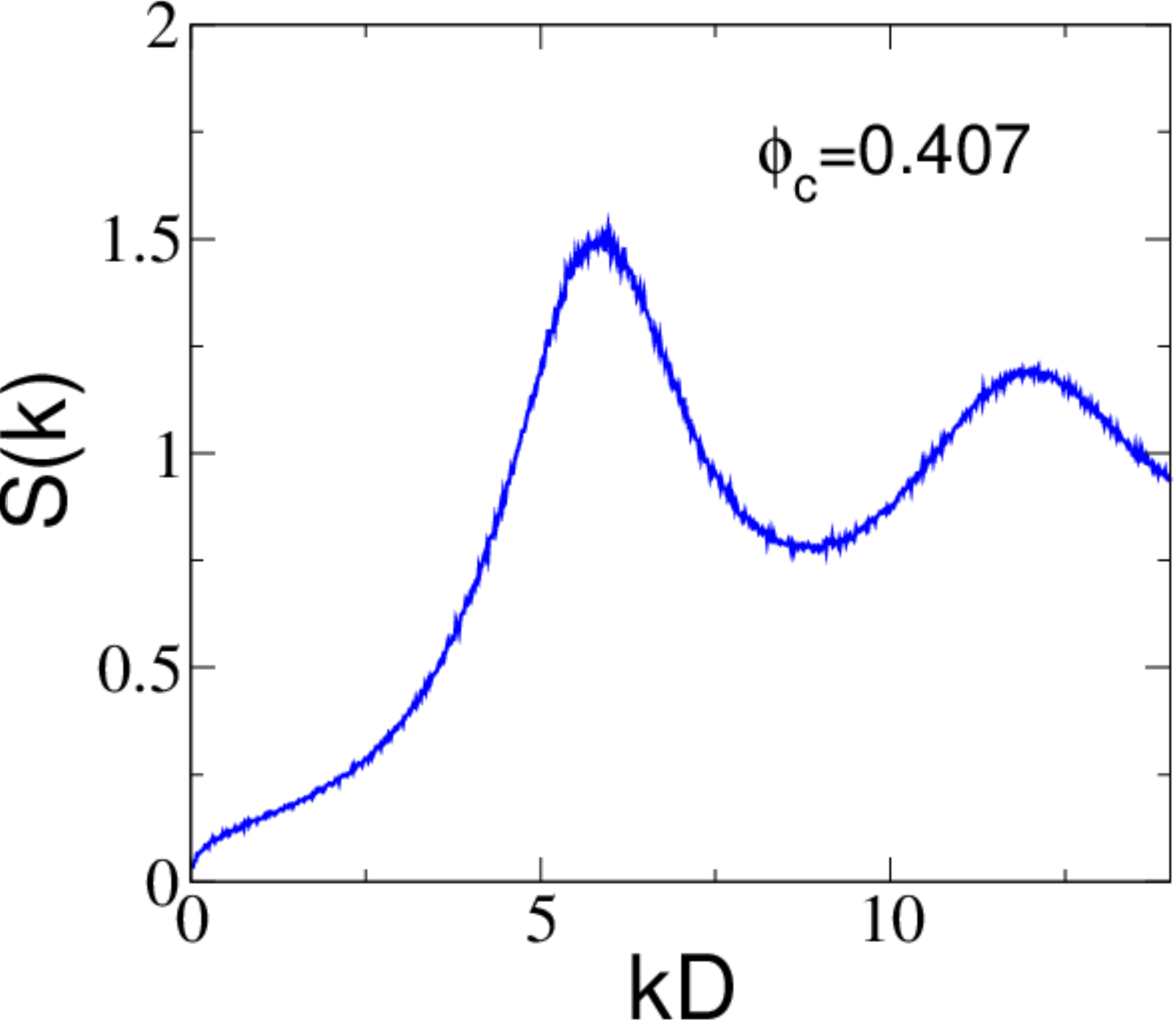}}
\centerline{\includegraphics[height=2.7in,keepaspectratio,clip=]{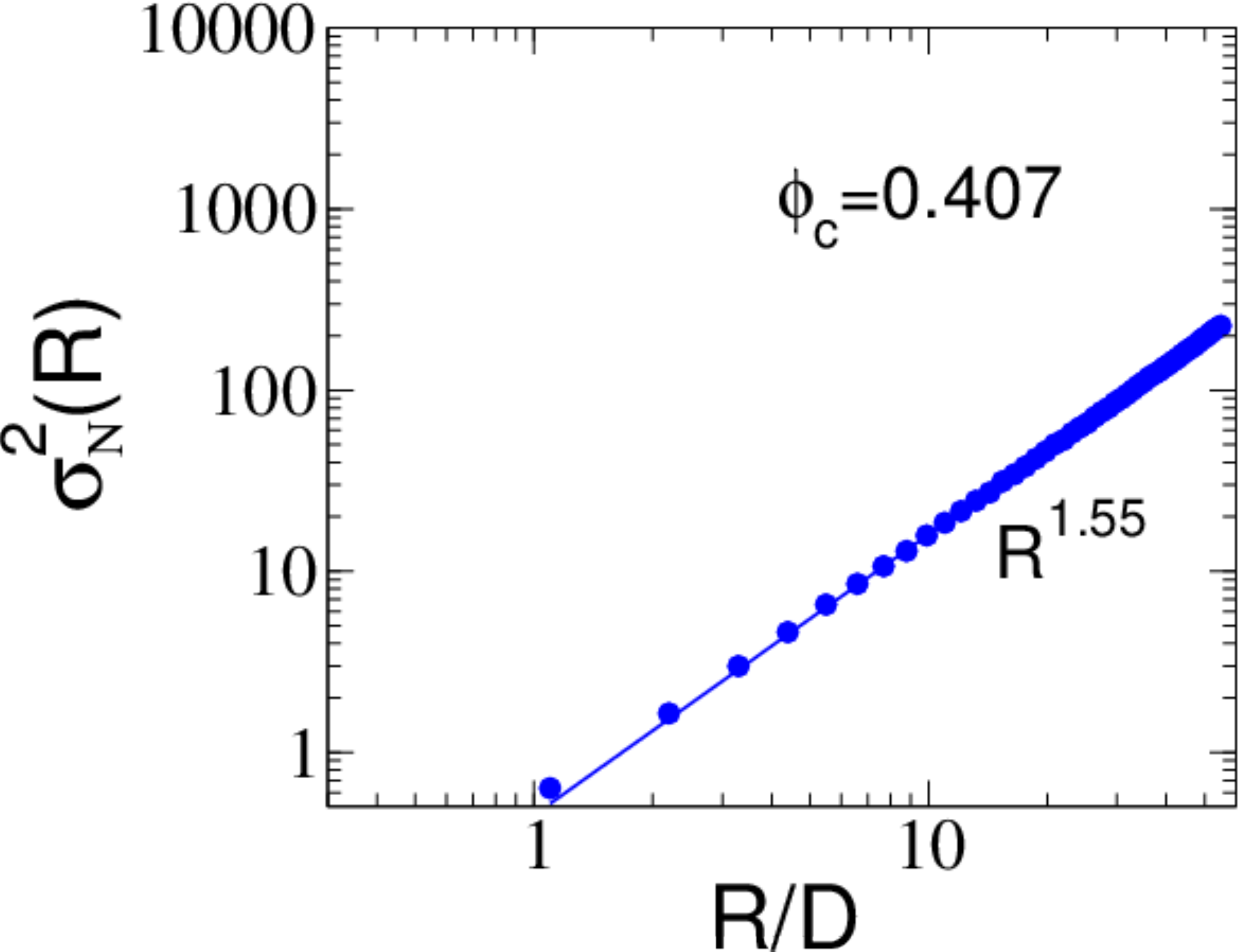}}
\caption{Top panel: A portion of a system of 100,000 circular disks of diameter $D$ at the critical state with packing
fraction $\phi_c=0.407$ as generated from the random organization model \cite{Chaik08}.
Middle panel: Corresponding structure factor $S(k)$
versus dimensionless wavenumber $kD$. Bottom panel: Corresponding logarithm of the number variance
$\sigma^2_{_N}(R)$ versus the logarithm of the window radius $R$, which shows a hyperuniform scaling of $R^{1.55}$,
and hence belongs in class III; see Eq. (\ref{sigma-N-asy}). }
\label{rand-org}
\end{figure}

A particularly simple version of this model takes the influence region 
to be a $d$-dimensional sphere \cite{Tj15}.  Because of the isotropy of this shape, the system evolves to a statistically isotropic 
distribution of particles. We have generated realizations of 100,000 particles in two dimensions that start
from relative dense but {\it unsaturated} nonoverlapping random-sequential-addition of circular disks of diameter
$a$ \cite{Fe80,Ta00,Zh13b} but with an influence circle of radius $D > a$
to allow particle overlaps (active particles).   For a given value of $D$, 
the amplitude of a kick given to an active particle is randomly and uniformly chosen between $0$ and $D/4$, and the system dynamics are followed until the steady-state
conditions are established. The diameter is allowed to systematically increase in order to identify  the critical packing fraction $\phi_c$,
at which point the system is a packing, i.e., identical hard circular disks of diameter $D$.  
A portion of a system at a critical absorbing state  with $\phi_c=0.407$ is shown in Fig. \ref{rand-org}.
(This value of the critical packing fraction is consistent with the one reported
in Ref. \cite{Tj15} with this maximum displacement size.)
The corresponding structure factor and number variance are also depicted in Fig. \ref{rand-org}.
Both quantities reveal that the critical state is effectively hyperuniform and belongs
to class III [cf. Eq. (\ref{sigma-N-asy})], since $\sigma^2_{_N}(R) \sim R^{1.55}$.
This is consistent with the results reported in Ref. \cite{He15}.
Note that, as in the case of the generation of MRJ particle
packings (Sec. \ref{slowing}), a precise numerical determination of the critical point of 
absorbing-state models is nontrivial due to a critical slowing down of the system.

The discovery that periodically sheared low-Reynolds number non-Brownian systems can self-organize into a correlated absorbing state \cite{Chaik08}  has led to a flurry of activity on periodically-driven systems. Superconducting vortices in an oscillating magnetic field \cite{Re09}, driven glassy systems \cite{Na14}, compressed/expanded foams, emulsions in oscillating flow \cite{We15} have all been shown to have absorbing states. It has recently been found that periodically sheared frictional granular systems have an absorbing-state transition as well \cite{Ro15}. Moreover, at criticality, these absorbing-state systems are hyperuniform. Remarkably, it has been discovered that 
off criticality, ``kicking" or reactivating these systems makes them more hyperuniform with  exponents 
that are comparable to those found for MRJ packings, i.e., the structure factor tends to $S(k) \sim k$ in the small wavenumber
limit \cite{He17b}, and hence they belong to class II hyperuniform systems. Interestingly, a recent numerical
study has shown that {\it chaotically} driven suspensions can lead to hyperuniform  states \cite{We17}.

These models have been extended by  introducing an additional symmetry 
beyond particle conservation, namely, when two particles interact, stochastic kicks are given that conserve their center of mass \cite{He17b}. 
It was found that the active states are hyperuniform with a greater
suppression of large-scale density fluctuations  than those in previous random organization models; in fact, they
belong to class I, as defined by  (\ref{sigma-N-asy}). Large-scale fluctuations are determined by a competition between a noise term that generates fluctuations, 
and a deterministic term that reduces them.

\subsubsection{Interpretation of the Ideal MRJ Sphere Packing as a Critical-Absorbing State}

Interestingly, an ideal MRJ sphere packing can be interpreted  as a  critical-absorbing state
associated with a nonequilibrium absorbing phase transition.
As usual, one must define the active and inactive particles as well as the dynamics.
We propose the following absorbing-state models for disordered strictly jammed sphere packings.
To begin, imagine an initial random configuration of a dilute concentration of $N$ identical hard spheres that will
undergo interparticle collisions via molecular dynamics within a fundamental simulation cell that is allowed to deform and {\it shrink} (on average)
as a function of time in the spirit of an adaptive-shrinking cell scheme \cite{To09b,To09c}.
(Note that conceptually this is opposite to the modified event-driven molecular dynamics
in which the particles are allowed to grow in a deforming volume-preserving simulation box that is not allowed
to shrink in volume \cite{Do05a,Do05b}.) Active particles are those that can undergo collisions
with themselves (or the container boundary in the case of a hard-wall container)
and inactive particles are those that are frozen or at least locally jammed at any particular instant
of time. The dynamics cease when the system reaches an absorbing state in which
all of the particles are inactive, i.e., when the entire packing is strictly jammed,
and hence achieves a critical absorbing state, but only from below. Notably,  this implies that
such critical absorbing states cannot possess rattlers because rattlers are active
due to collisions with the rattlers cages. A rattler-free 
critical strictly-jammed absorbing state that is  maximally random
harkens back to the ideal MRJ state alluded to in Sec. \ref{ideal-MRJ}, which
would be hyperuniform according to the hyperuniformity-jamming conjecture
(see Sec. \ref{conjecture}).  Depending on the initial conditions (spatial configuration and density) and simulation
box deformations, the final 
(jammed) critical absorbing states can have a spectrum of degrees of ordering, including,
for example, the maximally dense fcc lattice sphere packing.

\section{Natural Disordered Hyperuniform Systems}
\label{natural}

It has recently been discovered that disordered hyperuniformity can confer to biological systems
optimal or nearly optimal  functionality, including the avian retina
\cite{Ji14} and the immune system \cite{Ma15}, which are briefly reviewed
here. It is likely that there are many other natural disordered hyperuniform materials or systems 
that have yet to be found.

\subsection{Avian Photoreceptor Cells} 
\label{avian}

Biology has recently taught us that disordered hyperuniform point patterns
offer desirable color-sensing characteristics \cite{Ji14}. 
The purpose of a visual system is to sample light in such a way as
to provide an animal with actionable knowledge of its surroundings
that will permit it to survive and reproduce \cite{Pu11}. 
Often, this goal is achieved most effectively by a highly regular
two-dimensional  array of cone cells that evenly sample
incoming light to produce an accurate representation of the visual
scene. Cone cells are one of three types of photoreceptor cells in the retina of vertebrates
that are responsible for color vision and function best in relatively bright light, as opposed to rod cells
that are better able to detect dim light.  According to classical sampling theory \cite{Sh49,Pe62}, 
the optimal arrangement of a two-dimensional array of light detectors is the
triangular lattice. Indeed, studies suggest that any deviation from perfectly regularity
will cause deterioration in the
quality of the image produced by a retina \cite{Fr77}.
Accordingly, many species have evolved an optimal or nearly optimal sampling
arrangement of their photoreceptors, including the insect compound eye \cite{Re76,Lu11}, many
teleost fish \cite{Ly57,En63,Ra04} and some reptiles \cite{Du66}.

Diurnal birds have one of the most sophisticated cone visual
systems of any vertebrate, consisting of five cone types: four  single cones
(violet, blue, green and red) that mediate color vision and
double cones involved in luminance detection \cite{Ha01}; see Fig. \ref{cones-1}. Despite the presence of numerous
evolutionary specializations in the avian eye, the overall
arrangement of bird cone photoreceptors is not  ordered
but rather is irregular \cite{Mor70,Kr10}. The five avian cone
types exist as five independent, spatial patterns, all embedded
within a single monolayered epithelium.  Each cell
type of this {\it multicomponent} system is maximally sensitive to
visible light in a limited but different range of wavelengths. 
Given the acute vision of birds and the utility of the perfect triangular-lattice arrangement of
photoreceptors for color sensing \cite{Fr77}, the presence of disorder in
the spatial arrangement of avian cone patterns had been puzzling.

\begin{figure}[H]
\begin{center}
{\includegraphics[  width=3.5in,
    keepaspectratio]{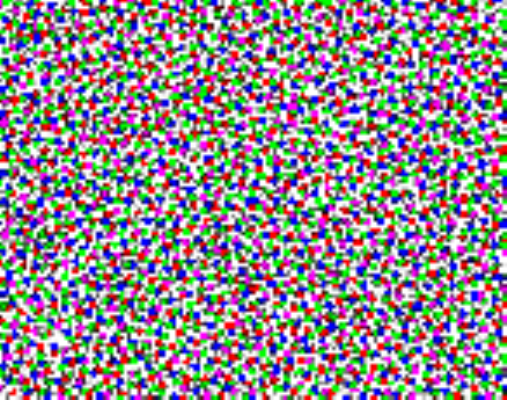}}
\caption{Voronoi tessellation representation of  the spatial distribution of the 
five types of light-sensitive  cones in the chicken retina: violet, blue, green, red
and white (representing double cones). Courtesy of Joseph Corbo and Timothy Lau, Washington University in St. Louis.}
\label{cones-1}
\end{center}
\end{figure}

\begin{figure}[H]
\begin{center}
{\includegraphics[  width=5in,
    keepaspectratio,clip=]{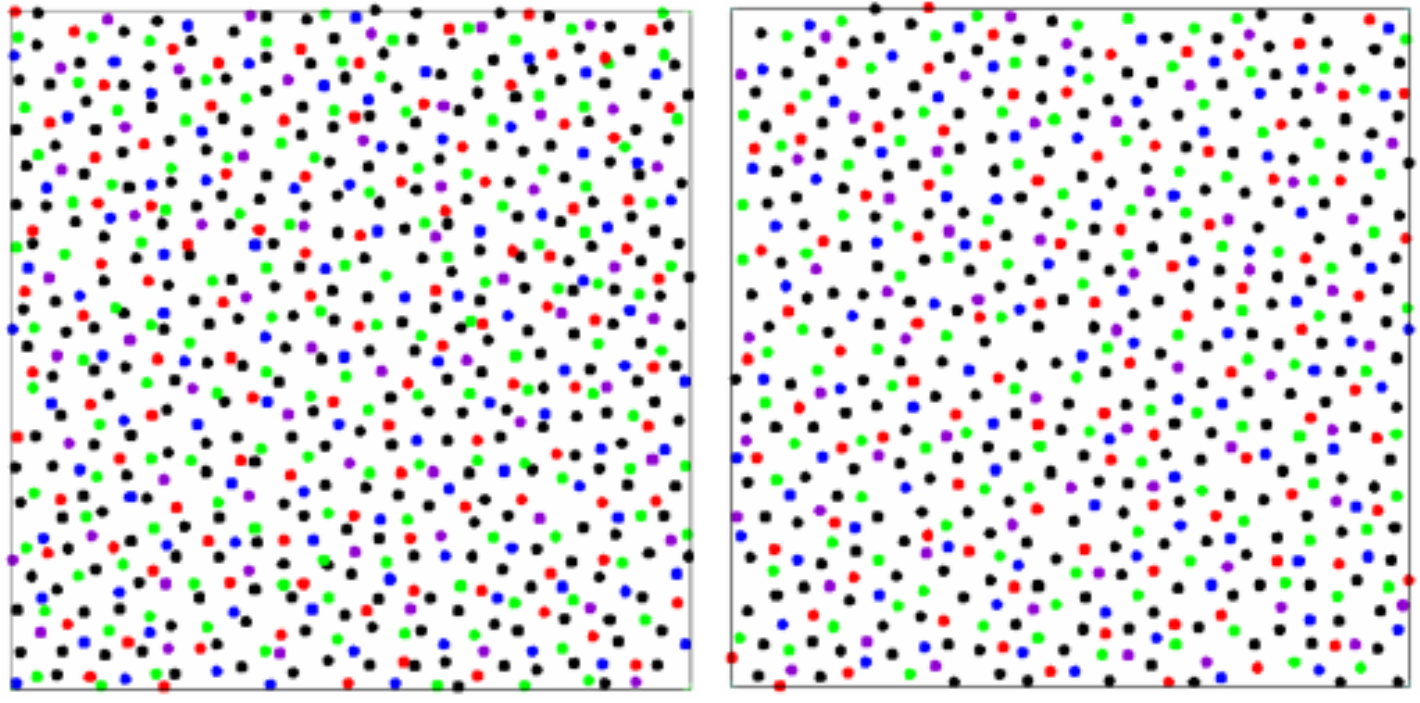}}
\caption{Left panel: Experimentally obtained configurations
representing the spatial arrangements of the five chicken cone photoreceptors.
The colored dots (enlarged  for visualization purposes) represent the centers of the cells. 
Here blacks dots represent the double cones. 
Right panel: Simulated point configurations representing
the spatial arrangements of chicken cone photoreceptors.
These figures are taken from Ref. \cite{Ji14}.}
\label{cones-2}
\end{center}
\end{figure}

Recently, images of large  two-dimensional arrays of chicken cone photoreceptors ($\sim 5000$ cells per data set)
were spatially analyzed and characterized using a host of
sensitive statistical correlation functions and other microstructural descriptors \cite{Ji14}; see
the left panel of Fig. \ref{cones-2}. Among other quantities, the local number variances 
and the structure factors were ascertained. It was found that the disordered photoreceptor patterns are 
hyperuniform, a property that  had previously not been  identified in any living organism \cite{Ji14}.
Remarkably, in a departure from
any known physical system, the patterns of both the total population and the individual cell types are simultaneously hyperuniform,
as shown by the corresponding structure factors depicted in  Fig. \ref{avian-S}. 
Such patterns are called ``multihyperuniform" because multiple distinct subsets of the overall point 
pattern are themselves hyperuniform \cite{Ji14}. This singular property implies that if any 
such distinct subset is removed from the overall population, the remaining pattern is still hyperuniform.
Multihyperuniform disordered structures could have implications for the
design of materials with novel physical properties and therefore may represent a fruitful area for future research.

To model the avian photoreceptor patterns, Jiao et al. \cite{Ji14} considered a statistical-mechanical cell model with
with two types of effective cell-cell interactions: isotropic
short-range hard-core repulsions between any pair of cells
and isotropic longer-ranged soft-core repulsions between pairs
of like-cells.  The local-energy minimizing configurations of such a many-particle
interacting system were simulated and shown to quantitatively capture, with high accuracy, the
unique spatial characteristics of avian photoreceptor patterns, including multihyperuniformity \cite{Ji14}.
Figure \ref{avian-S} shows that there is excellent agreement between the structure factors
obtained from the aforementioned statistical-mechanical cell model
and their experimental counterparts for the total population and the individual cell types.
This epithelium system stands in contrast to epithelia in mammalian  skin,
which has been shown not be hyperuniform \cite{Ch16}, since the latter, unlike
the former, must be very  pliable to deformations.

\begin{figure}[H]
\begin{center}
\includegraphics[height=10cm,keepaspectratio]{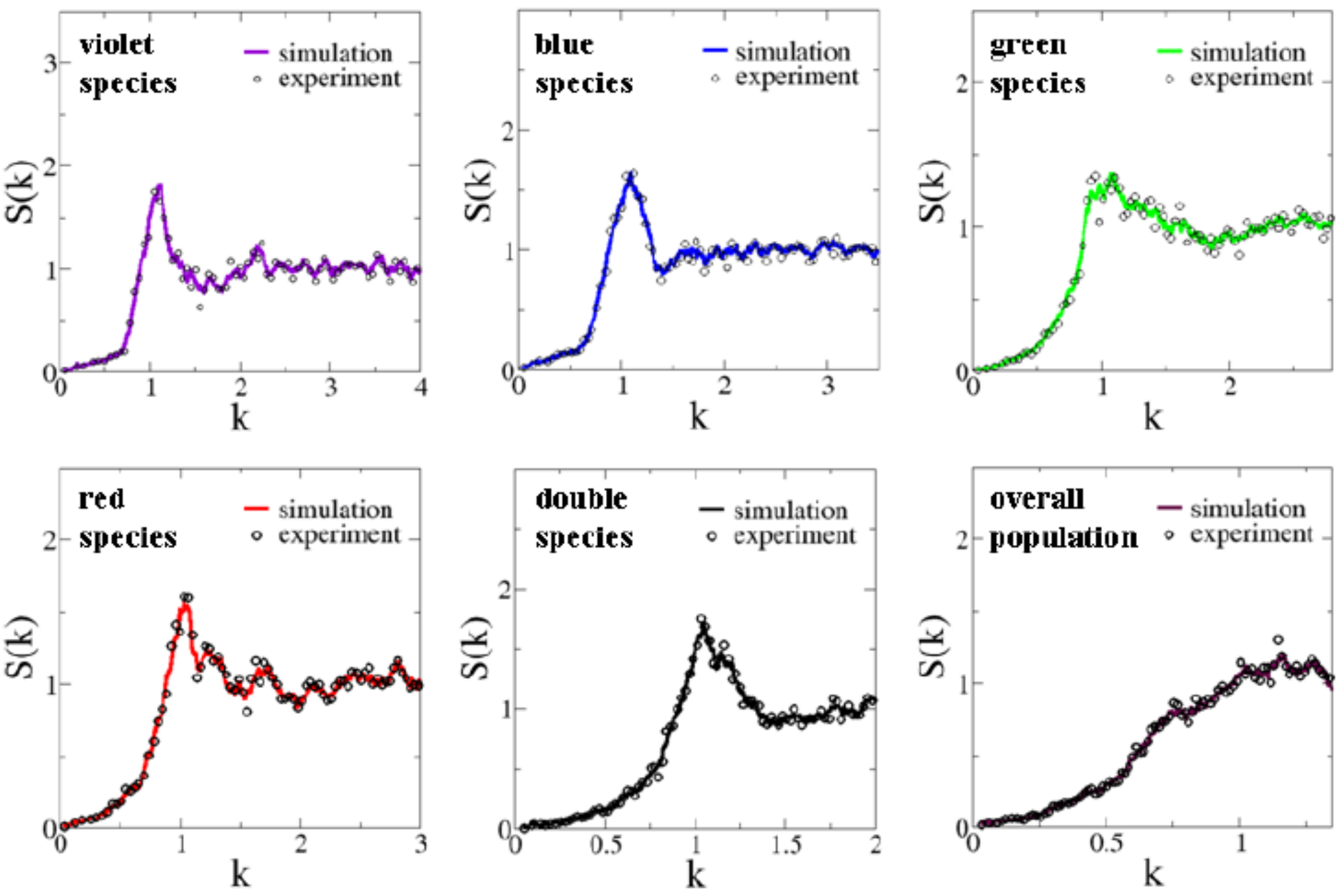} \\
\end{center}
\caption{Comparison of the structure factors $S(k)$ of the
experimentally obtained and simulated point configurations
representing the spatial arrangements of the individual cell types
and total population, as presented in Ref. \cite{Ji14}. Multihyperuniformity is manifested in the fact
that the individual cone types and the total population
are hyperuniform.} 
\label{avian-S}
\end{figure}

\subsection{ Receptor Organization in the Immune System}

A well-adapted immune system should be tuned to the pathogenic environment to reduce
the cost of infections to the organism. Mayer {\it et al.} \cite{Ma15} devised a general theoretical
framework to predict the optimal repertoire of lymphocyte receptors that minimizes the cost of infections contracted
from a given distribution of pathogens.  The model assumes that receptor repertoire is bounded in size
and that receptors are ``cross-reactive" (each antigen binds many receptors;
each receptor binds many antigens) and the cost of an infection
increases with time. An important question that they investigated 
is whether the cross-reactivity generically drives the optimal receptor
distribution to cluster into peaks. They found 
that the optimal repertoire is strongly peaked on a discrete
forest of receptors and showed that the width of these peaks
decreases as numerical precision is increased, suggesting that in a
continuous limit the optimum consists of a weighted sum of
Dirac delta functions, i.e., distinct, discretely spaced receptors in
different amounts in the shape space of the antigens. The peaks tended to repel each other and to organize
into local tiling patterns. Remarkably, the optimal distribution of the peaks
in shape space of the antigens were demonstrated to be disordered and hyperuniform, as determined by the
structure factor at small wavenumbers. In biological terms, hyperuniformity
means that the distribution of receptor peaks provides a much
more uniform coverage of the antigen space than if the peaks were
positioned randomly according to a Poisson distribution. 
As methods are developed to better
characterize pathogenic landscapes and receptor cross-reactivity,
the predictions for the composition of optimal repertoires derived
from this theoretical framework may be directly compared with experiments.

\section{Generalizations of the Hyperuniformity Concept}
\label{GEN}

Given the fundamental as well as practical importance of disordered hyperuniform systems elucidated thus far, it is natural to explore
generalizations of the hyperuniformity notion and its consequences. Recently, the hyperuniformity concept has been generalized 
to treat fluctuations in the interfacial area in multiphase heterogeneous media and surface-area driven
evolving microstructures, random scalar fields, random vector fields, and statistically anisotropic
many-particle systems and two-phase media \cite{To16a}. The relevant mathematical underpinnings are briefly reviewed
and illustrative examples are provided.  In the instances of random vector fields and statistically anisotropic
structures, it is crucial to note that the standard definition of hyperuniformity must be generalized 
so that it accounts for the dependence of the relevant spectral functions on the direction in which the origin in Fourier space
is approached. These recent generalizations of hyperuniformity are expected to provide scientists new avenues
to understand a very broad range of phenomena across a variety of fields through the hyperuniformity
``lens."

\subsection{Fluctuations in the Interfacial Area}
\label{interface}

Interfacial-area fluctuations are of importance whenever interfaces determine the underlying physical
phenomena. This includes flow in porous media \cite{To02a,Sa03}, diffusion and reaction in porous media (including 
nuclear magnetic resonance (NMR) relaxation processes) \cite{To02a,To90e,Wi91,Mit92}, and  surface-energy
driven coarsening phenomena, such as those that occur in
spinodal decomposition and morphogenesis \cite{Ca58,Sw77}.

The global specific surface $s$ (interface area per unit volume) 
is a one-point correlation function that is independent of position for  a statistically homogeneous 
two-phase system in $\mathbb{R}^d$ \cite{To02a}. On the other hand, the specific surface fluctuates at a local level. 
It is straightforward to show that the local specific-surface variance associated with
a $d$-dimensional spherical window of radius $R$  is given by \cite{To16a} 
\begin{eqnarray}
\sigma_{_S}^2(R) = \frac{1}{s^2 v_1(R)} \int_{\mathbb{R}^d} \chi_{_S}(\mathbf{r}) \alpha_2(r; R) d\mathbf{r},
\label{s-var-1}
\end{eqnarray}
where $\chi_{_S}(\mathbf{r})$ is the autocovariance function associated with the interface
indicator function, $r=|\bf r|$, $\alpha_2(r;R)$ is the scaled intersection volume, defined by
(\ref{alpha}) and we have invoked statistical homogeneity. Again, application of Parseval's theorem to (\ref{s-var-1})
yields the alternative  Fourier representation of $\sigma_{_S}^2(R)$ in terms of the
spectral density ${\tilde \chi}_{_S}({\bf k})$:
\begin{eqnarray}
\sigma_{_S}^2(R) = \frac{1}{s^2 v_1(R)(2\pi)^d} \int_{\mathbb{R}^d} {\tilde \chi}_{_S}(\mathbf{k}) {\tilde \alpha}_2(k; R) d\mathbf{k}.
\label{s-var-2}
\end{eqnarray}

A two-phase system is hyperuniform with respect to surface-area fluctuations if the spectral density ${\tilde \chi}_{_S}({\bf k})$
obeys the condition
\begin{eqnarray}
\lim_{|\mathbf{k}|\rightarrow 0}\tilde{\chi}_{_S}(\mathbf{k}) = 0,
\label{hyper-3}
\end{eqnarray}
which implies the sum rule
\begin{equation}
\int_{\mathbb{R}^d} \chi_{_S}({\bf r}) d{\bf r}=0.
\label{sum-3}
\end{equation}
This hyperuniformity property  is equivalent to requiring that the surface-area variance  $\sigma_{S}^2(R)$ for large $R$ goes to zero more rapidly  than $R^{-d}$, which is the same condition as that for the volume-fraction variance  discussed in the Introduction.
Using precisely the same analysis as for point configurations \cite{To03a,Za09,Za11b},
it is simple to show that three different hyperuniform scaling regimes
arise from (\ref{s-var-2}) when the surface-area spectral density
goes to zero with the power-law form ${\tilde \chi}_{_S}({\bf k}) \sim |{\bf k}|^\alpha$:
\begin{eqnarray}  
\sigma^2_{_S}(R) \sim \left\{
\begin{array}{lr}
R^{-(d+1)}, \quad \alpha >1\\
R^{-(d+1)} \ln R, \quad \alpha = 1 \qquad (R \rightarrow \infty).\\
R^{-(d+\alpha)}, \quad 0 < \alpha < 1
\end{array}\right.
\label{sigma-S-asy}
\end{eqnarray}
Note that these scaling forms are exactly the same as those for volume-fraction fluctuations
[cf.  (\ref{sigma-V-asy})].

For a packing of identical $d$-dimensional spheres in $\mathbb{R}^d$, the spectral density ${\tilde \chi}_{_S}({\bf k})$
is related to the structure factor $S({\bf k})$ associated with the sphere centers as follows \cite{To16a}:
\begin{equation}
{\tilde \chi}_{_S}({\bf k})=\rho\, {\tilde m}_s^2(k;a) S({\bf k}),
\label{chi_S-S}
\end{equation}
where 
\begin{equation}
{\tilde m}_s(k;a)=\left(\frac{2\pi a}{k}\right)^{d/2} k \, J_{d/2-1}(ka)
\end{equation}
is the Fourier transform of the interface indicator function $m_s(r;a)$ for a sphere
of radius $a$ \cite{To16a}.
From formula (\ref{chi_S-S}), it immediately follows that if the underlying
point process is hyperuniform and/or stealthy,  then
the spectral density ${\tilde \chi}_{_S}({\bf k})$ inherits the same hyperuniformity property (\ref{hyper-3}) \cite{To16a}.
Corresponding formulas for the autocovariance function and the associated spectral
density for packings of hard spheres with a continuous
or discrete size distribution were previously obtained \cite{To02a,Lu91}
and collected in a more recent article \cite{To16a} to analyze their hyperuniformity properties.

It is of interest to compare volume-fraction and surface-area fluctuations for the same two-phase system.
The left panel of Fig. \ref{specs} shows the two spectral functions, ${\tilde \chi}_{_V}({\bf k})$ and ${\tilde \chi}_{_S}({\bf k})$,
for the step-function $g_2$-invariant packing process in three dimensions at the terminal density $\rho_c=3/(4\pi)$ \cite{To03a};
see also Sec. \ref{invariant} 
The right panel of  Fig.  \ref{specs} depicts the corresponding local variances for the same system, as obtained
from these spectral functions, and relations (\ref{phi-var-2}) and (\ref{s-var-2}).
Notice that the surface-area spectral function exhibits stronger and longer-ranged behavior
compared to the volume-fraction spectral function, indicating that the former
is a more sensitive microstructural descriptor.
While both local variances decay like $R^{-4}$ for large $R$,
the surface-area variance is  larger than its volume-fraction
counterpart at a fixed value of $R$.

\begin{figure}
\begin{center}
\includegraphics[  width=3.in, keepaspectratio,clip=]{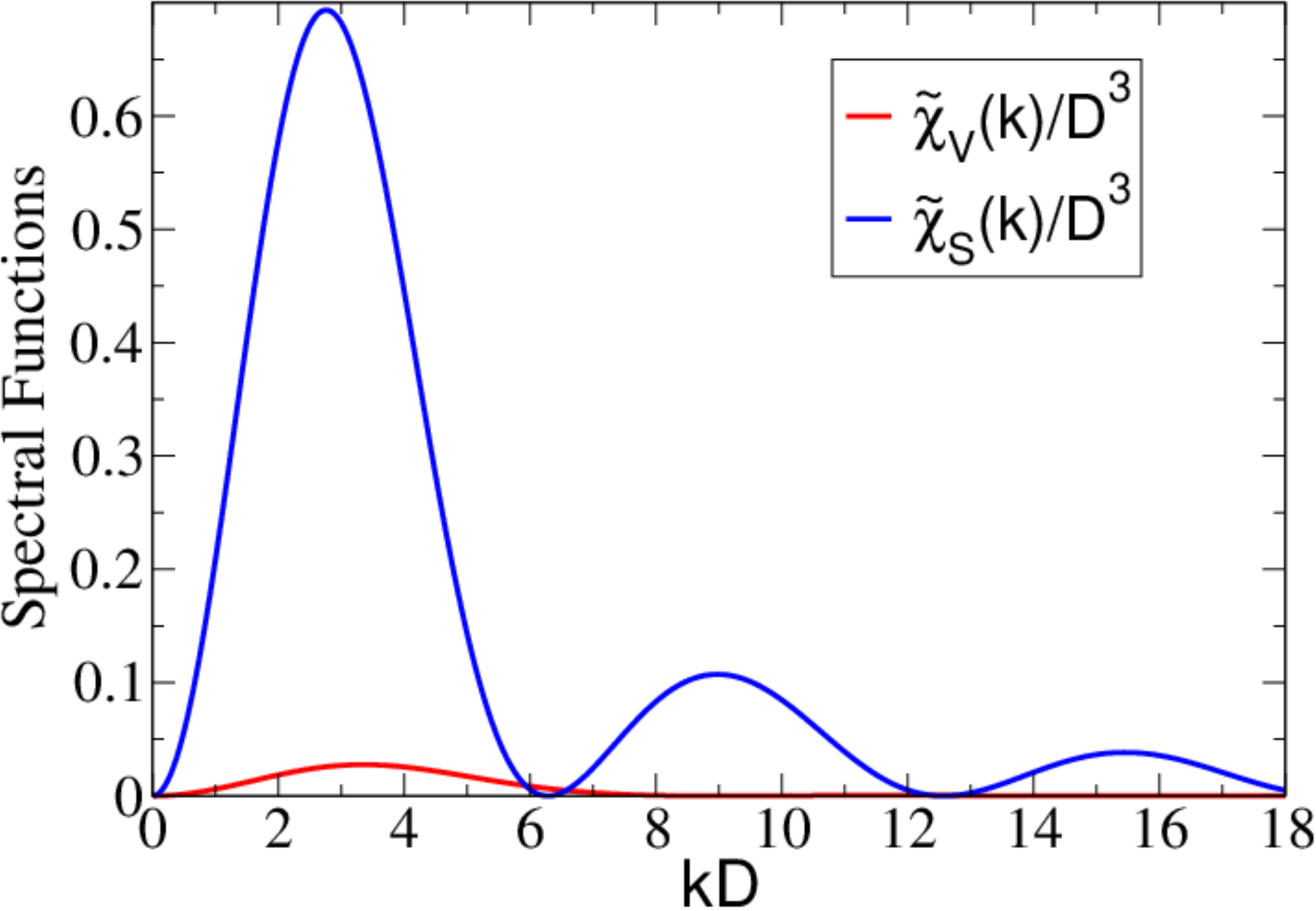} \hspace{0.2in}
\includegraphics[  width=3.in, keepaspectratio,clip=]{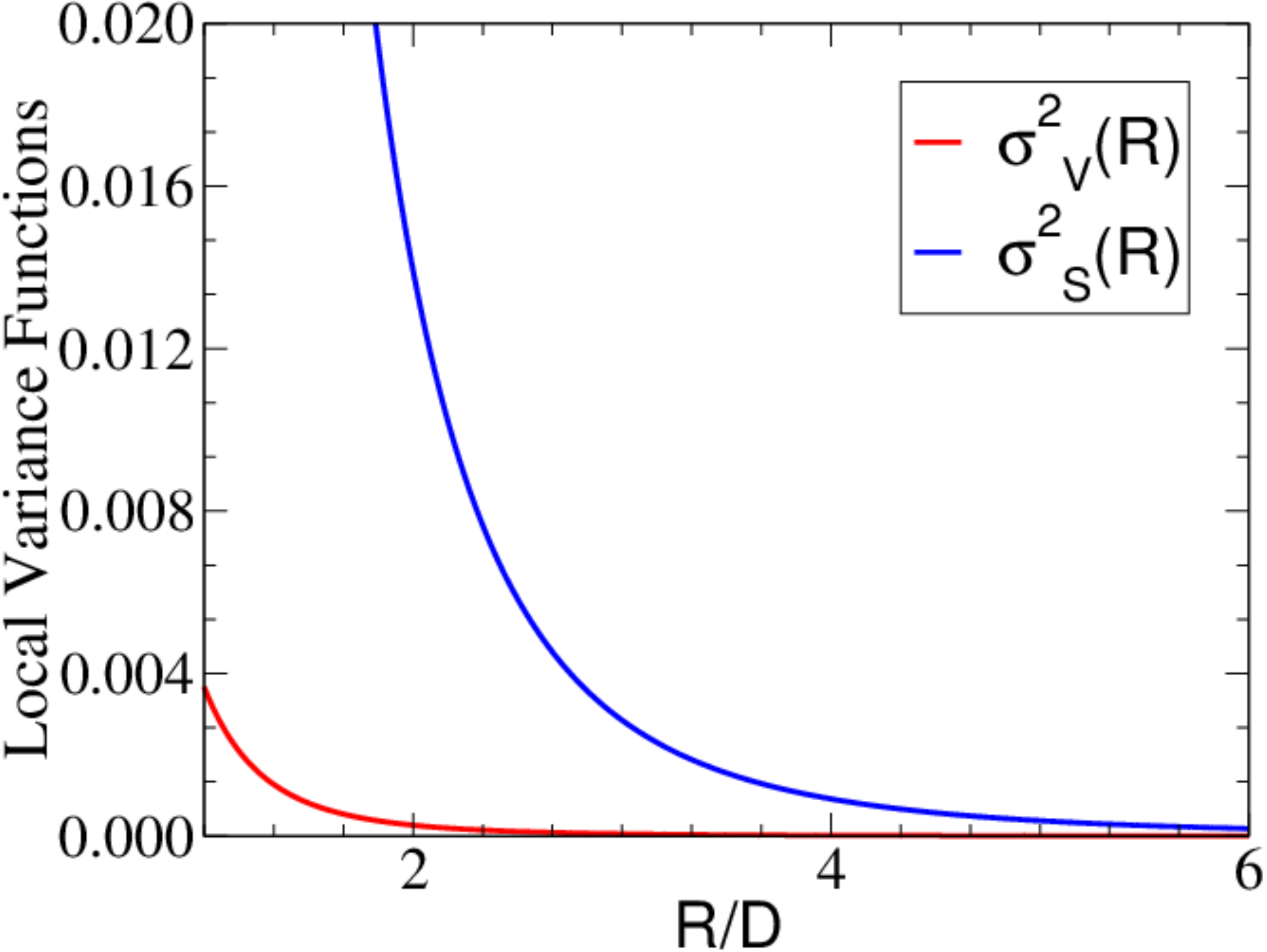}
\caption{Left panel: Comparison of the two hyperuniform spectral functions ${\tilde \chi}_{_V}(k)$ (lower curve) and ${\tilde \chi}_{_S}(k)$
versus wavenumber $k$ for a packing of identical spheres corresponding to the step-function $g_2$-invariant process in three dimensions at the
hyperuniform terminal density $\rho_c=3/(4\pi)$, as obtained in Ref. \cite{To03a}. Here $D$ is the diameter of a hard sphere.
Right panel: Corresponding volume-fraction variance $\sigma^2_{_V}(R)$ (lower curve) and
surface-area  variance  $\sigma^2_{_S}(R)$ versus window sphere radius $R$.
}
\label{specs}
\end{center}
\end{figure}

\subsection{Random Scalar Fields}

The hyperuniformity concept has been recently generalized to characterize fluctuations associated with 
random scalar fields in $\mathbb{R}^d$.  Such fields can arise in a variety of physical contexts, including concentration and temperature
fields in heterogeneous and porous media \cite{To02a,Sa03} as well as in turbulent  flows \cite{Ba59,Mo75},
laser speckle patterns \cite{Pi88,Wi08,Dog15,Dib16}, and temperature fluctuations associated
with the cosmic microwave background \cite{Pe93,Kom03}. Other example include spatial patterns that arise
in biological and chemical systems that have been theoretically described by, for example,
Cahn-Hilliard \cite{Ca58} and Swift-Hohenberg equations \cite{Sw77}.

Consider a statistically homogeneous random scalar field $F(\bf x)$ in $\mathbb{R}^d$ that is real-valued with an autocovariance function
\begin{equation}
\psi({\bf r})= \Bigg\langle    \Big(F({\bf x}_1)- \langle F({\bf x}_1)\rangle\Big) \, \Big(F({\bf x}_2) -  \langle F({\bf x}_2)\rangle\Big) \,\Bigg\rangle,
\label{spec-field}
\end{equation}
where we have invoked the statistical homogeneity of the field, since ${\bf r}={\bf x}_2 -{\bf x}_1$, which is a $d$-dimensional vector. We assume that
the associated spectral density ${\tilde \psi}({\bf k})$ (Fourier
transform of the autocovariance) exists. The hyperuniformity condition
is simply that the nonnegative spectral density obeys the
small-wavenumber condition:
\begin{equation}
\lim_{|{\bf k}| \rightarrow {\bf 0}} {\tilde \psi}({\bf k})=0,
\label{hyp-field}
\end{equation}
which implies the sum rule
\begin{equation}
\int_{\mathbb{R}^d} \psi({\bf r}) d{\bf r}=0.
\end{equation}
The local variance associated with fluctuations in the field within a spherical window
of radius $R$, denoted by $\sigma_{_F}^2(R)$,
is related to the autocovariance function or spectral function  in the usual way:
\begin{eqnarray}
\sigma^2_{_F}(R)&=& \frac{1}{v_1(R)} \int_{\mathbb{R}^d} \psi(\mathbf{r}) \alpha_2(r; R) d\mathbf{r},\nonumber \\
&=&\frac{1}{ v_1(R)(2\pi)^d} \int_{\mathbb{R}^d} {\tilde \psi}({\bf k})
{\tilde \alpha}_2(k;R) d{\bf k}.
\label{local-scalar}
\end{eqnarray}

It is interesting to note that one can construct models of two-phase or multiphase heterogeneous systems
via level cuts of random scalar fields by choosing appropriate threshold levels to define the
phases \cite{Berk87,Be91,Te91,Cr91,Bl93,Ro95}.
This approach is particularly useful in modeling
{\it bicontinuous} media (two-phase media in which each phase percolates),
such as microemulsions \cite{Berk87}, carbonate rocks \cite{Cr91},
Vycor glass \cite{Cr91}, amorphous alloys, \cite{Ro95} and aerogels \cite{Ro97}.
To derive a hyperuniform two-phase
medium from a thresholded random field $F({\bf r})$, the field must possess the special
correlations required to yield an autocovariance function $\chi_{_V}({\bf r})$ that satisfies the
 rule (\ref{sum-2}) \cite{To16a}.

The study of hyperuniformity in random scalar fields is in its very early stages.
Some recent results are worth noting. For example, it was proved that a class of random scalar fields derived from 
underlying hyperuniform point configurations are themselves hyperuniform \cite{To16a}. Ma and Torquato \cite{Ma17} formulated
 methods to explicitly construct hyperuniform scalar fields and, by thresholding them, to ascertain whether the resulting two-phase random media were 
hyperuniform \cite{Ma17}. They specifically considered spatial patterns generated from Gaussian random fields, the Cahn-Hilliard equation for spinodal decomposition, and the Swift-Hohenberg equation for pattern formation. Gaussian random fields that have been used to model a variety of systems, including the microwave background radiation \cite{Pe93,Kom03}, heterogeneous materials \cite{To02a} and laser speckle fields \cite{Pi88,Wi08,Dog15,Dib16}.
The time-evolving patterns that arise from spinodal decomposition via the Cahn-Hilliard description \cite{Ca58}, which are ubiquitous in chemistry 
and biological systems. The Swift-Hohenberg equations that describe thermal convection in hydrodynamics \cite{Sw77} as well as a general model of emergent pattern formation \cite{Cr09}.

Ma and Torquato demonstrated that it is straightforward to construct Gaussian random scalar fields that are hyperuniform. They also numerically studied the time evolution of spinodal decomposition patterns modeled by the Cahn-Hilliard equation and showed that these patterns are hyperuniform in the scaling regime.
Figure  \ref{spinodal} shows a snapshot of such a hyperuniform two-dimensional realization as well as the corresponding spectral density
${\tilde \psi}(k)$. Consistent with hyperuniform behavior, the corresponding local variance $\sigma^2_{_F}(a)$ was shown to decay like $1/a^3$ for large $a$,
where $a$ is the side length of a square window. Moreover, they also characterized labyrinth-like patterns generated by the Swift-Hohenberg equation, and found they are effectively hyperuniform. A toy ``polycrystal" model was introduced to explain the features of the local field fluctuations and spectral densities of these labyrinth-like patterns. Finally, they 
showed that thresholding (level-cutting) a hyperuniform scalar field to produce a two-phase random medium can easily destroy the hyperuniformity of the progenitor scalar field. In particular,  a thresholded disordered Gaussian random field is generally not hyperuniform. Several guidelines were provided to achieve effectively hyperuniform two-phase media obtained from thresholded scalar fields. 

In summary,  hyperuniformity can emerge in scalar fields,
and the quantification of long-wavelength  scalar field fluctuations provide useful ways to characterize the degree of global order of scalar fields and hence enables the classification of wide class of spatial patterns.  These theoretical results are expected to guide experimentalists to synthesize new classes of hyperuniform materials with novel physical properties via coarsening processes and using state-of-the-art techniques, such as stereolithography and 3D printing.

\begin{figure}[H]
\centering
\includegraphics[width=2.8in]{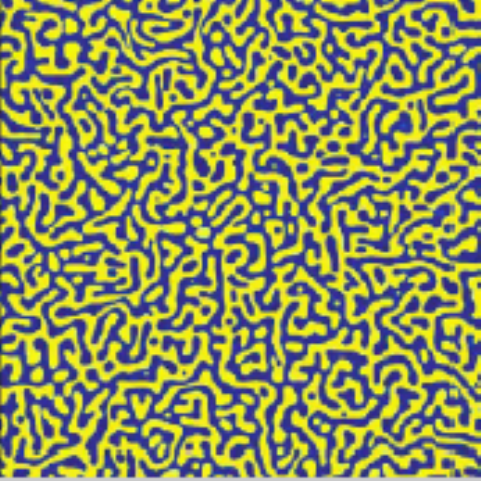}\hspace{0.2in}\includegraphics[width=3.4in]{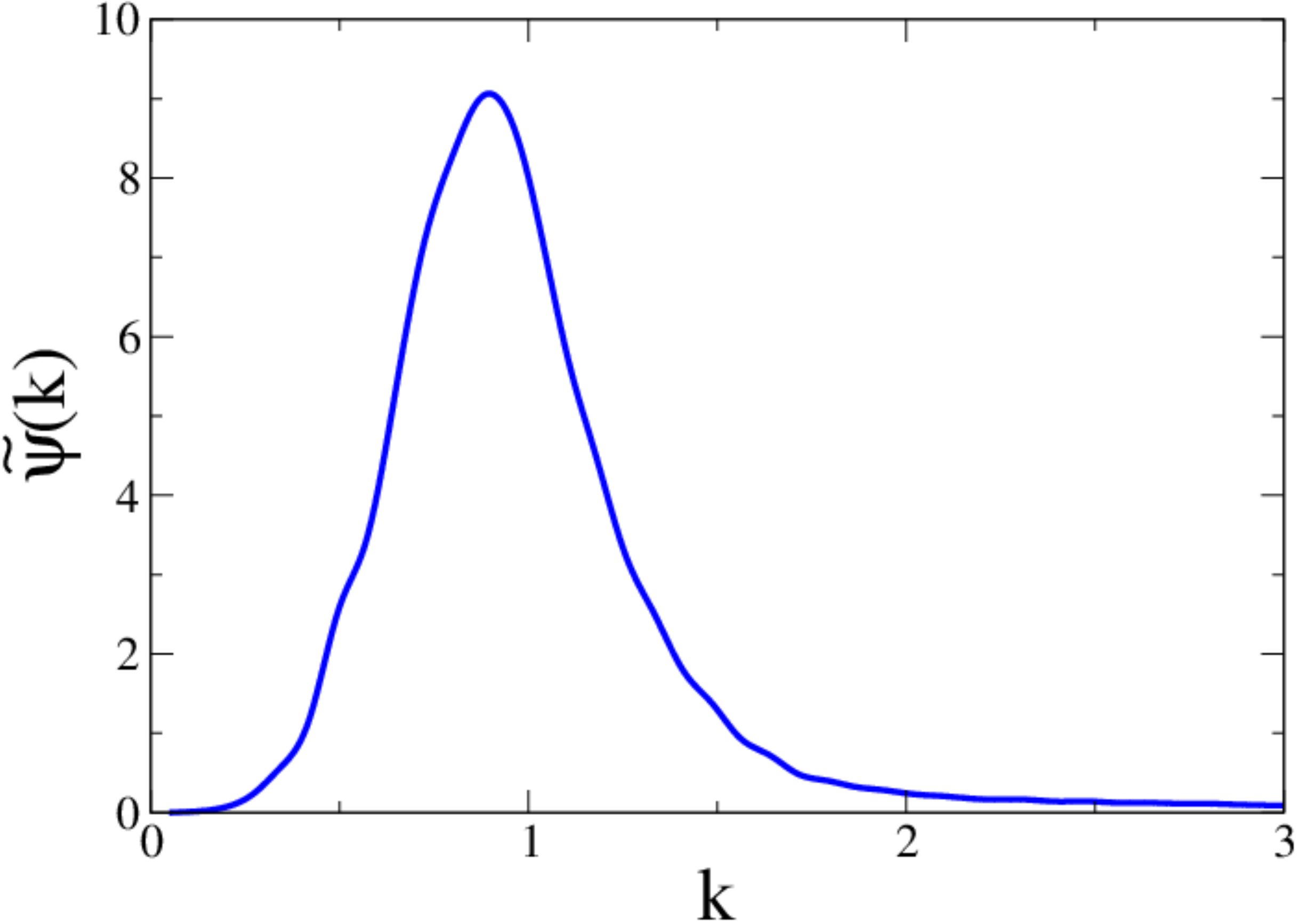}
\caption{Snapshot of a two-dimensional system undergoing a coarsening spinodal-decomposition
process (left panel) as obtained from a  numerical simulation of the Cahn-Hilliard equation reported in Ref. \cite{Ma17}. 
At this instant of time in the scaling regime, the system is hyperuniform, as evidenced by a scaled spectral density
function ${\tilde \psi}(k)$, which is seen to vanish as the dimensionless wavenumber tends to zero (right panel). The reader
is referred to Ref. \cite{Ma17} for the manner in which  ${\tilde \psi}(k)$ and $k$ are scaled.}
\label{spinodal}
\end{figure}

It should be noted that when simulating random fields on the computer or when
extracting them from experimentally obtained images,
one must inevitably treat discrete or digitized  renditions of the fields.
The ``pixels" or "voxels" (smallest components of the digitized systems
in two and three dimensions, respectively) take on
gray-scale intensities that span the intensity range
associated with the continuous field. Thus, the discrete versions
of relations (\ref{spec-field}) and (\ref{local-scalar}) are to be applied
in such instances; see, for example, Ref. \cite{Bl93}.

\subsection{Random Vector Fields}

The hyperuniformity concept has been recently generalized to characterize fluctuations
associated with random vector fields \cite{To16a}. This was done in the context of  divergence-free random vector fields
for simplicity,  but the basic ideas
apply to more general vector fields. Excellent physical examples within this class of vectors fields
occur in heterogeneous  media, including
divergence-free heat, current or mass flux fields, divergence-free electric displacement
fields associated with dielectrics, divergence-free magnetic induction fields, and divergence-free low-Reynolds-number velocity
fields \cite{To02a,Sa03}. Incompressible  turbulent flow fields provide yet
other very well-known set of examples \cite{Ba59,Mo75}.

Consider a statistically homogeneous divergence-free (solenoidal) random vector field ${\bf u}({\bf x})$ in $\mathbb{R}^d$
that is real-valued with zero mean.
A key quantity is the autocovariance function $\Psi_{ij}({\bf r})$ ($i,j=1,2,\ldots,d$)
associated with the vector field ${\bf u}({\bf x})$, which is a  second-rank
tensor field defined by
\begin{equation}
\Psi_{ij}({\bf r})= \langle    u_i({\bf x}) u_j({\bf x}+{\bf r}) \rangle,
\label{auto}
\end{equation}
where we have invoked the statistical homogeneity of the field.
Let ${\tilde \Psi}_{ij}({\bf k})$ denote the spectral density tensor, i.e., the Fourier transform of
the autocovariance tensor (\ref{auto}). The real-valued spectral density tensor
is positive semi-definite, i.e., for an arbitrary real vector $\bf a$,
\begin{equation}
a_i {\tilde \Psi}_{ij}({\bf k}) a_j \ge 0, \qquad \mbox{for all} \; {\bf k},
\end{equation}
where Einstein indicial summation notation is implied.  

From the theory of turbulence of an incompressible fluid \cite{Ba59,Mo75}, it is well known that if an arbitrary
divergence-free vector field ${\bf u}({\bf x})$ is also isotropic, then the spectral
density tensor must take the following general form:
\begin{equation}
 {\tilde \Psi}_{ij}({\bf k})=\left(\delta_{ij} -\frac{k_ik_j}{k^2}\right) {\tilde \psi}(k),
\label{spec-tensor}
\end{equation}
where $\delta_{ij}$ is the Kronecker delta or identity tensor,
and ${\tilde \psi}(k)$ is a nonnegative scalar radial function of the wavenumber $k=|\bf k|$.
A random vector field is isotropic if all of its associated $n$-point correlation
functions are independent of translations, rotations and reflections of the
coordinates. Note that the trace of ${\tilde \Psi}_{ij}({\bf k})$ is trivially related
to ${\tilde \psi(k)}$, i.e.,
\begin{equation}
{\tilde \Psi}_{ii}({\bf k})=(d-1) {\tilde \psi}(k),
\end{equation}
and so we see that
\begin{equation}
{\tilde \Psi}_{ii}({\bf k}={\bf 0})=(d-1) {\tilde \psi}(k=0)=\int_{\mathbb{R}^d} \Psi_{ii}({\bf r}) d{\bf r}
\end{equation}
and
\begin{equation}
\Psi_{ii}({\bf r}={\bf 0})=\frac{(d-1)}{(2\pi)^d} \int_{\mathbb{R}^d} {\tilde \psi}(k) d{\bf k}.
\end{equation}

Now if the radial function ${\tilde \psi}(k)$ is continuous but positive at $k=0$ (not hyperuniform), it immediately follows
 from
the form (\ref{spec-tensor}) that the spectral tensor can only vanish (i.e., be hyperuniform)
in {\it certain directions}. For example, the component ${\tilde \Psi}_{11}({\bf k})$ is
zero for $k=k_1$ (all wave vectors along the $k_1$-axis) and the component ${\tilde \Psi}_{12}({\bf k})$ is
zero whenever $k_1=0$ or $k_2=0$. The fact that the value of  ${\tilde \Psi}_{11}({\bf k})$
depends on the direction in which the origin is approached means that it is nonanalytic at $\bf k=0$.
On the other hand, if ${\tilde \psi}(k)$ is hyperuniform and continuous at $k=0$,
then each component of ${\tilde \Psi}_{ij}({\bf k})$ will inherit the radial hyperuniformity
of  ${\tilde \psi}(k)$, and hence is independent of the direction in which
the origin is approached. For example, consider the situation in which ${\tilde \psi}(k)$
admits the following small-wavenumber expansion
\begin{equation}
{\tilde \psi}(k) = a_1 |{\bf k}|^\alpha + {o}(|{\bf k}|^\alpha),
\label{exp}
\end{equation}
where $a_1$ and $\alpha$ are positive constants and $o$ signifies higher order terms. Note that whenever $\alpha$ is a noninteger
or odd integer, ${\tilde \psi}(k)$ is a nonanalytic function at the origin due to a derivative discontinuity.
(An analytic radial function would admit an expansion in even powers of the wavenumber only.)
For any $\alpha >0$, substitution of  (\ref{exp}) in $(\ref{spec-tensor})$ reveals that the
spectral tensor is radially hyperuniform near ${\bf k=0}$ such that  it vanishes as $|{\bf k}|^\alpha$.

Thus we see that one needs an even more general hyperuniformity concept
in the case of a spectral tensor, namely, one in which hyperuniformity
depends on the direction in which the origin is approached in Fourier
space. Let ${\bf k}_Q$ represent a  $d$-dimensional unit vector emanating
from the origin $\bf k=0$. The field is said to be  hyperuniform  for a particular component $i=I$ and $j=J$ of the spectral
tensor of a vector field (isotropic or not) in the direction ${\bf k}_Q$ if
\begin{equation}
\lim_{t \rightarrow {0}} {\tilde \Psi}_{IJ}(t{\bf k}_Q)={0},
\label{HYPER}
\end{equation}
where $t$ is a scalar parameter.
Note that there are many different unit vectors (directions) for a particular spectral tensor that can satisfy
this condition, whether this set is countable, or it is uncountable because
these unit vectors can occur in a continuous range of directions. Moreover, if the condition (\ref{HYPER})
applies independent of the direction of the unit vector,  then
it reduces to the standard spectral definition of hyperuniformity.

To illustrate the hyperuniformity concept in the context of a divergence-free isotropic
vector field, consider the
following hyperuniform radial function:
\begin{equation}
{\tilde \psi}(k)= c(d) (ka) \exp(-(ka)^2),
\label{radial}
\end{equation}
where
\begin{equation}
c(d)= \frac{\Gamma(d/2)\, a^d}{2^d \pi^{d/2} \Gamma((d+1)/2)},
\end{equation}
where $a$ is a characteristic length scale.
In any dimension, this nonnegative spectral function corresponds to a realizable scalar field  
associated with an autocovariance function such that $\psi(r=0)=1$.
For visual purposes, we examine the two-dimensional outcome when (\ref{radial}) is substituted into
the spectral tensor (\ref{spec-tensor}).  Figure \ref{tensor} shows three components of this
symmetric tensor and the radial function ${\tilde \psi}(k)$. The hyperuniformity
property in a compact region around the origin for all components is readily visible.

\begin{figure}[H]
\begin{center}
\includegraphics*[  width=2.8in,clip=keepaspectratio]{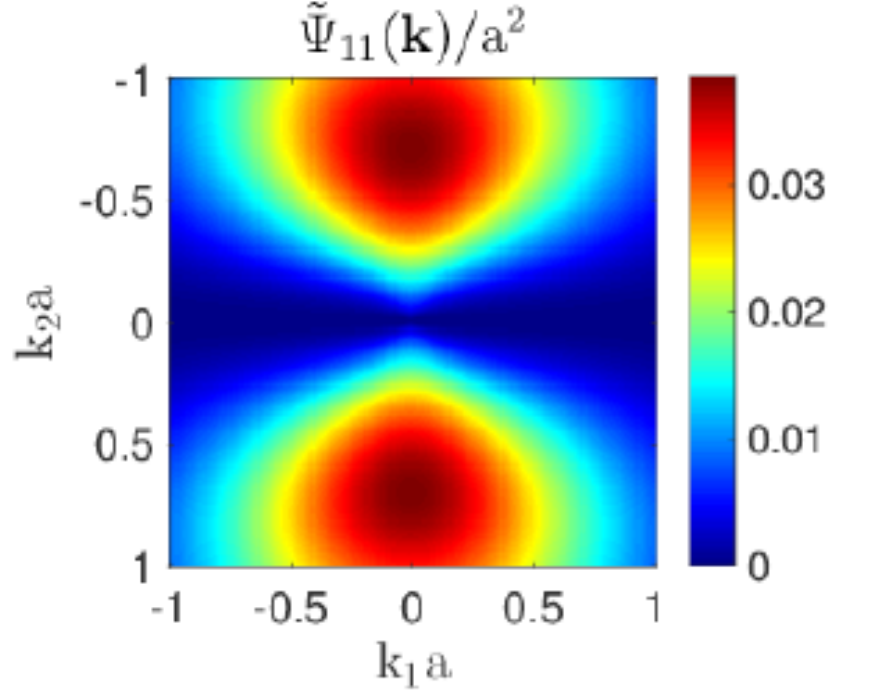}
\includegraphics*[  width=2.8in,clip=keepaspectratio]{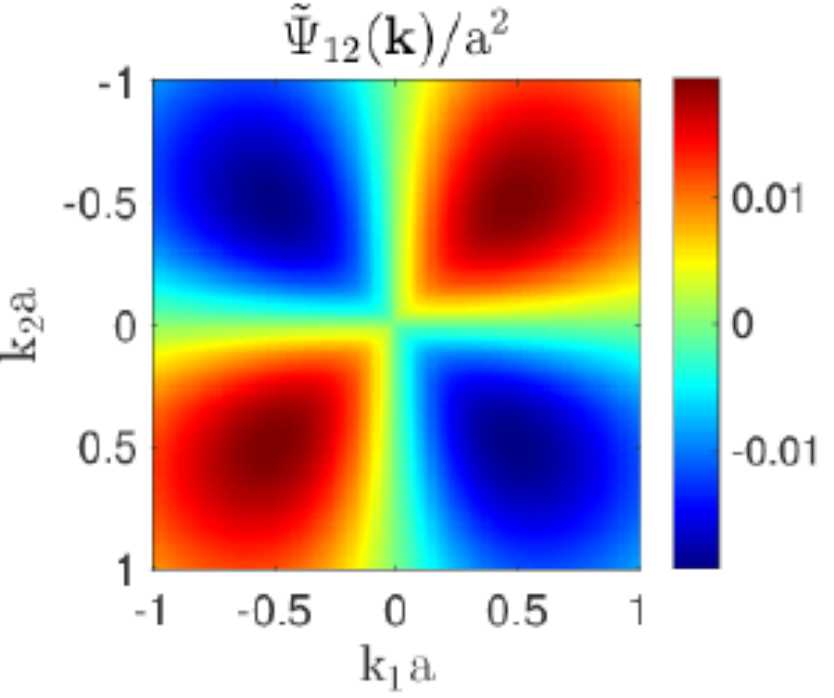}
\includegraphics*[  width=2.8in,clip=keepaspectratio]{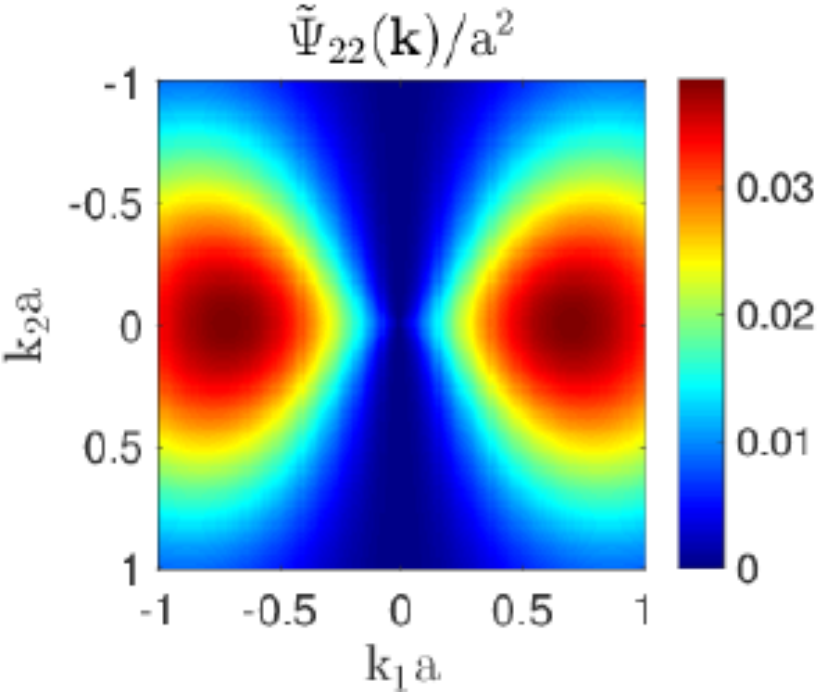}\vspace{0.2in}
\includegraphics*[  width=2.5in,clip=keepaspectratio]{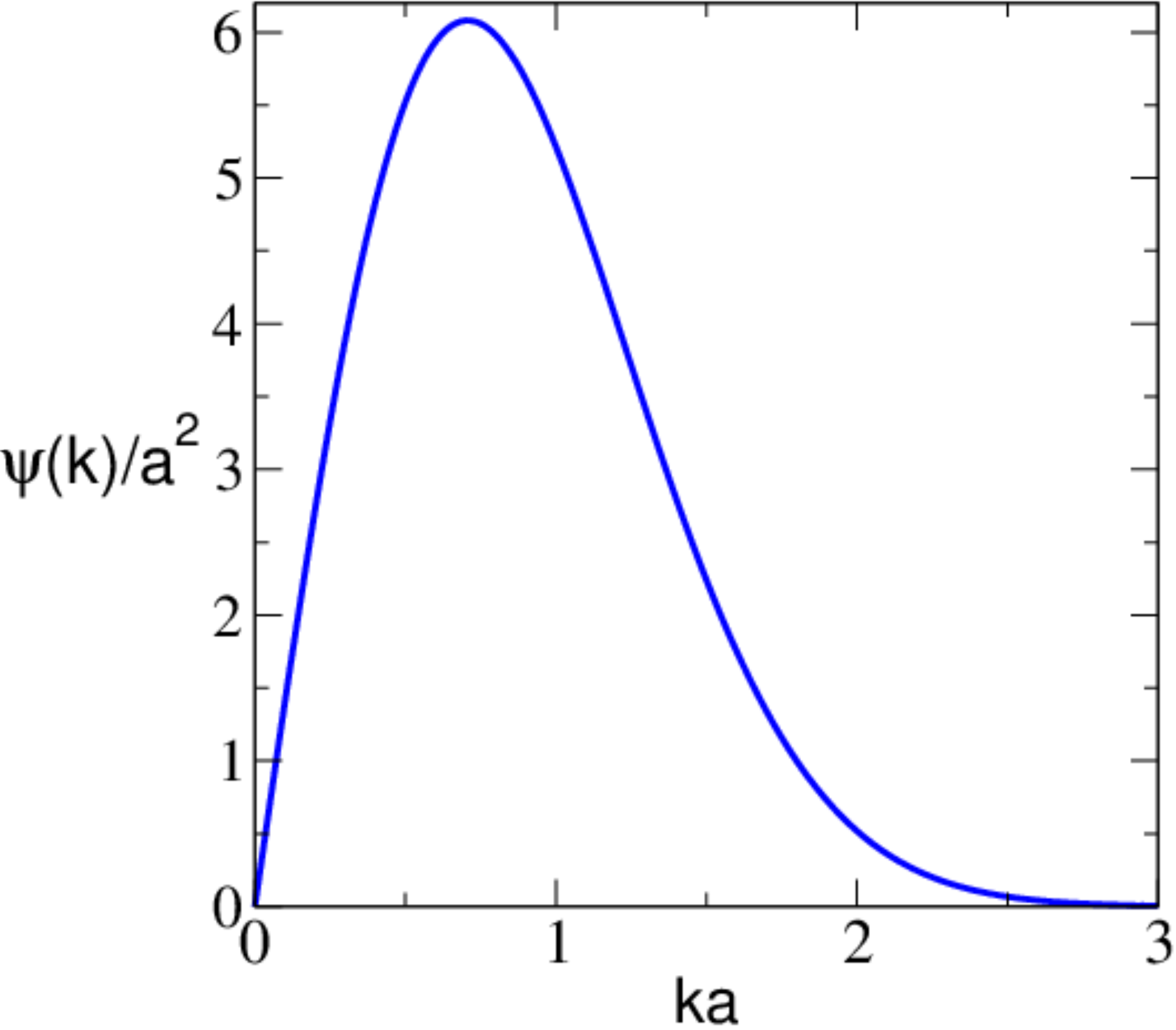}
\caption{Spectral patterns for the tensor components
of a divergence-free isotropic vector field in $\mathbb{R}^2$ generated
from the radial function $\psi(k)$, defined by (\ref{radial}) with $d=2$, which is
depicted on the right side of  the bottom panel. Unlike the nonnegative 11- and 22-components,
the 12-component can be both positive and negative, and so its color map
indicating zero intensity (darkest shade) is different from those for the diagonal components. These figures are taken from Ref. \cite{To16a}.}
\label{tensor}
\end{center}
\end{figure}

\vspace{-0.2in}

\subsection{Statistical Anisotropic Structures}
\label{anisotropy}

Other classes of disordered systems in which ``directional" hyperuniformity is relevant
include many-particle and heterogeneous systems that are statistically
anisotropic, but  statistically homogeneous. In such instances,
the spectral function conditions (\ref{hyper}), (\ref{hyper-2}) and (\ref{hyper-3}) should be replaced
with the following ones, respectively:
\begin{equation}
\lim_{t \rightarrow 0} S(t{\bf k}_Q) = 0,
\label{Hyper}
\end{equation}
\begin{eqnarray}
\lim_{t \rightarrow 0}\tilde{\chi}_{_V}(t\mathbf{k}_Q) = 0,
\label{Hyper-2}
\end{eqnarray}
\begin{eqnarray}
\lim_{t \rightarrow 0}\tilde{\chi}_{_S}(t\mathbf{k}_Q) = 0,
\label{Hyper-3}
\end{eqnarray}
where the unit vector ${\bf k}_Q$ is defined immediately above relation (\ref{HYPER}).

To illustrate the implications of such generalizations, the collective-coordinate optimization technique
outlined in Sec. \ref{STEALTHY} was employed to create a many-particle system that is
hyperuniform in only certain directions in Fourier space \cite{To16a}.
Specifically, the structure factor was constrained
to be exactly zero within a lemniscate region around
the origin ${\bf k = 0}$, implying that this entire region is stealthy
(scattering is completely suppressed) but is
hyperuniform in only certain directions; see the top panel of
Fig. \ref{lemniscate}. Indeed, this anisotropic structure factor is attainable
by like-linear ``filamentary" chains of particles that run more or less horizontally 
(bottom panel of Fig. \ref{lemniscate}), which corresponds to the ground state
of the associated bounded long-ranged anisotropic (directional) pair potential.
These ground states will possess exotic physical properties, as described in Sec. \ref{properties}.

\begin{figure}[H]
\begin{center}

$\begin{array}{c}\\
\includegraphics[width=0.50\textwidth]{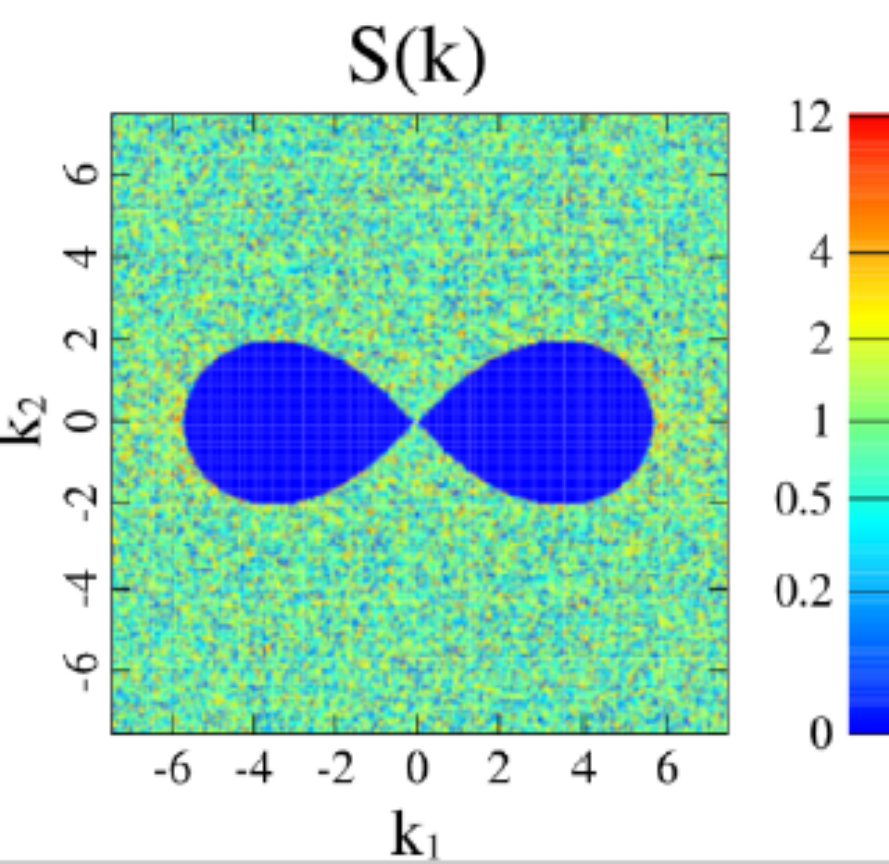} \\
\includegraphics[width=0.35\textwidth]{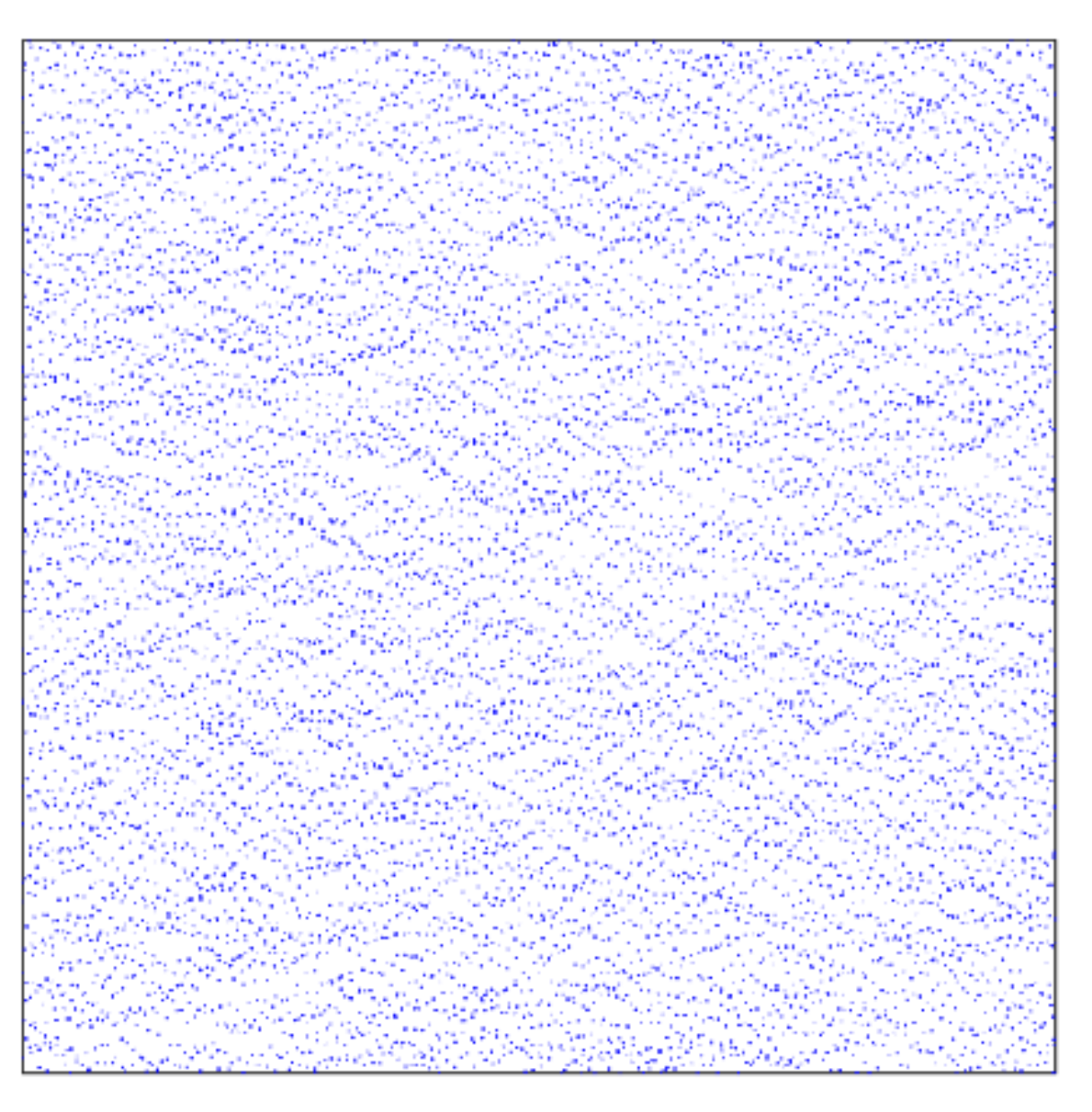} \\

\end{array}$
\end{center}
\caption{ Top panel: A targeted scattering pattern showing a lemniscate region around
the origin in which the scattering intensity is exactly zero (darkest shade). This ``stealthy" pattern clearly
shows
that hyperuniformity depends on the direction in which the origin $\bf k=0$
is approached. Bottom panel: A statistically anisotropic ground-state configuration of 10,000 particles
that corresponds
to the unusual scattering pattern shown in the top panel, which is generated using the collective-coordinate
optimization procedure \cite{Uc04b,Zh15a,Zh15b} in a square simulation box under periodic boundary conditions.
These figures are taken from Ref. \cite{To16a}.}
\label{lemniscate}

\end{figure}

\vspace{-0.15in}

For structurally anisotropic systems, one can choose an appropriately shaped nonspherical
window occupying region $\Omega$ with fixed orientation that maximizes sensitivity of fluctuations 
in certain directions \cite{To16a}. Figure \ref{nematic} schematically depicts
a  statistically homogeneous, anisotropic nematic liquid crystal configuration
of particles and an appropriate window shape and orientational distribution to distinguish ``directional" fluctuations associated
with either the centroidal positions of the particles, volume fraction,  or interfacial
area of the particles. It is clear that window sampling in the direction indicated
in Fig. \ref{nematic} will produce fluctuations that are different from those obtained by sampling in the
orthogonal direction.

\begin{figure}[H]
\begin{center}
\includegraphics[  width=2.5in, keepaspectratio,clip=]{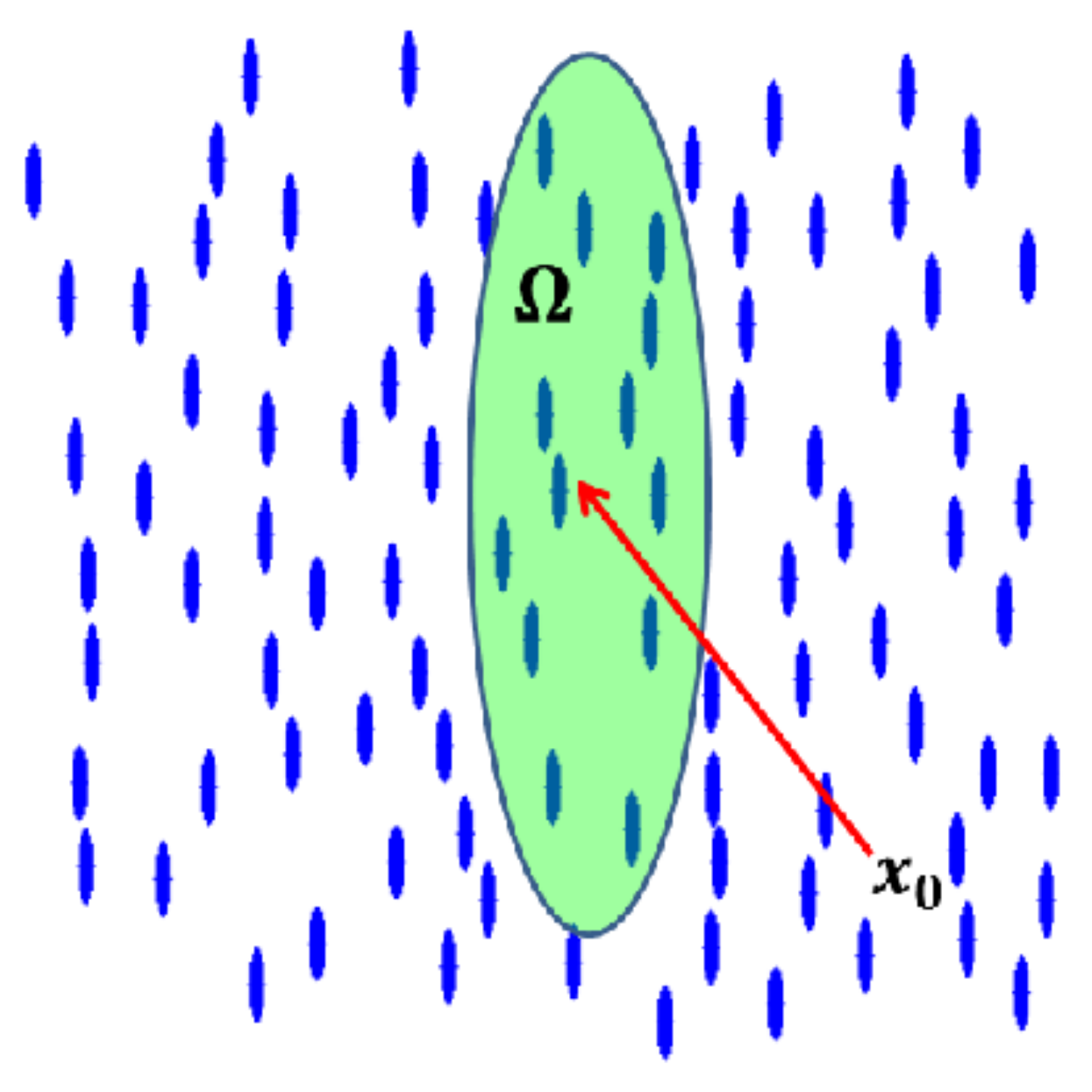}
\caption{ Schematic illustration of a statistically homogeneous and anisotropic nematic liquid crystal configuration
taken from Ref. \cite{To16a}.
An appropriately shaped window that occupies region $\Omega$ is also shown. Here $\bf x_0$ denotes
both the centroidal position and orientation of the window, the latter of which
is chosen generally from a prescribed probability distribution that depends on the specific
structure of interest. }
\label{nematic}
\end{center}
\end{figure}

Many-particle systems that respond to external fields are frequently described
by anisotropic structure factors and thus represent a class of systems
where directional hyperuniformity can potentially arise. Huang, Wang and Holm \cite{Hu05} have carried out molecular dynamics
simulations of colloidal ferrofluids subjected to external fields that  capture the salient
structural features observed in corresponding experimental systems as measured
by the structure factor.  In these systems, structural anisotropy arises
due to the formation of particle chains that tend to align in the direction
of the applied magnetic field; see Fig. \ref{ferro} 
taken from Ref. \cite{Hu05}. It is apparent that the system is effectively hyperuniform in the vertical direction.

\begin{figure}[H]
\centerline{\includegraphics[  width=2.2in, keepaspectratio,clip=]{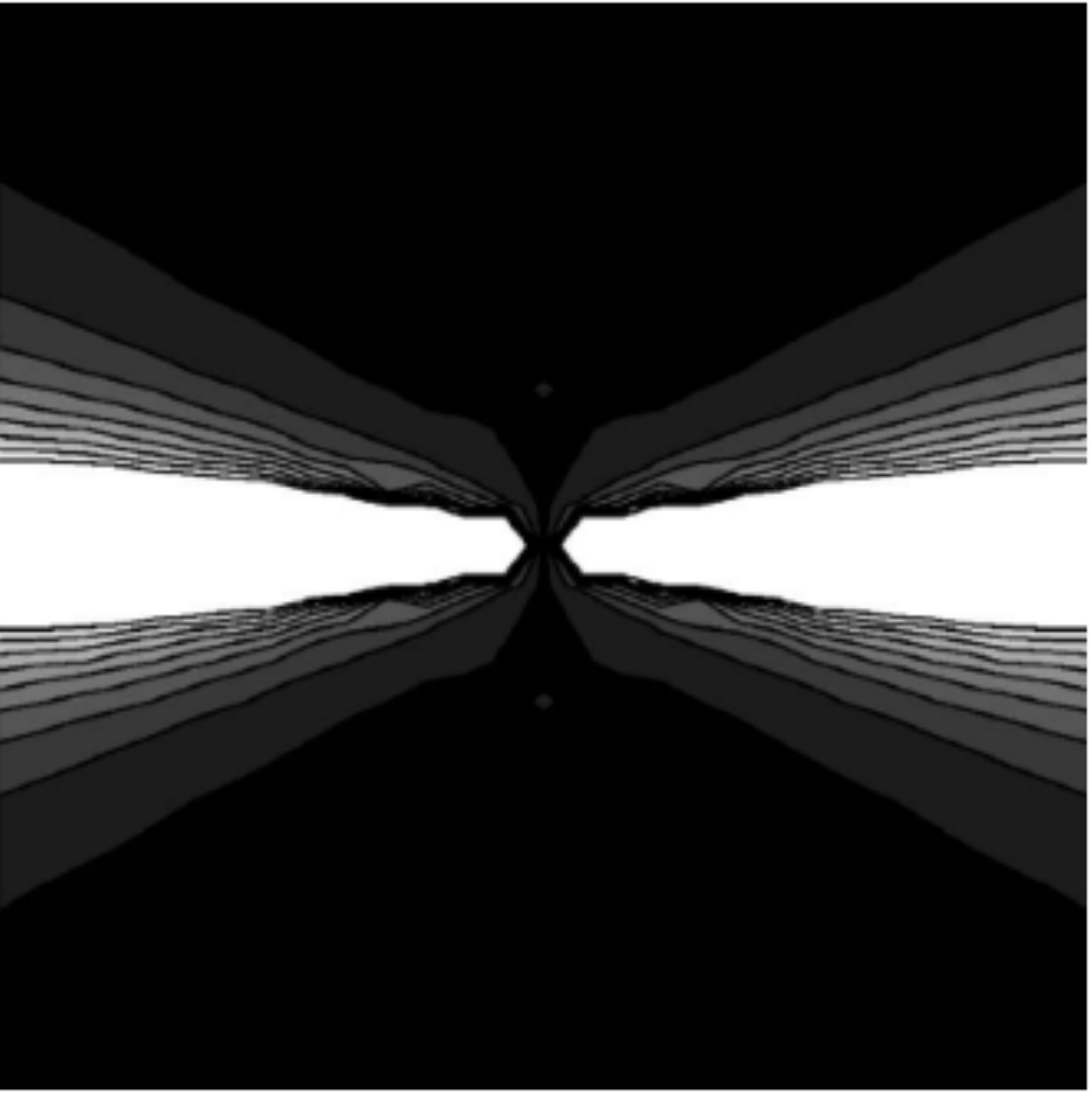}}
\caption{Anisotropic structure factor of a colloidal ferrofluid in the plane in which the particle chains
align, as obtained from  Fig. 6 of Ref. \cite{Hu05}. Dark and light regions indicate low and high intensities,
respectively. Observe that depending on the direction in which the origin
is approached, the structure factor can exhibit effective hyperuniformity.}
\label{ferro}
\end{figure}

Long-range hydrodynamic correlations among settling particles in a viscous liquid lead to complex 
many-body dynamics, exhibiting very large density fluctuations and large-scale dynamic structures when the particles
are spheres or other centrally symmetric objects \cite{Ca85,Ham88,Ni95,Se97}.  Goldfriend, Diamant and  Witten studied \cite{Go17}
the over damped sedimentation of non-Brownian objects of irregular shape using fluctuating hydrodynamics. They showed
that the anisotropic response of the objects to flow, caused by their tendency to align with gravity, directly suppresses 
density and velocity fluctuations in certain directions with varying intensities. This allows the suspension to avoid the anomalous fluctuations predicted for suspensions of spheres and other centrally symmetric
particles. The suppression of density fluctuations in suspensions with irregularly-shaped particles
leads to  structures characterized by directional hyperuniformity.

\subsection{Spin Systems}

We note in passing that the hyperuniformity concept has recently been generalized to spin systems,
including a capability to  construct disordered stealthy hyperuniform spin configurations as classical ground states  \cite{Ch16a}.
 The discovered exotic disordered spin ground
states, which are distinctly different from spin glasses \cite{Me87},
are the spin analogs of disordered stealthy hyperuniform
many-particle ground states \cite{Uc04b,Ba08,To15} that have been shown to
be endowed with novel bandgap and wave characteristics \cite{Fl09b,Man13b,Yu15,De16,Le16,Fr16,Gk17,Wu17,Ch18,Kl18}.
The implications and significance of the existence of such disordered
spin ground states warrant further study, including
whether their bulk physical properties and excited states, like their many-particle
system counterparts, are singularly remarkable, and can be experimentally realized.

\section{Novel Bulk Physical Properties of Disordered Hyperuniform Materials}
\label{properties}

We noted in the Introduction that disordered hyperuniform states of matter 
lie between a crystal and liquid in that  they are like perfect crystals in the way they suppress large-scale density fluctuations and yet are like liquids or glasses in that they are statistically isotropic with no Bragg peaks. 
These unusual attributes can endow such materials with novel equilibrium and nonequilibrium
bulk physical properties. Isotropic disordered hyperuniform materials may offer
advantages over periodic materials with high crystallographic symmetries when the latter
possess bulk physical properties that have undesirable directional dependence.

While our understanding of the bulk properties of disordered hyperuniform materials
is nascent and thus represents a fertile area for future research, evidence
is beginning to emerge that such materials have important practical
and technological implications. Here we briefly review some of these developments.

\subsection{Thermodynamic Properties}

Interparticle interactions in many-particle hyperuniform systems in thermal equilibrium must necessarily be long-ranged
in order to achieve the requisite strong suppression of density fluctuations at
large length scales. The anomalous  thermodynamic properties of classical one-component plasmas,
which must be hyperuniform at all temperatures due to overall charge neutrality with a rigid
background of opposite charge, are well-documented
\cite{Ba78,Ba80,Na80,Ja81,Han13} and hence will not be reviewed here.
We saw in Sec. \ref{STEALTHY} that the ground states of isotropic stealthy pair potentials with a structure
factor $S({\bf k})$ that is constrained to be zero in a spherical region
around the origin in reciprocal space  (see Fig. \ref{patterns}) are remarkably disordered and highly degenerate
provided that $\chi <1/2$ \cite{Ba08,To15}. Such pair interactions are bounded, oscillating, long-ranged
functions and the resulting energy landscape is such that the disordered ground states
cannot resist shear deformations. This feature distinguishes them from
crystalline ground states associated with steep repulsive interactions (e.g.,
Lennard-Jones and hard-sphere potentials) whose shear moduli are positive.
It has been shown that stealthy systems at low positive temperatures
possess anomalous equilibrium  properties, which again is  attributed to the topography of the
underlying energy landscape \cite{Ba09a,Ba09b}.

\subsection{Wave Characteristics}

By mapping relatively large 2D disordered stealthy hyperuniform particle configurations, obtained via the 
collective-coordinate optimization procedure \cite{Uc04b,Ba08}, to certain 2D trivalent dielectric
networks via a Delaunay centroidal tessellation \cite{Fl09b} (see left panel of Fig. \ref{PBG}) what was thought to
be impossible at the time  became possible, namely, the first disordered
solids to have large {\it complete} (both polarizations and all directions) photonic band gaps comparable in
size to those in photonic crystals, but with the additional advantage of perfect isotropy \cite{Fl09b}.
Under the constraint of statistical isotropy, it is the largest degree of stealthiness 
($\chi$ nearly equal to 0.5) with an accompanying substantial degree
of short-range order 
that is responsible for what appears to be the maximal complete band-gap size in disordered hyperuniform dielectric networks.
(We have seen in Sec. \ref{STEALTHY} that short-range order increases with increasing $\chi$ 
in the disordered regime $0 \le \chi \le 1/2$ \cite{Uc04b,Ba08,Zh15a}.) 
This computational study enabled the design and  fabrication of disordered cellular solids with unprecedented 
waveguide geometries unhindered by crystallinity and anisotropy, and robust to defects \cite{Fl13,Man13b}; see Fig. \ref{PBG}.
Such materials are thus suitable for various applications, including   lasers, sensors, 
and optical microcircuits \cite{Man13b}. Florescu, Steinhardt and Torquato \cite{Fl09b}
suggested that disordered hyperuniform solid networks had ramifications for
electronic and phononic band gaps in disordered materials, which indeed has been borne
out \cite{He13,Xie13,Gk17}; see also Sec. \ref{silicon}. A recent 2D numerical study by Froufe-P{\'e}rez {\it et al.} \cite{Fr16}
of disordered photonic solids suggested that short-range
order and hyperuniformity were more important than stealthiness in
determining band gap formation, but the systems examined were relatively small
and its not clear that the same conclusions would be drawn for larger
systems, as explained immediately below.

A critical theoretical question is whether a complete isotropic photonic band gap 
persists in a disordered solid network in the infinite-system-size (thermodynamic) limit.
Since most disordered systems, including many hyperuniform varieties, contain
arbitrarily large ``holes" (see Sec. \ref{holes}), such inhomogeneities would seem to preclude
them from possessing complete photonic band gaps in the thermodynamic limit.
On the other hand, disordered stealthy hyperuniform systems cannot tolerate
arbitrarily large holes \cite{Zh17a}, which we conjecture here to be a necessary condition
to have a complete photonic band gap in the thermodynamic limit. 
This is precisely why one must be very careful in drawing conclusions about
band gap size from numerical simulations because hole formation is an extremely rare event 
that is virtually impossible to see in the very limited simulation box sizes and number of configurations,
especially those  that make it is computationally feasible to solve Maxwell's electromagnetic equations.

\begin{figure}[H]
\begin{center}
\includegraphics[  width=6in, keepaspectratio,clip=]{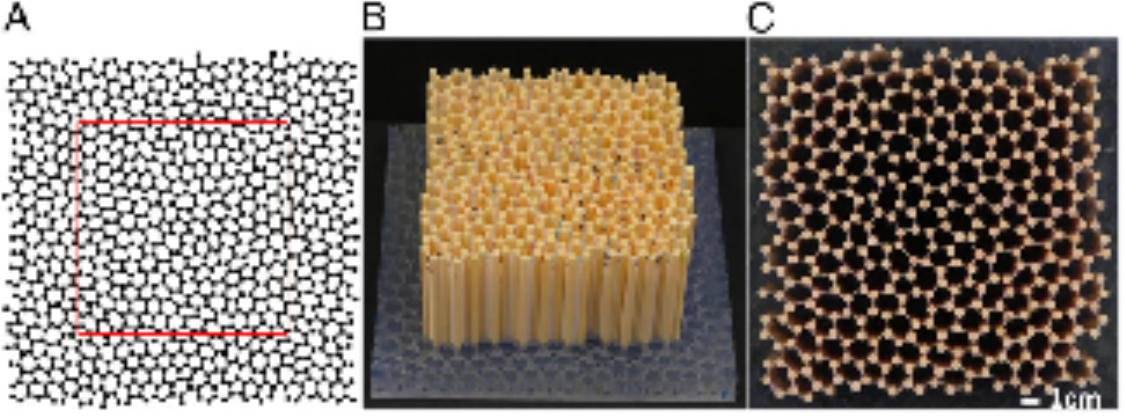}
\caption{Computational design of a 2D stealthy hyperuniform disordered network
solid (cross-section) and corresponding fabricated three-dimensional structure built of ``cylinders" and
"walls". (A)  Delaunay centroidal tessellation of a stealthy hyperuniform point configurations decorated with
cylinders and walls \cite{Fl09b}. The area enclosed in the red box is the structure used in the experimental study. 
Side view (B) and top view (C) of the stealthy hyperuniform disordered structure used in the experiment, assembled with Al$_2$O$_3$ cylinders and walls \cite{Man13b}. This figure is taken from Ref. \cite{Man13b}. }
\label{PBG}
\end{center}
\end{figure}

The propagation of electromagnetic waves in hyperuniform two-phase dielectric media are predicted to be lossless
in the long-wavelength limit. This prediction is derived from the corresponding ``strong-contrast" expansions for the 
frequency-dependent effective dielectric constant 
$\varepsilon_e$  \cite{Re08a}. Through lowest-order in the perturbation expansion
as well as an accurate approximations for $\varepsilon_e$
based on a resummation of  the series the imaginary part of the effective
dielectric constant depends on the volume integral over the autocovariance function
$\chi_{_V}({\bf r})$, defined by (\ref{Auto}). This means that the imaginary
part, which accounts for attenuation (losses) due
to incoherent multiple scattering in a typical disordered two-phase system,
is identically zero for any hyperuniform material (to an excellent approximation) due to the
sum rule (\ref{sum-2}) and hence waves can propagate
without any dissipation. This suggests that
``stealthy" disordered hyperuniform two-phase dielectric materials (to an excellent approximation)  
must be dissipativeless for a wider range of wavelengths \cite{Ch18}. Recent calculations of the frequency-dependent
effective dielectric constant have been carried out for designed phase-inversion-symmetric disordered hyperuniform composites and
disordered stealthy dispersions in two dimensions  \cite{Ch18} as well as dispersions of MRJ spheres
in three dimensions \cite{Kl18}.  It is notable that another recent study \cite{Wu17}
has demonstrated that a Lunesburg lens (which focuses a plane wave or transforms
a circular wave from a point source to a plane wave) based on a disordered hyperuniform design
has superior radiation properties compared to those of previously reported metamaterial
designs.

It is noteworthy that a recent computational study
by Leseur, Pierrat and Carminati \cite{Le16}  has demonstrated that high-density
materials made of non-absorbing subwavelength stealthy hyperuniform scatterers 
can be made transparent. More precisely, under very general conditions, such materials can be transparent for a range of wavelengths at densities 
for which an uncorrelated disordered material would be opaque due to multiple scattering.

As discussed in Sec. \ref{avian},  Nature has shown us a disordered {\it multihyperuniform} design in the avian retina
for acute color sensing  \cite{Ji14}. This strongly suggests that
the performance of color displays and sensors could be improved by the use of
multihyperuniform color patterns, and hence represents a potentially technologically important
area for future development. 


\subsection{Transport Properties}

Recently, stealthy disordered hyperuniform point configurations have been mapped to two-phase
media by circumscribing each point with a possibly overlapping sphere of a common radius $a$: the
``particle" and ``void" phases are taken to be the space interior and exterior to the spheres, respectively,
and their transport properties were computed \cite{Zh16b}.
The effective diffusion coefficient of point particles diffusing in the void phase as well as static and time-dependent characteristics associated with diffusion-controlled reactions in which the particles are perfectly absorbing traps for diffusing reactants
were considered.  
It was shown that  these transport as well as certain geometrical and topological
properties of two-phase media derived from decorated stealthy ground states are distinctly different
from those of equilibrium hard-sphere systems and spatially uncorrelated overlapping spheres.
As the extent of short-range order increases, stealthy disordered two-phase media can attain nearly
maximal effective diffusion coefficients over a broad range of volume fractions while also maintaining
isotropy, and therefore may have practical applications in situations where ease of transport is
desirable. 

The Fourier-space based construction procedure described in Sec. \ref{design} has recently been employed
to design both phase-inversion-symmetric composites and disordered stealthy hyperuniform dispersions in a matrix
with desirable effective conductivities \cite{Ch18}.
The stealthy dispersions were shown to possess nearly optimal effective conductivities,
while being fully statistically isotropic. 

In the context of type-II superconductors, it was shown that vortices with hyperuniform pinning site 
geometries exhibit  enhanced pinning, enabling higher critical currents \cite{Thien17}, which is robust over a wide range of parameters. 
The stronger pinning arises in the
hyperuniform arrays due to the suppression of pinning density fluctuations (greater
global uniformity), allowing higher pin
occupancy and the reduction of ``weak links" that lead to easy flow channeling.

Interestingly, it has been shown theoretically that there is a  relationship between a dynamical exponent associated
with the long-time diffusion coefficient in disordered porous media is related to large-scale
volume-fraction fluctuations and, specifically, the exponent $\alpha$ defined by (\ref{spec-asy}) \cite{Bur15}.
This link, which enables one to distinguish hyperuniform from
nonhyperuniform systems,  was recently verified experimentally using diffusion nuclear magnetic resonance methods \cite{Pa17}.

\subsection{Mechanical Properties}

By virtue of the similarities between the optimal microstructures for the effective electrical (thermal) conductivity
and effective bulk and shear moduli of two-phase composites \cite{To02a,Mi02}, certain decorations of disordered stealthy
hyperuniform point patterns are expected to  have nearly optimal effective elastic moduli in the same
way that they have nearly optimal effective conductivities \cite{Ch18}. Moreover, the suppression
of volume-fraction  fluctuations in disordered hyperuniform particulate composites is known to suppress
crack propagation within the matrix phase \cite{To00a}. Thus, appropriately designed disordered
hyperuniform composites could have desirable mechanical failure characteristics. Indeed, in a recent study
by Xu {\it et al.} \cite{Xu17},   designed  hyperuniform
composites were shown to possess a significantly higher brittle fracture strength than 
nonhyperuniform ones.

\subsection{Structurally Anisotropic Materials}

Structural anisotropic many-particle systems and heterogeneous materials
generally possess directional-dependent physical properties, including optical, 
transport, acoustic and elastic behaviors. Such materials with directional hyperuniformity
(see Sec. \ref{anisotropy}) are expected to possess unusual physical properties, but their potential
for technological applications has yet to be explored. The collective-coordinate optimization
procedure described in Sec. \ref{anisotropy} provides a systematic procedure to design hyperuniform
anisotropic materials. Such tunability could have technological relevance for manipulating light and sound waves 
in media in ways heretofore not thought to be possible.
Interestingly, although anisotropic stealthy ground-state configurations, such as 
one shown in Fig. \ref{lemniscate},  cannot support shear (for similar reasons
as in their isotropic counterparts \cite{Zh15b}), they are generally elastically anisotropic
because the stress tensor is {\it asymmetric}. Indeed, such configurations generally will possess internal
force couples that resist out-of-plane torques \cite{To16a}.
Directional structural hyperuniformity raises the interesting possibility that there
may exist disordered many-particle systems in equilibrium that at positive
temperature $T$ are incompressible in certain directions  and compressible in other directions
-- a highly unusual situation \cite{To16a}.


\section{Effect of Imperfections on Hyperuniform States}
\label{defects}

In the same way that there is no infinitely large perfect crystal in practice due to the inevitable existence of imperfections, such as point defects (e.g., vacancies), dislocations and thermal excitations, there is no infinitely large perfect disordered hyperuniform system, whether in thermal equilibrium or not.
For instance, in both equilibrium and nonequilibrium systems, point defects can arise and consequently destroy the hyperuniformity, albeit in some cases to a small degree.
In addition, we have seen that any  compressible ($\kappa_T >0$) equilibrium system  
cannot be hyperuniform at positive temperatures due
to thermal excitations, which follows from the compressibility relation (\ref{comp}) \cite{To15,To16a}.
Moreover, it can be difficult to attain perfect hyperuniformity in disordered nonequilibrium systems
due to a \textit{critical slowing down}, as discussed in Secs. \ref{slowing} and \ref{slowing-2}.

Thus, in both experiments and numerical simulations, it is important to understand the extent to which the hyperuniformity can be degraded 
or destroyed due to such imperfections in an otherwise hyperuniform system.
Interestingly, the presence of imperfections can significantly alter the properties of materials.
Specifically, point defects and dislocations can substantially enhance the mechanical strength of metals \cite{Di86, As76}.
Moreover, impurities can have a dramatic influence on the electronic properties of crystalline
solids, e.g., color centers \cite{As76, Ki05}, doped semiconductors
\cite{As76, Ki05}, and the Kondo effect \cite{Ko64, An70}.

In what follows, we discuss how the degree of hyperuniformity of a perfect hyperuniform system, either
ordered or disordered, is affected by the introduction of imperfections, such as point vacancies, particle displacements, and thermal excitations, following closely the recent work of Kim and Torquato \cite{Ki18a}.
We start with an initial hyperuniform point process at number density $\rho$ that has a
structure factor $S_0({\bf k})$ and a  pair-correlation function $g_2^{(0)} ({\bf r})$.

Point vacancies in a crystal refers to missing particles at the crystal sites that are fully occupied by particles in the perfect crystal.
In many cases, the presence of point defects induces elastic deformations in the crystal, resulting in asymmetric diffuse scattering near Bragg peaks, which is called Huang diffuse scattering \cite{Hu47, De73b}.
If one neglects the effect of deformations, the structure factor of a defective crystal with uncorrelated point vacancies is given by
\begin{equation}\label{eq:Sk point defects}
S({\bf k}) = p + (1-p) S_0 ({\bf k}),
\end{equation}
where $p$ is the concentration of vacancies.
It is noteworthy that this relation is also applicable to disordered point processes, whether they are 
hyperuniform or not; 
see Refs. \cite{De73b} and \cite{Ki18a} for various derivations of this equation.
We see that even a small concentration of defects can destroy hyperuniformity, even if to a small degree.

Now consider a perturbed point configuration in which the position of the 
$i$th particle ${\bf r}_i$ ($i=1,2,\cdots$) in an initial hyperuniform 
point configuration is displaced to a new  position ${\bf r}_i + {\bf u}({\bf r}_i)$.
When the initial point configuration is a lattice, we call the perturbed system a {\it perturbed lattice} \cite{We80}, which is also known as the {\it shuffled lattice} \cite{Ga02,To03a,Ga04b,Ga04}.
Welberry \textit{et al.} \cite{We80} investigated perturbed lattices with Gaussian perturbations in the context of distorted crystals, and Gabrielli \cite{Ga04b} and Gabrielli, Joyce and Torquato \cite{Ga08}
studied this subject more generally in the context of hyperuniformity.

The structure factor of a perturbed point process, whether it is initially ordered
or disordered, can be written as \cite{We80,Ga04b}
\begin{equation}\label{eq:Sk_perturbed lattice_general}
S({\bf k}) = 1+\rho \int d {{\bf r}} e^{-i{\bf k \cdot r}}g_2^{(0)} 
({\bf r})\hat{\phi}({\bf k};{\bf r}),
\end{equation}
where $\hat{\phi}({\bf k};{\bf r}) \equiv \int d{{\bf u}}d{{\bf v}} 
\exp({i{\bf k}\cdot ({\bf u}-{\bf v})} )f_2({{\bf u},{\bf v};{\bf r}})$ and
$f_2({{\bf u},{\bf v};{\bf r}})d{\bf u}d{\bf v}$ is the joint probability that two particles, initially separated by ${\bf r}$, are displaced by ${\bf u}$ within a volume element $d{\bf u}$ and ${\bf v}$ 
within $d{\bf v}$, respectively.
Note that relation  (\ref{eq:Sk_perturbed lattice_general}) is valid whether the initial configuration is hyperuniform or not.
If the displacements are uncorrelated and identically distributed, the associated joint 
probability density function is the product of two singlet probability densities, i.e., 
$f_2({\bf u},{\bf v};{\bf r}) = f_1({\bf u}) f_1({\bf v})$, where $f_1({\bf u})$ is the singlet probability density function of a displacement $\bf u$. 
Thus, (\ref{eq:Sk_perturbed lattice_general}) simplifies as follows \cite{Ga04b, Ga04}:
\begin{equation}\label{eq:Sk_perturbed lattice_uncorrelated}
S({\bf k}) = 1-|\tilde{f}_1({\bf k})|^2        + |\tilde{f}_1({\bf k})|^2 S_0 
({\bf k}),
\end{equation}
where $\tilde{f}_1({\bf k})$ is the Fourier transform of $f_1({\bf u})$.
The fact that $\tilde{f}_1({\bf k}) \to 1$ as $|{\bf k}|\to  0$ enables  one to conclude from
(\ref{eq:Sk_perturbed lattice_general}) that perturbing a hyperuniform point process preserves the hyperuniformity of the original system if the displacements  are uncorrelated. However, in all cases, the resulting growth
rate of the number variance cannot be slower than that of the original hyperuniform system. For example,
depending upon whether the variance of $f_1({\bf u})$ exists, an original class I hyperuniform
system can remain class I or may change to class II or III.

Let us now suppose that the displacements $\bf u$ are isotropic and two of its orthogonal
components are uncorrelated when $d\geq2$. 
It then follows that $\hat{\phi}({\bf k};{\bf r})$ in (\ref{eq:Sk_perturbed lattice_general})  
for small $|{\bf k}|$ must be  given asymptotically by
\begin{equation}\label{eq:G function}
\hat{\phi}({{\bf k};{\bf r}}) = 1 +  |{\bf k}|^2 \left[G({{\bf r}})-G({{\bf 0}})\right] + {\cal O}(|{\bf k}|^4),
\end{equation}
where $G({\bf r}) \equiv \left<{u_{1}({\bf r}+{\bf r}_0)u_{1}({\bf r}_0)}\right>$ is the 
displacement-displacement correlation function, $\left< {\bf u}\right> = {\bf 0}$ due to the 
isotropy of ${\bf u}$, and thus $G({{\bf 0}}) =\left<{|{\bf u}|^2}\right>/d$ if $\left<{|{\bf u}|^2}\right>$ exists.
Using (\ref{eq:G function}), the structure factor $S({\bf k})$ in (\ref{eq:Sk_perturbed lattice_general})  for small $|\bf k|$ is given approximately by \cite{Ga04b}
\begin{equation}\label{eq:Sk_perturbed lattice_correlated}
S({{\bf k}}) \approx \left[|{\bf k}|^2 G({{\bf 0}})+ \left(1-|{\bf k}|^2G({{\bf 0}})\right) S_0({{\bf k}})         \right] +\rho  |{\bf k}|^2\tilde{G}({\bf k})+\rho |{\bf k}|^2 \int 
d{{\bf r}}e^{-i{\bf k}\cdot {\bf r}} h_0({{\bf r}}) G({{\bf r}})  ,
\end{equation}
where ${h_0}({\bf r}) \equiv g_2^{(0)}({\bf r}) -1$ and $\tilde{G}({\bf k})$ is the Fourier transform of $G({\bf r})$.
Note that since $ |{\tilde f}_1({\bf k})|^2 \approx 1-|{\bf k}|^2 \langle |{\bf u}|^2 \rangle/d = 1-|{\bf k}|^2 G({\bf 0})$ 
for small $|{\bf k}|$ if $\langle |{\bf u}|^2 \rangle$ exists, 
the terms in the square brackets in (\ref{eq:Sk_perturbed lattice_correlated}) correspond to
the right-hand side of relation (\ref{eq:Sk_perturbed lattice_uncorrelated}), and the 
next two terms in (\ref{eq:Sk_perturbed lattice_correlated}) describe contributions from correlated displacements.
Gabrielli \cite{Ga04b} pointed out that when $\tilde{G}({\bf k})\sim |\bf k|^{\beta}$ for small $|\bf k|$, the perturbed point process is hyperuniform if $\beta>-2$.

Structure factors of crystals in thermal equilibrium have been extensively studied in 
the past by utilizing approximations for $\hat{\phi}({\bf k}; {\bf r})$ in  (\ref{eq:Sk_perturbed lattice_general}).
For instance, the Debye-Waller factor $\exp(- \left<({\bf q}\cdot{\bf u})^2\right>)$ 
\cite{As76}, a major contribution to diffuse scattering in harmonic crystals, is obtained from the terms
$1- |{\bf k}|^2 G({\bf 0})$ in (\ref{eq:G function}).
However, previous studies mainly focused on diffuse scattering near Bragg peaks rather than on the small-wavenumber behavior. 
Thus, it had not been quantitatively understood how thermal excitations effect the behavior of 
the structure factor for small wavenumbers and hence how they precisely destroy the hyperuniformity of ground-state crystals as the temperature is increased.
Recently, Kim and Torquato \cite{Ki18a} addressed this question by deriving the structure factor of a classical harmonic hypercubic lattice in thermal equilibrium; in short, they showed that
\begin{equation}\label{eq:Sk_phonon}
S({\bf k}) \approx \left<|{{\bf k}\cdot {\tilde{ {\bf u}}({\bf k}, t)}|^2}\right> 
= |{\bf k}|^2 \frac{k_B T}{m {\omega^2 
_{\parallel}}({\bf k})}~~~~(|{\bf k}|a \ll 1),
\end{equation}
where $m$ is the mass of a single particle, $a$ is lattice constant, $\tilde{{\bf u}}({\bf k},t)$ refers to
the normal coordinates of a harmonic crystal \cite{Chaik95} at a wave vector $\bf k$ 
[essentially the Fourier components of a small displacement ${\bf u}({\bf r}, t)$ 
of a particle initially located at $\bf{r}$]
and ${\omega_{\parallel}}({\bf k})$ is the angular frequency of $\tilde{{\bf u}}({\bf k},t)$ with a longitudinal polarization.
In the small-wavenumber limit, (\ref{eq:Sk_phonon}) recovers the compressibility relation (\ref{comp}) with an isothermal compressibility that is given by
\begin{equation}\label{eq:isothermal compressibility}
\kappa_T = \left(K a^{2-d}\right)^{-1},
\end{equation}
where $K$ is the spring constant between nearest neighbors. Relation (\ref{eq:isothermal compressibility}) has the same form for other lattices but with different constant parameters \cite{Ki18a}.
We note in passing that equilibrated 3D hard-spheres systems obtained
via molecular dynamics on approach
to the fcc jamming point along the stable crystal branch exhibits the same
structure factor scaling as in (\ref{eq:Sk_phonon}) \cite{At16b}.

Interestingly, harmonic crystals in thermal equilibrium can be viewed as correlated perturbed lattices. 
We learned that uncorrelated perturbed systems, which contribute only to the one-point statistics, cannot arise long-wavelength density fluctuations.
However, correlated perturbations can influence the two-point statistics in a point process;
if the correlations are sufficiently long-ranged, i.e., $\tilde{G}({\bf k}) \sim |{\bf k}|^{-2}$ 
for small $|\bf k|$, then the hyperuniformity of an initial point process can be destroyed \cite{Ga04b} because the contribution $|{\bf k}|^2 \tilde{G}({\bf k})$ from ``correlations'' in (\ref{eq:Sk_perturbed lattice_correlated}) remains a constant for small wavenumbers.
Displacement-displacement correlation functions of classical harmonic crystals in thermal equilibrium 
exactly satisfy this condition, regardless of the space dimension $d$ \cite{Ki18a}.
 For low dimensions ($d\leq 2$), this implies that $\left<{|{\bf u}|^2}\right>$ diverges, or equivalently the Debye-Waller factor vanishes, which is consistent with previous theoretical studies \cite{Im79}.


\section{Nearly Hyperuniform Systems}
\label{nearly}

Quantifying large-scale density fluctuations of a many-particle system is of interest, regardless
of whether it is hyperuniform. For example, the relative magnitudes of the volume and surface-area coefficients in the asymptotic
expansion of the local number variance $\sigma^2_{_N}(R)$  contain crucial information about the large-scale structure
of a many-body system. This was borne out in a recent study concerning amorphous ices and transitions between their different forms
\cite{Mar17}. Moreover, understanding structural and physical properties of  a system as it approaches a hyperuniform state
or whether  near hyperuniformity in a system is signaling crucial large-scale structural changes appears
to be fundamentally important. In what follows, we briefly review some developments along these lines.

\subsection{Growing Length Scale on Approach to a Hyperuniform State}
\label{growing}

In Sec. \ref{inverted}, we saw that the volume integral of the direct correlation
function $c({\bf r})$, i.e., $\tilde{c}({\bf k=0})$, diverges to -$\infty$ at a
hyperuniform state, whether in equilibrium or not, where $\tilde{c}({\bf k})$ is the
Fourier transform of $c({\bf r})$. This volume integral is expected
to have a large magnitude for any system that is near a hyperuniform state.
Thus, it is natural to define the following length scale \cite{Ho12b,Ma13a}:
\begin{equation}
\xi_c \equiv \left[ -\tilde{c}({\bf k=0}) \right]^{1/d},
\label{xi}
\end{equation}
to herald the anomalous suppression of large-scale density fluctuations. 
In the case of overcompressed nonequilibrium hard-sphere systems,
$\xi_c$ grows with increasing density well before the critical
hyperuniform state is reached \cite{Ho12b,At16b}. For well-known atomic glass formers, 
this length scale is able to distinguish subtle structural
differences between glassy and liquid states and grows upon supercooling,
even as the temperature is lowered past the glass transition temperature
and even if these states are only nearly hyperuniform \cite{Ma13a}.
Moreover, in all of these cases \cite{Ho12b,Ma13a}, as $\xi_c$ grew, so did the nonequilibrium
index $X$, defined by (\ref{XX}). Interestingly, using a revised cell theory, it has been
shown that jamming results in a percolation transition described by a static diverging length scale that is proportional to $\xi_c$ \cite{Con17}.

\subsection{Amorphous Silicon and Electronic Band Gaps}
\label{silicon}

The development of accurate structural models of amorphous silicon (a-Si) and other 
tetrahedrally coordinated solids has been an active area of research for many
decades \cite{Za32,Wea71,Za83,Mo02}, but many challenges and questions remain. The structure of a-Si
is approximated well by continuous random network (CRN) models \cite{Za32,Za83}, which are fully four-coordinated,
isotropic disordered networks that contain primarily five, six, and seven atom rings, while maintaining nearly perfect local
tetrahedral order. The concept of nearly hyperuniform network (NHN) structures has been proposed
\cite{He13} as alternatives to the CRN models for amorphous tetrahedrally coordinated solids, such as a-Si,
especially after annealing. A NHN has been defined as an amorphous tetrahedral network whose structure factor 
$S({\bf k})$ at $\bf k=0$ is smaller than the liquid value at the melting temperature. As a practical matter,
one is interested in cases where $S({\bf k= 0})$ is substantially
less, by 50\% or more, which implies a substantial reduction
in the large-scale density fluctuations and runs counter to the
limitations imposed by the ``frozen-liquid" paradigm of a glass \cite{Deg10}.
Using a novel implementation of the Stillinger-Weber potential for the interatomic
interactions \cite{St85b}, Henja {\it et al.} \cite{He13} showed
 that the energy landscape for a spectrum of NHNs includes a sequence of local minima
with an increasing degree of hyperuniformity that is significantly below the frozen-liquid
value and that correlates with other measurable features in
the structure factor  at intermediate and large wavenumbers. 
Among other results, it was shown  the degree of hyperuniformity
correlates with the width of the electronic band
gap, i.e., in terms of the ``hyperuniformity parameter" $H$, defined by relation (\ref{Hyp}),
this means  that this width increases as $H$ decreases.

On the experimental side, results of highly sensitive transmission X-ray scattering
measurements have been reported on nearly fully dense high-purity
a-Si samples for the purpose of determining
their degree of hyperuniformity \cite{Xie13}.  Annealing
was observed to increase the degree of hyperuniformity in a-Si
where it was found that $S(0) = 0.0075 (\pm 0.0005)$, which is significantly below
the suggested lower bound 
based on computational studies of CRN models \cite{Deg10}, but consistent with the recently proposed
NHN picture of a-Si \cite{He13}. Increasing hyperuniformity
was shown to be correlated with narrowing of the first diffraction
peak and the extension of the range of oscillations in the pair distribution
function.

\subsection{Polymeric Materials: Liquid State and Glass Formation}

Here we briefly describe some work that demonstrates that  models of
polymers in equilibrium and of glassy polymer systems can be made, under certain conditions,
to be nearly hyperuniform and, in some cases, nearly stealthy and hyperuniform.

\subsubsection{Equilibrium Liquids}

\begin{figure}[bthp]
\centerline{\includegraphics[  width=3in, keepaspectratio,clip=]{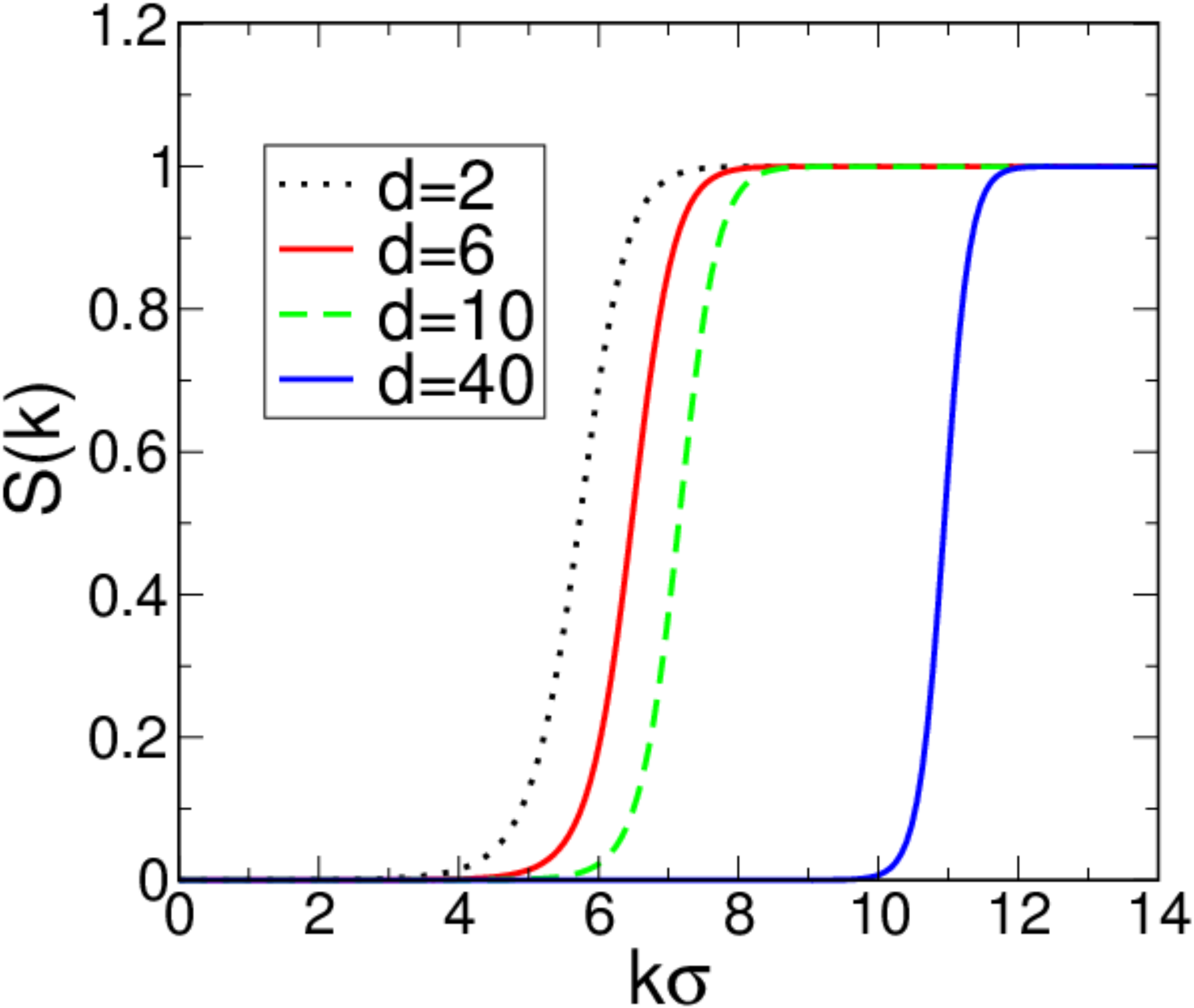}}
\caption{ Structure factor $S(k)$ in the MFA for the GCM at selected  $d$
with $\beta \epsilon=100$ and $\rho \sigma^d=36/\pi$, as obtained from relation (\ref{MFA})
given in Ref. \cite{Za08}.}
\label{GCM}
\end{figure}

The Gaussian-core model (GCM), in which particles interact with a
a purely repulsive Gaussian pair potential, $v_2(r)=\epsilon \exp[-(r/\sigma)^2]$, was introduced by Stillinger
\cite{St76} as a simple model to mimic effective pair interactions between the centers of mass of two polymer chains.
Since then, the GCM  has been investigated by many groups \cite{La00,Za08,Ik11,Co17b}.
Zachary, Stillinger and Torquato \cite{Za08} studied the liquid states of this model in $\mathbb{R}^d$ in various approximations
and arbitrary dimensions. Among other results, they employed  the $d$-dimensional generalization
of the structure factor in the {\it mean-field-approximation} (MFA) for the GCM to quantify its large-scale density fluctuations
in the high-density limit:
\begin{equation}
S(k) = \frac{1}{1+\lambda_d \exp\left[- \displaystyle\frac{( k \sigma)^2}{4}\right]},
\label{MFA}
\end{equation}
where  $\lambda_d= \beta\epsilon\rho \sigma^d \pi^{d/2}$ is a dimensionless parameter that depends on dimension,
$\beta=1/(k_B T)$ is an inverse temperature, and $k_B$ is Boltzmann's constant.
This approximation was first derived by  Lang et al. in three dimensions \cite{La00} and shown by them to be in excellent
agreement with computer simulations. The MFA structure factor (\ref{MFA}) is plotted in Fig. \ref{GCM}
for selected dimensions at fixed dimensionless temperature $\beta \epsilon=100$ and fixed dimensionless
density $\rho \sigma^d=36/\pi$. It is seen that the systems are not only nearly
hyperuniform but they are nearly stealthy for a range of wavenumbers that increases
as $d$ increases. The degree of hyperuniformity increases as $d$ increases. Thus, the GCM in the high-density 
regime are examples of disordered systems in equilibrium that 
at positive temperatures are very nearly hyperuniform and stealthy. To understand this more quantitatively, consider
the first two terms in the series expansion of the MFA structure factor (\ref{MFA}) about $k=0$ for any $d$:
\begin{equation}
S(k)=\frac{1}{1+\lambda_d}+ \frac{\lambda_d}{4(1+\lambda_d)^2} (k\sigma)^2 +{\cal O}(k^4).
\end{equation}
Thus, for large $\lambda_d$, $S(0)\sim  \lambda_d^{-1}$. In particular, fixing $\beta \epsilon=100$ and $\rho \sigma^d=36/\pi$, as
in the cases plotted in Fig. \ref{GCM}, we find that $S(0) \approx 2.8 \times 10^{-4},  2.8 \times 10^{-5},  2.9 \times 10^{-6}$,
and $ 1.0 \times 10^{-13}$ for $d=2,6,10$ and $40$, respectively.
Ikeda and Miyazaki studied the high-density equilibrium behavior of the GCM in three dimensions via molecular dynamics \cite{Ik11}.
Figure 5 in this paper shows $S(k)$ at four densities for liquids at the declining freezing temperature,
with an insert indicating strong convergence to hyperuniformity [i.e., $S(0)=0$] upon compression,
which is consistent with the aforementioned results.

\subsubsection{Glassy Systems}

\begin{figure}[H]
\centerline{\includegraphics[  width=3.7in, keepaspectratio,clip=]{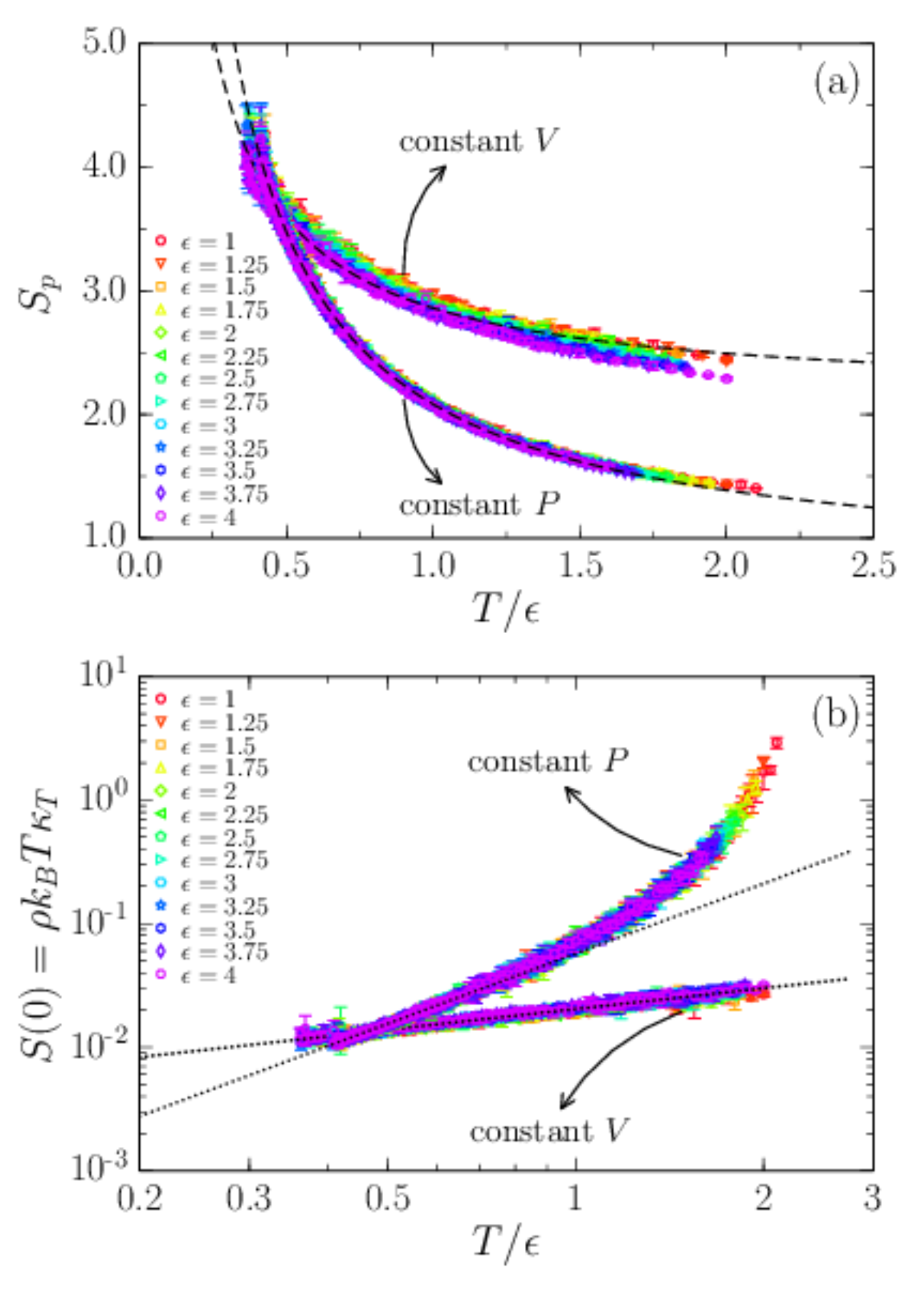}}
\caption{ (a) Height $S_p$ of the first peak of $S(k)$ and (b)
$S(0)$ as a function of $T/\epsilon$ for various $\epsilon$ at constant volume $V$ and constant pressure $P$. Dashed and dotted
lines in (a) and (b), respectively, are fits to the data. This is Fig. 7 of Ref. \cite{Xu16}.}
\label{douglas}
\end{figure}

Douglas and coworkers explored the utility of information in the static structure factor
$S(k)$ to quantify molecular jamming processes that underlie glass-formation in polymeric materials.  
Their molecular dynamics simulations indicated that polymeric materials (bulk materials composed 
of flexible polymers \cite{Xu16}  and nano particle fluids with chains 
grafted onto their surfaces \cite{Chr17}) are prototypical ``soft" materials 
because the many internal degrees of freedom within the molecules allow for rearrangements at 
a molecular scale that naturally permit the approach to a hyperuniform solid-like state. In particular, they 
found that the onset temperature of glass formation $T_A$ may be estimated by a 
Hansen-Verlet freezing criterion, corresponding to the peak height in the first peak in $S(k)$.  
In physical terms, when this peak is large the particles have begun to be locally ``jammed." 
However, this localization is transient in equilibrium liquids because the inertia of the polymer 
segments ultimately allows them to escape from their ``cages." The temperature $T_A$ 
is the temperature at which transient particle localization first emerges.
They also found that polymer fluids progressively approach a hyperuniform glassy 
state upon cooling and that the temperature range demarcating the end of glass 
formation process in measurements may be estimated from the magnitude of the 
``hyperuniformity” parameter $H$ \cite{At16a}, defined by relation (\ref{Hyp}).

Figure \ref{douglas} shows  structure-factor characteristics as a function of temperature $T$, normalized energy $\epsilon$, a
parameter measuring the strength of the intermolecular cohesive interaction, which was first presented in Ref. \cite{Xu16}. 
Data is shown for both constant volume $V$ and  constant pressure $P$. From this universal curve, the temperature at which $H = {\cal O}(10^{-3})$ 
can be estimated to be about $T/\epsilon  = 0.3$ at constant pressure, a value that is comparable to the characteristic Vogel-Fulcher-Tammann temperature at which the viscosity of the cooled liquid extrapolates to infinity based on 
measurements performed above the glass-transition temperature. Simulations of nanoparticle fluids in 
which the nanoparticles have layers of grafted polymer chains also led to fluid materials 
that are effectively hyperuniform at low temperatures, proving further evidence that ``soft” molecular 
fluids in which the molecules have many internal degrees of freedom that can adjust under 
local jamming conditions are ideal materials for actually achieving effectively hyperuniform 
materials \cite{Chr17}. One may thus estimate both the ``onset" and ``end" of 
the glass-formation process from thermodynamic properties of glass-forming liquids, at least under 
constant pressure conditions. On approaching the glassy state, the liquid evidently shows an 
opposite tendency in comparison to approaching a liquid critical point, in the sense that the 
compressibility tends to vanish rather than diverge. Douglas and coworkers emphasize that while the 
evidence points to a thermodynamic transition underlying glass formation, no phase transition 
involving the singular changes in the free energy arises as temperature or density is varied.


\section{Discussion and Outlook}
\label{conclusions}

Hyperuniform systems are characterized by an unusually large suppression of 
density fluctuations at long wavelengths (large length scales) compared to those in ordinary 
disordered systems, such as typical fluids and amorphous solids.
The hyperuniformity concept provides a unified means to classify and structurally characterize crystals, quasicrystals and special
disordered systems. While this review article focused attention on disordered hyperuniform systems,
we have seen that problems involving the quantification of the hyperuniformity
of the ordered kinds, crystals and quasicrystals, have deep connections
to problems in pure mathematics (e.g., number theory and discrete geometry) 
and theoretical physics (e.g., low-temperature states of matter and integrable quantum systems). 

Disordered hyperuniform systems and their manifestations were largely unknown in the scientific community about a decade and a half ago, but this survey reveals that
such systems arise in a plethora of contexts across the physical, materials, chemical, mathematical, engineering and biological sciences,
including disordered ground states,  glass formation, jamming, high-pressure states
of matter, Coulomb systems, spin systems, photonic and electronic band structure, localization
of waves and excitations, self-organization, fluid dynamics, number theory, stochastic point processes,
integral and stochastic geometry, the immune system, and photoreceptor cells \cite{Do05d,Za11a,Ji11c,Ch14a,Kl16,At16a,Si09,Be11,Ku11,Dr15,Le14,Ji14,Ja15,He15,We15,Tj15,Su15,Ch16a,He17a,He17b,We17,Kw17,Ma15,To08b,Uc04b,Ba08,Ba09a,To15,Zh15a,Zh15b,Xu16,Zh16a,To16a,To16b,Chr17,Xu17,Zh17a,Zh17b,Mar17,Wu17,Ch18,Kl18,Fl09b,Fl13,Man13a,Man13b,Ha13,Ma16,Zh16,De15,De16,Zi15b,Le16,We15,Yu15,Fr16,Fr17,Gk17,Zh16b,He13,Xie13,Ho12b,Ma13a}.
Remarkably, they  can be obtained via equilibrium or nonequilibrium routes,
and come in both quantum-mechanical and classical varieties.
Such isotropic amorphous states of matter are poised at exotic critical points, and
are like perfect crystals in the way they suppress large-scale density fluctuations,
a special type of long-range order, and yet are like liquids or glasses in that they are statistically isotropic with no Bragg peaks and hence lack any conventional long-range order. It was shown that these unusual attributes   
endow such materials with novel  physical properties, including
photonic, phononic, electronic, transport and mechanical properties that are
only beginning to be discovered, both computationally and experimentally.
Designer hyperuniform materials have already been fabricated and tested,
but their full potential in technological applications has yet to be realized.

We have also seen that the hyperuniformity concept has illuminated the importance of 
studying the  large-scale structural correlations in amorphous systems, regardless of whether they are hyperuniform,
via the quantification of the long-wavelength behavior of their spectral functions or corresponding density fluctuations.  
Understanding structural and physical properties of  a system as it approaches a hyperuniform state
or whether  near hyperuniformity  is signaling crucial large-scale structural changes in a system is
fundamentally important and should lead to new insights about condensed phases of matter.
The hyperuniformity concept has suggested a ``nonequilibrium index" for glasses and novel correlation functions
from which one can extract relevant growing length scales as a function of temperature
as a liquid is supercooled below its glass transition temperature.

While the notion of hyperuniformity was introduced in the context
of density fluctuations associated with point configurations and subsequently
of volume-fraction fluctuations in heterogeneous materials \cite{Za09}, it   has very recently been broadened along four different directions \cite{To16a}. 
This includes generalizations to treat fluctuations
in the interfacial area  in heterogeneous media and surface-area driven
evolving microstructures, random scalar fields (e.g., spinodal decomposition), random vector fields (e.g., velocity fields  in turbulence
and transport in random media), and statistically anisotropic
many-particle systems (e.g., structures that respond to external fields) and two-phase media. In the instances of random vector fields and statistically anisotropic
structures, the standard definition of hyperuniformity must be generalized such that it accounts
for the dependence of the relevant spectral functions on the direction in which the origin in Fourier space
is approached (nonanalyticities at the origin). 

Another obvious and untapped generalization involves
hyperuniform  stochastic processes in {\it time}. A temporal hyperuniform process 
is defined by a spectral function that tends to zero as the {\it frequency} tends to zero.
Ascertaining hyperuniformity in time would involve spectral analysis of time series, which could be   applied to real-valued continuous data
or discrete  data \cite{Pr81}.
Such processes arise in host of contexts and applications, including signal processing, control theory, communication theory, 
statistics, astrophysics, econometrics, quantitative finance, seismology, meteorology, and geophysics.

The connections of hyperuniform states of matter to many different areas of fundamental science  appear to be profound and yet our theoretical
understanding of them is only in its infancy. Can one identify general organizing principles that
drive a system to a hyperuniform state? When a system is at a disordered hyperuniform state, 
what does it say about the system and the underlying process leading to it?  Is it possible to devise a non-perturbative statistical-mechanical
theory of stealthy ground states for arbitrary $\chi$ that goes beyond the existing small-$\chi$
perturbation scheme \cite{To15}? Is the  hyperuniformity correlation length scale 
based on the direct correlation function  that grows with decreasing temperature
upon supercooling a liquid \cite{Ho12b,Ma13a} related to corresponding growing  point-to-set length scales  \cite{Be12,Hock12,Ch13}?
While relatively small hyperuniform materials have been primarily fabricated at millimeter length scales
using 3D printing technologies, a fabrication challenge is to design interactions in soft-matter and molecular systems 
to self-organize  to produce large samples of disordered hyperuniform materials
with heterogeneity length scales at the micron or smaller level. There similar
challenges on the computational side. For example, can  new algorithms and/or optimization procedures 
be devised that efficiently generate 
very large samples of high-quality disordered hyperuniform systems.
When a biological systems is poised at a disordered hyperuniform state, what
are the concomitant biological advantages being conferred?
This list represents  a very small fraction of the open fascinating questions in the field.

Judging from the explosive pace of published articles on disordered hyperuniform systems
over the last several years, prospects for exciting future theoretical and experimental developments
in this field are very promising. It would not be surprising that by the time
this review article is published, such advances will have been made.

\section*{Acknowledgement}

I am deeply grateful to Frank Stillinger, Obioma Uche, Aleksandar Donev, Robert Batten, Andrea Gabrielli,
Michael Joyce, Antonello Scardicchio,  Paul Steinhardt, Paul Chaikin, Marian Florescu,
Weining Man, Yang Jiao, Chase Zachary,  Adam Hopkins, {\'E}tienne Marcotte, Gabrielle Long, Sjoerd Roorda,
Joseph Corbo, Robert DiStasio, Eli Chertkov, Ge Zhang, Duyu Chen, Jaeuk Kim, 
Zheng Ma, Steven Atkinson, Remi Dreyfus, Arjun Yodh, Jianxiang Tian, Joshua Socolar, Erdal Oguz, Chaney Lin, Roberto Car, Fausto Martelli, Nicholas Giovambattista, Enrqiue Lomba and Jean-Jacques Weis
with whom I have collaborated on topics described in this review article. I am very thankful to Jaeuk Kim, Duyu Chen, Zheng Ma, Michael Klatt, Jack Douglas,
Fausto Martelli,  Roberto Car, Mathhew De Courcy-Ireland and Yang Jiao for comments
that greatly improved this article. The author's work on hyperuniformity
since 2003 has been supported by various grants from the Office of Basic Energy Sciences and 
the National Science Foundation.

\appendix\section{Gauss Circle Problem and Its Generalizations }
\label{gauss}

 How does the number of points
contained within some prescribed  domain (window) $\Omega$ in which ${\bf x}_0$ (centroidal position)
is fixed grow as the size of $\Omega$ is uniformly increased? 
A classic number-theoretic problem involving a specific version of this question 
asks how many lattice points of a square lattice at unit density lie inside or on a circle
of radius $R$ centered at a lattice point, denoted by  $N(R)$, which 
amounts to finding all of the integer solutions of $n_1^2+n_2^2 \le R^2$ \cite{Ke48,Ke53b}.
(This problem is directly related to the determination of the number of energy levels less than some fixed energy 
in integrable quantum systems \cite{Ble93}).
Clearly, $N(R)$ asymptotically approaches the window area $\pi R^2$ for large $R$.
The {\it discrepancy}  $E_2(R) \equiv N(R) -\pi R^2$ grows with $R$ in apparently ``chaotic" 
manner; see  Fig. \ref{Gauss}.  Using elementary considerations, Gauss showed that
the error $|E_2(R)|$ is bounded from above by some constant times the circumference of the circle, i.e.,
proportional to $R$. Since that time, many investigators have sought to improve the
bound, a notoriously difficult task. Expressing the error as  $|E_2(R)| \le  C R^t$, the best known bounds on the
exponent $t$ are $1/2 < t \le 131/208=0.6298 \ldots$, where $C$
is a constant; see Ref. \cite{Hu03} and references therein.

\begin{figure}[bthp]
\centerline{\psfig{file=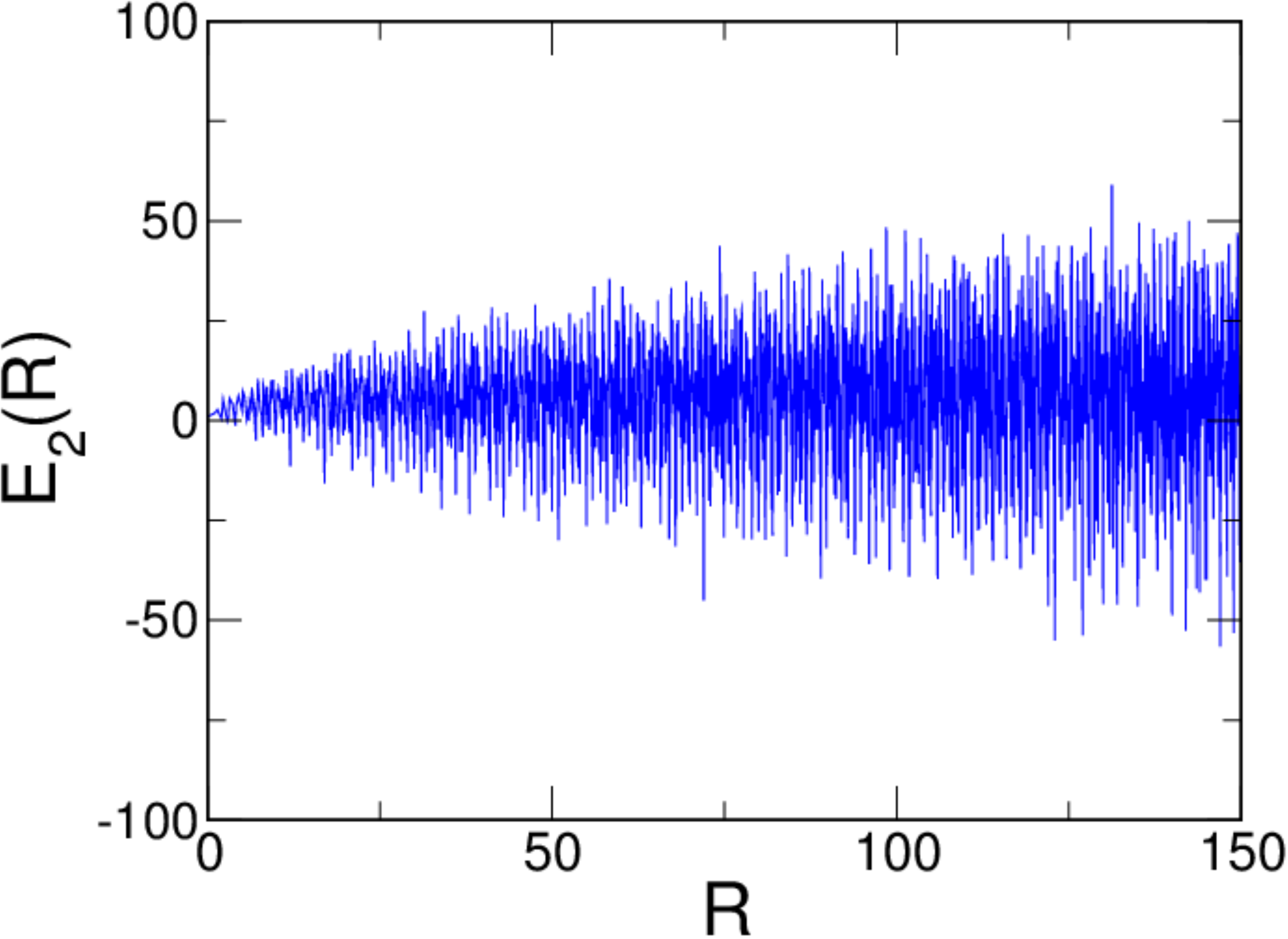,width=3.0in,clip=}}
\caption{The discrepancy  $E_2(R)\equiv N(R)-\pi R^2$ versus $R$ 
for the square lattice at unit number density using a circular window
of radius $R$ centered on a lattice point.}
\label{Gauss}
\end{figure}

A $d$-dimensional generalization of this question asks for
the number of  points of a $d$-dimensional lattice at unit density contained within and on
a $d$-dimensional spherical window of radius $R$, $N(R)$, which asymptotically
approaches the volume of a sphere of radius $R$, $v_1(R)$. An exact expression
for the discrepancy  $E_d(R) \equiv N(R) - v_1(R)$ in dimension $d$ is immediately obtained
from relation (\ref{N-per}) for $N(R;{\bf x}_0)$, where $N(R) \equiv N(R;{\bf x}_0=0)$,
yielding
\begin{equation}
E_d(R) = (2\pi R)^{d/2} \sum_{\bf q \neq 0} \frac{J_{d/2}(qR)}{q^{d/2}}.
\label{E1}
\end{equation}
Assuming the hypercubic lattice $\mathbb{Z}^d$ 
and considering the large-$R$ asymptotic limit, we can replace the Bessel function by the first term of
its asymptotic expansion and obtain an estimate for the error:
\begin{equation}
|E_d(R)| = \frac{R^{(d-1)/2}}{\pi}\Bigg|\sum_{n \ge  1} \frac{ z_m(n)
\cos\Big[2\pi \sqrt{n} R-\pi(d+1)/4\Big]}{n^{(d+1)/4}}\Bigg| + 
{\cal O}\left(R^{(d-2)/2}\right)
\label{E2}
\end{equation}
where $z_m(n)$ is the number of ways of writing $n$ as a sum of $m$ squares.
Note that whereas the correction term of order $R^{(d-2)/2}$ involves an absolutely
convergent sum, the first sum in (\ref{E2}) is conditionally convergent.
Improving on the elementary upper bound that $|E_d(R)| \le C R^{d-1}$ for $d \ge 2$ rests 
on obtaining sharper $R$-dependent estimates of the first sum.  That such estimates
are most difficult to obtain in dimensions two and three is related to the fact that every positive integer cannot
always be written as a sum of two and three squares, respectively.  
(The best known solution in three dimensions is   $|E_3(R)| \le  C R^{t+\epsilon}$,
where $t=17/14=1.21428\ldots$ and $\epsilon>0$ is arbitrarily small \cite{Ar08}.) Lagrange
showed that every positive integer can be written as the sum of at most four squares 
and so one expects that it become easier to estimate the error when $d \ge 4$. Indeed,
the problem is settled when $d \ge 5$, where its known that $|E_d(R)| =  {\cal O}(R^{d-2})$;
see Ref. \cite{Ts00} and references therein.




\begin{thebibliography}{100}
\expandafter\ifx\csname url\endcsname\relax
  \def\url#1{\texttt{#1}}\fi
\expandafter\ifx\csname urlprefix\endcsname\relax\def\urlprefix{URL }\fi
\expandafter\ifx\csname href\endcsname\relax
  \def\href#1#2{#2} \def\path#1{#1}\fi

\bibitem{Ve75}
D.~J. Vezzetti, A new derivation of some fluctuation theorems in statistical
  mechanics, J. Math. Phys. 16 (1975) 31--33.

\bibitem{Zi77}
R.~M. Ziff, On the bulk distribution functions and fluctuation theorems, J.
  Math. Phys. 18 (1977) 1825--1831.

\bibitem{La80}
L.~D. Landau, E.~M. Lifshitz, Statistical Physics, Pergamon Press, New York,
  1980.

\bibitem{Chaik95}
P.~M. Chaikin, T.~C. Lubensky, Principles of Condensed Matter Physics,
  Cambridge University Press, New York, 1995.

\bibitem{Tr98b}
T.~M. Truskett, S.~Torquato, P.~G. Debenedetti, Density fluctuations in
  many-body systems, Phys. Rev. E 58 (1998) 7639--7380.

\bibitem{Han13}
J.~P. Hansen, I.~R. McDonald, Theory of Simple Liquids, 4th Edition, Academic
  Press, New York, 2013.

\bibitem{Wi65}
B.~Widom, Equation of state in the neighborhood of the critical point, J. Chem.
  Phys. 43 (1965) 3898--3905.

\bibitem{Ka66}
L.~P. Kadanoff, Scaling laws for {I}sing models near {$T_c$}, Physics 2 (1966)
  263--272.

\bibitem{Fi67}
M.~E. Fisher, The theory of equilibrium critical phenomena, Rep. Prog. Phys. 30
  (1967) 615.

\bibitem{St71}
H.~E. Stanley, Introduction to Phase Transitions and Critical Phenomena, Oxford
  University Press, New York, 1971.

\bibitem{Wi74}
K.~G. Wilson, J.~Kogut, The renormalization group and the $\epsilon$ expansion,
  Phys. Rep. 12 (1974) 75--199.

\bibitem{Bi92}
J.~J. Binney, N.~J. Dowrick, A.~J. Fisher, M.~E.~J. Newman, The Theory of
  Critical Phenomena: An Introduction to the Renormalization Group, Oxford
  University Press, Oxford, England, 1992.

\bibitem{Berry86}
M.~V. Berry, M.~Robnik, Statistics of energy levels without time-reversal
  symmetry: {A}haronov-{B}ohm chaotic billiards, J. Phys. A: Math. \& Gen. 19
  (1986) 649.

\bibitem{Me91}
M.~L. Metha, Random Matrices, Academic Press, New York, 1991.

\bibitem{Bl93}
R.~Blumenfeld, S.~Torquato, Coarse-graining procedure to generate and analyze
  heterogeneous materials: {T}heory, Phys. Rev. E 48 (1993) 4492--4500.

\bibitem{Wa02}
A.~Wax, C.~Yang, V.~Backman, K.~Badizadegan, C.~W. Boone, R.~R. Dasari, M.~S.
  Feld, Cellular organization and substructure measured using angle-resolved
  low-coherence interferometry, Biophys. J. 82 (2002) 2256 -- 2264.

\bibitem{Pe93}
P.~J.~E. Peebles, Principles of Physical Cosmology, Princeton University Press,
  Princeton, 1993.

\bibitem{Ga05}
A.~Gabrielli, F.~S. Labini, M.~Joyce, L.~Pietronero, Statistical Physics for
  Cosmic Structures, Springer-Verlag, New York, 2005.

\bibitem{Sch08}
F.~C. Grozema, S.~Tonzani, Y.~A. Berlin, G.~C. Schatz, L.~D.~A. Siebbeles,
  M.~A. Ratner, Effect of structural dynamics on charge transfer in {DNA}
  hairpins, J. Am. Chem. Soc. 130 (2008) 5157--5166.

\bibitem{Ye03}
R.~Chang, A.~Yethiraj, Strongly charged flexible polyelectrolytes in poor
  solvents: Molecular dynamics simulations with explicit solvent, J. Chem.
  Phys. 118 (2003) 6634--6647.

\bibitem{Mu06}
Z.~Ou, M.~Muthukumar, Entropy and enthalpy of polyelectrolyte complexation:
  {L}angevin dynamics simulations, J. Chem. Phys. 124 (2006) 154902.

\bibitem{Be07}
L.~Berthier, G.~Biroli, J.-P. Bouchaud, W.~Kob, K.~Miyazaki, D.~R. Reichman,
  Spontaneous and induced dynamic fluctuations in glass formers. {I}. {G}eneral
  results and dependence on ensemble and dynamics, J. Chem. Phys. 126 (2007)
  184503.

\bibitem{Chand99}
K.~Lum, D.~Chandler, J.~D. Weeks, Hydrophobicity at small and large length
  scales, J. Phys. Chem. B 103 (1999) 4570--4577.

\bibitem{Sc99}
A.~M. Kulkarni, A.~P. Chatterjee, K.~S. Schweizer, C.~F. Zukoski, Depletion
  interactions in the protein limit: Effects of polymer density fluctuations,
  Phys. Rev. Lett. 83 (1999) 4554--4557.

\bibitem{Wa96}
S.~Warr, J.-P. Hansen, Relaxation of local density fluctuations in a fluidized
  granular medium, Europhys. Lett. 36 (1996) 589.

\bibitem{Ji11d}
Y.~Jiao, H.~Berman, T.-R. Kiehl, S.~Torquato, Spatial organization and
  correlations of cell nuclei in brain tumors, PloS one 6 (2011) e27323.

\bibitem{To03a}
S.~Torquato, F.~H. Stillinger, Local density fluctuations, hyperuniform
  systems, and order metrics, Phys. Rev. E 68 (2003) 041113.

\bibitem{Ga02}
A.~Gabrielli, M.~Joyce, F.~S. Labini, Glass-like universe: {R}eal-space
  correlation properties of standard cosmological models, Phys. Rev. D 65
  (2002) 083523.

\bibitem{Sh84}
D.~Shechtman, I.~Blech, D.~Gratias, J.~W. Cahn, Metallic phase with long-range
  orientational order and no translational symmetry, Phys. Rev. Lett. 53 (1984)
  1951--1953.

\bibitem{Le84}
D.~Levine, P.~J. Steinhardt, Quasicrystals: {A} new class of ordered
  structures, Phys. Rev. Lett. 53 (1984) 2477--2480.

\bibitem{Lev86}
D.~Levine, P.~J. Steinhardt, Quasicrystals. {I}. {D}efinition and structure,
  Phys. Rev. B 34 (1986) 596.

\bibitem{Za09}
C.~E. Zachary, S.~Torquato, Hyperuniformity in point patterns and two-phase
  heterogeneous media, J. Stat. Mech.: Theory \& Exp. (2009) P12015.

\bibitem{Og17}
E.~C. {O{\u g}uz}, J.~E.~S. {Socolar}, P.~J. {Steinhardt}, S.~{Torquato},
  {Hyperuniformity of Quasicrystals}, Phys. Rev. B 95 (2017) 054119.

\bibitem{To08b}
S.~Torquato, A.~Scardicchio, C.~E. Zachary, Point processes in arbitrary
  dimension from {F}ermionic gases, random matrix theory, and number theory, J.
  Stat. Mech.: Theory Exp. (2008) P11019.

\bibitem{Le00}
D.~Levesque, J.-J. Weis, J.~Lebowitz, Charge fluctuations in the
  two-dimensional one-component plasma, J. Stat. Phys. 100 (2000) 209--222.

\bibitem{Do05d}
A.~Donev, F.~H. Stillinger, S.~Torquato, Unexpected density fluctuations in
  disordered jammed hard-sphere packings, Phys. Rev. Lett. 95 (2005) 090604.

\bibitem{Sk06}
M.~Skoge, A.~Donev, F.~H. Stillinger, S.~Torquato, Packing hyperspheres in
  high-dimensional {E}uclidean spaces, Phys. Rev. E 74 (2006) 041127.

\bibitem{Za11a}
C.~E. Zachary, Y.~Jiao, S.~Torquato, Hyperuniform long-range correlations are a
  signature of disordered jammed hard-particle packings, Phys. Rev. Lett. 106
  (2011) 178001.

\bibitem{Ji11c}
Y.~Jiao, S.~Torquato, Maximally random jammed packings of {P}latonic solids:
  {H}yperuniform long-range correlations and isostaticity, Phys. Rev. E 84
  (2011) 041309.

\bibitem{Ho12b}
A.~B. Hopkins, F.~H. Stillinger, S.~Torquato, Nonequilibrium static diverging
  length scales on approaching a prototypical model glassy state, Phys. Rev. E
  86 (2012) 021505.

\bibitem{Ch14a}
D.~Chen, Y.~Jiao, S.~Torquato, Equilibrium phase behavior and maximally random
  jammed state of truncated tetrahedra, J. Phys. Chem. B 118 (2014) 7981--7992.

\bibitem{Kl16}
M.~A. Klatt, S.~Torquato, Characterization of maximally random jammed sphere
  packings. {II}. {Co}rrelation functions and density fluctuations, Phys. Rev.
  E 94 (2016) 022152.

\bibitem{At16a}
S.~Atkinson, G.~Zhang, A.~B. Hopkins, S.~Torquato, Critical slowing down and
  hyperuniformity on approach to jamming, Phys. Rev. E 94 (2016) 012902.

\bibitem{At16b}
S.~Atkinson, F.~H. Stillinger, S.~Torquato, Static structural signatures of
  nearly jammed disordered and ordered hard-sphere packings: {D}irect
  correlation function, Phys. Rev. E 94 (2016) 032902.

\bibitem{Si09}
L.~E. Silbert, M.~Silbert, Long-wavelength structural anomalies in jammed
  systems, Phys. Rev. E 80 (2009) 041304.

\bibitem{Be11}
L.~Berthier, P.~Chaudhuri, C.~Coulais, O.~Dauchot, P.~Sollich, Suppressed
  compressibility at large scale in jammed packings of size-disperse spheres,
  Phys. Rev. Lett. 106 (2011) 120601.

\bibitem{Ku11}
R.~Kurita, E.~R. Weeks, Incompressibility of polydisperse random-close-packed
  colloidal particles, Phys. Rev. E 84 (2011) 030401.

\bibitem{Dr15}
R.~{Dreyfus}, Y.~{Xu}, T.~{Still}, L.~A. {Hough}, A.~G. {Yodh}, S.~{Torquato},
  {Diagnosing hyperuniformity in two-dimensional, disordered, jammed packings
  of soft spheres}, Phys. Rev. E 91 (2015) 012302.

\bibitem{Ri17}
J.~Ricouvier, R.~Pierrat, R.~Carminati, P.~Tabeling, P.~Yazhgur, Optimizing
  hyperuniformity in self-assembled bidisperse emulsions, Phys. Rev. Lett. 119
  (2017) 208001.

\bibitem{Le14}
I.~{Lesanovsky}, J.~P. {Garrahan}, {Out-of-equilibrium structures in strongly
  interacting Rydberg gases with dissipation}, Phys. Rev. A 90 (2014) 011603.

\bibitem{Ja15}
R.~L. Jack, I.~R. Thompson, P.~Sollich, Hyperuniformity and phase separation in
  biased ensembles of trajectories for diffusive systems, Phys. Rev. Lett. 114
  (2015) 060601.

\bibitem{He15}
D.~{Hexner}, D.~{Levine}, {Hyperuniformity of critical absorbing states}, Phys.
  Rev. Lett. 114 (2015) 110602.

\bibitem{We15}
J.~H. Weijs, R.~Jeanneret, R.~Dreyfus, D.~Bartolo, Emergent hyperuniformity in
  periodically driven emulsions, Phys. Rev. Lett. 115 (2015) 108301.

\bibitem{Tj15}
E.~Tjhung, L.~Berthier, Hyperuniform density fluctuations and diverging dynamic
  correlations in periodically driven colloidal suspensions, Phys. Rev. Lett.
  114 (2015) 148301.

\bibitem{He17a}
D.~Hexner, D.~Levine, {Noise, Diffusion, and Hyperuniformity}, Phys. Rev. Lett.
  118 (2017) 020601.

\bibitem{He17b}
D.~Hexner, P.~M. Chaikin, D.~Levine, Enhanced hyperuniformity from random
  reorganization, Proc. Nat. Acad. Sci. 114 (2017) 4294–--4299.

\bibitem{We17}
J.~H. Weijs, D.~Bartolo, Mixing by unstirring: {H}yperuniform dispersion of
  interacting particles upon chaotic advection, Phys. Rev. Lett. 119 (2017)
  048002.

\bibitem{Kw17}
S.~Kwon, J.~M. Kim, Hyperuniformity of initial conditions and critical decay of
  a diffusive epidemic process belonging to the manna class, Phys. Rev. E 96
  (2017) 012146.

\bibitem{Ji14}
Y.~Jiao, T.~Lau, H.~Hatzikirou, M.~Meyer-Hermann, J.~C. Corbo, S.~Torquato,
  Avian photoreceptor patterns represent a disordered hyperuniform solution to
  a multiscale packing problem, Phys. Rev. E 89 (2014) 022721.

\bibitem{Ma15}
A.~Mayer, V.~Balasubramanian, T.~Mora, A.~M. Walczak, How a well-adapted immune
  system is organized, Proc. Nat. Acad. Sci. 112 (2015) 5950--5955.

\bibitem{Fe56}
R.~P. Feynman, M.~Cohen, Energy spectrum of the excitations in liquid helium,
  Phys. Rev. 102 (1956) 1189--1204.

\bibitem{Uc04b}
O.~U. Uche, F.~H. Stillinger, S.~Torquato, Constraints on collective density
  variables: Two dimensions, Phys. Rev. E 70 (2004) 046122.

\bibitem{Ba08}
R.~D. Batten, F.~H. Stillinger, S.~Torquato, Classical disordered ground
  states: {S}uper-ideal gases, and stealth and equi-luminous materials, J.
  Appl. Phys. 104 (2008) 033504.

\bibitem{Ba09a}
R.~D. Batten, F.~H. Stillinger, S.~Torquato, Novel low-temperature behavior in
  classical many-particle systems, Phys. Rev. Lett. 103 (2009) 050602.

\bibitem{To15}
S.~{Torquato}, G.~{Zhang}, F.~H. {Stillinger}, {Ensemble Theory for Stealthy
  Hyperuniform Disordered Ground States}, Phys. Rev. X 5 (2015) 021020.

\bibitem{Zh15a}
G.~Zhang, F.~Stillinger, S.~Torquato, Ground states of stealthy hyperuniform
  potentials: {I}. {E}ntropically favored configurations, Phys. Rev. E 92
  (2015) 022119.

\bibitem{Zh15b}
G.~Zhang, F.~Stillinger, S.~Torquato, Ground states of stealthy hyperuniform
  potentials: {II}. {S}tacked-slider phases, Phys. Rev. E 92 (2015) 022120.

\bibitem{Zh16a}
G.~Zhang, F.~H. Stillinger, S.~Torquato, The perfect glass paradigm: Disordered
  hyperuniform glasses down to absolute zero, Sci. Rep. 6.

\bibitem{Zh17a}
G.~Zhang, F.~H. Stillinger, S.~Torquato, {Can Exotic Disordered ``Stealthy''
  Particle Configurations Tolerate Arbitrarily Large Holes?}, Soft Matter 13
  (2017) 6197.

\bibitem{Zh17b}
G.~{Zhang}, F.~H. {Stillinger}, S.~{Torquato}, {Classical many-particle systems
  with unique disordered ground states}, Phys. Rev. E 96 (2017) 042146.

\bibitem{Mon73}
H.~L. Montgomery, The pair correlation of zeros of the zeta function, Amer.
  Math. Soc. (1973) 181--193.

\bibitem{Dy70}
F.~J. Dyson, Correlations between eigenvalues of a random matrix, Comm. Math.
  Phys. 19 (1970) 235--250.

\bibitem{To06d}
S.~Torquato, O.~U. Uche, F.~H. Stillinger, Random sequential addition of hard
  spheres in high {E}uclidean dimensions, Phys. Rev. E 74 (2006) 061308.

\bibitem{Zh13b}
G.~Zhang, S.~Torquato, Precise algorithm to generate random sequential addition
  of hard hyperspheres at saturation, Phys. Rev. E 88 (2013) 053312.

\bibitem{To16b}
S.~Torquato, Disordered hyperuniform heterogeneous materials, J. Phys.: Cond.
  Mat 28 (2016) 414012.

\bibitem{Xu16}
W.-S. Xu, J.~F. Douglas, K.~F. Freed, Influence of cohesive energy on the
  thermodynamic properties of a model glass-forming polymer melt,
  Macromolecules 49 (2016) 8341--8354.

\bibitem{To16a}
S.~Torquato, Hyperuniformity and its generalizations, Phys. Rev. E 94 (2016)
  022122.

\bibitem{Ch16a}
E.~Chertkov, R.~A. DiStasio, G.~Zhang, R.~Car, S.~Torquato, Inverse design of
  disordered stealthy hyperuniform spin chains, Phys. Rev. B 93 (2015) 064201.

\bibitem{Chr17}
A.~Chremos, J.~F. Douglas, Particle localization and hyperuniformity of
  polymer-grafted nanoparticle materials, Annalen der Physik 529.

\bibitem{Mar17}
F.~{Martelli}, S.~{Torquato}, N.~{Giovambattista}, R.~{Car}, {Large-scale
  structure and hyperuniformity of amorphous ices}, Phys. Rev. Lett. 119 (2017)
  136002.

\bibitem{Xu17}
Y.~Xu, S.~Chen, P.-E. Chen, W.~Xu, Y.~Jiao, Microstructure and mechanical
  properties of hyperuniform heterogeneous materials, Phys. Rev. E 96 (2017)
  043301.

\bibitem{Wu17}
B.-Y. Wu, X.-Q. Sheng, Y.~Hao, Effective media properties of hyperuniform
  disordered composite materials, PloS one 12 (2017) e0185921.

\bibitem{Ch18}
D.~{Chen}, S.~{Torquato}, {Designing Disordered Hyperuniform Two-Phase
  Materials with Novel Physical Properties}, Acta Materialia 142 (2018)
  152--161.

\bibitem{Kl18}
M.~A. {Klatt}, S.~{Torquato}, {Characterization of Maximally Random Jammed
  Sphere Packings. {III}. Transport and Electromagnetic Properties via
  Correlation Functions}, Phys. Rev. E 97 (2018) 012118.

\bibitem{Fl09b}
M.~Florescu, S.~Torquato, P.~J. Steinhardt, Designer disordered materials with
  large complete photonic band gaps, Proc. Nat. Acad. Sci. 106 (2009)
  20658--20663.

\bibitem{Fl13}
M.~Florescu, P.~J. Steinhardt, S.~Torquato, Optical cavities and waveguides in
  hyperuniform disordered photonic solids, Phys. Rev. B 87 (2013) 165116.

\bibitem{Man13b}
W.~Man, M.~Florescu, E.~P. Williamson, Y.~He, S.~R. Hashemizad, B.~Y.~C. Leung,
  D.~R. Liner, S.~Torquato, P.~M. Chaikin, P.~J. Steinhardt, Isotropic band
  gaps and freeform waveguides observed in hyperuniform disordered photonic
  solids, Proc. Nat. Acad. Sci. 110 (2013) 15886--15891.

\bibitem{Man13a}
W.~Man, M.~Florescu, K.~Matsuyama, P.~Yadak, G.~Nahal, S.~Hashemizad,
  E.~Williamson, P.~Steinhardt, S.~Torquato, P.~Chaikin, Photonic band gap in
  isotropic hyperuniform disordered solids with low dielectric contrast, Opt.
  Express 21 (2013) 19972--19981.

\bibitem{Ha13}
J.~Haberko, N.~Muller, F.~Scheffold, Direct laser writing of three dimensional
  network structures as templates for disordered photonic materials, Phys. Rev.
  A 88 (2013) 043822.

\bibitem{Ma16}
T.~Ma, H.~Guerboukha, M.~Girard, A.~D. Squires, R.~A. Lewis, M.~Skorobogatiy,
  3{D} printed hollow-core terahertz optical waveguides with hyperuniform
  disordered dielectric reflectors, Adv. Optical Mater. 4 (2016) 2085--2094.

\bibitem{Zh16}
W.~Zhou, Z.~Cheng, B.~Zhu, X.~Sun, H.~K. Tsang, Hyperuniform disordered network
  polarizers, IEEE J. Selected Topics in Quantum Elec. 22 (2016) 288--294.

\bibitem{De15}
C.~De~Rosa, F.~Auriemma, C.~Diletto, R.~Di~Girolamo, A.~Malafronte,
  P.~Morvillo, G.~Zito, G.~Rusciano, G.~Pesce, A.~Sasso, Toward hyperuniform
  disordered plasmonic nanostructures for reproducible surface-enhanced {R}aman
  spectroscopy, Phys. Chem. Chem. Phys. 17 (2015) 8061--8069.

\bibitem{De16}
R.~Degl'Innocenti, Y.~D. Shah, L.~Masini, A.~Ronzani, A.~Pitanti, Y.~Ren, D.~S.
  Jessop, A.~Tredicucci, H.~E. Beere, D.~A. Ritchie, Hyperuniform disordered
  terahertz quantum cascade laser, Sci. Reports 6 (2016) 19325.

\bibitem{Zi15b}
G.~Zito, G.~Rusciano, G.~Pesce, A.~Malafronte, R.~Di~Girolamo, G.~Ausanio,
  A.~Vecchione, A.~Sasso, Nanoscale engineering of two-dimensional disordered
  hyperuniform block-copolymer assemblies, Phys. Rev. E 92 (2015) 050601.

\bibitem{Le16}
O.~{Leseur}, R.~{Pierrat}, R.~{Carminati}, {High-density hyperuniform materials
  can be transparent}, Optica 3 (2016) 763--767.

\bibitem{Yu15}
S.~{Yu}, X.~{Piao}, J.~{Hong}, N.~{Park}, {Bloch-like wave dynamics in
  disordered potentials based on supersymmetry}, Nat. Mater. 6 (2015) 8269.

\bibitem{Fr16}
L.~S. Froufe-P\'erez, M.~Engel, P.~F. Damasceno, N.~Muller, J.~Haberko, S.~C.
  Glotzer, F.~Scheffold, Role of short-range order and hyperuniformity in the
  formation of band gaps in disordered photonic materials, Phys. Rev. Lett. 117
  (2016) 053902.

\bibitem{Fr17}
L.~S. {Froufe-P{\'e}rez}, M.~{Engel}, J.~{Jos{\'e} S{\'a}enz}, F.~{Scheffold},
  {Transport Phase Diagram and Anderson Localization in Hyperuniform Disordered
  Photonic Materials}, Proc. Nat. Acad. Sci. 114 (2017) 9570–--9574.

\bibitem{Gk17}
G.~Gkantzounis, T.~Amoah, M.~Florescu, Hyperuniform disordered phononic
  structures, Phys. Rev. B 95 (2017) 094120.

\bibitem{Zh16b}
G.~Zhang, F.~H. Stillinger, S.~Torquato, {Transport, Geometrical and
  Topological Properties of Stealthy Disordered Hyperuniform Two-Phase
  Systems}, J. Chem. Phys 145 (2016) 244109.

\bibitem{He13}
M.~Hejna, P.~J. Steinhardt, S.~Torquato, Nearly hyperuniform network models of
  amorphous silicon, Phys. Rev. B 87 (2013) 245204.

\bibitem{Xie13}
R.~Xie, G.~G. Long, S.~J. Weigand, S.~C. Moss, T.~Carvalho, S.~Roorda,
  M.~Hejna, S.~Torquato, P.~J. Steinhardt, Hyperuniformity in amorphous silicon
  based on the measurement of the infinite-wavelength limit of the structure
  factor, Proc. Nat. Acad. Sci. 110 (2013) 13250--13254.

\bibitem{Ma13a}
{\' E}.~Marcotte, F.~H. Stillinger, S.~Torquato, Nonequilibrium static growing
  length scales in supercooled liquids on approaching the glass transition, J.
  Chem. Phys. 138 (2013) 12A508.

\bibitem{Lu07}
V.~Lubchenko, P.~G. Wolynes, Theory of structural glasses and supercooled
  liquids, Ann. Rev. Phys. Chem. 58 (2007) 235--266.

\bibitem{Sc07}
K.~S. Schweizer, Dynamical fluctuation effects in glassy colloidal suspensions,
  Current Opinion Coll. Inter. Sc. 12 (2007) 297 -- 306.

\bibitem{Chand10}
D.~Chandler, J.~P. Garrahan, Dynamics on the way to forming glass: Bubbles in
  space-time, Ann. Rev. Phys. Chem. 61 (2010) 191--217.

\bibitem{Be12}
L.~Berthier, W.~Kob, Static point-to-set correlations in glass-forming liquids,
  Phys. Rev. E 85 (2012) 011102.

\bibitem{Hock12}
G.~M. Hocky, T.~E. Markland, D.~R. Reichman, Growing point-to-set length scale
  correlates with growing relaxation times in model supercooled liquids, Phys.
  Rev. Lett. 108 (2012) 225506.

\bibitem{Ch13}
B.~Charbonneau, P.~Charbonneau, G.~Tarjus, Geometrical frustration and static
  correlations in hard-sphere glass formers, J. Chem. Phys. 138.

\bibitem{Con17}
A.~{Coniglio}, M.~{Pica Ciamarra}, T.~{Aste}, {Universal behaviour of the glass
  and the jamming transitions in finite dimensions}, Soft Matter 13 (2017)
  8766--8771.

\bibitem{To02a}
S.~Torquato, Random Heterogeneous Materials: Microstructure and Macroscopic
  Properties, Springer-Verlag, New York, 2002.

\bibitem{Sa03}
M.~Sahimi, Heterogeneous Materials {I}: {L}inear Transport and Optical
  Properties, Springer-Verlag, New York, 2003.

\bibitem{Lu90a}
B.~L. Lu, S.~Torquato, Photographic granularity-- {M}athematical formulation
  and effect of impenetrability of grains, J. Opt. Soc. Am. A 7 (1990)
  717--724.

\bibitem{Lu90b}
B.~L. Lu, S.~Torquato, Local volume fraction fluctuations in heterogeneous
  media, J. Chem. Phys. 93 (1990) 3452--3459.

\bibitem{Qu97b}
J.~Quintanilla, S.~Torquato, Local volume fraction fluctuations in random
  media, J. Chem. Phys. 106 (1997) 2741--2751.

\bibitem{Qu99}
J.~Quintanilla, S.~Torquato, Local volume fraction fluctuations in periodic
  heterogeneous media, J. Chem. Phys. 110 (1999) 3215--3219.

\bibitem{Za11c}
C.~E. Zachary, Y.~Jiao, S.~Torquato, Hyperuniformity, quasi-long-range
  correlations, and void-space constraints in maximally random jammed particle
  packings. {I}. {P}olydisperse spheres, Phys. Rev. E 83 (2011) 051308.

\bibitem{Za11d}
C.~E. Zachary, Y.~Jiao, S.~Torquato, Hyperuniformity, quasi-long-range
  correlations, and void-space constraints in maximally random jammed particle
  packings. {II}. {A}nisotropy in particle shape, Phys. Rev. E 83 (2011)
  051309.

\bibitem{Ch15}
D.~Chen, S.~Torquato, Confined disordered strictly jammed binary sphere
  packings, Phys. Rev. E 92 (2015) 062207.

\bibitem{Di18}
R.~A. DiStasio, G.~Zhang, F.~H. Stillinger, S.~Torquato, Rational design of
  stealthy hyperuniform patterns with tunable orderIn preparation.

\bibitem{St95}
D.~Stoyan, W.~S. Kendall, J.~Mecke, Stochastic Geometry and Its Applications,
  2nd Edition, Wiley, New York, 1995.

\bibitem{Sc08}
A.~Scardicchio, F.~H. Stillinger, S.~Torquato, Estimates of the optimal density
  of sphere packings in high dimensions,, J. Math. Phys. 49 (2008) 043301.

\bibitem{Le73}
A.~Lenard, Correlation functions and the uniqueness of the state in classical
  statistical mechanics, Commun. Math. Phys. 30 (1973) 35--44.

\bibitem{To06b}
S.~Torquato, F.~H. Stillinger, New conjectural lower bounds on the optimal
  density of sphere packings, Experimental Math. 15 (2006) 307--331.

\bibitem{Co93}
J.~H. Conway, N.~J.~A. Sloane, Sphere Packings, Lattices and Groups,
  Springer-Verlag, New York, 1998.

\bibitem{To82b}
S.~Torquato, G.~Stell, Microstructure of two-phase random media: {I}. {T}he
  $n$-point probability functions, J. Chem. Phys. 77 (1982) 2071--2077.

\bibitem{To83a}
S.~Torquato, G.~Stell, Microstructure of two-phase random media: {II}. {T}he
  {M}ayer--{M}ontroll and {K}irkwood--{S}alsburg hierarchies, J. Chem. Phys. 78
  (1983) 3262--3272.

\bibitem{De49}
P.~Debye, A.~M. Bueche, Scattering by an inhomogeneous solid, J. Appl. Phys. 20
  (1949) 518--525.

\bibitem{De57}
P.~Debye, H.~R. Anderson, H.~Brumberger, Scattering by an inhomogeneous solid.
  {II}. {T}he correlation function and its applications, J. Appl. Phys. 28
  (1957) 679--683.

\bibitem{Ma80}
P.~Martin, T.~Yalcin, The charge fluctuations in classical {C}oulomb systems,
  J. Stat. Phys. 22 (1980) 435--463.

\bibitem{Ke53b}
D.~G. Kendall, R.~A. Rankin, On the number of points of a given lattice in a
  random hypersphere, Quart. J. Math. Oxford 4 (1953) 178--189.

\bibitem{He06}
L.~Heinrich, H.~Schmidt, V.~Schmidt, Central limit theorems for {P}oisson
  hyperplane tessellations, Annals Applied Prob. 16 (2006) 919--950.

\bibitem{Hen13}
L.~Heinrich, M.~Spiess, Central limit theorems for volume and surface content
  of stationary {P}oisson cylinder processes in expanding domains, Adv. in
  Appl. Probab. 45 (2013) 312--331.

\bibitem{Ga08}
A.~Gabrielli, M.~Joyce, S.~Torquato, Tilings of space and superhomogeneous
  point processes, Phys. Rev. E 77 (2008) 031125.

\bibitem{Za11b}
C.~E. Zachary, S.~Torquato, Anomalous local coordination, density fluctuations,
  and void statistics in disordered hyperuniform many-particle ground states,
  Phys. Rev. E 83 (2011) 051133.

\bibitem{Ab17}
L.~D. Abreu, J.~M. Pereira, J.~L. Romero, S.~Torquato, The {W}eyl-{H}eisenberg
  ensemble: {H}yperuniformity and higher landau levels, J. Stat. Mech.: Th. and
  Exper. 2017 (2017) 043103.

\bibitem{Ga04b}
A.~Gabrielli, Point processes and stochastic displacement fields, Phys. Rev. E
  70 (2004) 066131.

\bibitem{Ga04}
A.~Gabrielli, S.~Torquato, Voronoi and void statistics for superhomogeneous
  point processes, Phys. Rev. E 70 (2004) 041105.

\bibitem{Ki18a}
J.~Kim, S.~Torquato, Effect of imperfections on the hyperuniformity of
  many-body systems, Phys. Rev. BIn press.

\bibitem{Ai01}
M.~Aizenman, S.~Goldstein, J.~Lebowitz, Bounded fluctuations and translation
  symmetry breaking in one-dimensional particle systems, J. Stat. Phys. 103
  (2001) 601--618.

\bibitem{Beck87}
J.~Beck, Irregularties of distribution {I}., Acta Mathematica 159 (1987) 1--49.

\bibitem{Ma17}
Z.~Ma, S.~Torquato, Random scalar fields and hyperuniformity, J. Appl. Phys.
  121 (2017) 244904.

\bibitem{Uc06b}
O.~U. Uche, S.~Torquato, F.~H. Stillinger, Collective coordinates control of
  density distributions, Phys. Rev. E 74 (2006) 031104.

\bibitem{Re67}
L.~Reatto, G.~V. Chester, Phonons and the properties of a {B}ose system, Phys.
  Rev. 155 (1967) 88--100.

\bibitem{Ga03}
A.~Gabrielli, B.~Jancovici, M.~Joyce, J.~L. Lebowitz, L.~Pietronero, F.~S.
  Labini, Generation of primordial cosmological perturbations from statistical
  mechanical models, Phys. Rev. D 67 (2003) 043506.

\bibitem{Ja81}
B.~Jancovici, Exact results for the two-dimensional one-component plasma, Phys.
  Rev. Lett. 46 (1981) 386--388.

\bibitem{Lo17}
E.~{Lomba}, J.~J. {Weis}, S.~{Torquato}, {Disordered hyperuniformity in
  two-component non-additive hard disk plasmas}, Phys. Rev. E 96 (2017) 062126.

\bibitem{Lo18a}
E.~{Lomba}, J.-J. {Weis}, S.~{Torquato}, {Disordered multihyperuniformity
  derived from binary plasmas}, Phys. Rev. E 97 (2018) 010102(R).

\bibitem{Ki17}
J.~Kim, S.~Torquato, {Effect of Window Shape on the Detection of
  Hyperuniformity via the Local Number Variance}, J. Stat. Mech.: Th. and
  Exper. 2017 (2017) 013402.

\bibitem{Sa06}
P.~Sarnak, A.~Str{\" o}mbergsson, Minima of {E}pstein's zeta function and
  heights of flat tori, Inventiones Math. 165 (2006) 115--151.

\bibitem{Ra53}
R.~A. Rankin, A minimum problem for the {E}pstein zeta function, Proc. Glasg.
  Math. Assoc. 1 (1953) 149--158.

\bibitem{En64}
V.~Ennola, On a problem about the {E}pstein zeta function, Proc. Camb. Philos.
  Soc. 60 (1964) 855--875.

\bibitem{Vi17}
M.~{Viazovska}, {The sphere packing problem in dimension 8}, Annal Math. 185
  (2017) 991--1015.

\bibitem{Co17}
H.~{Cohn}, A.~{Kumar}, S.~D. {Miller}, D.~{Radchenko}, M.~{Viazovska}, {The
  sphere packing problem in dimension 24}, Annal Math. 183 (2017) 1017--1033.

\bibitem{So86}
J.~E.~S. Socolar, P.~J. Steinhardt, Quasicrystals. {II}. {U}nit-cell
  configurations, Phys. Rev. B 34 (1986) 617.

\bibitem{Li17}
C.~Lin, P.~J. Steinhardt, S.~Torquato, Hyperuniformity variation with
  quasicrystal local isomorphism class, J. Phys.: Cond. Matter 29 (2017)
  204003.

\bibitem{Ha05}
T.~C. Hales, A proof of the {K}epler conjecture, Ann. Math. 162 (2005)
  1065--1185.

\bibitem{To07}
S.~Torquato, F.~H. Stillinger, Toward the jamming threshold of sphere packings:
  {T}unneled crystals, J. Appl. Phys. 102 (2007) 093511, {E}rratum, {\bf 103},
  129902 {(}2008{)}.

\bibitem{To10d}
S.~Torquato, Reformulation of the covering and quantizer problems as ground
  states of interacting particles, Phys. Rev. E 82 (2010) 056109.

\bibitem{Ke48}
D.~G. Kendall, On the number of lattice points inside a random oval, Quart. J.
  Math. (Oxford) 19 (1948) 1--24.

\bibitem{Be01}
J.~Beck, Randomness in lattice point problems, Discrete Math. 229 (2001) 29 --
  55.

\bibitem{Ken63}
M.~G. Kendall, P.~A.~P. Moran, Geometrical Probability, Griffin, London, 1963.

\bibitem{Ma89}
B.~Mat\'{e}rn, Precision of area estimation: a numerical study, J. Microsc. 153
  (1989) 269--284.

\bibitem{Chi17}
A.~T. Chieco, R.~Dreyfus, D.~J. Durian, Characterizing pixel and point patterns
  with a hyperuniformity disorder length, Phys. Rev. E 96 (2017) 032909.

\bibitem{Ye98a}
C.~L.~Y. Yeong, S.~Torquato, Reconstructing random media, Phys. Rev. E 57
  (1998) 495--506.

\bibitem{Ji09b}
Y.~Jiao, F.~H. Stillinger, S.~Torquato, A superior descriptor of random
  textures and its predictive capacity, Proc. Nat. Acad. Sci. 106 (2009)
  17634--17639.

\bibitem{To83b}
S.~Torquato, G.~Stell, Microstructure of two-phase random media: {III}. {T}he
  $n$-point matrix probability functions for fully penetrable spheres, J. Chem.
  Phys. 79 (1983) 1505--1510.

\bibitem{To85b}
S.~Torquato, G.~Stell, Microstructure of two-phase random media: {V}. {T}he
  $n$-point matrix probability functions for impenetrable spheres, J. Chem.
  Phys. 82 (1985) 980--987.

\bibitem{Lu91}
B.~L. Lu, S.~Torquato, General formalism to characterize the microstructure of
  polydispersed random media, Phys. Rev. A 43 (1991) 2078--2080.

\bibitem{To90d}
S.~Torquato, B.~Lu, Rigorous bounds on the fluid permeability: {E}ffect of
  polydispersivity in grain size, Phys. Fluids A 2 (1990) 487--490.

\bibitem{Or14}
L.~S. Ornstein, F.~Zernike, Accidental deviations of density and opalescence at
  the critical point of a single substance, Proc. Akad. Sci. (Amsterdam) 17
  (1914) 793--806.

\bibitem{Jo64}
M.~D. Johnson, P.~Hutchinson, N.~H. March, Ion-ion oscillatory potentials in
  liquid metals, Proc. R. Soc. Lond. A 282 (1964) 283--302.

\bibitem{Stell77}
G.~Stell, Fluids with long-range forces: {T}oward a simple analytic theory, in:
  B.~J. Berne (Ed.), Statistical Mechanics, Part A, Plenum Press, New York,
  1977, pp. 47--82.

\bibitem{Hu87}
K.~Huang, Statistical Mechanics, John Wiley, New York, 1987.

\bibitem{Fo96}
P.~J. Forrester, B.~Jancovici, G.~T{\'e}llez, Universality in some classical
  {C}oulomb systems of restricted dimension, J. Stat. Phys. 84 (1996) 359--378.

\bibitem{To02b}
S.~Torquato, Statistical description of microstructures, Ann. Rev. Mater. Res.
  32 (2002) 77--111.

\bibitem{Cr03}
J.~R. Crawford, S.~Torquato, F.~H. Stillinger, Aspects of correlation function
  realizability, J. Chem. Phys. 119 (2003) 7065--7074.

\bibitem{Ba78}
M.~Baus, On the compressibility of a one-component plasma, J. Physics A: Math.
  \& Gen. 11 (1978) 2451.

\bibitem{Na80}
E.~J. M.~Navet, M.~R. Feix, Virial pressure of the classical one-component
  plasma, J. Physique Lett. 41 (1980) 69--73.

\bibitem{Ch80}
P.~Choquard, P.~Favre, C.~Gruber, On the equation of state of classical
  one-component systems with long-range forces, J. Stat. Phys. 23 (1980)
  405--442.

\bibitem{Ba80}
M.~Baus, J.-P. Hansen, Statistical mechanics of simple {C}oulomb systems, Phys.
  Rep. 59 (1980) 1--94.

\bibitem{Se14}
S.~{Serfaty}, {Ginzburg-{L}andau Vortices, {C}oulomb Gases, and Renormalized
  Energies}, J. Stat. Phys. 154 (2014) 660--680.

\bibitem{Sa15}
E.~Sandier, S.~Serfaty, 2{D} {C}oulomb gases and the renormalized energy, Ann.
  Probab. 43 (2015) 2026--2083.

\bibitem{Se15}
S.~Serfaty, Coulomb gases and {G}inzburg-{L}andau vortices, Zurich Lecture
  Notes in Mathematics, Eur. Math. Soc.

\bibitem{Leb16}
T.~Lebl{\'e}, Logarithmic, {C}oulomb and {R}iesz energy of point processes, J.
  Stat. Phys. 162 (2016) 887--923.

\bibitem{Pe18}
M.~Petrache, S.~R. Nodari, Equidistribution of jellium energy for {C}oulomb and
  {R}iesz interactions, Constructive Approximation 47 (2018) 163--210.

\bibitem{Sa02}
H.~Sakai, F.~H. Stillinger, S.~Torquato, Equi-$g(r)$ sequences of systems
  derived from the square-well potential, J. Chem. Phys. 117 (2002) 297--307.

\bibitem{Fa91}
Y.~Fan, J.~K. Percus, D.~K. Stillinger, F.~H. Stillinger, Constraints on
  collective density variables: One dimension, Phys. Rev. A 44 (1991)
  2394--2402.

\bibitem{Ba09b}
R.~D. Batten, F.~H. Stillinger, S.~Torquato, Interactions leading to disordered
  ground states and unusual low-temperature behavior, Phys. Rev. E 80 (2009)
  031105.

\bibitem{Ma13}
S.~Martis, {\' E}.~Marcotte, F.~H. Stillinger, S.~Torquato, Exotic ground
  states of directional pair potentials via collective-density variables, J.
  Stat. Phys. 150 (2013) 414.

\bibitem{As76}
N.~W. Ashcroft, D.~N. Mermin, Solid State Physics, Thomson Learning, Toronto,
  1976.

\bibitem{Bac10}
A.~B\'acsi, A.~Virosztek, Local density of states and {F}riedel oscillations in
  graphene, Phys. Rev. B 82 (2010) 193405.

\bibitem{Ha14}
C.~Hanel, C.~N. Likos, R.~Blaak, Effective interactions between multilayered
  ionic microgels, Materials 7 (2014) 7689--7705.

\bibitem{Su05}
A.~S{\"u}t{\H{o}}, Crystalline ground states for classical particles, Phys.
  Rev. Lett. 95 (2005) 265501.

\bibitem{Hough09}
J.~B. Hough, M.~Krishnapur, Y.~Peres, B.~Vir{\'a}g, Zeros of {G}aussian
  analytic functions and determinantal point processes, Vol.~51, American
  Mathematical Society, Providence, RI, 2009.

\bibitem{Gh17}
S.~{Ghosh}, J.~L. {Lebowitz}, {Generalized stealthy hyperuniform processes :
  {Ma}ximal rigidity and the bounded holes conjecture}, ArXiv e-prints\href
  {http://arxiv.org/abs/1707.04328} {\path{arXiv:1707.04328}}.

\bibitem{Ka48}
W.~Kauzmann, The nature of the glassy state and the behavior of liquids at low
  temperatures., Chem. Rev. 43 (1948) 219--256.

\bibitem{To00a}
S.~Torquato, Modeling of physical properties of composite materials, Int. J.
  Solids Structures 37 (2000) 411--422.

\bibitem{To10c}
S.~Torquato, F.~H. Stillinger, Jammed hard-particle packings: {F}rom {K}epler
  to {B}ernal and beyond, Rev. Mod. Phys. 82 (2010) 2633.

\bibitem{To08a}
S.~Torquato, F.~H. Stillinger, New duality relations for classical ground
  states,, Phys. Rev. Lett. 100 (2008) 020602.

\bibitem{To11b}
S.~Torquato, C.~E. Zachary, F.~H. Stillinger, Duality relations for the
  classical ground states of soft-matter systems, Soft Matter 7 (2011) 3780.

\bibitem{De95}
P.~G. De~Gennes, J.~Prost, The physics of liquid crystals, Oxford University
  Press, Oxford, England, 1995.

\bibitem{Wi51}
E.~P. Wigner, On the statistical distribution of the widths and spacings of
  nuclear resonance levels, in: Proc. Cambridge Philos. Soc., Vol.~47, 1951,
  pp. 790--798.

\bibitem{La55}
A.~M. Lane, R.~G. Thomas, E.~P. Wigner, Giant resonance interpretation of the
  nucleon-nucleus interaction, Phys. Rev. 98 (1955) 693--701.

\bibitem{Me60}
M.~L. Mehta, On the statistical properties of the level-spacings in nuclear
  spectra, Nuclear Phys. 18 (1960) 395--419.

\bibitem{Mc60}
R.~McWeeny, Some recent advances in density matrix theory, Rev. Mod. Phys. 32
  (1960) 335--369.

\bibitem{Dy62a}
F.~J. Dyson, Statistical theory of the energy levels of complex systems. {I},
  J. Math. Phys. 3 (1962) 140--156.

\bibitem{Dy62b}
F.~J. Dyson, Statistical theory of the energy levels of complex systems. {II},
  J. Math. Phys. 3 (1962) 157--165.

\bibitem{Dy62c}
F.~J. Dyson, Statistical theory of the energy levels of complex systems. {III},
  J. Math. Phys. 3 (1962) 166--175.

\bibitem{Dy63}
M.~L. Metha, F.~J. Dyson, Statistical theory of the energy levels of complex
  systems. {IV}, J. Math. Phys. 4 (1963) 713--719.

\bibitem{Mo73}
T.~Mori, K.~Tanaka, Average stress in matrix and average elastic energy of
  materials with misfitting inclusions, Acta Metall. 21 (1973) 571--574.

\bibitem{Od87}
A.~M. Odlyzko, On the distribution of spacings between zeros of the zeta
  function, Mathematics of Computation 48 (1987) 273--308.

\bibitem{Rud96}
Z.~Rudnick, P.~Sarnak, Zeros of principal {$L$}-functions and random matrix
  theory, Duke Math. J. 81 (1996) 269--322.

\bibitem{Fe98}
R.~P. Feynman, Statistical Mechanics, Westview Press, Boulder, Colorado, 1998.

\bibitem{Ka99}
N.~Katz, P.~Sarnak, Zeroes of zeta functions and symmetry, Bull. Am. Math. Soc.
  36 (1999) 1--26.

\bibitem{Fo10}
P.~J. Forrester, Log-gases and random matrices, Princeton University Press,
  Princeto, New Jersey, 2010.

\bibitem{Ta11}
T.~Tao, V.~Vu, Random matrices: {U}niversality of local eigenvalue statistics,
  Acta mathematica 206 (2011) 127--204.

\bibitem{Be85}
M.~V. Berry, Semiclassical theory of spectral rigidity, Proc. R. Soc. Lond. A.
  400 (1985) 229--251.

\bibitem{Gu98}
T.~Guhr, A.~M{\"u}ller-Groeling, H.~A. Weidenm{\"u}ller, Random-matrix theories
  in quantum physics: common concepts, Phys. Rep. 299 (1998) 189--425.

\bibitem{Ba14}
V.~Balasubramanian, M.~Berkooz, S.~F. Ross, J.~Sim{\'o}n, Black holes,
  entanglement and random matrices, Classical Quantum Gravity 31 (2014) 185009.

\bibitem{Cot17}
J.~S. Cotler, G.~Gur-Ari, M.~Hanada, J.~Polchinski, P.~Saad, S.~H. Shenker,
  D.~Stanford, A.~Streicher, M.~Tezuka, {Black Holes and Random Matrices}, J.
  High Energy Phys. 2017 (2017) 118.

\bibitem{La99}
L.~Laloux, P.~Cizeau, J.-P. Bouchaud, M.~Potters, Noise dressing of financial
  correlation matrices, Phys. Rev. Lett. 83 (1999) 1467--1470.

\bibitem{Pl02}
V.~Plerou, P.~Gopikrishnan, B.~Rosenow, L.~A.~N. Amaral, T.~Guhr, H.~E.
  Stanley, Random matrix approach to cross correlations in financial data,
  Phys. Rev. E 65 (2002) 066126.

\bibitem{Maj09}
S.~N. Majumdar, C.~Nadal, A.~Scardicchio, P.~Vivo, Index distribution of
  {G}aussian random matrices, Phys. Rev. Lett. 103 (2009) 220603.

\bibitem{Ru96}
M.~A. Rutgers, J.~H. Dunsmuir, J.~Z. Xue, W.~B. Russel, P.~M. Chaikin,
  Measurement of the hard-sphere equation of state using screened charged
  polystyrene colloids, Phys. Rev. B 53 (1996) 5043--5046.

\bibitem{Ma75}
G.~Matheron, Random Sets and Integral Geometry, Wiley, New York, 1975.

\bibitem{So00}
A.~Soshnikov, Determinantal random point fields, Russian Mathematical Surveys
  55 (2000) 923--975.

\bibitem{Bu93}
R.~Burton, R.~Pemantle, Local characteristics, entropy and limit theorems for
  spanning trees and domino tilings via transfer-impedances, Ann. Probab. 21
  (1993) 1329--1371.

\bibitem{Jo04}
K.~Johansson, Determinantal processes with number variance saturation, Comm.
  Math. Phys. 252 (2004) 111--148.

\bibitem{Pe05}
Y.~Peres, B.~Vir{\'a}g, Zeros of the iid {G}aussian power series: {A}
  conformally invariant determinantal process, Acta Mathematica 194 (2005)
  1--35.

\bibitem{Cos04}
O.~Costin, J.~Lebowitz, On the construction of particle distributions with
  specified single and pair densities, J. Phys. Chem. B. 108 (2004)
  19614--19618.

\bibitem{De08}
J.~De~Coninck, F.~Dunlop, T.~Huillet, On the correlation structure of some
  random point processes on the line, Physica A 387 (2008) 725--744.

\bibitem{Sc09}
A.~Scardicchio, C.~E. Zachary, S.~Torquato, Statistical properties of
  determinantal point processes in high-dimensional {E}uclidean spaces, Phys.
  Rev. E 79 (2009) 041108,.

\bibitem{Mi05}
H.~Minkowski, Diskontinuit{\" a}tsbereich f{\" u}r arithmetische
  {{\"A}}quivalenz, J. {r}eine {a}ngew. {M}ath. 129 (1905) 220--274.

\bibitem{To06a}
S.~Torquato, F.~H. Stillinger, Exactly solvable disordered sphere-packing model
  in arbitrary-dimensional {E}uclidean spaces, Phys. Rev. E 73 (2006) 031106.

\bibitem{Ho06}
J.~B. Hough, M.~Krishnapur, Y.~Peres, B.~Vir{\'a}g, Determinantal processes and
  independence, Prob. Surveys 3~(206-229).

\bibitem{Gi65}
J.~Ginibre, Statistical ensembles of complex, quaternion, and real matrices, J.
  Math. Phys. 6 (1965) 440--449.

\bibitem{Laugh87}
R.~B. Laughlin, Elementary theory: {T}he incompressible quantum fluid, in: The
  Quantum Hall Effect, Springer, 1987, pp. 233--301.

\bibitem{Ca99}
C.~Cavazzoni, G.~L. Chiarotti, S.~Scandolo, E.~Tosatti, M.~Bernasconi,
  M.~Parrinello, Superionic and metallic states of water and ammonia at giant
  planet conditions, Science 283~(5398) (1999) 44--46.

\bibitem{Su15}
J.~Sun, B.~K. Clark, S.~Torquato, R.~Car, The phase diagram of high-pressure
  superionic ice, Nature Comm. 6 (2015) 8156.

\bibitem{Bo79}
J.~B. Boyce, B.~A. Huberman, Superionic conductors: {T}ransitions, structures,
  dynamics, Phys. Reports 51 (1979) 189--265.

\bibitem{Mak09}
R.~Makiura, T.~Yonemura, T.~Yamada, M.~Yamauchi, R.~Ikeda, H.~Kitagawa,
  K.~Kato, M.~Takata, Size-controlled stabilization of the superionic phase to
  room temperature in polymer-coated agi nanoparticles, Nature Mater. 8 (2009)
  476--480.

\bibitem{Hai13}
A.~{Haimi}, H.~{Hedenmalm}, {The Polyanalytic {G}inibre Ensembles}, J. Stat.
  Phys. 153 (2013) 10--47.

\bibitem{Do04d}
A.~Donev, F.~H. Stillinger, P.~M. Chaikin, S.~Torquato, Unusually dense crystal
  ellipsoid packings, Phys. Rev. Lett. 92 (2004) 255506.

\bibitem{Pa10}
G.~Parisi, F.~Zamponi, Mean field theory of hard sphere glasses and jamming,
  Rev. Mod. Phys. 82 (2010) 789---845.

\bibitem{Co98}
R.~Connelly, K.~Bezdek, A.~Bezdek, Finite and uniform stability of sphere
  packings, Discrete Comput. Geom. 20 (1998) 111--130.

\bibitem{To00b}
S.~Torquato, T.~M. Truskett, P.~G. Debenedetti, Is random close packing of
  spheres well defined?, Phys. Rev. Lett. 84 (2000) 2064--2067.

\bibitem{To01b}
S.~Torquato, F.~H. Stillinger, Multiplicity of generation, selection, and
  classification procedures for jammed hard-particle packings, J. Phys. Chem. B
  105 (2001) 11849--11853.

\bibitem{Oh02}
C.~S. O'Hern, S.~A. Langer, A.~J. Liu, S.~R. Nagel, Random packings of
  frictionless particles, Phys. Rev. Lett. 88 (2002) 075507.

\bibitem{Oh03}
C.~S. O'Hern, L.~E. Silbert, A.~J. Liu, S.~R. Nagel, Jamming at zero
  temperature and zero applied stress: The epitome of disorder, Phys. Rev. E 68
  (2003) 011306.

\bibitem{Do04b}
A.~Donev, I.~Cisse, D.~Sachs, E.~A. Variano, F.~H. Stillinger, R.~Connelly,
  S.~Torquato, P.~M. Chaikin, Improving the density of jammed disordered
  packings using ellipsoids, Science 303 (2004) 990--993.

\bibitem{Ma05}
W.~Man, A.~Donev, F.~H. Stillinger, M.~Sullivan, W.~B. Russel, D.~Heeger,
  S.~Inati, S.~Torquato, P.~M. Chaikin, {Experiments on Random Packing of
  Ellipsoids}, Phys. Rev. Lett. 94 (2005) 198001.

\bibitem{Wy05}
M.~Wyart, L.~E. Silbert, S.~R. Nagel, T.~A. Witten, Effects of compression on
  the vibrational modes of marginally jammed solids, Phys. Rev. E 72 (2005)
  051306.

\bibitem{Ma08}
C.~Song, P.~Wang, H.~A. Makse, A phase diagram for jammed matter, Nature 453
  (2008) 629--632.

\bibitem{Mail09}
M.~Mailman, C.~F. Schreck, C.~S. O'Hern, B.~Chakraborty, Jamming in systems
  composed of frictionless ellipse-shaped particles, Phys. Rev. Lett. 102
  (2009) 255501.

\bibitem{Ma09}
R.~Mari, F.~Krzakala, J.~Kurchan, Jamming versus glass transitions, Phys. Rev.
  Lett. 103 (2008) 025701.

\bibitem{Ja13}
R.~Jadrich, K.~S. Schweizer, Equilibrium theory of the hard sphere fluid and
  glasses in the metastable regime up to jamming. {I}. {T}hermodynamics, J.
  Chem. Phys. 139.

\bibitem{To03c}
S.~Torquato, A.~Donev, F.~H. Stillinger, Breakdown of elasticity theory for
  jammed hard-particle packings: Conical nonlinear constitutive theory, Int. J.
  Solids Structures 40 (2003) 7143--7153.

\bibitem{Do04a}
A.~Donev, S.~Torquato, F.~H. Stillinger, R.~Connelly, Jamming in hard sphere
  and disk packings, J. Appl. Phys. 95 (2004) 989--999.

\bibitem{Do05c}
A.~Donev, S.~Torquato, F.~H. Stillinger, Pair correlation function
  characteristics of nearly jammed disordered and ordered hard-sphere packings,
  Phys. Rev. E 71 (2005) 011105: 1--14.

\bibitem{To09b}
S.~Torquato, Y.~Jiao, Dense packings of the {P}latonic and {A}rchimedean
  solids, Nature 460 (2009) 876--881.

\bibitem{To09c}
S.~Torquato, Y.~Jiao, Dense polyhedral packings: {P}latonic and {A}rchimedean
  solids, Phys. Rev. E 80 (2009) 041104.

\bibitem{Be60}
J.~D. Bernal, Geometry and the structure of monatomic liquids, Nature 185
  (1960) 68--70.

\bibitem{Ber65}
J.~D. Bernal, The geometry of the structure of liquids, in: T.~J. Hughel (Ed.),
  Liquids: structure, properties,solid interactions, Elsevier, New York, 1965,
  pp. 25--50.

\bibitem{At14}
S.~Atkinson, F.~H. Stillinger, S.~Torquato, Existence of isostatic, maximally
  random jammed monodisperse hard-disk packings, Proc. Nat. Acad. Sci. 111
  (2014) 18436--18441.

\bibitem{Ed94}
S.~F. Edwards, The role of entropy in the specification of a powder, in:
  A.~Mehta (Ed.), Granular Matter, Springer-Verlag, New York, 1994, pp.
  121--140.

\bibitem{Kla14}
M.~A. Klatt, S.~Torquato, Characterization of maximally random jammed sphere
  packings: Voronoi correlation functions, Phys. Rev. E 90 (2014) 052120.

\bibitem{Lu90}
B.~D. Lubachevsky, F.~H. Stillinger, Geometric properties of random disk
  packings, J. Stat. Phys. 60 (1990) 561--583.

\bibitem{Ik15}
A.~Ikeda, L.~Berthier, Thermal fluctuations, mechanical response, and
  hyperuniformity in jammed solids, Phys. Rev. E 92 (2015) 012309.

\bibitem{Wu15}
Y.~Wu, P.~Olsson, S.~Teitel, Search for hyperuniformity in mechanically stable
  packings of frictionless disks above jamming, Phys. Rev. E 92 (2015) 052206.

\bibitem{Ik17}
A.~Ikeda, L.~Berthier, G.~Parisi, Large-scale structure of randomly jammed
  spheres, Phys. Rev. E 95 (2017) 052125.

\bibitem{St03}
F.~H. Stillinger, S.~Torquato, H.~Sakai, Lattice-based random jammed
  configurations for hard particles, Phys. Rev. E 67 (2003) 031107.

\bibitem{Do05a}
A.~Donev, S.~Torquato, F.~H. Stillinger, Neighbor list collision-driven
  molecular dynamics for nonspherical hard particles: {I}. {A}lgorithmic
  details, J. Comput. Phys. 202 (2005) 737--764.

\bibitem{Do05b}
A.~Donev, S.~Torquato, F.~H. Stillinger, Neighbor list collision-driven
  molecular dynamics for nonspherical hard particles: {II}. {A}pplications to
  ellipses and ellipsoids,, J. Comput. Phys. 202 (2005) 765--793.

\bibitem{Do04c}
A.~Donev, S.~Torquato, F.~H. Stillinger, R.~Connelly, A linear programming
  algorithm to test for jamming in hard-sphere packings, J. Comput. Phys. 197
  (2004) 139--166.

\bibitem{Cha10}
P.~{Chaudhuri}, L.~{Berthier}, S.~{Sastry}, {Jamming Transitions in Amorphous
  Packings of Frictionless Spheres Occur over a Continuous Range of Volume
  Fractions}, Phys. Rev. Lett. 104 (2010) 165701.

\bibitem{To10e}
S.~Torquato, Y.~Jiao, Robust algorithm to generate a diverse class of dense
  disordered and ordered sphere packings via linear programming, Phys. Rev. E
  82 (2010) 061302.

\bibitem{At13}
S.~Atkinson, F.~H. Stillinger, S.~Torquato, Detailed characterization of
  rattlers in exactly isostatic, strictly jammed sphere packings, Phys. Rev. E
  88 (2013) 062208.

\bibitem{Ji10b}
Y.~Jiao, F.~H. Stillinger, S.~Torquato, Distinctive features arising in
  maximally random jammed packings of superballs, Phys. Rev. E 81 (2010)
  041304.

\bibitem{Ti15}
J.~Tian, Y.~Xu, Y.~Jiao, S.~Torquato, A geometric-structure theory for
  maximally random jammed packings, Sci. Reports 5 (2015) 16722.

\bibitem{De10}
G.~W. Delaney, P.~W. Cleary, The packing properties of superellipsoids,
  Europhys. Lett. 89 (2010) 34002.

\bibitem{Li08}
S.-X. Li, J.~Zhao, X.~Zhou, Numerical simulation of random close packing with
  tetrahedra, Chinese Phys. Lett. 25 (2008) 1724.

\bibitem{Gl09}
A.~Haji-Akbari, M.~Engel, A.~S. Keys, X.~Zheng, R.~G. Petschek,
  P.~Palffy-Muhoray, S.~C. Glotzer, Disordered, quasicrystalline and
  crystalline phases of densely packed tetrahedra, Nature 462 (2009) 773--777.

\bibitem{Ja10}
A.~Jaoshvili, A.~Esakia, M.~Porrati, P.~M. Chaikin, Experiments on the random
  packing of tetrahedral dice, Phys. Rev. Lett. 104 (2010) 185501.

\bibitem{Sm10}
K.~C. Smith, M.~Alam, T.~S. Fisher, Athermal jamming of soft frictionless
  {P}latonic solids, Phys. Rev. E 82 (2010) 051304.

\bibitem{Bak10}
J.~Baker, A.~Kudrolli, Maximum and minimum stable random packings of {P}latonic
  solids, Phys. Rev. E 82 (2010) 061304.

\bibitem{Sp83}
H.~Spohn, Long range correlations for stochastic lattice gases in a
  non-equilibrium steady state, J. Phys. A: Math. \& Gen. 16 (1983) 4275.

\bibitem{Ga90}
P.~L. Garrido, J.~L. Lebowitz, C.~Maes, H.~Spohn, Long-range correlations for
  conservative dynamics, Phys. Rev. A 42 (1990) 1954.

\bibitem{Ch91}
Z.~Cheng, P.~L. Garrido, J.~L. Lebowitz, J.~L. Vall{\'e}s, Long-range
  correlations in stationary nonequilibrium systems with conservative
  anisotropic dynamics, Europhys. Lett. 14 (1991) 507.

\bibitem{De07}
B.~Derrida, Non-equilibrium steady states: fluctuations and large deviations of
  the density and of the current, J. Stat. Mech.: Theory \& Exp. 2007 (2007)
  P07023.

\bibitem{Bo08}
T.~Bodineau, B.~Derrida, V.~Lecomte, F.~van Wijland, Long range correlations
  and phase transitions in non-equilibrium diffusive systems, J. Stat. Phys.
  133 (2008) 1013--1031.

\bibitem{Hi00}
H.~Hinrichsen, Non-equilibrium critical phenomena and phase transitions into
  absorbing states, Adv. Phys. 49 (2000) 815--958.

\bibitem{Lu04}
S.~L{\"u}beck, Universal scaling behavior of non-equilibrium phase transitions,
  Int. J. Mod. Phys. B 18 (2004) 3977--4118.

\bibitem{He08}
M.~Henkel, H.~Hinrichsen, S.~L{\"u}beck, Non-Equilibrium Phase Transitions -
  Volume 1: Absorbing Phase Transitions, Springer, New York, 2008.

\bibitem{Pi05}
D.~J. Pine, J.~P. Gollub, J.~F. Brady, A.~M. Leshansky, Chaos and threshold for
  irreversibility in sheared suspensions, Nature 438 (2005) 997--1000.

\bibitem{Chaik08}
C.~Laurent, P.~M. Chaikin, J.~P. Gollub, D.~J. Pine, Random organization in
  periodically driven systems, Nature Phys. 4 (2008) 420--424.

\bibitem{Ro00}
J.~N. Roux, Geometric origin of mechanical properties of granular materials,
  Phys. Rev. E 61 (2000) 6802--6836.

\bibitem{Le04}
S.-B. Lee, Universality class of the conserved {M}anna model in one dimension,
  Phys. Rev. E 89 (2014) 060101.

\bibitem{Fe80}
J.~Feder, Random sequential adsorption, J. Theor. Biol. 87 (1980) 237--254.

\bibitem{Ta00}
J.~Talbot, G.~Tarjus, P.~R. Van~Tassel, P.~Viot, From car parking to protein
  adsorption: {A}n overview of sequential adsorption processes, Colloids and
  Surfaces A 165 (2000) 287--324.

\bibitem{Re09}
C.~Reichhardt, C.~J.~O. Reichhardt, Random organization and plastic depinning,
  Phys. Rev. Lett. 103 (2009) 168301.

\bibitem{Na14}
K.~H. Nagamanasa, S.~Gokhale, A.~K. Sood, R.~Ganapathy, Experimental signatures
  of a nonequilibrium phase transition governing the yielding of a soft glass,
  Phys. Rev. E 89 (2014) 062308.

\bibitem{Ro15}
J.~R. Royer, P.~M. Chaikin, Precisely cyclic sand: {S}elf-organization of
  periodically sheared frictional grains, Proc. Nat. Acad. Sci. 112 (2015)
  49--53.

\bibitem{Pu11}
D.~Purves, R.~B. Lotto, Why we see what we do redux: A wholly empirical theory
  of vision, Sinauer Associates, Sunderland, Massachusetts, 2011.

\bibitem{Sh49}
C.~E. Shannon, Communication in the presence of noise, Proceedings of the IRE
  37 (1949) 10--21.

\bibitem{Pe62}
D.~P. Petersen, D.~Middleton, Sampling and reconstruction of
  wave-number-limited functions in {N}-dimensional {E}uclidean spaces, Info.
  Control 5 (1962) 279--323.

\bibitem{Fr77}
A.~S. French, A.~W. Snyder, D.~G. Stavenga, Image degradation by an irregular
  retinal mosaic, Bio. Cybernetics 27 (1977) 229--233.

\bibitem{Re76}
D.~F. Ready, T.~E. Hanson, S.~Benzer, Development of the drosophila retina, a
  neurocrystalline lattice, Develop. Bio. 53 (1976) 217--240.

\bibitem{Lu11}
D.~K. Lubensky, M.~W. Pennington, B.~I. Shraiman, N.~E. Baker, A dynamical
  model of ommatidial crystal formation, Proc. Nat. Acad. Sci. 108 (2011)
  11145--11150.

\bibitem{Ly57}
A.~H. Lyall, Cone arrangements in teleost retinae, J. Cell Sci. 3 (1957)
  189--201.

\bibitem{En63}
K.~Engstro{\" o}m, Cone types and cone arrangements in teleost retinae1, Acta
  Zoologica 44 (1963) 179--243.

\bibitem{Ra04}
P.~A. Raymond, L.~K. Barthel, A moving wave patterns the cone photoreceptor
  mosaic array in the zebrafish retina, Int. J. Develop. Bio. 48 (2004)
  935--945.

\bibitem{Du66}
R.~F. Dunn, Studies on the retina of the gecko coleonyx variegatus: Ii. the
  rectilinear visual cell mosaic, J. Ultrastructure Res. 16 (1966) 672--684.

\bibitem{Ha01}
N.~S. Hart, The visual ecology of avian photoreceptors, Prog. Retinal Eye Res.
  20 (2001) 675--703.

\bibitem{Mor70}
V.~B. Morris, Symmetry in a receptor mosaic demonstrated in the chick from the
  frequencies, spacing and arrangement of the types of retinal receptor, J.
  Comparative Neurology 140 (1970) 359--397.

\bibitem{Kr10}
Y.~A. Kram, S.~Mantey, J.~C. Corbo, Avian cone photoreceptors tile the retina
  as five independent, self-organizing mosaics, PloS One 5 (2010) e8992.

\bibitem{Ch16}
D.~Chen, W.-Y. Aw, D.~Devenport, S.~Torquato, Structural characterization and
  statistical-mechanical model of epidermal patterns, Biophys. J. 111 (2016)
  2534--2545.

\bibitem{To90e}
S.~Torquato, Relationship between permeability and diffusion-controlled
  trapping constant of porous media, Phys. Rev. Lett. 64 (1990) 2644--2646.

\bibitem{Wi91}
D.~J. Wilkinson, D.~L. Johnson, L.~M. Schwartz, Nuclear magnetic relaxation in
  porous media: {T}he role of the mean lifetime $\tau(\rho,d)$, Phys. Rev. B 44
  (1991) 4960--4971.

\bibitem{Mit92}
P.~P. Mitra, P.~N. Sen, Effects of microgeometry and surface relaxation on nmr
  pulsed-field-gradient experiments: Simple pore geometries, Phys. Rev. B 45
  (1992) 143.

\bibitem{Ca58}
J.~W. Cahn, J.~E. Hilliard, Free energy of a nonuniform system. {I}.
  {I}nterfacial free energy, J. Chem. Phys. 28 (1958) 258--267.

\bibitem{Sw77}
J.~Swift, P.~C. Hohenberg, Hydrodynamic fluctuations at the convective
  instability, Phys. Rev. A 15 (1977) 319--328.

\bibitem{Ba59}
G.~K. Batchelor, The Theory of Homogeneous Turbulence, Cambridge University
  Press, Cambridge, England, 1959.

\bibitem{Mo75}
A.~S. Monin, A.~M. Yaglom, Statistical Fluid Mechanics: Mechanics of
  Turbulence, Vol.~2, MIT Press, Cambridge, Massachusetts, 1975.

\bibitem{Pi88}
D.~J. Pine, D.~A. Weitz, P.~M. Chaikin, E.~Herbolzheimer, Diffusing wave
  spectroscopy, Phys. Rev. Lett. 60 (1988) 1134--1137.

\bibitem{Wi08}
D.~S. Wiersma, The physics and applications of random lasers, Nature Phys. 4
  (2008) 359--367.

\bibitem{Dog15}
A.~Dogariu, R.~Carminati, Electromagnetic field correlations in
  three-dimensional speckles, Phys. Rep. 559 (2015) 1--29.

\bibitem{Dib16}
D.~{Di Battista}, D.~{Ancora}, M.~{Leonetti}, G.~{Zacharakis}, {From amorphous
  speckle pattern to reconfigurable Bessel beam via wavefront shaping}, ArXiv
  e-prints\href {http://arxiv.org/abs/1511.04964} {\path{arXiv:1511.04964}}.

\bibitem{Kom03}
E.~Komatsu, A.~Kogut, M.~R. Nolta, C.~L. Bennett, M.~H. i, G.~Hinshaw,
  N.~Jarosik, M.~Limon, S.~S. Meyer, L.~Page, D.~N. Spergel, G.~S. Tucker,
  L.~Verde, E.~Wollack, E.~L. Wright, First-year {W}ilkinson microwave
  anisotropy probe ({WMAP}) observations: {T}ests of {G}aussianity, Astrophys.
  J. Suppl. Series 148 (2003) 119.

\bibitem{Berk87}
N.~F. Berk, Scattering properties of a model bicontinuous structure with a well
  defined length scale, Phys. Rev. Lett. 58 (1987) 2718--2721.

\bibitem{Be91}
N.~F. Berk, Scattering properties of the leveled-wave model of random
  morphologies, Phys. Rev. A 44 (1991) 5069--5079.

\bibitem{Te91}
M.~Teubner, Level surfaces of {G}aussian random fields and microemulsions,
  Europhys. Lett. 14 (1991) 403--408.

\bibitem{Cr91}
P.~A. Crossley, L.~M. Schwartz, J.~R. Banavar, Image-based models of porous
  media-- {A}pplication to vycor glass and carbonate rocks, Appl. Phys. Lett.
  59 (1991) 3553--3555.

\bibitem{Ro95}
A.~P. Roberts, M.~Teubner, Transport properties of heterogeneous materials
  derived from {G}aussian random fields: {B}ounds and simulation, Phys. Rev. E
  51 (1995) 4141--4154.

\bibitem{Ro97}
A.~P. Roberts, Morphology and thermal conductivity of model organic aerogels,
  Phys. Rev. E 55 (1997) R1286--R1289.

\bibitem{Cr09}
M.~Cross, H.~Greenside, Pattern formation and dynamics in nonequilibrium
  systems, Cambridge University Press, Cambridge, England, 2009.

\bibitem{Hu05}
J.~P. Huang, Z.~W. Wang, C.~Holm, Computer simulations of the structure of
  colloidal ferrofluids, Phys. Rev. E 71 (2005) 061203.

\bibitem{Ca85}
R.~E. Caflisch, J.~H. Luke, Variance in the sedimentation speed of a
  suspension, Phys. Fluids 28 (1985) 759--760.

\bibitem{Ham88}
J.~M. Ham, G.~M. Homsy, Hindered settling and hydrodynamic dispersion in
  quiescent sedimenting suspensions, Int. J. Multiphase flow 14 (1988)
  533--546.

\bibitem{Ni95}
H.~Nicolai, E.~Guazzelli, Effect of the vessel size on the hydrodynamic
  diffusion of sedimenting spheres, Phys. Fluids 7 (1995) 3--5.

\bibitem{Se97}
P.~N. Segr\`e, E.~Herbolzheimer, P.~M. Chaikin, Long-range correlations in
  sedimentation, Phys. Rev. Lett. 79 (1997) 2574--2577.

\bibitem{Go17}
T.~Goldfriend, H.~Diamant, T.~A. Witten, Screening, hyperuniformity, and
  instability in the sedimentation of irregular objects, Phys. Rev. Lett. 118
  (2017) 158005.

\bibitem{Me87}
M.~M{\'e}zard, G.~Parisi, M.~Virasoro, Spin glass theory and beyond: An
  Introduction to the Replica Method and Its Applications, Vol.~9, World
  Scientific Publishing, 1987.

\bibitem{Re08a}
M.~C. Rechtsman, S.~Torquato, Effective dielectric tensor for electromagnetic
  wave propagation in random media, J. Appl. Phys. 103 (2008) 084901.

\bibitem{Thien17}
Q.~L. Thien, D.~McDermott, C.~J. Reichhardt, C.~Reichhardt, Enhanced pinning
  for vortices in hyperuniform substrates and emergent hyperuniform vortex
  states, Phys. Rev. B 96 (2017) 094516.

\bibitem{Bur15}
L.~M. Burcaw, E.~Fieremans, D.~S. Novikov, Mesoscopic structure of neuronal
  tracts from time-dependent diffusion, NeuroImage 114 (2015) 18--37.

\bibitem{Pa17}
A.~Papaioannou, D.~S. Novikov, E.~Fieremans, G.~S. Boutis, Observation of
  structural universality in disordered systems using bulk diffusion
  measurement, Phys. Rev. E 96 (2017) 061101.

\bibitem{Mi02}
G.~W. Milton, The Theory of Composites, Cambridge University Press, Cambridge,
  England, 2002.

\bibitem{Di86}
G.~E. Dieter, Mechanical metallurgy, 3rd Edition, McGraw-Hill, New York, 1986.

\bibitem{Ki05}
C.~Kittel, Introduction to solid state physics, 8th Edition, Wiley, New York,
  2005.

\bibitem{Ko64}
J.~Kondo, Resistance minimum in dilute magnetic alloys, Prog. Theor. Phys. 32
  (1964) 37--49.

\bibitem{An70}
P.~W. Anderson, A poor man's derivation of scaling laws for the kondo problem,
  J. Phys. C: Solid State Phys. 3 (1970) 2436.

\bibitem{Hu47}
K.~Huang, X-ray reflexions from dilute solid solutions, Proc. R. Soc. Lond. A
  190 (1947) 102--117.

\bibitem{De73b}
P.~H. Dederichs, The theory of diffuse x-ray scattering and its application to
  the study of point defects and their clusters, J. Phys. F: Met. Phys. 3
  (1973) 471.

\bibitem{We80}
T.~R. Welberry, G.~H. Miller, C.~E. Carroll, Paracrystals and growth-disorder
  models, Acta Crystallogr., Sect. A: Found. Crystallogr. 36 (1980) 921--929.

\bibitem{Im79}
Y.~Imry, Long-range order in two dimensions, Crit. Rev. Solid State Mater. Sci.
  8 (1979) 157--174.

\bibitem{Za32}
W.~H. Zachariasen, The atomic arrangement in glass, J. Am. Chem. Soc. 54 (1932)
  3841--3851.

\bibitem{Wea71}
D.~Weaire, M.~F. Thorpe, Electronic properties of an amorphous solid. i. a
  simple tight-binding theory, Phys. Rev. B 4 (1971) 2508.

\bibitem{Za83}
R.~Zallen, The Physics of Amorphous Solids, Wiley, New York, 1983.

\bibitem{Mo02}
N.~Mousseau, G.~T. Barkema, S.~M. Nakhmanson, Recent developments in the study
  of continuous random networks, Phil. Mag. B 82 (2002) 171--183.

\bibitem{Deg10}
A.~M.~R. De~Graff, M.~F. Thorpe, The long-wavelength limit of the structure
  factor of amorphous silicon and vitreous silica, Acta Crystallography Sec. A
  66 (2010) 22--31.

\bibitem{St85b}
F.~H. Stillinger, T.~A. Weber, Computer simulation of local order in condensed
  phases of silicon, Phys. Rev. B 31 (1985) 5262--5271.

\bibitem{Za08}
C.~E. Zachary, F.~H. Stillinger, S.~Torquato, Gaussian-core model phase diagram
  and pair correlations in high {E}uclidean dimensions, J. Chem. Phys. 128
  (2008) 224505.

\bibitem{St76}
F.~H. Stillinger, Phase transitions in the {G}aussian core system, J. Chem.
  Phys. 65 (1976) 3968.

\bibitem{La00}
A.~Lang, C.~N. Likos, M.~Watzlawek, H.~Low{\" e}n, Fluid and solid phases of
  the {G}aussian core model, J. Phys. Cond. Matter 12 (2000) 5087--5108.

\bibitem{Ik11}
A.~Ikeda, K.~Miyazaki, Thermodynamic and structural properties of the high
  density {G}aussian core model, J. Chem. Phys. 135.

\bibitem{Co17b}
H.~{Cohn}, M.~{de Courcy-Ireland}, {The {G}aussian core model in high
  dimensions}, ArXiv e-prints\href {http://arxiv.org/abs/1603.09684}
  {\path{arXiv:1603.09684}}.

\bibitem{Pr81}
M.~B. Priestley, Spectral Analysis and Time Series, Academic Press, New York,
  1981.

\bibitem{Ble93}
P.~M. Bleher, F.~J. Dyson, J.~L. Lebowitz, Non-{G}aussian energy level
  statistics for some integrable systems, Phys. Rev. Lett. 71 (1993)
  3047--3050.

\bibitem{Hu03}
M.~N. Huxley, Exponential sums and lattice points {III}, Proc. Lond. Math. Soc.
  87 (2003) 591--609.

\bibitem{Ar08}
L.~G. Arkhipova, Number of lattice points in a sphere, Moscow Univ. Math. Bull.
  63 (2008) 214--215.

\bibitem{Ts00}
K.-M. Tsang, Counting lattice points in the sphere, Bull. Lond. Math. Soc. 32
  (2000) 679--688.

\end{thebibliography}
\end{document}